\documentclass[12pt,preprint]{aastex}

\usepackage{lineno}
\usepackage{times}
\usepackage{natbib}
\usepackage{pdflscape}
\def\F{{\it Fermi}-LAT }
\def\SZ{{\it Suzaku }}

\shorttitle{}
\shortauthors{Takeuchi et al.}

\begin{document}

\title{Multiband Diagnostics of Unidentified 1FGL Sources with Suzaku and Swift X-ray Observations}

\author{Y.~Takeuchi\altaffilmark{1}, J.~Kataoka\altaffilmark{1}, K.~Maeda\altaffilmark{1}, Y.~Takahashi\altaffilmark{1}, T.~Nakamori\altaffilmark{2}, M.~Tahara\altaffilmark{1}}

\email{uto\_of\_take@suou.waseda.jp}

\altaffiltext{1}{Research Institute for Science and Engineering, Waseda University, 3-4-1 Okubo, Shinjuku, Tokyo, 169-8555, Japan.}
\altaffiltext{2}{Department of Physics, Faculty of Science, Yamagata University, 990-8560}

\begin{abstract}
We have analyzed all the archival X-ray data of 134 unidentified (unID) gamma-ray sources 
listed in the first {\it Fermi/LAT} (1FGL) catalog and subsequently followed up 
by {\it Swift/XRT}. 
We constructed the spectral energy distributions (SEDs) from radio to gamma-rays
for each X-ray source detected, and tried to pick up unique objects that display
anomalous spectral signatures. 
In these analysis, we target all the 1FGL unID sources, using updated data from 
the second {\it Fermi/LAT} (2FGL) catalog on their LAT position and spectra. 
We found several potentially interesting objects, particularly
three sources, 1FGL\,J0022.2-1850, 1FGL\,J0038.0+1236 and 1FGL\,J0157.0-5259, which were then more deeply 
observed with \SZ as a part of an AO7 program in 2012. We successfully detected an X-ray counterpart 
for each source whose X-ray spectra were well fitted by a single power-law function. The positional 
coincidence with a bright radio counterpart (currently identified as AGN) in the 2FGL error circles
suggests these are definitely the X-ray emission from the same AGN, but their SEDs show a wide 
variety of behavior. In particular, the SED of 1FGL\,J0038.0+1236 is difficult to be explained 
by conventional emission models of blazars. The source 1FGL\,J0022.2-1850 may be in a transition state between a low-frequency peaked
BL Lac and a high-frequency peaked BL Lac object, and 1FGL\,J0157.0-5259 could be a 
rare kind of extreme blazar. We discuss the possible nature of these three sources observed with \SZ, 
together with the X-ray identification results and SEDs of all 134 sources observed with {\it Swift}/XRT. 

\end{abstract}

\keywords{galaxies:active - gamma rays:general - radiation mechanism:general - X-rays:general}

\section{Introduction}

Since the successful launch of the {\it Fermi} Gamma-ray Space Telescope in 2008 June, we now have a 
new opportunity to study gamma-ray emission from different types of high energy sources with 
much improved sensitivity and localization capabilities than with the EGRET instrument onboard 
the {\it Compton Gamma-Ray Observatory} (CGRO). With the field of view 
covering 20\% of the sky at every moment (five times larger than EGRET), and its improved sensitivity 
(by more than an order of magnitude with respect to EGRET), the Large Area Telescope 
\citep[LAT;][]{LAT} aboard {\it Fermi} surveys the entire sky each day down to photon 
flux levels of $F_{>100MeV}$ $\simeq$ few $\times 10^{-7} {\rm ph\,cm^{-2}\,s^{-1}}$. 
The number of detected gamma-ray sources has increased, with the 2nd \F Catalog \citep[2FGL;][]{2FGL}
containing 1873 gamma-ray sources in the 100 MeV to 100 GeV range, while 271 objects were previously listed in the 
3rd EGRET Catalog \citep[3EG;][]{3EG}. More than 1,000 gamma-ray 
sources included in the 2FGL are proposed to be associated with active galactic nuclei (AGNs) and 
87 sources with pulsars \citep[PSRs;][]{PSRs}, including 21 millisecond pulsars (MSPs).
Other associations included supernova remnants \citep[SNRs;][]{SNRs}, low-mass/high-mass 
X-ray binaries \citep{XB09}, pulsar wind nebulae \citep{PWN10}, normal and starburst galaxies 
\citep{SBG10}, and the giant lobes of a radio galaxy \citep{RG10}. 

However, no obvious counterparts at longer wavelength have been found for as much 
as 31\% of the 2FGL \F objects so that several hundreds of GeV sources currently remain unassociated 
with any known astrophysical systems. In other words the nature of unassociated gamma-ray 
sources are still one of the major puzzles, and the mystery has never been solved yet.
Fortunately, an improved localization capabilities of the \F (typical 95\% confidence 
radii $r_{95}\sim0\degr.1-0\degr.2$, and even 0$\degr$.005 - 0$\degr$.01 for the brightest 
sources; \citep{2FGL}), when compared to that of EGRET (typical $r_{95}\sim0\degr.4-0\degr.7$), 
enables more effective follow-up studies at radio, optical, and X-ray frequencies, 
which can help to unravel the nature of the unID gamma-ray emitters. 
Indeed for example, a lot of {\it Fermi} sources were identified using WISE IR data \citep{dab13,mas13}.

In this context, we started a new project to investigate the nature of unID \F objects 
through X-ray follow-up observations with the XIS sensor onboard the \SZ X-ray satellite (see section 2).
For example, the results of the first-year campaign conducted in \SZ AO-4 (2009) were presented
in \citep{mae11}. In this campaign, the X-ray counterpart for one of the brightest unassociated 
\F objects, 1FGL\,J1231.1-1410 (also detected by EGRET as 3EG\,J1234-1318 and EGR\,J1231-1412), 
was discovered for the first time. The X-ray spectrum was well fitted by a blackbody with an 
additional power-law component, supporting the recent identification of this source with a  
MSP. In the second-year campaign (AO-5), another seven unID \F sources were subsequently 
observed with \SZ \citep{tak12}. In particular, this paper presented a convenient method to classify the objects 
into ``AGN-like'' and ``PSR-like'' by comparing their multiwavelength properties with those of 
known AGNs and pulsars. In the third-year (AO-6), 1FGL\,J2339.7-0531 (Yatsu et al. 2013 in prep; also 
Romani \& Shaw 2011) and 1FGL\,J1311.7-3429 \citep{rom12,kat12} were intensively monitored with 
a total exposure time of 200 ksec. Both sources are now suggested to be ``black widow'' MSP systems and 
newly categorized as ``radio-quiet'' MSPs.
As these projects show, the X-ray follow-up observations especially using \SZ 
provided various fruitful results to clarify the nature of unassociated gamma-ray sources, and 
was able to find a new type of gamma-ray emitter.

To complete a series of X-ray follow-up programs described above, we further carried out the 
analysis of all the archival 
X-ray data of 134 unID gamma-ray sources in the 1FGL catalog of point sources 
\citep[1FGL;][]{1FGL} with {\it Swift}/XRT. Note that, all these 134 sources have been 
detected in the 2FGL catalog, hence the updated data on there LAT position and spectra 
were available from the 2nd $Fermi$/LAT catalog so are used throughout this work. This allowed us to construct the SEDs 
(SEDs) of each objects from radio to gamma-rays (see Section\,2 and appendix) for the first time.
Note that we target all the 1FGL unID sources that satisfy our selections (see Section\,2), 
using updated/improved information from the 2FGL catalog on their LAT positions and spectra in this paper.
Moreover, three sources that displayed potentially interesting SEDs, 1FGL\,J0022.2-1850 
(or 2FGL J0022.2-1853), 1FGL\,J0038.0+1236 (or 2FGL J0037.8+1238) and 1FGL\,J0157.0-5259 
(or 2FGL J0157.2-5259), were deeply observed with \SZ as a part of AO7 campaign in 2012.
In the 2FGL catalog, both 1FGL\,J0022.2-1850 and 1FGL\,J0157.0-5259 are categorized as active galaxies of 
uncertain type (agu), while 1FGL\,J0038.0+1236 was classified as a BL Lac type of blazar (bzb) 
based on the positional coincidences to sources observed at another wavelength. But as we see below, 
the unique SEDs of these three objects are not well understood as conventional blazars. 
In Section\,2 we first describe the analysis of the 134 1FGL unID sources with {\it Swift}. Subsequently in Section\,3, deep
\SZ follow-up observations of the selected three sources, 1FGL\,J0022.2-1850, 1FGL\,J0038.0+1236 and 1FGL\,J0157.0-5259 
are shown. The results of the analysis are given in Section\,4. The discussion and summary 
are presented in Section\,5 and 6, respectively.

\section{{\it Swift} analysis of 134 1FGL unID sources}
\subsection{Observation and Data Reduction}

{\it Swift} \citep{Geh04} is a gamma-ray observatory launched on 2004 November 20. The primary goal of this mission 
is to explore and follow-up the gamma-ray burst, but high mobility and sensitivity to localize sources 
especially using the XRT \citep{bur05} and UVOT \citep{rom05}, makes it also viable to follow-up unID gamma-ray objects discovered by \F.
In fact, {\it Swift} follow-up observations helped study many unID \F sources \citep{cog11,kei11,ran11,the11,che12,kon12}. 
Here we tried to perform a systematic and uniform analysis of unID gamma-ray sources observed thus far with 
{\it Swift} using the archival data. The selection criteria is given as follows; (1) categorized as unID 
sources in 1FGL catalog, (2) localized at high Galactic latitude $|b|$ $>$ 10$^{\circ}$, (3) observational data 
are in public at the time of October 2011, (4) the positional center of {\it Swift} field-of-view is within 
12 arcmin from the 1FGL sources. Among 630 unID sources listed in the 1FGL catalog, this selection yields 134 sources which we analyzed here.
In the reduction and the analysis of the {\it Swift}/XRT and UVOT data, HEADAS software version 6.11 and the most recent 
calibration databases (CALDB) as of 2011 October 20 were used. We did not use BAT data because most sources are 
not bright enough to be detected within short exposure of typically 10 ksec or less. In the XRT analysis, we only use 
the PC mode data, while only image data taken from photometry observation is used in the UVOT analysis. 

\subsection{XRT Analysis}
Two types of XRT archival data can be obtained from the {\it Swift} Data Center, Level 1 and 2.
Level 2 cleaned data have gone through the standard pipeline process;
however, we calibrated the Level 1 data ourselves in a way 
recommended by {\it Swift} team\footnote{The SWIFT XRT Data Reduction Guide:http://heasarc.nasa.gov/docs/swift/analysis/xrt\_swguide\_v1\_2.pdf}. Particularly, we selected the good time interval (GTI) from the Level 1 data using \texttt{xrtpipeline}.
In this process, we only changed the default \texttt{xrtpipeline} selection in the temperature of CCDs from
"$\leqq -47$" to "$\leqq -50$", where the former is the default value.
In the XRT image analysis, we tried to detect the X-ray counterparts of 1FGL unID sources,
and localize each source. First, we extracted X-ray images in the energy range 
of $0.3 - 10$ keV using \texttt{xselect}. Next by using \texttt{ximage}, we searched for ``possible'' 
X-ray sources which were $>$ 3 $\sigma$ confidence level in photon statistics against background.
The position of these sources are determined with a typical accuracy of $\sim 5''$ using 
\texttt{xrtcentroid}. The results of X-ray source detection is listed in Table 6 
(see, appendix). In the appendix, we also show the XRT images corresponding to each 
of the 134 1FGL unID gamma-ray sources indicating also the 2FGL error ellipses (Figure\,11).
In these images, bright radio sources and X-ray sources listed in The ROSAT All-Sky Survey Bright Source Catalogue \citep[1RXS;][]{vog99}
corresponding to the XRT sources are also plotted as magenta crosses. 
Finally, if the X-ray sources were detected were within the 2FGL 95$\%$ position error ellipse,
we performed X-ray spectral analysis for those sources. We note that very bright X-ray sources with more than 0.6 c/s
should cause serious pile-up effects in the XRT CCD, however, there were no such bright sources 
in our analyzed sample. 

In the XRT spectral analysis, PHA files were extracted from event files with \texttt{xselect}
and exposure maps were made using \texttt{xrtexpomap}. We made auxiliary response 
files (ARFs) using \texttt{xrtmkarf}, while we used the current redistribution matrix files (RMFs) in CALDB.
We extracted photons from circles with 30 arcsec radii around the source positions as the source regions,
and we set the concentric rings centered at the source positions with radius 30-180 arcsec as the background regions.
In the case when some background sources appear in the field, or overall regions cannot be fitted in the CCD, 
we simply remove the region surrounding these background sources, or the region outside the CCD chip. 
If there is no source detected above $3\sigma$ inside the 2FGL error ellipse, we derived the 
upper limit assuming the 2FGL error ellipses as the source region of corresponding X-ray flux. 
The results of spectral analysis are included in the SEDs given in the appendix.
We note that X-ray spectral data were binned at two different ways according to the source 
brightness;
(a)binned with \texttt{grppha} so that at least 20 photons are included in each bin, 
(b)divide X-ray spectral data (0.5-10.0 keV) logarithmically into 5 bin. 
When the source have more than 40 counts, we used (a), but otherwise we used (b).

\subsection{UVOT Analysis}

We performed the analysis of UVOT data only when the X-ray counterpart of 1FGL source was found 
in the {\it Swift}/XRT field of view. {\it Swift}/UVOT archival data has six types of filters ($v$, $b$, $u$, $uvw1$, $uvw2$ and 
$uvm2$), with each filter providing different wavelength data. When each filter included more than one observation, 
the images and exposure maps were summed using \texttt{uvotimsum}. Using \texttt{uvotdetect}, we detected 
the sources which have high signal-to-noise ratio ($> 3 \sigma$). We set the circles around those sources 
with radius ~5 arcsec as the sources regions if any source was found in the 90\% error ellipse of the {\it Swift}/XRT source.
Then, circles with radius 30'' as the background regions were taken from the area where no sources are found.
Finally, we obtained the magnitude of each filter using \texttt{uvotsource}.
The correction of Galactic extinction were performed following the way described in Cardelli et al. (1998).

\subsection{Multiwavelength Analysis}
To construct SEDs of each 1FGL sources, we used not only the {\it Swift}/XRT and {\it Swift}/UVOT flux 
data analyzed in this paper, but gamma-ray fluxes listed in the 2FGL catalog, and radio fluxes mostly from 
the NED database and W3Browse based on a variety of catalogs. We searched for radio counterparts associated with 
the XRT or UVOT objects which were mentioned above in the HEASARC/Master Radio Catalog 
\footnote{http://heasarc.nasa.gov/W3Browse/all/radio.html}, (includes 
the NRAO VLA Sky Survey \citep[NVSS;][1.4 GHz]{NVSS},
the FIRST Survey Catalog of 1.4-GHz Radio Sources \citep[FIRST;][1.4 GHz]{wh97},
Sydney University Molonglo Sky Survey Source Catalog \citep[SUMSS;][843 MHz]{SUMSS},
VLA Low-Frequency Sky Survey Discrete Source Catalog \citep[VLSS;][74 MHz]{co07},
the Westerbork in the Southern Hemisphere Survey \citep[WISH;][]{de02},
Australia Telescope 20-GHz Survey Catalog \citep[AT20G;][]{mu10},
the Green Bank 6-cm Catalog of Radio Sources \citep[GB6;][]{gr96},
and
Parkes-MIT-NRAO Southern, Tropical, Equatorial and Zenith Survey \citep[PMN;][]{wr94})
and those radio fluxes are added in the SEDs.
Note that if no corresponding radio sources were found in the 2FGL 95\% error ellipses,
we obtained the upper limit from the most bright radio source in the error ellipse.
Likewise, if no source was found in the error ellipse but some sources were found outside of 
the region, we used the sensitivity limits of those sources as an upper limit.
Finally resultant SEDs and flux values of each wavelength are given in the appendix (Figure 12, Table 5 and 6).

\section{\SZ analysis of three 1FGL unID sources}
\subsection{Observation and Data Reduction}

We observed three \F objects  that exhibited potentially interesting or anomalous
SEDs (which is difficult to be explained by standard emission models of blazars, i.e., 
SSC or external Compton models as explained in \citet{fos97}) with the \SZ X-ray astronomy
satellite \citep{mit07}. These were denoted in
the 1FGL catalog as 1FGL\,J0022.2-1850, 1FGL\,J0038.0+1236,
and 1FGL\,J0157.0-5259, and in the 2FGL catalog as 
2FGL\,J0022.2-1853, 2FGL\,J0037.8+1238, and 2FGL\,J0157.2-5259, respectively.
The \SZ observation logs are summarized in Table\,1.
The observations were performed with XIS which consists of
four CCD cameras each placed in the focal plane of the
X-ray Telescope \citep[XRT;][]{ser07}, and with the Hard
X-ray Detector (HXD) which consists of Si PIN photo-diodes
(HXD-PIN) and GSO scintillation counters (HXD-GSO) \citep[HXD;][]{kok07,tak07}.

One of the XIS sensors (XIS 1) has a back-illuminated
(BI) CCD, while the other three (XISs 0, 2, and 3)
utilize front-illuminated (FI) CCDs. However, because
of an anomaly in 2006 November, the operation of XIS2
was terminated. Hence, here we use only the three remaining 
CCDs. The XIS was operated in the normal full-frame
clocking mode with the $3 \times 3$ or $5 \times 5$
editing mode.
We analyzed the screened XIS data, reduced using the \SZ software version 1.2.
The screening was based on the following criteria:
(1) only {\it ASCA}-grade 0, 2, 3, 4, 6 events were accumulated,
while hot and flickering pixels were removed from the XIS image using the \texttt{sisclean} script \citep{day98},
(2) the time interval after the passage of South Atlantic Anomaly was greater than 60\,s,
(3) the object was at least $5^{\circ}$ and $20^{\circ}$ above the rim of the Earth (ELV) during night and day, respectively.
In addition, we also selected the data with a cutoff rigidity (COR) larger than 6\,GV.
In the reduction and the analysis of the \SZ XIS data, HEADAS software version 6.12 and a calibration databases (CALDB) released on 2009 September 25 were used.
The XIS cleaned event data-set was obtained in the combined $3 \times 3$ and $5 \times 5$ edit modes using \texttt{xselect}.

The HXD data were also processed in a standard way as follows.
First, we obtained the appropriate version 2.0 "tuned" non-X-ray background file (NXB) for this observation. 
Because the HXD background file has a time variation, we made a new GTI file
to match the Good Time Interval (GTI) between observation data and NXB data using \texttt{mgtime}.
Next, using this new GTI file, We generated time-averaged HXD spectra with \texttt{xselect}.
These were then dead time corrected using \texttt{hxddtcor} script.
Epoch appropriate response files for XIS-nominal
pointing were downloaded from the \SZ CALDB website.
The contribution from the Cosmic X-ray Background (CXB) was
simulated following a recipe provided by the HXD team \footnote{See \texttt{http://heasarc.gsfc.nasa.gov/docs/suzaku/analysis/pin\_cxb.html}}.

\subsection{Image Analysis}
We extracted the XIS images within the photon energy range of $0.4 - 10$\,keV from only the two FI CCDs (XIS 0, XIS 3).
In the image analysis, we excluded calibration sources at the corner of the CCD chips. The images of the NXB were obtained from the night Earth data using \texttt{xisnxbgen} \citep{taw08}. Since the exposure times for the original data were different from that of NXB, we calculated the appropriate exposure-corrected original and NXB maps using \texttt{xisexpmapgen} \citep{ish07}. The corrected NXB images were next subtracted from the corrected original images. In addition, we simulated flat sky images using \texttt{xissim} \citep{ish07}, and applied a vignetting correction. All the images obtained with XIS0 and XIS3 were combined and re-binned by a factor of 4 \citep[CCD pixel size $24$\,$\mu$m\,$\times 24$\,$\mu$m, so that $1024 \times 1024$ pixels cover an $18' \times 18'$ region on the sky;][]{koy07}. Throughout these processes, we performed vignetting correction for all the images. Finally, the images were smoothed with a Gaussian function with $\sigma = 0.24'$. Note that the apparent features at the edge of these exposure corrected images are undoubtedly spurious due to low exposure in those regions.

\subsection{Spectral Analysis}
In the spectral analysis of XIS, we analyzed the three target sources as point sources, based on the result of our image analysis(See Section 5). Source regions for spectral analysis indicated by inner green circles were selected around each detected X-ray sources within the error ellipse of a gamma-ray emitters. The corresponding background regions were indicated by outer green ellipse after the removal of source regions. 
Moreover, if the X-ray sources other than target source was found, we excluded the region around those sources from the region for spectral analysis. We extracted the spectra from each source regions with \texttt{xselect} for each CCD (XIS 0, XIS 1, XIS 3). Next, we made redistribution matrix files (RMFs) and auxiliary response files (ARFs) using \texttt{xisrmfgen} and \texttt{xissimarfgen} \citep{ish07}, respectively. 
In addition, we used the new contamination files ae\_xi0\_contami\_20120711.fits, ae\_xi1\_contami\_20120711.fits and ae\_xi3\_contami\_20120711.fits \footnote{See : \texttt{http://byakko.scphys.kyoto-u.ac.jp:31415/xis/XIS\_Suzaku/2012\_MeetRep
     /20120711\_wada\_CALDB/}}, because response function of XIS0 is imperfect for recent observations.
Using these RMFs and ARFs, the corrected spectrum about energy response and the effective area of XIS were obtained.
Finally, Spectral analysis and model fitting were performed with \texttt{xspec} version 12.7.0.
In the spectral analysis of HXD, We also subtracted the NXB and CXB to obtain the HXD-PIN spectrum, then we performed model fitting together with XIS spectrum. 

\section{Results of \SZ observations}
We show in this section the results of X-ray image and spectral analysis for each object observed with \SZ.
Since we didn't detect any time-variability for each source in \SZ exposures, the results of timing analysis are 
not shown in this paper. We detected successfully significant signals from the X-ray counterpart of 1FGL\,J0157.0-5259 with HXD/PIN,
but below the sensitivity limit of the HXD/GSO, while other two sources were too faint to be detected either with 
HXD/PIN or GSO. Therefore, as for HXD analysis about 1FGL\,J0157.0-5259 in this paper, we only use the data from HXD/PIN.

\subsection{1FGL\,J0022.2-1850}
In our \SZ observations, we detected one X-ray point source (RA, DEC)=(5$\degr$.540, -18$\degr$.894) within the updated 2FGL error ellipse corresponding to 1FGL\,J0022.2-1850. Figure 1 shows the corresponding X-ray image of 
1FGL\,J0022.2-1850, as described in Section 4.1. 
The radio source NVSS\,J003750+123818 appears to be the counterpart of 1FGL\,J0022.2-1850, as indicated by the magenta cross at the 
center of this X-ray source (See section 5). Moreover, one unknown point source is detected within the background region 
for spectral analysis (because the central source is very bright, it is difficult to see this source in Fig 1).

In Figure 4, the X-ray spectrum of the \SZ source, which we argue is the most likely counterpart of 1FGL\,J0022.2-1850, is shown. 
The XIS spectra is given for the energy range $0.6 - 7.5$\,keV. In the spectral analysis, the target X-ray source is 
so bright that we selected a source region assuming radii 3$\arcmin$ (a typical half-power diameter of the XRT is 
2$\arcmin$:\citep{ser07}). Meanwhile, we excluded one contaminating field X-ray point source detected by \SZ, assuming the 
source region radii 2$\arcmin$ from the source and background regions for spectral analysis described in Section 4.2. 
The spectrum is well fitted by a single power-law continuum with a photon index, $\Gamma = 2.43 \pm 0.03$, 
moderated by the Galactic absorption only. The Galactic hydrogen column density was fixed as $N_{\rm H} = 2.02 
\times 10^{20}$\,cm$^{-2}$ \citep{dic90}. The value of $\chi^{2}$\,/\,d.o.f\,$= 46.81/46$ indicated that this 
is a satisfactory model for 1FGL\,J0022.2-1850. The details of the fitting results are summarized in Table\,2.

\subsection{1FGL\,J0038.0+1236}
One X-ray point source (RA, DEC)=(9$\degr$.472, 12$\degr$.639) was found with {\it Suzaku} within the improved 2FGL error ellipse corresponding to 1FGL\,J0038.0+1236. 
The corresponding X-ray image made through the way described in Section 4.1 is shown in Figure\,2.
Moreover, one unknown point source is detected within the background region for spectral analysis.
The radio source, NVSS\,J00209-185332 (shown by magenta cross; see section 6), is coincident with the X-ray position and we propose this to be the most likely counterpart of 1FGL\,J0038.0+1236. 

Figure\,2 shows the X-ray spectrum of the point source detected by \SZ near the center of the 2FGL error ellipse of 
1FGL\,J0038.0+1236 ($0.6 - 7.5$\,keV). The source region was selected with radii 3$\arcmin$, because the target 
source is too bright not to cover entire region of the emission from target source with radii 2$\arcmin$. When we 
selected the background region, the region from X-ray contaminant source was excluded with radii 2$\arcmin$ in order to not subtract 
too much as background. The spectrum could be well fitted by a single power-law continuum with $\Gamma = 
2.76 \pm 0.17$, moderated by the Galactic absorption only. The Galactic hydrogen column density was fixed as $N_{\rm H} = 
5.38 \times 10^{20}$\,cm$^{-2}$ \citep{dic90}. The value of $\chi^{2}$\,/\,d.o.f\,$= 85.54/70$ indicates that this is 
a satisfactory model for 1FGL\,J0038.0+1236. The details of the fitting results are summarized in Table\,3.

\subsection{1FGL\,J0157.0-5259}
We succeeded in detecting a bright X-ray point source with {\it Suzaku} within the 2FGL error ellispe corresponding to 1FGL\,J0157.0-5259.
Figure 3 shows the corresponding X-ray image (see Section 4.1). The X-ray source is located at (RA, DEC)=
(29$\degr$.253, -53$\degr$.035), as shown in Figure\,3. The position of the radio source, SUMSS\,J015657-530157, is shown by 
a magenta cross (see section 5). In this observation, we did not find any other contamination source like in the above two observations.

The X-ray spectrum (XIS + HXD) of the \SZ source, which we propose to be the most likely counterpart of 
1FGL\,J0157.0-5259, is shown in Figure 6 within the energy range $0.6 - 40.0$\,keV (XIS $0.4-10.0$\,keV, 
HXD $15.0-40.0$\,keV). In this spectral analysis, we set the extraction region to be a radius of 4$\arcmin$ to encircle this bright source. 
On the other hand, we set the background region with the more larger radius, and 
the location of the center of background region displaced from the center of target source not to be over the region 
covering by CCD. The spectrum could be well fitted by a single power-law continuum with
$\Gamma = 1.85 \pm 0.01$, moderated by the Galactic absorption only. The Galactic hydrogen column 
density was fixed as $N_{\rm H} = 2.70 \times 10^{20}$\,cm$^{-2}$ \citep{dic90}. The value of 
$\chi^{2}$\,/\,d.o.f\,$= 3184.74/2757$ indicates that this is a satisfactory model for 1FGL\,J0157.0-5259. 
The details of the fitting results are summarized in Table\,4.

\section{Discussion}
In the uniform analysis of archival {\it Swift} data, we found several objects which seemed to display
anomalous SEDs that are not typical of AGNs or PSRs. Then we performed \SZ X-ray follow-up observations 
of three such sources to more accurately determine the SEDs of each object (Figures 7, 8, and 9). 
Different fluxes between {\it Swift}/XRT and \SZ/{\it XIS} seen in these SEDs indicate that these objects 
should have temporal variability that was not seen within the individual shorter \SZ exposures. 
Moreover, thanks to the good sensitivity and long exposure of the \SZ data, we have additional hints 
to reveal the nature of each source as discussed below.
 
An X-ray source found within the updated 2FGL error ellipse of 1FGL\,J0022.2-1850 is positionally 
consistent with the radio source NVSS\,J00209-185332 found in the NVSS catalog \citep{NVSS} (See Table\,2).
Moreover, infrared counterpart source WISE\,J002209.25-185334.7 located at (RA, DEC)=(5$\degr$.5385563, -18$\degr$.8929772) was found in
the Wide Field Infrared Survey Explorer (WISE) All-Sky Release \citep{wri10}.
The SED of 1FGL\,J0022.2-1850/WISE\,J002209.25-185334.7/NVSS\,J00209-185332, 
including our {\it Suzaku}/XIS data and derived XRT and 
UVOT fluxes from {\it Swift}, are shown in Figure\,7. From the relatively high radio flux and flat X-ray 
spectrum obtained with {\it Swift}/XRT, this object is likely to be a low-frequency peaked BL Lac (LBL). 
However, during our \SZ observation, the X-ray spectrum was observed to be substantially steeper, more typical 
of a high-frequency peaked BL Lac (HBL). Moreover, the flat GeV gamma-ray spectrum is typical of HBLs like Mrk 421 
and Mrk 501, rather than a LBL. Considering the Cherenkov Telescope Array (CTA) which is an initiative to built 
a next generation observatory for very-high energy gamma-rays will have an improved sensitivity by 
an order of magnitude with respect to current instruments ($\sim 10^{-14} {\rm erg/cm^{2}/s}$ above a few TeV), 
the upward shape of the \F spectrum suggests the source could be detected also in TeV energy in the near future.

In the case of 1FGL\,J0038.0+1236, the location of a X-ray counterpart discovered in our \SZ observations 
is consistent with NVSS\,J003750+123818 described in Table\,2, and each optical counterpart 
SDSS\,J003750.88+123819.9 (classified as GALAXY) located at (RA, DEC)=(9$\degr$.462, 12$\degr$.638875)
and infrared counterpart WISE\,J003750.87+123819.9 located at (RA, DEC)=(9$\degr$.4619958, 12$\degr$.6388878),
were also found in the Sloan Digital Sky Survey (SDSS) catalog \citep{SDSS} and WISE catalog respectively.
The constructed radio to X-ray SED together 
with the {\it Swift} XRT/UVOT and the LAT spectrum is shown in Figure\,8. Since the X-ray spectrum appears very steep, this 
source seems to be associated with a HBL, while the steep gamma-ray spectrum observed with \F favors a FSRQ origin of this source.
While optical and ultraviolet fluxes are extremely bright, this could be due to a
contribution of soft photons from the host galaxy as seen in some blazar spectra
(see e.g., the SED of Mrk 501; \citet{kat99}).
These results show that this source is difficult to be explained by conventional leptonic models of blazars 
(i.e., SSC or external Compton models). 

In the case of 1FGL\,J0157.0-5259, our {\it Suzaku}/XIS observations revealed the presence of a quite 
bright X-ray counterpart in the LAT error circle, and since the hard X-ray fluxes of this source are very high, 
we could also obtain data from {\it Suzaku}/HXD. At the position of this \SZ X-ray source, the radio counterpart 
SUMSS J015657-530157 was found in the SUMSS catalog \citep{SUMSS}(Table\,2).
The broad-band SED of 1FGL\,J0157.0-5259/SUMSS\,J015657-530157 with our {\it Suzaku}/XIS,HXD and derived 
{\it Swift}/XRT,UVOT data is presented in Figure\,9. Since the X-ray fluxes are connected by a straight line with 
the radio fluxes, these fluxes are seemed to be explained by synchrotron radiation. The peak frequency of the synchrotron 
spectrum is very high ($>$10 keV) suggesting that the source could be one a rare type of "extreme" blazar like 
Mrk 501 in the historical high state \citep{ghi04}. If this source is an "extreme" blazar, observations at TeV 
energies or more deep observations with \F may detect significant signal in the future.

Finally, we made the similar figure described in Takahashi et al. (2012) for the 1FGL 134 unID objects in which we performed 
follow-up observations with {\it Swift}/XRT (Figure\.10). This figure presents a comparison of the AGNs (aqua), 
PSRs (green), and unassociated sources (red) which classified in the 2FGL catalog in the X-ray to gamma-ray 
flux ratios versus radio to gamma-ray ratios plane. The three sources we observed with \SZ in this paper are shown 
in black stars. Apparently, that these sources are situated in the typical AGN region of this diagnostic plane. 
It is noteworthy that, 1FGL\,J0157.0-5259 is at the right edge of this typical AGN region, 
and this means the X-ray and gamma-ray flux ratios of this source is quite high, such that $\sim$ 2, 
which is consistent with our speculation that the source is an extreme HBL-type blazar. 

\section{Summary}
In this paper we reported on the results of X-ray follow-up observations of three unID gamma-ray sources 
detected by \F instrument which indicate anomalous SEDs.
We have successfully detected X-ray counterparts of 1FGL\,J0022.2-1850, 1FGL\,J0038.0+1236 and 1FGL\,J0157.0-5259 
using \SZ. The characteristics of each object are summarized below. We also note that these objects display
temporal variability in X-rays, as indicated by the different X-ray fluxes measured by {\it Suzaku}/XIS 
and {\it Swift}/XRT (see, Figure\,7, 8, 9). The X-ray spectrum of the discovered \SZ counterpart of 
1FGL\,J0022.2-1850 is well fitted by single power-law model with $\Gamma = 2.43 \pm 0.03$. 
The spectral shape obtained with \SZ (in X-ray) and \F (in gamma ray) suggest the source 
is typical of a HBL-type blazar, but previous $Swift$ observations rather show it was similar to the LBL-type blazar. 
The source is potentially a TeV emitter that could be detected in the near future. 
In the case of 1FGL\,J0038.0+1236, the X-ray spectrum is well fitted by single power-law model with a photon index, 
$\Gamma = 2.76 \pm 0.17$. At first glance, this source also seems to be classified as a HBL because the X-ray spectrum
seen in Figure\,8 appear very steep. However, its steep gamma-ray spectrum observed with \F favors a FSRQ origin for this source. 
These results show that this source is difficult to be explained by standard emission models of blazars, i.e., 
SSC or external Compton models. In the case of 1FGL\,J0157.0-5259, the \SZ X-ray spectrum obtained from XIS and 
HXD are well fitted by single power-law model with a photon index, $\Gamma = 1.85 \pm 6.66 \times 10^{-3}$. 
From the multiwavelength analysis shown in Figure\,9, the peak frequency of synchrotron spectrum is very 
high ($\sim$10 keV) suggesting that source could be one of "extreme" blazar like Mrk 501 in the historical high state.

\acknowledgments
We would like to thank C.  C. Cheung for useful 
comments that helped to improve the organization of the manuscript.

\clearpage

\begin{table}
\begin{center}
\caption{{\it Suzaku} XIS observation logs.}
\label{tbl-1}
\begin{tabular}{ccccc}
\tableline
Target Name & R.A. [deg] & Dec. [deg] & Exposure [ks] & Obs. Start (UT) \\
\tableline\tableline
1FGL\,J0022.2-1850 & 5.5540 & -18.9060 & 34.2 & 2012-May.30 12:57:00 \\
1FGL\,J0038.0+1236 & 9.4627 & 12.6391 & 18.8 & 2012-Jun.29 23:56:00 \\
1FGL\,J0157.0-5259 & 29.3640 & -53.0280 & 12.1 & 2012-May.28 16:19:00 \\
\tableline
\end{tabular}
\end{center}
\end{table}

\begin{table}
\begin{center}
\caption{Radio counterpart sources for each target.}
\label{tbl-2}
\begin{tabular}{ccccc}
\tableline
Target Name & R.A. [deg] & Dec. [deg] & $F_{\rm 1.4\,GHz}$ [mJy] & $F_{\rm 843\,MHz}$ [mJy] \\
\tableline\tableline
NVSS J00209-185332 & 5.5381667 & -18.892444 & $22.1 \pm 0.8$ & - \\
NVSS J003750+123818 & 9.461875 & 12.638556 & $75.1 \pm 2.3$ & - \\
SUMSS J015657-530157 & 29.240833 & -53.032778 & - & $43.4 \pm 1.5$ \\
\tableline
\end{tabular}
\end{center}
\end{table}

\begin{table}[m]
\small
\caption{Fitting parameters for each {\it Suzaku} target (the power-law model).}
\label{tbl-1}
\scalebox{0.9}{
\begin{tabular}{cccccc}
\tableline
Target Name & $N_{\rm H}$ [cm$^{-2}$] & $\Gamma$ & $\chi^2$/d.o.f & $P(\chi^2)$ & $F_{\rm 0.6 - 7.5\,keV}$ [erg\,cm$^{-2}$\,s$^{-1}$] \\
\tableline\tableline
1FGL\,J0022.2-1850 & $2.02 \times 10^{20}$ (fixed) & $2.43 \pm 0.03$ & $46.81/46$ & $4.39 \times 10^{-1}$ & $2.12 \times 10^{-12}$\\
1FGL\,J0038.0+1236 & $5.38 \times 10^{20}$ (fixed) & $2.76 \pm 0.17$ & $85.54/70$ & $9.98 \times 10^{-2}$ & $2.11 \times 10^{-13}$\\
1FGL\,J0157.0-5259 & $2.70 \times 10^{20}$ (fixed) & $1.85 \pm 6.66 \times 10^{-3}$ & $3184.74/2757$ & $1.95 \times 10^{-8}$ & $3.21 \times 10^{-11}$\\
\tableline
\end{tabular}
}
\end{table}

\clearpage

\begin{figure}
\begin{center}
\includegraphics[width=150mm]{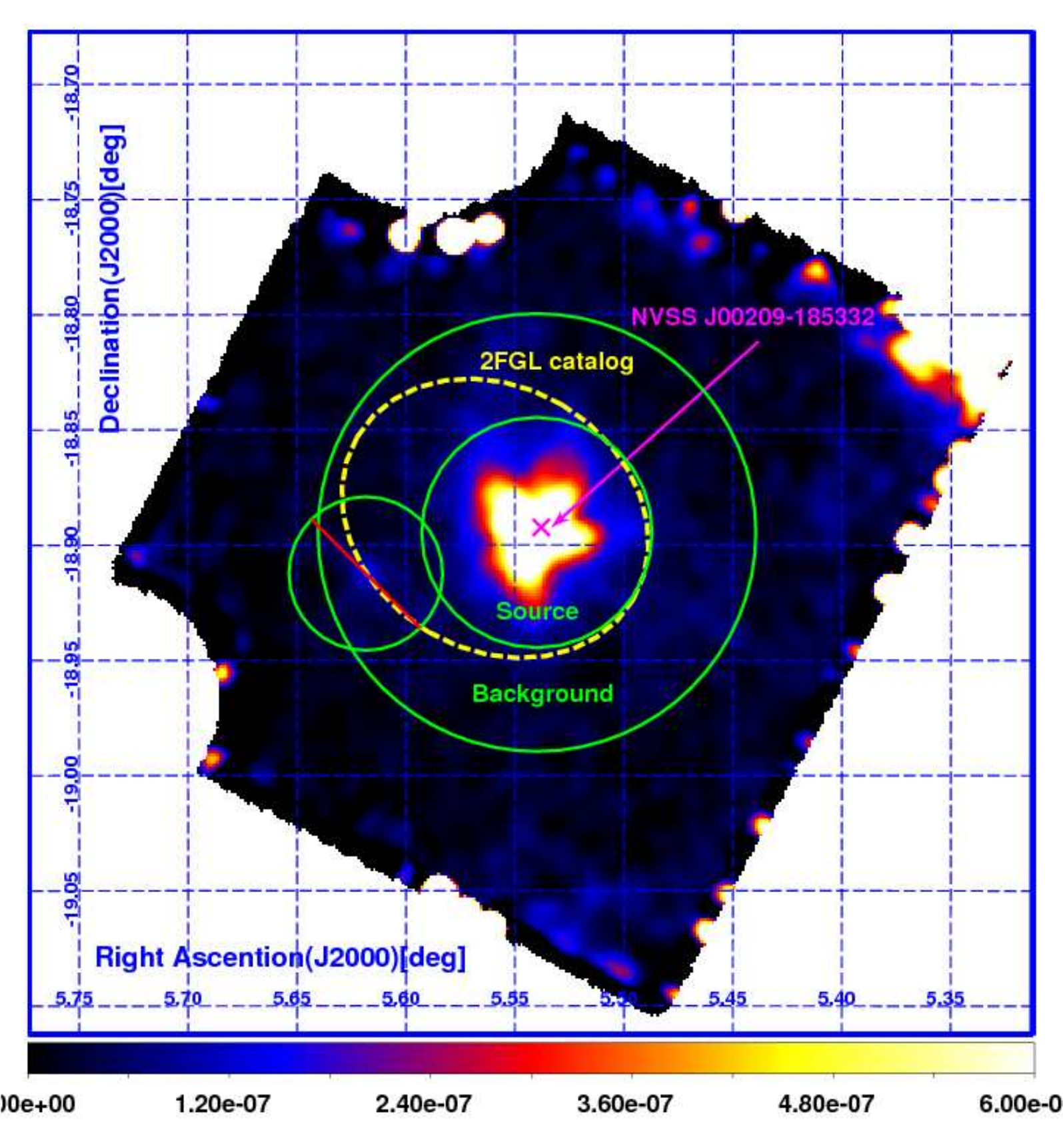}\\
\caption{\SZ X-ray image of 1FGL\,J0022.2-1850 (vignetting and exposure corrections applied). Data from XIS 0 and XIS 3 are summed in the $0.4 - 10$\,keV energy band. The magenta cross denote the position of radio counterpart source NVSS\,J00209-185332, the yellow dotted ellipse denotes the 95$\%$ position error of 2FGL\,J0022.2-1853, and the inner green ellipse denote the \SZ source extraction region, the outer green ellipse without inner source region denote the \SZ background extraction region.}
\label{fig-2}
\end{center}
\end{figure}

\clearpage

\begin{figure}
\begin{center}
\includegraphics[width=150mm]{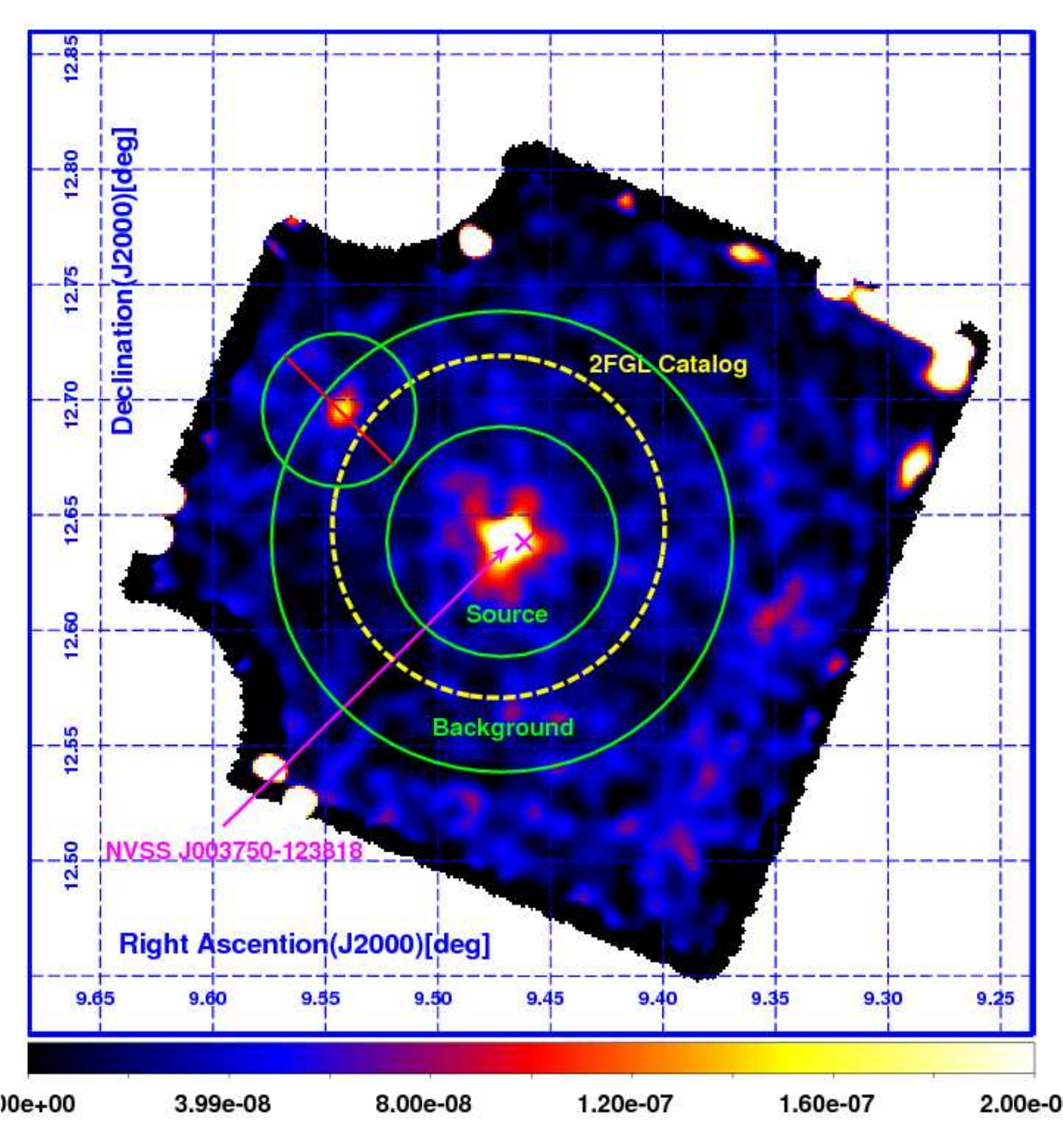}\\
\caption{Same as Figure\,1, but target source is 1FGL\,J0038.0+1236 and radio counterpart is NVSS\,J003750+123818.}
\label{fig-3}
\end{center}
\end{figure}

\clearpage

\begin{figure}
\begin{center}
\includegraphics[width=150mm]{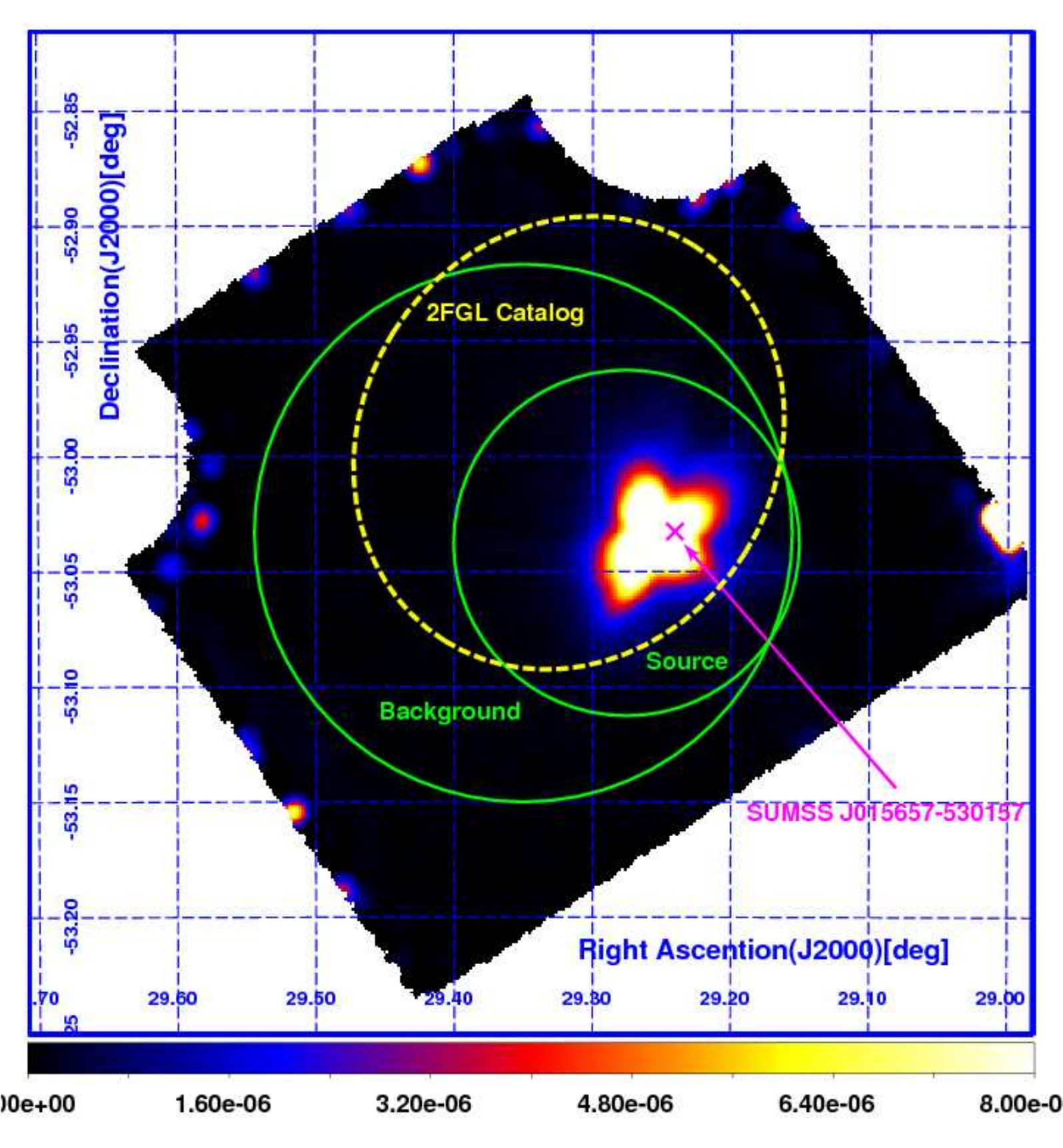}\\
\caption{Same as Figure\,1, but target source is 1FGL\,J0157.0-5259 and radio counterpart is SUMSS\,J015657-530157.}
\label{fig-4}
\end{center}
\end{figure}

\clearpage

\begin{figure}
\begin{center}
\includegraphics[width=120mm,angle=-90]{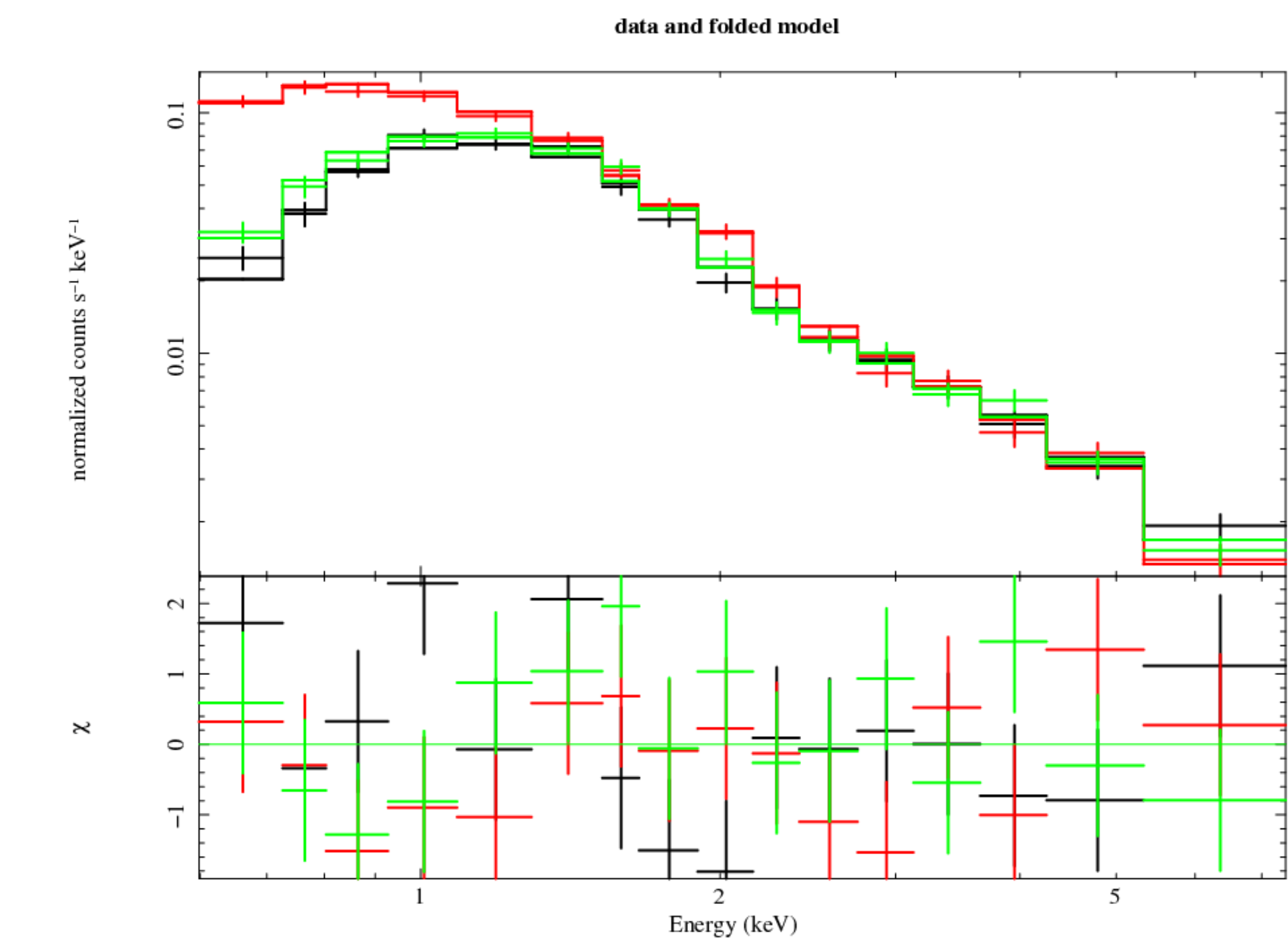}\\
\caption{\SZ XIS spectra of 1FGL\,J0022.2-1850 in the photon energy range $0.4 - 10$\,keV, fitted by the model \texttt{wabs + power-law}. Spectra of XIS0, XIS1, XIS3 are shown in black, red, and green, respectively.}
\label{fig-5}
\end{center}
\end{figure}

\clearpage

\begin{figure}
\begin{center}
\includegraphics[width=120mm,angle=-90]{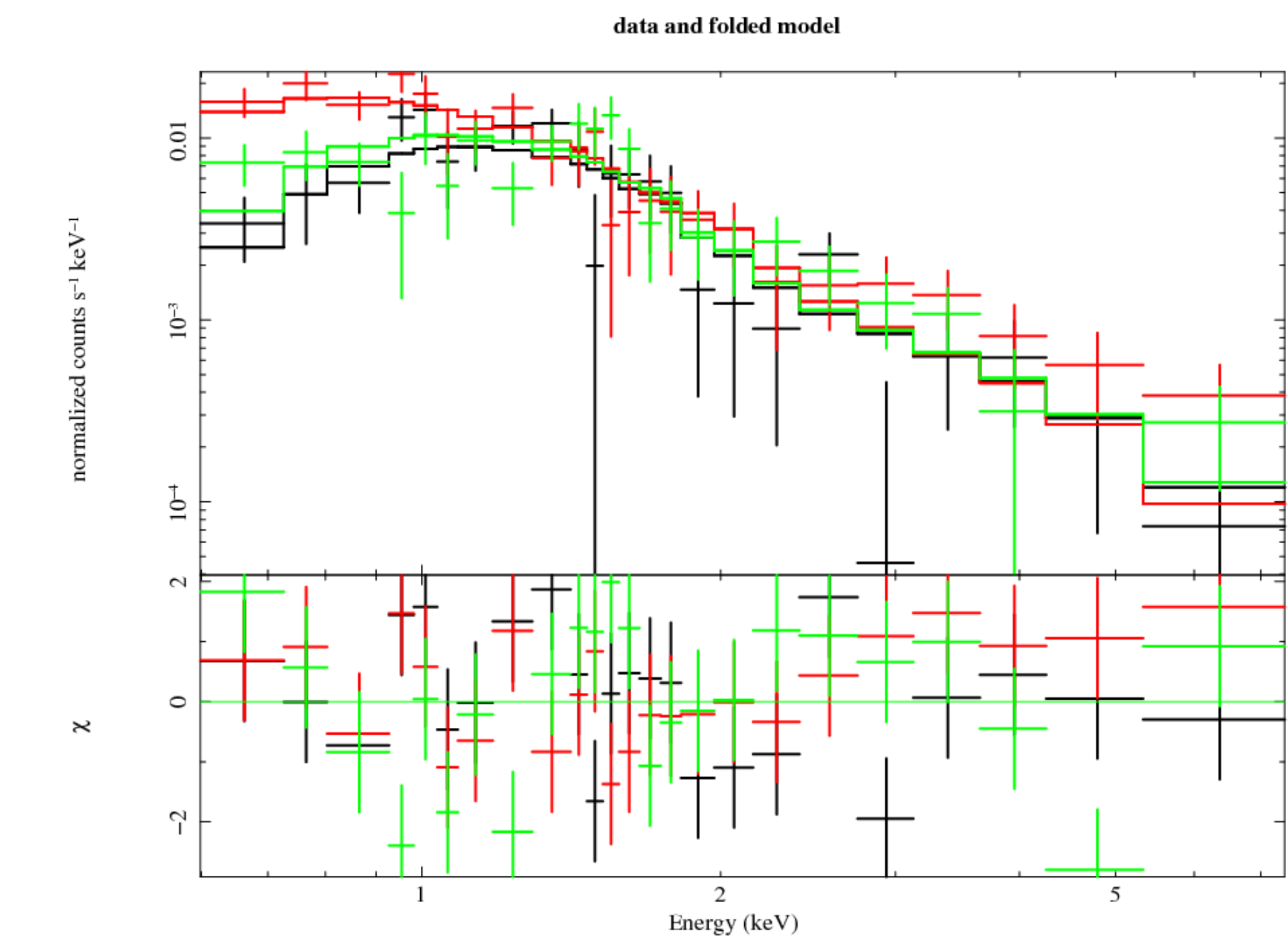}\\
\caption{\SZ XIS spectra of 1FGL\,J0038.0+1236 in the photon energy range $0.4 - 10$\,keV, fitted by the model \texttt{wabs + power-law}. Spectra of XIS0, XIS1, XIS3 are shown in black, red, and green, respectively.}
\label{fig-6}
\end{center}
\end{figure}

\clearpage

\begin{figure}
\begin{center}
\includegraphics[width=120mm,angle=-90]{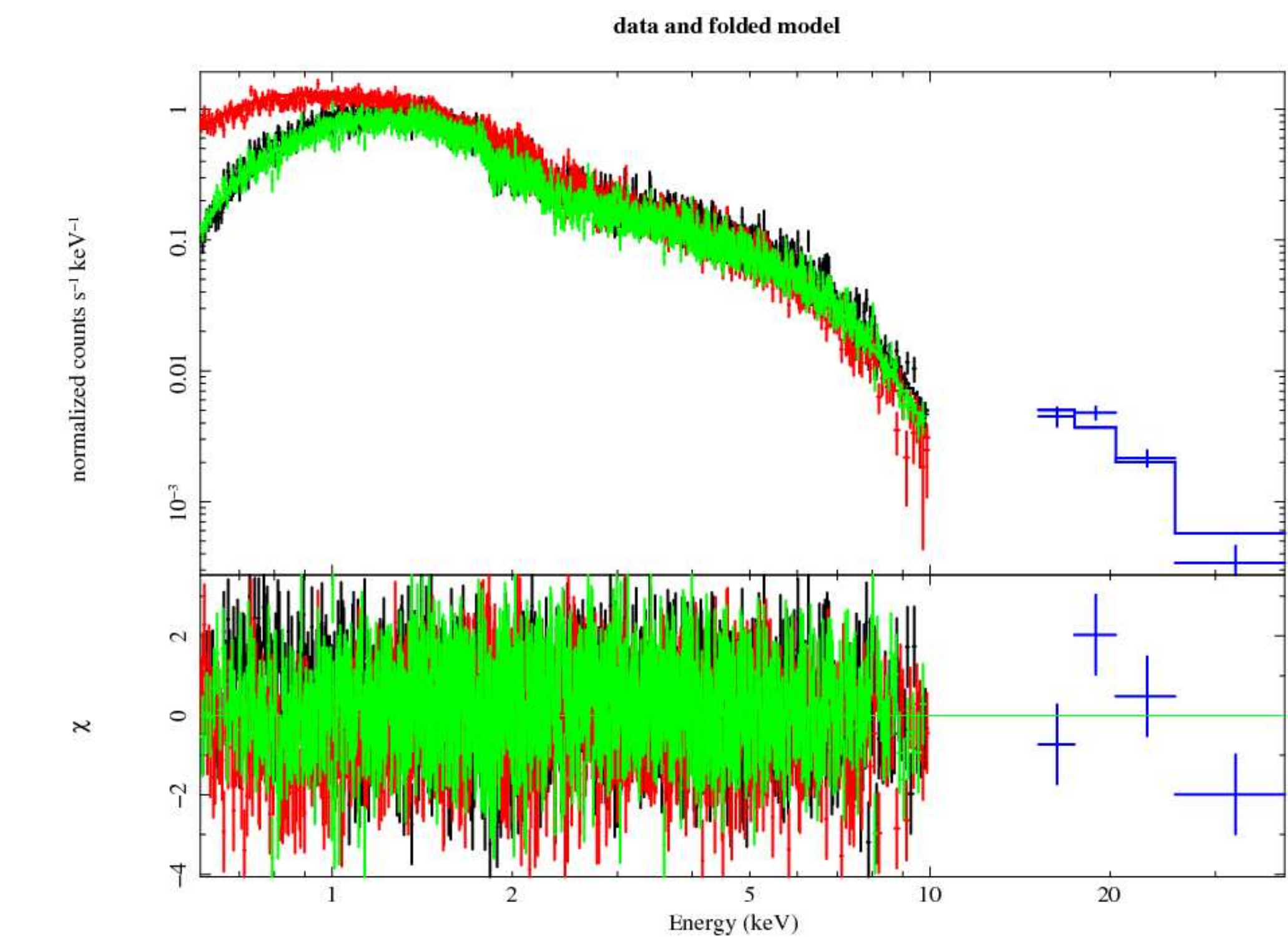}\\
\caption{\SZ XIS spectra of 1FGL\,J0157.0-5259 in the photon energy range $0.4 - 40$\,keV (XIS $0.4-10$\,keV, HXD $15-40$\,keV), fitted by the model \texttt{wabs + power-law}. Spectra of XIS0, XIS1, XIS3, HXD are shown in black, red, green, and blue respectively.}
\label{fig-7}
\end{center}
\end{figure}

\clearpage

\begin{figure}
\begin{center}
\includegraphics[scale=0.80]{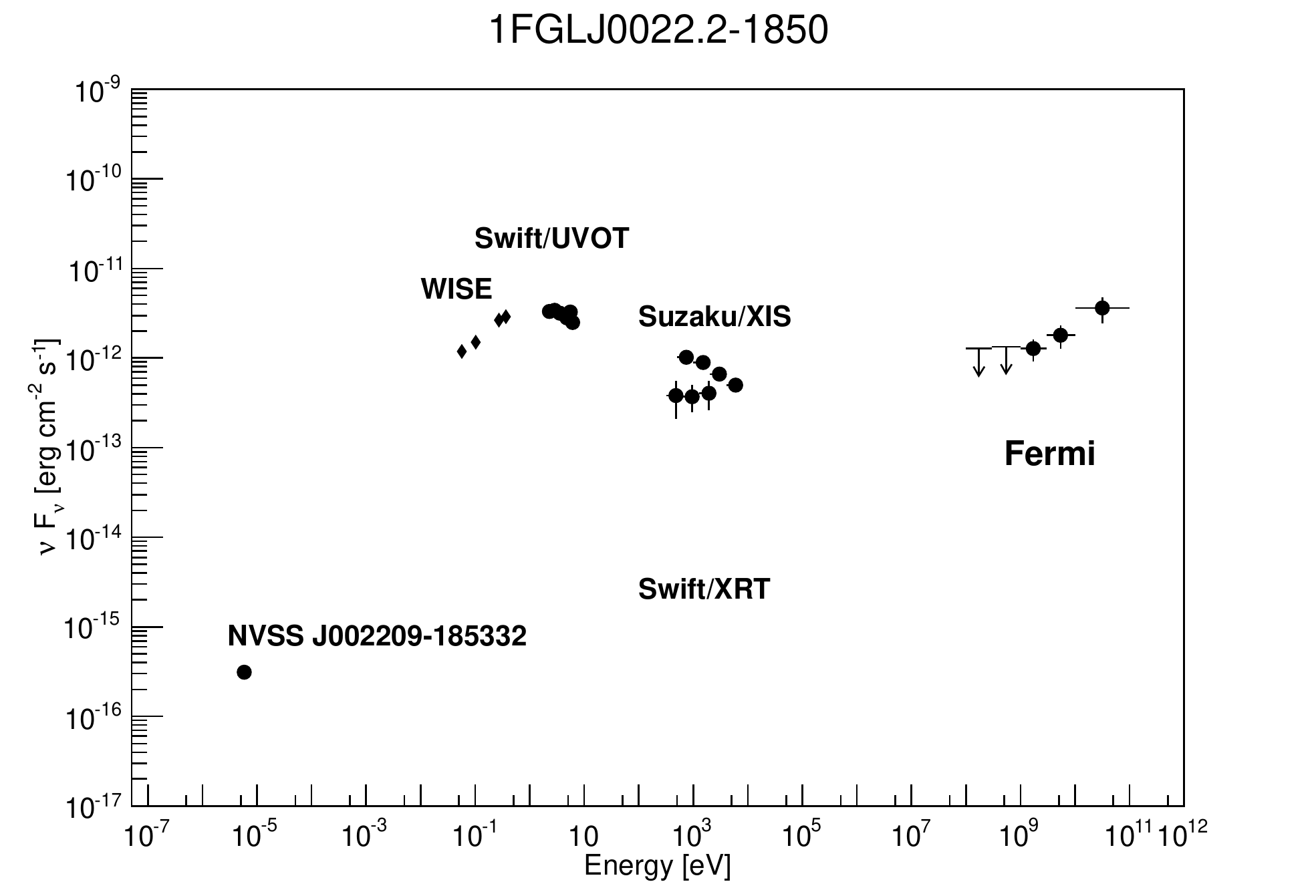}\\
\caption{Broad-band SEDs of 1FGL\,J0022.2-1850. The radio fluxes (1.4 GHz) for this source are taken from NVSS\,J00209-185332 as radio counterpart. The infrared fluxes (3.4$\micron$, 4.6$\micron$, 12$\micron$ and 22$\micron$) are taken from the WISE catalog. The optical/UV fluxes were derived from the {\it Swift}/UVOT observations (this work). The X-ray are fluxes taken from the {\it Suzaku}/XIS and {\it Swift}/XRT observations (this work). Finally, the gamma-ray data points are taken from the 2FGL catalog \citep{2FGL}.}
\label{fig-8}
\end{center}
\end{figure}

\clearpage

\begin{figure}
\begin{center}
\includegraphics[scale=0.80]{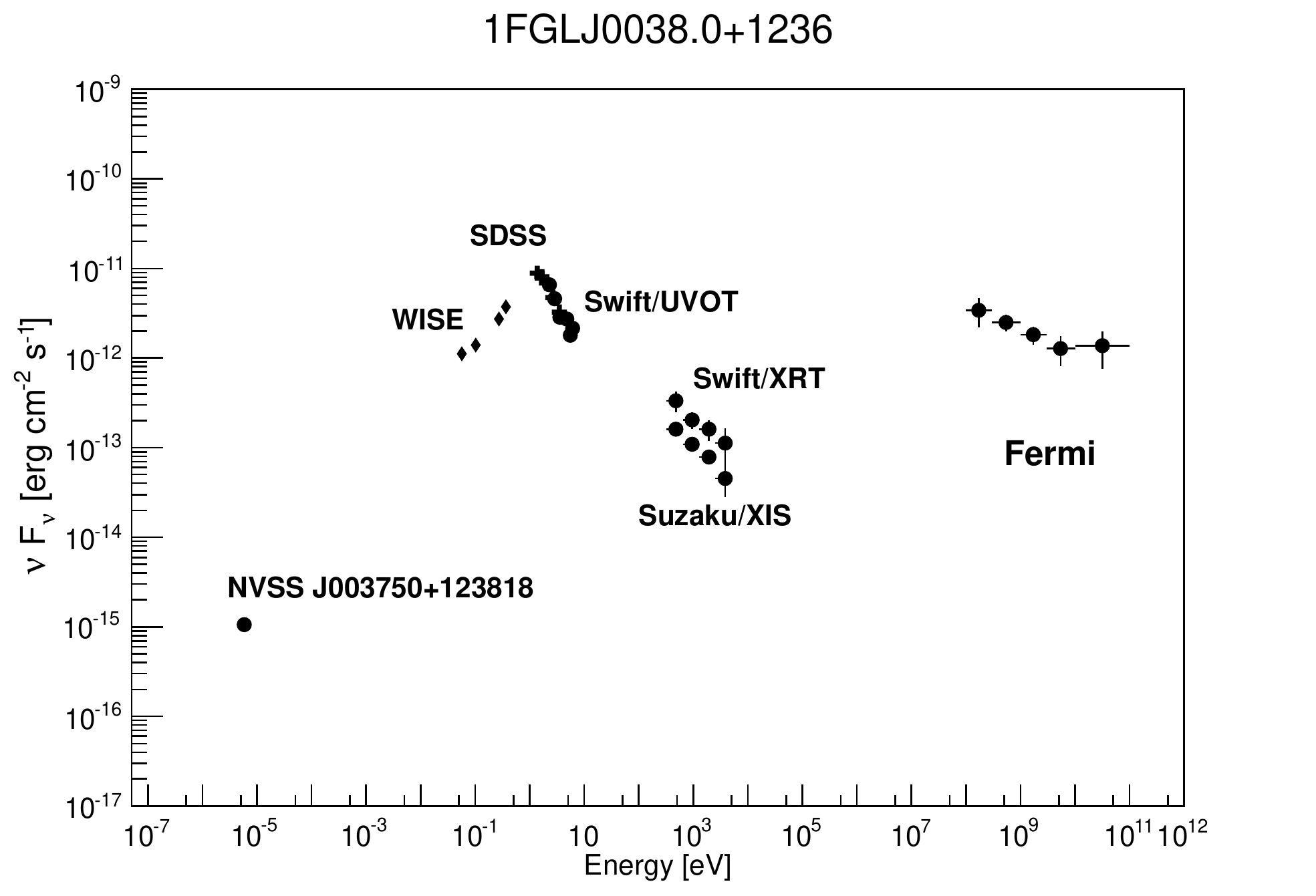}\\
\caption{Broad-band SEDs of 1FGL\,J0038.0+1236. The radio fluxes (1.4 GHz) for this source are taken from NVSS\,J003750*123818 as radio counterpart. The infrared fluxes (3.4$\micron$, 4.6$\micron$, 12$\micron$ and 22$\micron$) are taken from the WISE catalog.
The optical/UV fluxes were derived from the {\it Swift}/UVOT observations (this work) and the SDSS catalog. The X-ray fluxes are taken from the {\it Suzaku}/XIS and {\it Swift}/XRT observations (this work). Finally, the gamma-ray data points are taken from the 2FGL catalog \citep{2FGL}.}
\label{fig-9}
\end{center}
\end{figure}

\clearpage

\begin{figure}
\begin{center}
\includegraphics[scale=0.80]{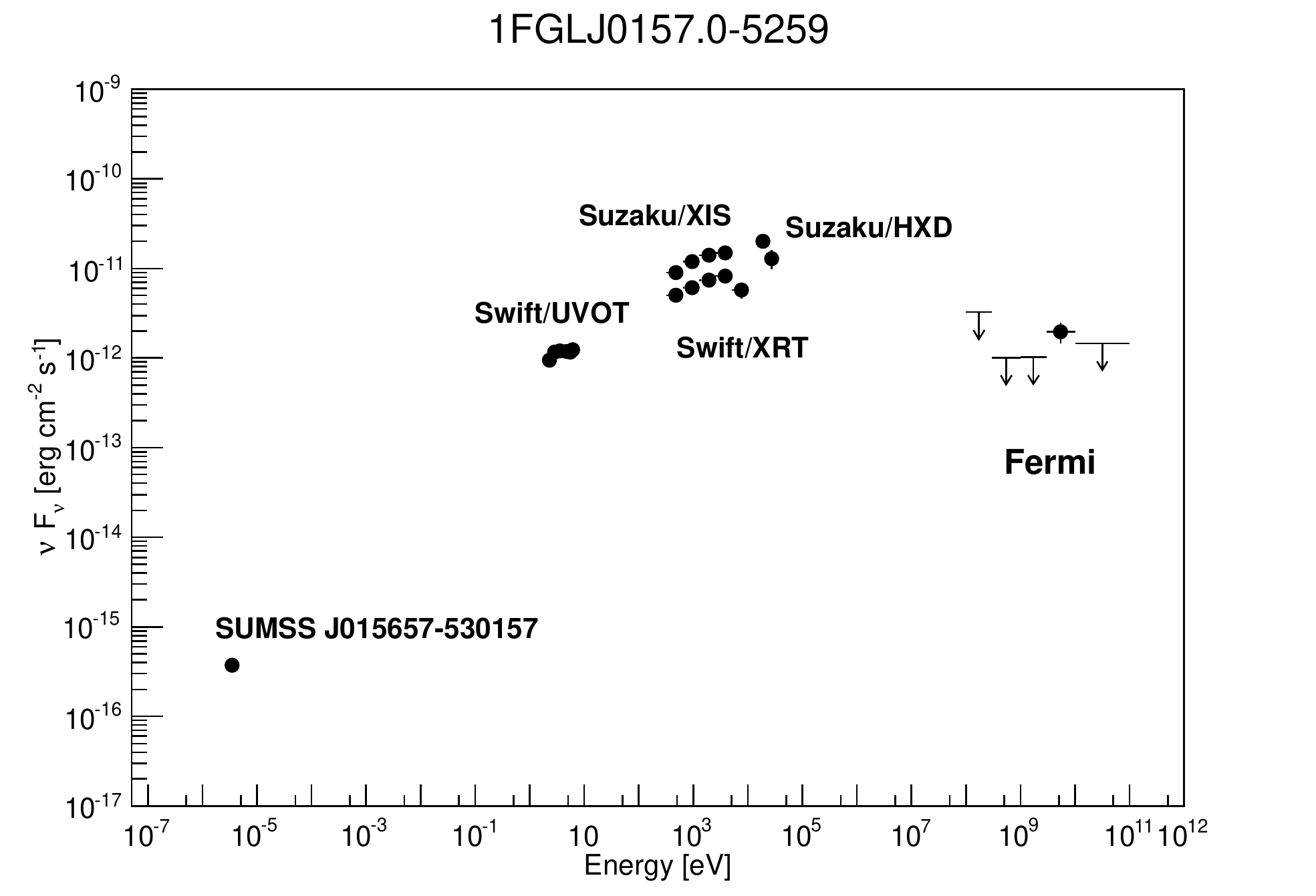}\\
\caption{Broad-band SEDs of 1FGL\,J0157.0-5259. The radio fluxes (843 MHz) for this source a re taken from SUMSS\,J015657-530157 as radio counterpart source. The optical/UV fluxes were derived from the {\it Swift}/UVOT observations. The X-ray fluxes taken from the {\it Suzaku}/XIS and {\it Swift}/XRT observations (this work). The hard X-ray fluxes are from the {\it Suzaku}/HXD observations (this work). Finally, the gamma-ray data points are taken from the 2FGL catalog \citep{2FGL}.}
\label{fig-10}
\end{center}
\end{figure}

\clearpage

\begin{figure}
\includegraphics[scale=0.8]{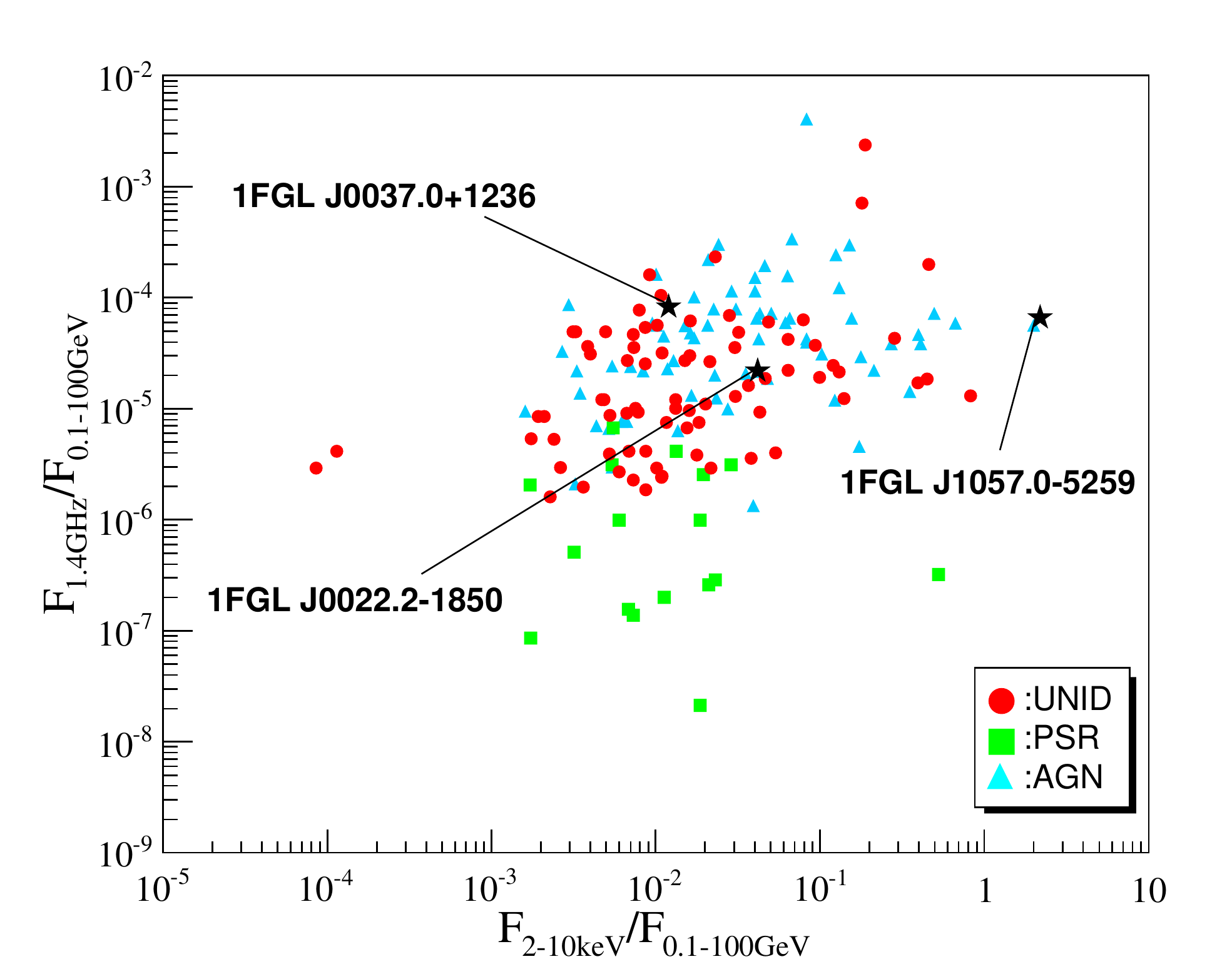}\\
\caption{X-ray to gamma-ray flux ratios versus radio to gamma-ray ratios for 1FGL 134 unID objects performed follow-up observations with Swift/XRT. Red circles show the flux ratios of the \F sources which are still unID in 2FGL catalog. Aqua triangles indicate the data plots of AGN listed in 2FGL catalog. Green squares represent the data points of sources that are associated with PSRs listed in 2FGL catalog. Three black stars show the target sources in this paper, 1FGL\,J0022.2-1850, 1FGL\,J0038.0+1236, and 1FGL\,J0157.0-5259, respectively.}
\label{fig-1}
\end{figure}

\clearpage

\begin{appendix}
\chapter{{\it Swift} Observed 1FGL unID sources}
\begin{figure}[m]
 \begin{center}
  \begin{minipage}{0.32\hsize}
    \begin{center}
      {\small (1) 1FGL\,J0001.9--4158} \\
      \includegraphics[width=52mm]{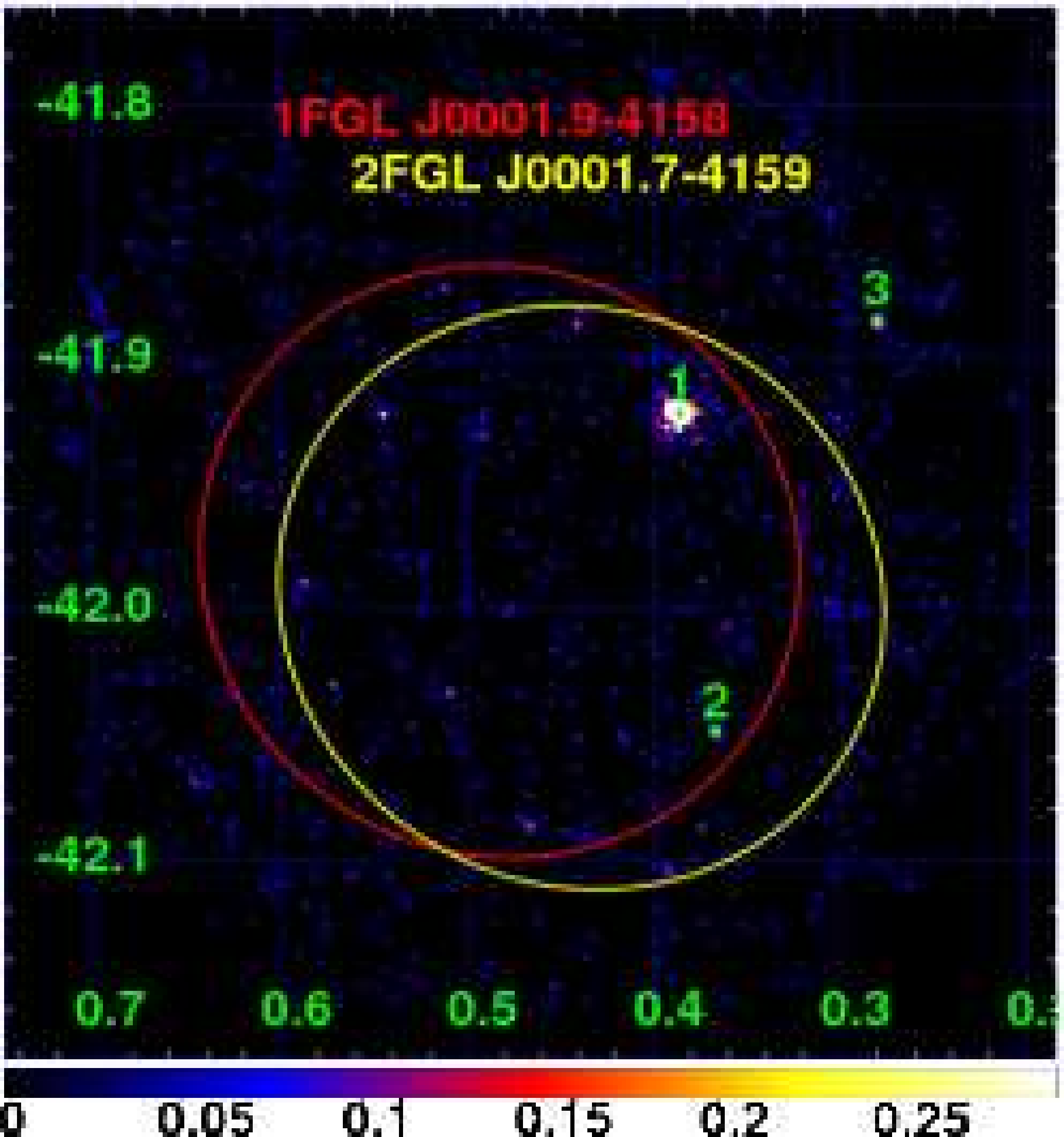}
    \end{center} 
  \end{minipage}
  \begin{minipage}{0.32\hsize}
    \begin{center}
      {\small (2) 1FGL\,J0009.1$+$5031} \\
      \includegraphics[width=52mm]{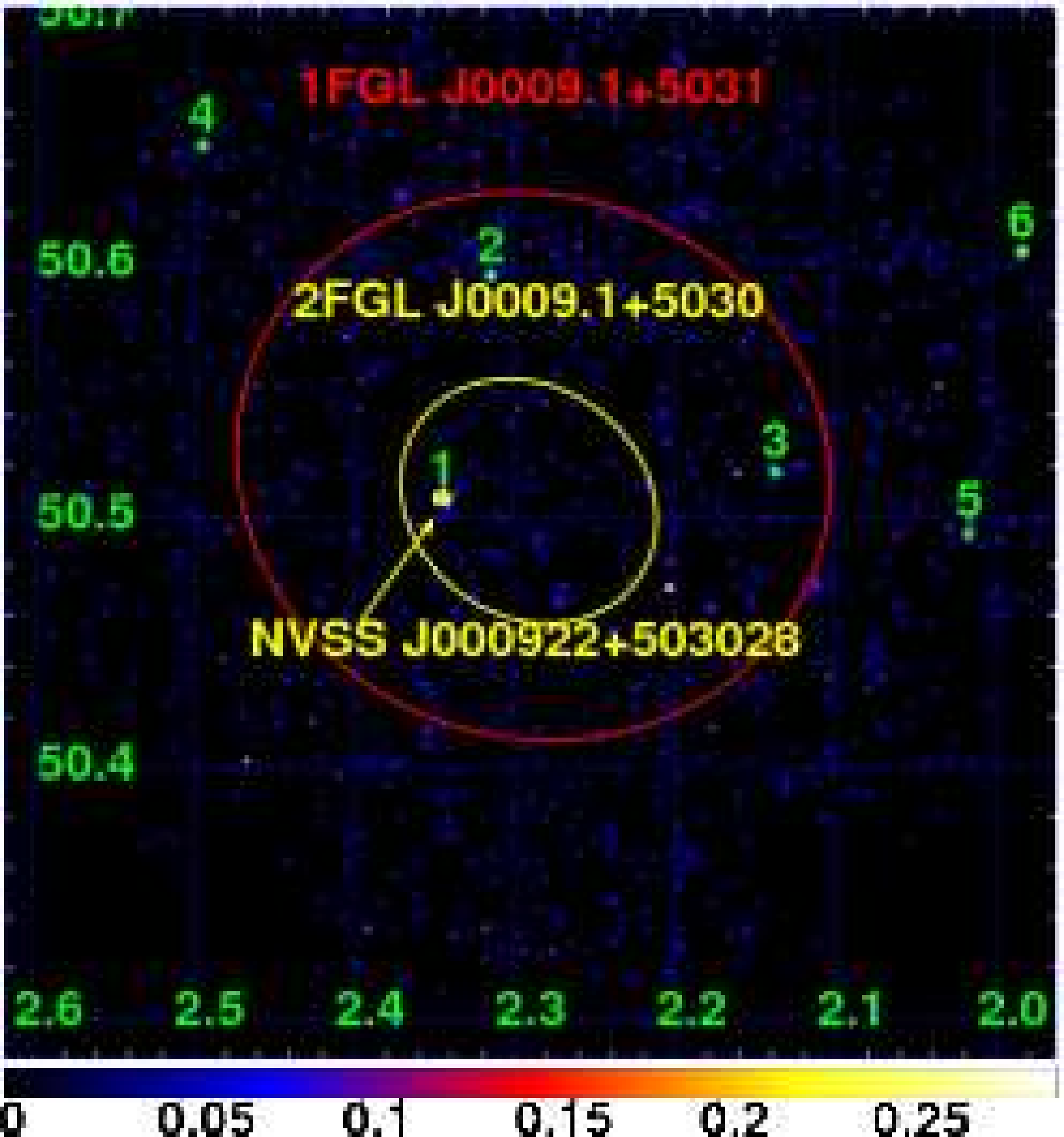}
    \end{center}
  \end{minipage}
  \begin{minipage}{0.32\hsize}
    \begin{center}
      {\small (3) 1FGL\,J0022.2--1850} \\
      \includegraphics[width=52mm]{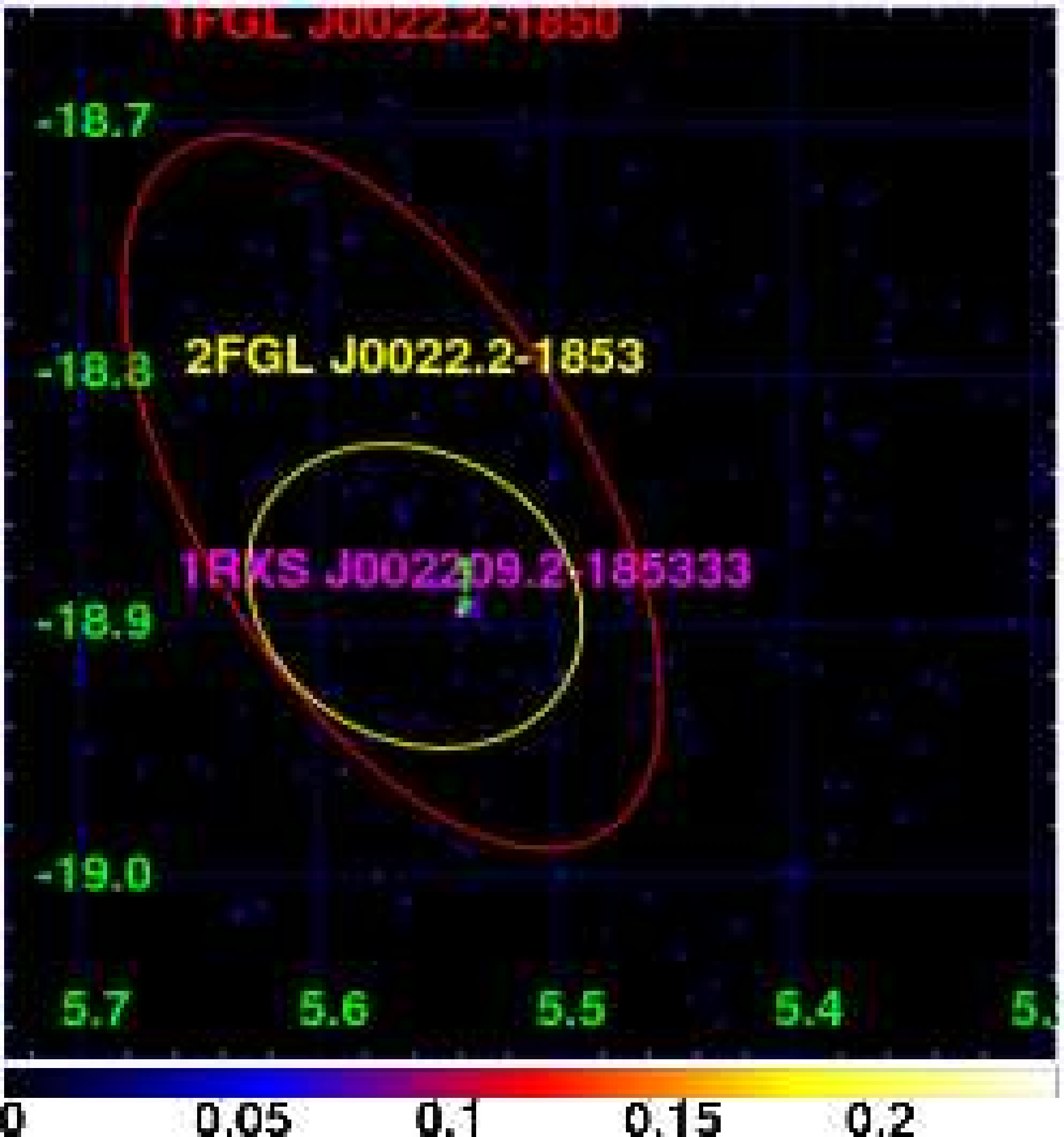}
    \end{center}
  \end{minipage}
  \begin{minipage}{0.32\hsize}
    \begin{center}
      {\small (4) 1FGL\,J0023.5$+$0930} \\
      \includegraphics[width=52mm]{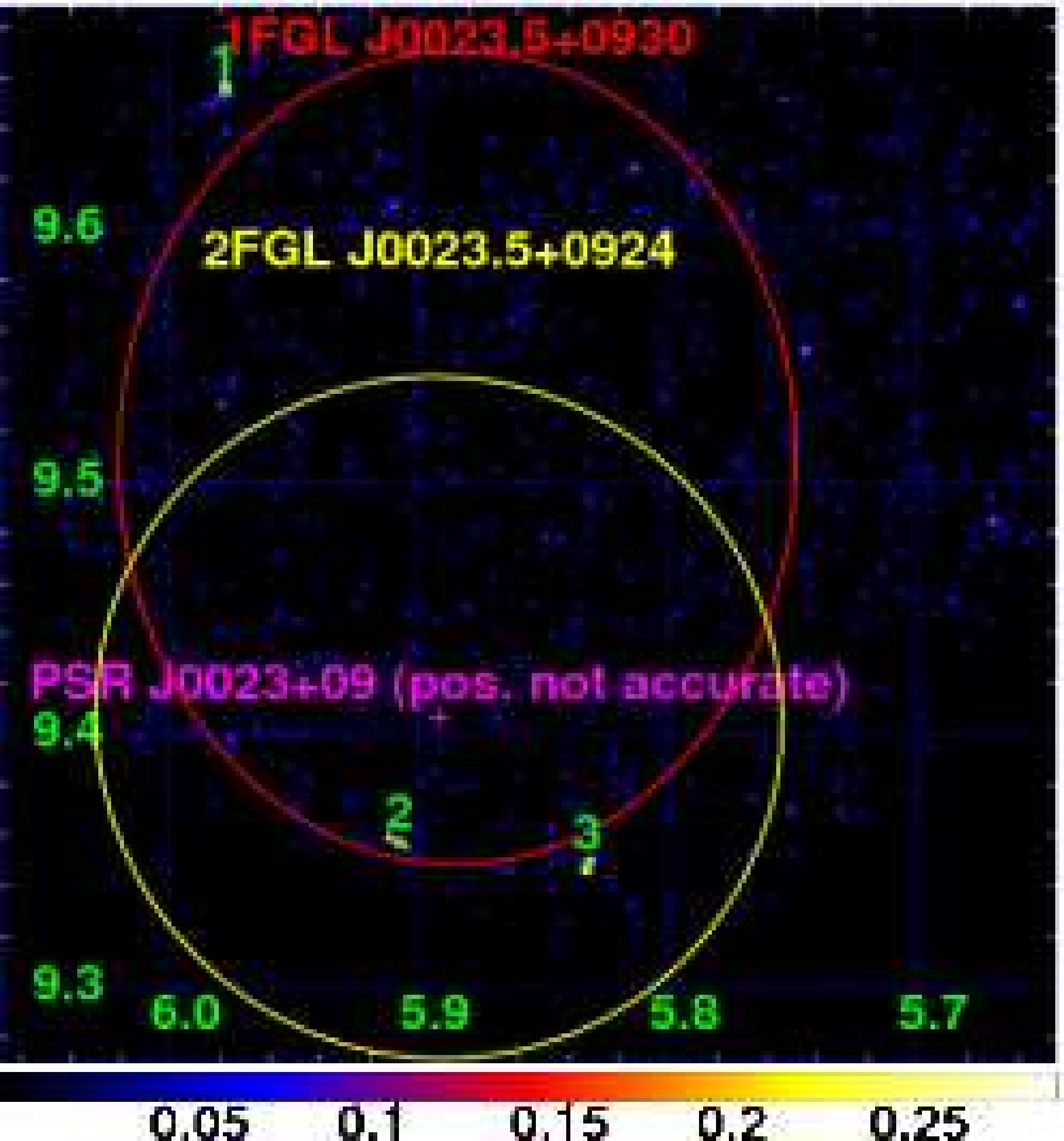}
    \end{center} 
  \end{minipage}
  \begin{minipage}{0.32\hsize}
    \begin{center}
      {\small (5) 1FGL\,J0030.7$+$0724} \\
      \includegraphics[width=52mm]{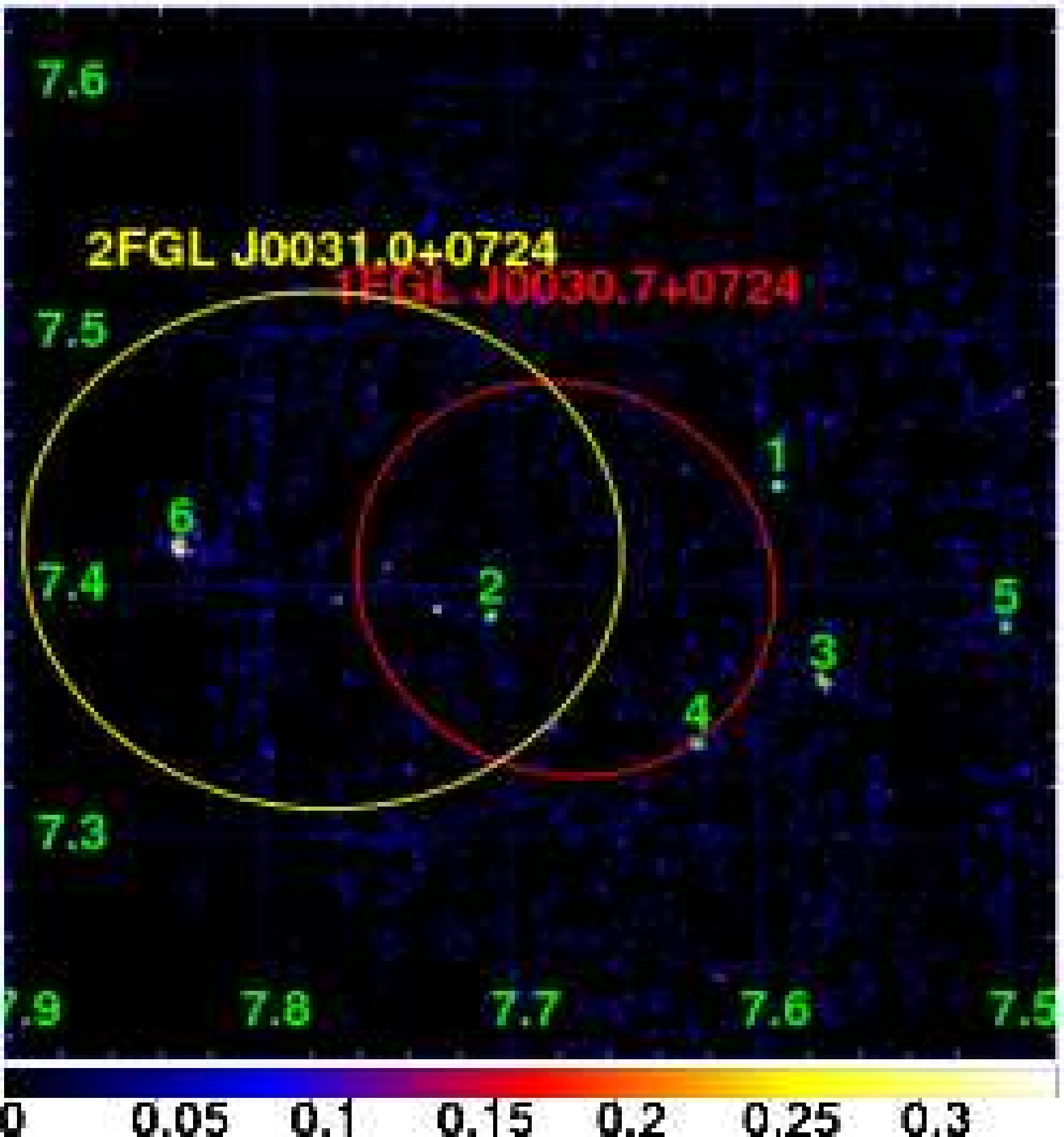}
    \end{center}
  \end{minipage}
  \begin{minipage}{0.32\hsize}
    \begin{center}
      {\small (6) 1FGL\,J0032.7--5519} \\
      \includegraphics[width=52mm]{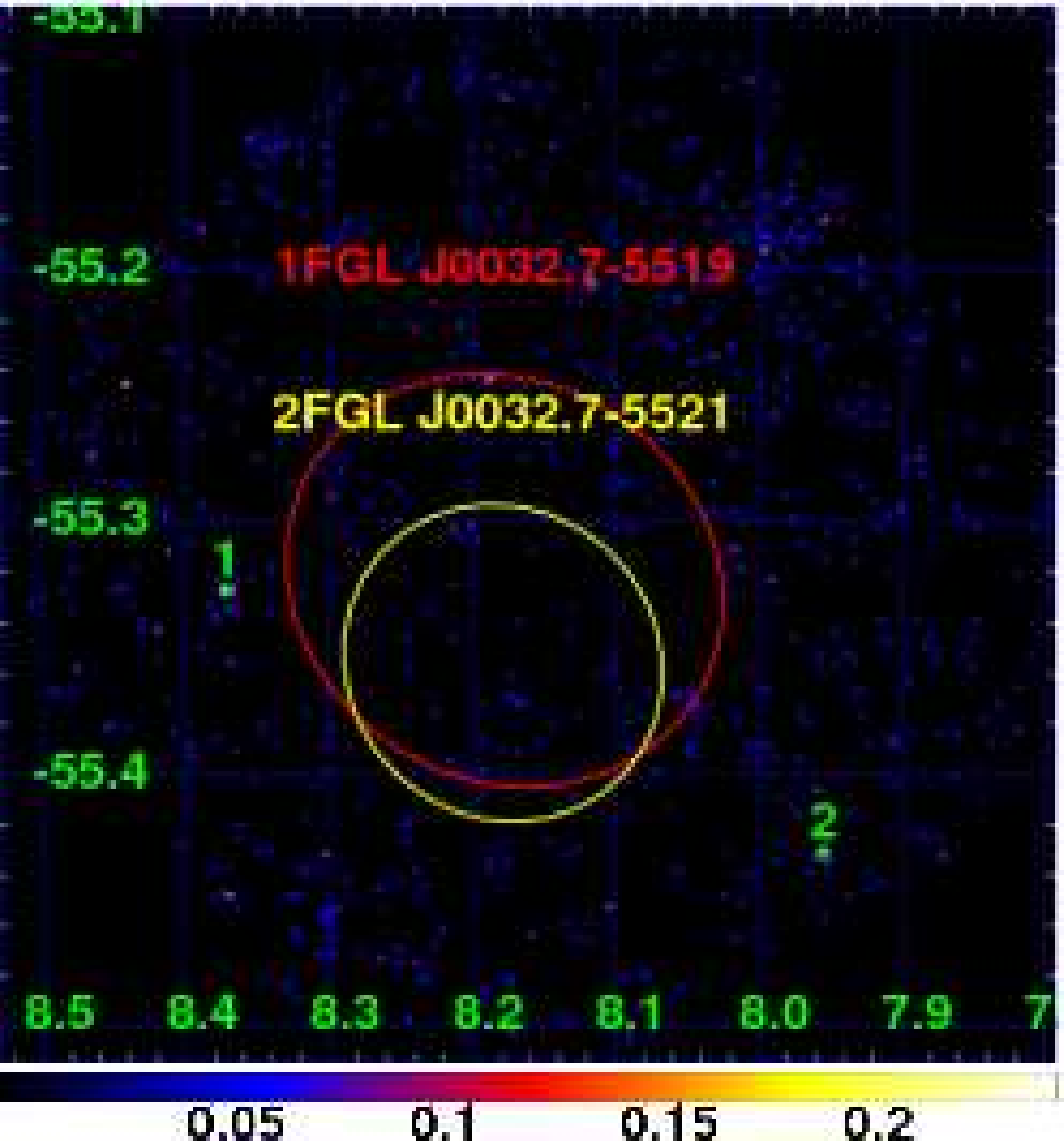}
    \end{center}
  \end{minipage}
  \begin{minipage}{0.32\hsize}
    \begin{center}
      {\small (7) 1FGL\,J0038.0$+$1236} \\
      \includegraphics[width=52mm]{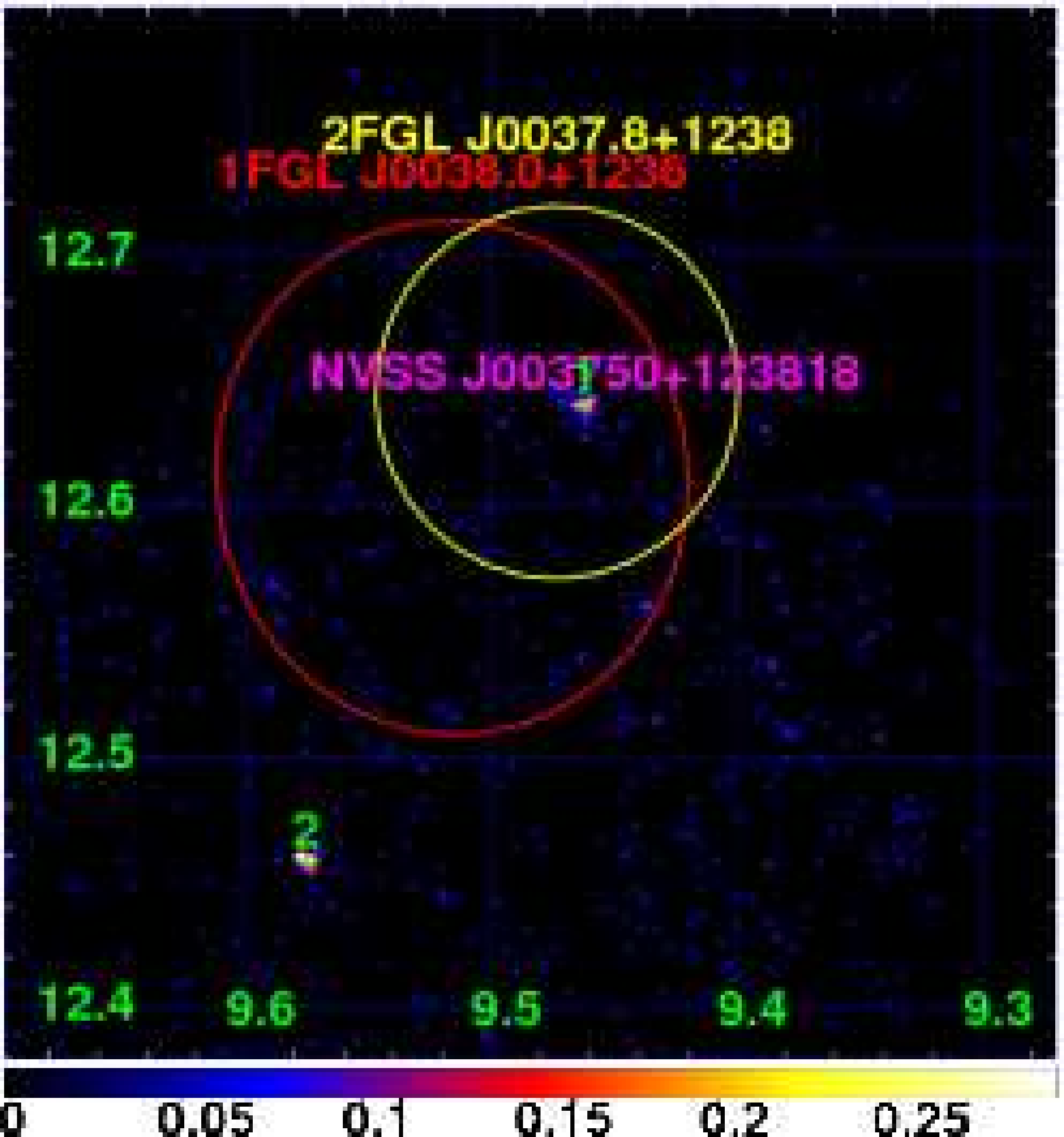}
    \end{center} 
  \end{minipage}
  \begin{minipage}{0.32\hsize}
    \begin{center}
      {\small (8) 1FGL\,J0039.2$+$4331} \\
      \includegraphics[width=52mm]{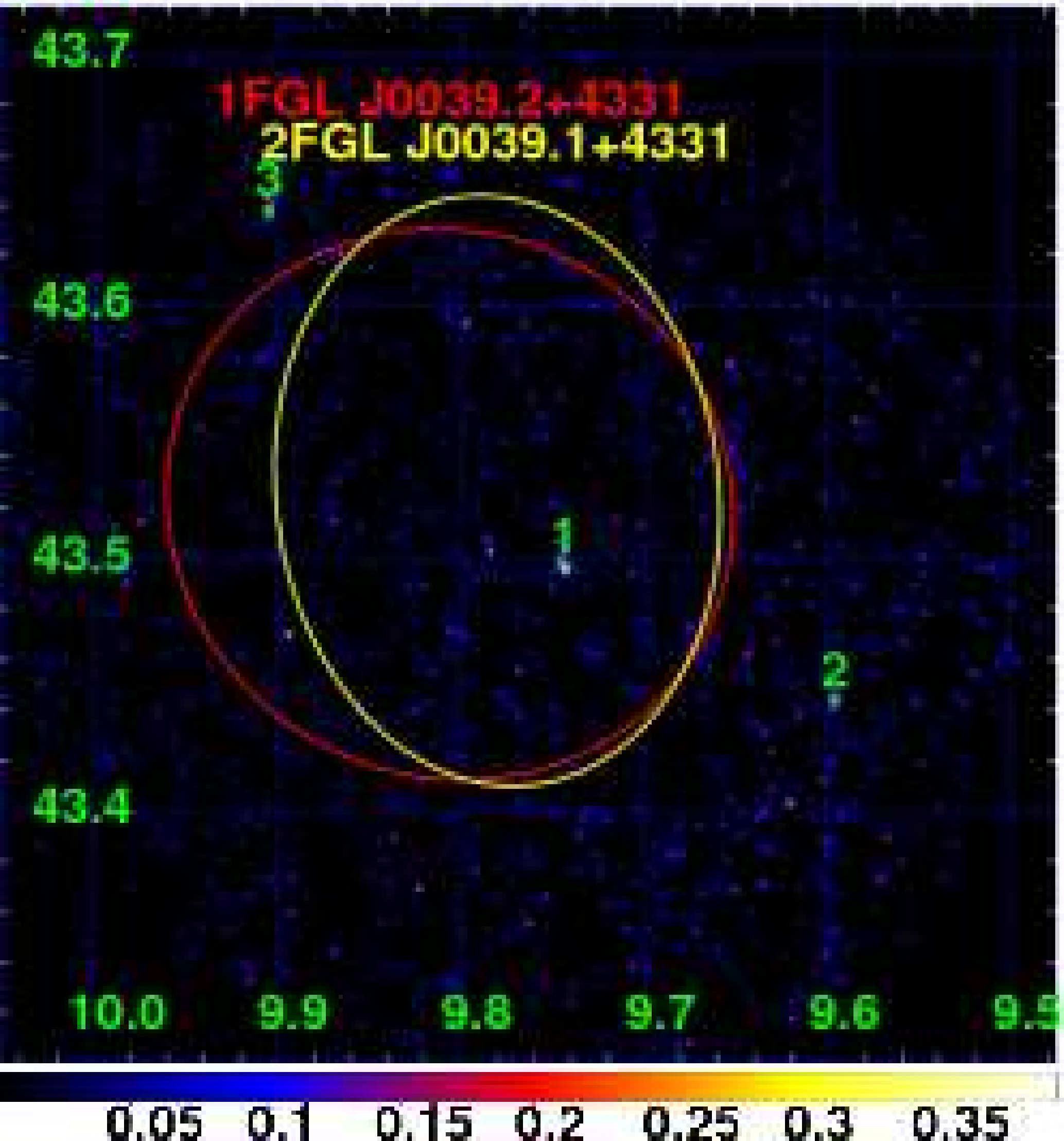}
    \end{center}
  \end{minipage}
  \begin{minipage}{0.32\hsize}
    \begin{center}
      {\small (9) 1FGL\,J0051.4--6242} \\
      \includegraphics[width=52mm]{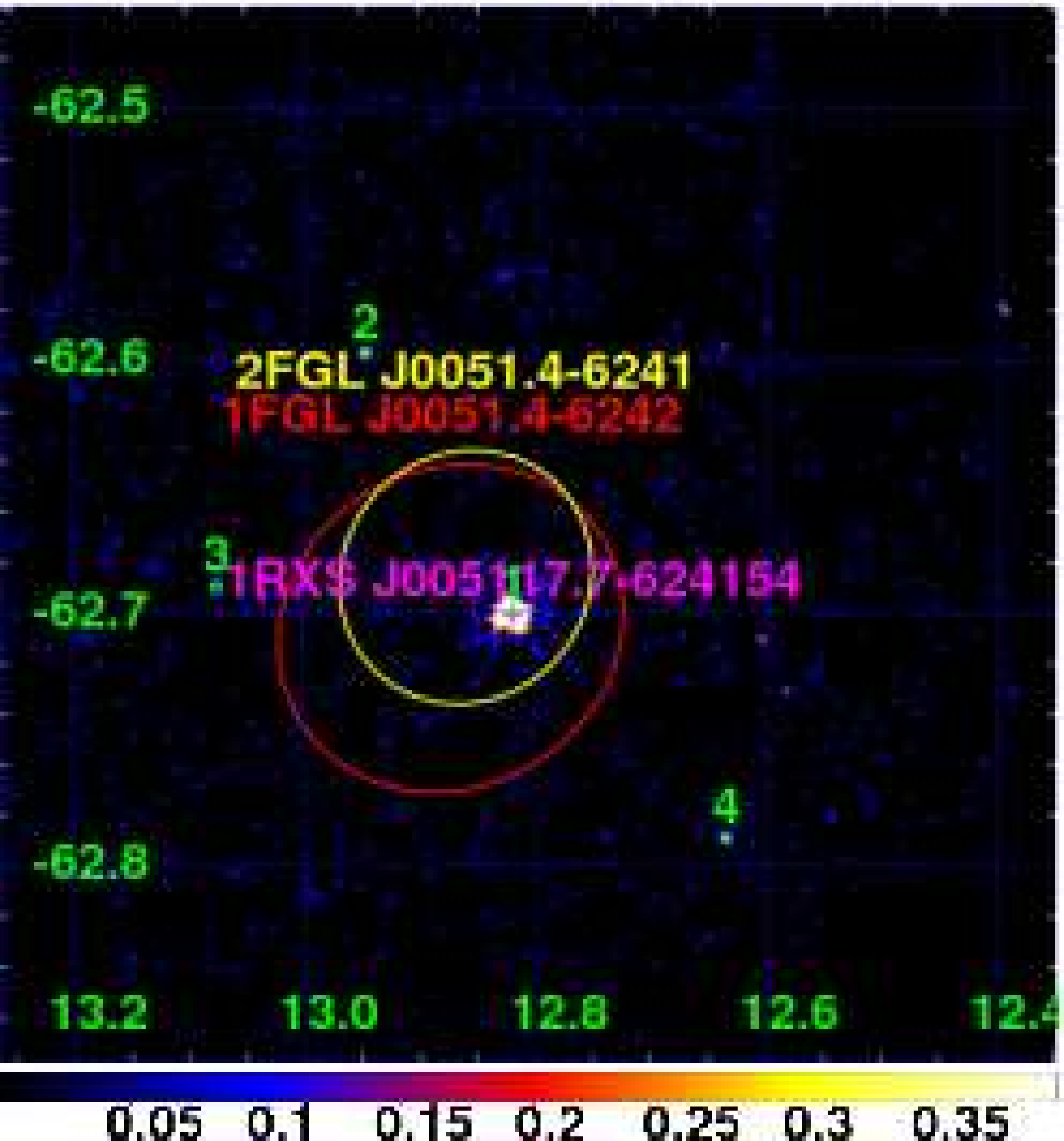}
    \end{center}
  \end{minipage}
  \begin{minipage}{0.32\hsize}
    \begin{center}
      {\small (10) 1FGL\,J0054.9--2455} \\
      \includegraphics[width=52mm]{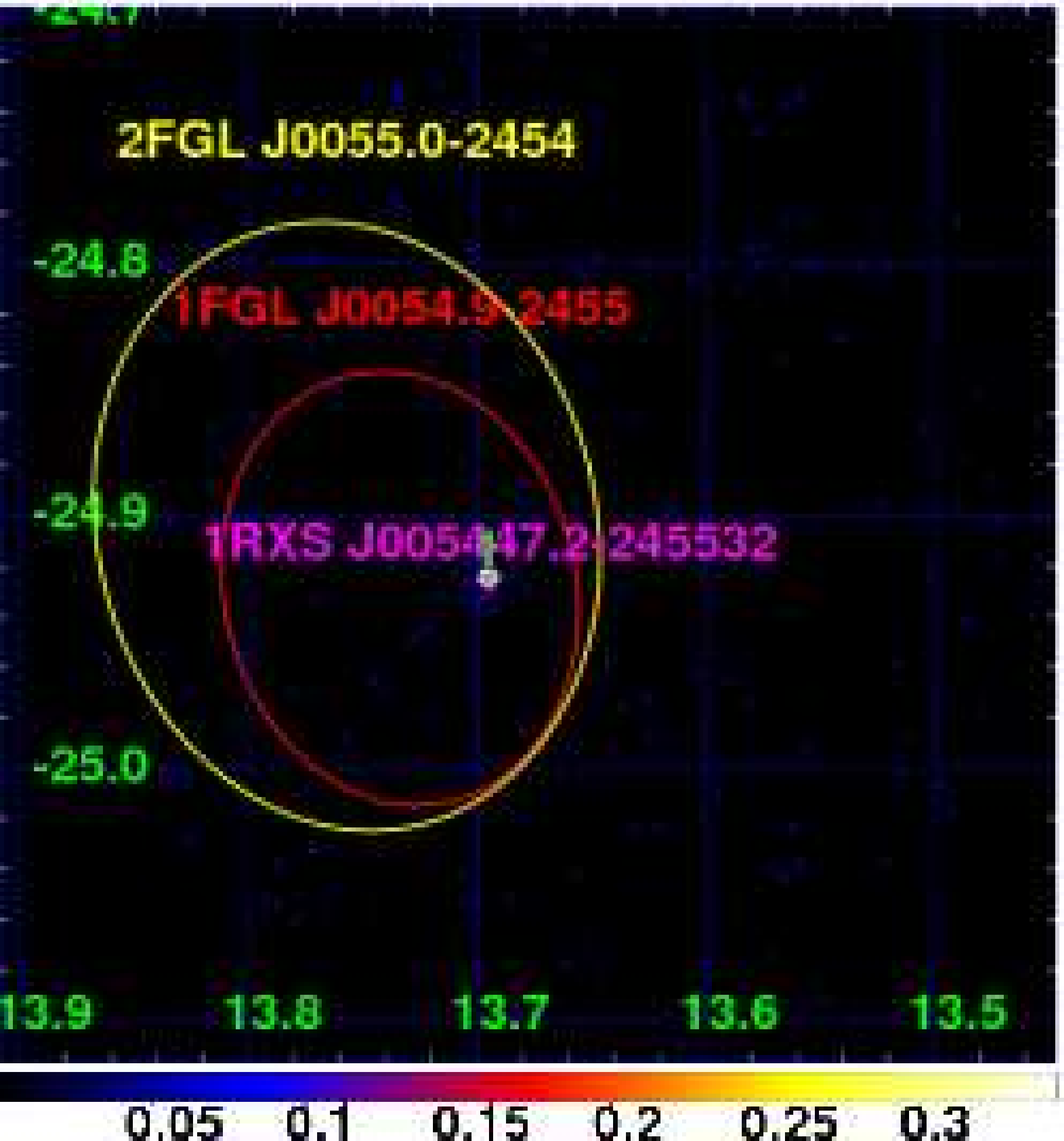}
    \end{center} 
  \end{minipage}
  \begin{minipage}{0.32\hsize}
    \begin{center}
      {\small (11) 1FGL\,J0101.0--6423} \\
      \includegraphics[width=52mm]{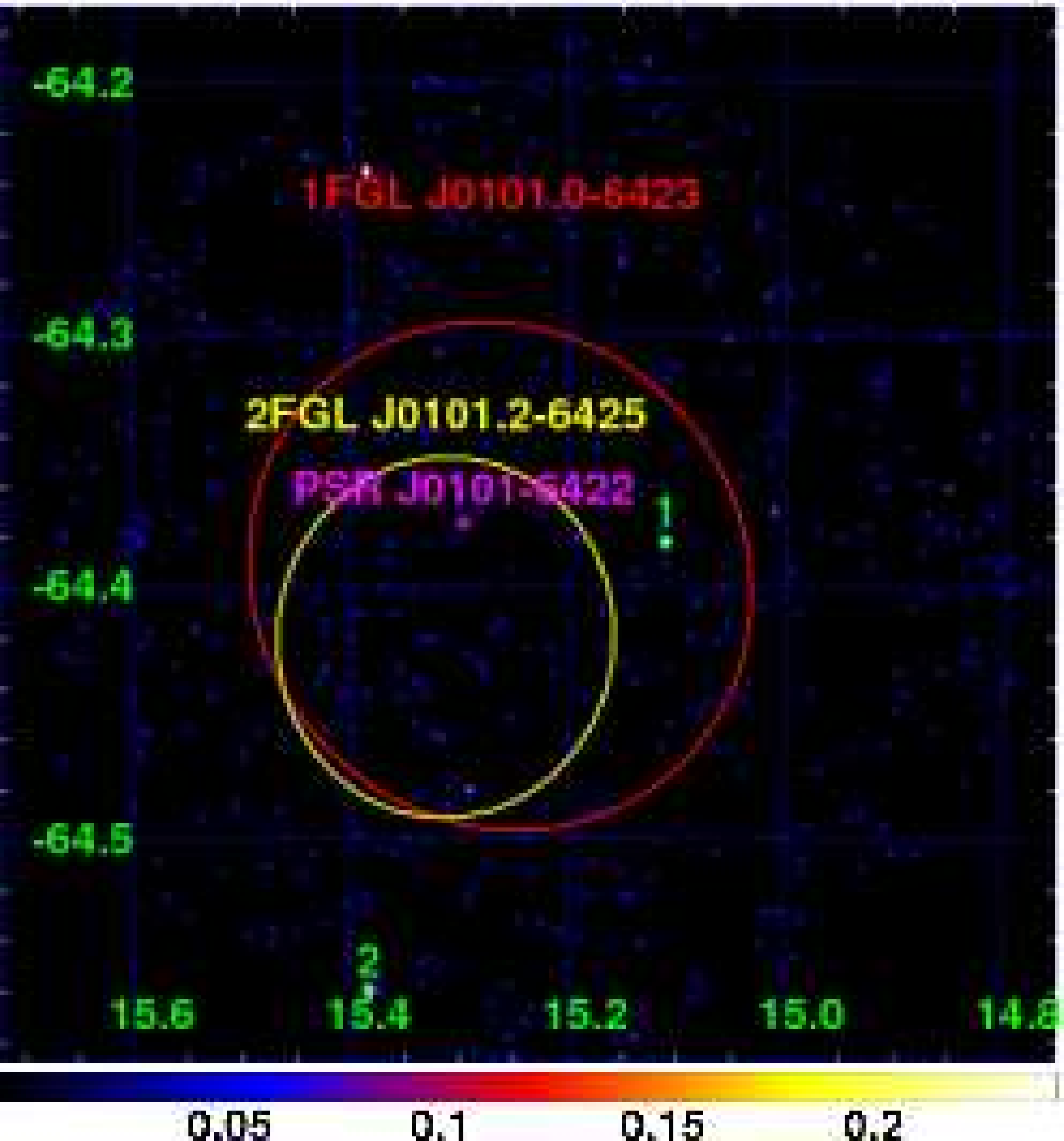}
    \end{center}
  \end{minipage}
  \begin{minipage}{0.32\hsize}
    \begin{center}
      {\small (12) 1FGL\,J0102.3$+$0942} \\
      \includegraphics[width=52mm]{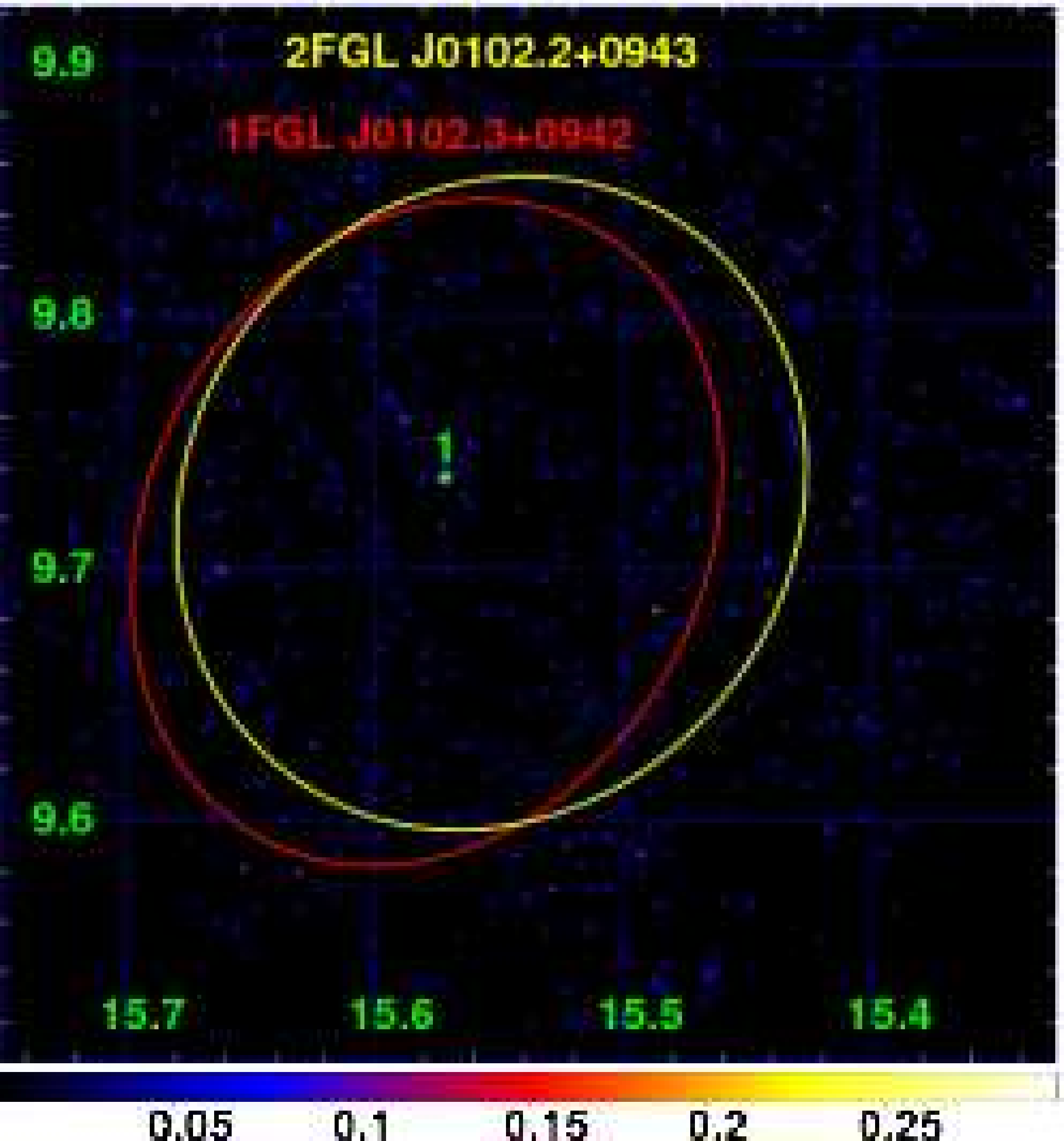}
    \end{center}  \end{minipage}
 \end{center}
\end{figure}
\clearpage
\begin{figure}[m]
 \begin{center}
  \begin{minipage}{0.32\hsize}
    \begin{center}
      {\small (13) 1FGL\,J0103.1$+$4840} \\
      \includegraphics[width=52mm]{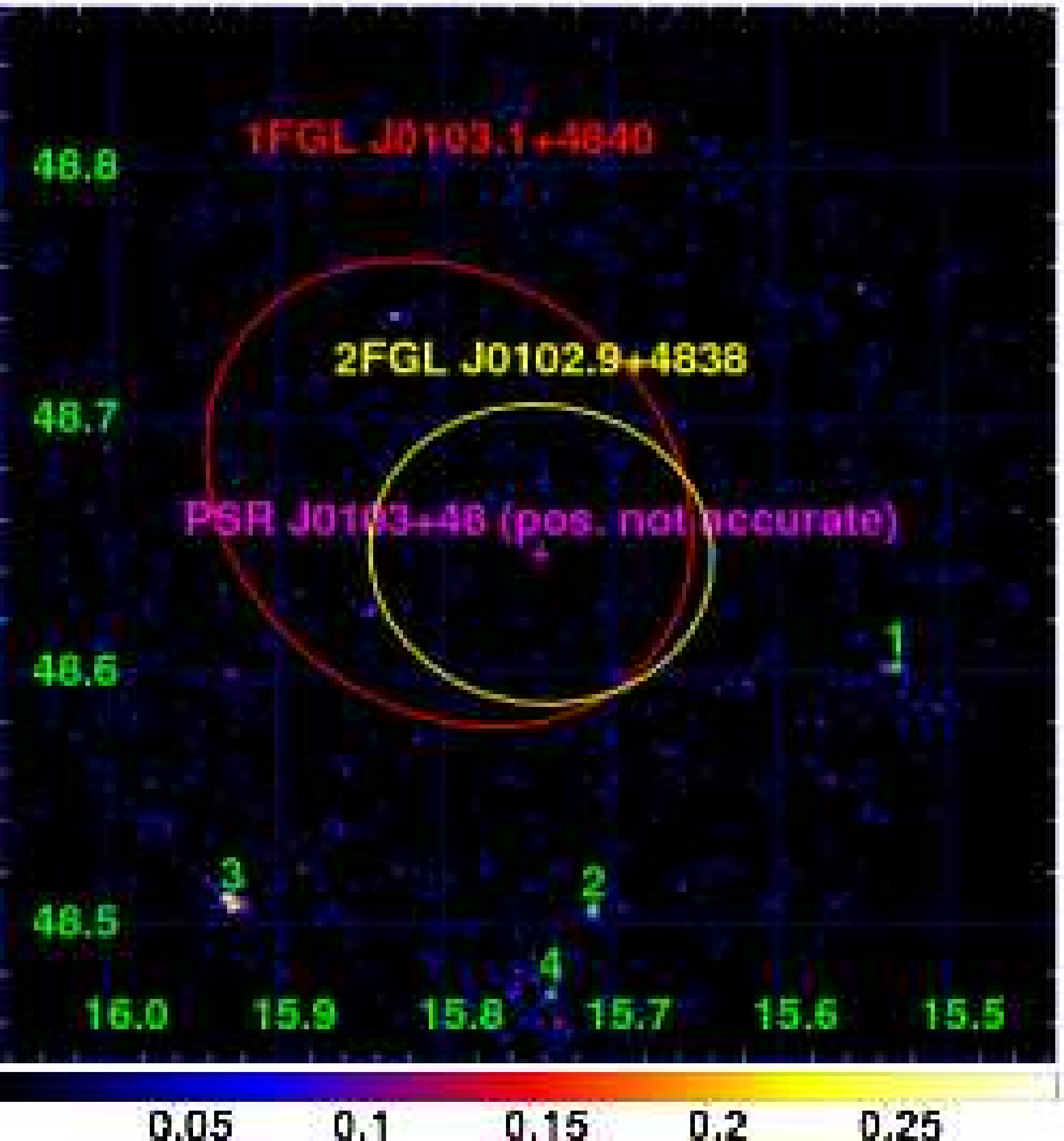}
    \end{center} 
  \end{minipage}
  \begin{minipage}{0.32\hsize}
    \begin{center}
      {\small (14) 1FGL\,J0106.7$+$4853} \\
      \includegraphics[width=52mm]{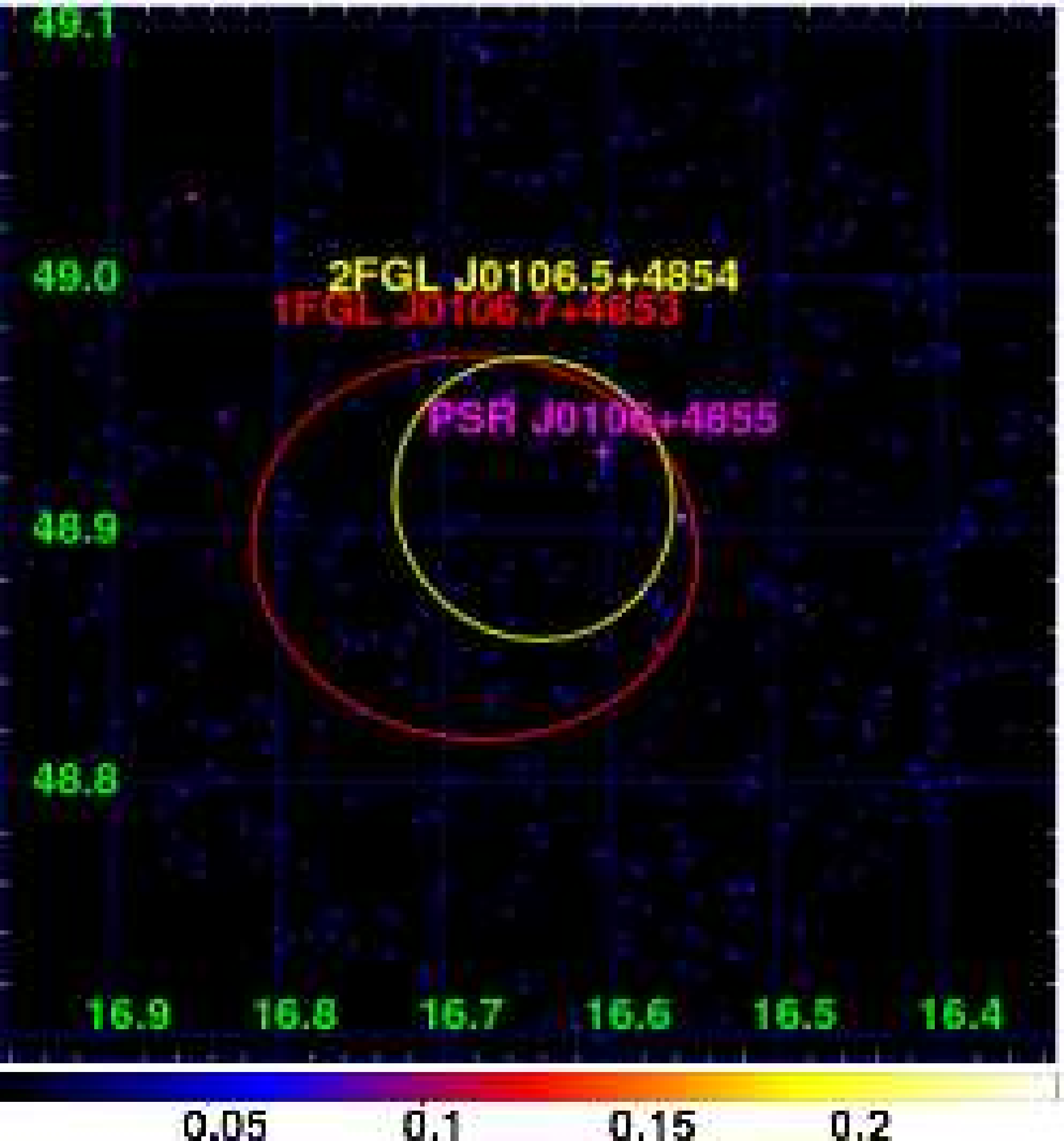}
    \end{center} 
  \end{minipage}
  \begin{minipage}{0.32\hsize}
    \begin{center}
      {\small (15) 1FGL\,J0115.7$+$0357} \\
      \includegraphics[width=52mm]{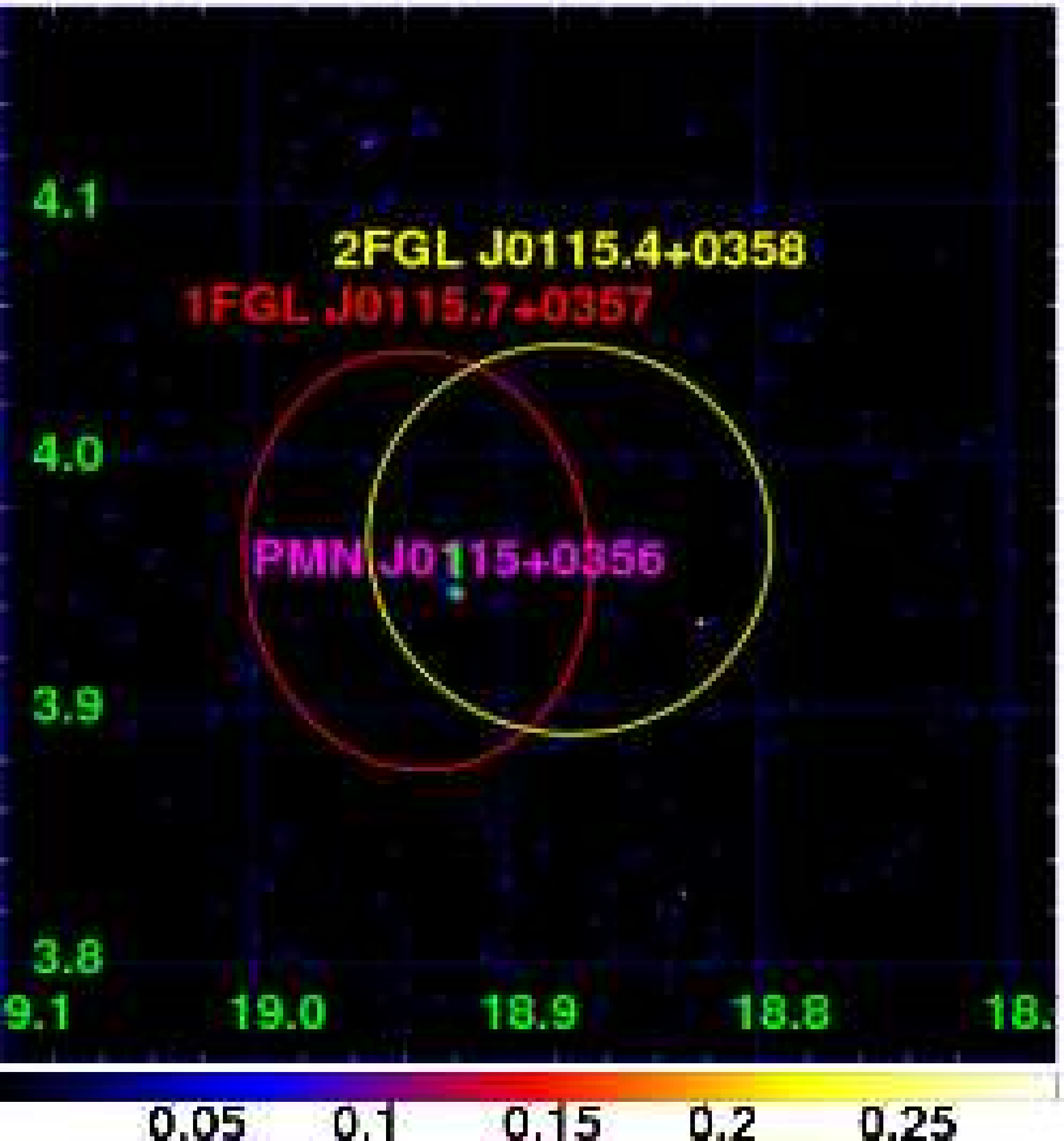}
    \end{center}
  \end{minipage}
  \begin{minipage}{0.32\hsize}
    \begin{center}
      {\small (16) 1FGL\,J0124.6--0616} \\
      \includegraphics[width=52mm]{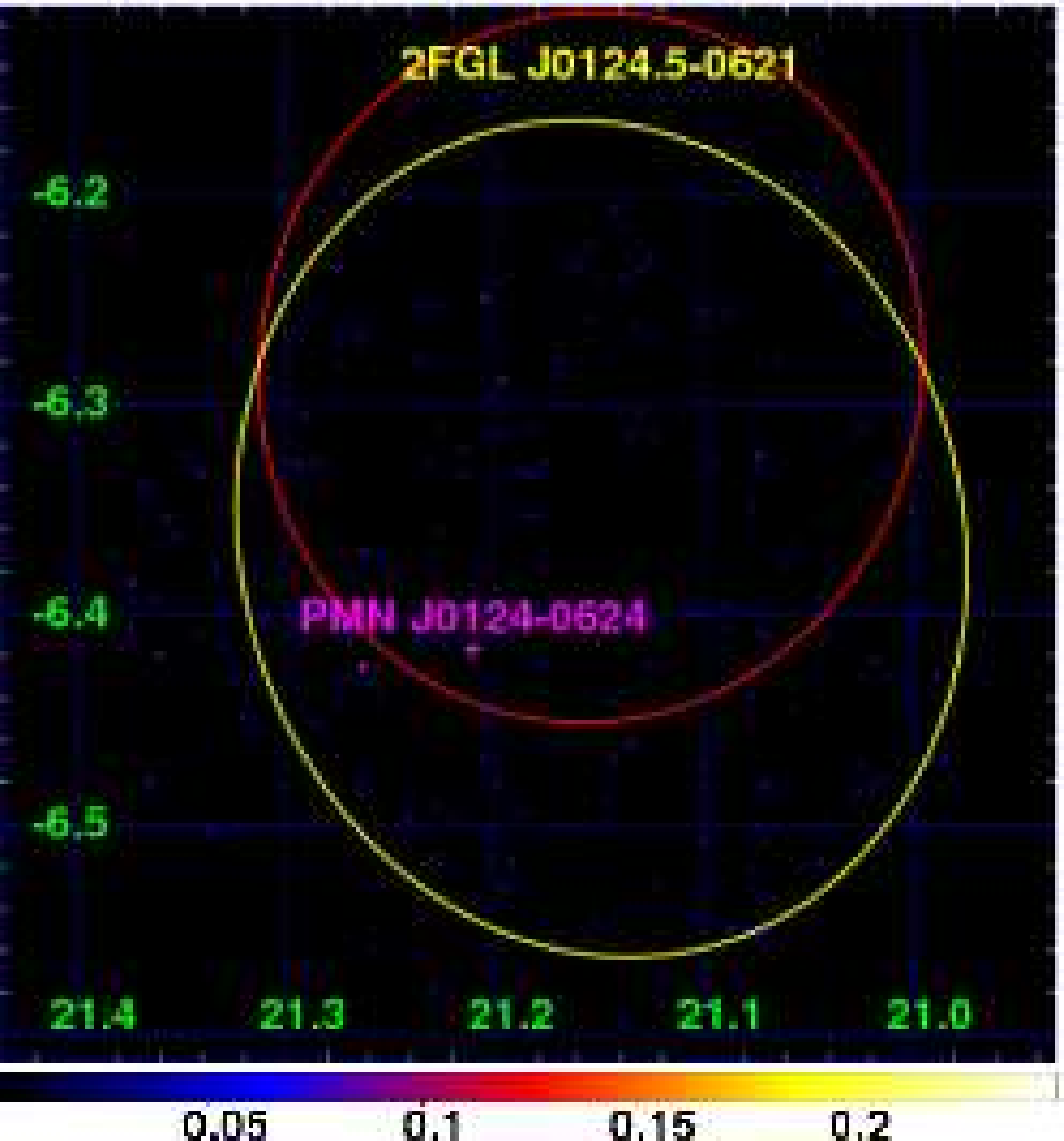}
    \end{center}
  \end{minipage}
  \begin{minipage}{0.32\hsize}
    \begin{center}
      {\small (17) 1FGL\,J0143.9--5845} \\
      \includegraphics[width=52mm]{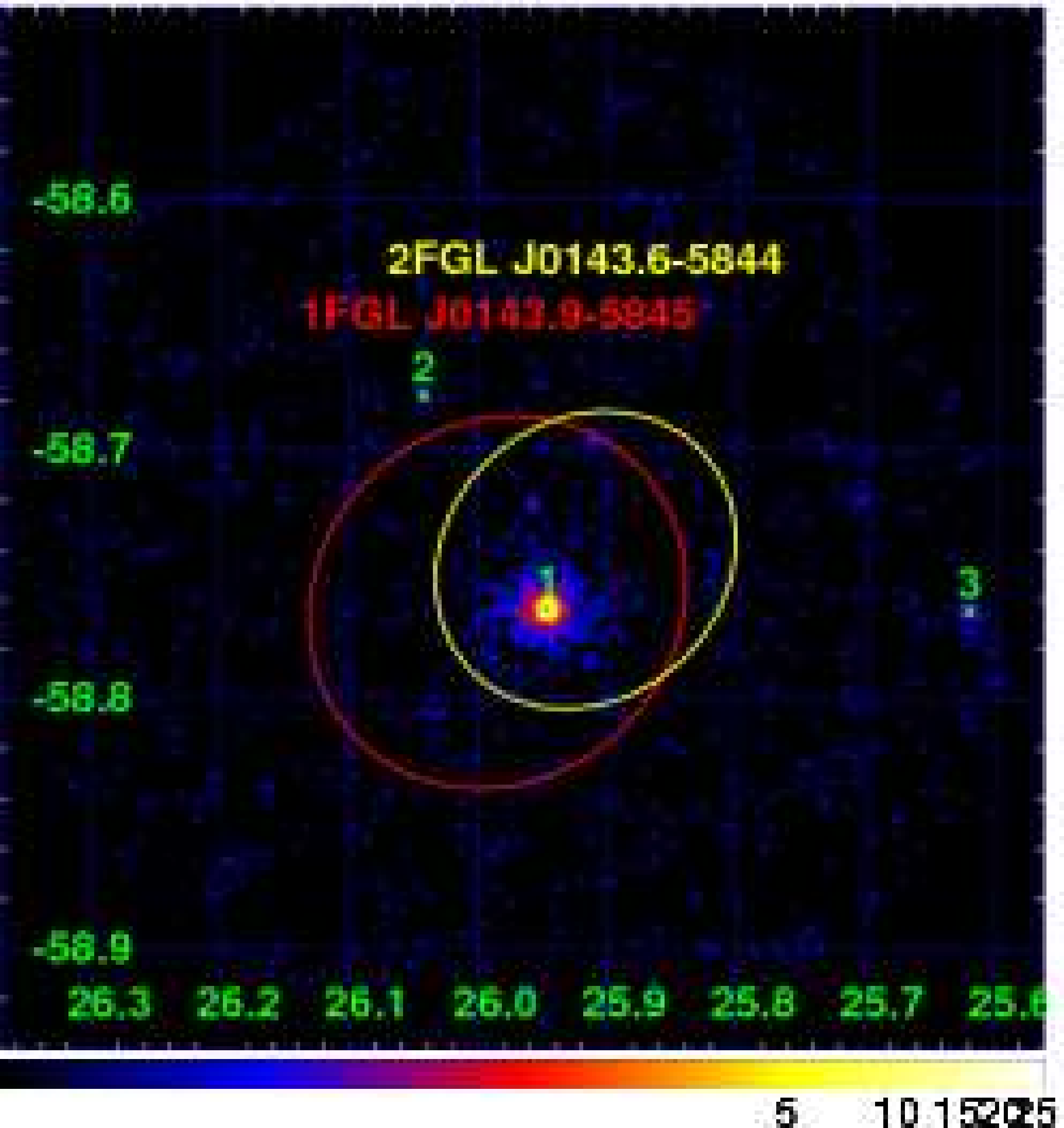}
    \end{center} 
  \end{minipage}
  \begin{minipage}{0.32\hsize}
    \begin{center}
      {\small (18) 1FGL\,J0157.0--5259} \\
      \includegraphics[width=52mm]{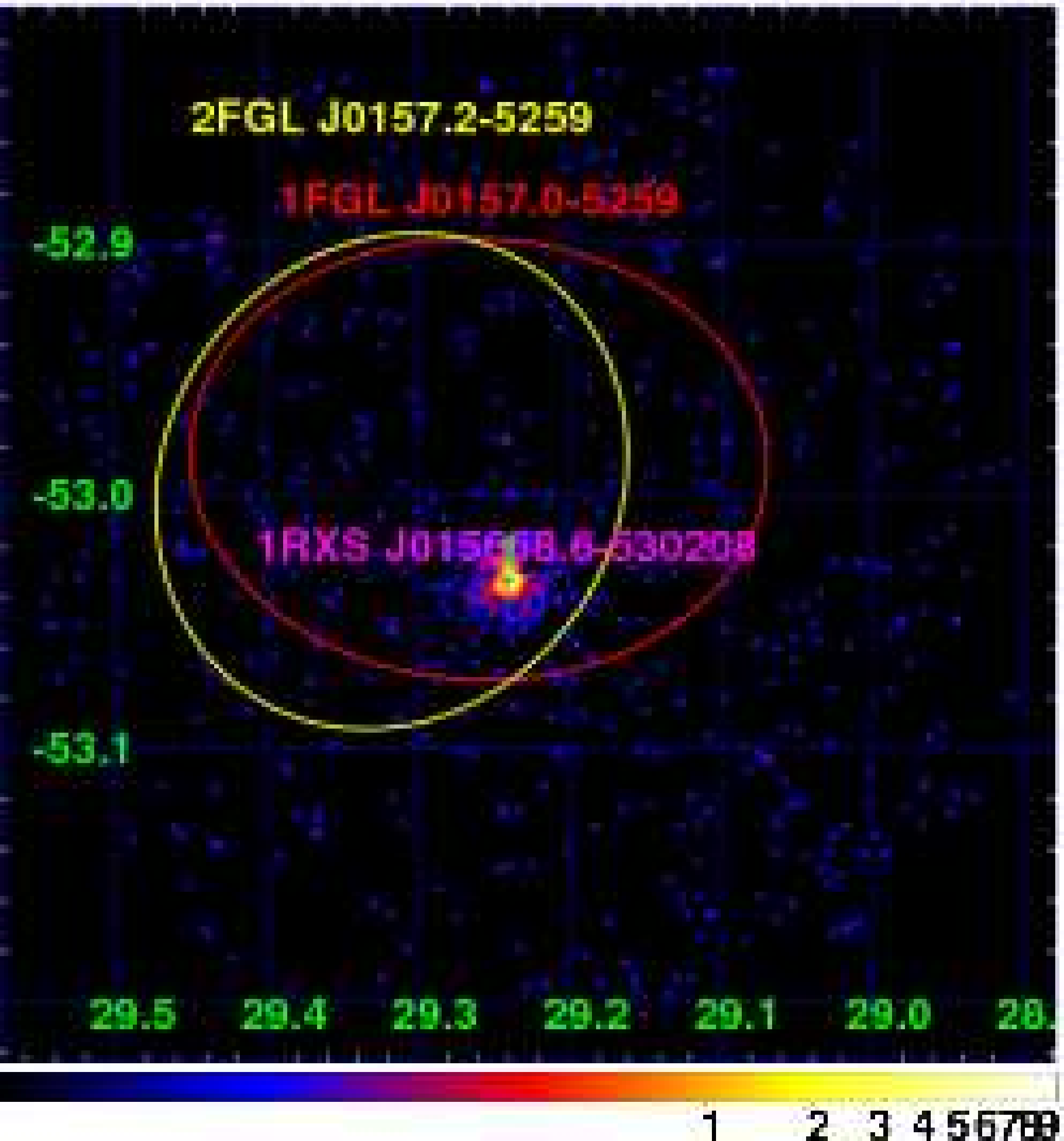}
    \end{center}
  \end{minipage}
  \begin{minipage}{0.32\hsize}
    \begin{center}
      {\small (19) 1FGL\,J0217.9--6630} \\
      \includegraphics[width=52mm]{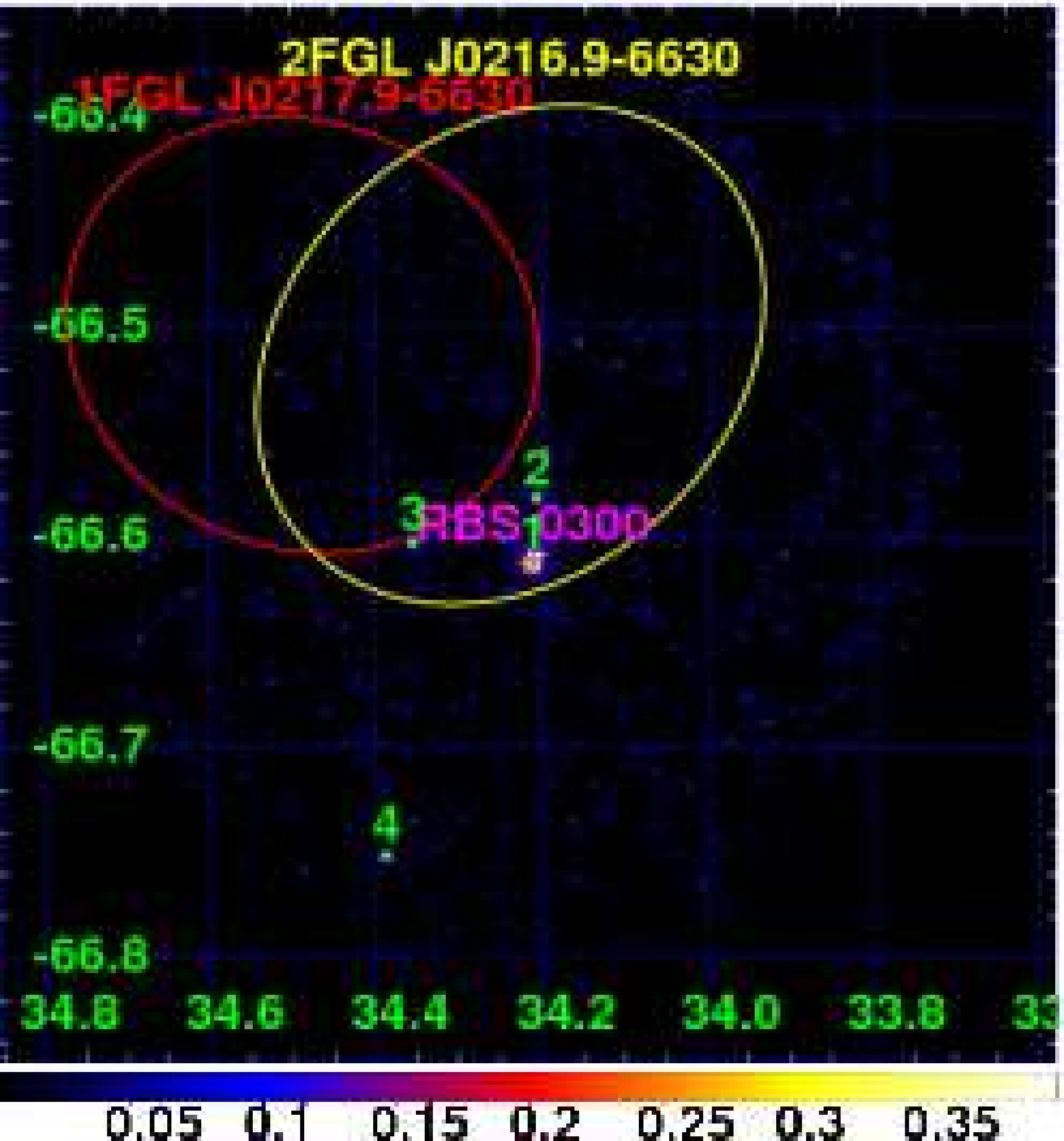}
    \end{center}
  \end{minipage}
  \begin{minipage}{0.32\hsize}
    \begin{center}
      {\small (20) 1FGL\,J0223.0--1118} \\
      \includegraphics[width=52mm]{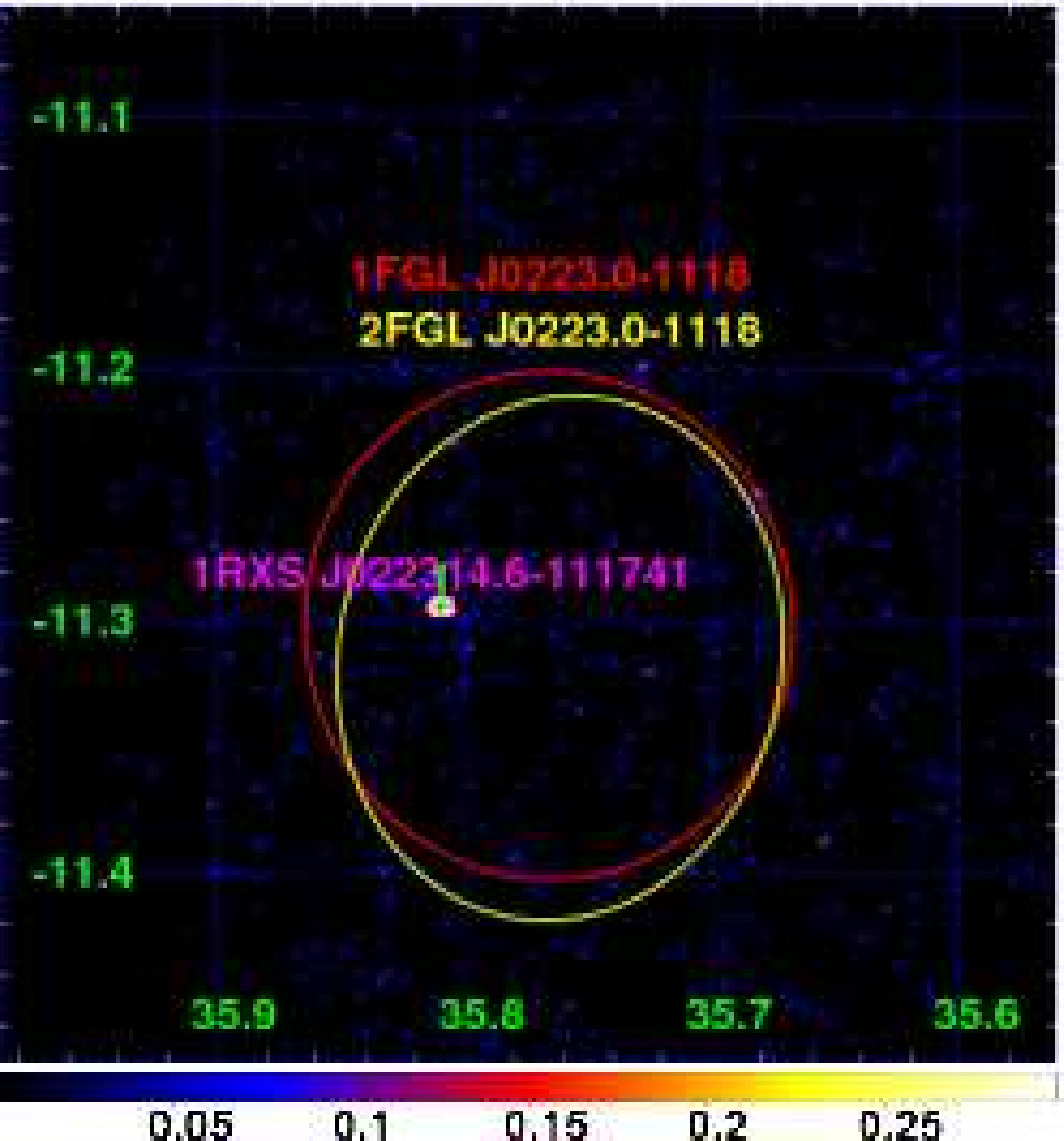}
    \end{center} 
  \end{minipage}
  \begin{minipage}{0.32\hsize}
    \begin{center}
      {\small (21) 1FGL\,J0226.3$+$0937} \\
      \includegraphics[width=52mm]{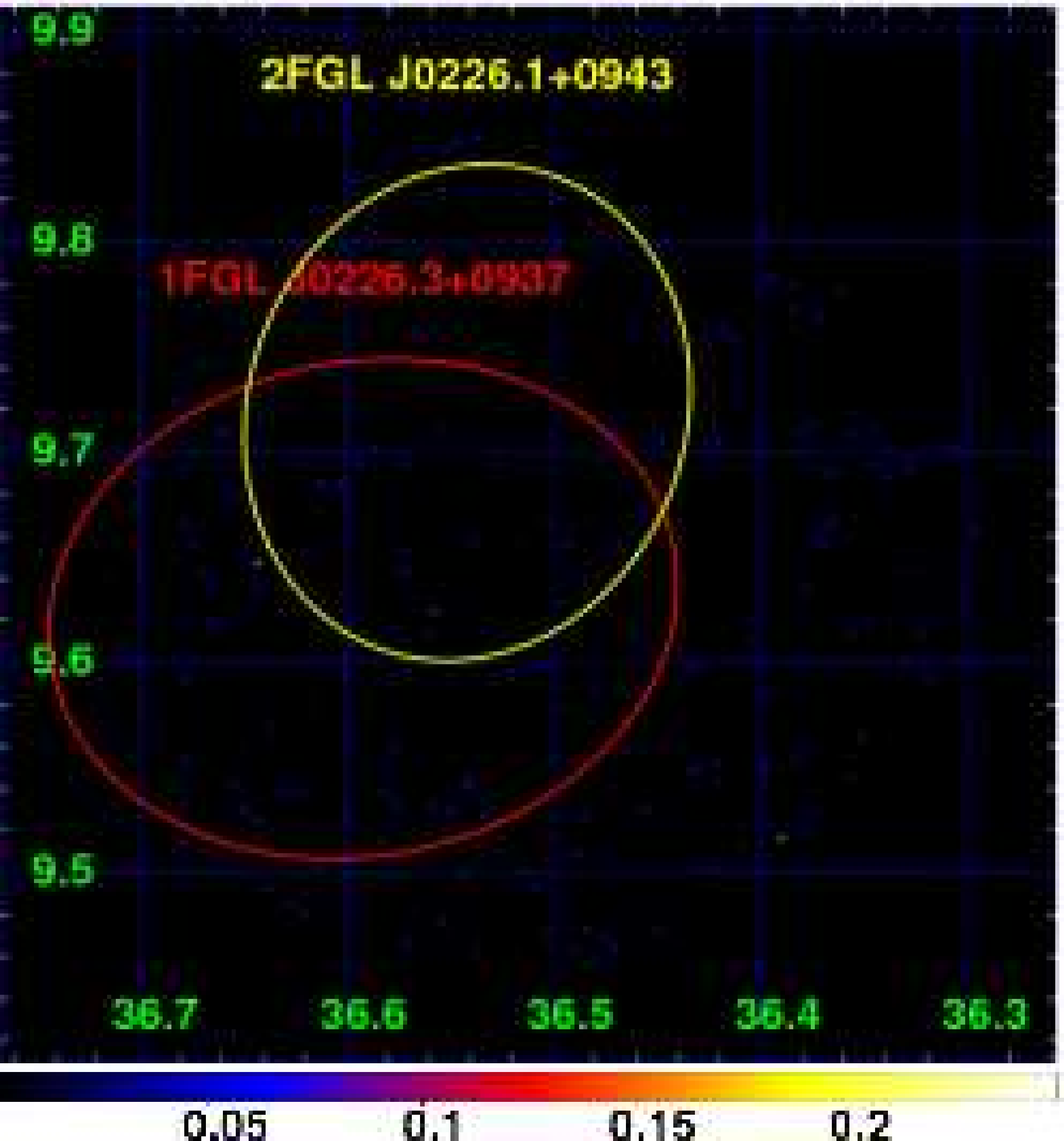}
    \end{center}
  \end{minipage}
  \begin{minipage}{0.32\hsize}
    \begin{center}
      {\small (22) 1FGL\,J0239.5$+$1324} \\
      \includegraphics[width=52mm]{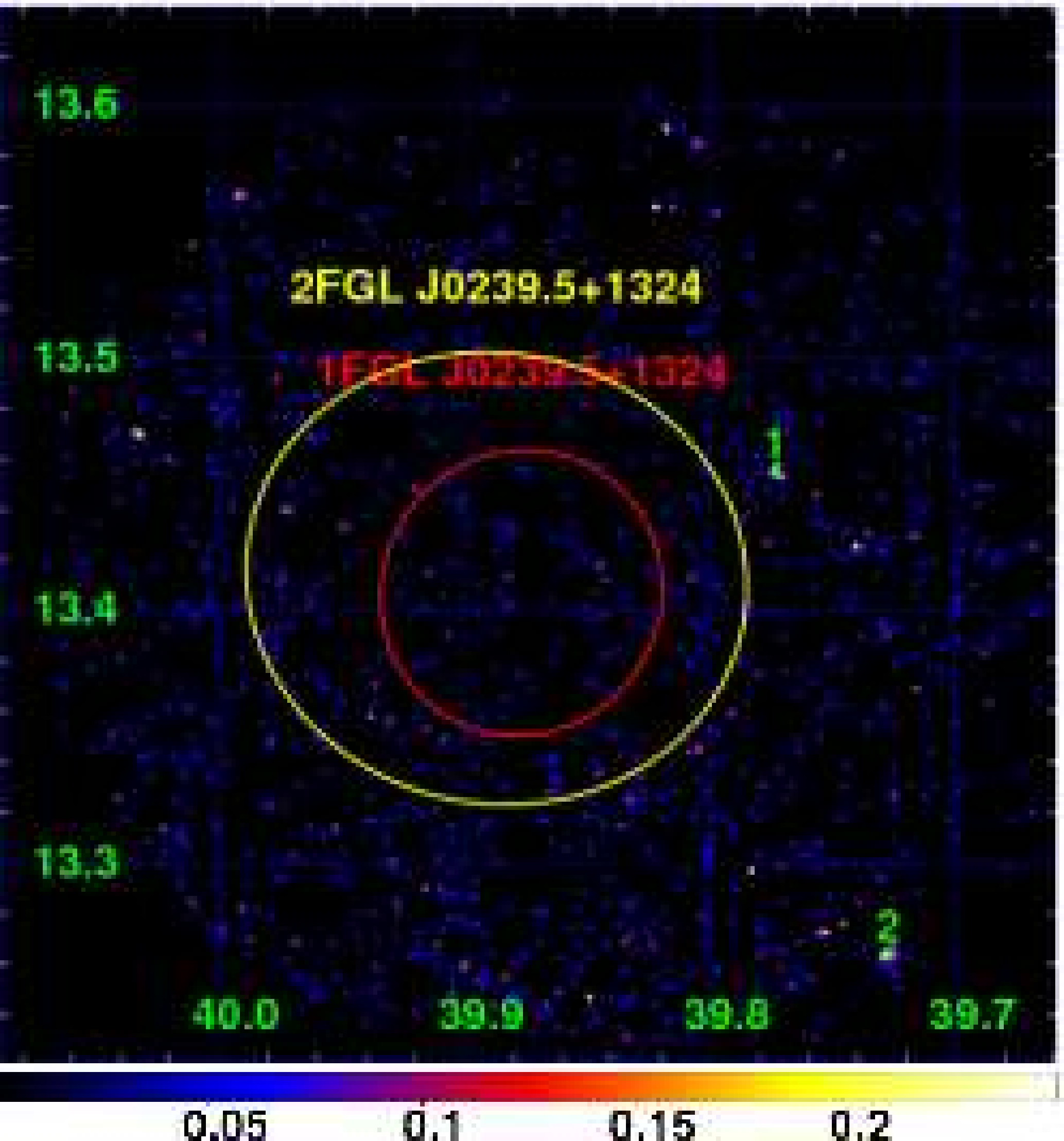}
    \end{center}
  \end{minipage}
  \begin{minipage}{0.32\hsize}
    \begin{center}
      {\small (23) 1FGL\,J0305.2--1601} \\
      \includegraphics[width=52mm]{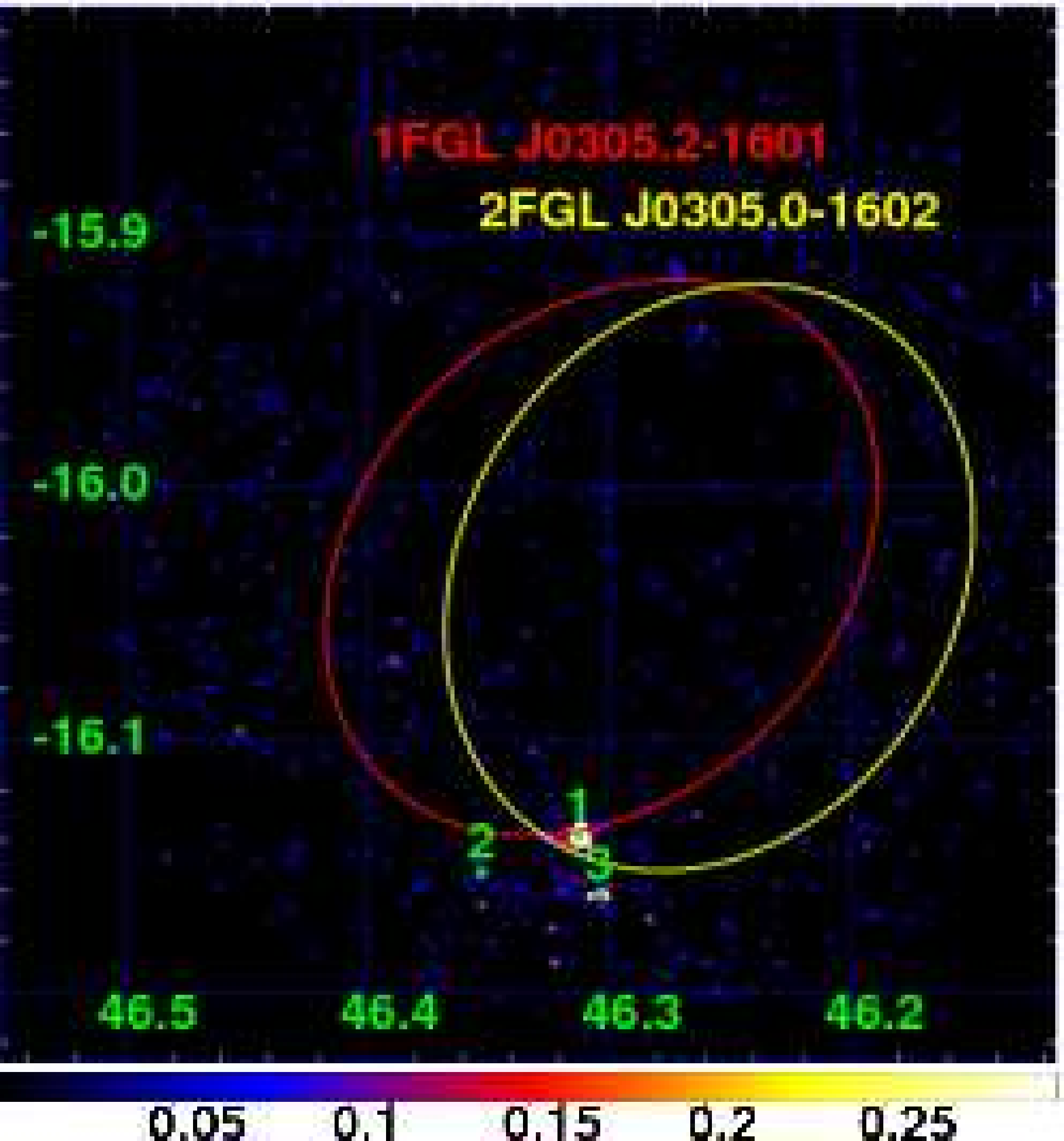}
    \end{center} 
  \end{minipage}
  \begin{minipage}{0.32\hsize}
    \begin{center}
      {\small (24) 1FGL\,J0308.6+7442} \\
      \includegraphics[width=52mm]{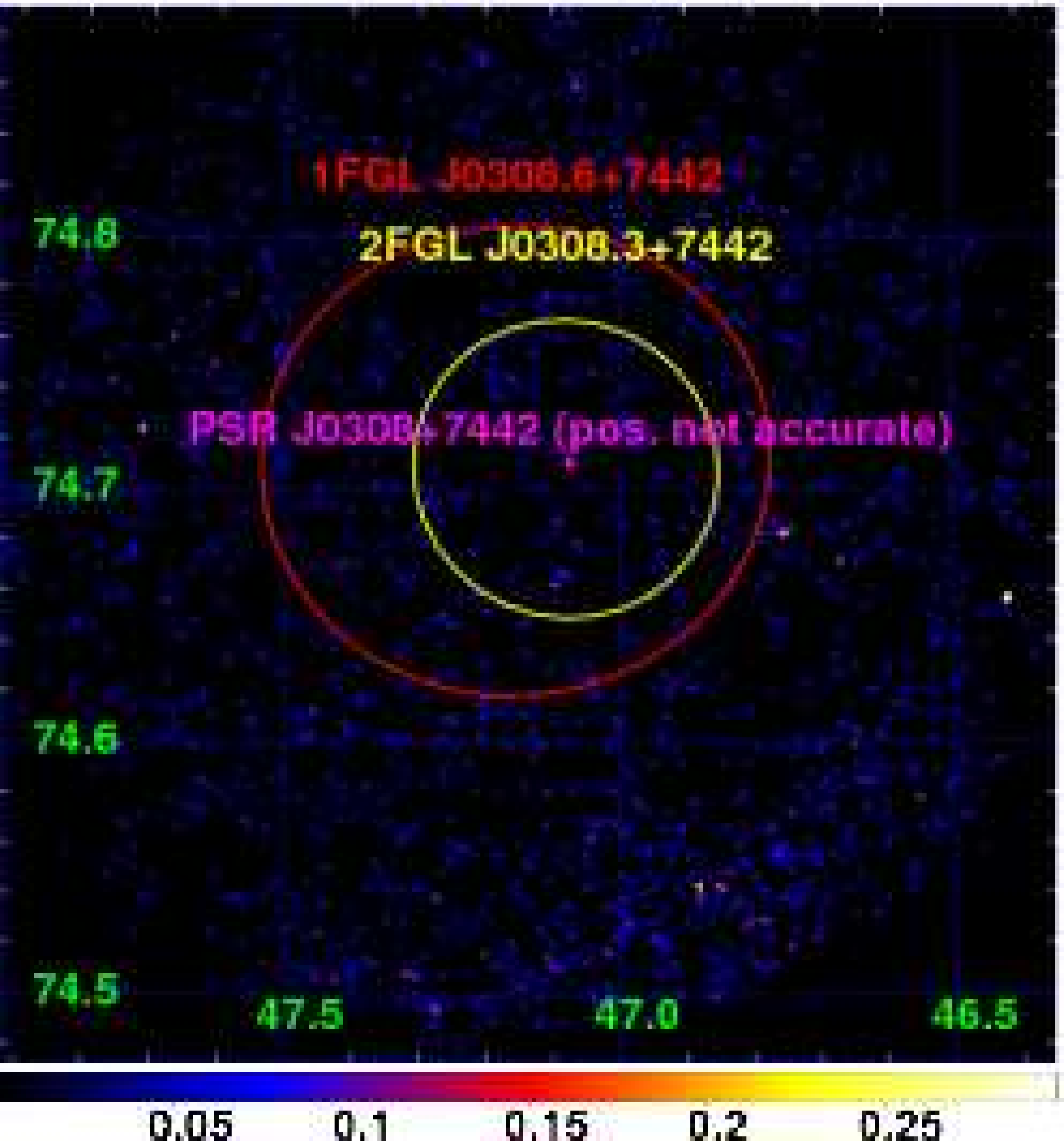}
    \end{center}
  \end{minipage}
 \end{center}
\end{figure}
\clearpage
\begin{figure}[m]
 \begin{center}
  \begin{minipage}{0.32\hsize}
    \begin{center}
      {\small (25) 1FGL\,J0316.3--6438} \\
      \includegraphics[width=52mm]{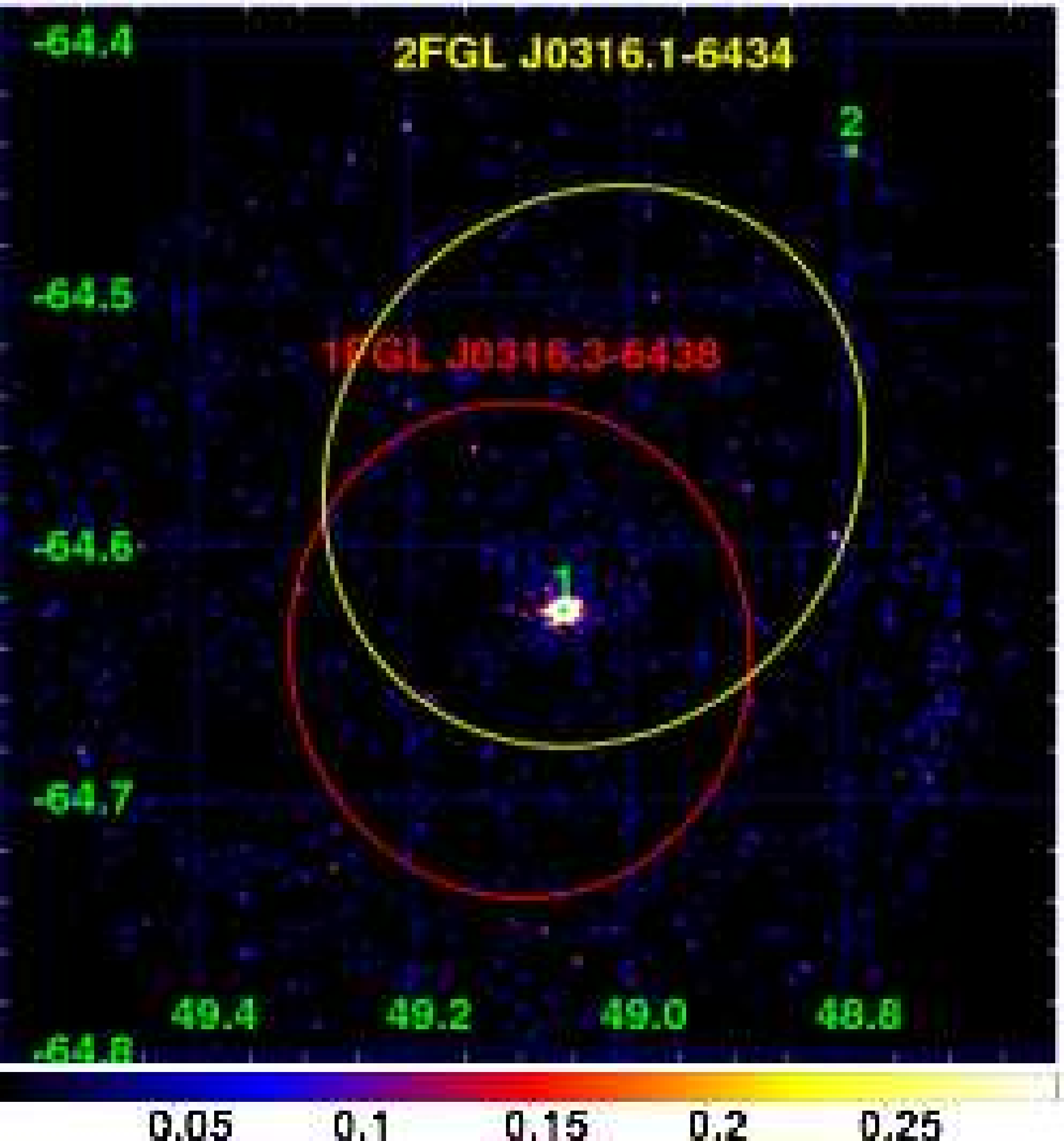}
    \end{center}
  \end{minipage}
  \begin{minipage}{0.32\hsize}
    \begin{center}
      {\small (26) 1FGL\,J0335.5--4501} \\
      \includegraphics[width=52mm]{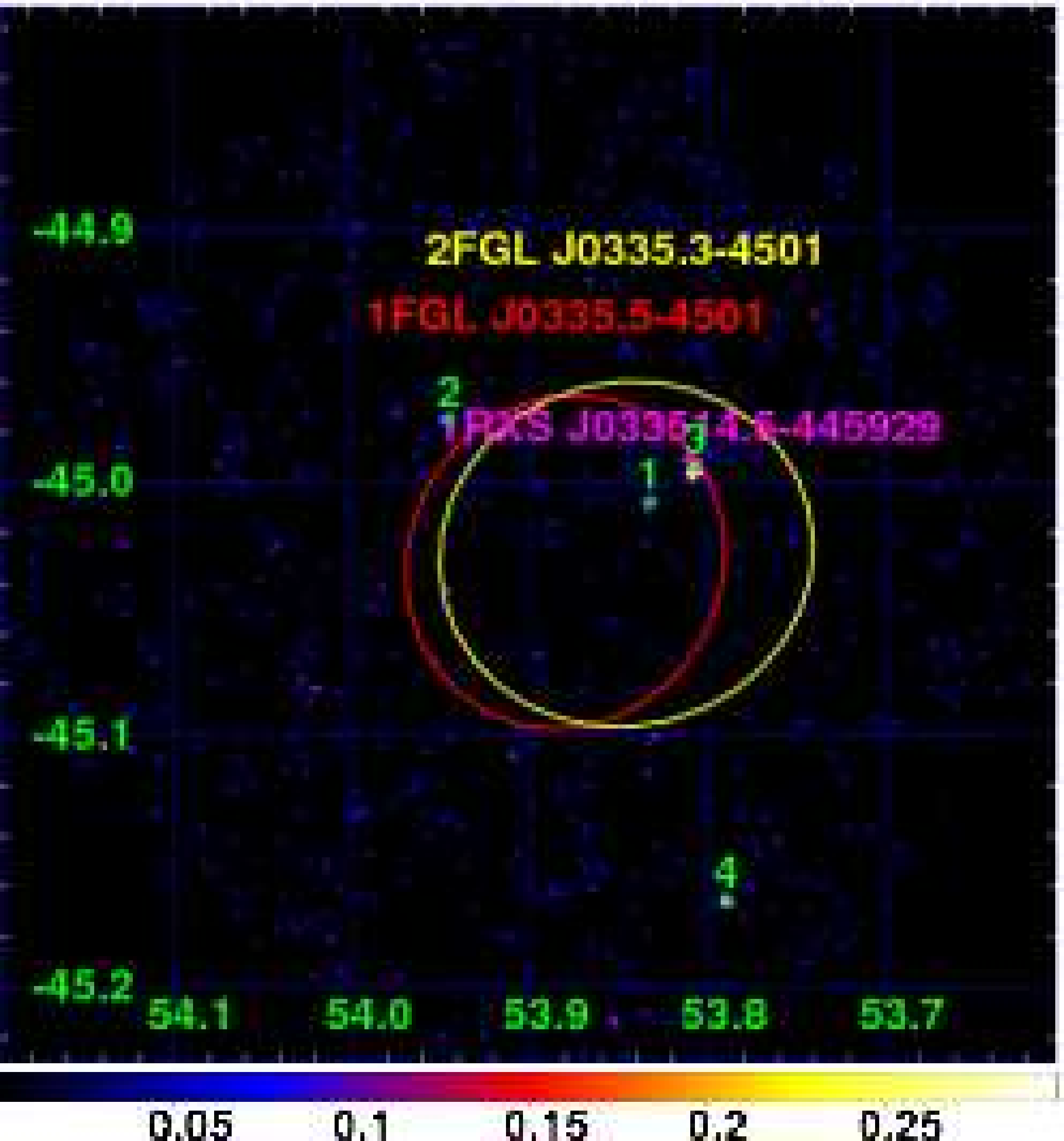}
    \end{center} 
  \end{minipage}
  \begin{minipage}{0.32\hsize}
    \begin{center}
      {\small (27) 1FGL\,J0340.4$+$4130} \\
      \includegraphics[width=52mm]{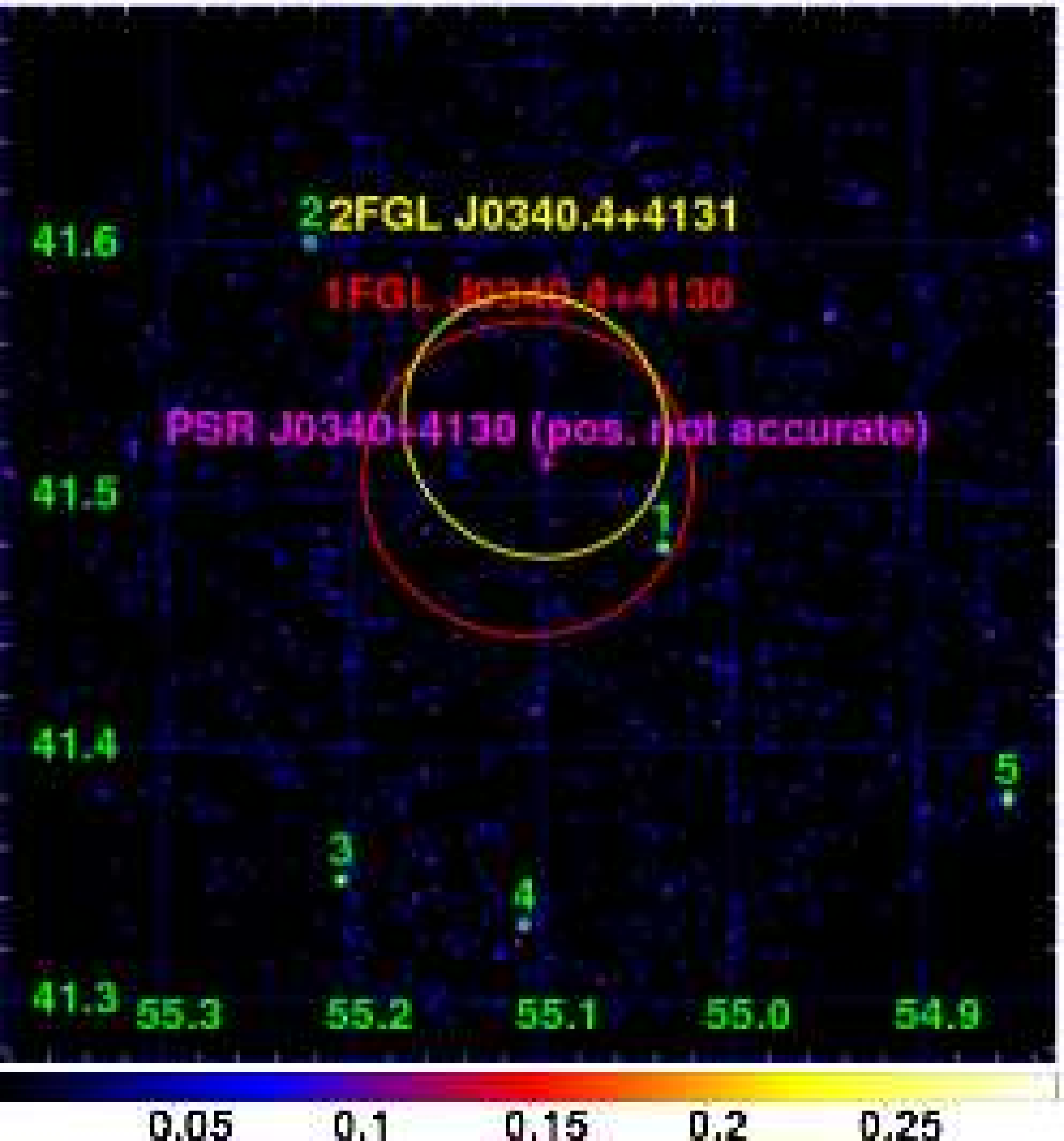}
    \end{center}
  \end{minipage}
  \begin{minipage}{0.32\hsize}
    \begin{center}
      {\small (28) 1FGL\,J0345.2--2355} \\
      \includegraphics[width=52mm]{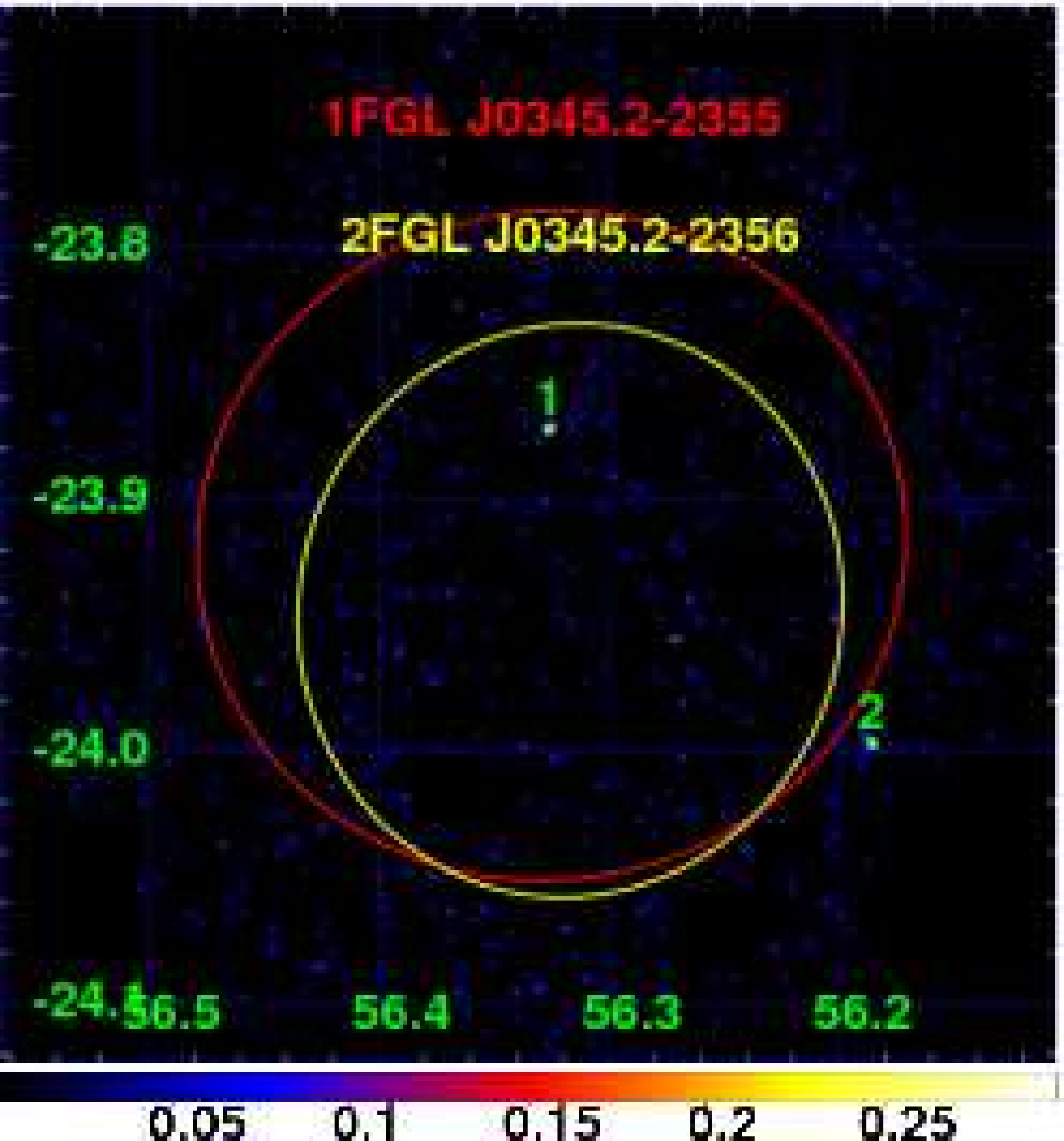}
    \end{center}
  \end{minipage}
  \begin{minipage}{0.32\hsize}
    \begin{center}
      {\small (29) 1FGL\,J0409.9--0357} \\
      \includegraphics[width=52mm]{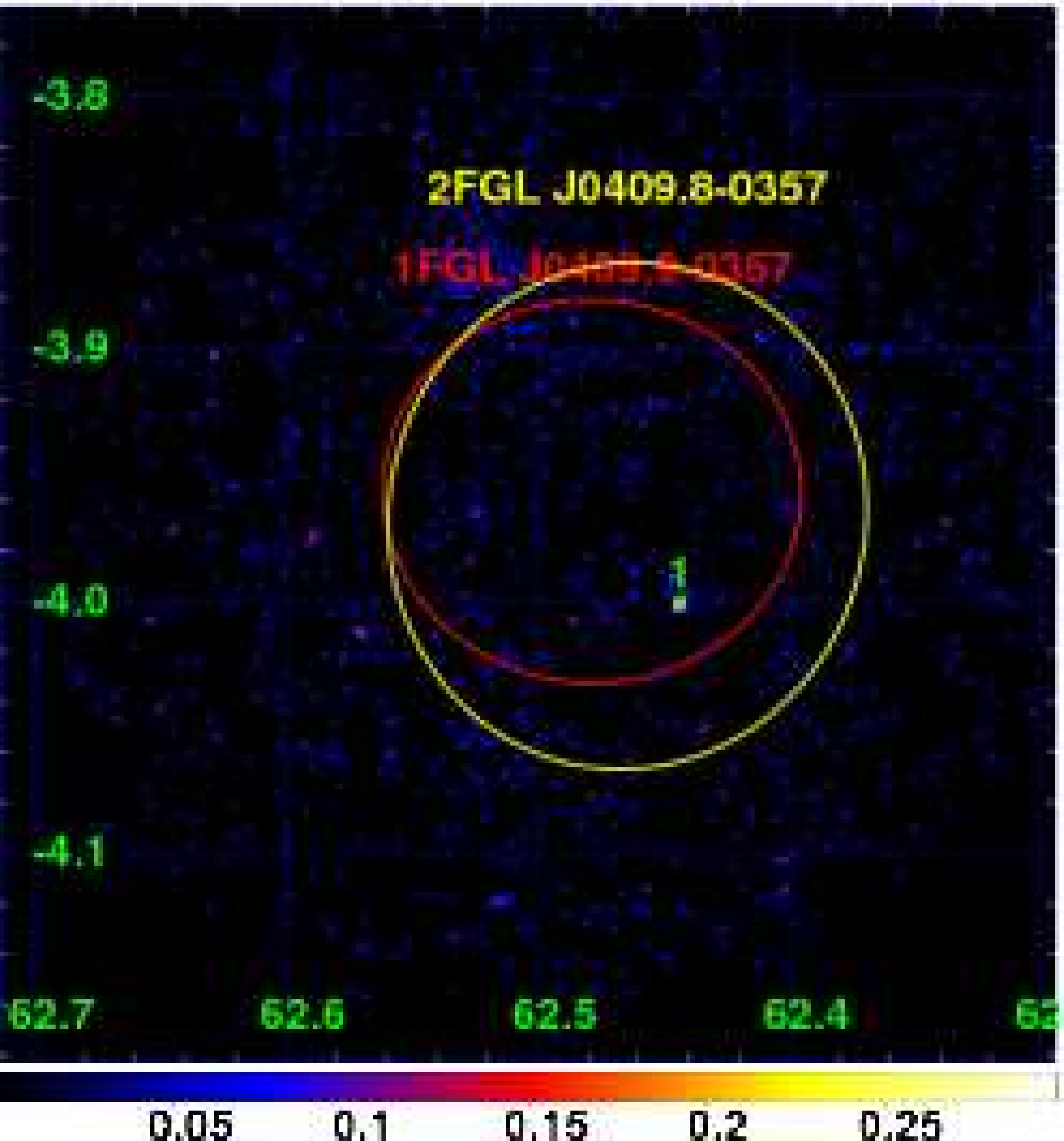}
    \end{center} 
  \end{minipage}
  \begin{minipage}{0.32\hsize}
    \begin{center}
      {\small (30) 1FGL\,J0439.8--1857} \\
      \includegraphics[width=52mm]{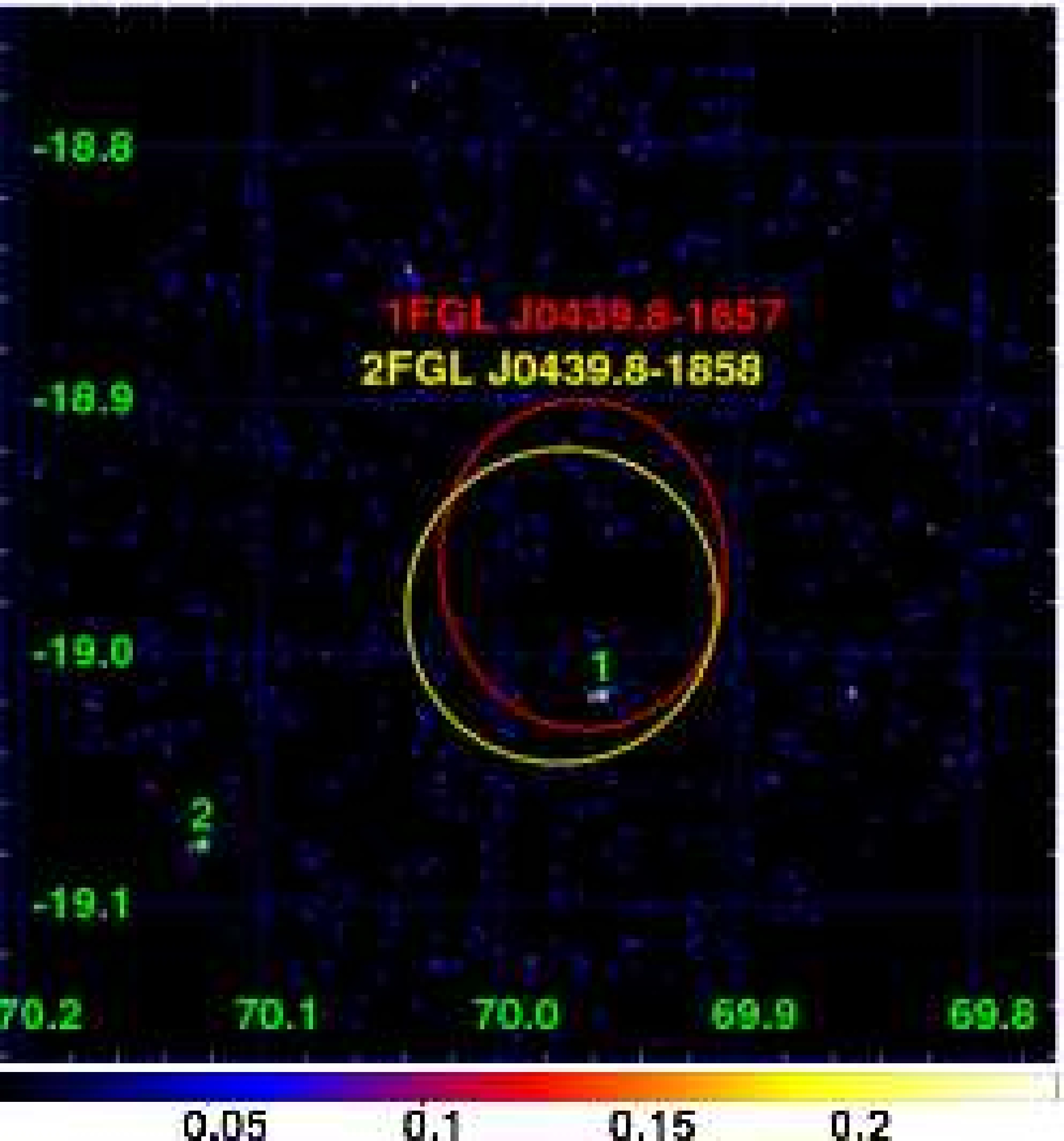}
    \end{center}
  \end{minipage}
  \begin{minipage}{0.32\hsize}
    \begin{center}
      {\small (31) 1FGL\,J0505.9$+$6121} \\
      \includegraphics[width=52mm]{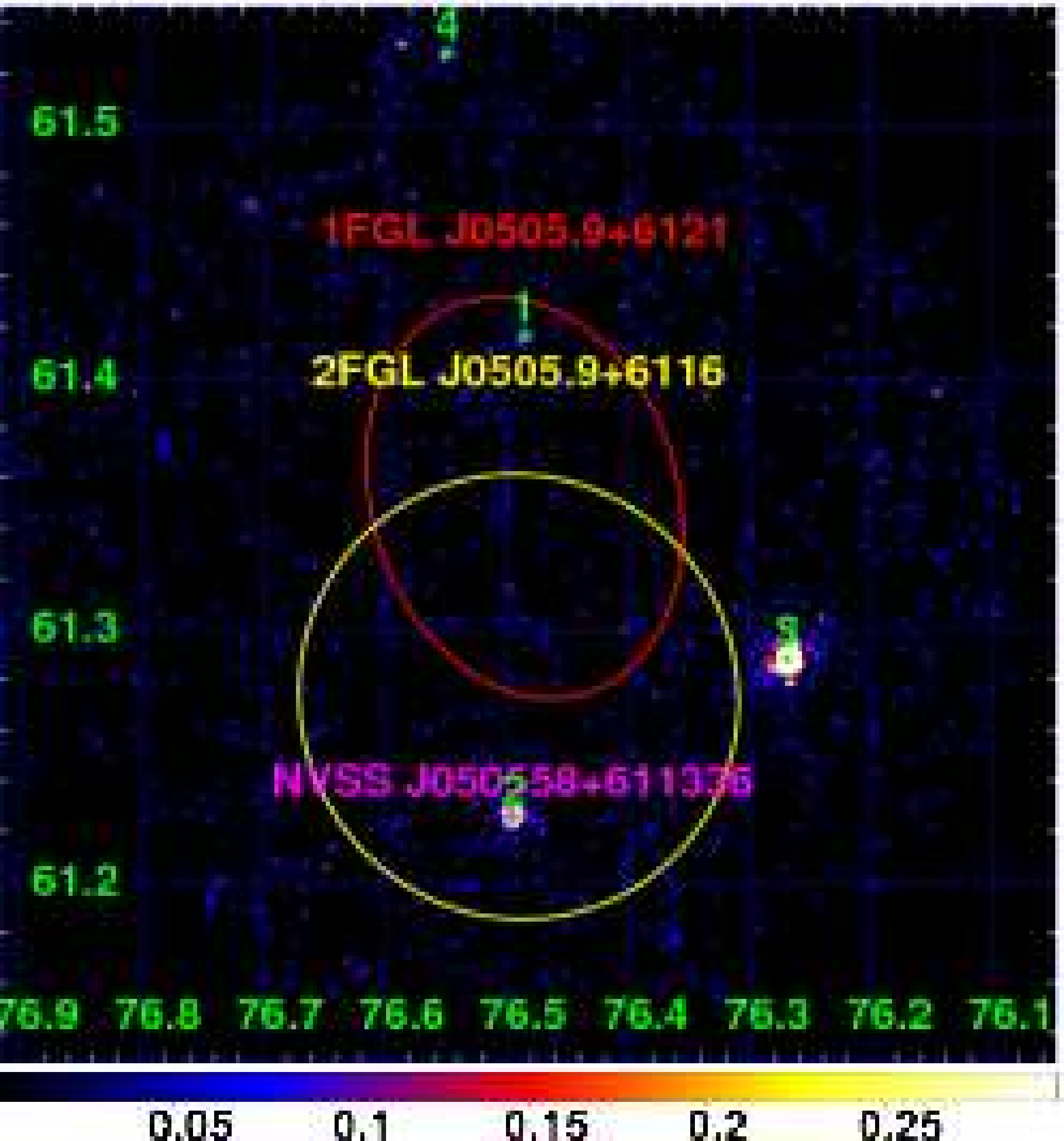}
    \end{center}
  \end{minipage}
  \begin{minipage}{0.32\hsize}
    \begin{center}
      {\small (32) 1FGL\,J0506.9--5435} \\
      \includegraphics[width=52mm]{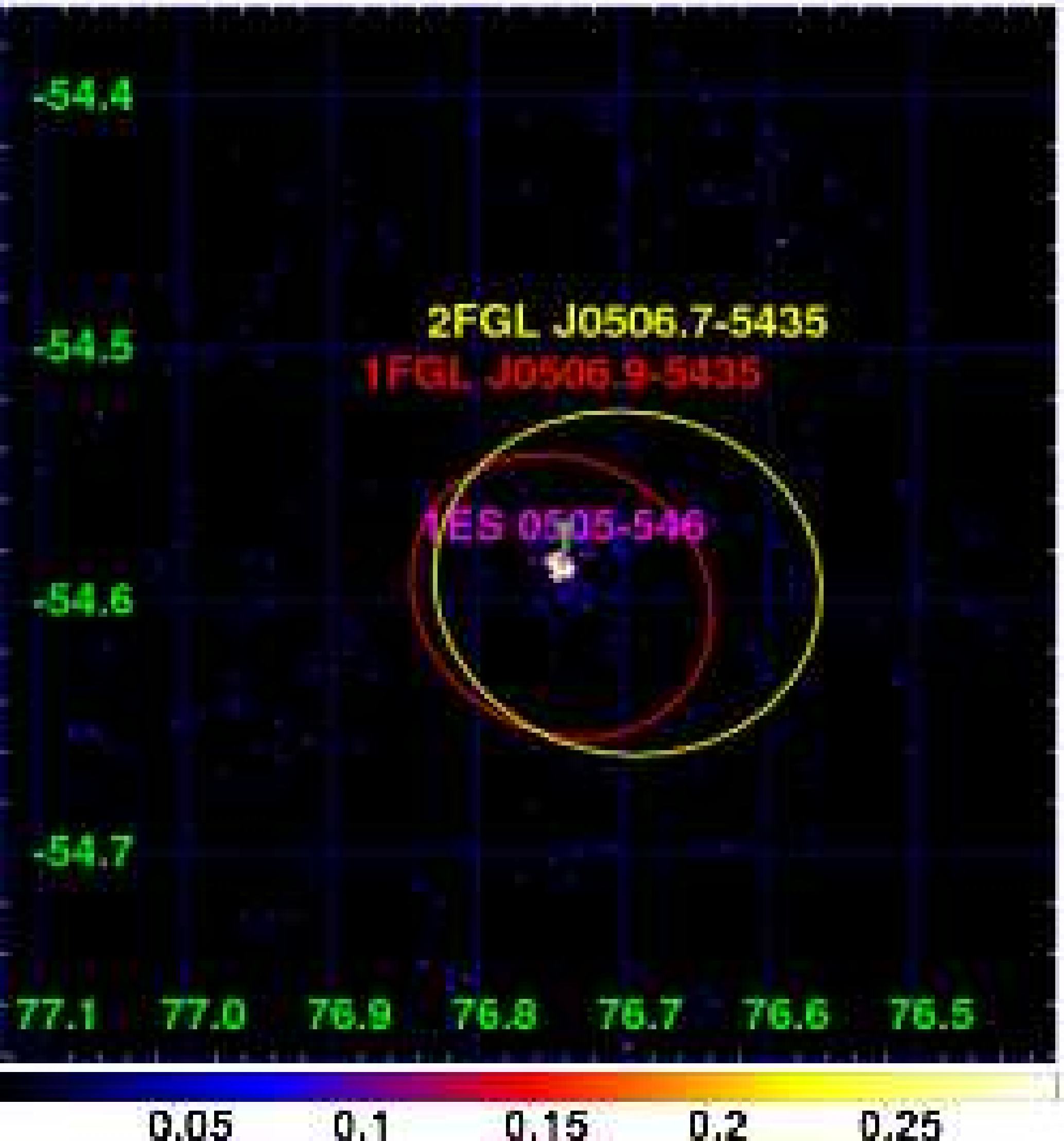}
    \end{center} 
  \end{minipage}
  \begin{minipage}{0.32\hsize}
    \begin{center}
      {\small (33) 1FGL\,J0515.9$+$1528} \\
      \includegraphics[width=52mm]{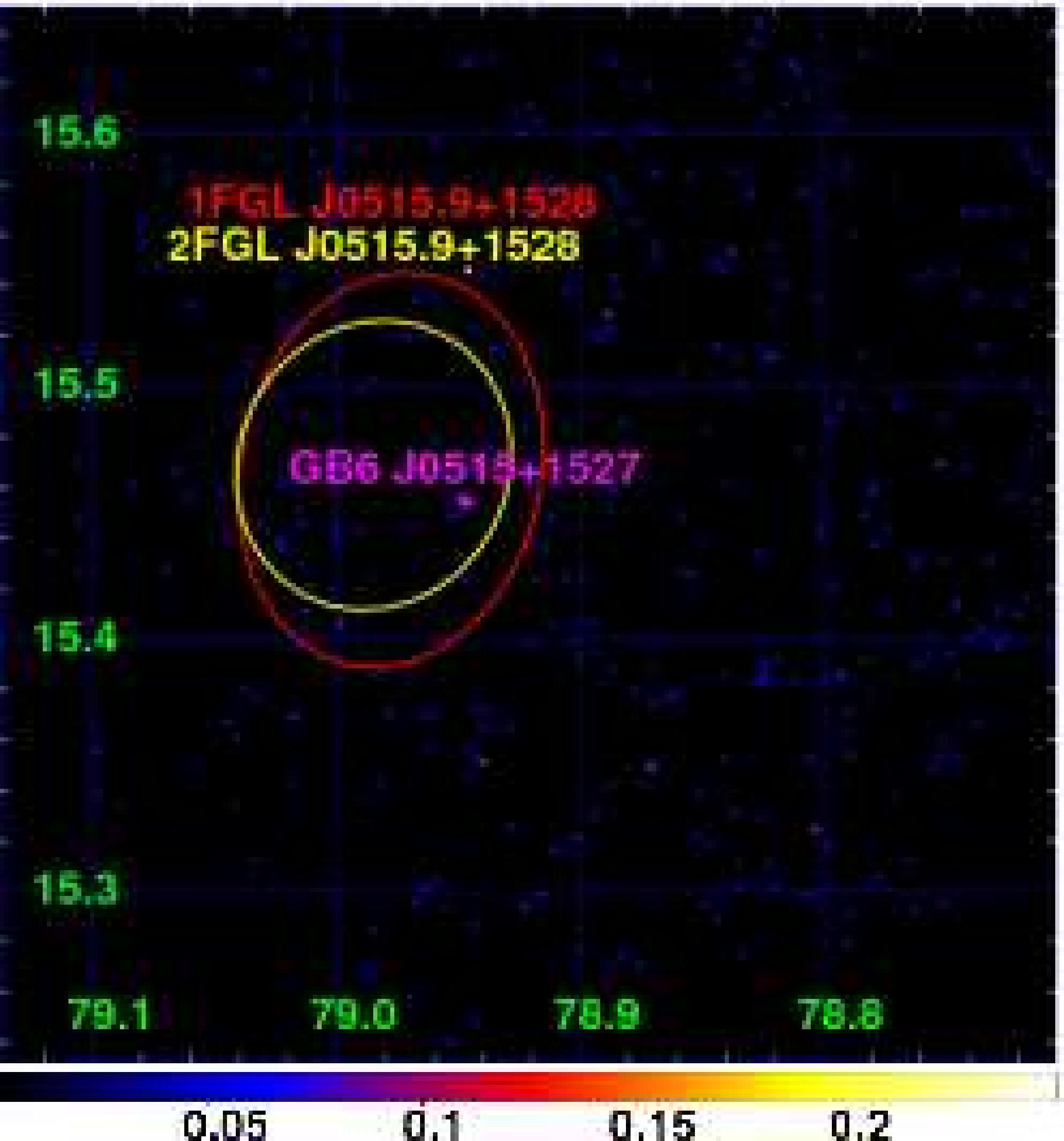}
    \end{center}
  \end{minipage}
  \begin{minipage}{0.32\hsize}
    \begin{center}
      {\small (34) 1FGL\,J0521.6$+$0103} \\
      \includegraphics[width=52mm]{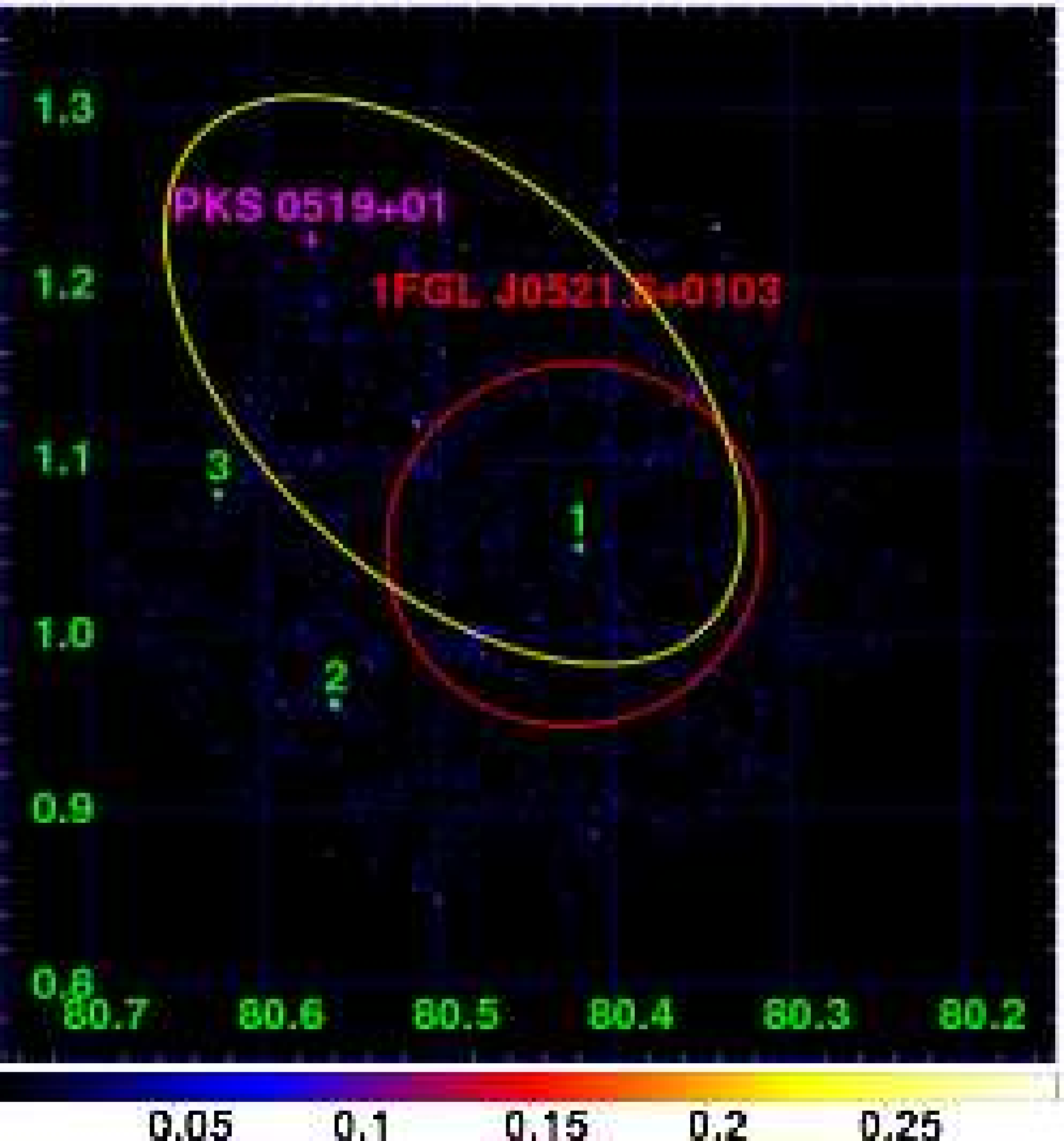}
    \end{center}
  \end{minipage}
  \begin{minipage}{0.32\hsize}
    \begin{center}
      {\small (35) 1FGL\,J0523.5--2529} \\
      \includegraphics[width=52mm]{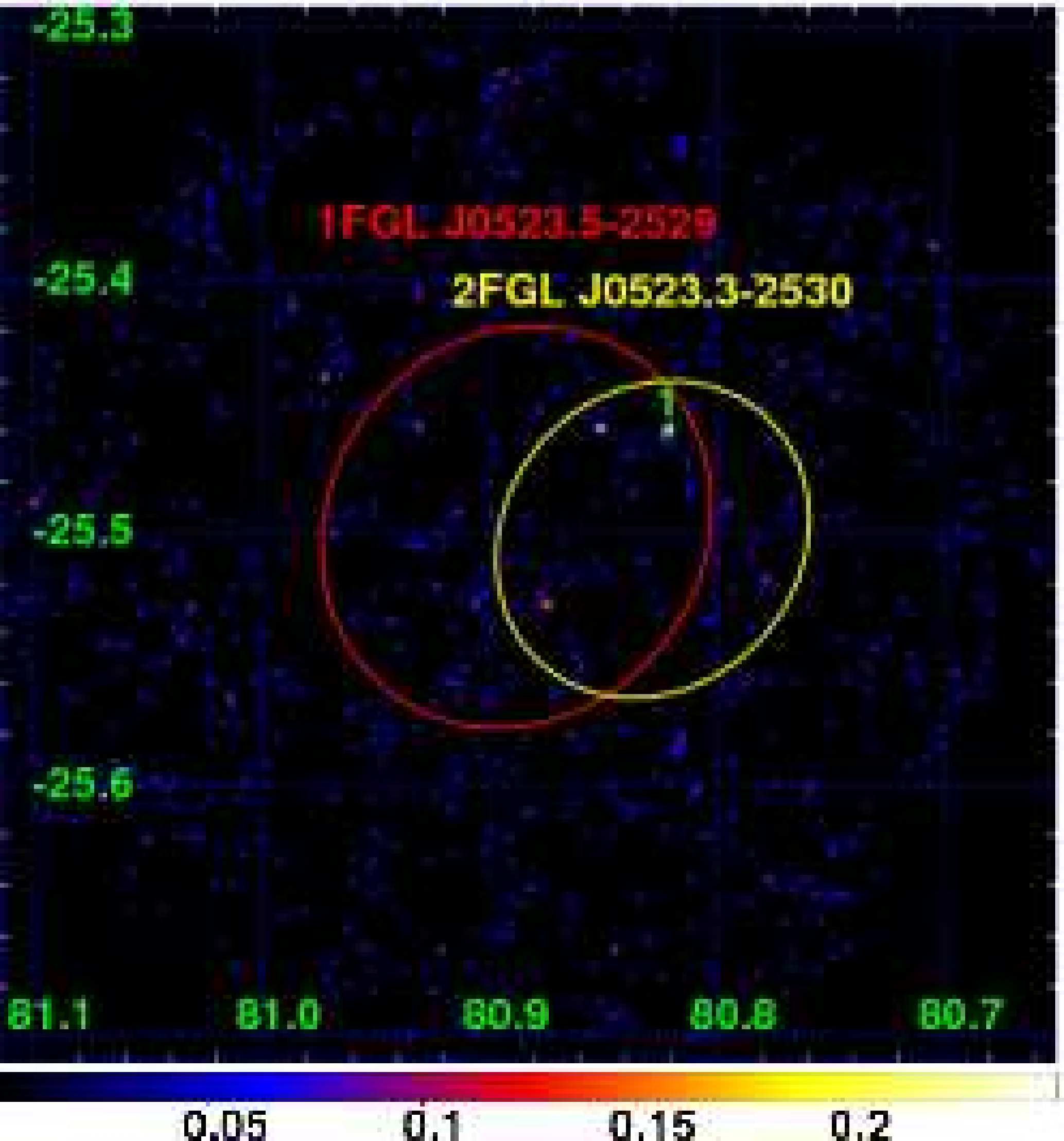}
    \end{center} 
  \end{minipage}
  \begin{minipage}{0.32\hsize}
    \begin{center}
      {\small (36) 1FGL\,J0533.9$+$6758} \\
      \includegraphics[width=52mm]{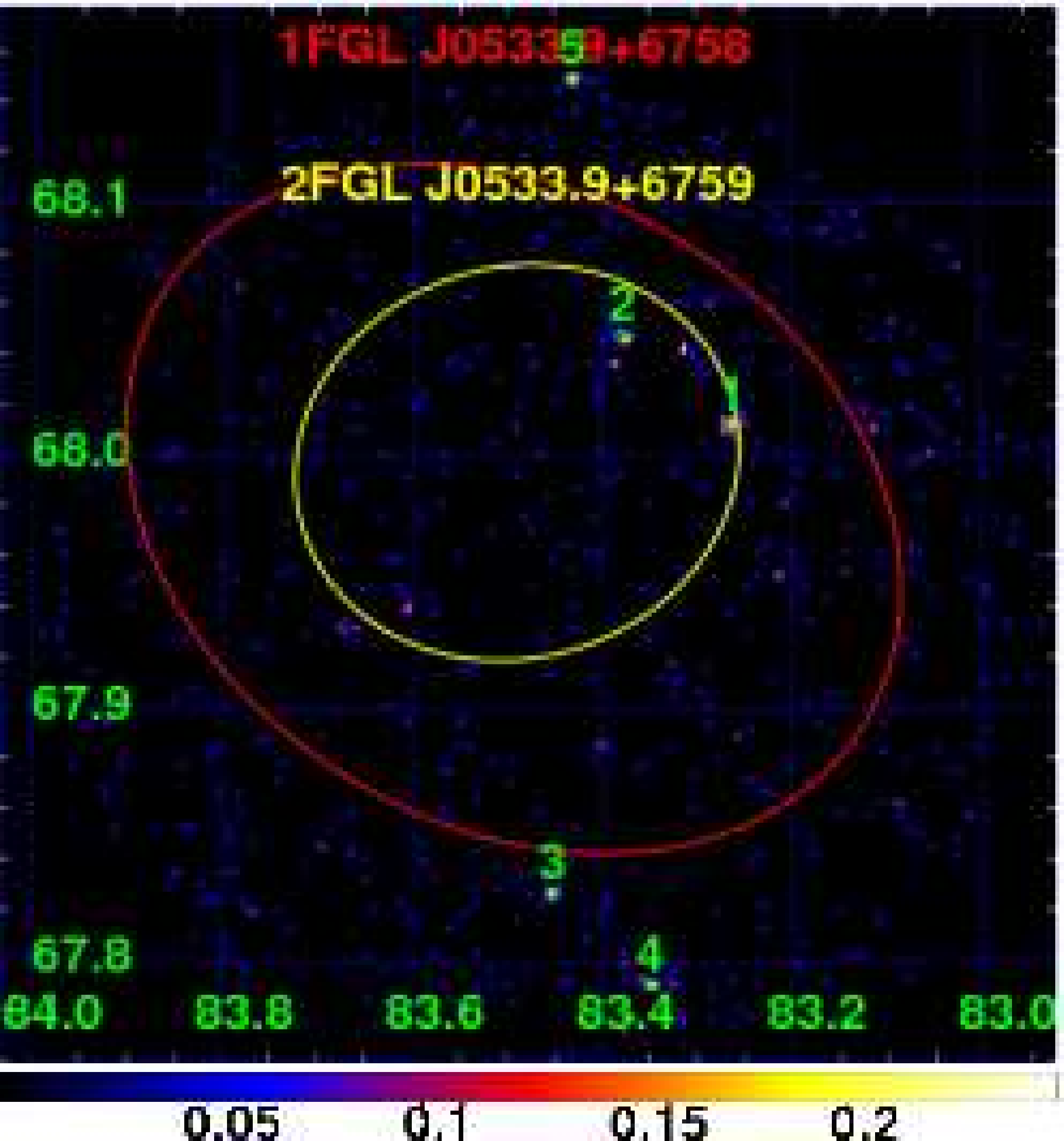}
    \end{center}
  \end{minipage}
 \end{center}
\end{figure}
\clearpage
\begin{figure}[m]
 \begin{center}
  \begin{minipage}{0.32\hsize}
    \begin{center}
      {\small (37) 1FGL\,J0537.7--5717} \\
      \includegraphics[width=52mm]{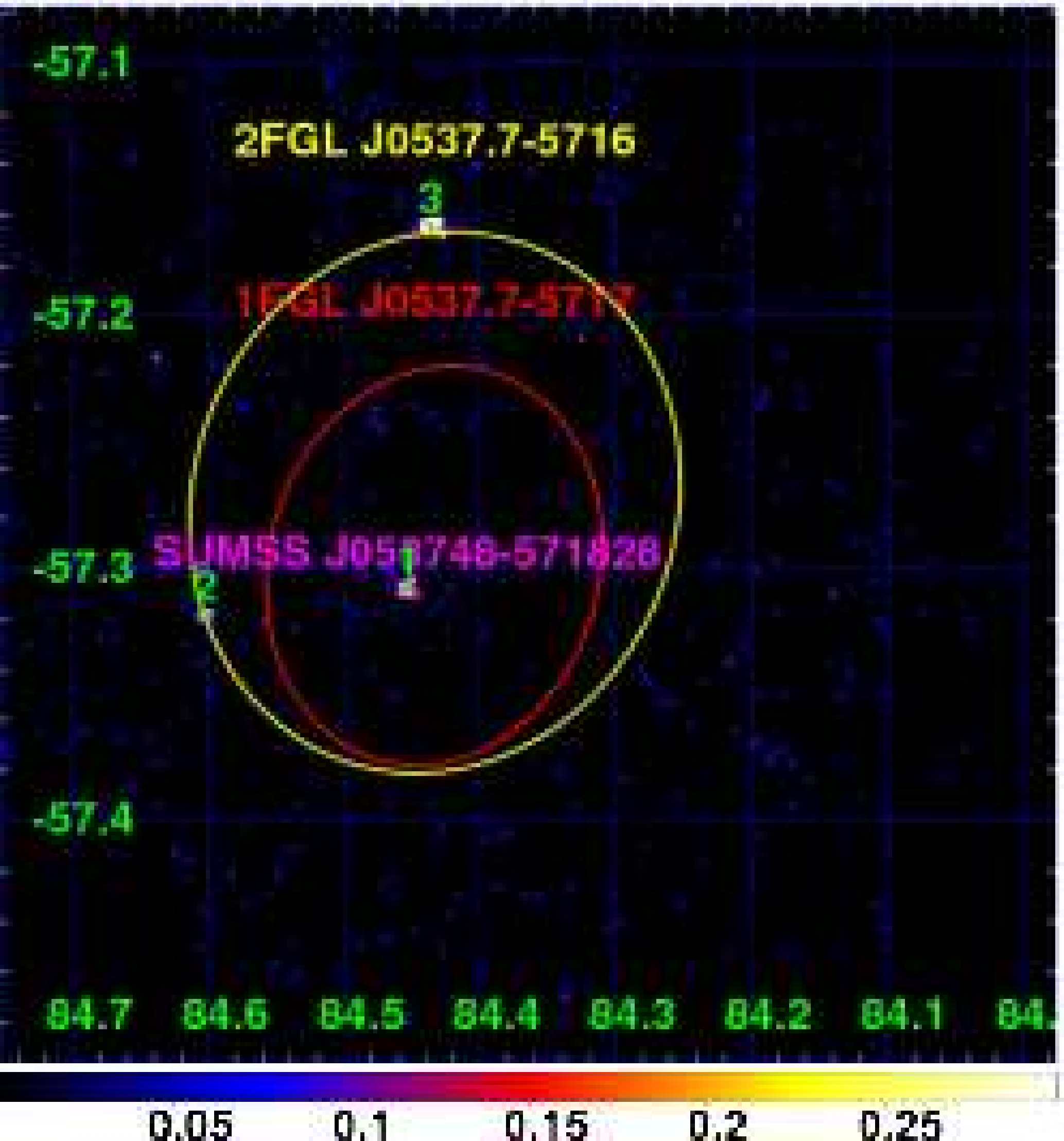}
    \end{center}
  \end{minipage}
  \begin{minipage}{0.32\hsize}
    \begin{center}
      {\small (38) 1FGL\,J0538.4--3910} \\
      \includegraphics[width=52mm]{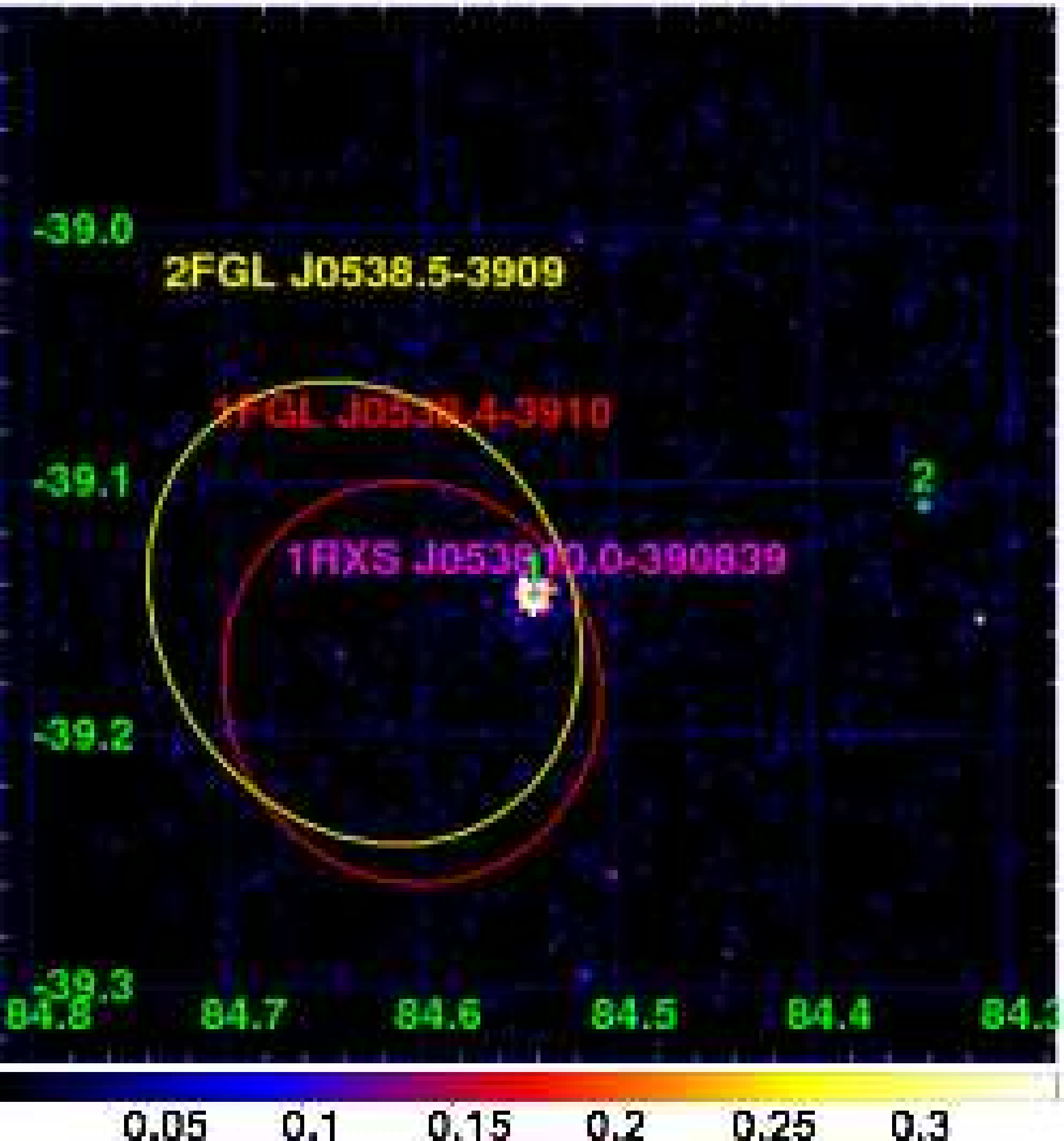}
    \end{center} 
  \end{minipage}
  \begin{minipage}{0.32\hsize}
    \begin{center}
      {\small (39) 1FGL\,J0545.6$+$6022} \\
      \includegraphics[width=52mm]{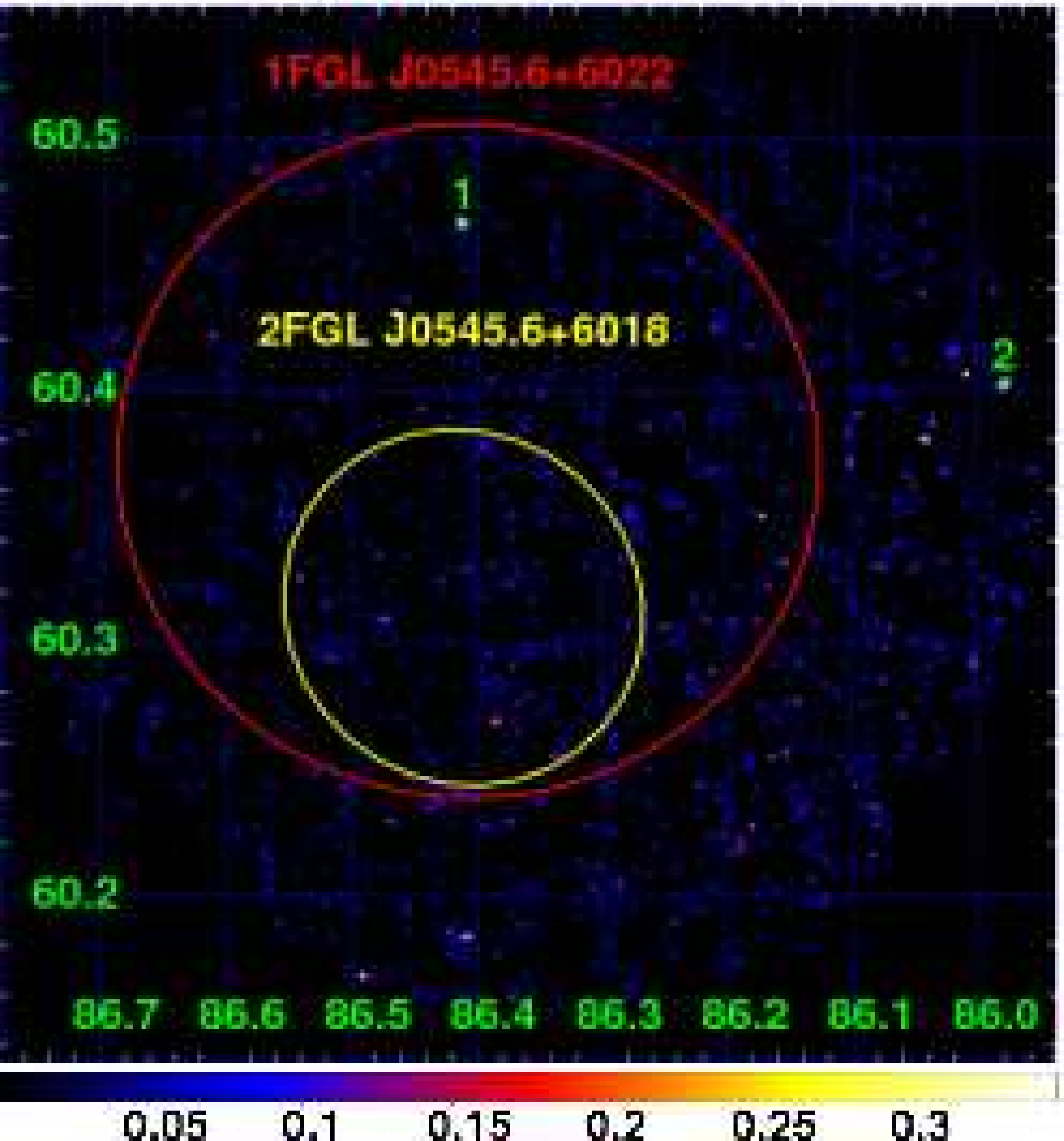}
    \end{center}
  \end{minipage}
  \begin{minipage}{0.32\hsize}
    \begin{center}
      {\small (40) 1FGL\,J0600.5--2006} \\
      \includegraphics[width=52mm]{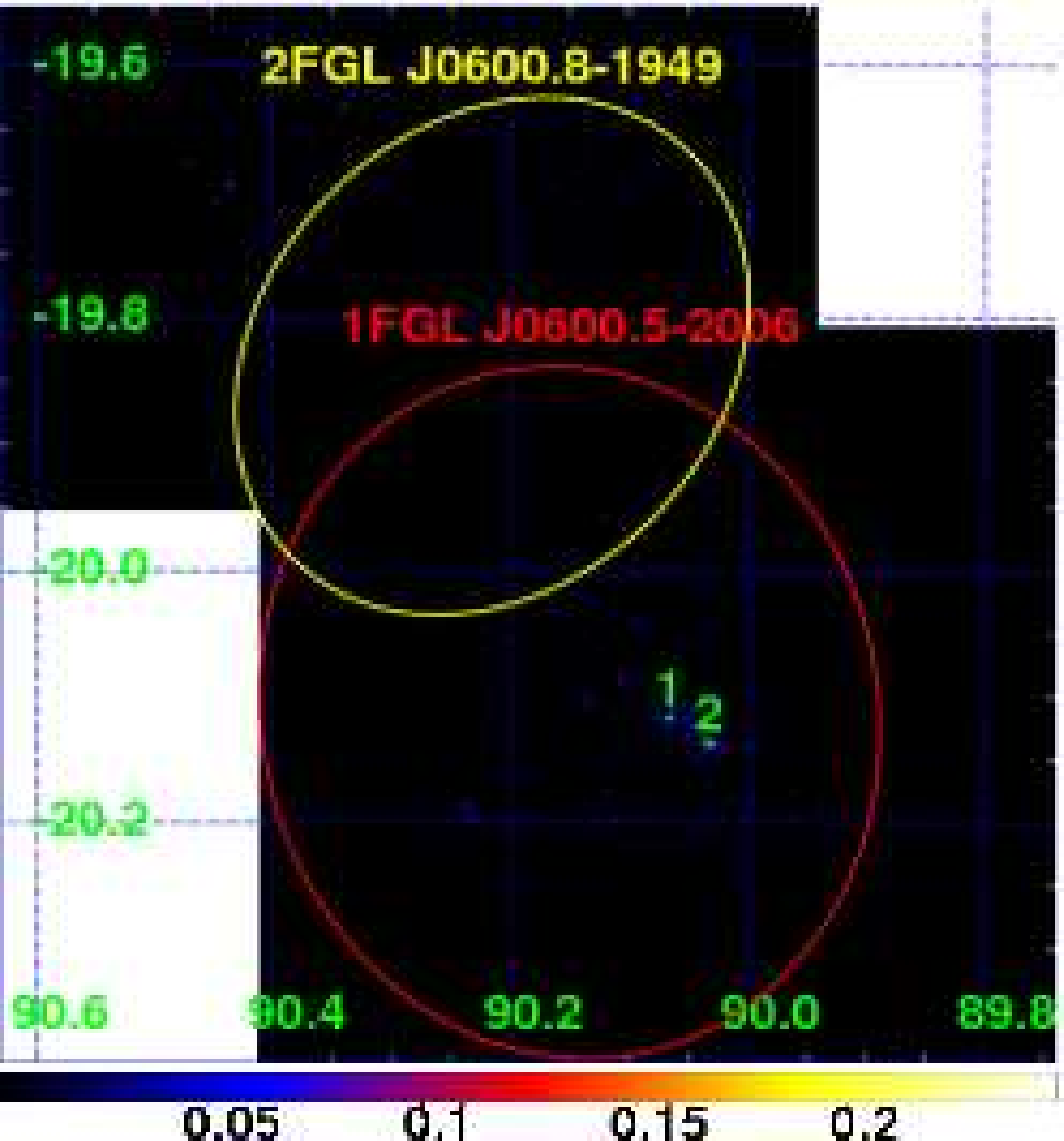}
    \end{center}
  \end{minipage}
  \begin{minipage}{0.32\hsize}
    \begin{center}
      {\small (41) 1FGL\,J0603.0--4012} \\
      \includegraphics[width=52mm]{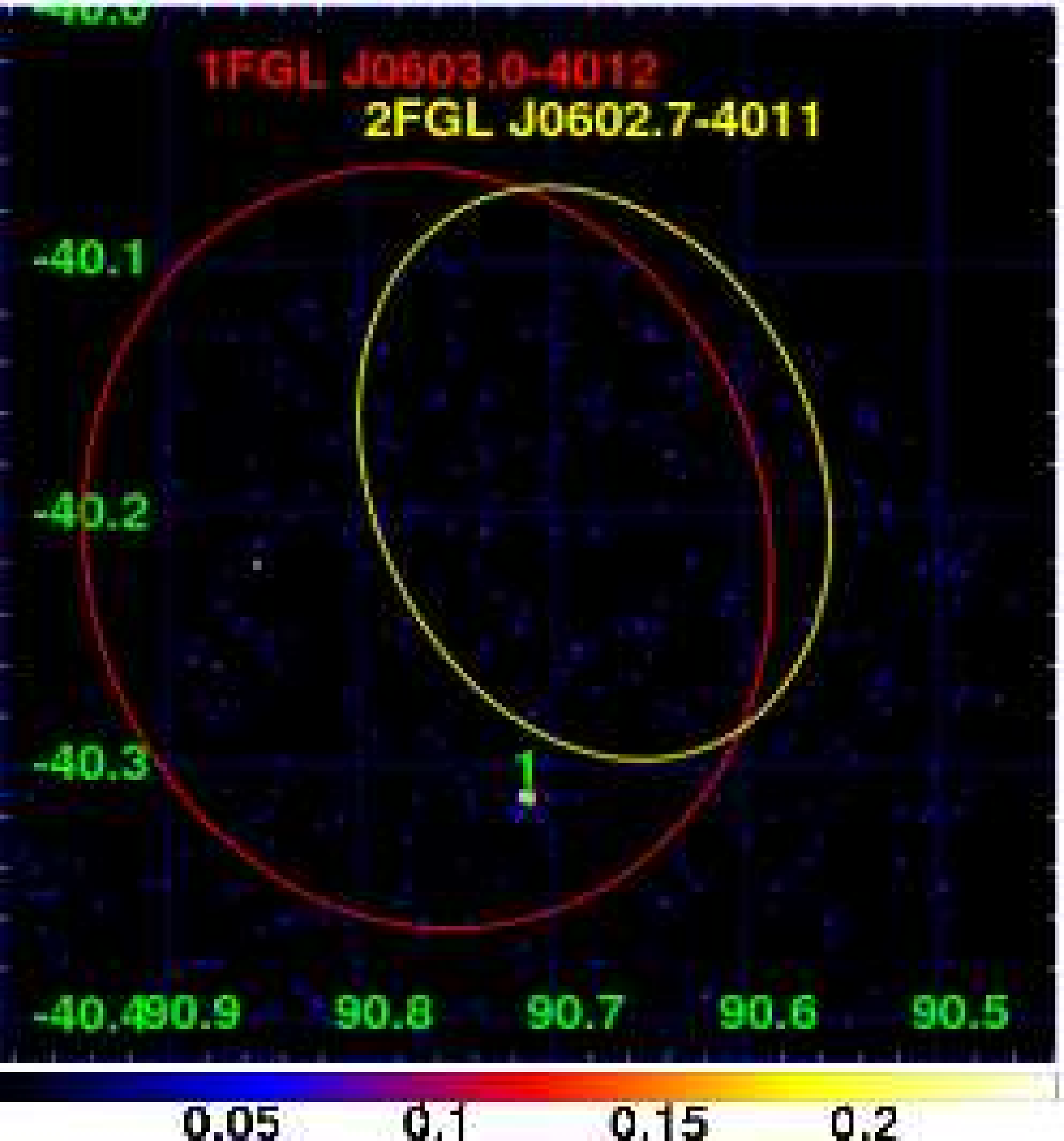}
    \end{center} 
  \end{minipage}
  \begin{minipage}{0.32\hsize}
    \begin{center}
      {\small (42) 1FGL\,J0604.2--4817} \\
      \includegraphics[width=52mm]{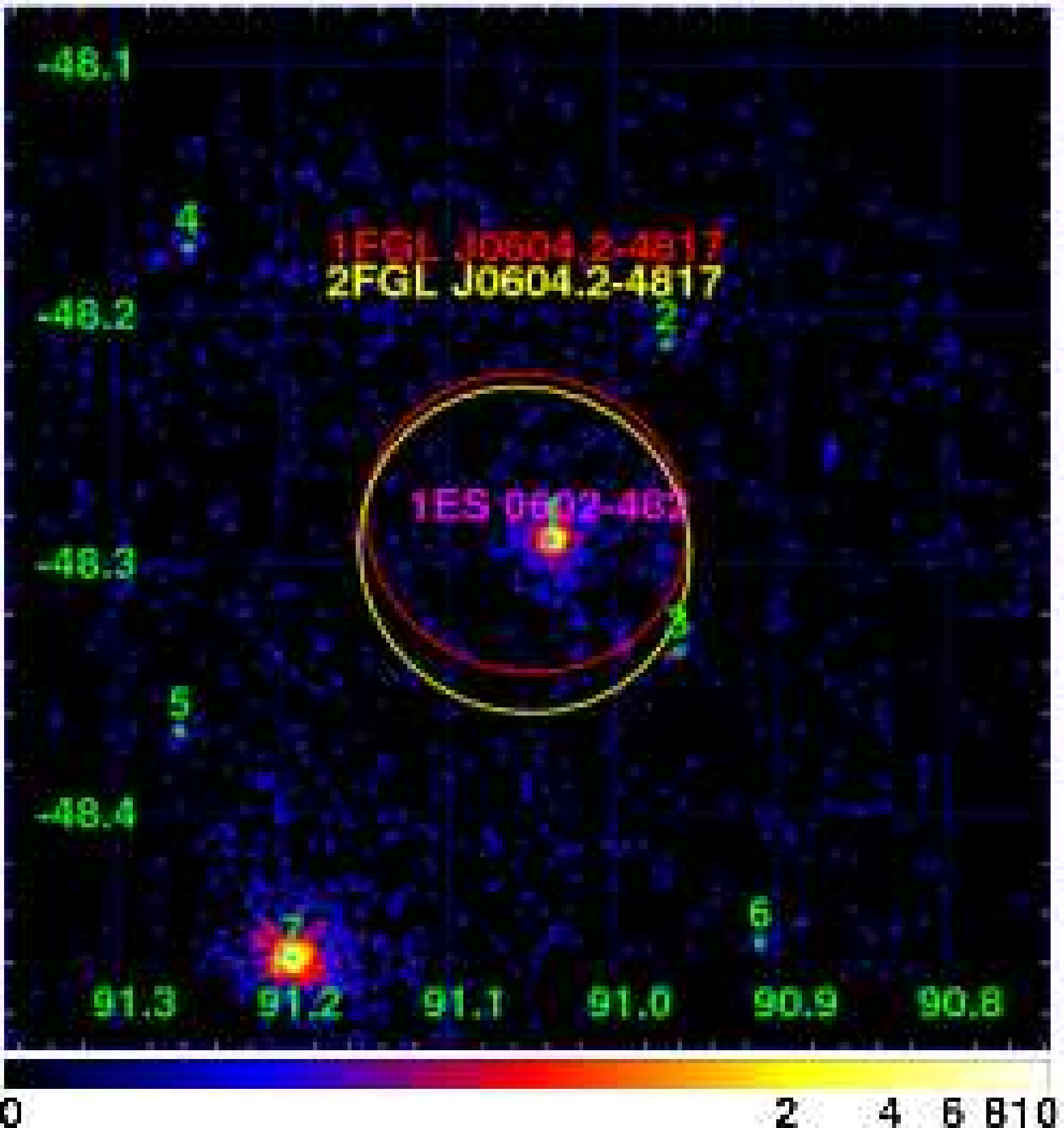}
    \end{center}
  \end{minipage}
  \begin{minipage}{0.32\hsize}
    \begin{center}
      {\small (43) 1FGL\,J0605.1$+$0005} \\
      \includegraphics[width=52mm]{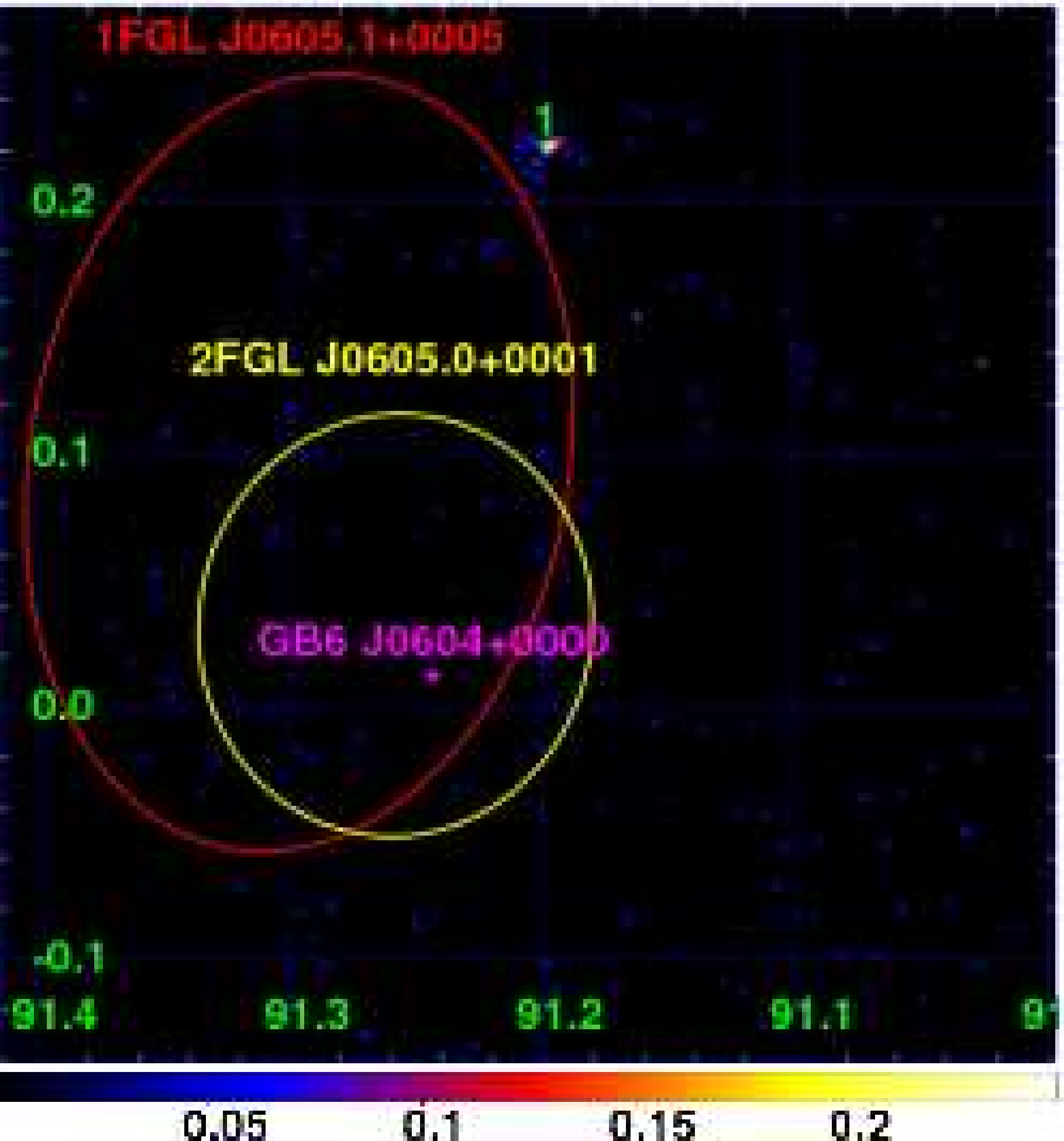}
    \end{center}
  \end{minipage}
  \begin{minipage}{0.32\hsize}
    \begin{center}
      {\small (44) 1FGL\,J0609.3--0244} \\
      \includegraphics[width=52mm]{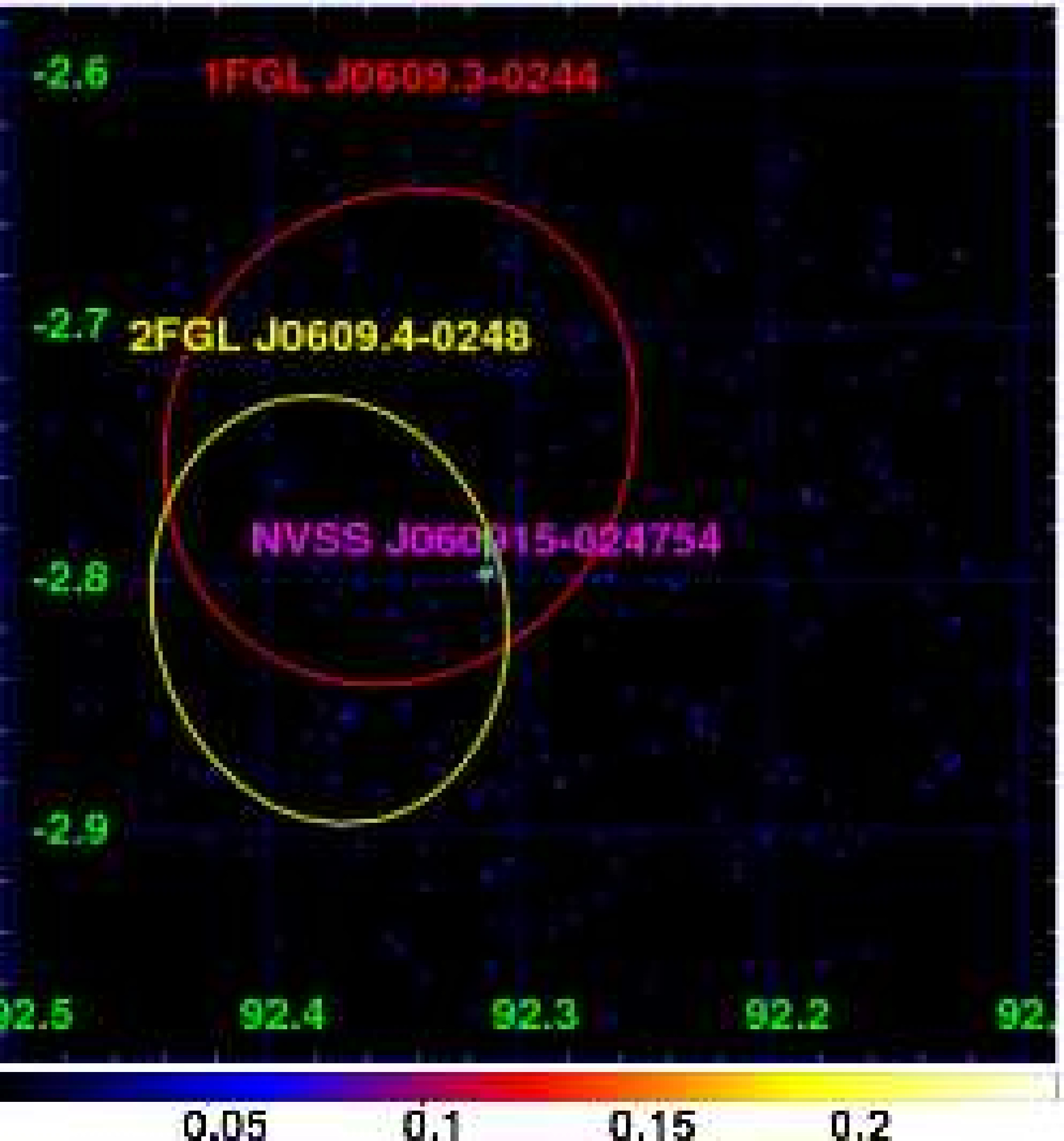}
    \end{center} 
  \end{minipage}
  \begin{minipage}{0.32\hsize}
    \begin{center}
      {\small (45) 1FGL\,J0622.2$+$3751} \\
      \includegraphics[width=52mm]{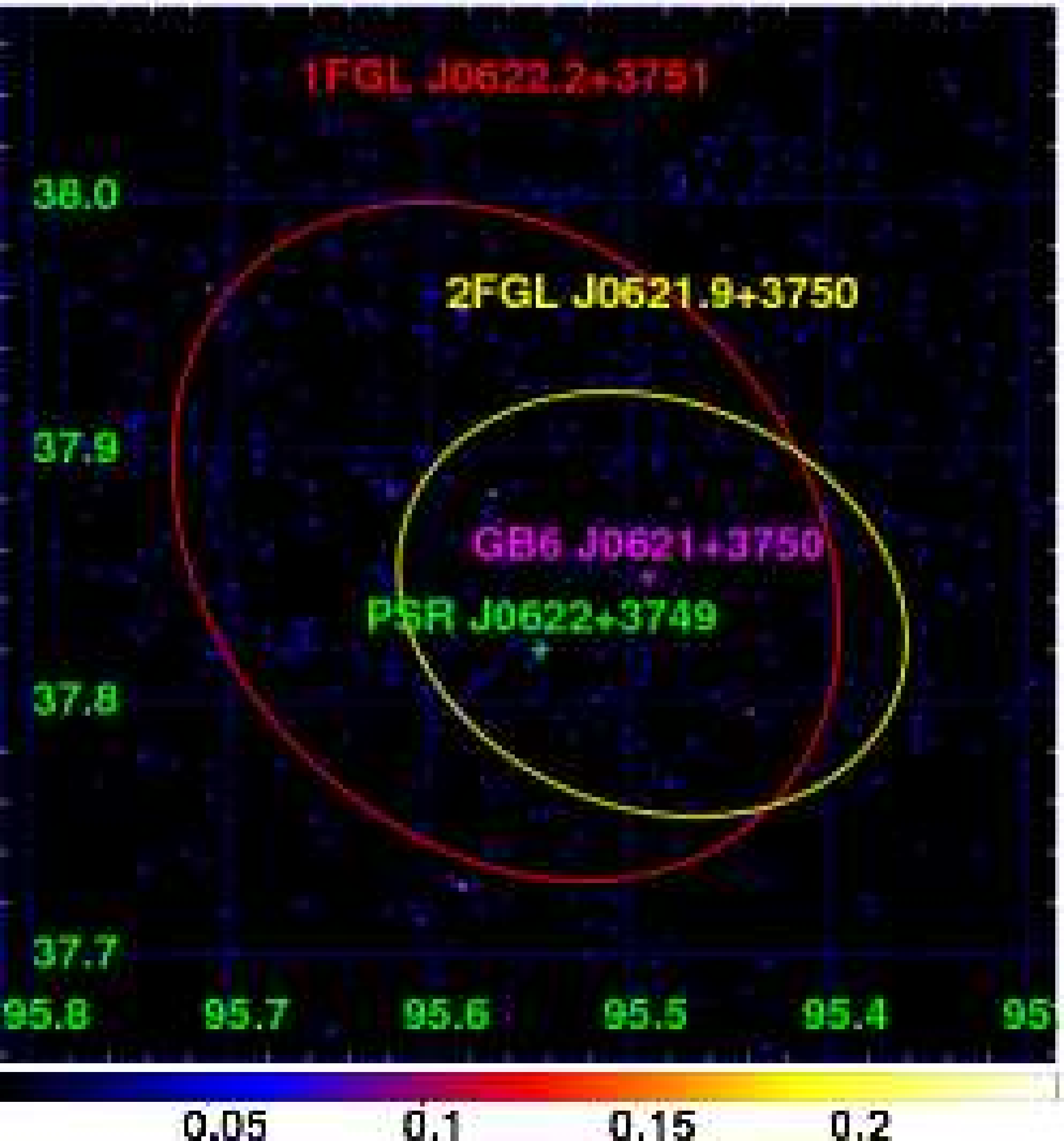}
    \end{center}
  \end{minipage}
  \begin{minipage}{0.32\hsize}
    \begin{center}
      {\small (46) 1FGL\,J0648.6--6052} \\
      \includegraphics[width=52mm]{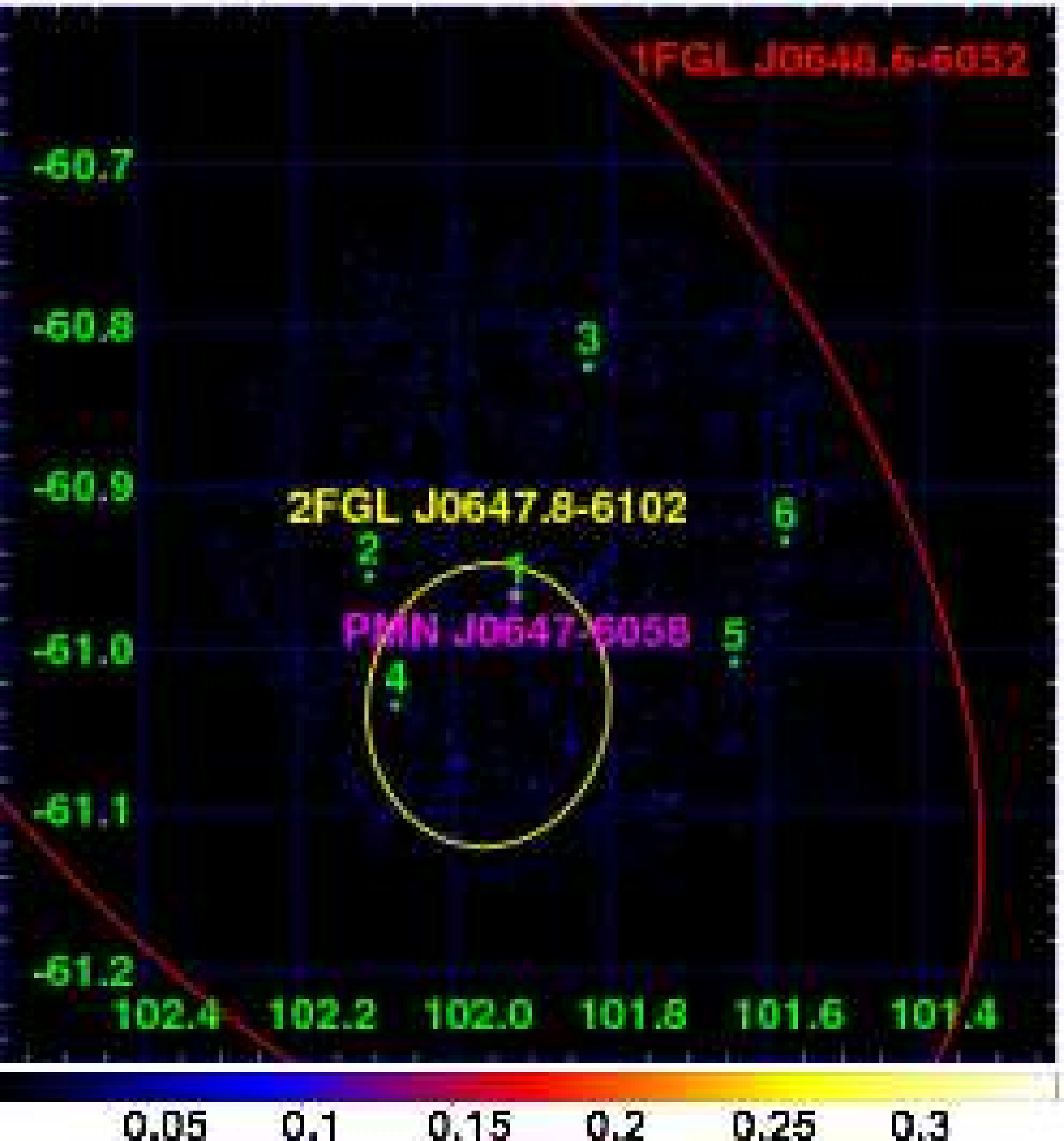}
    \end{center}
  \end{minipage}
  \begin{minipage}{0.32\hsize}
    \begin{center}
      {\small (47) 1FGL\,J0707.3$+$7742} \\
      \includegraphics[width=52mm]{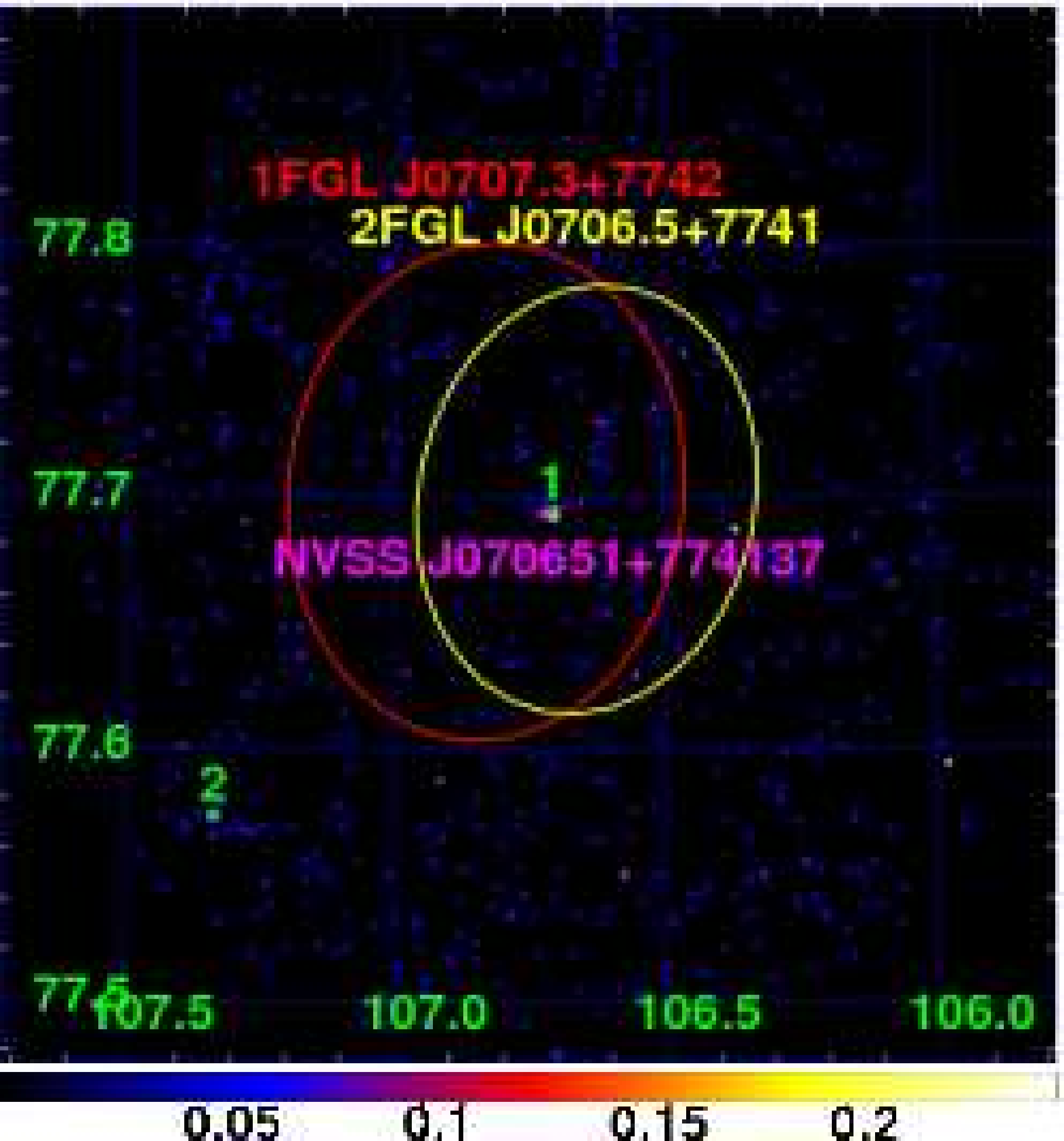}
    \end{center} 
  \end{minipage}
  \begin{minipage}{0.32\hsize}
    \begin{center}
      {\small (48) 1FGL\,J0718.8--4958} \\
      \includegraphics[width=52mm]{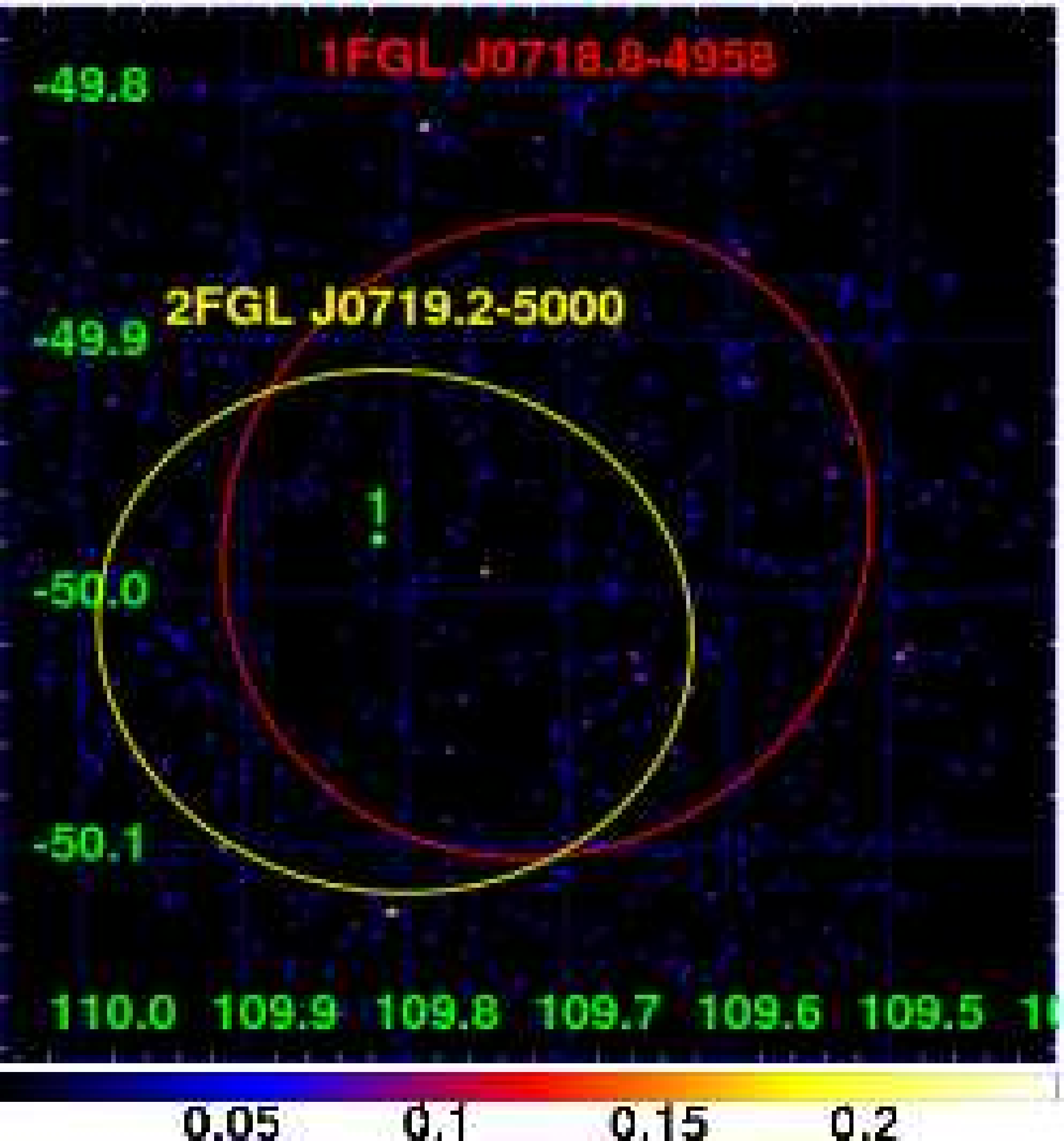}
    \end{center}
  \end{minipage}
 \end{center}
\end{figure}
\clearpage
\begin{figure}[m]
 \begin{center}
  \begin{minipage}{0.32\hsize}
    \begin{center}
      {\small (49) 1FGL\,J0803.1--0339} \\
      \includegraphics[width=52mm]{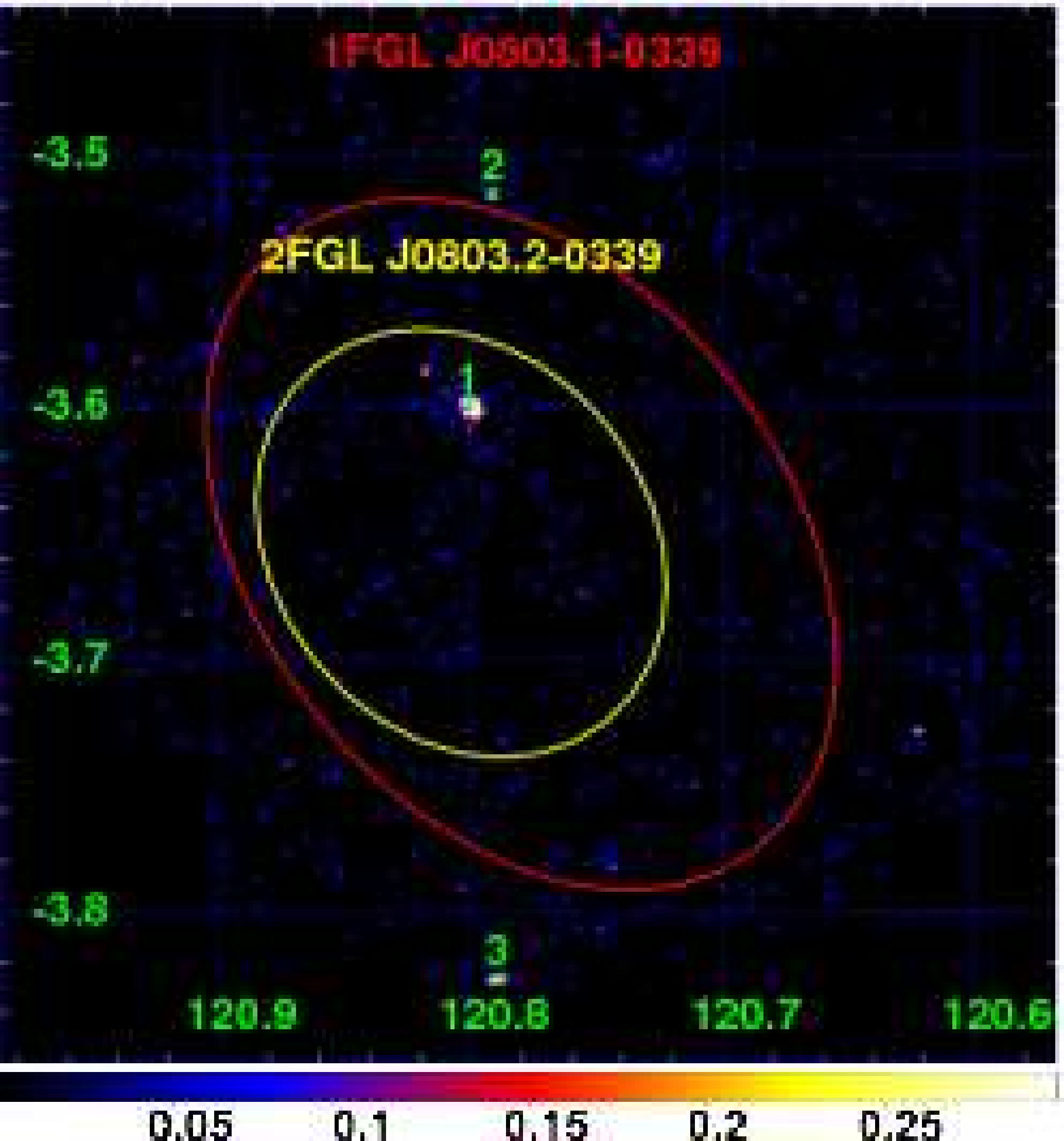}
    \end{center}
  \end{minipage}
  \begin{minipage}{0.32\hsize}
    \begin{center}
      {\small (50) 1FGL\,J0814.5--1011} \\
      \includegraphics[width=52mm]{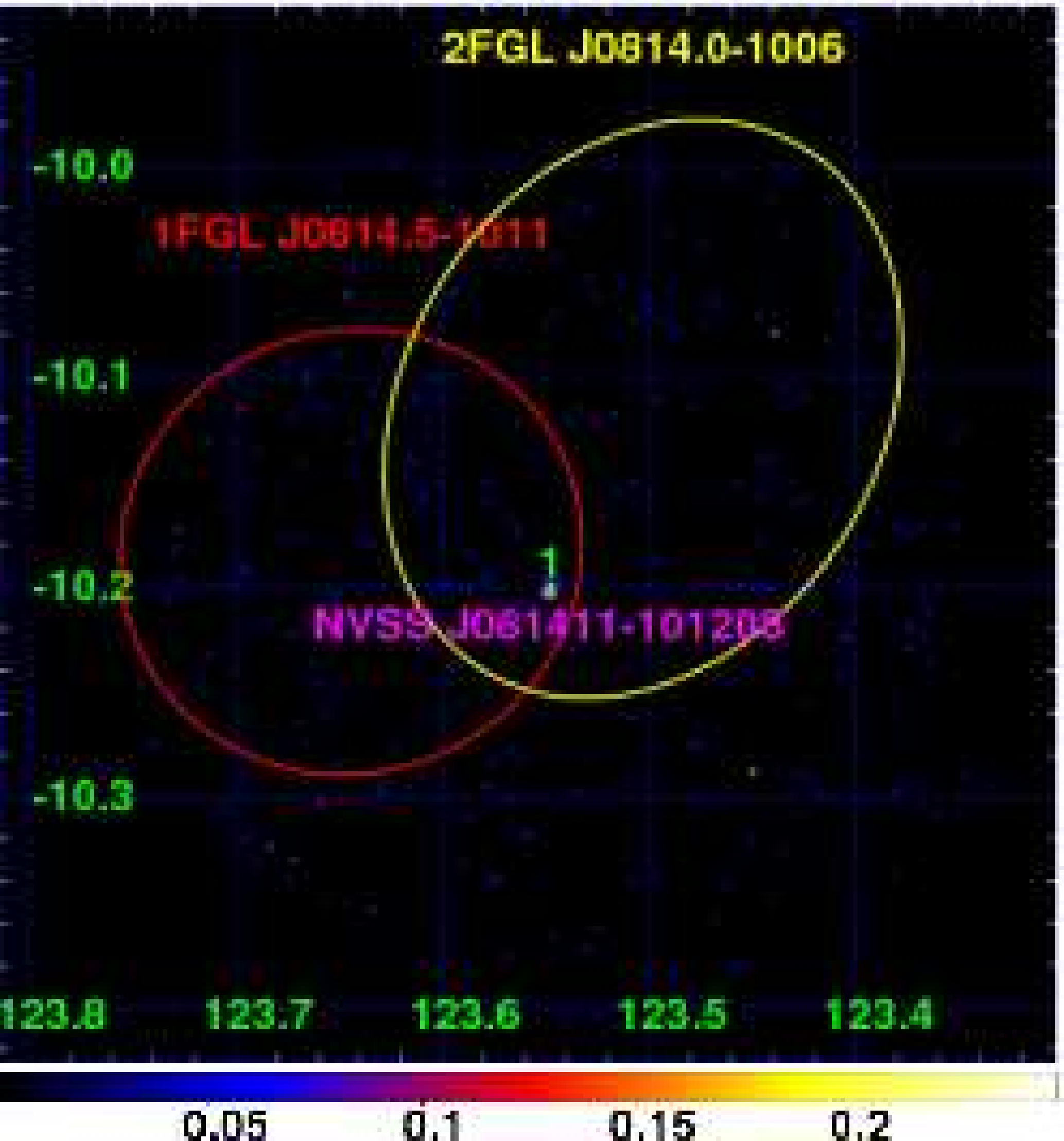}
    \end{center} 
  \end{minipage}
  \begin{minipage}{0.32\hsize}
    \begin{center}
      {\small (51) 1FGL\,J0843.4$+$6718} \\
      \includegraphics[width=52mm]{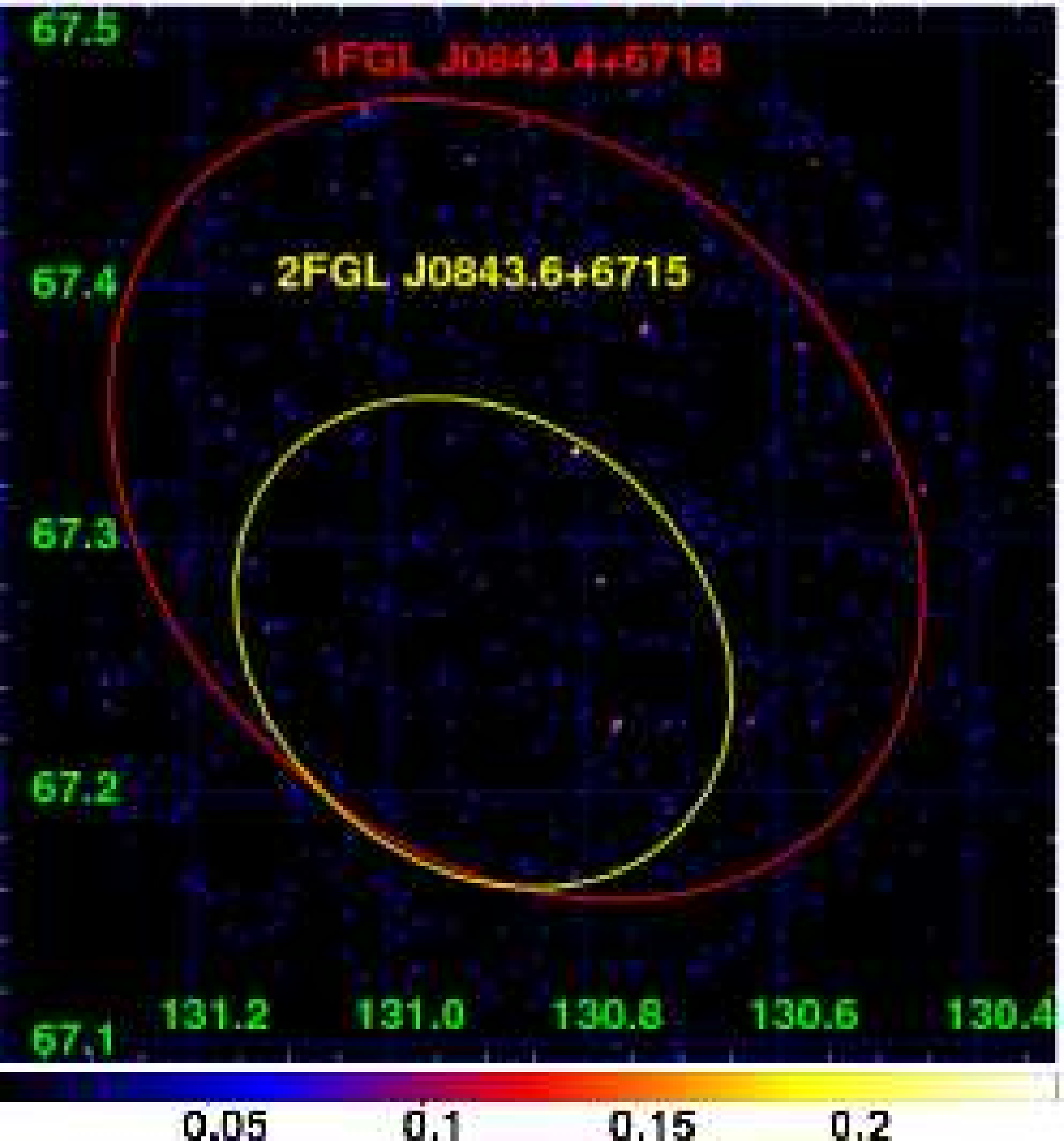}
    \end{center}
  \end{minipage}
  \begin{minipage}{0.32\hsize}
    \begin{center}
      {\small (52) 1FGL\,J0848.6$+$0504} \\
      \includegraphics[width=52mm]{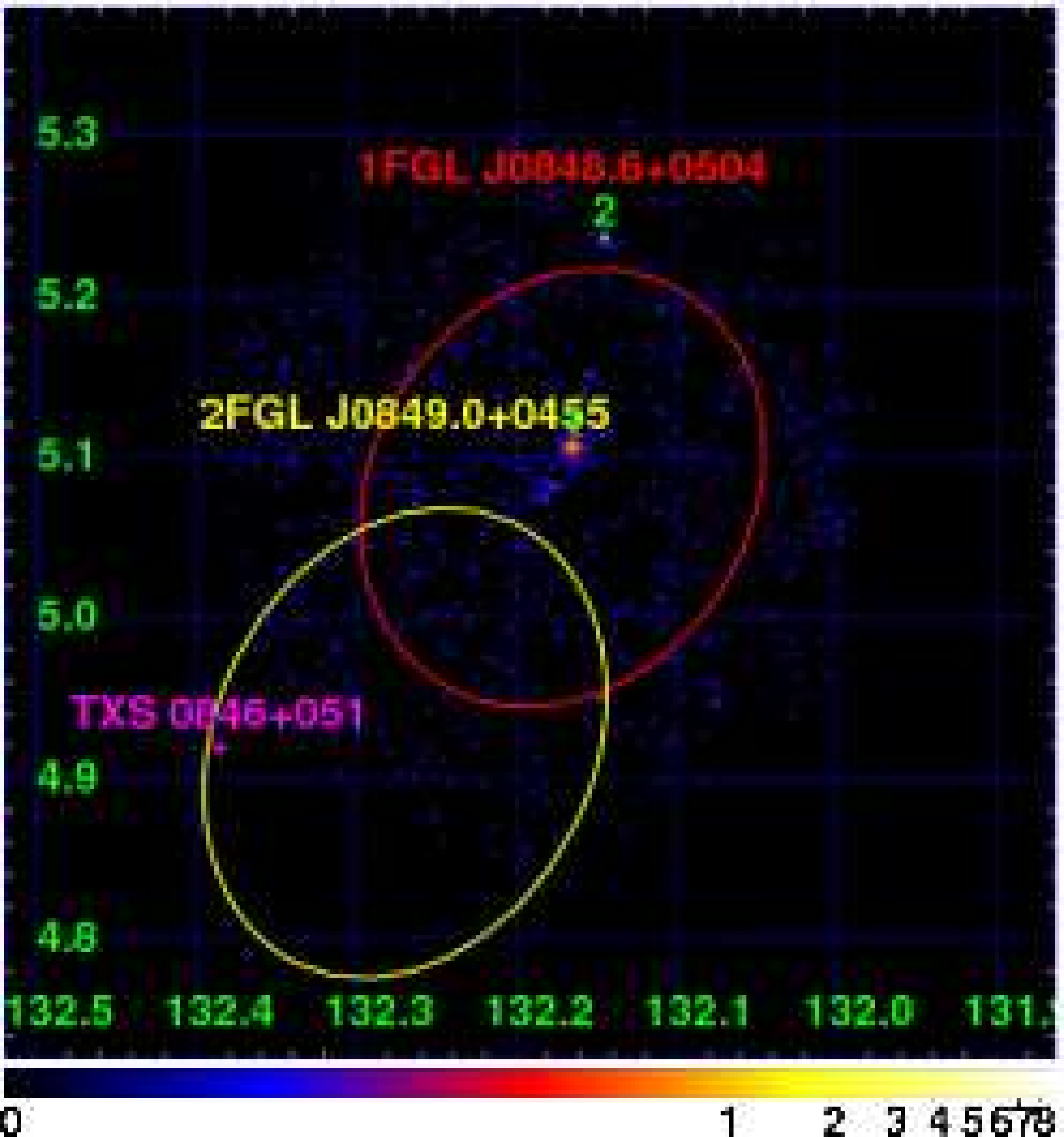}
    \end{center}
  \end{minipage}
  \begin{minipage}{0.32\hsize}
    \begin{center}
      {\small (53) 1FGL\,J0902.4$+$2050} \\
      \includegraphics[width=52mm]{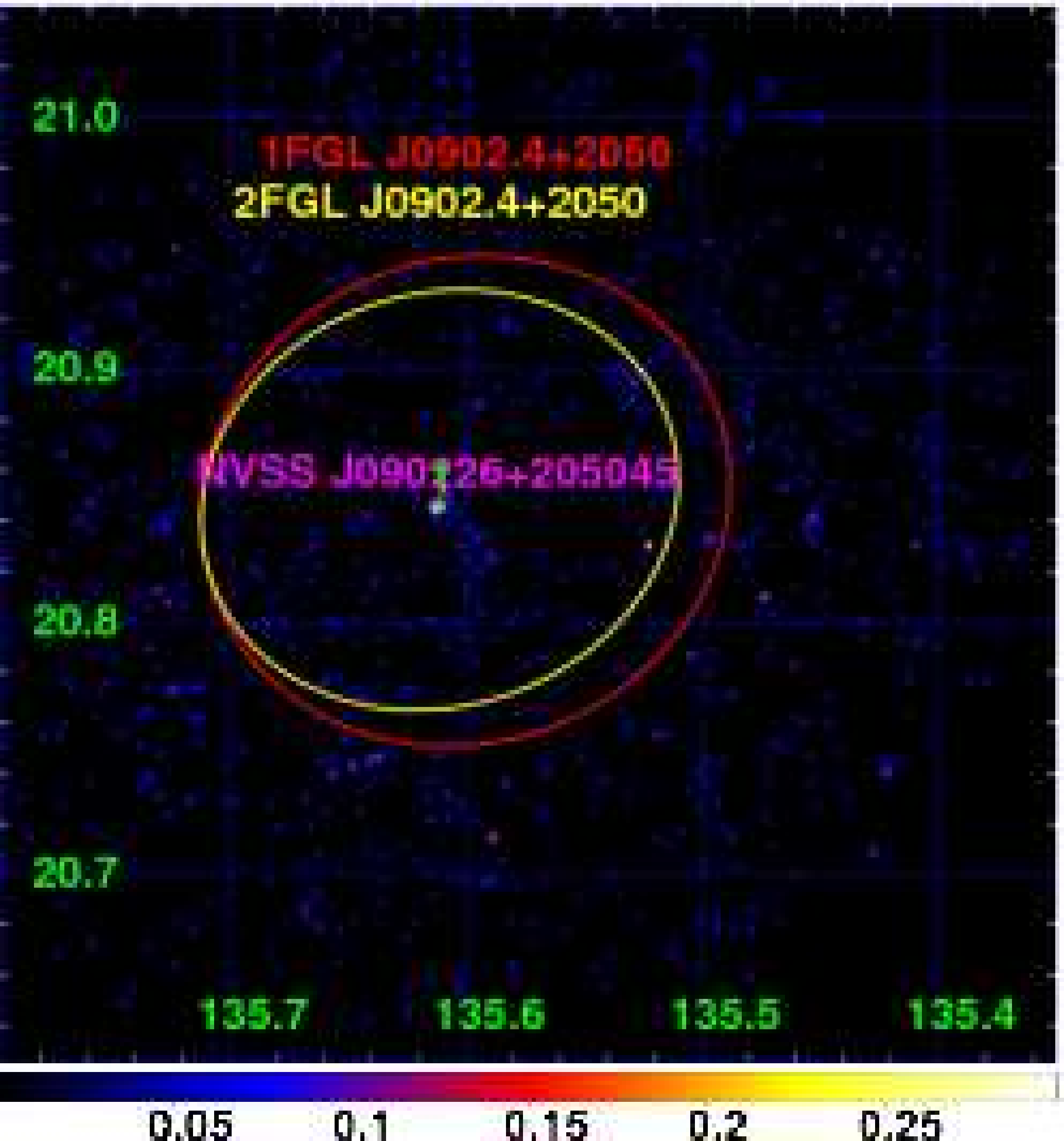}
    \end{center} 
  \end{minipage}
  \begin{minipage}{0.32\hsize}
    \begin{center}
      {\small (54) 1FGL\,J0906.4--0903} \\
      \includegraphics[width=52mm]{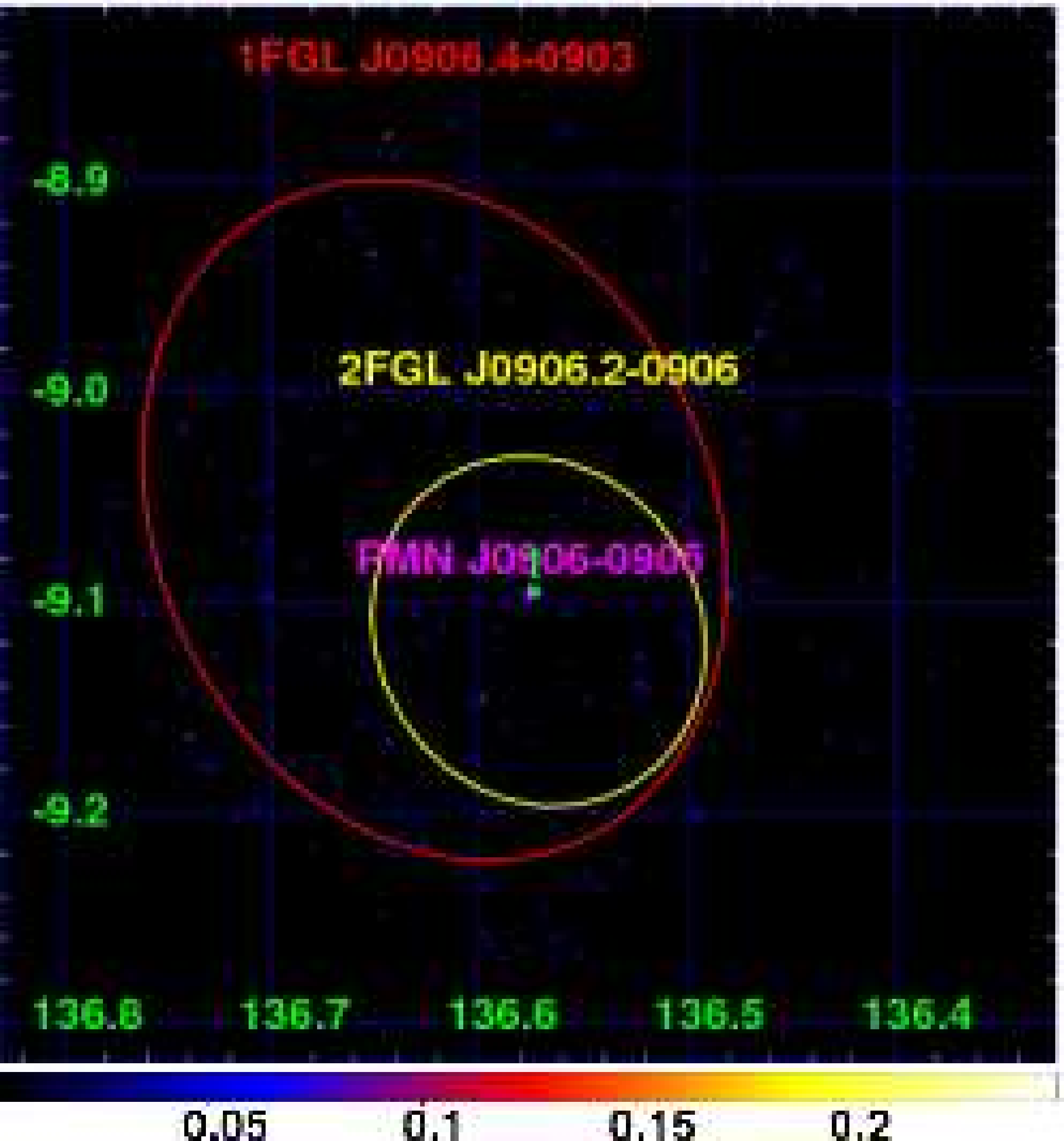}
    \end{center}
  \end{minipage}
  \begin{minipage}{0.32\hsize}
    \begin{center}
      {\small (55) 1FGL\,J0908.7--2119} \\
      \includegraphics[width=52mm]{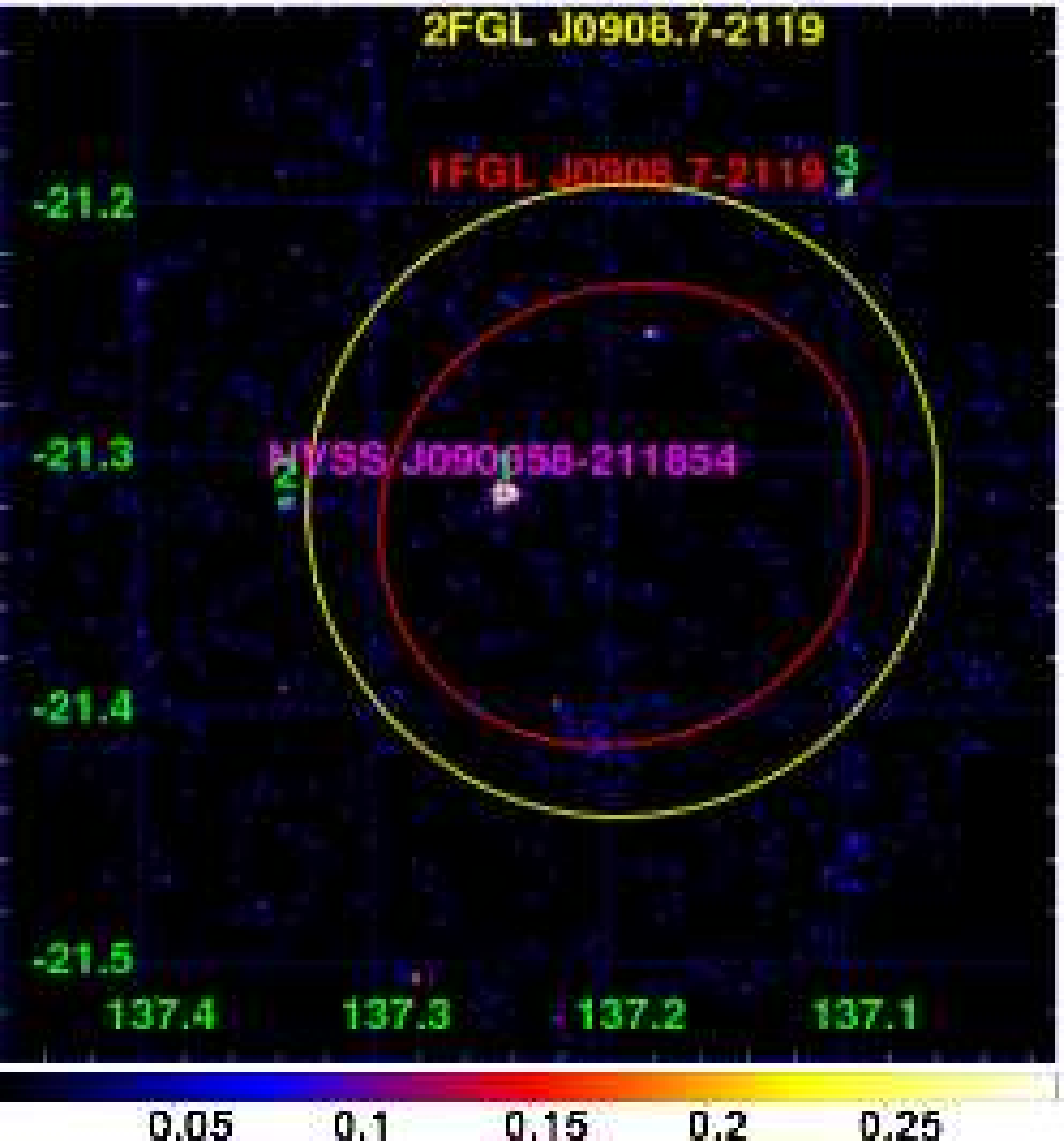}
    \end{center}
  \end{minipage}
  \begin{minipage}{0.32\hsize}
    \begin{center}
      {\small (56) 1FGL\,J0922.0$+$2337} \\
      \includegraphics[width=52mm]{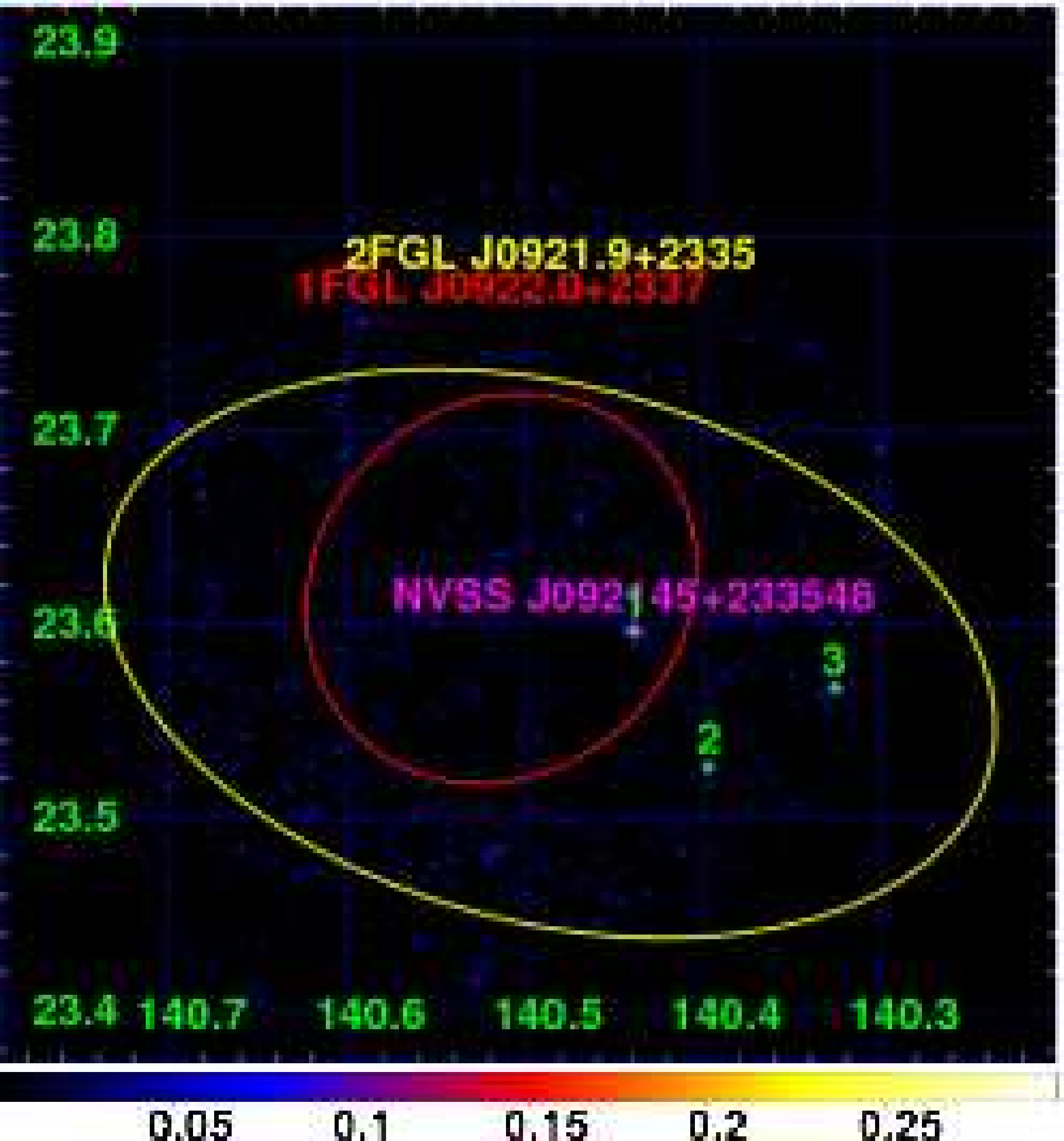}
    \end{center} 
  \end{minipage}
  \begin{minipage}{0.32\hsize}
    \begin{center}
      {\small (57) 1FGL\,J0940.2--7605} \\
      \includegraphics[width=52mm]{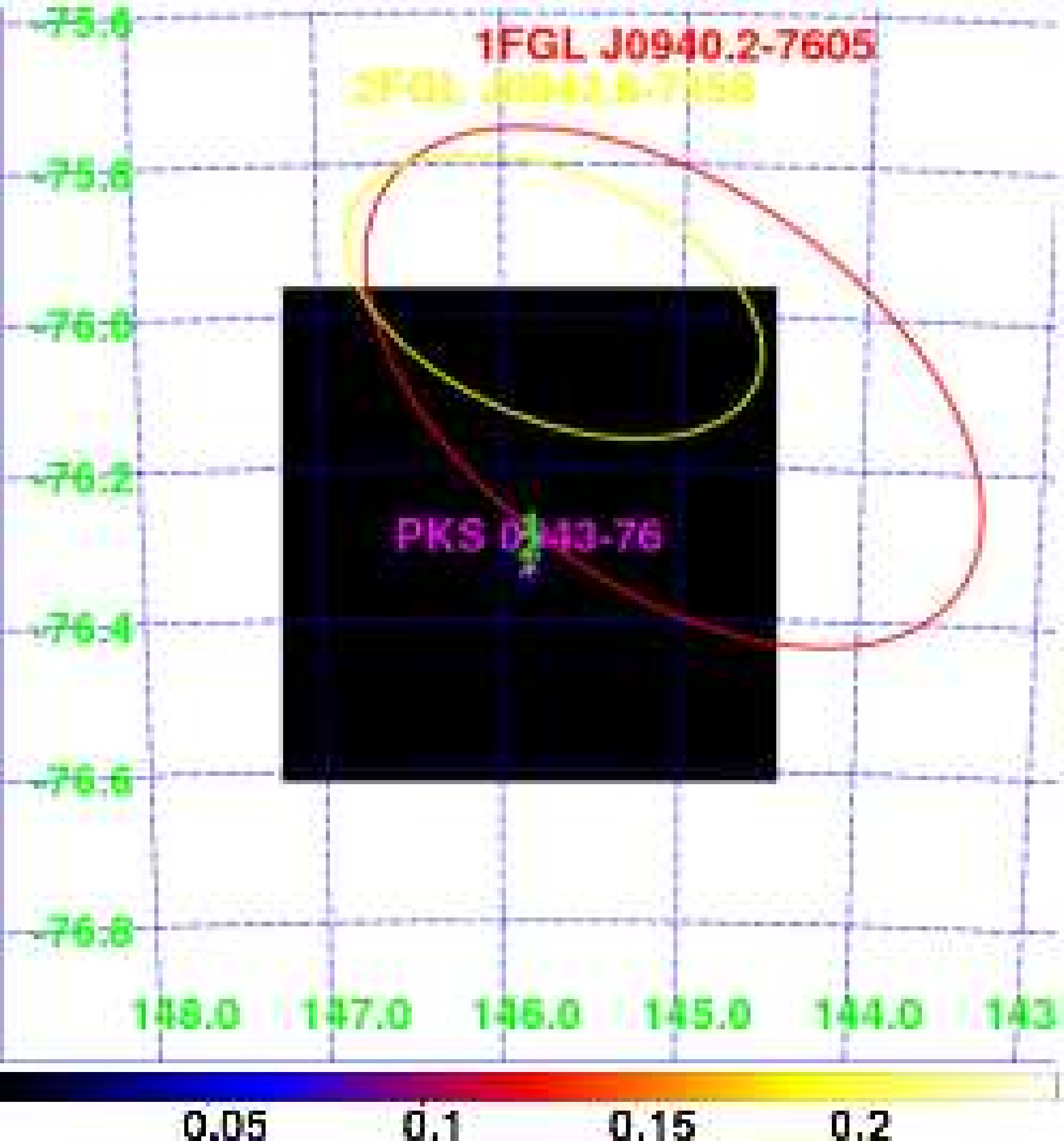}
    \end{center}
  \end{minipage}
  \begin{minipage}{0.32\hsize}
    \begin{center}
      {\small (58) 1FGL\,J0953.6--1505} \\
      \includegraphics[width=52mm]{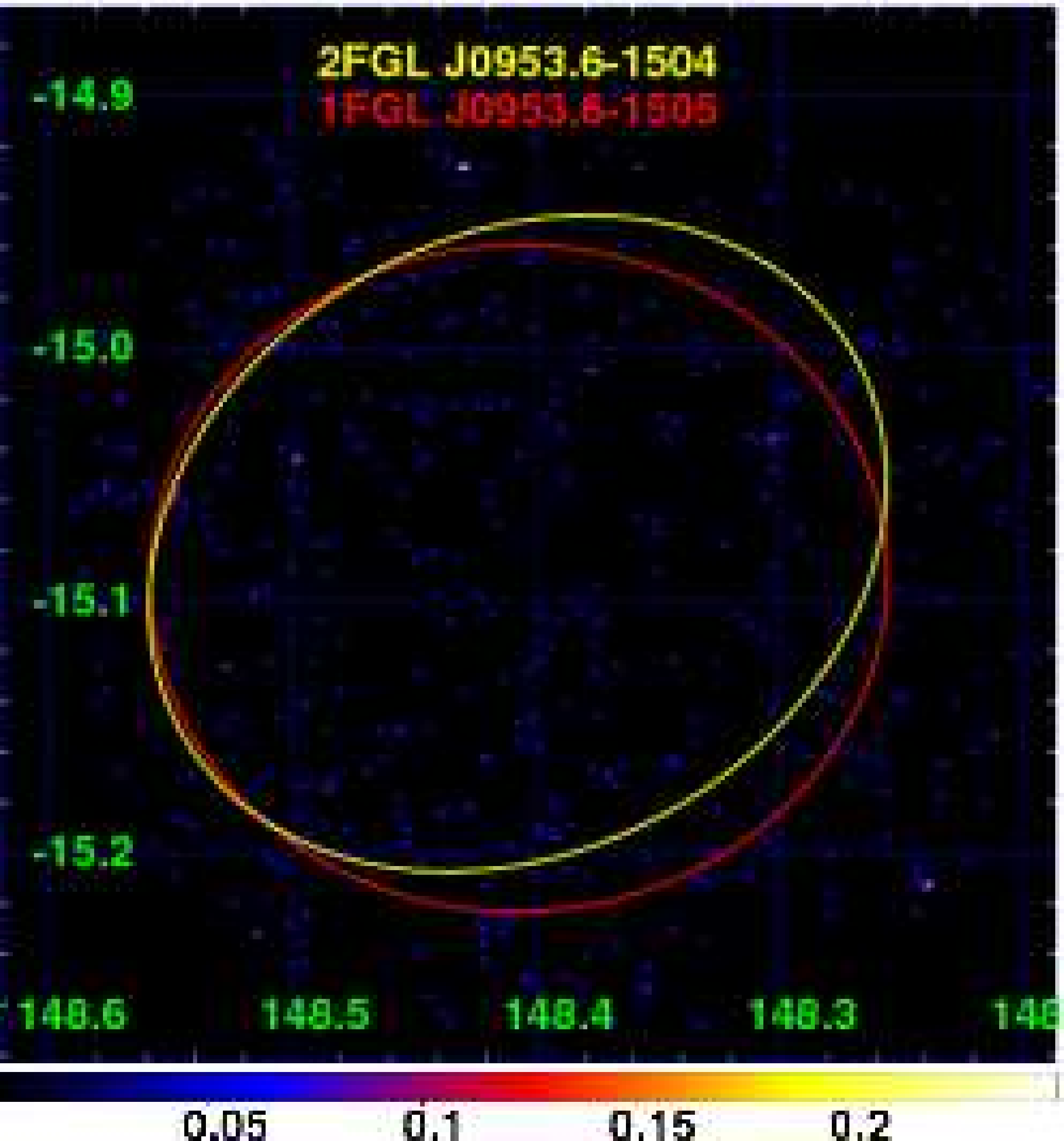}
    \end{center}
  \end{minipage}
  \begin{minipage}{0.32\hsize}
    \begin{center}
      {\small (59) 1FGL\,J0955.2--3949} \\
      \includegraphics[width=52mm]{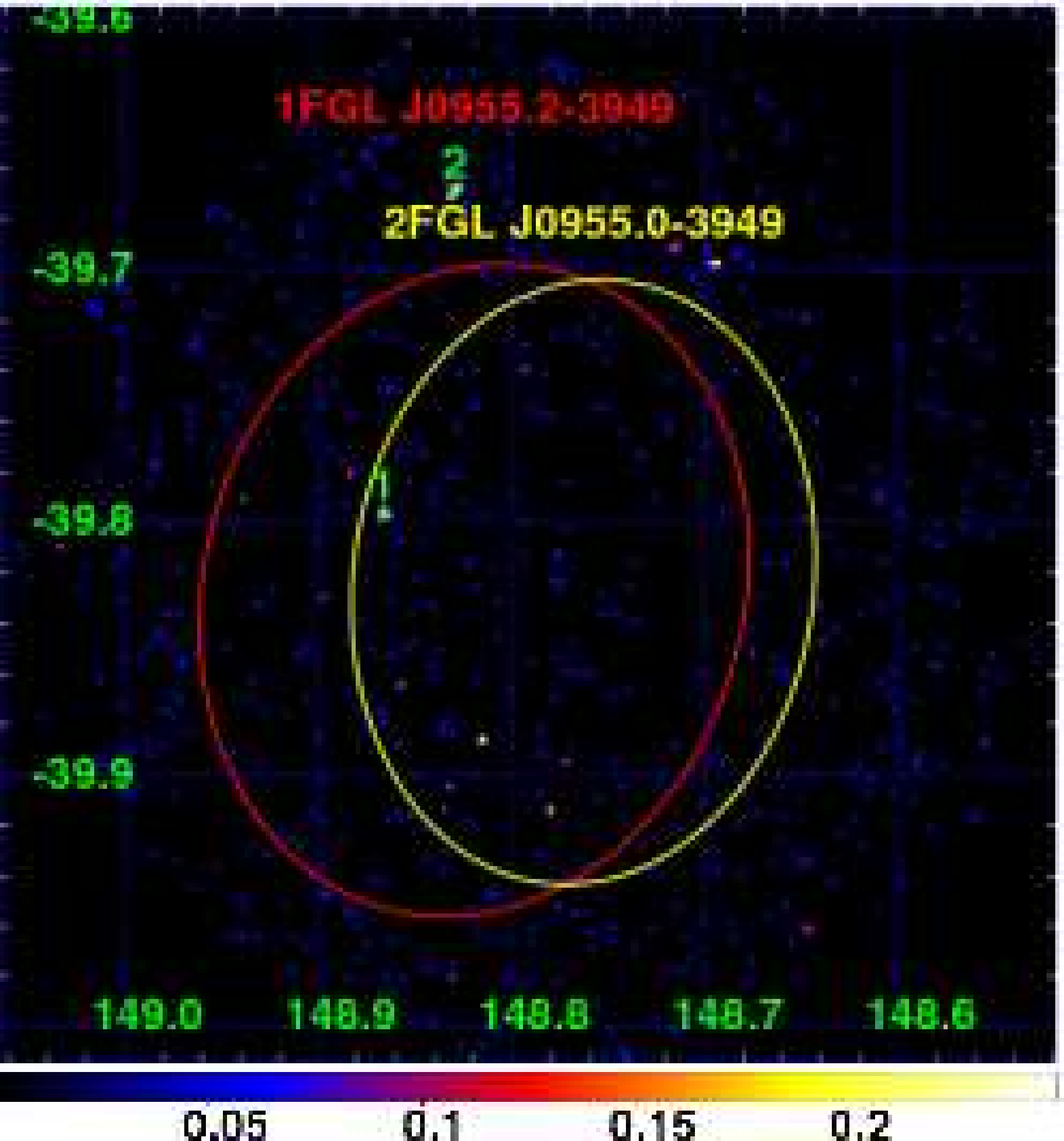}
    \end{center} 
  \end{minipage}
  \begin{minipage}{0.32\hsize}
    \begin{center}
      {\small (60) 1FGL\,J1040.5$+$0616} \\
      \includegraphics[width=52mm]{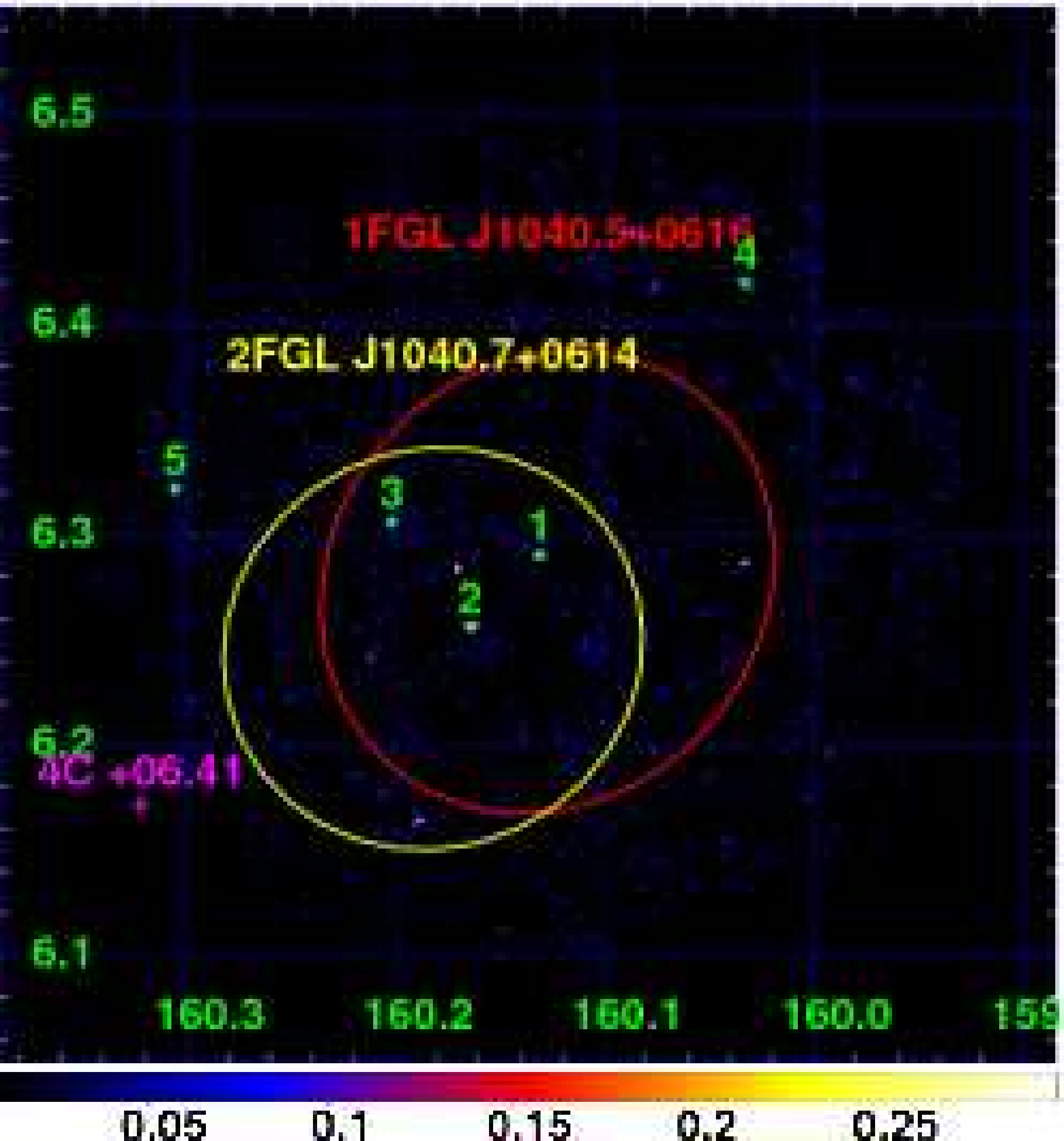}
    \end{center}
  \end{minipage}
 \end{center}
\end{figure}
\clearpage
\begin{figure}[m]
 \begin{center}
  \begin{minipage}{0.32\hsize}
    \begin{center}
      {\small (61) 1FGL\,J1048.7$+$2335} \\
      \includegraphics[width=52mm]{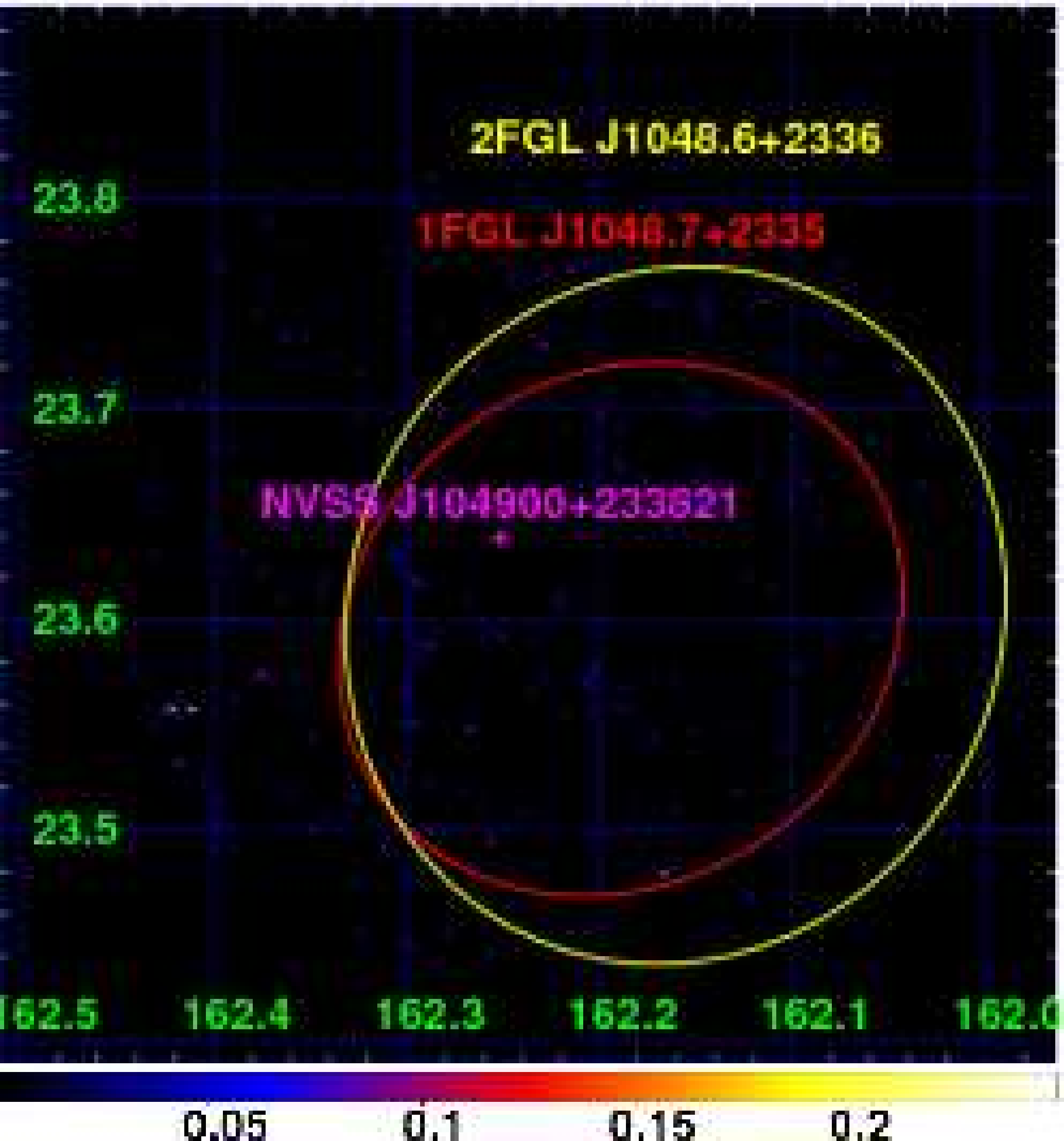}
    \end{center}
  \end{minipage}
  \begin{minipage}{0.32\hsize}
    \begin{center}
      {\small (62) 1FGL\,J1119.9--2205} \\
      \includegraphics[width=52mm]{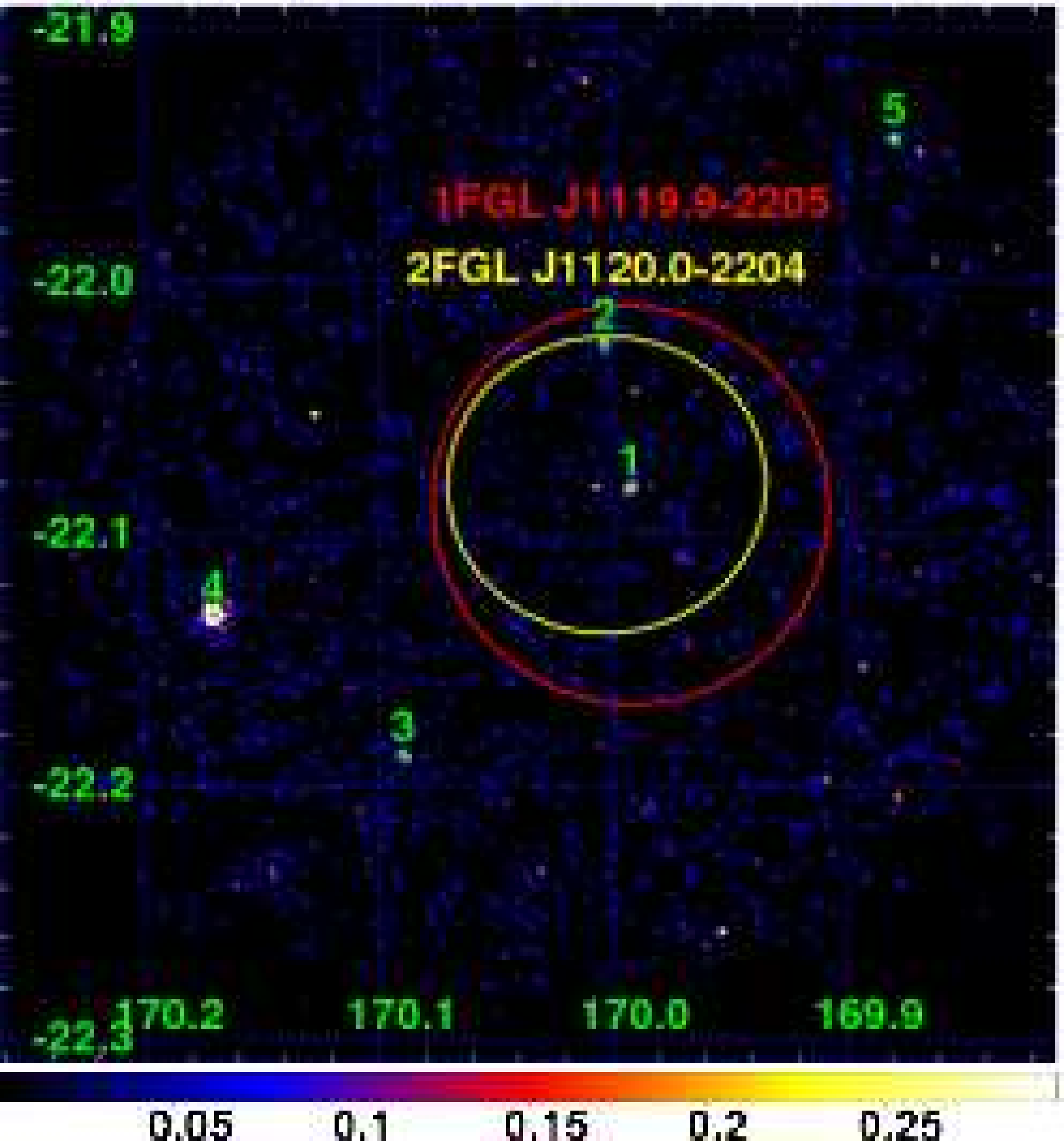}
    \end{center} 
  \end{minipage}
  \begin{minipage}{0.32\hsize}
    \begin{center}
      {\small (63) 1FGL\,J1124.4--3654} \\
      \includegraphics[width=52mm]{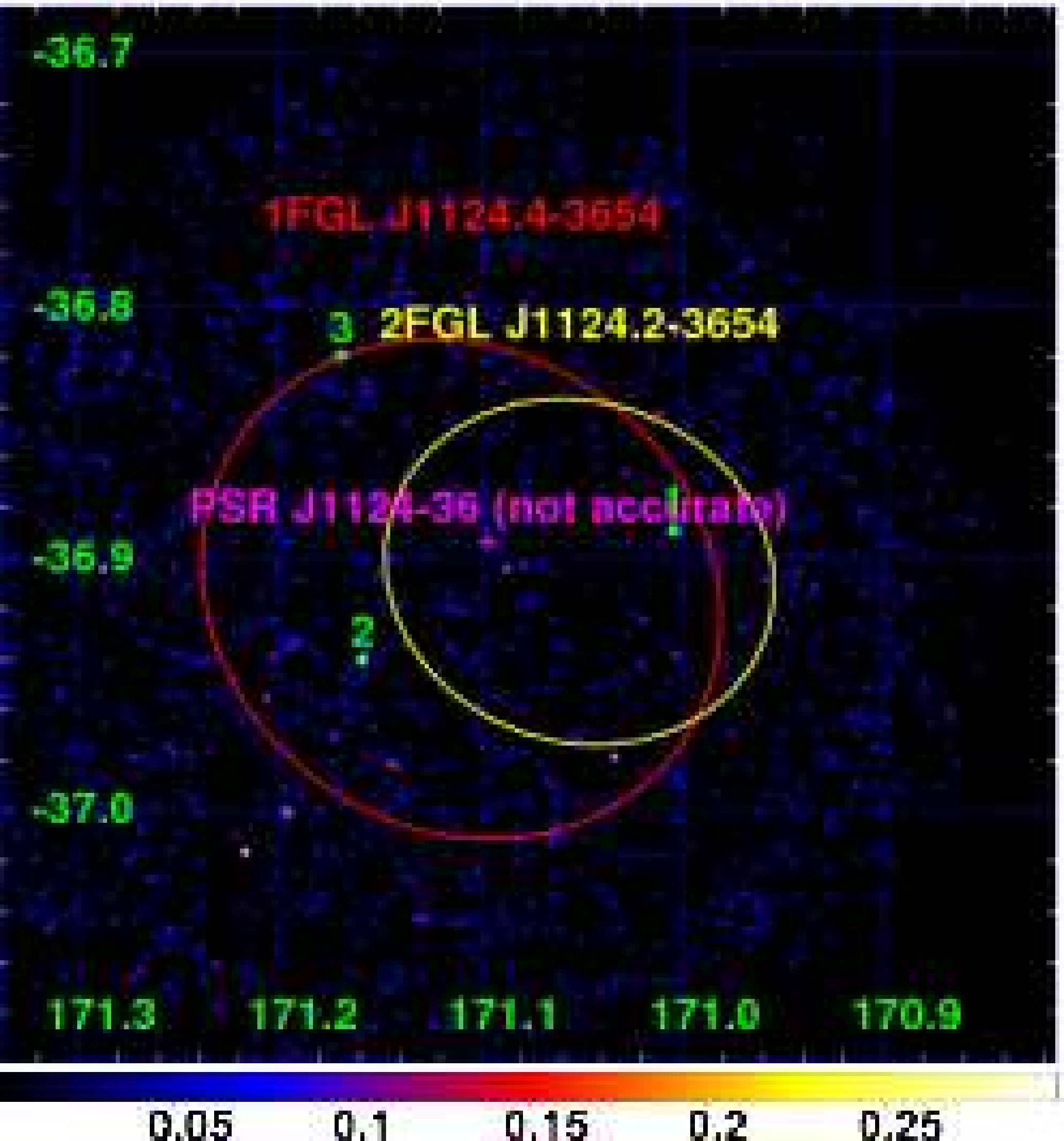}
    \end{center}
  \end{minipage}
  \begin{minipage}{0.32\hsize}
    \begin{center}
      {\small (64) 1FGL\,J1129.3$+$3757} \\
      \includegraphics[width=52mm]{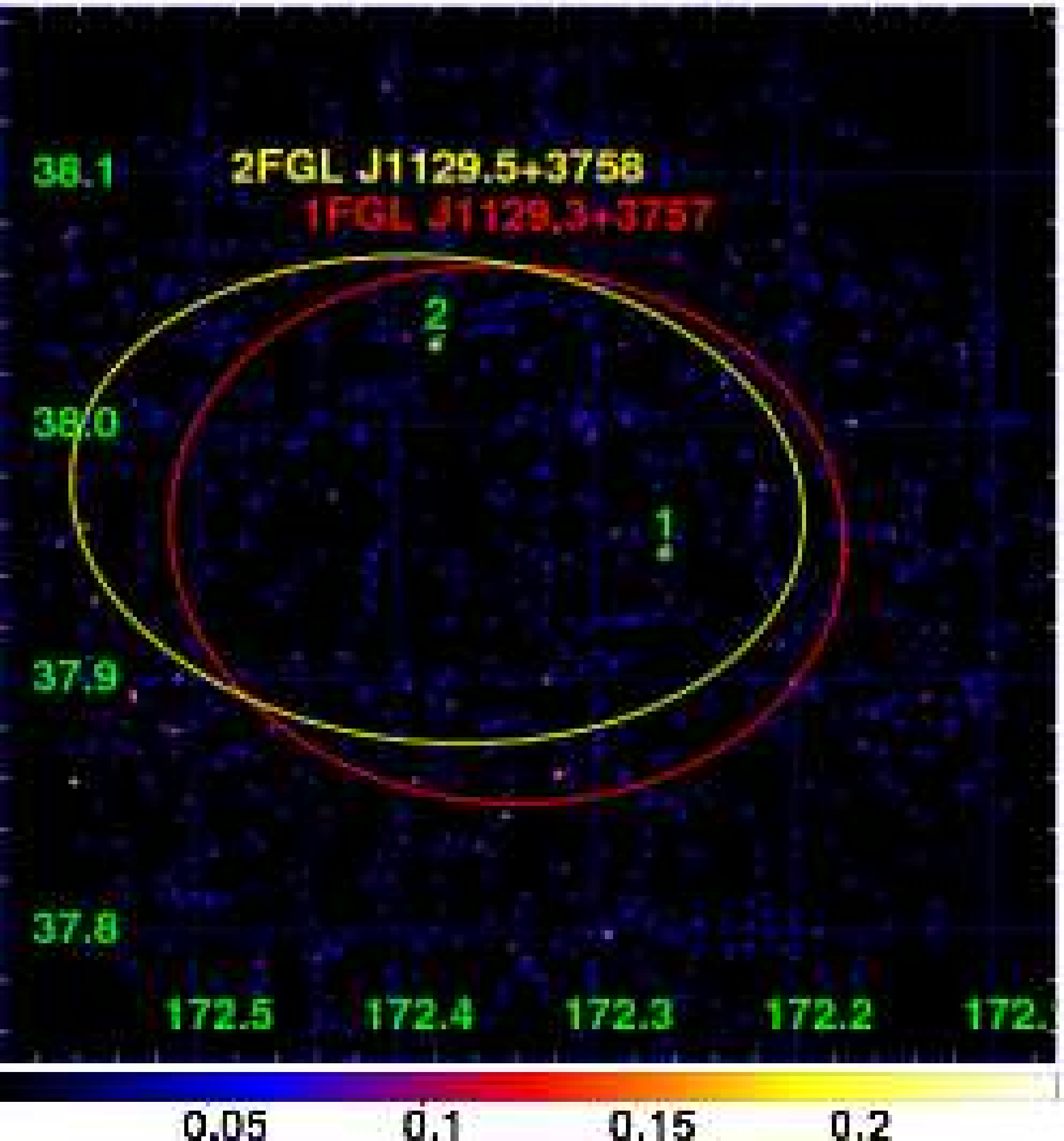}
    \end{center}
  \end{minipage}
  \begin{minipage}{0.32\hsize}
    \begin{center}
      {\small (65) 1FGL\,J1141.8--1403} \\
      \includegraphics[width=52mm]{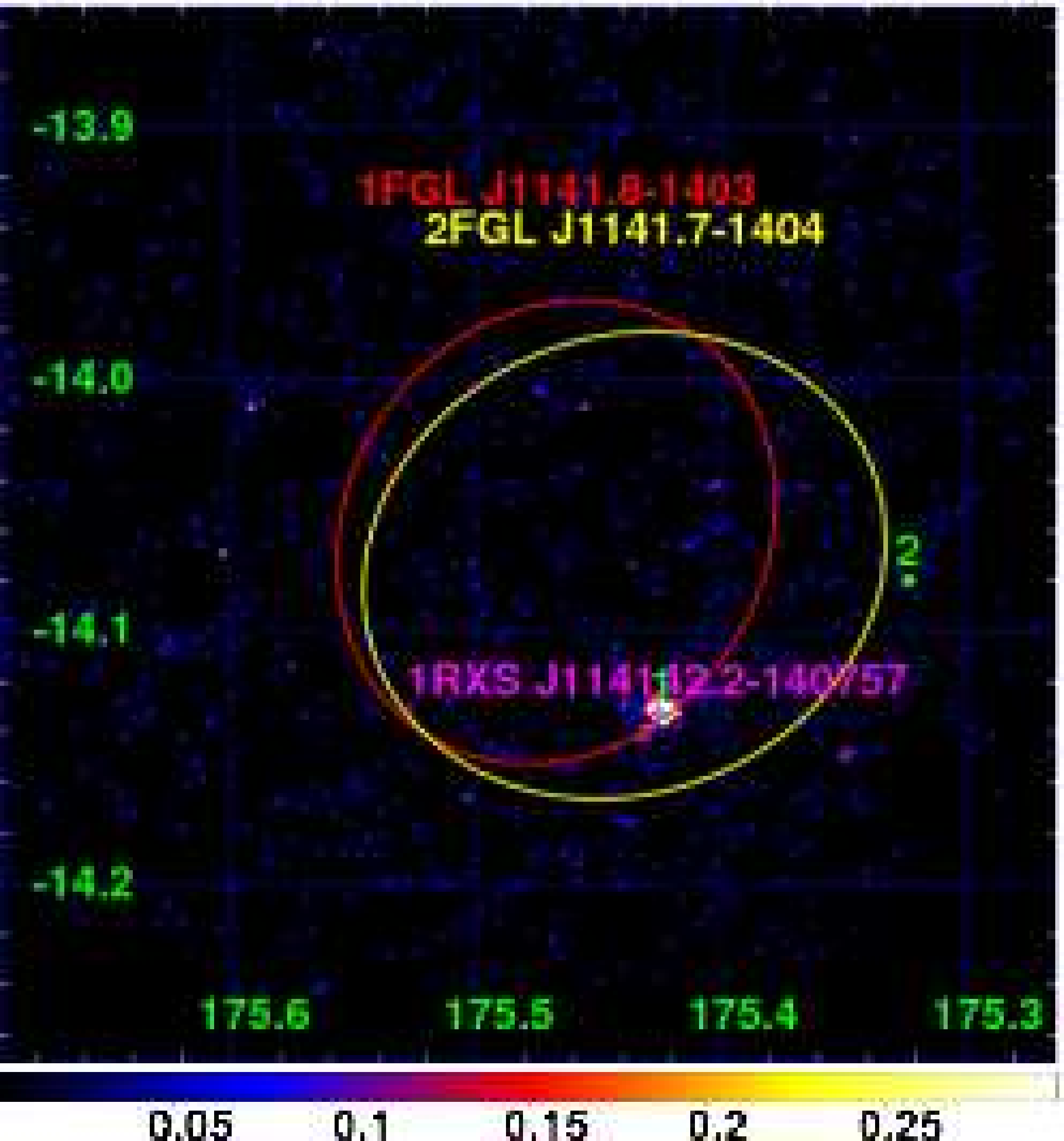}
    \end{center} 
  \end{minipage}
  \begin{minipage}{0.32\hsize}
    \begin{center}
      {\small (66) 1FGL\,J1159.8$+$0200} \\
      \includegraphics[width=52mm]{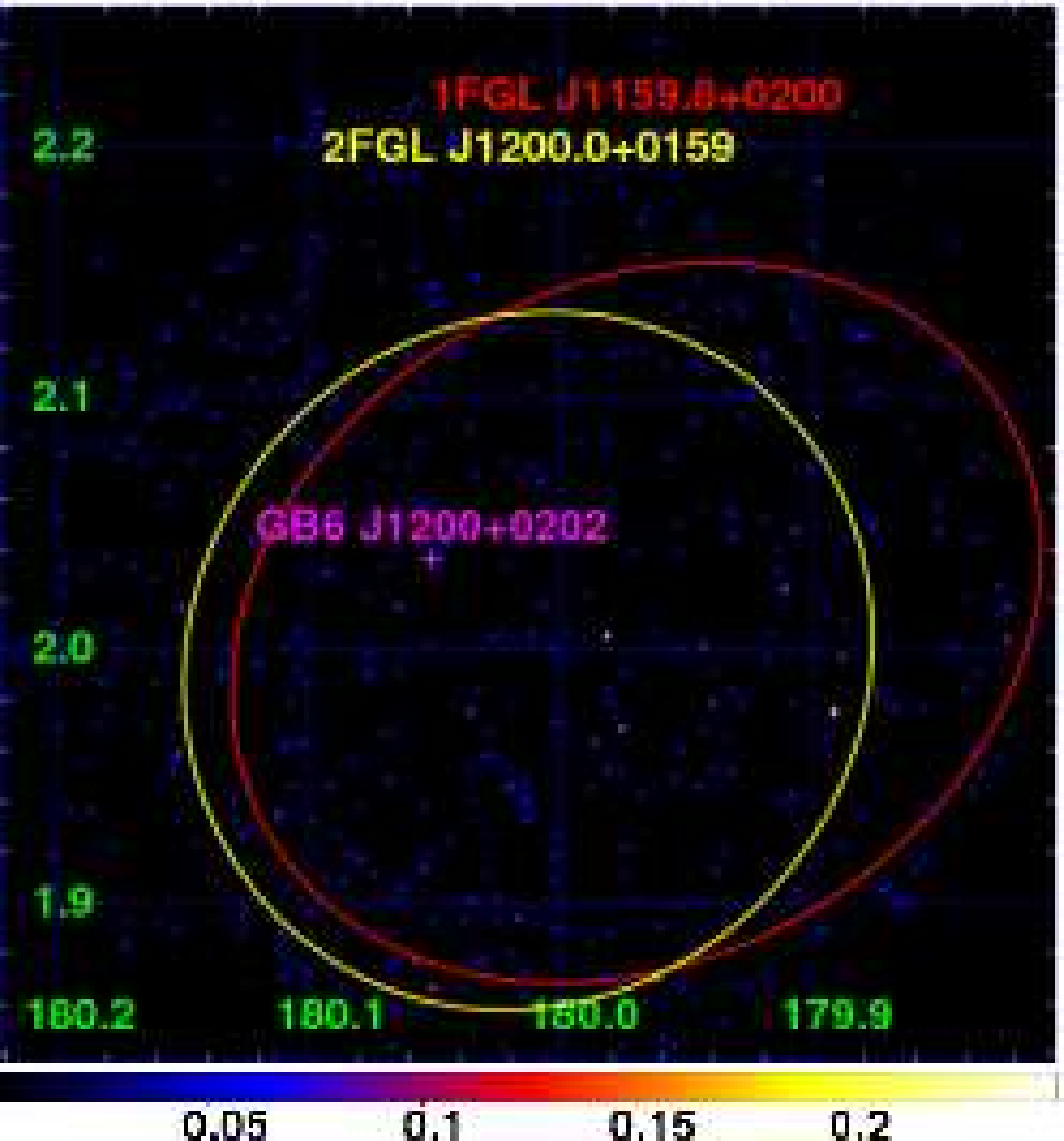}
    \end{center}
  \end{minipage}
  \begin{minipage}{0.32\hsize}
    \begin{center}
      {\small (67) 1FGL\,J1218.4--0128} \\
      \includegraphics[width=52mm]{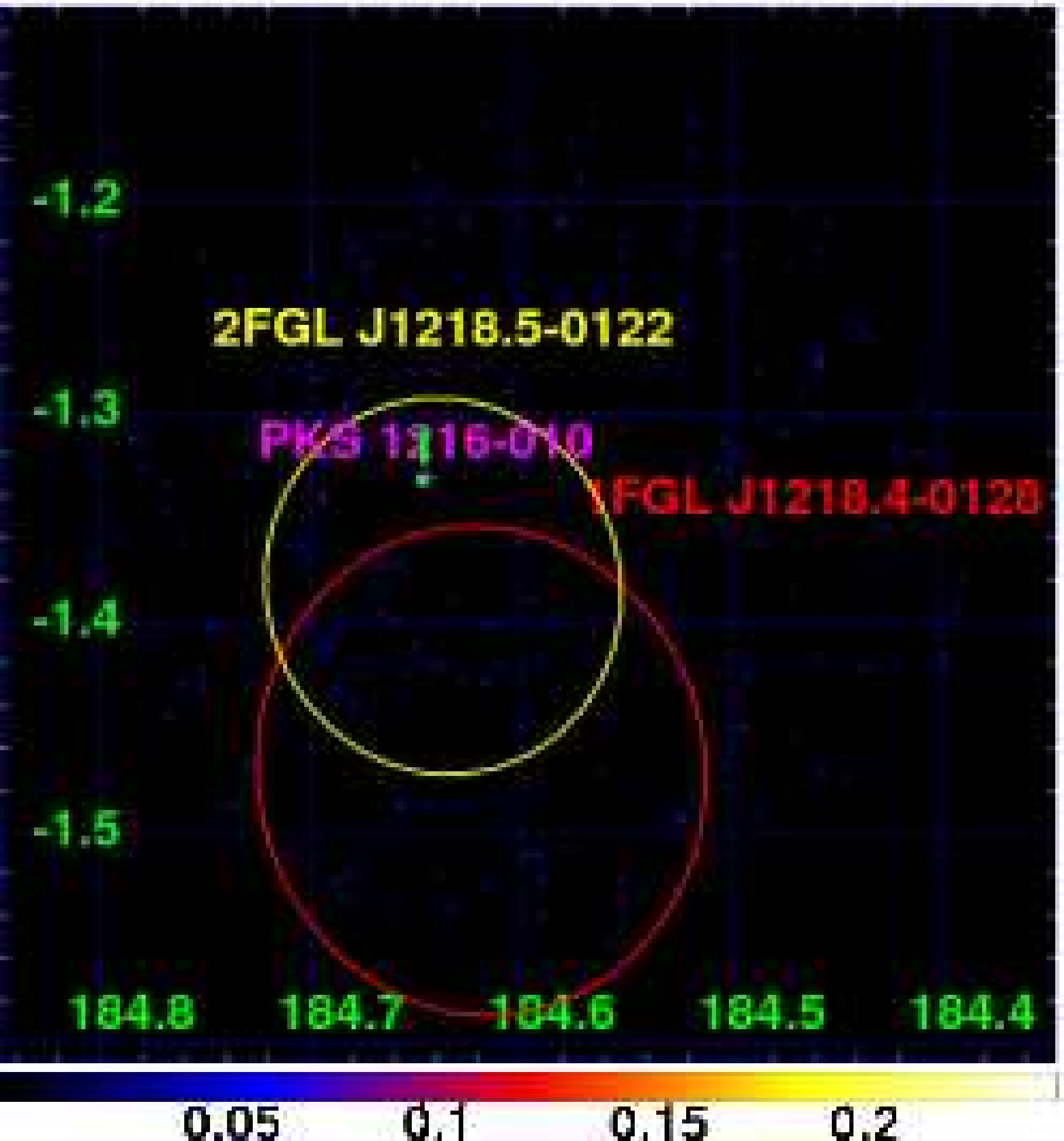}
    \end{center}
  \end{minipage}
  \begin{minipage}{0.32\hsize}
    \begin{center}
      {\small (68) 1FGL\,J1221.4--0635} \\
      \includegraphics[width=52mm]{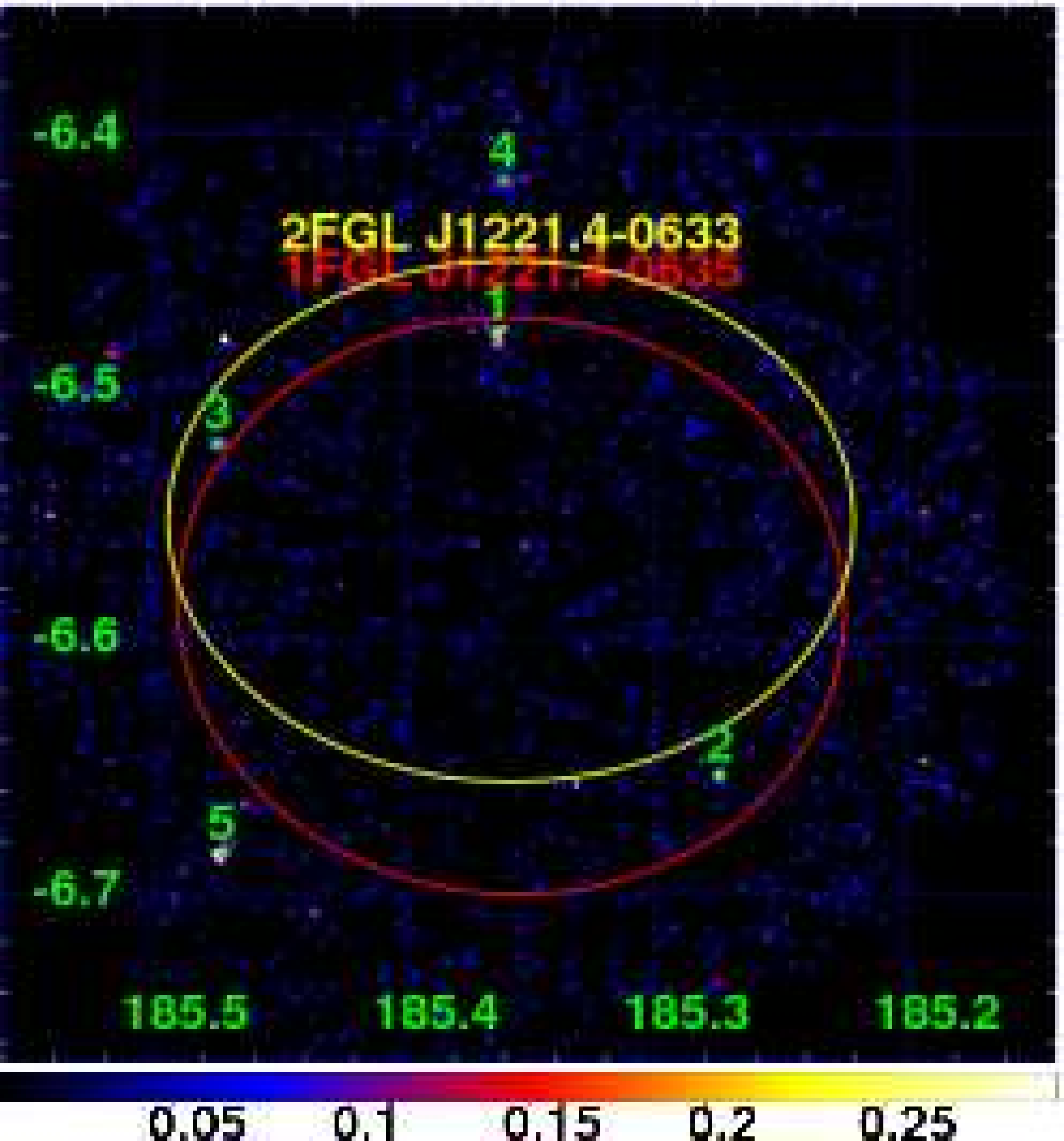}
    \end{center} 
  \end{minipage}
  \begin{minipage}{0.32\hsize}
    \begin{center}
      {\small (69) 1FGL\,J1226.0$+$2954} \\
      \includegraphics[width=52mm]{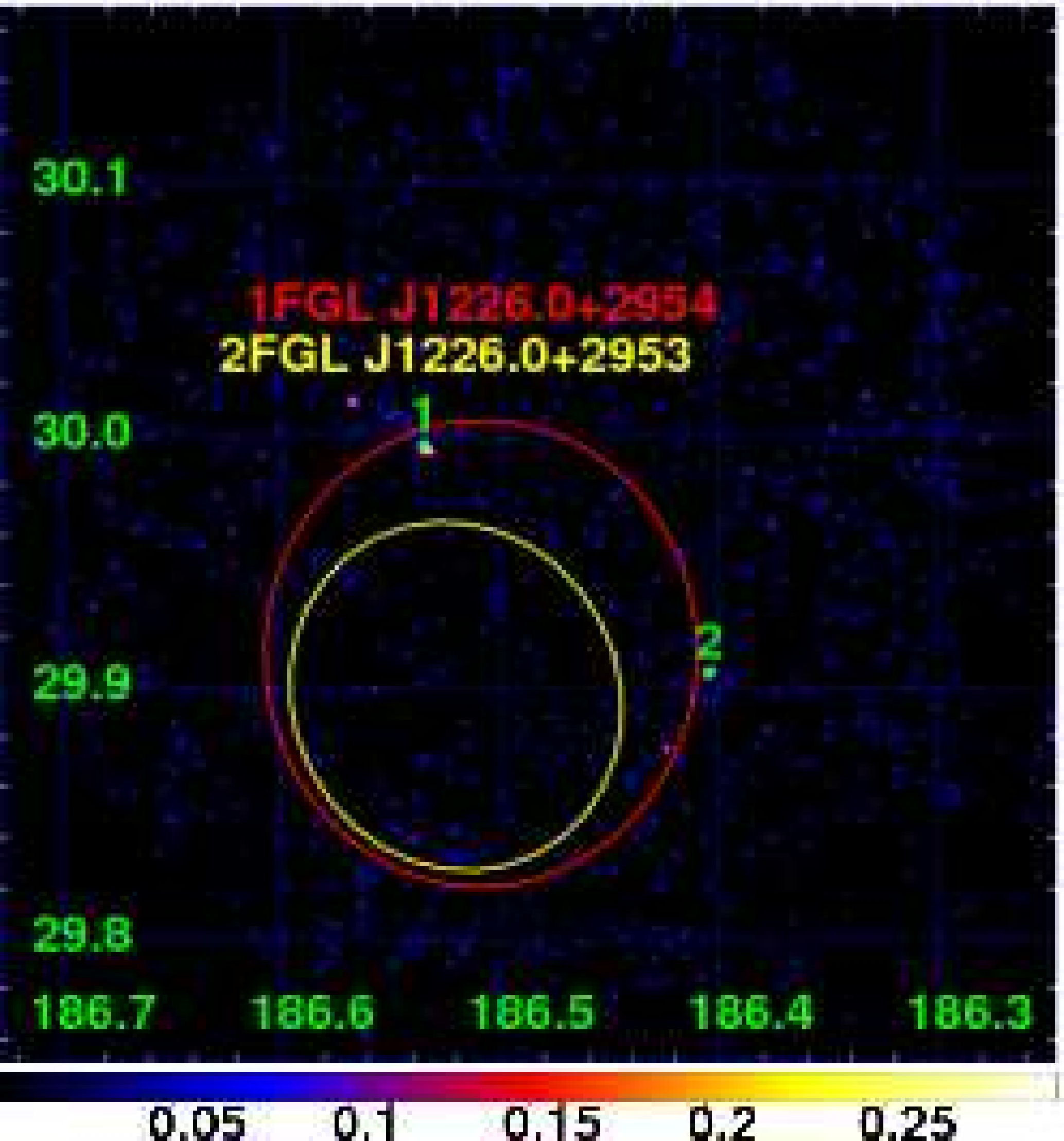}
    \end{center}
  \end{minipage}
  \begin{minipage}{0.32\hsize}
    \begin{center}
      {\small (70) 1FGL\,J1232.2--5118} \\
      \includegraphics[width=52mm]{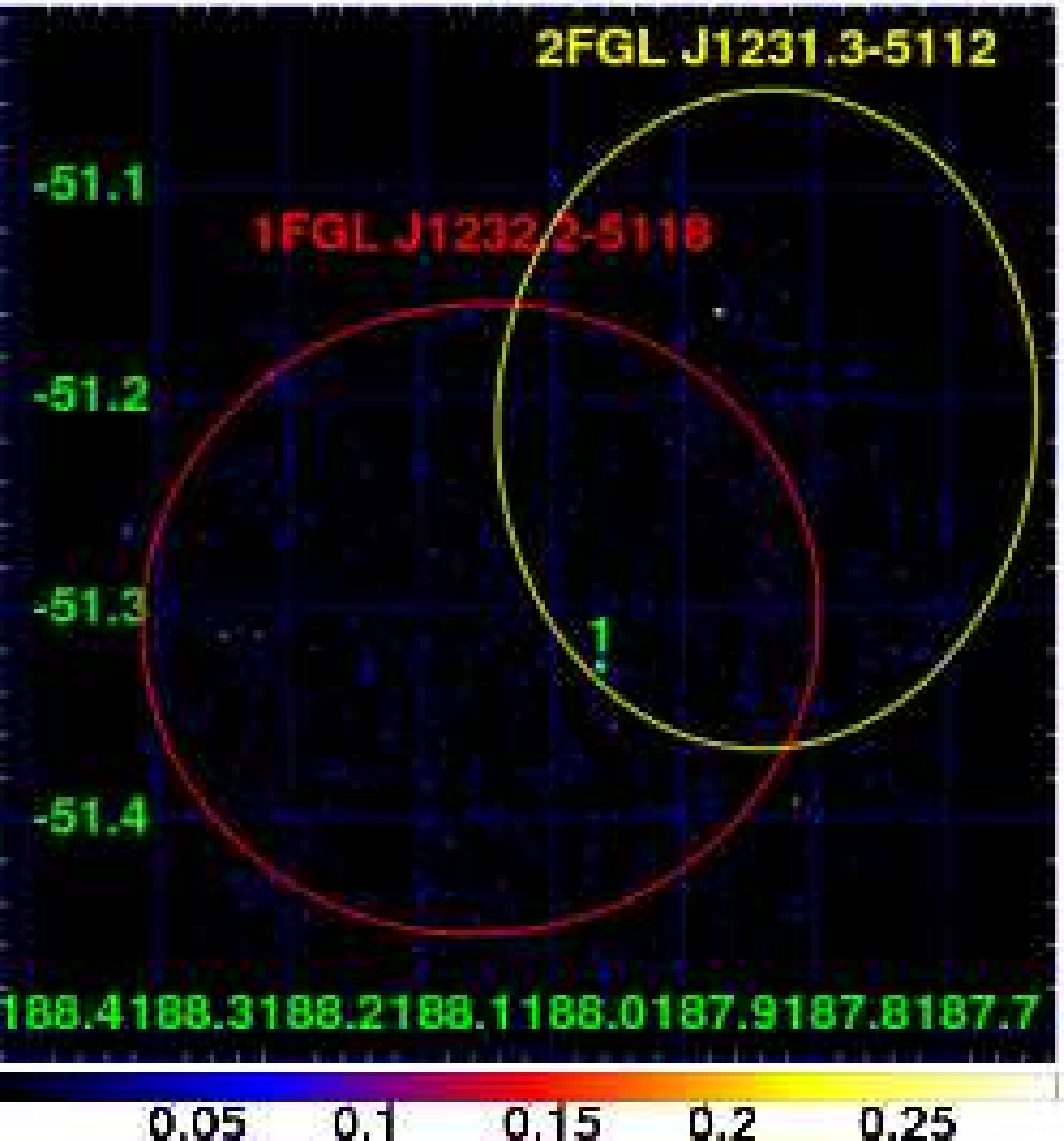}
    \end{center}
  \end{minipage}
  \begin{minipage}{0.32\hsize}
    \begin{center}
      {\small (71) 1FGL\,J1249.3--2812} \\
      \includegraphics[width=52mm]{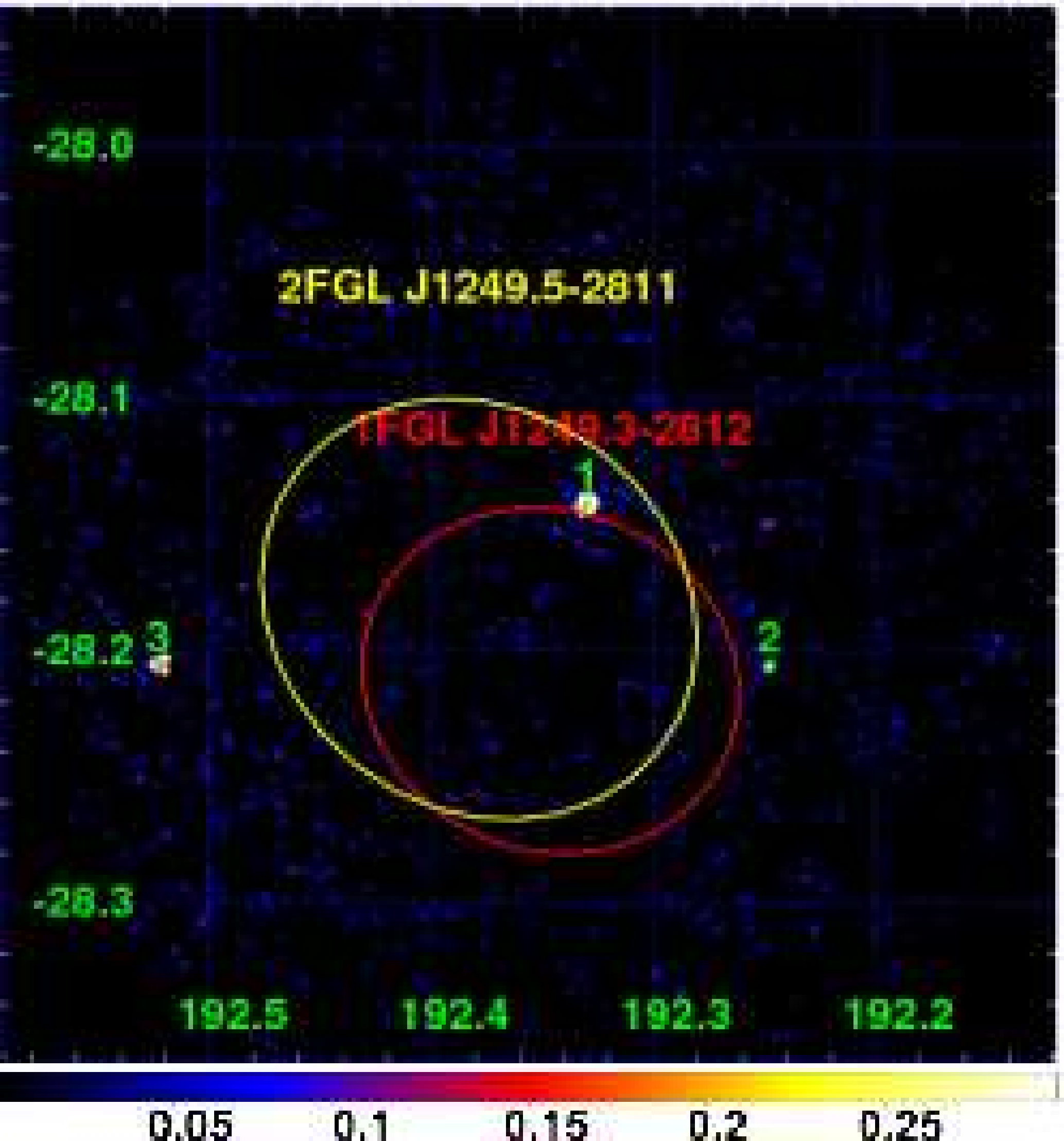}
    \end{center} 
  \end{minipage}
  \begin{minipage}{0.32\hsize}
    \begin{center}
      {\small (72) 1FGL\,J1251.3$+$1044} \\
      \includegraphics[width=52mm]{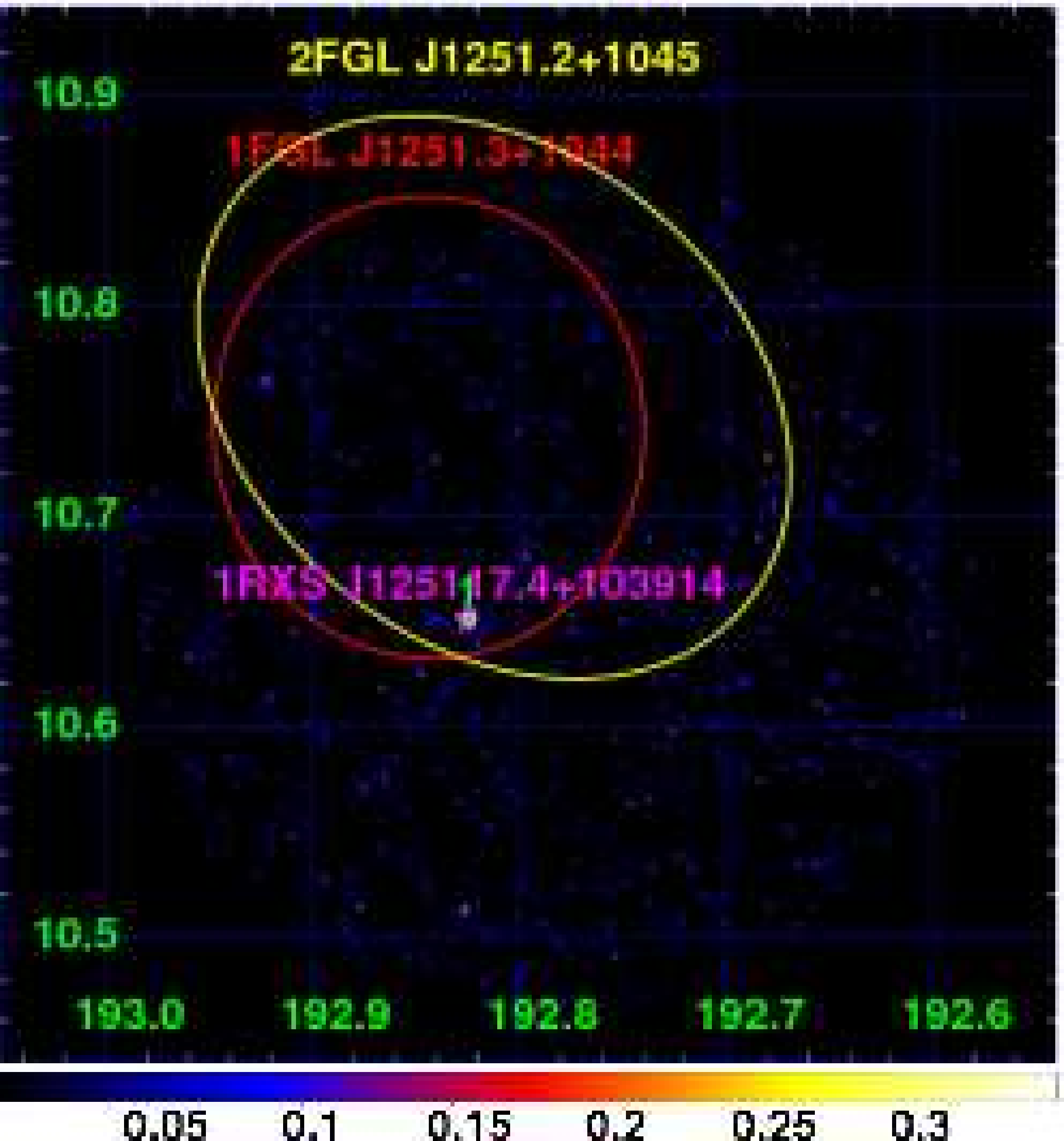}
    \end{center}
  \end{minipage}
 \end{center}
\end{figure}
\clearpage
\begin{figure}[m]
 \begin{center}
  \begin{minipage}{0.32\hsize}
    \begin{center}
      {\small (73) 1FGL\,J1254.4$+$2209} \\
      \includegraphics[width=52mm]{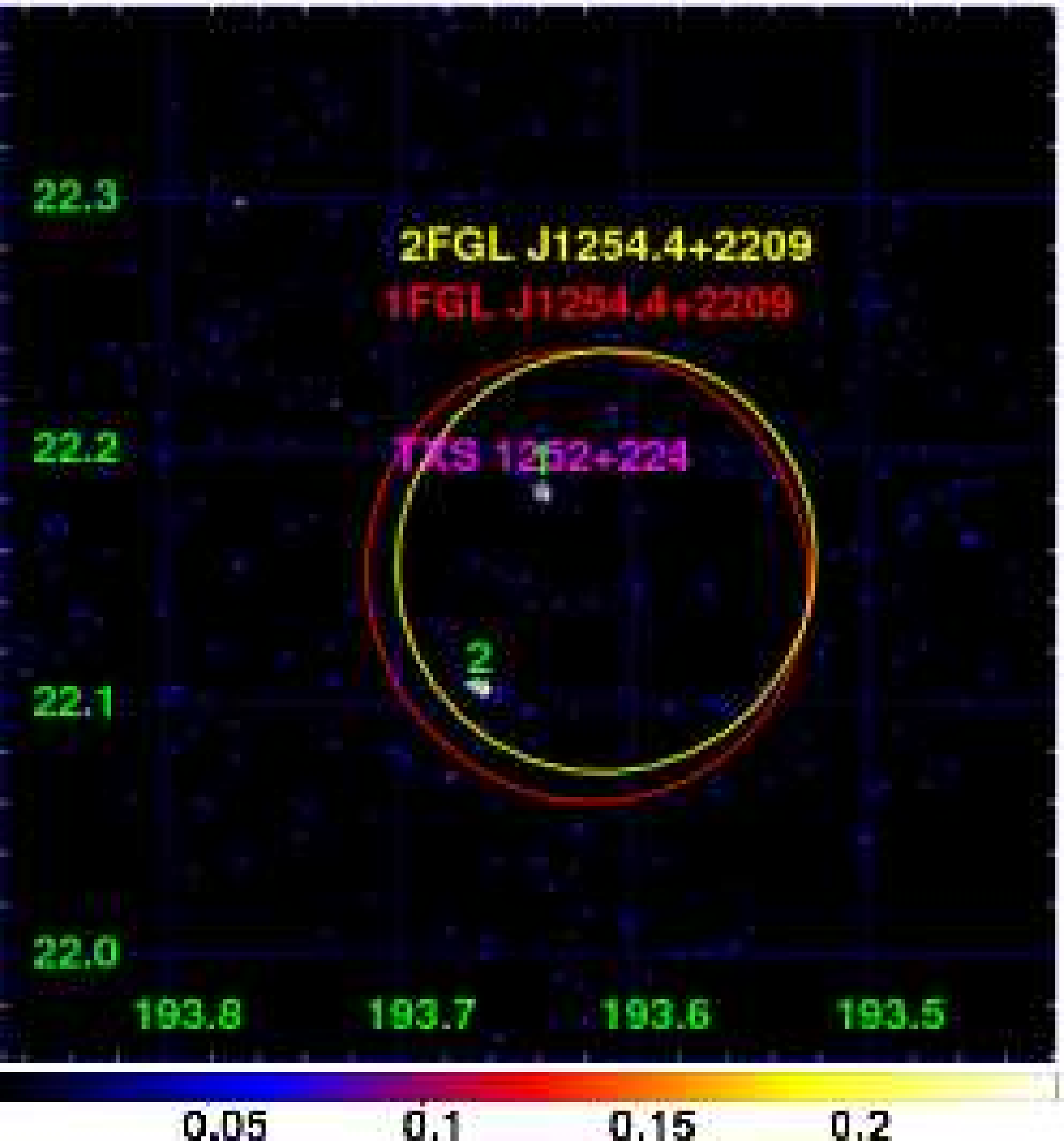}
    \end{center}
  \end{minipage}
  \begin{minipage}{0.32\hsize}
    \begin{center}
      {\small (74) 1FGL\,J1256.9$+$3650} \\
      \includegraphics[width=52mm]{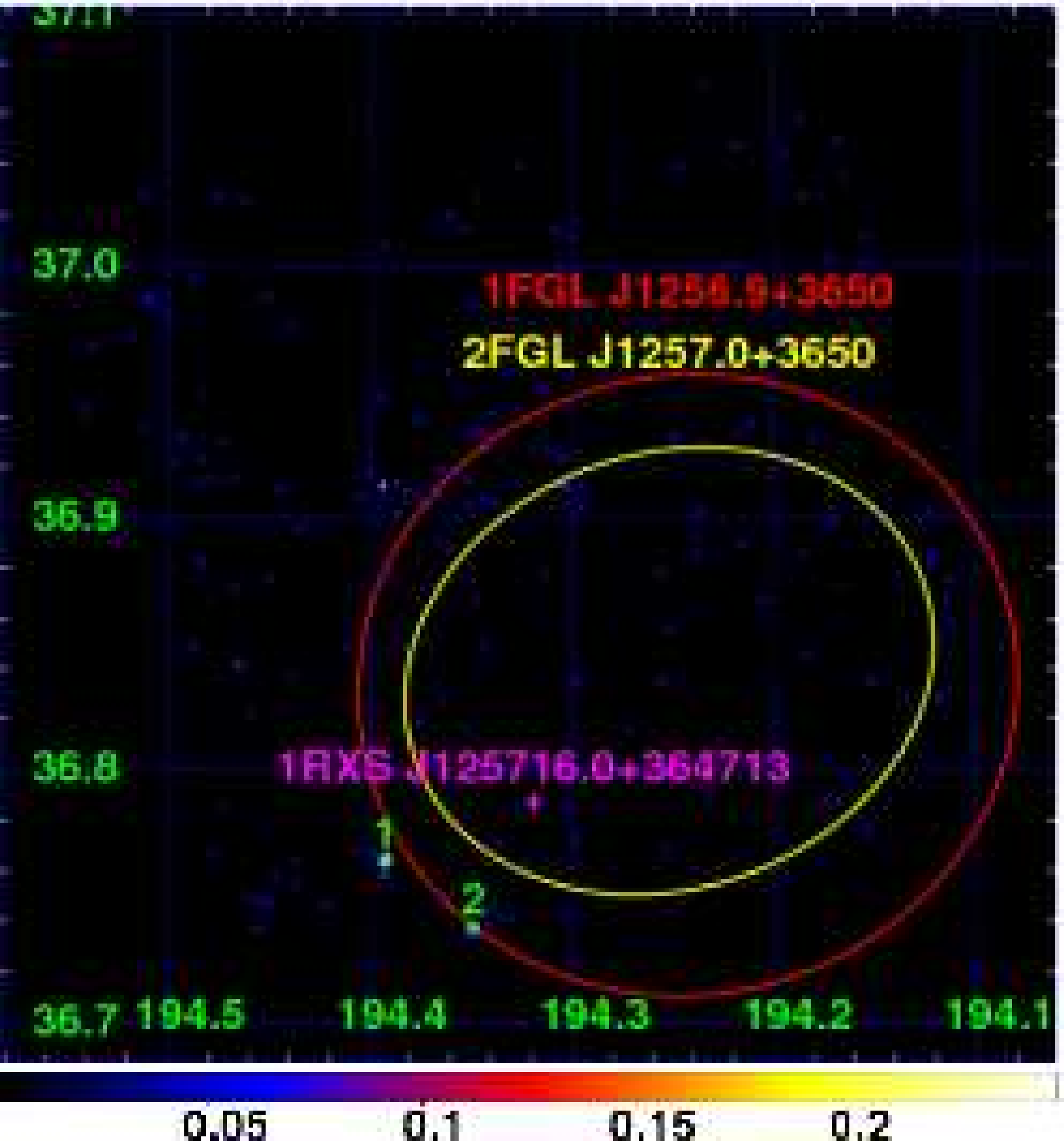}
    \end{center} 
  \end{minipage}
  \begin{minipage}{0.32\hsize}
    \begin{center}
      {\small (75) 1FGL\,J1301.8$+$0837} \\
      \includegraphics[width=52mm]{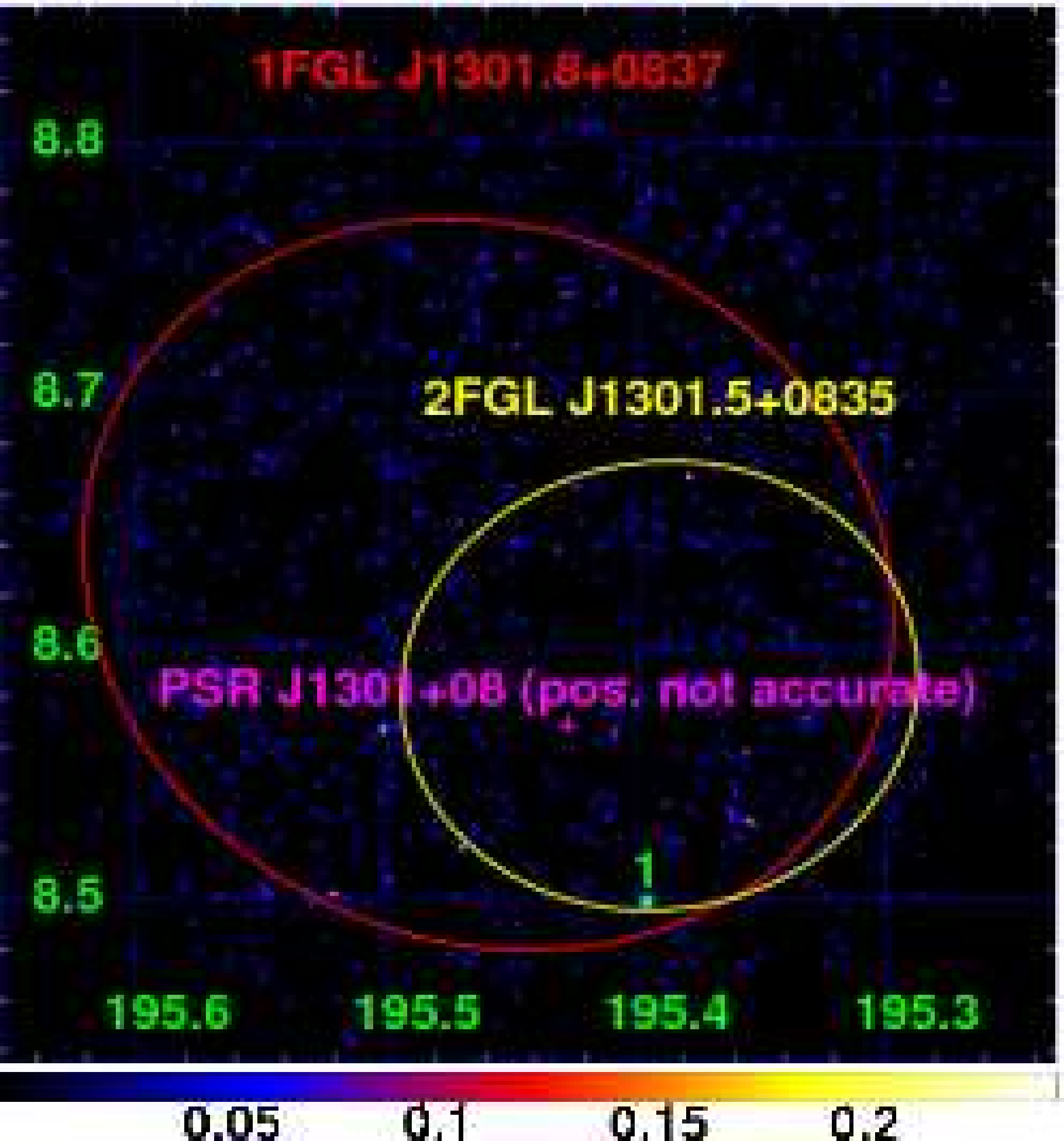}
    \end{center}
  \end{minipage}
  \begin{minipage}{0.32\hsize}
    \begin{center}
      {\small (76) 1FGL\,J1302.3--3255} \\
      \includegraphics[width=52mm]{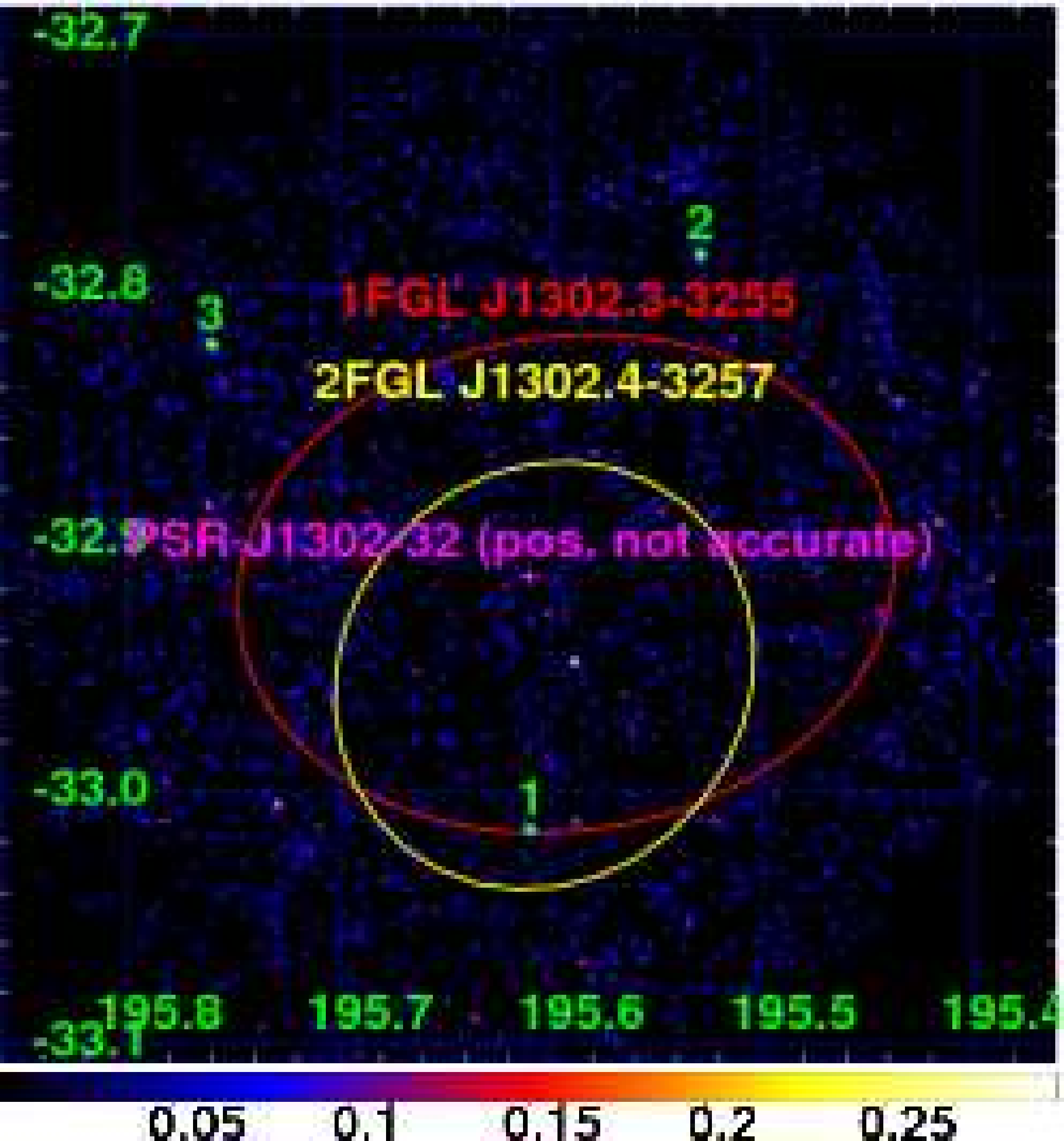}
    \end{center}
  \end{minipage}
  \begin{minipage}{0.32\hsize}
    \begin{center}
      {\small (77) 1FGL\,J1304.3--4352} \\
      \includegraphics[width=52mm]{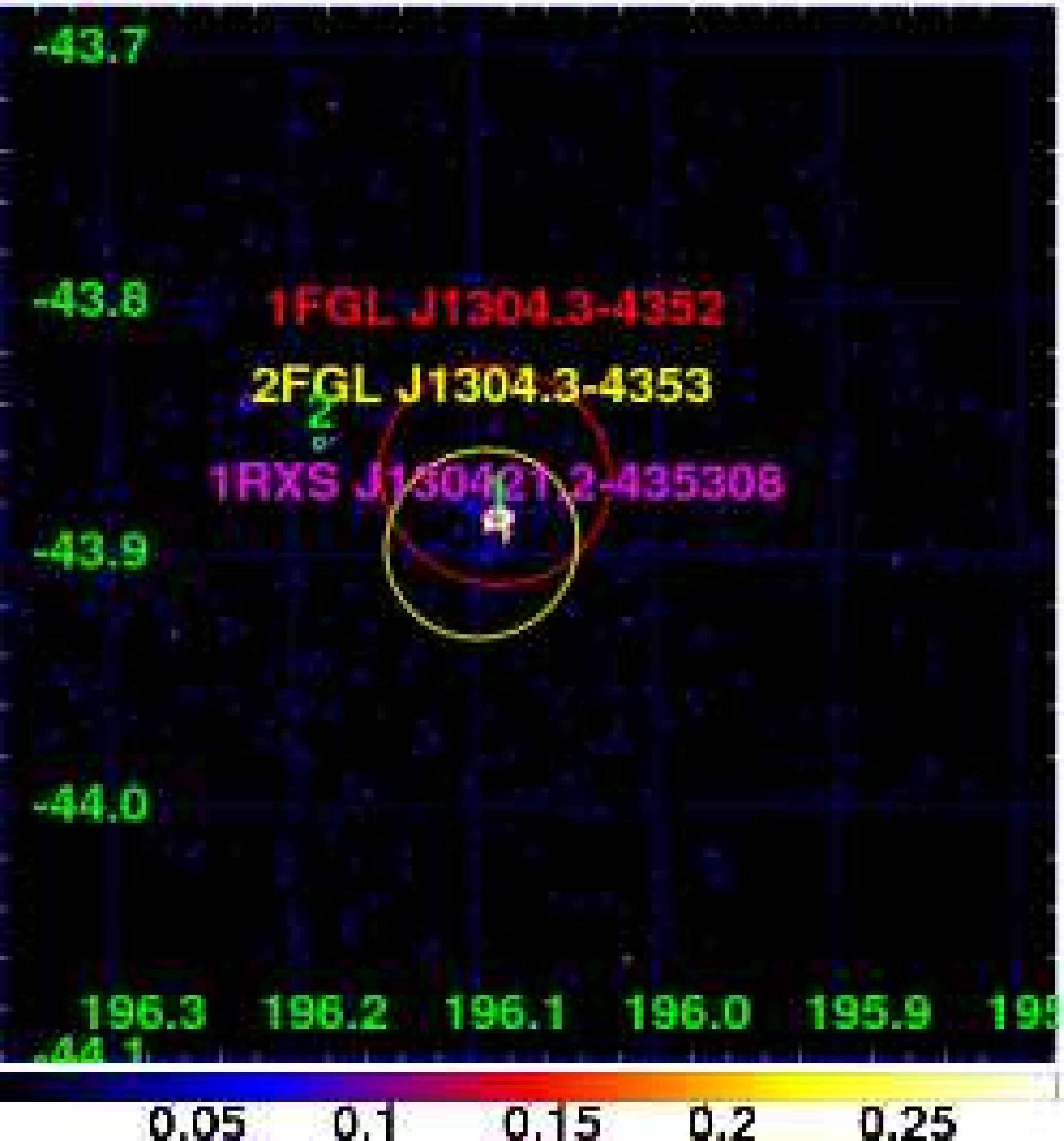}
    \end{center} 
  \end{minipage}
  \begin{minipage}{0.32\hsize}
    \begin{center}
      {\small (78) 1FGL\,J1307.0--4030} \\
      \includegraphics[width=52mm]{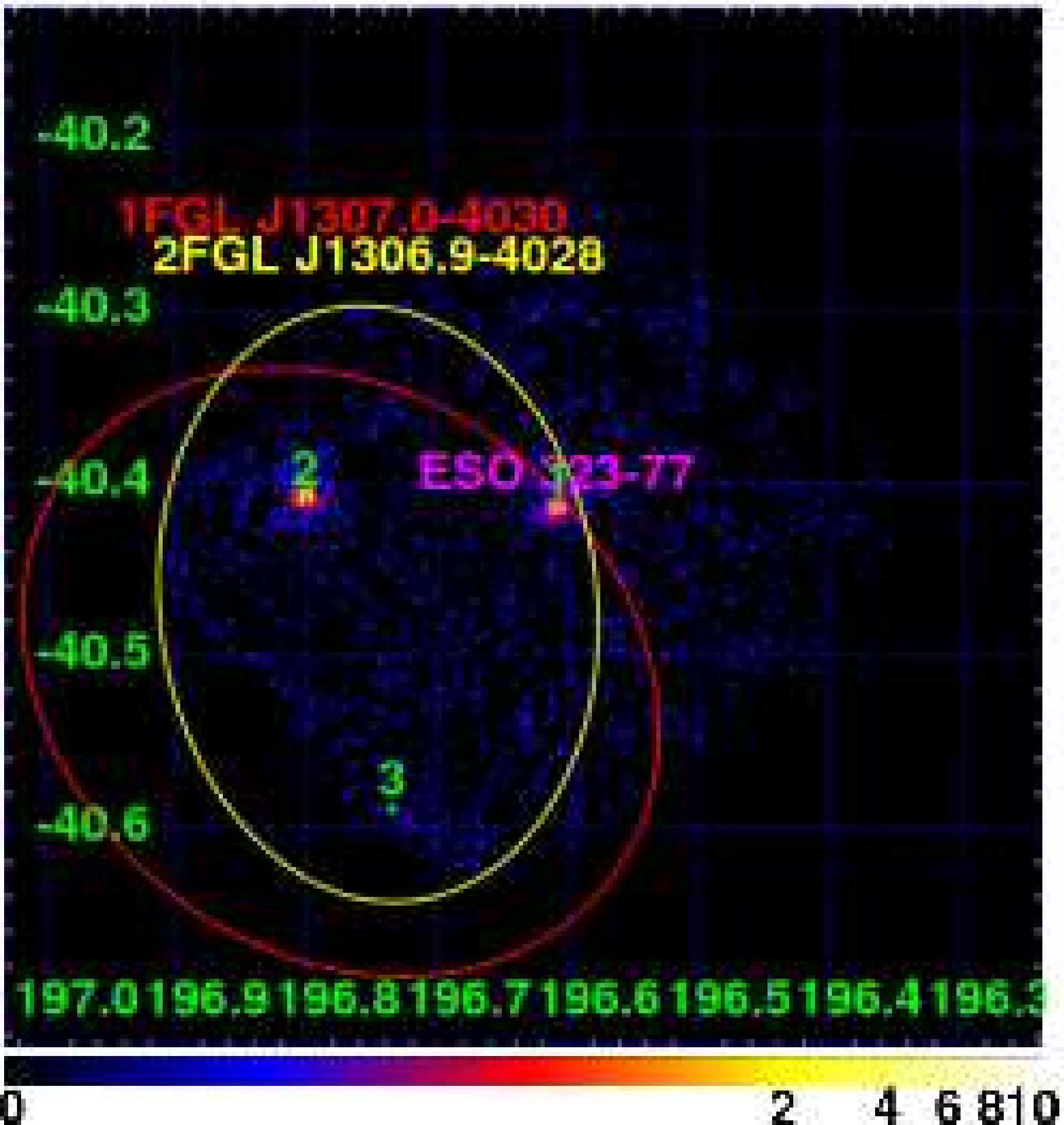}
    \end{center}
  \end{minipage}
  \begin{minipage}{0.32\hsize}
    \begin{center}
      {\small (79) 1FGL\,J1307.6--4259} \\
      \includegraphics[width=52mm]{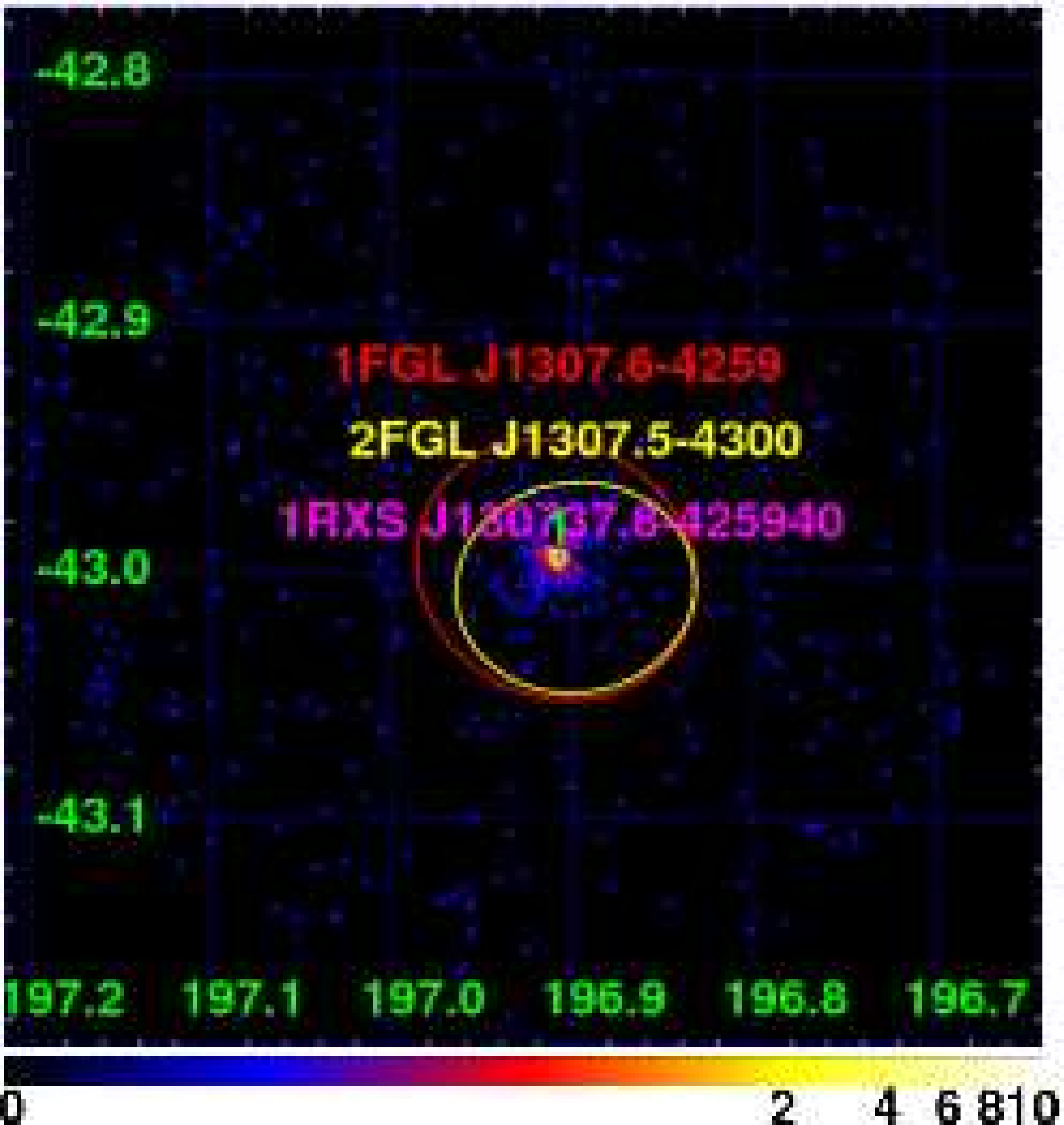}
    \end{center}
  \end{minipage}
  \begin{minipage}{0.32\hsize}
    \begin{center}
      {\small (80) 1FGL\,J1311.7--3429} \\
      \includegraphics[width=52mm]{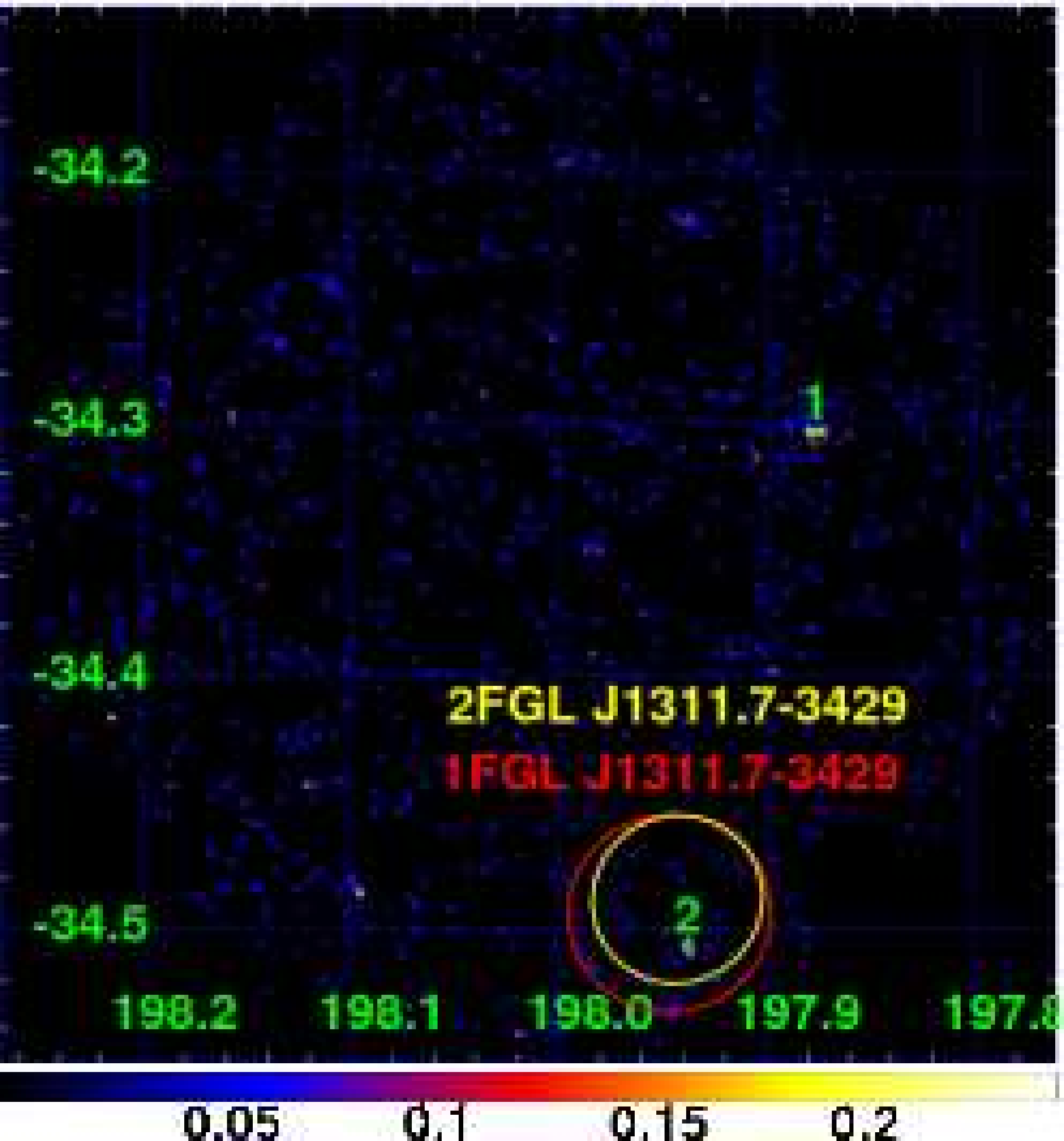}
    \end{center}
  \end{minipage}
  \begin{minipage}{0.32\hsize}
    \begin{center}
      {\small (81) 1FGL\,J1312.6$+$0048} \\
      \includegraphics[width=52mm]{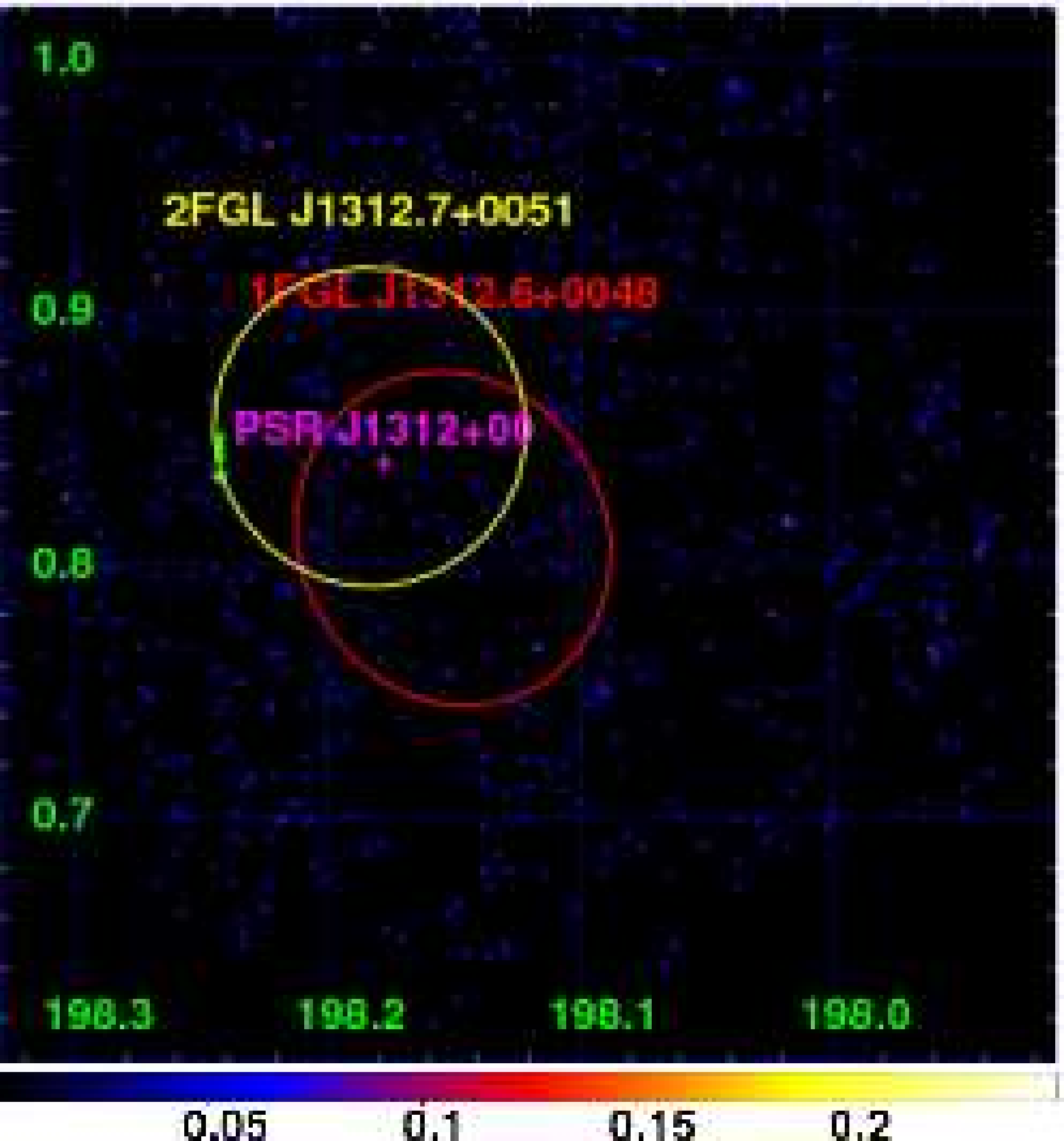}
    \end{center}
  \end{minipage}
  \begin{minipage}{0.32\hsize}
    \begin{center}
      {\small (82) 1FGL\,J1315.6--0729} \\
      \includegraphics[width=52mm]{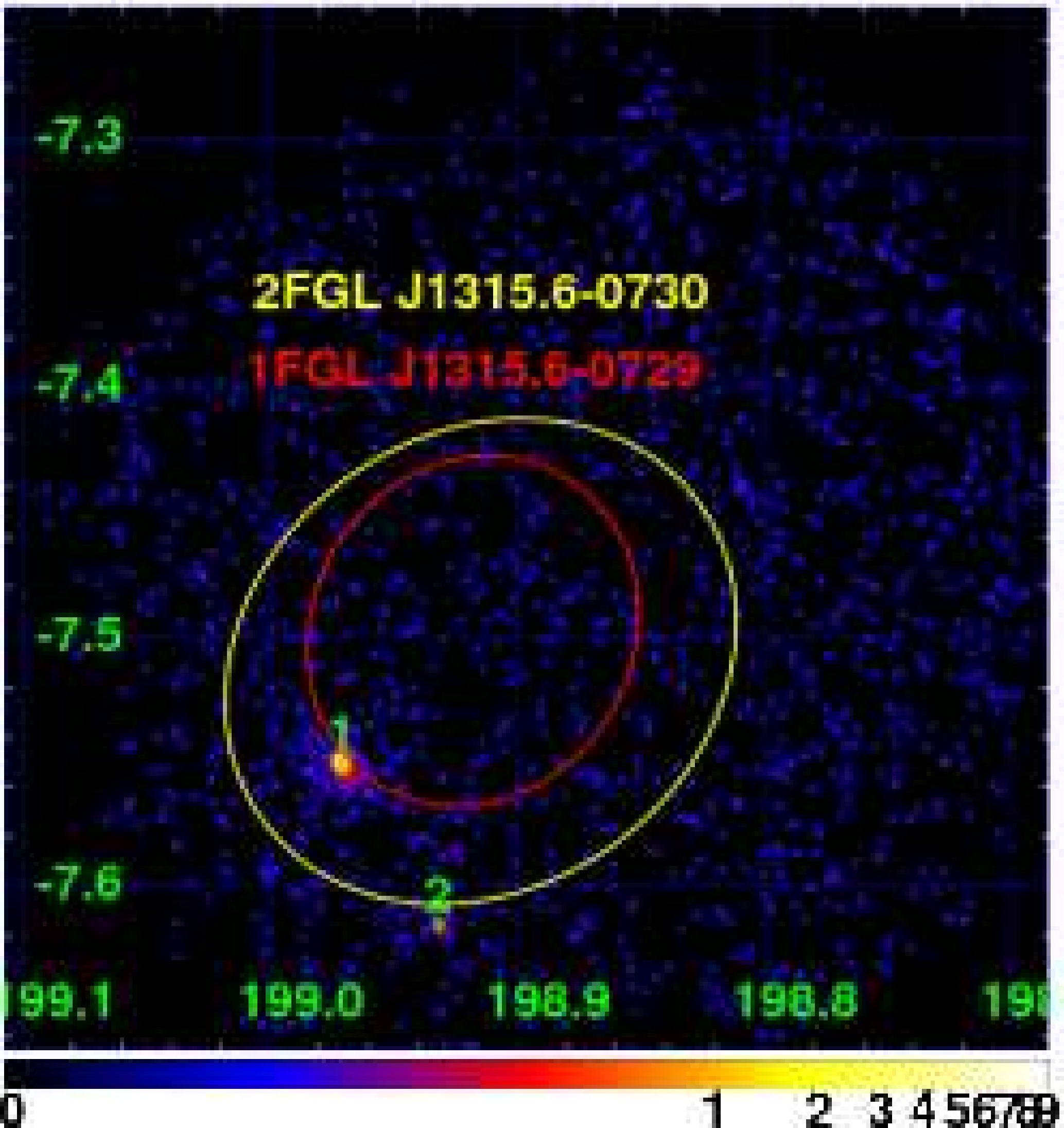}
    \end{center} 
  \end{minipage}
  \begin{minipage}{0.32\hsize}
    \begin{center}
      {\small (83) 1FGL\,J1328.2--4729} \\
      \includegraphics[width=52mm]{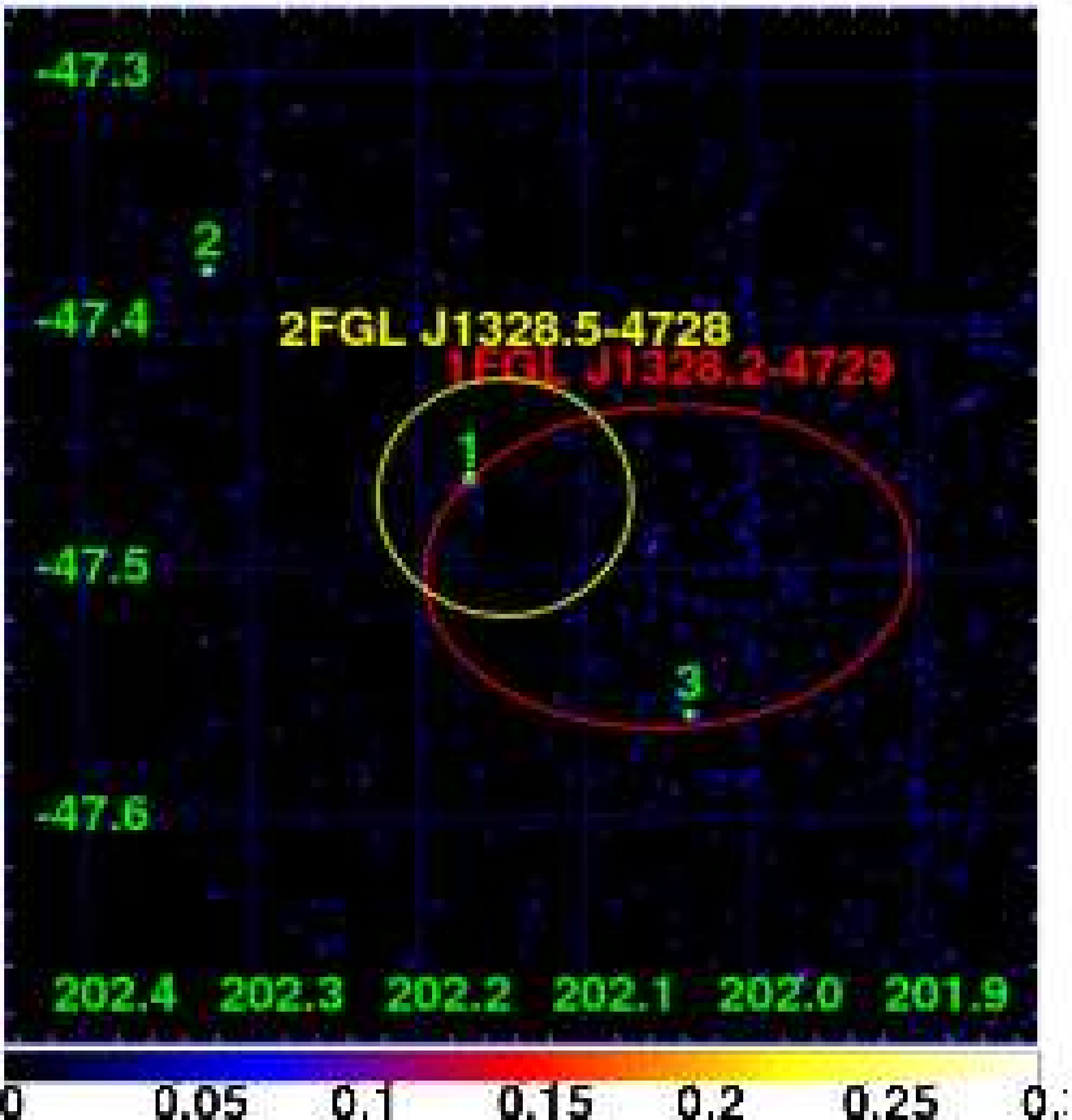}
    \end{center}
  \end{minipage}
  \begin{minipage}{0.32\hsize}
    \begin{center}
      {\small (84) 1FGL\,J1340.5--0413} \\
      \includegraphics[width=52mm]{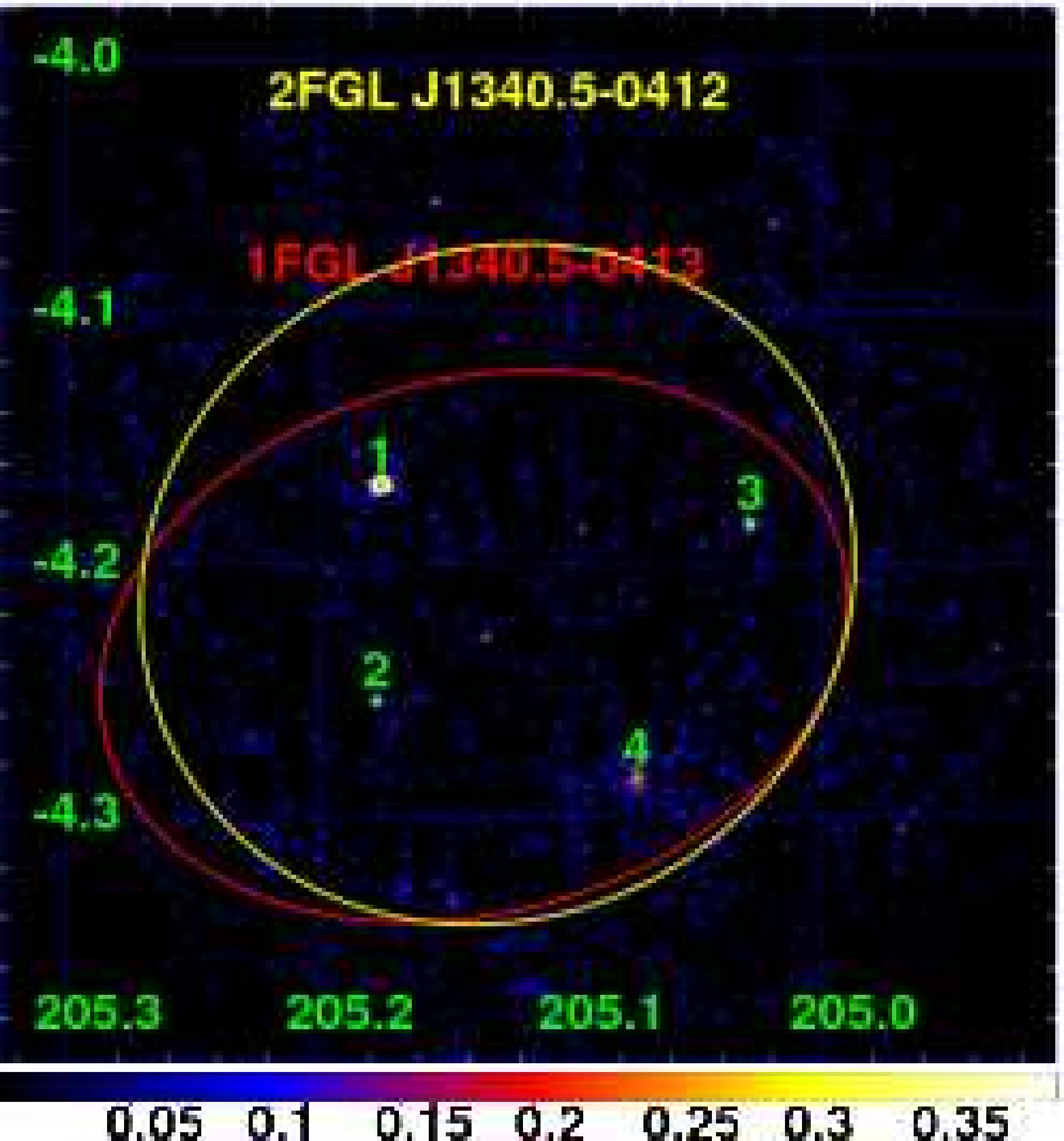}
    \end{center}
  \end{minipage}
 \end{center}
\end{figure}
\clearpage
\begin{figure}[m]
 \begin{center}
  \begin{minipage}{0.32\hsize}
    \begin{center}
      {\small (85) 1FGL\,J1406.2--2510} \\
      \includegraphics[width=52mm]{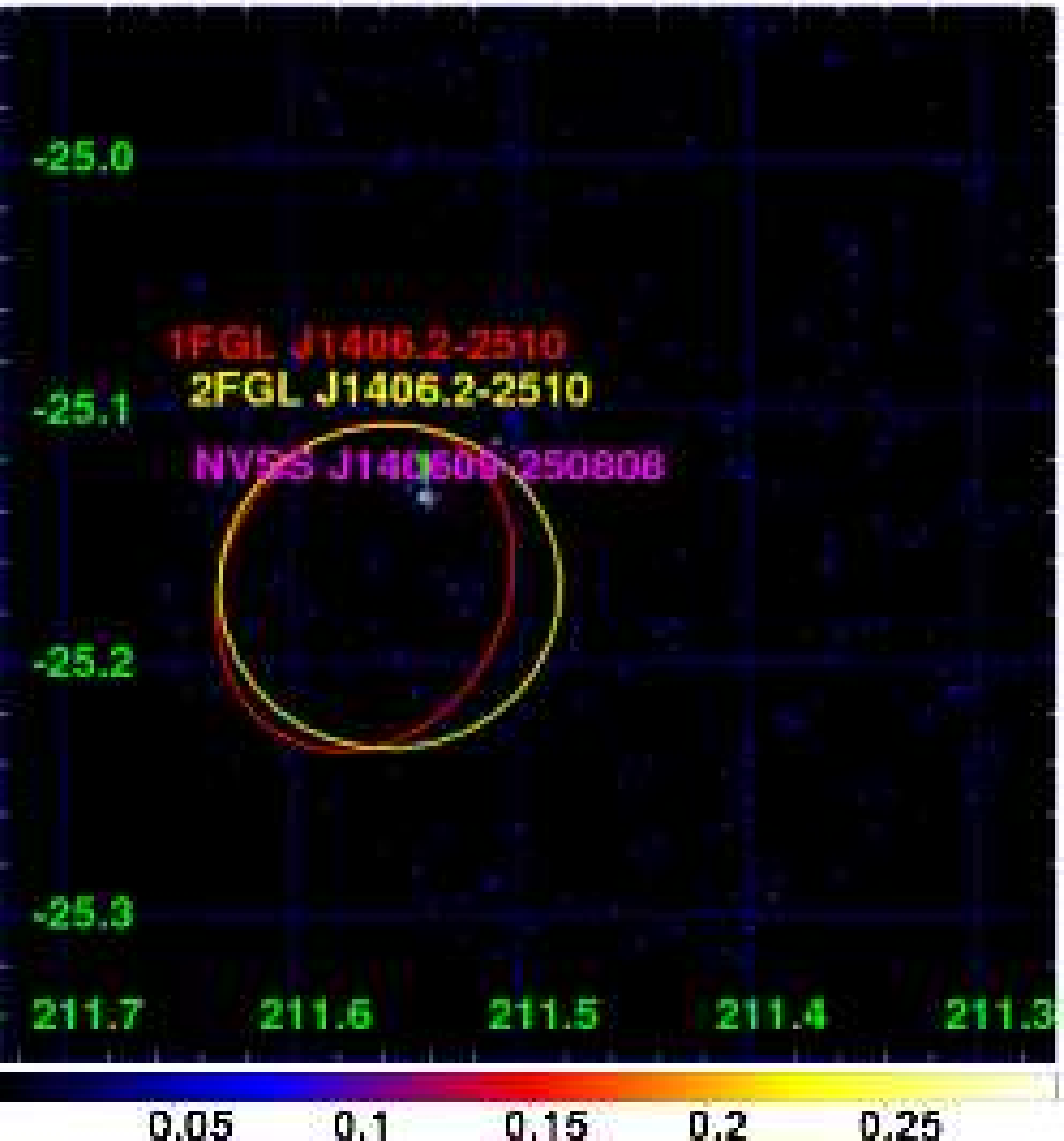}
    \end{center} 
  \end{minipage}
  \begin{minipage}{0.32\hsize}
    \begin{center}
      {\small (86) 1FGL\,J1419.7$+$7731} \\
      \includegraphics[width=52mm]{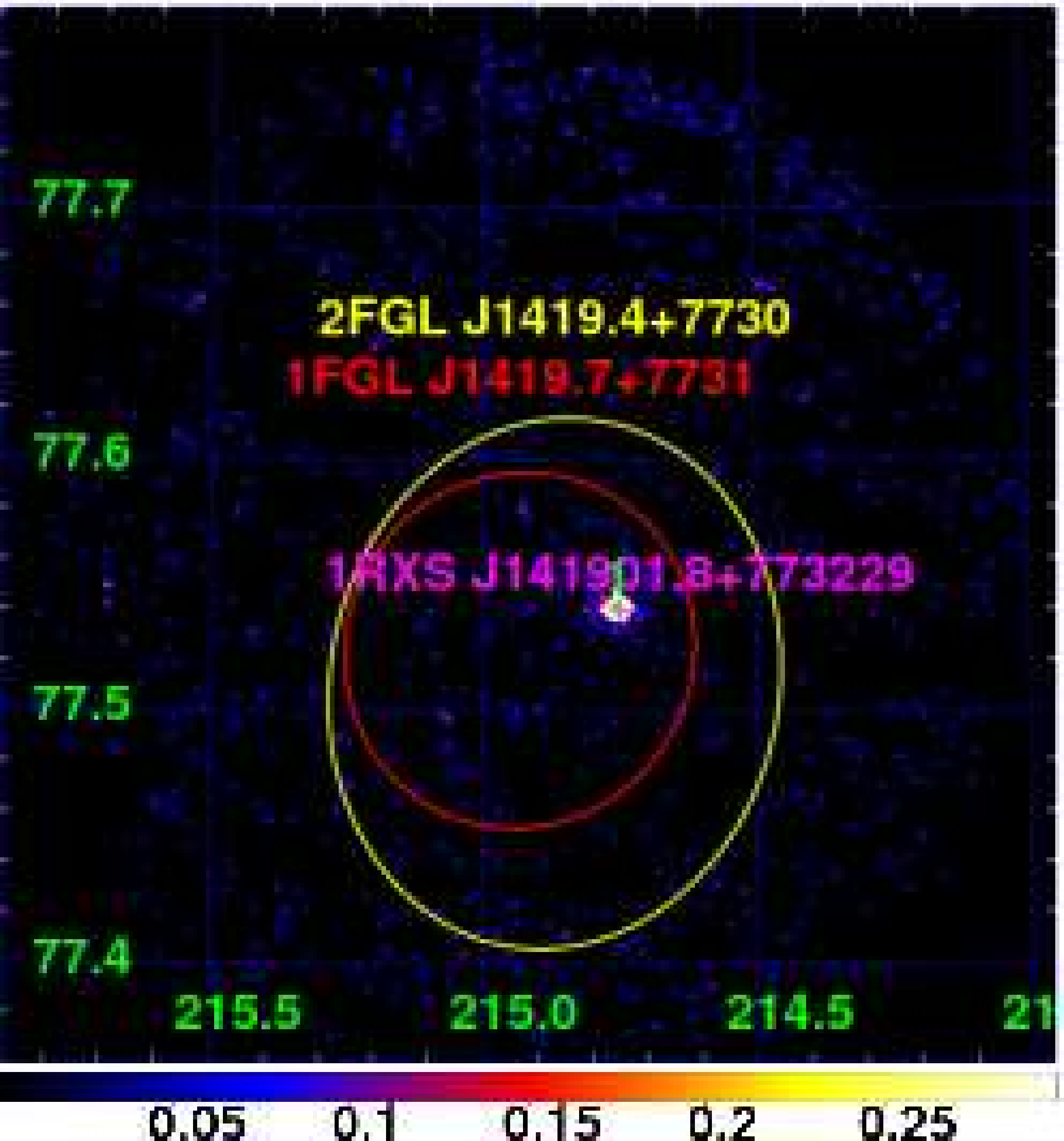}
    \end{center}
  \end{minipage}
  \begin{minipage}{0.32\hsize}
    \begin{center}
      {\small (87) 1FGL\,J1511.8--0513} \\
      \includegraphics[width=52mm]{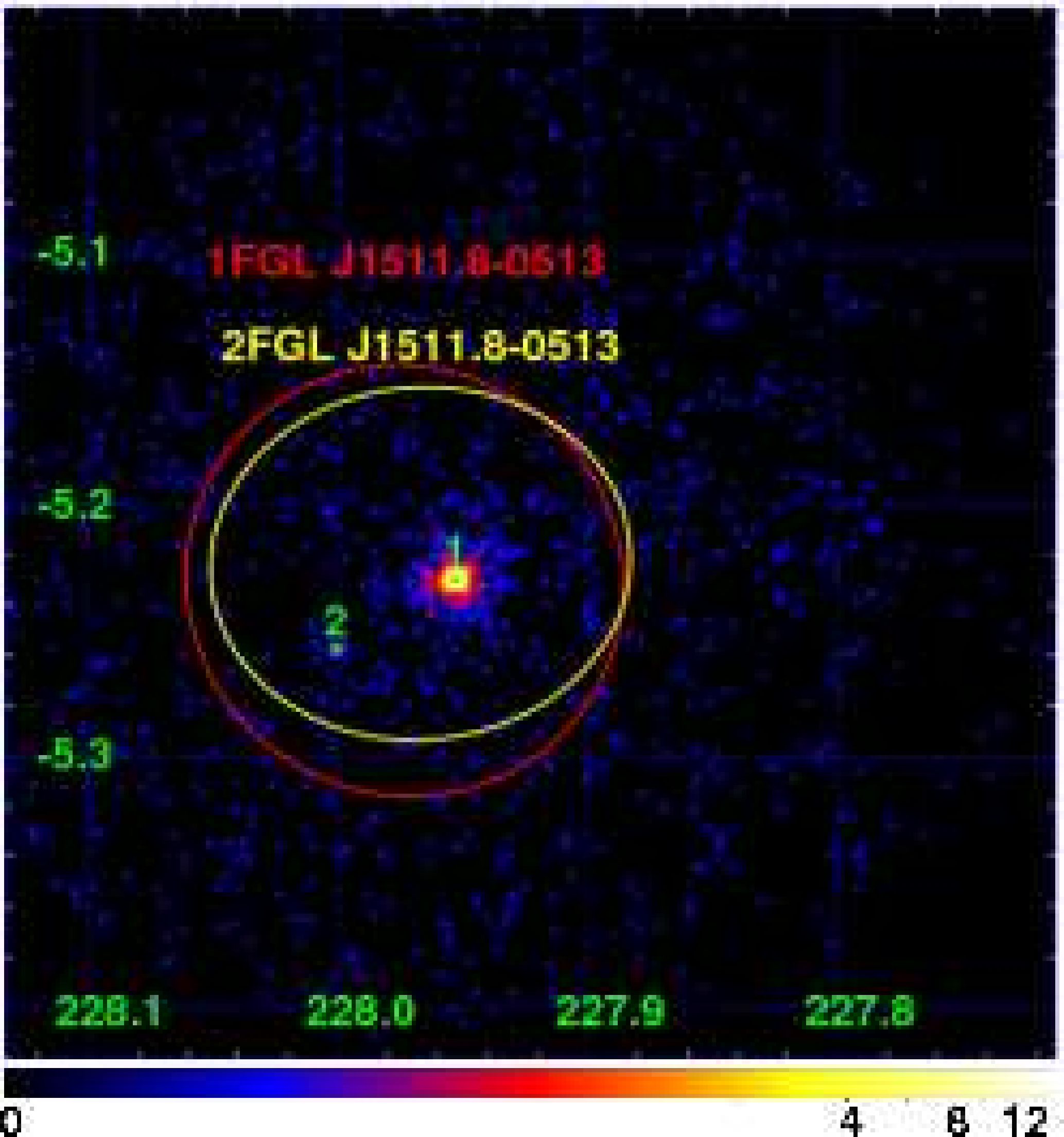}
    \end{center}
  \end{minipage}
  \begin{minipage}{0.32\hsize}
    \begin{center}
      {\small (88) 1FGL\,J1521.0--0350} \\
      \includegraphics[width=52mm]{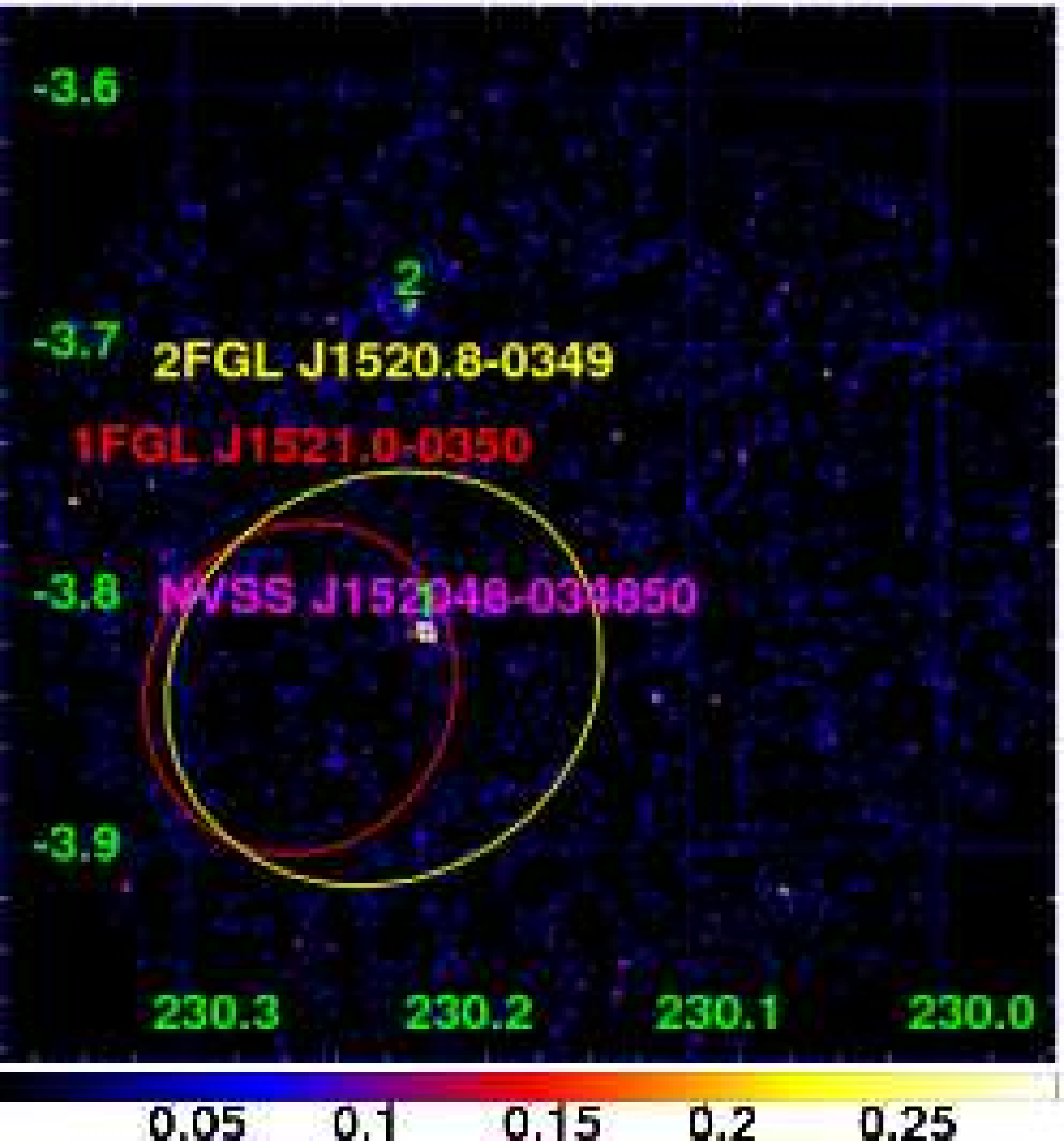}
    \end{center} 
  \end{minipage}
  \begin{minipage}{0.32\hsize}
    \begin{center}
      {\small (89) 1FGL\,J1539.0--3328} \\
      \includegraphics[width=52mm]{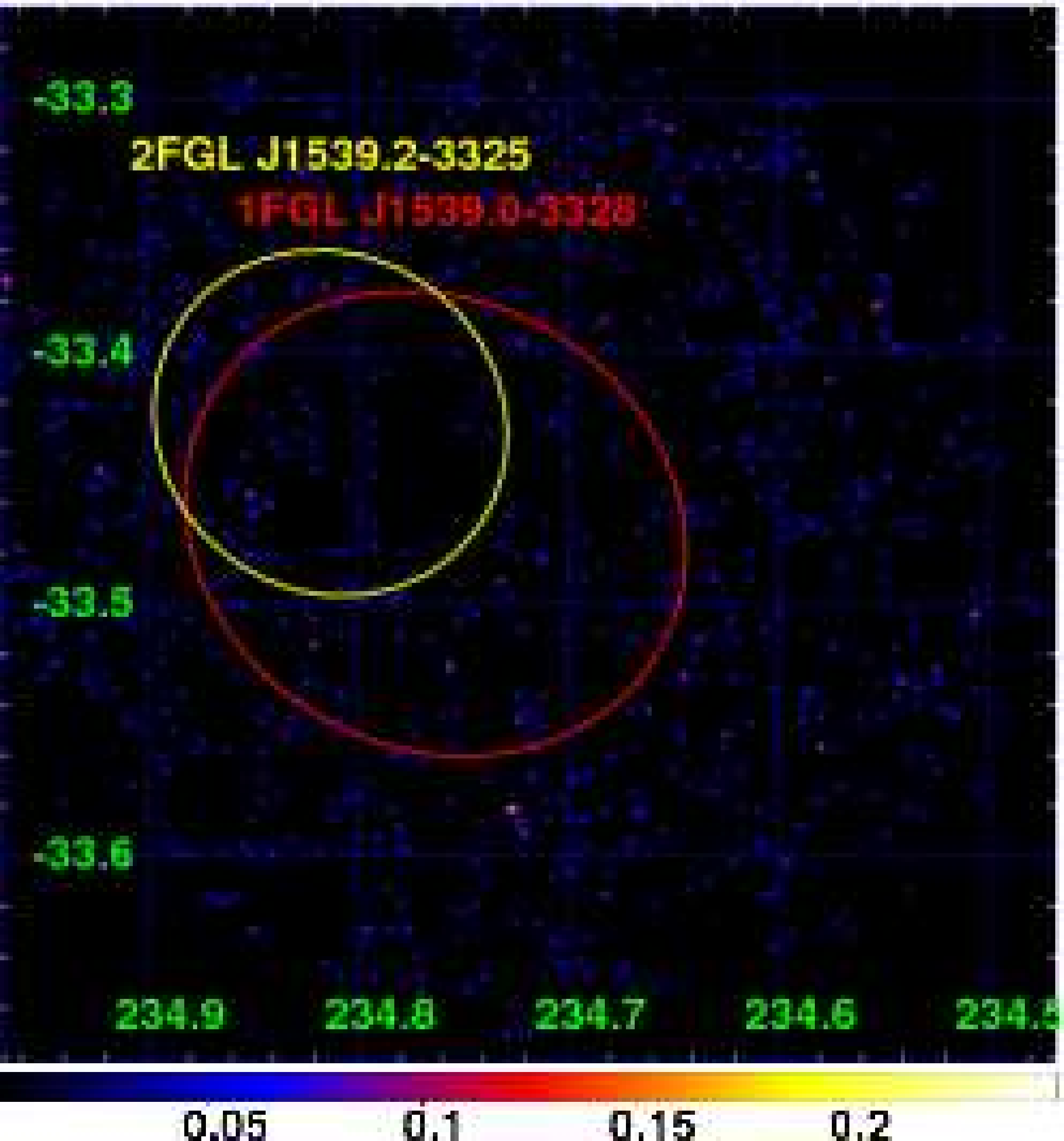}
    \end{center}
  \end{minipage}
  \begin{minipage}{0.32\hsize}
    \begin{center}
      {\small (90) 1FGL\,J1544.5--1127} \\
      \includegraphics[width=52mm]{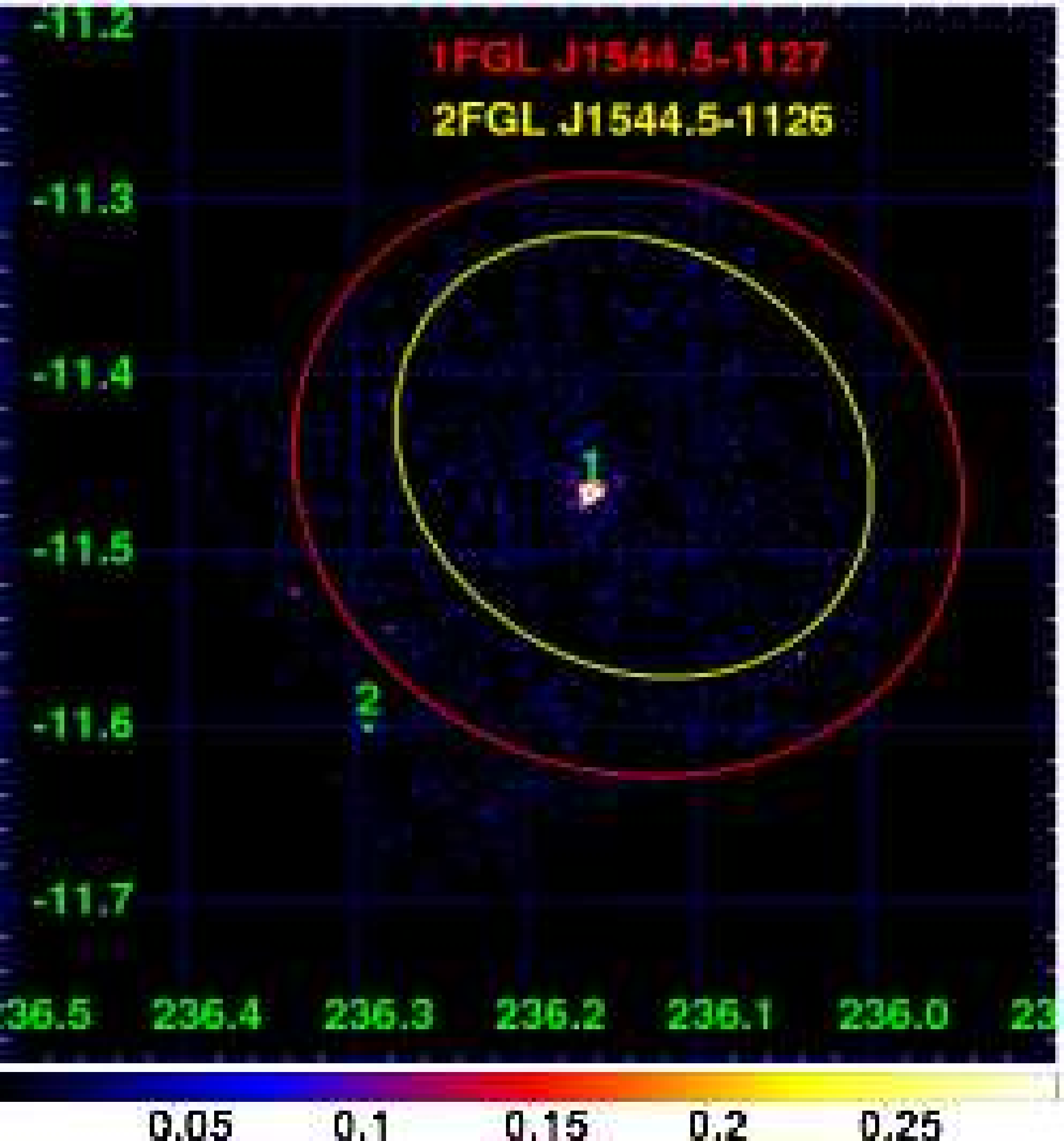}
    \end{center}
  \end{minipage}
  \begin{minipage}{0.32\hsize}
    \begin{center}
      {\small (91) 1FGL\,J1549.7--0659} \\
      \includegraphics[width=52mm]{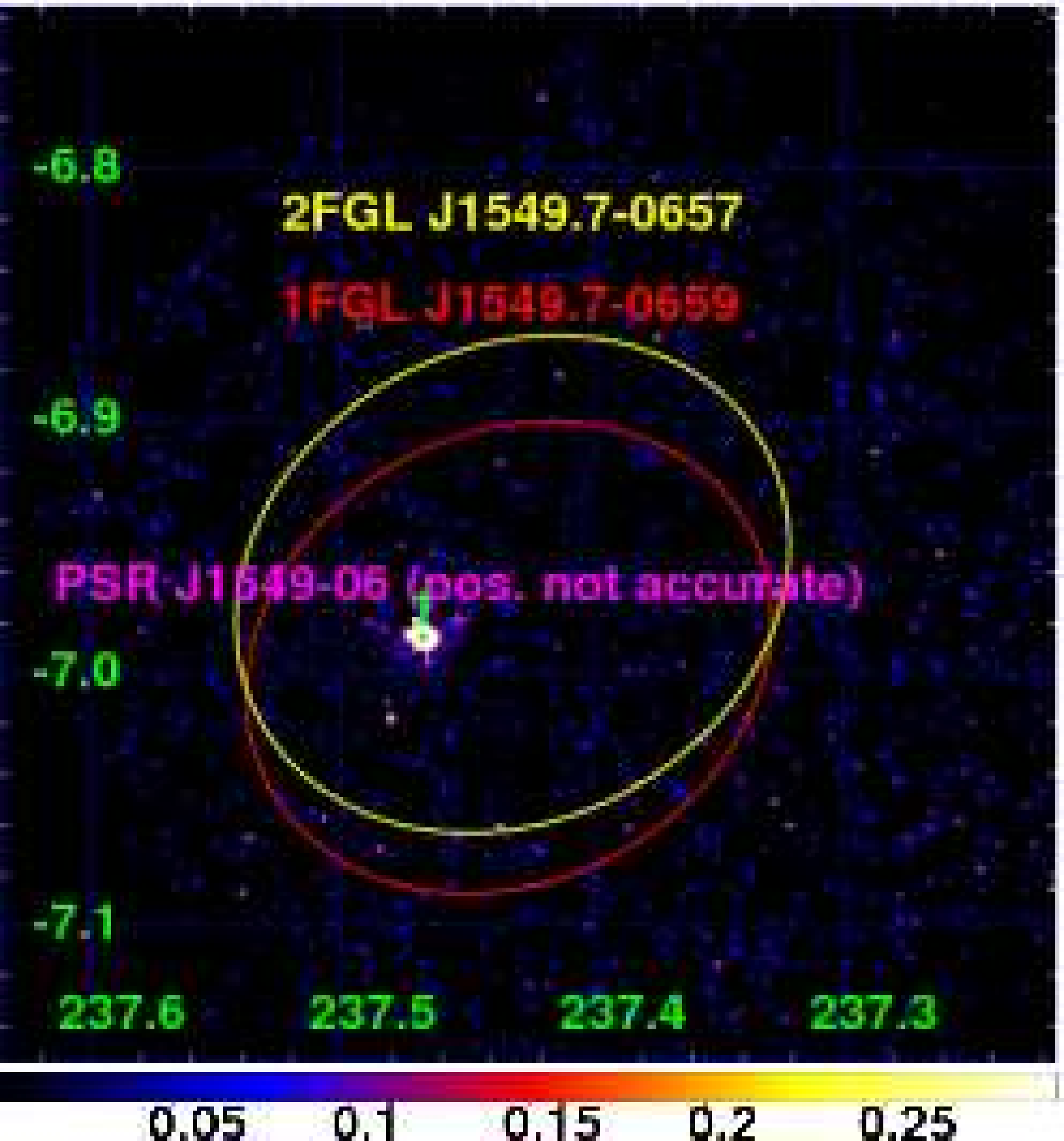}
    \end{center} 
  \end{minipage}
  \begin{minipage}{0.32\hsize}
    \begin{center}
      {\small (92) 1FGL\,J1625.3--0019} \\
      \includegraphics[width=52mm]{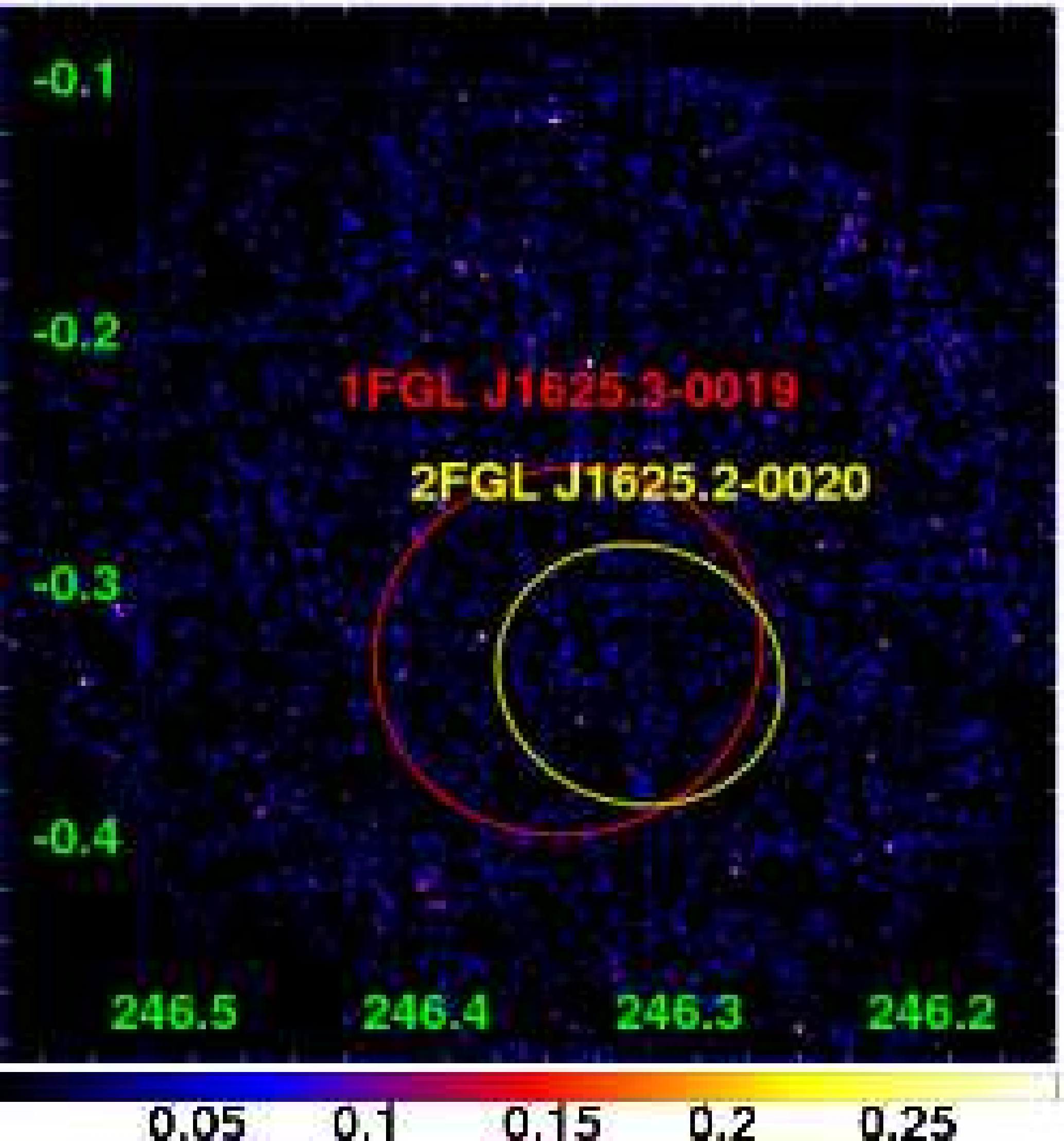}
    \end{center}
  \end{minipage}
  \begin{minipage}{0.32\hsize}
    \begin{center}
      {\small (93) 1FGL\,J1627.6$+$3218} \\
      \includegraphics[width=52mm]{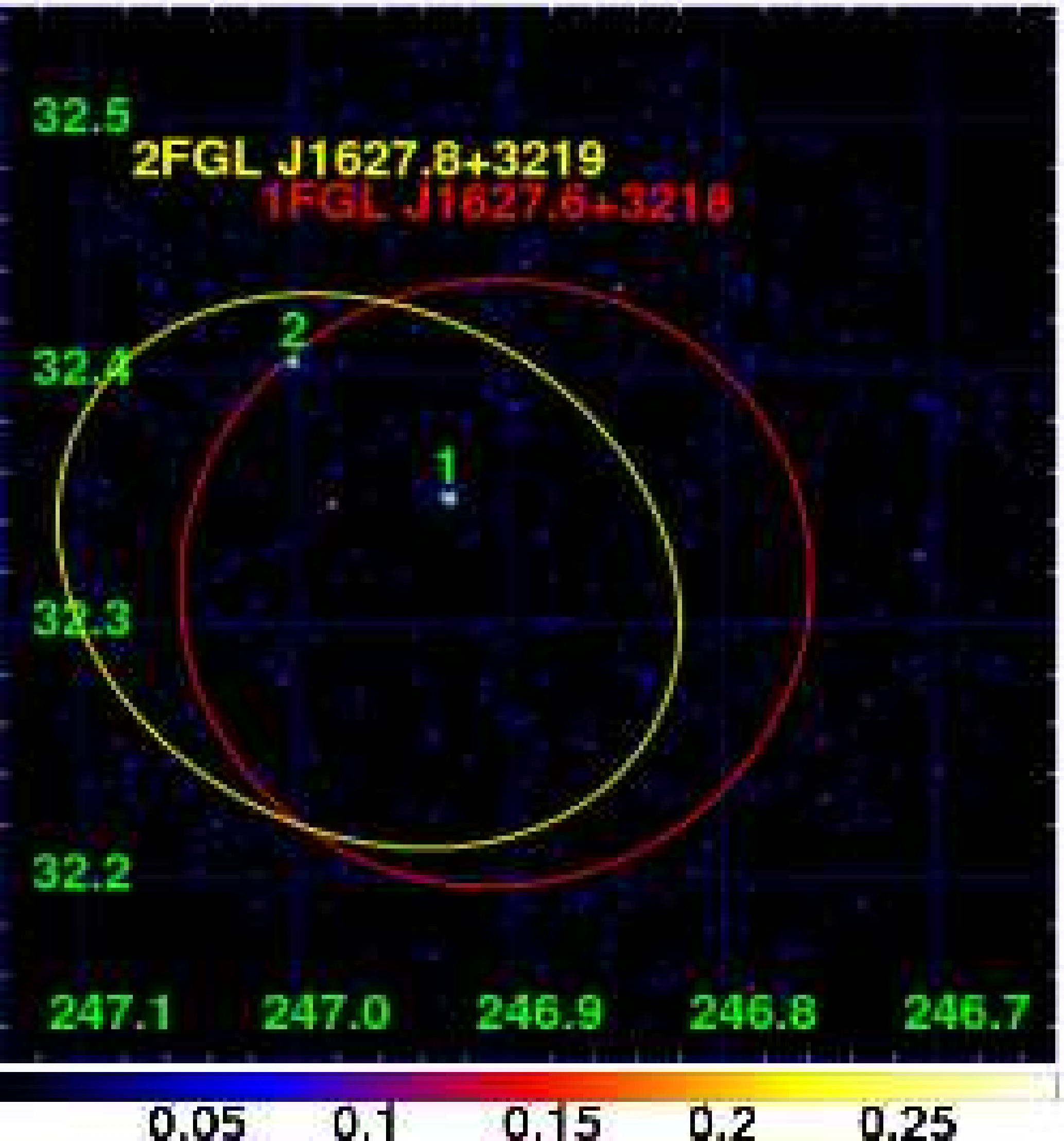}
    \end{center}
  \end{minipage}
  \begin{minipage}{0.32\hsize}
    \begin{center}
      {\small (94) 1FGL\,J1630.5$+$3735} \\
      \includegraphics[width=52mm]{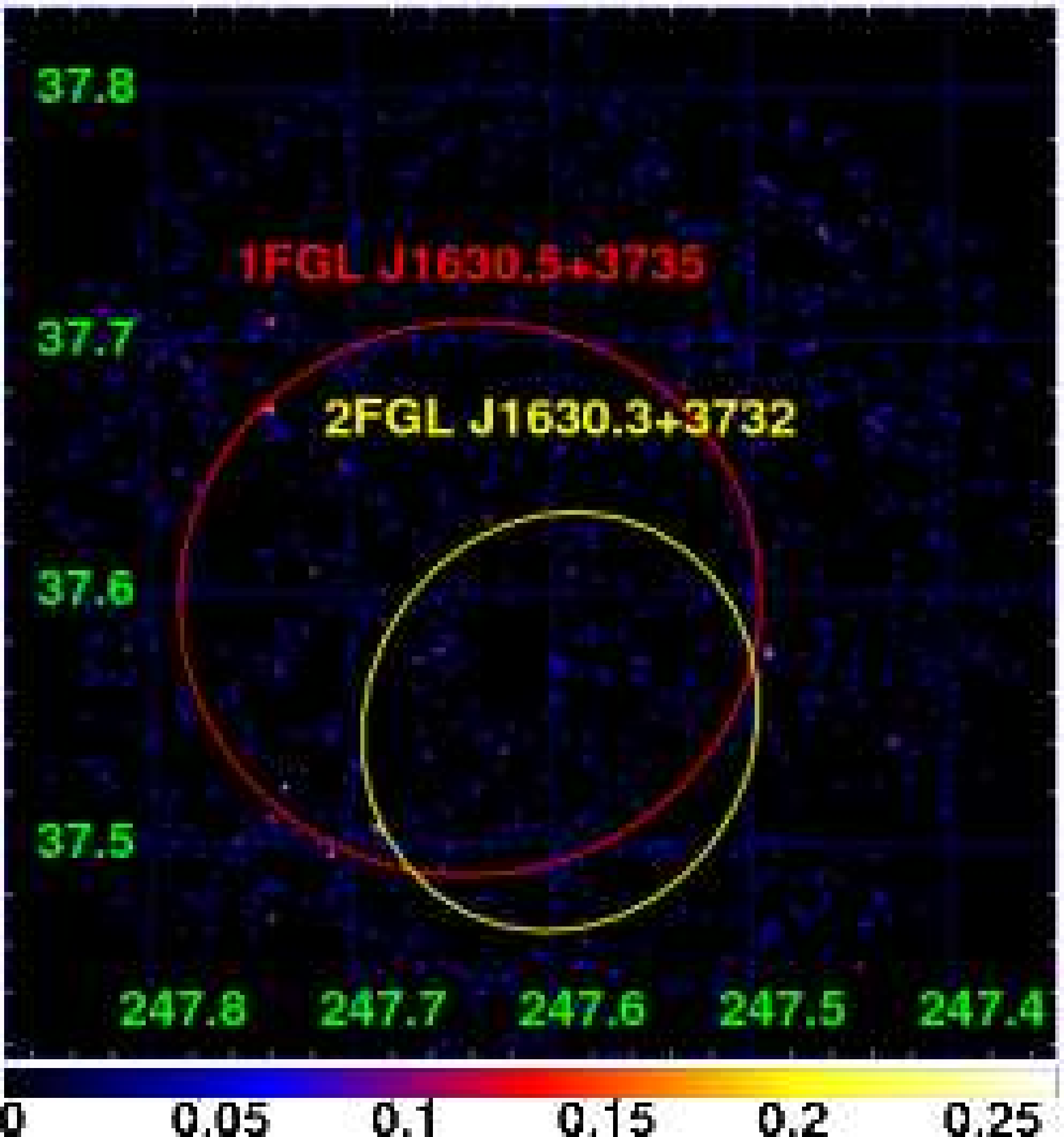}
    \end{center} 
  \end{minipage}
  \begin{minipage}{0.32\hsize}
    \begin{center}
      {\small (95) 1FGL\,J1653.6--0158} \\
      \includegraphics[width=52mm]{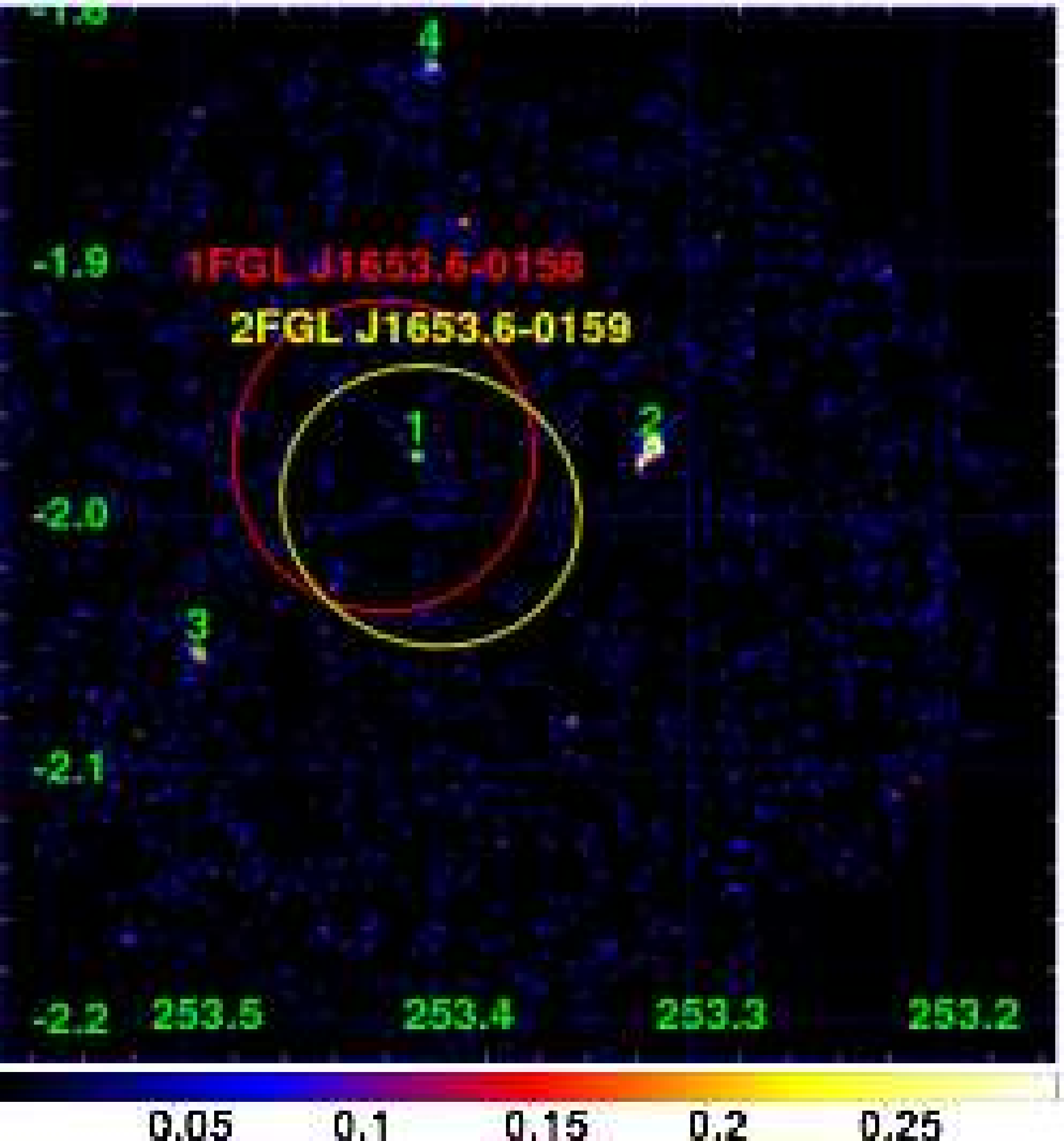}
    \end{center}
  \end{minipage}
  \begin{minipage}{0.32\hsize}
    \begin{center}
      {\small (96) 1FGL\,J1721.1$+$0713} \\
      \includegraphics[width=52mm]{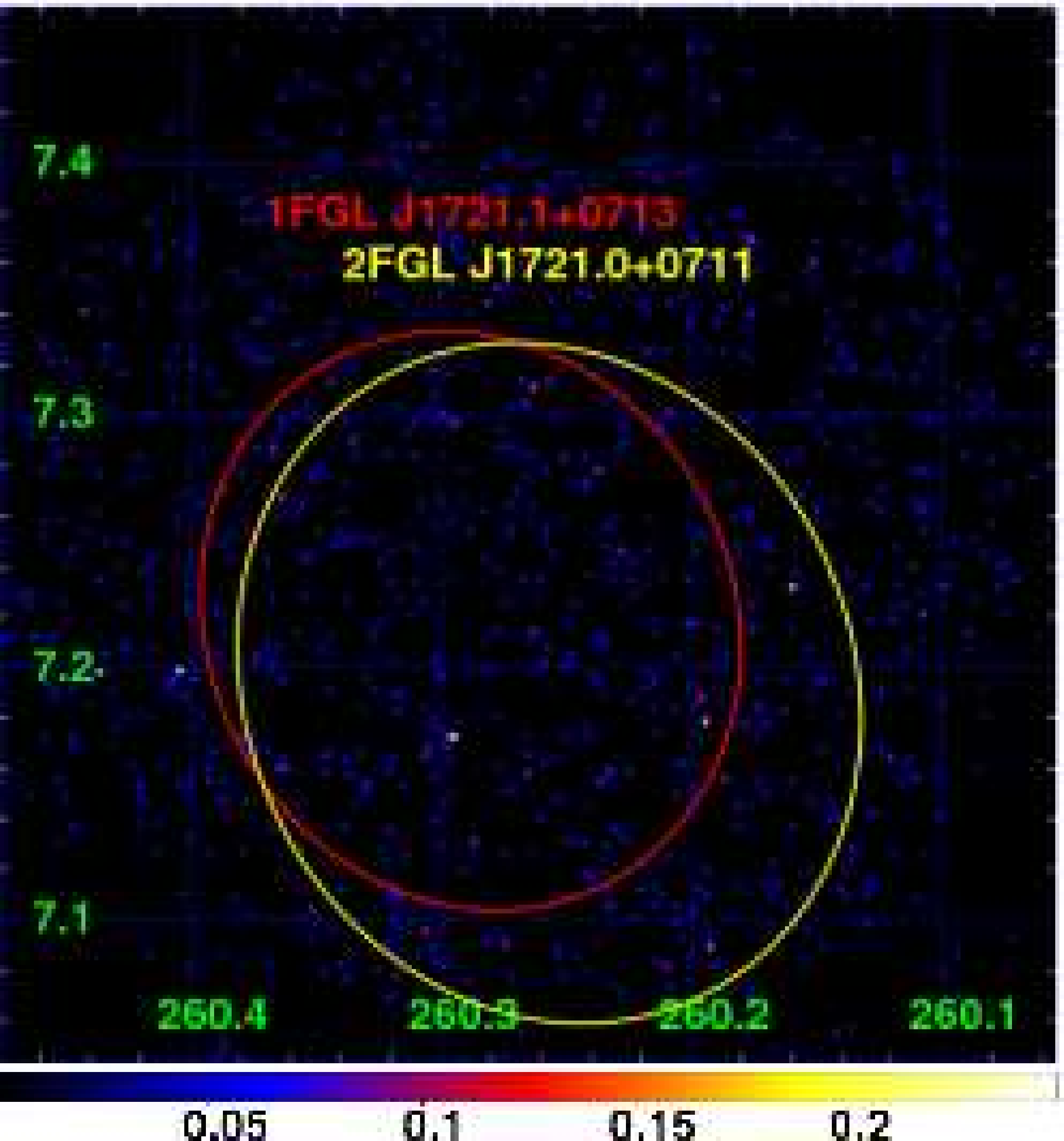}
    \end{center}
  \end{minipage}
 \end{center}
\end{figure}
\clearpage
\begin{figure}[m]
 \begin{center}
  \begin{minipage}{0.32\hsize}
    \begin{center}
      {\small (97) 1FGL\,J1739.4$+$8717} \\
      \includegraphics[width=52mm]{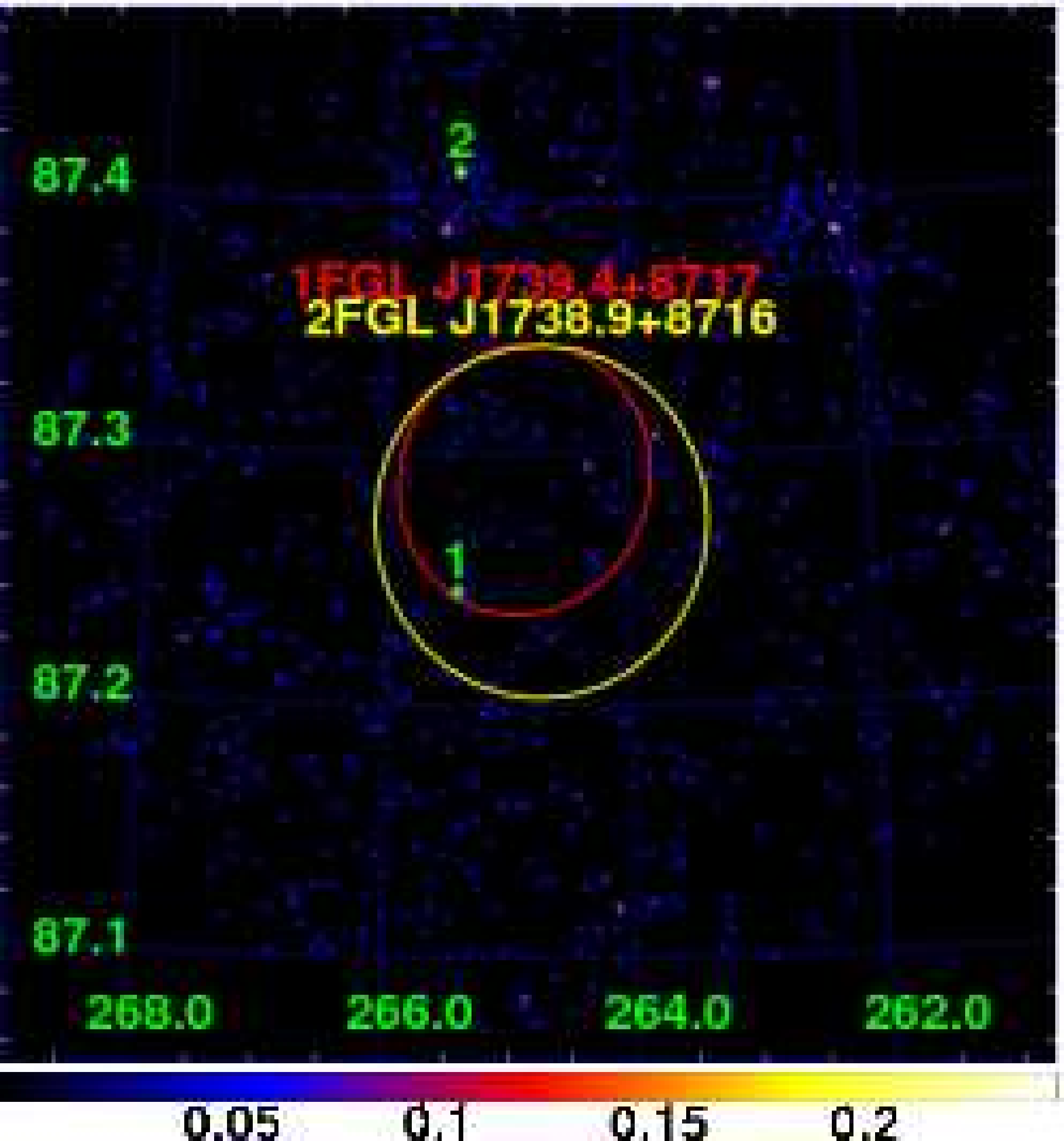}
    \end{center}
  \end{minipage}
  \begin{minipage}{0.32\hsize}
    \begin{center}
      {\small (98) 1FGL\,J1743.8--7620} \\
      \includegraphics[width=52mm]{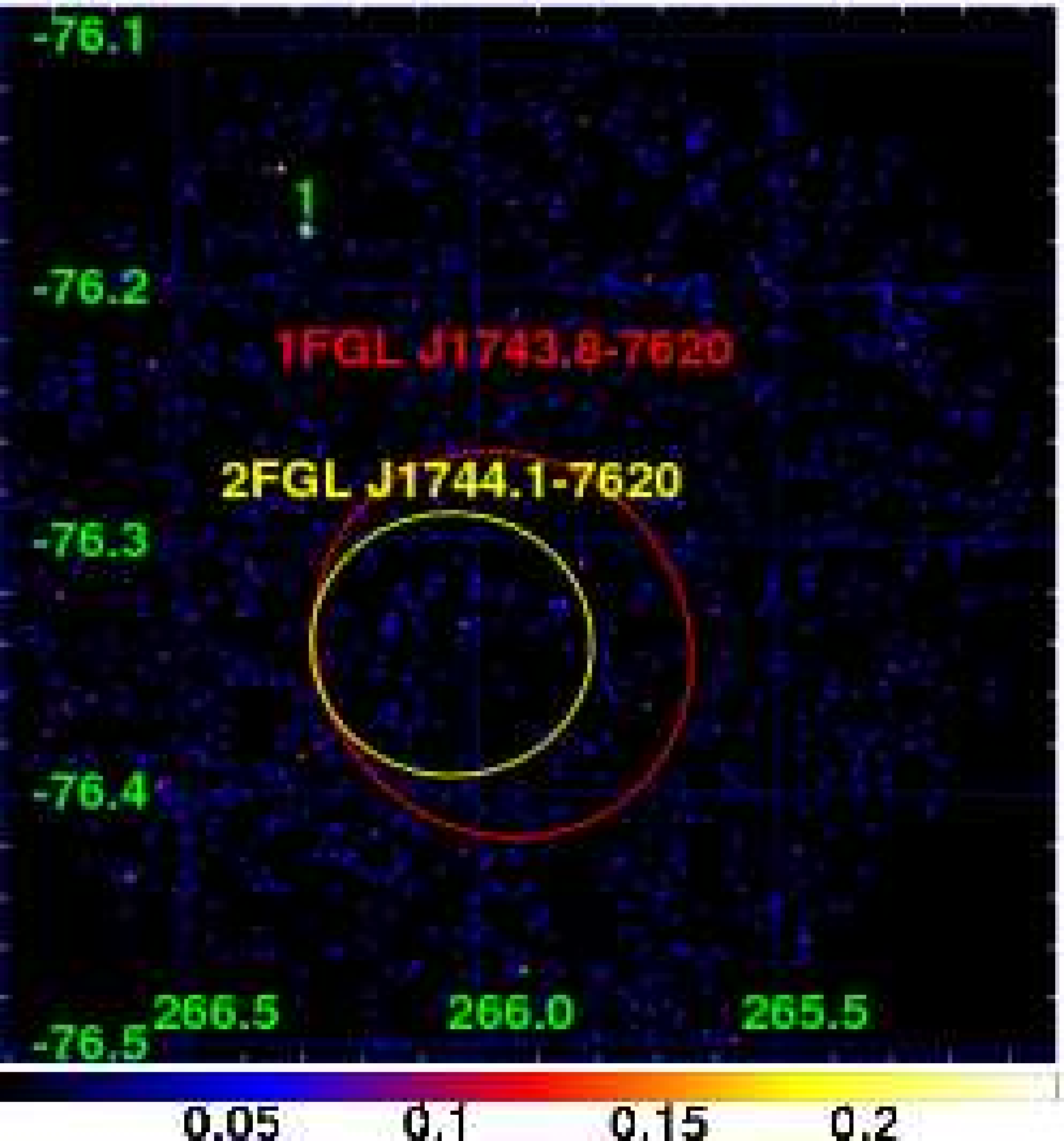}
    \end{center}
  \end{minipage}
  \begin{minipage}{0.32\hsize}
    \begin{center}
      {\small (99) 1FGL\,J1745.5$+$1018} \\
      \includegraphics[width=52mm]{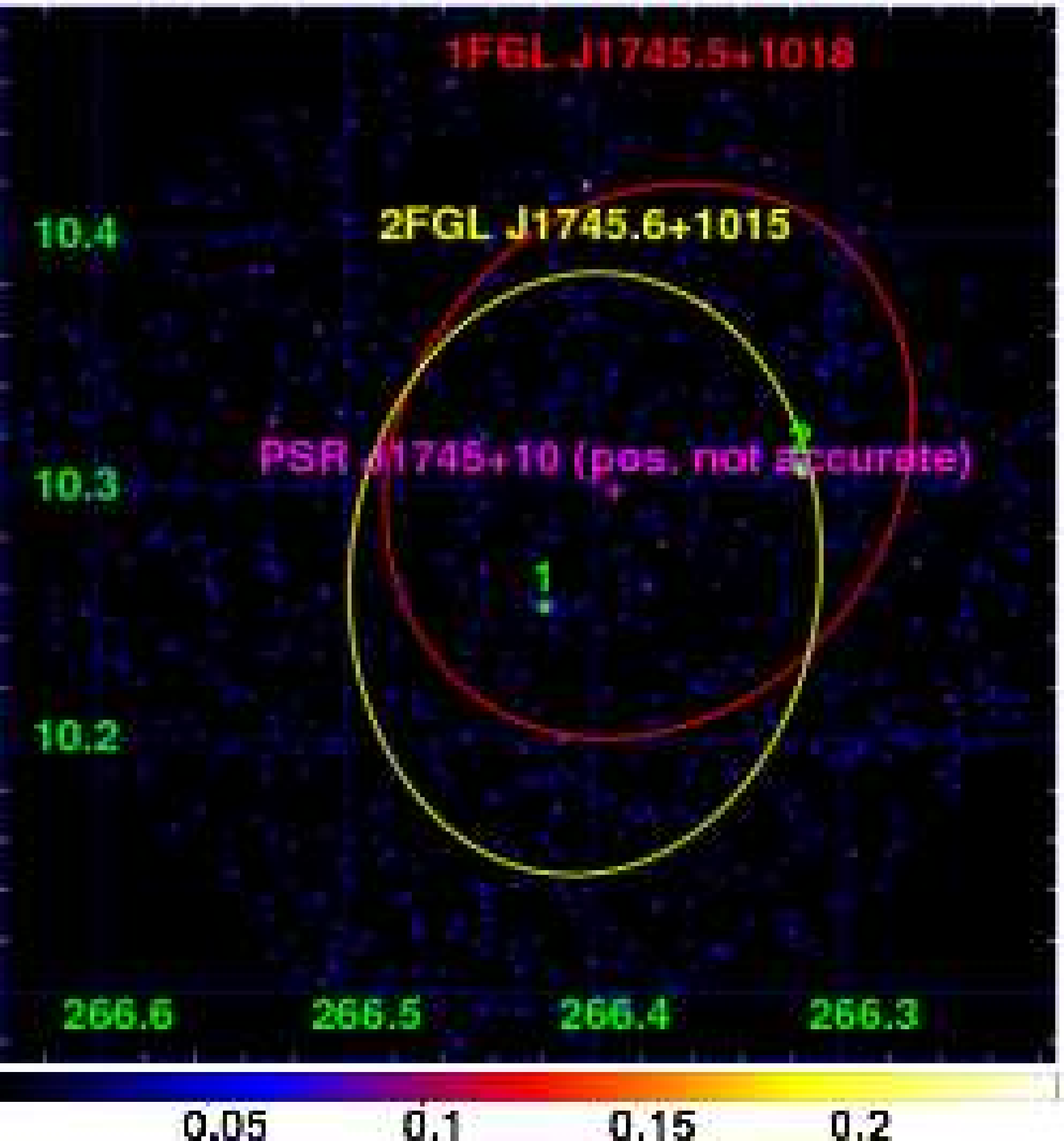}
    \end{center} 
  \end{minipage}
  \begin{minipage}{0.32\hsize}
    \begin{center}
      {\small (100) 1FGL\,J1754.0--5002} \\
      \includegraphics[width=52mm]{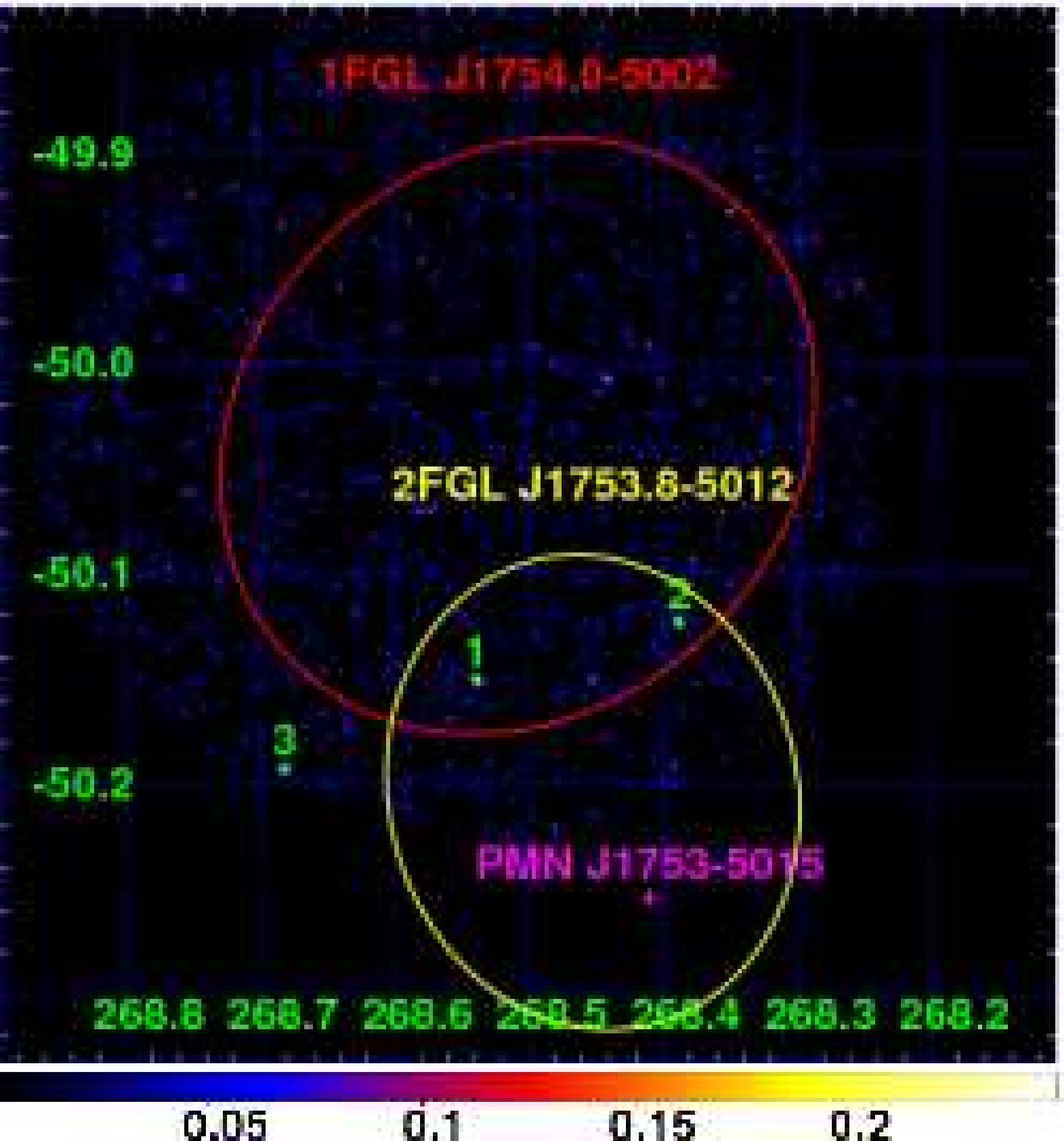}
    \end{center}
  \end{minipage}
  \begin{minipage}{0.32\hsize}
    \begin{center}
      {\small (101) 1FGL\,J1806.2$+$0609} \\
      \includegraphics[width=52mm]{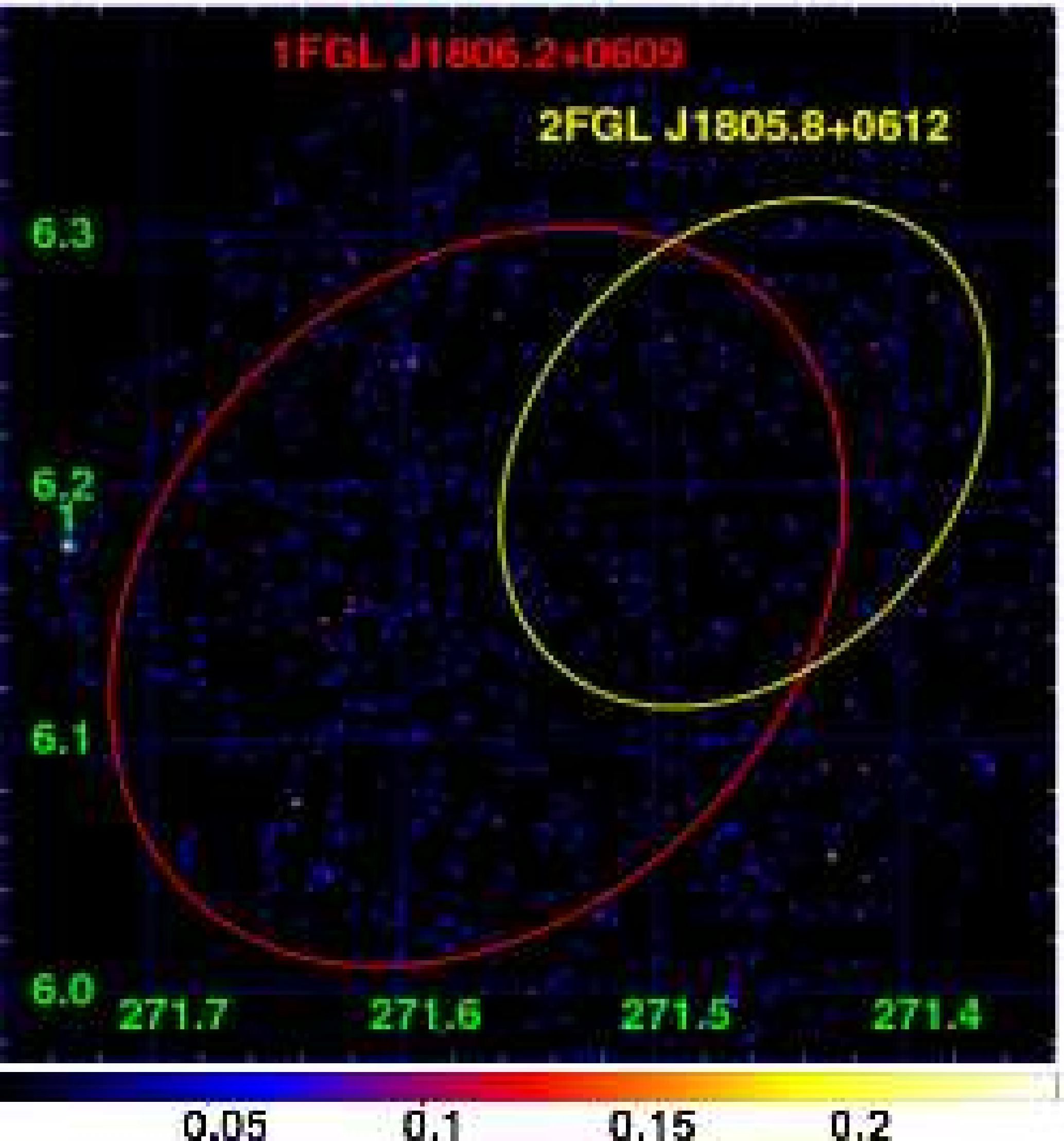}
    \end{center}
  \end{minipage}
  \begin{minipage}{0.32\hsize}
    \begin{center}
      {\small (102) 1FGL\,J1810.3$+$1741} \\
      \includegraphics[width=52mm]{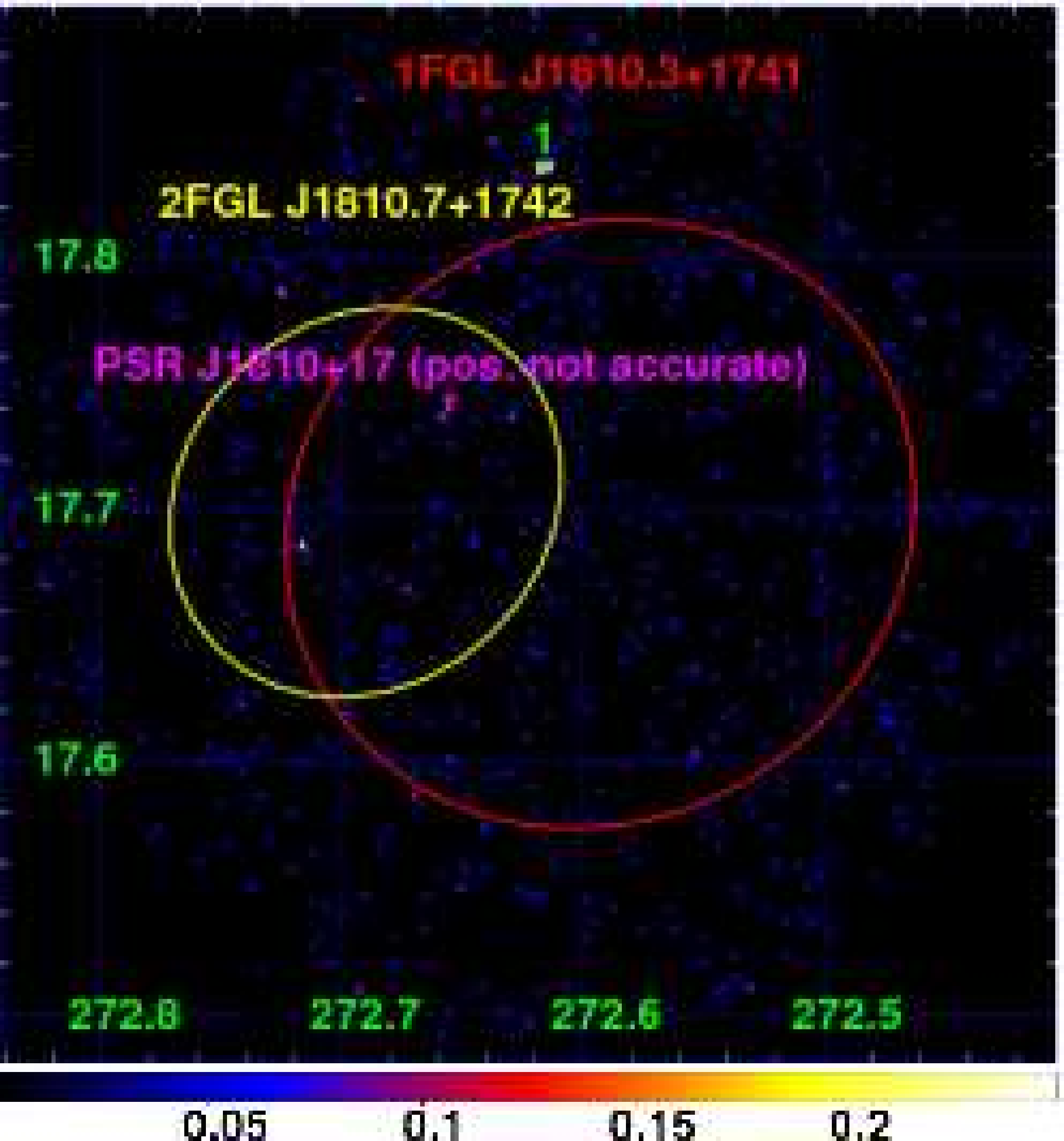}
    \end{center} 
  \end{minipage}
  \begin{minipage}{0.32\hsize}
    \begin{center}
      {\small (103) 1FGL\,J1816.7$+$4509} \\
      \includegraphics[width=52mm]{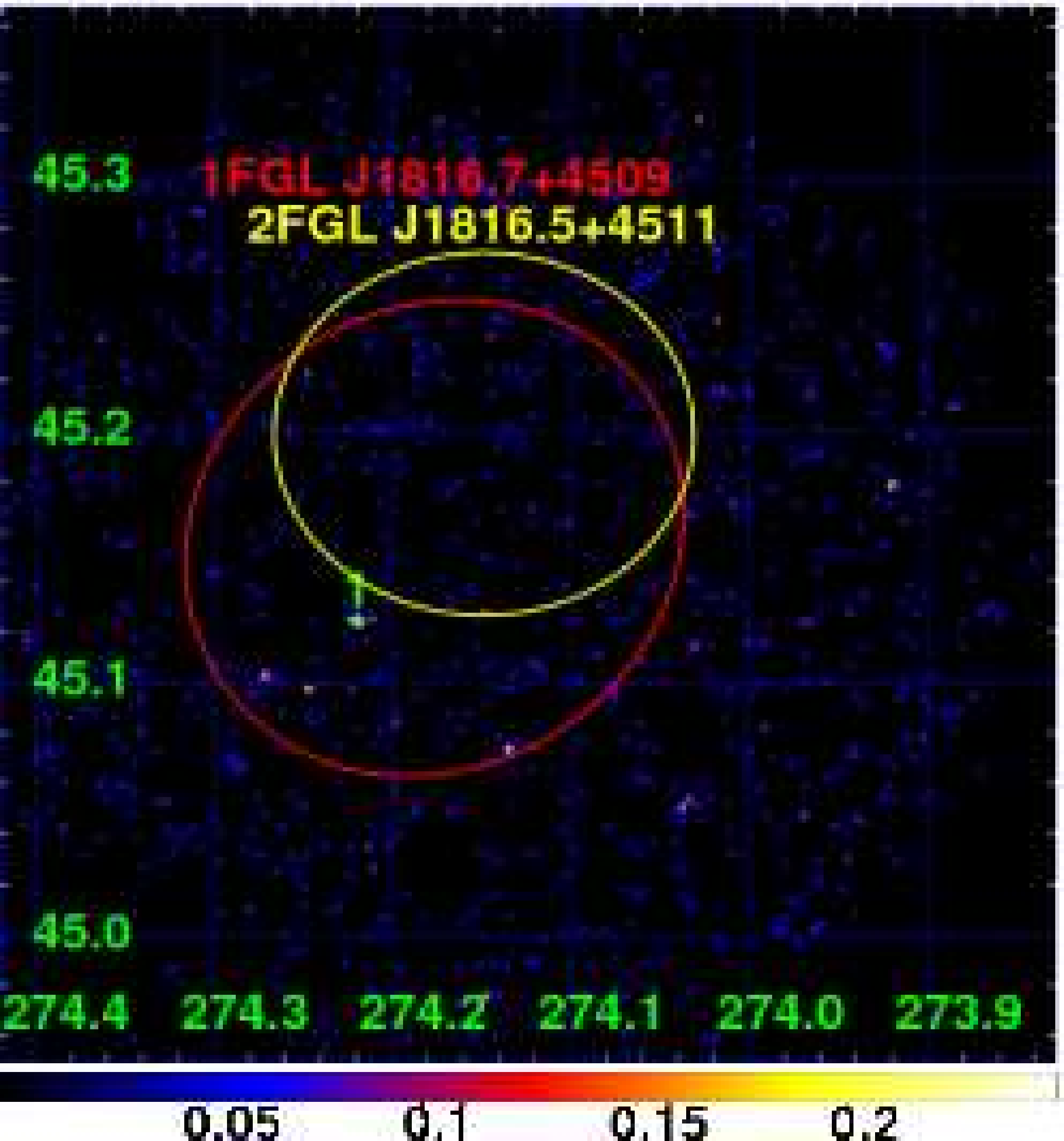}
    \end{center}
  \end{minipage}
  \begin{minipage}{0.32\hsize}
    \begin{center}
      {\small (104) 1FGL\,J1824.6$+$1013} \\
      \includegraphics[width=52mm]{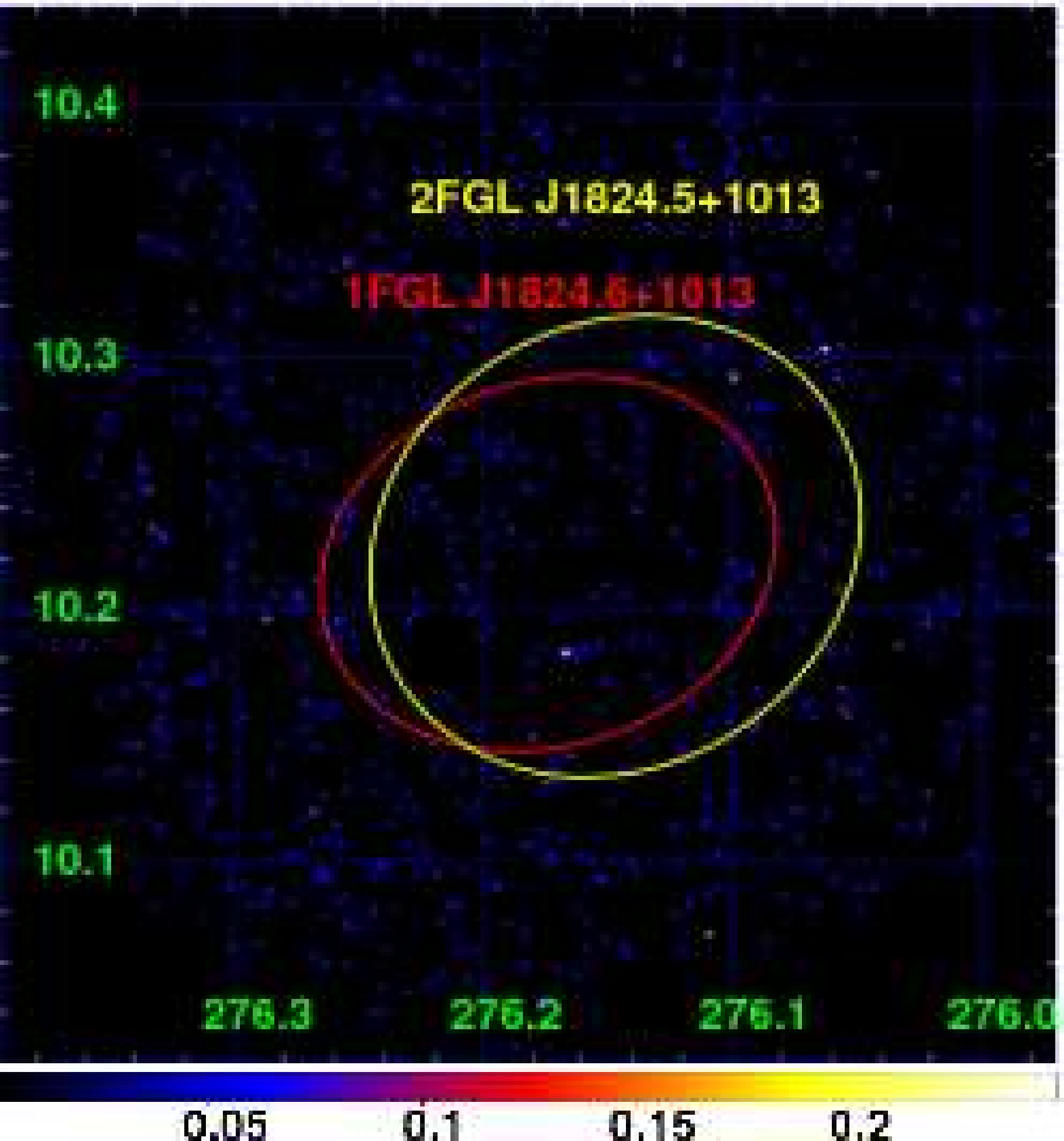}
    \end{center}
  \end{minipage}
  \begin{minipage}{0.32\hsize}
    \begin{center}
      {\small (105) 1FGL\,J1841.9$+$3220} \\
      \includegraphics[width=52mm]{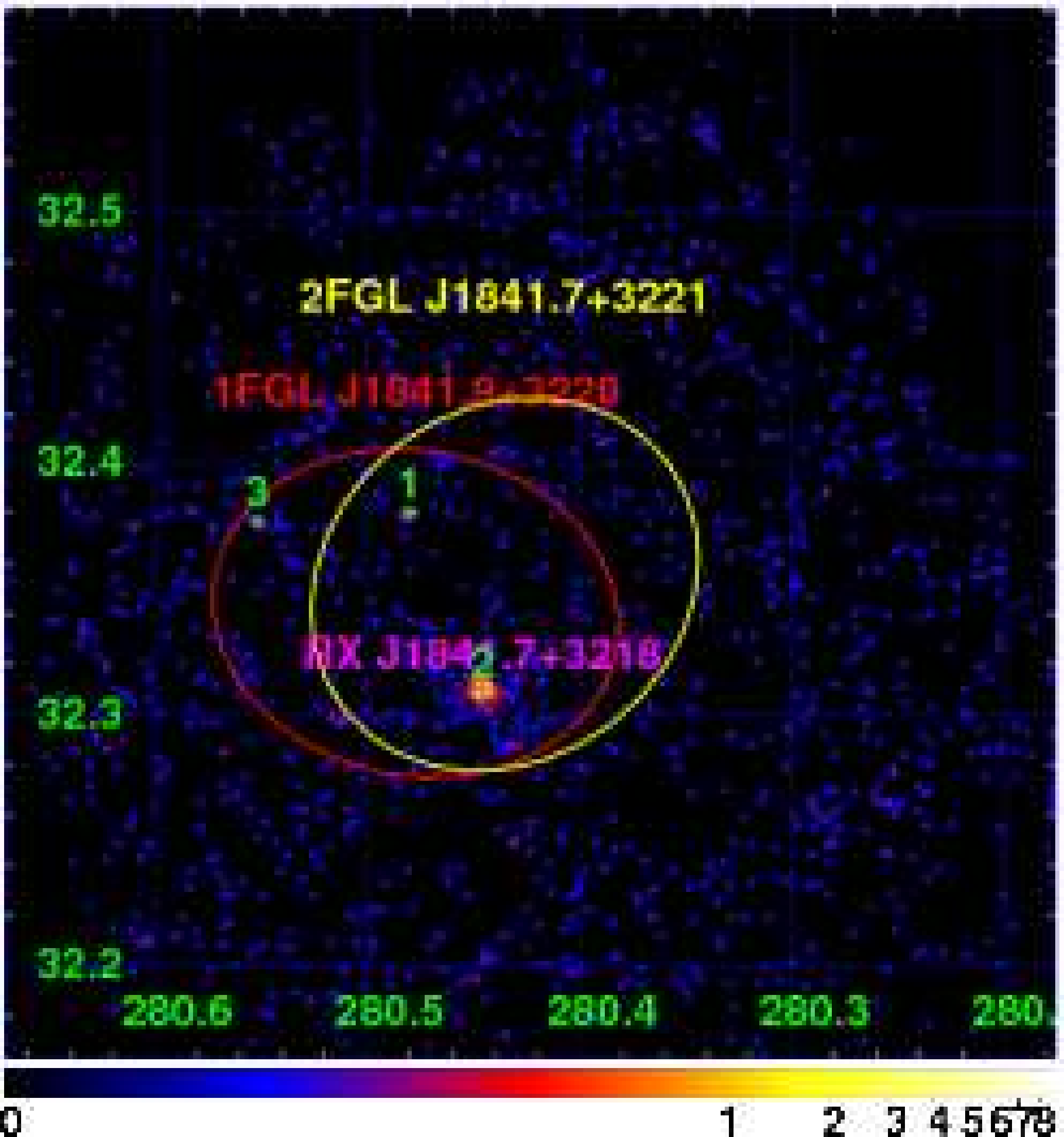}
    \end{center} 
  \end{minipage}
  \begin{minipage}{0.32\hsize}
    \begin{center}
      {\small (106) 1FGL\,J1842.3--5845} \\
      \includegraphics[width=52mm]{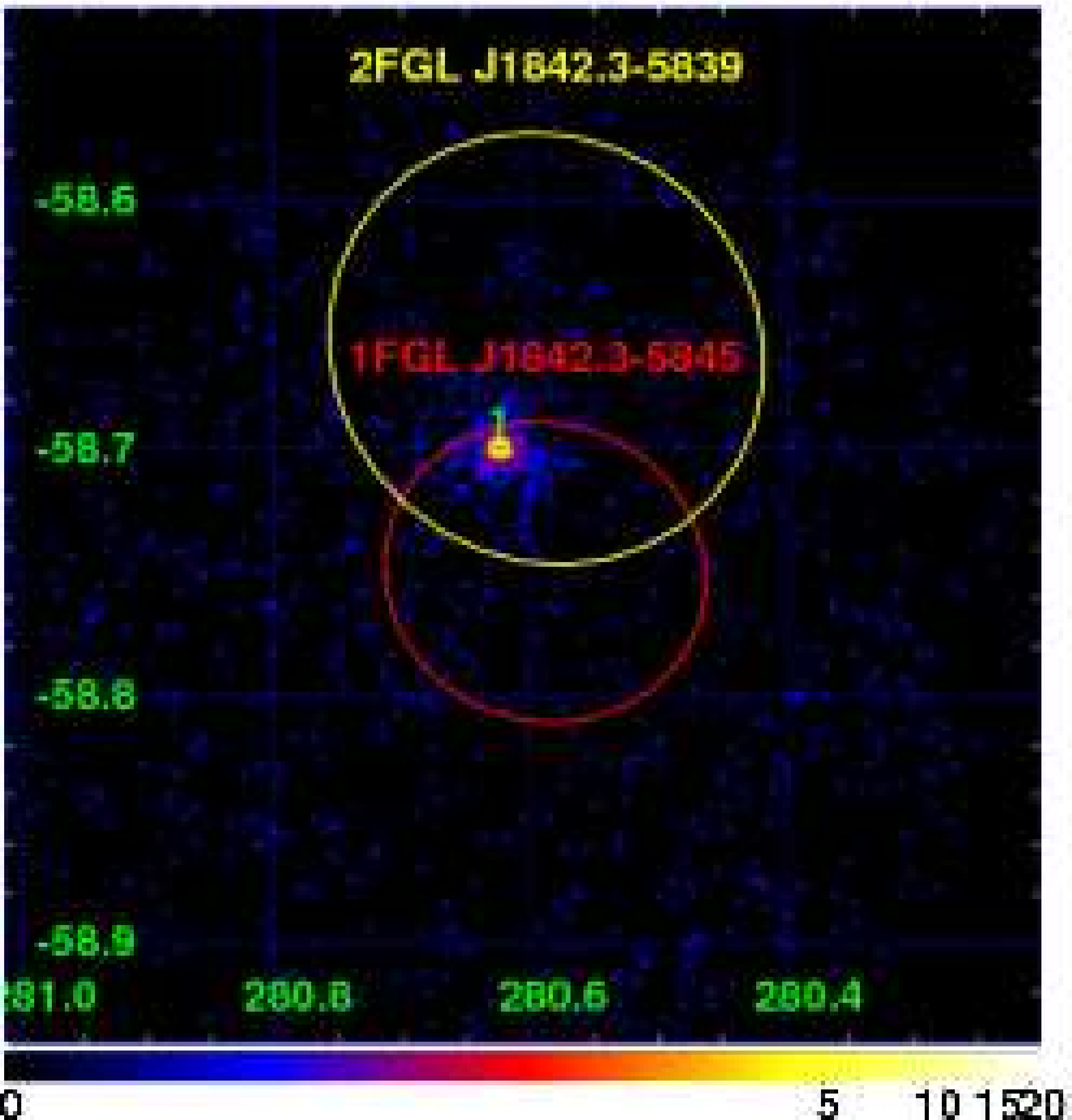}
    \end{center}
  \end{minipage}
  \begin{minipage}{0.32\hsize}
    \begin{center}
      {\small (107) 1FGL\,J1858.1--2218} \\
      \includegraphics[width=52mm]{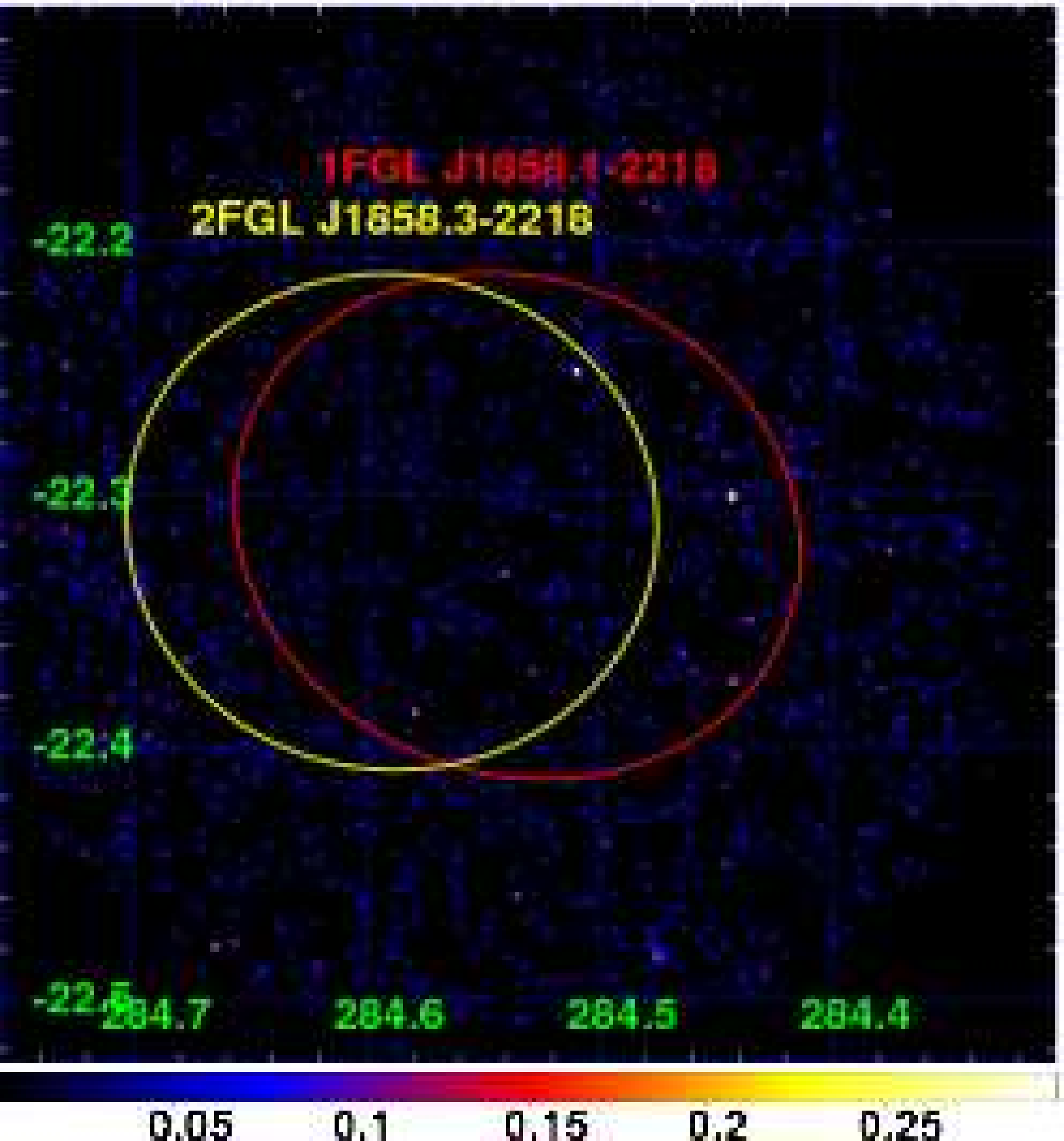}
    \end{center}
  \end{minipage}
  \begin{minipage}{0.32\hsize}
    \begin{center}
      {\small (108) 1FGL\,J1902.0--5110} \\
      \includegraphics[width=52mm]{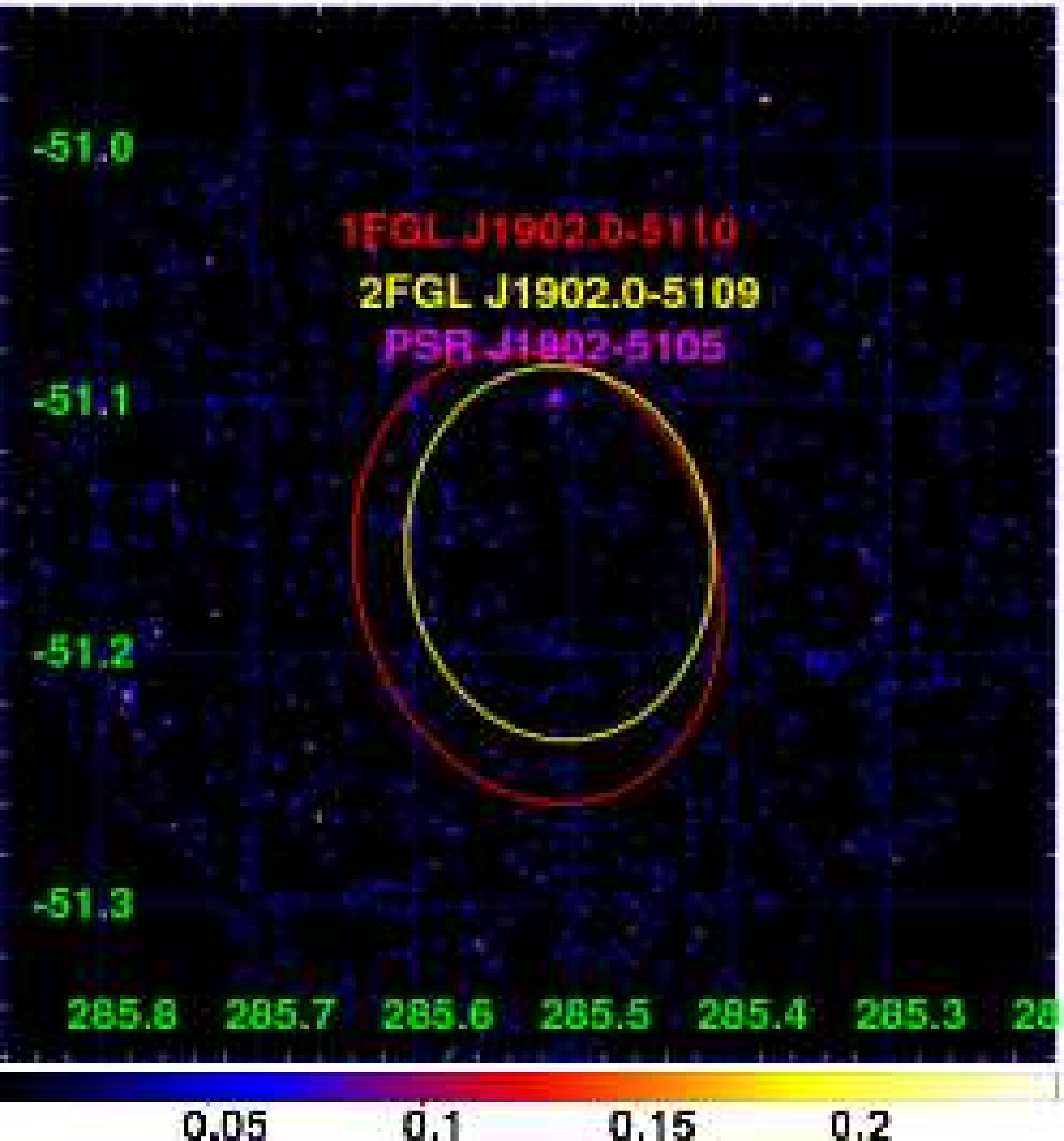}
    \end{center}
  \end{minipage}
 \end{center}
\end{figure}
\clearpage
\begin{figure}[m]
 \begin{center}
  \begin{minipage}{0.32\hsize}
    \begin{center}
      {\small (109) 1FGL\,J1916.9--3028} \\
      \includegraphics[width=52mm]{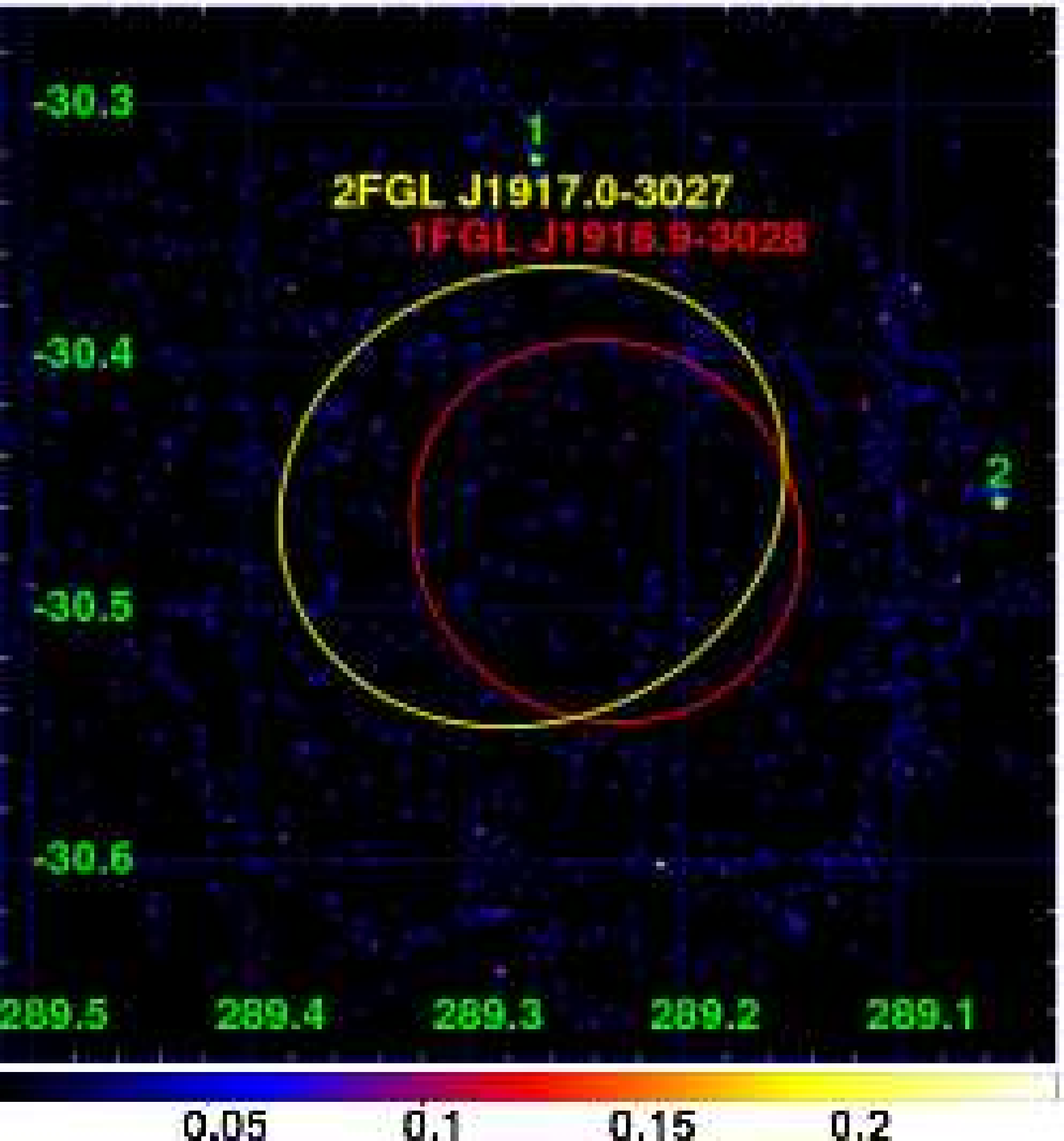}
    \end{center} 
  \end{minipage}
  \begin{minipage}{0.32\hsize}
    \begin{center}
      {\small (110) 1FGL\,J1926.8$+$6153} \\
      \includegraphics[width=52mm]{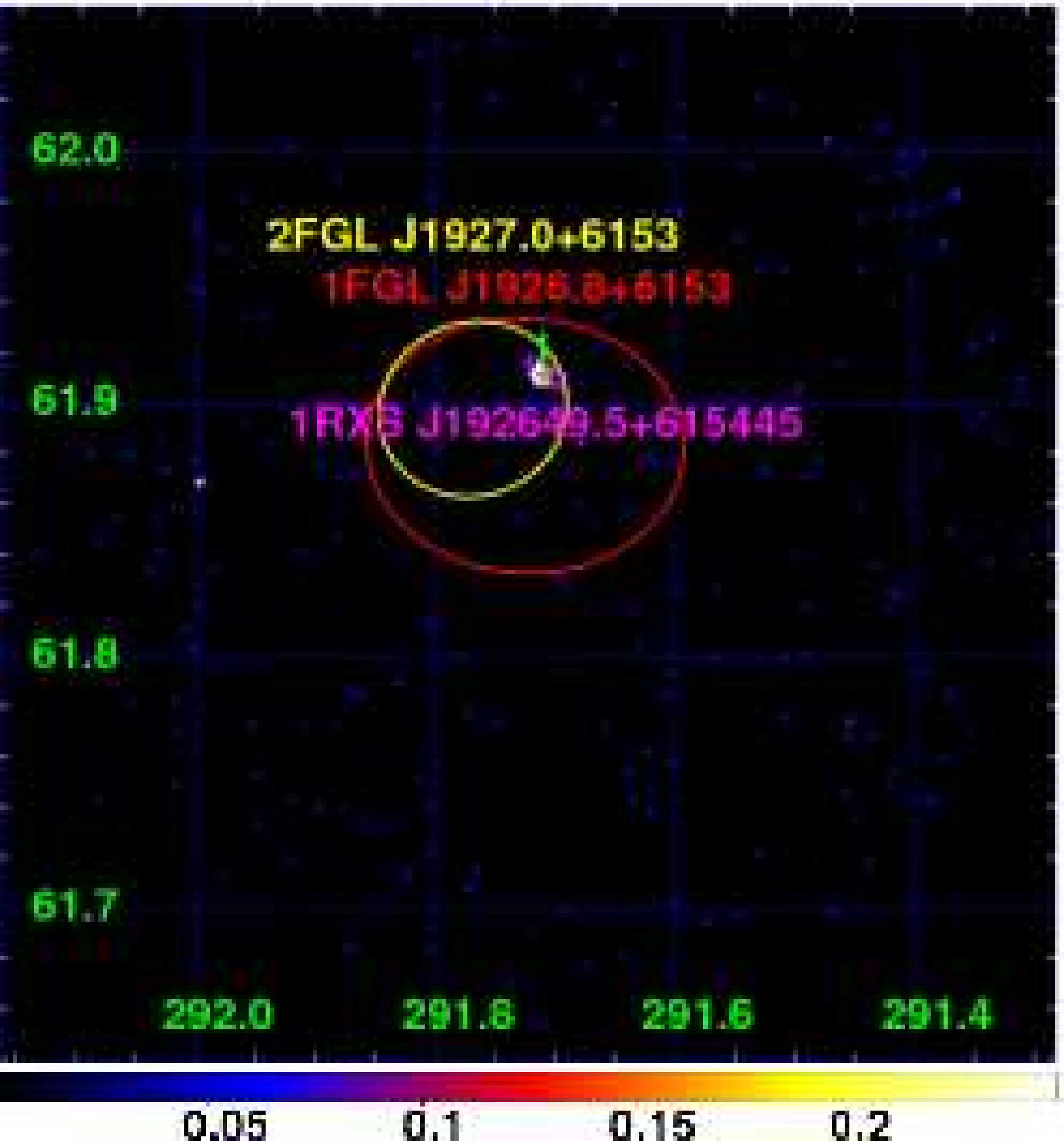}
    \end{center}
  \end{minipage}
  \begin{minipage}{0.32\hsize}
    \begin{center}
      {\small (111) 1FGL\,J1956.2--0238} \\
      \includegraphics[width=52mm]{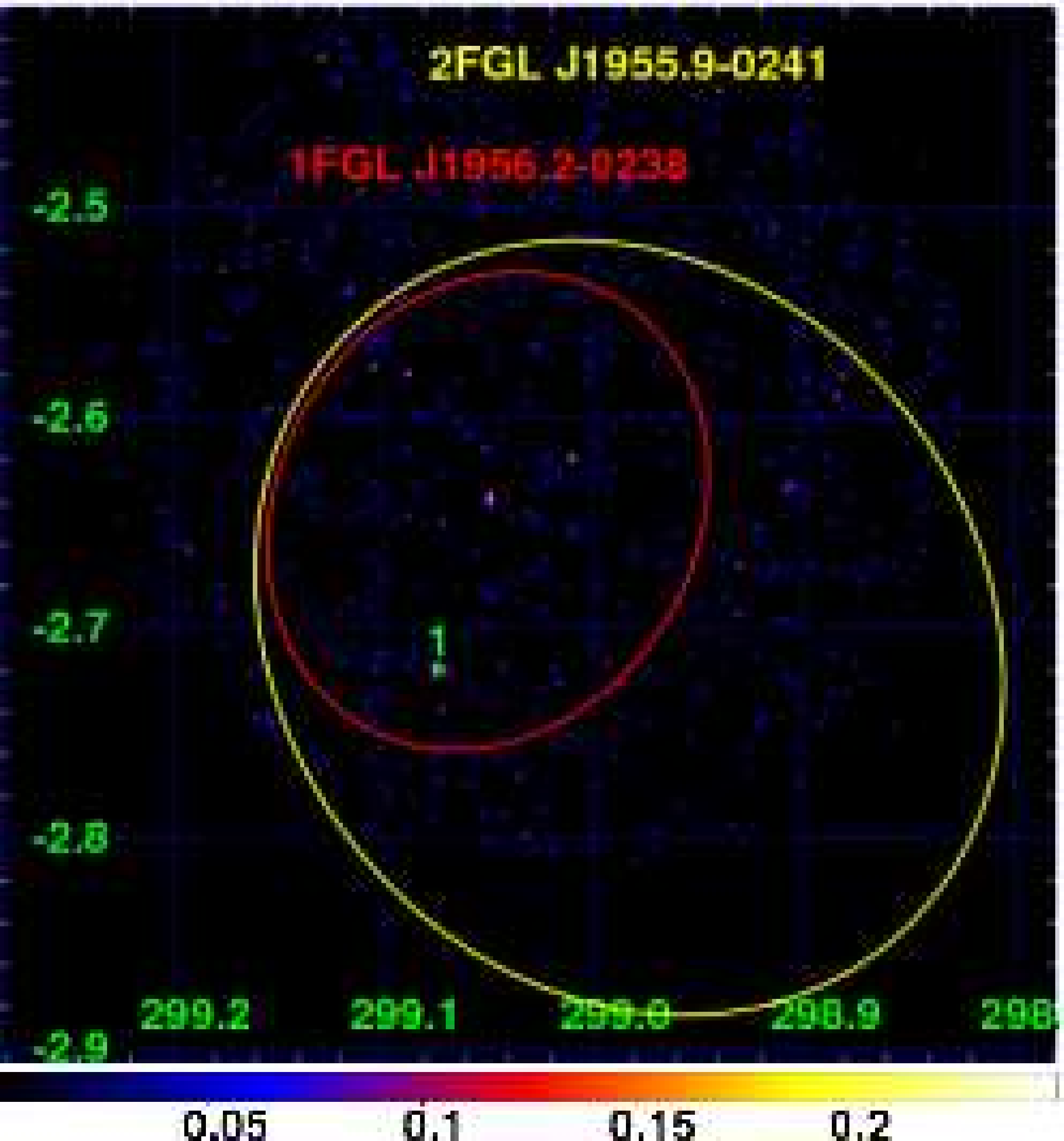}
    \end{center}
  \end{minipage}
  \begin{minipage}{0.32\hsize}
    \begin{center}
      {\small (112) 1FGL\,J1959.7--4730} \\
      \includegraphics[width=52mm]{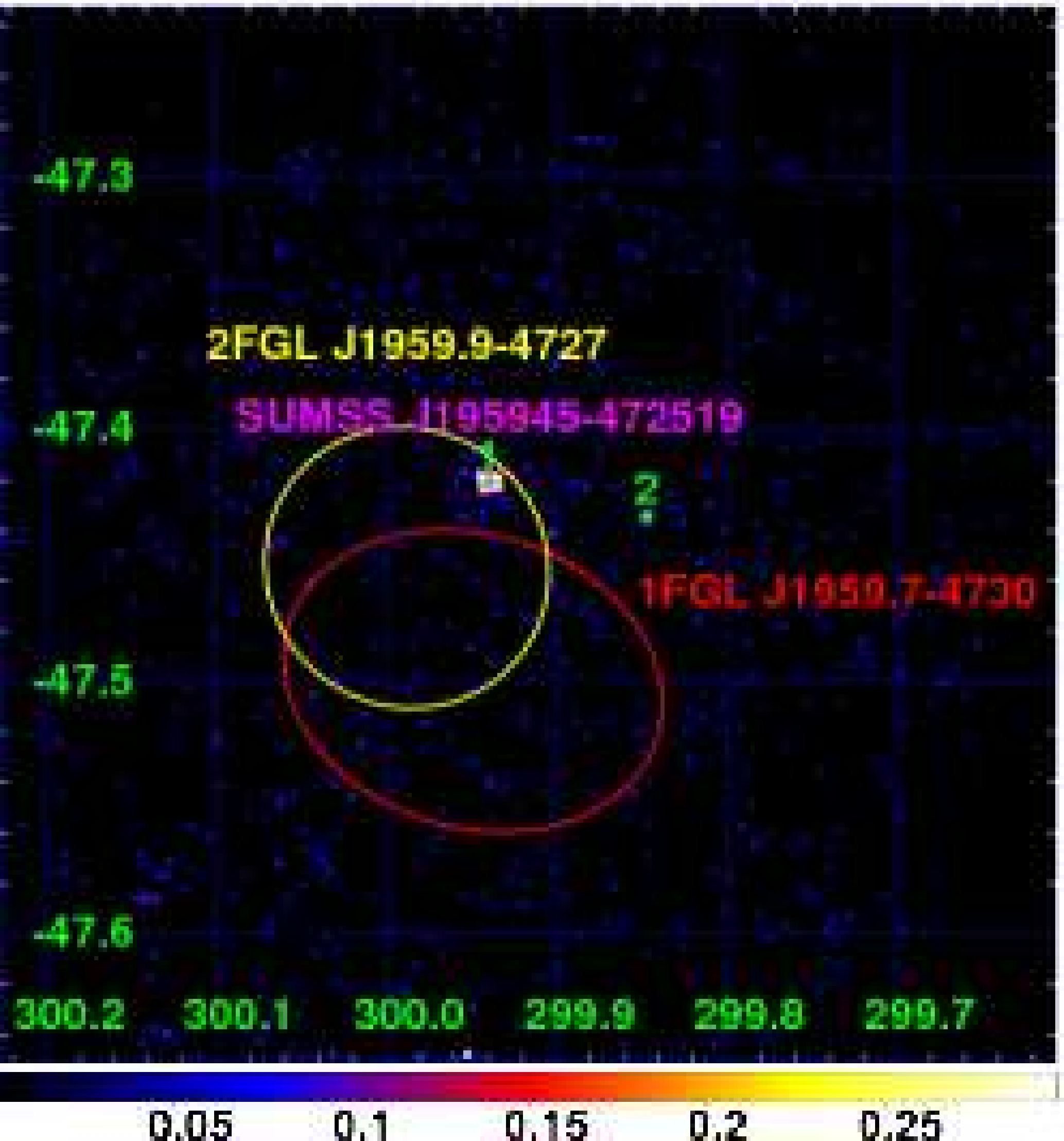}
    \end{center}
  \end{minipage}
  \begin{minipage}{0.32\hsize}
    \begin{center}
      {\small (113) 1FGL\,J2004.8$+$7004} \\
      \includegraphics[width=52mm]{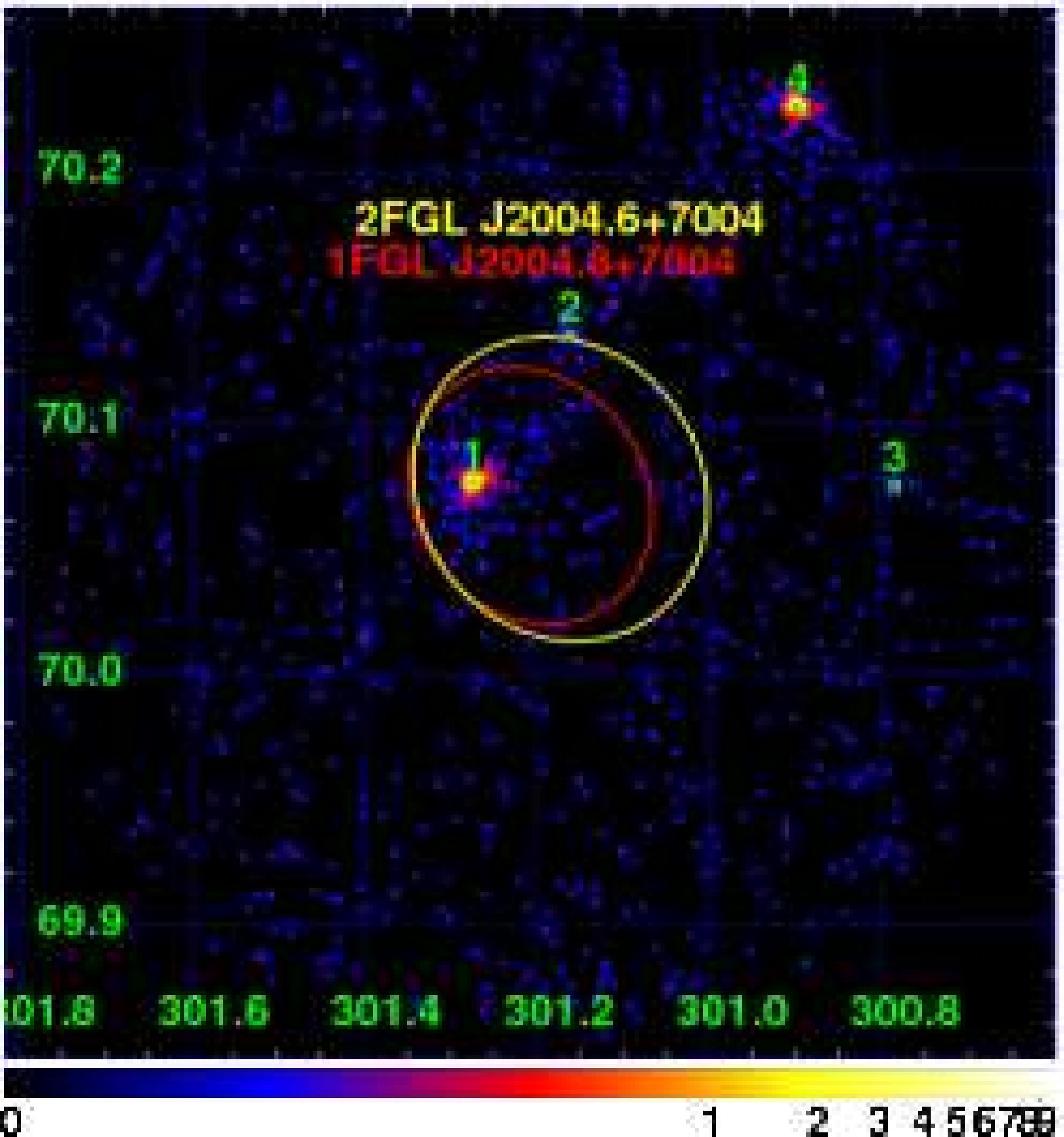}
    \end{center}
  \end{minipage}
  \begin{minipage}{0.32\hsize}
    \begin{center}
      {\small (114) 1FGL\,J2014.4$+$0647} \\
      \includegraphics[width=52mm]{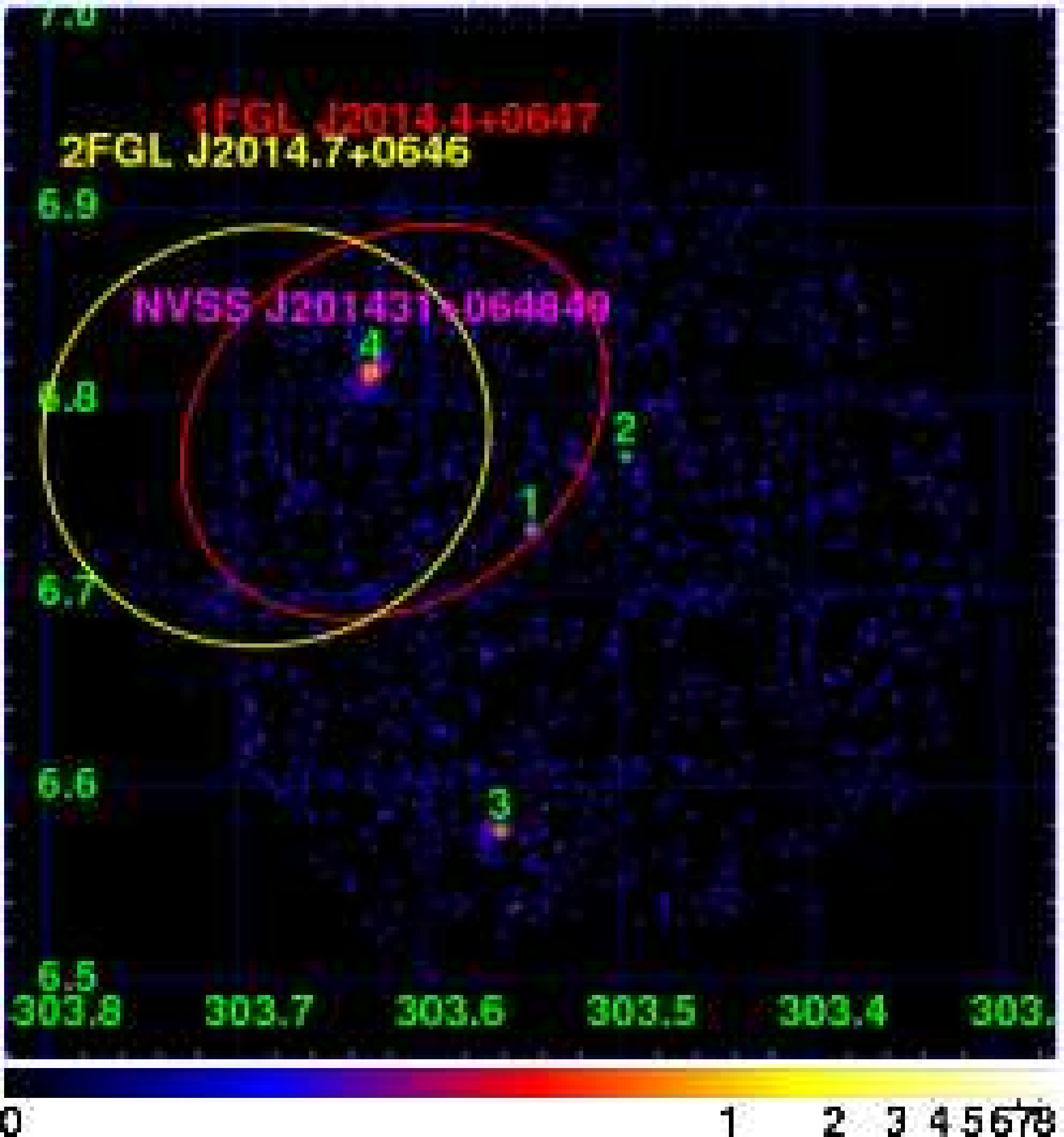}
    \end{center}
  \end{minipage}
  \begin{minipage}{0.32\hsize}
    \begin{center}
      {\small (115) 1FGL\,J2034.6--4202} \\
      \includegraphics[width=52mm]{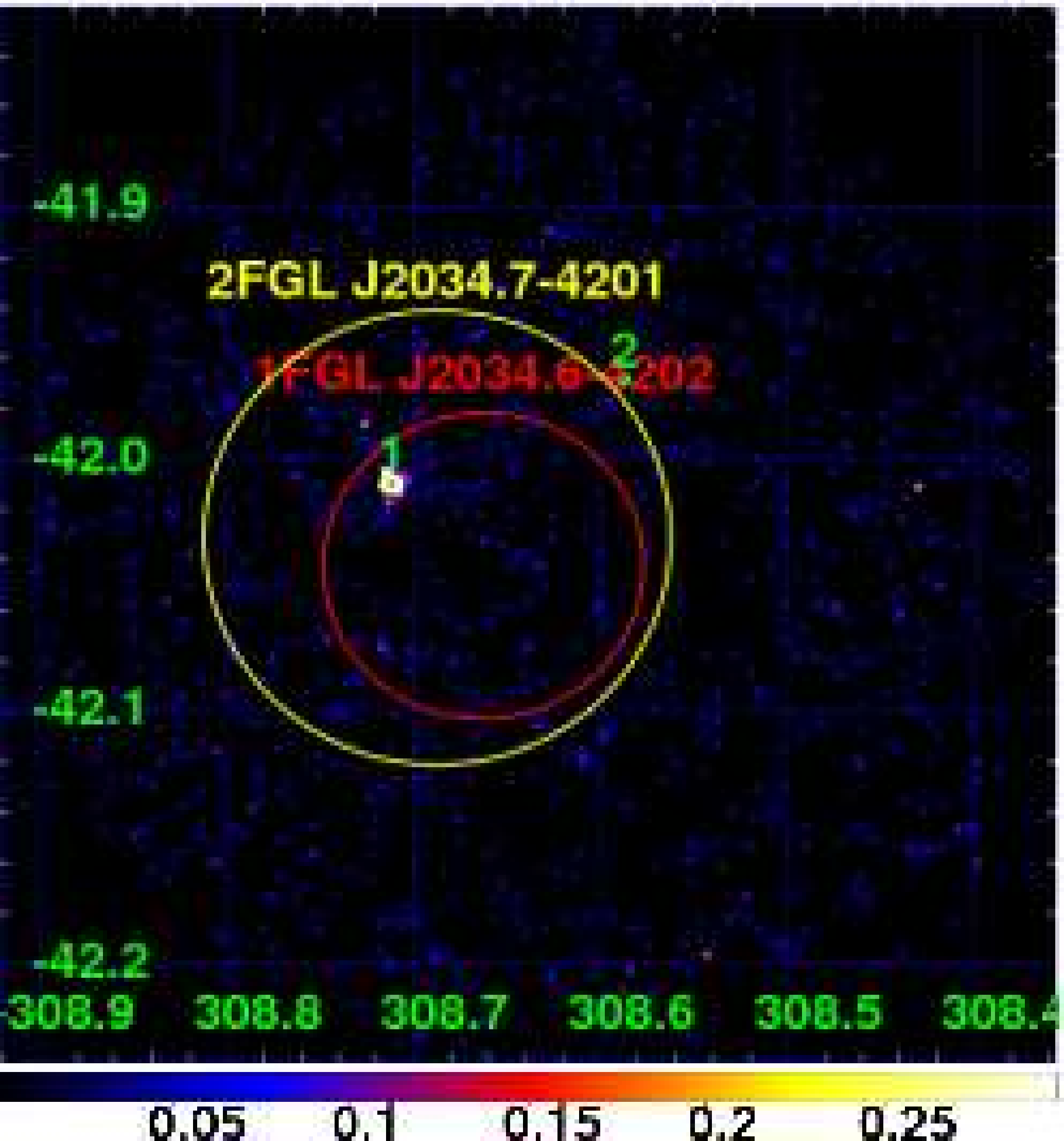}
    \end{center}
  \end{minipage}
  \begin{minipage}{0.32\hsize}
    \begin{center}
      {\small (116) 1FGL\,J2039.4--5621} \\
      \includegraphics[width=52mm]{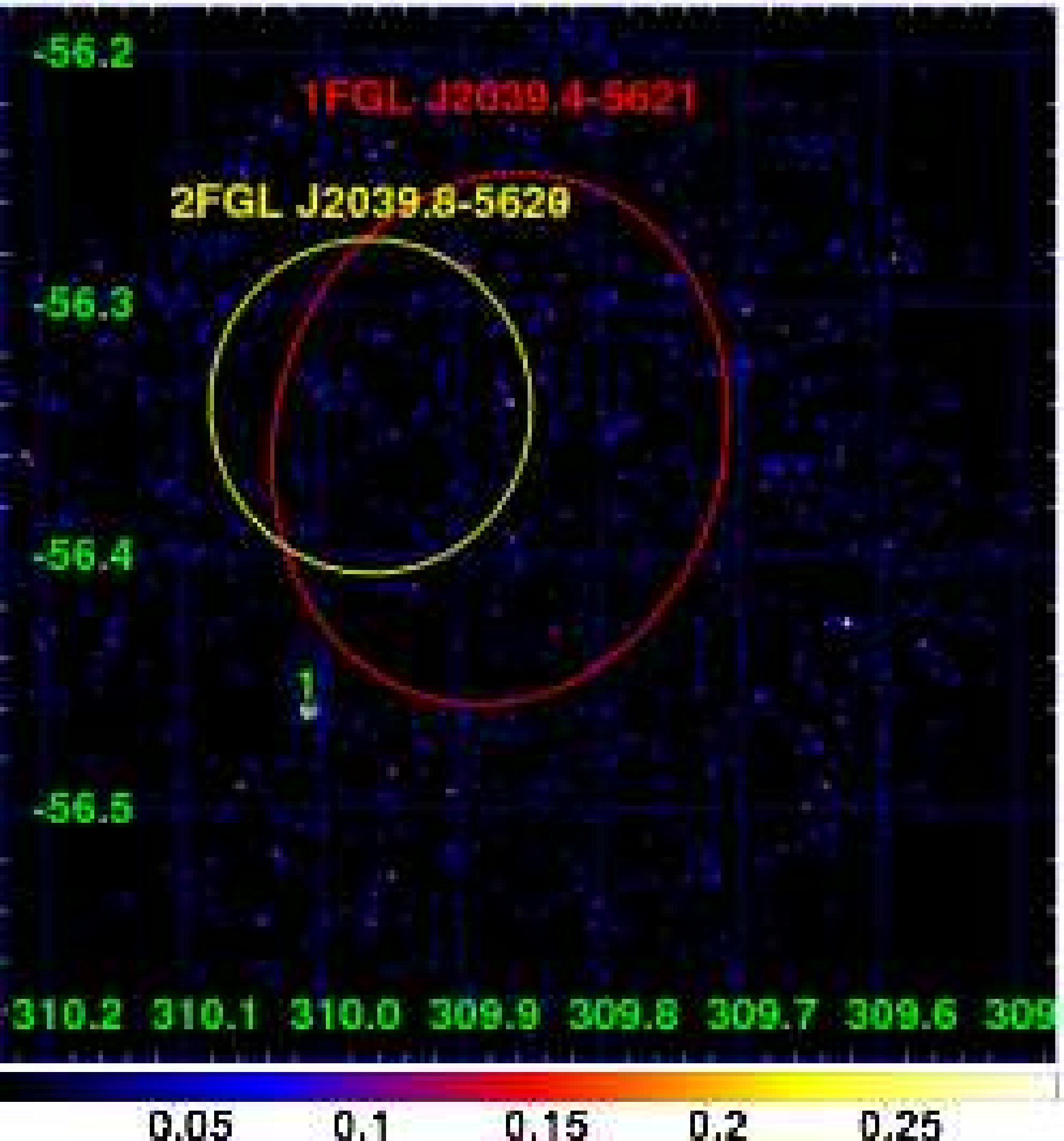}
    \end{center}
  \end{minipage}
  \begin{minipage}{0.32\hsize}
    \begin{center}
      {\small (117) 1FGL\,J2043.2$+$1709} \\
      \includegraphics[width=52mm]{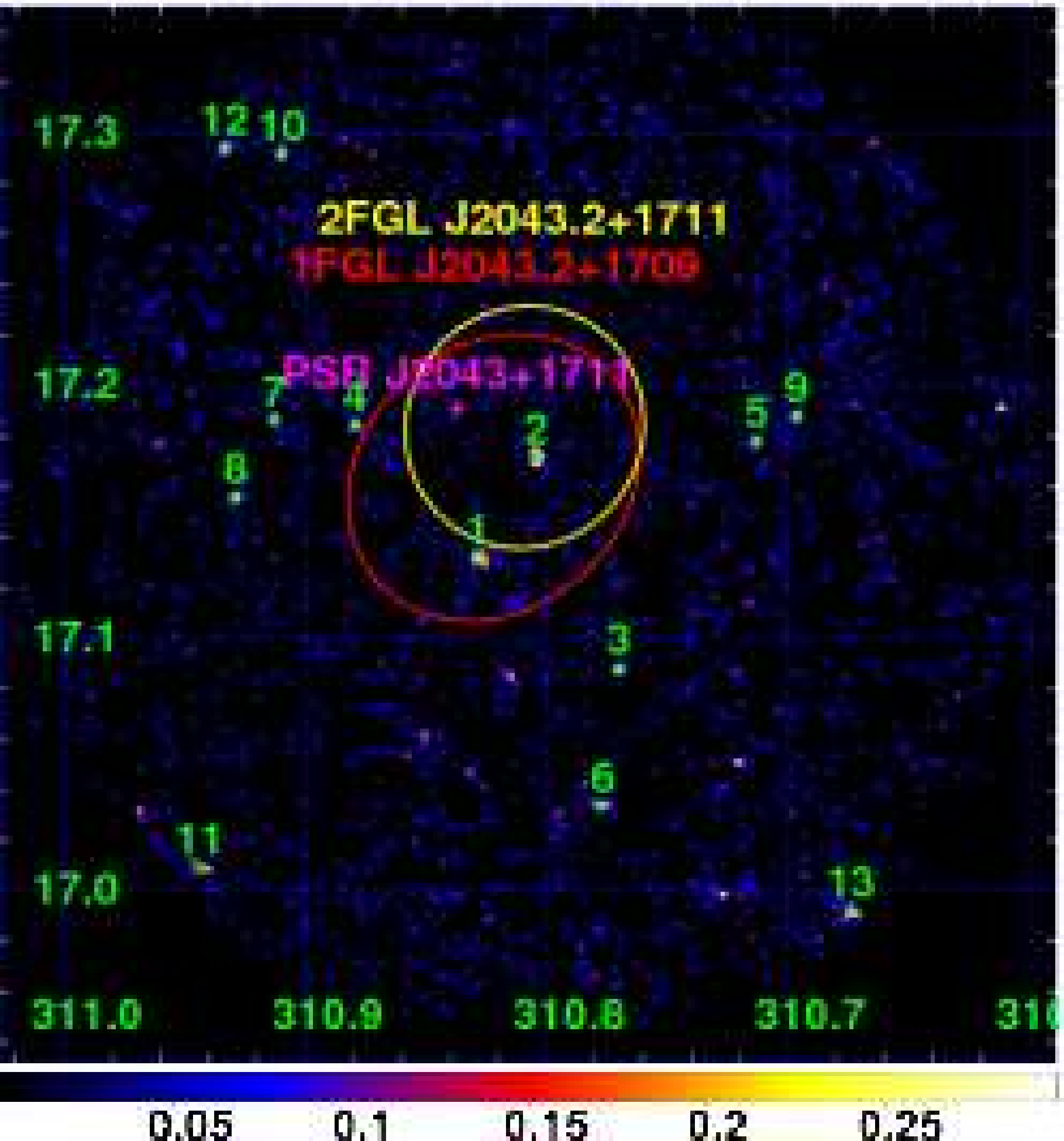}
    \end{center}
  \end{minipage}
  \begin{minipage}{0.32\hsize}
    \begin{center}
      {\small (118) 1FGL\,J2112.5--3044} \\
      \includegraphics[width=52mm]{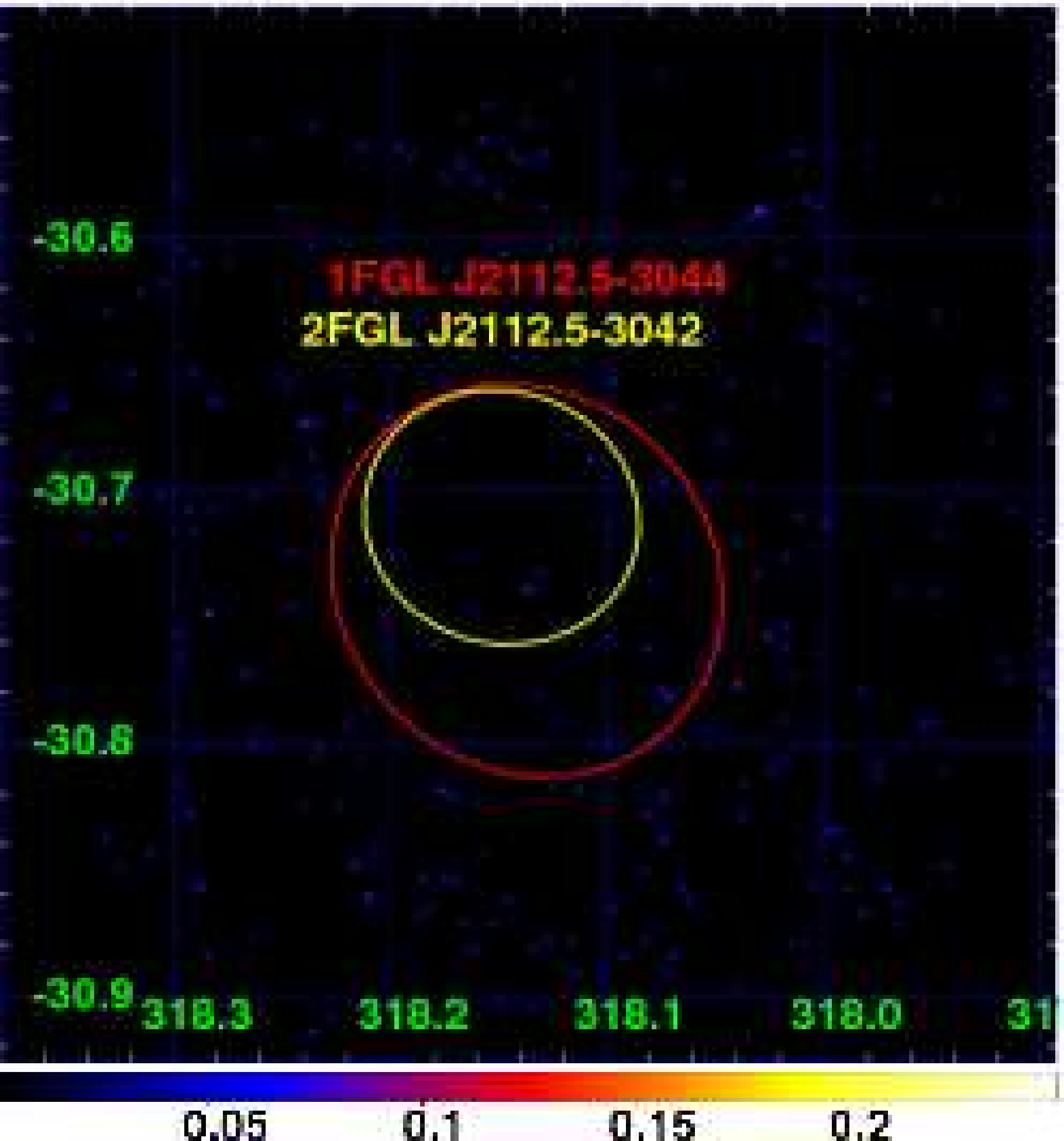}
    \end{center}
  \end{minipage}
  \begin{minipage}{0.32\hsize}
    \begin{center}
      {\small (119) 1FGL\,J2129.8--0427} \\
      \includegraphics[width=52mm]{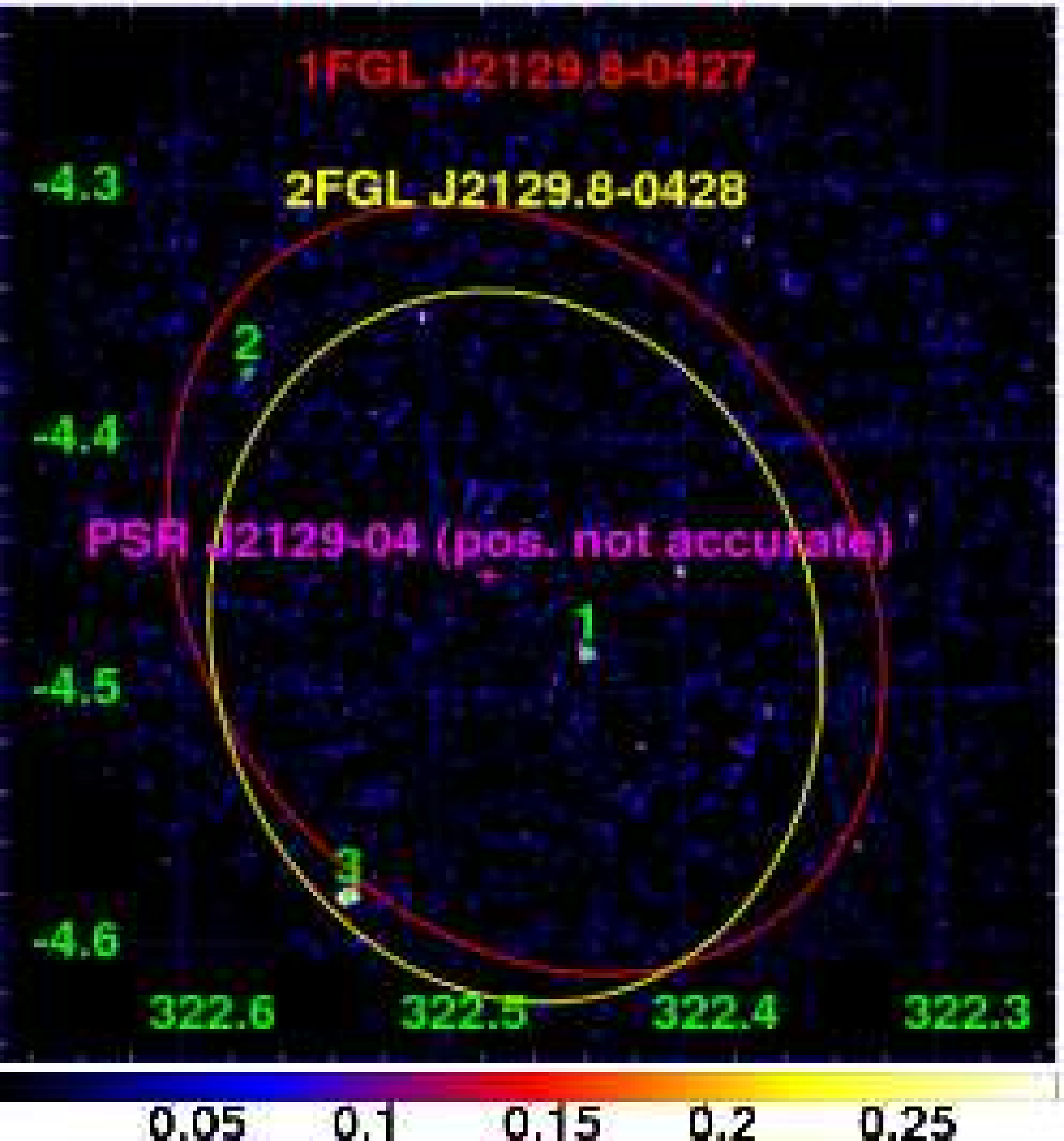}
    \end{center} 
  \end{minipage}
  \begin{minipage}{0.32\hsize}
    \begin{center}
      {\small (120) 1FGL\,J2134.5--2130} \\
      \includegraphics[width=52mm]{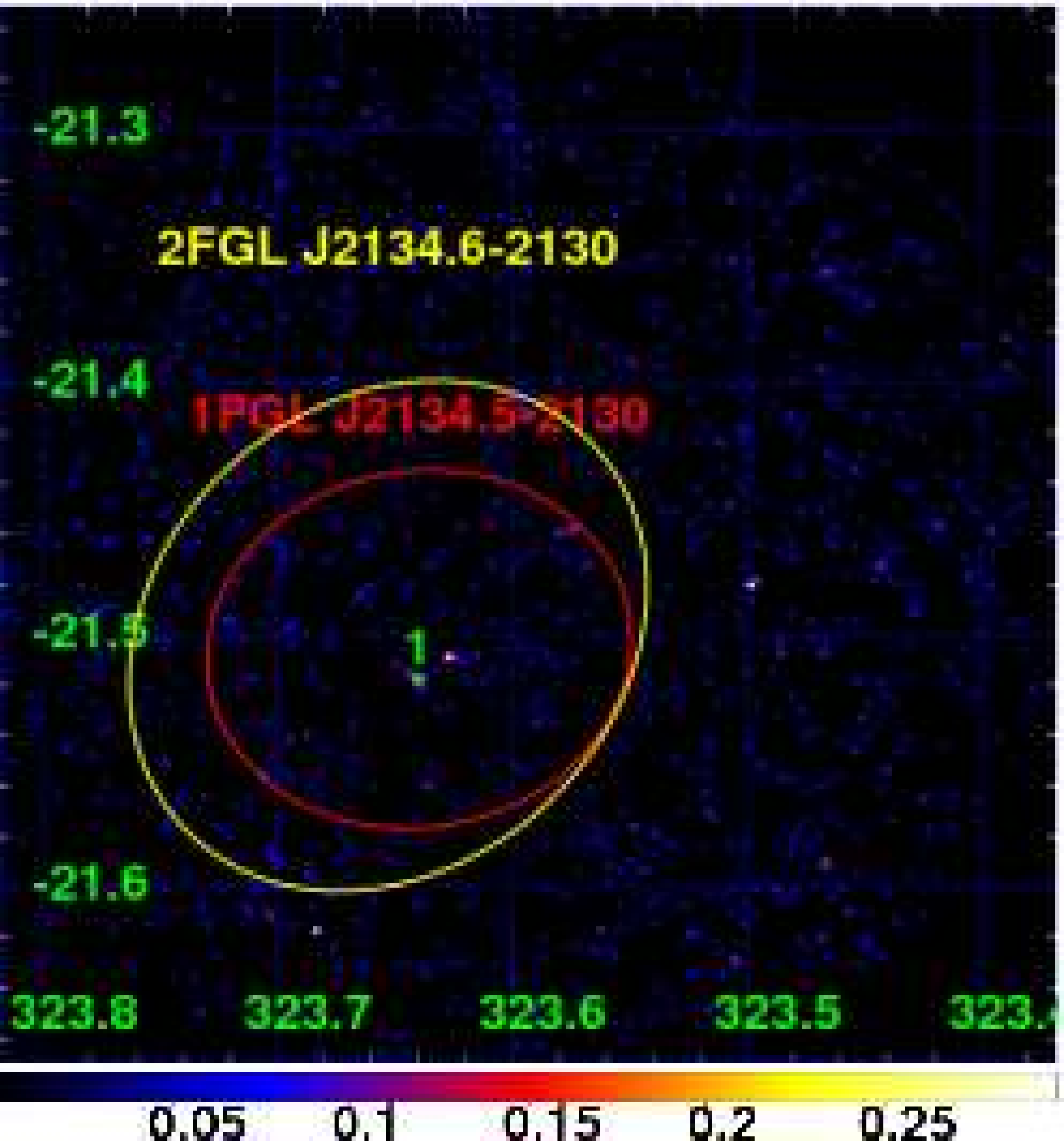}
    \end{center}
  \end{minipage}
 \end{center}
\end{figure}
\clearpage
\begin{figure}[m]
 \begin{center}
  \begin{minipage}{0.32\hsize}
    \begin{center}
      {\small (121) 1FGL\,J2223.3$+$0103} \\
      \includegraphics[width=52mm]{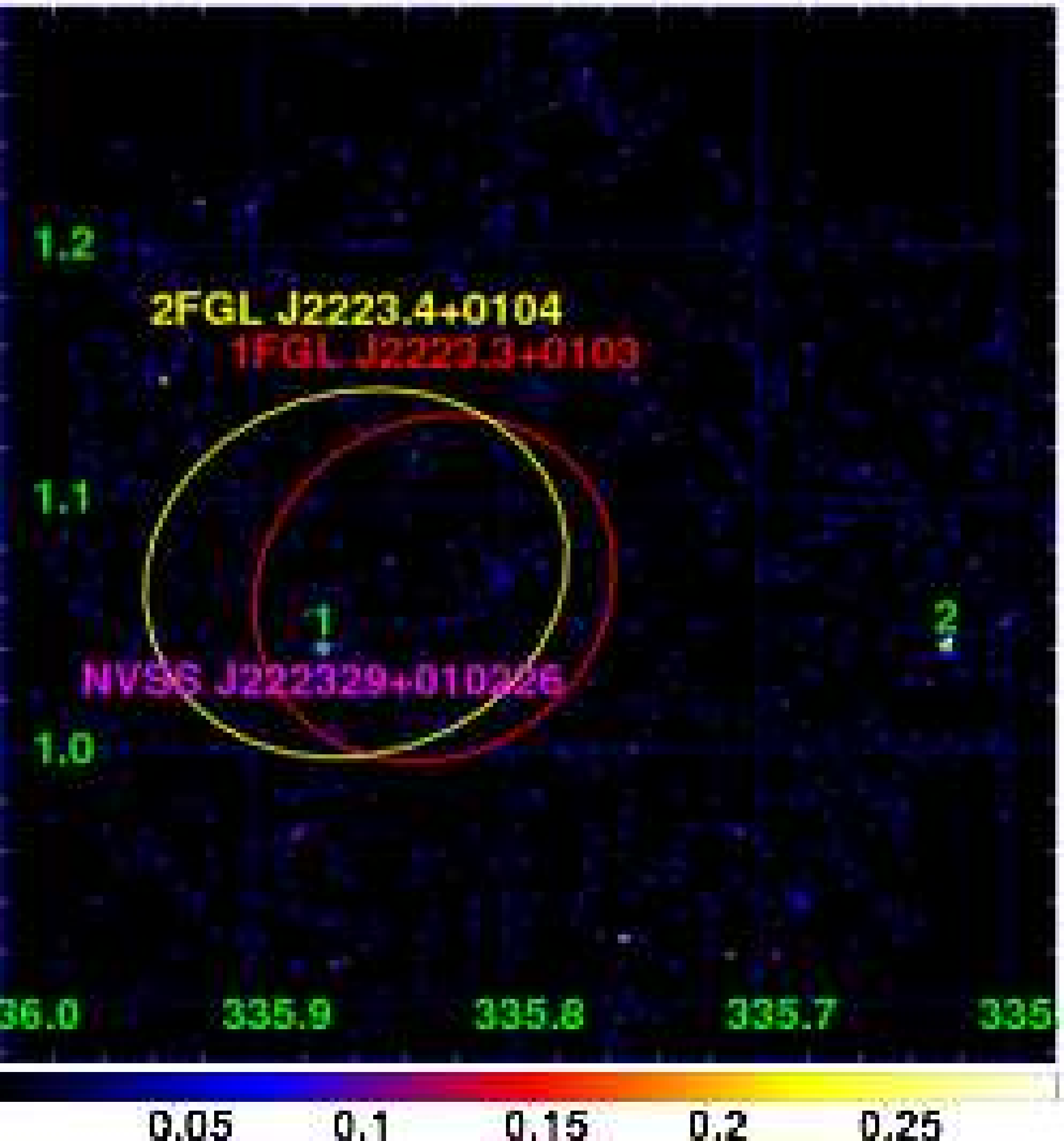}
    \end{center}
  \end{minipage}
  \begin{minipage}{0.32\hsize}
    \begin{center}
      {\small (122) 1FGL\,J2228.5--1633} \\
      \includegraphics[width=52mm]{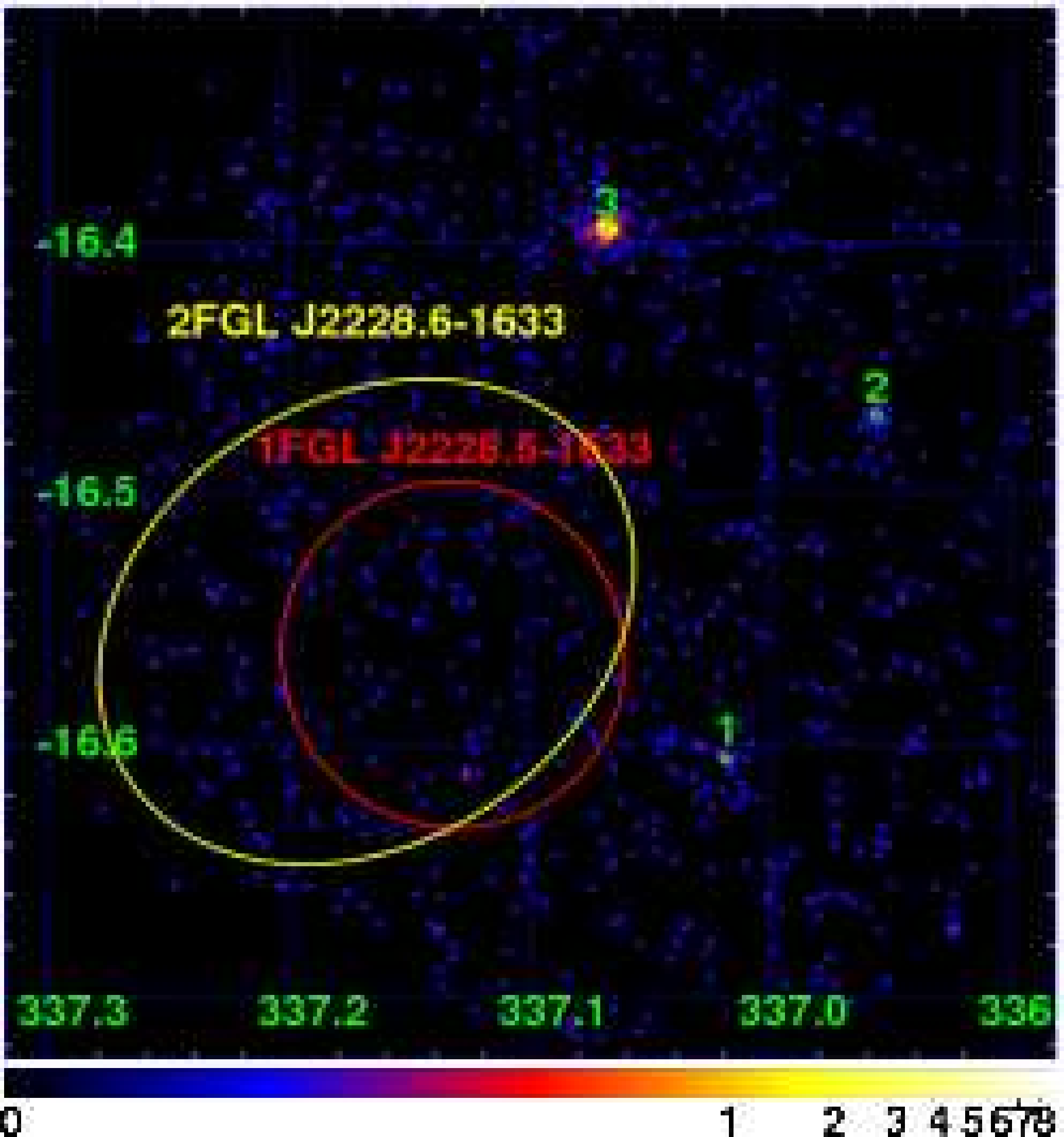}
    \end{center} 
  \end{minipage}
   \begin{minipage}{0.32\hsize}
    \begin{center}
      {\small (123) 1FGL\,J2243.4$+$4104} \\
      \includegraphics[width=52mm]{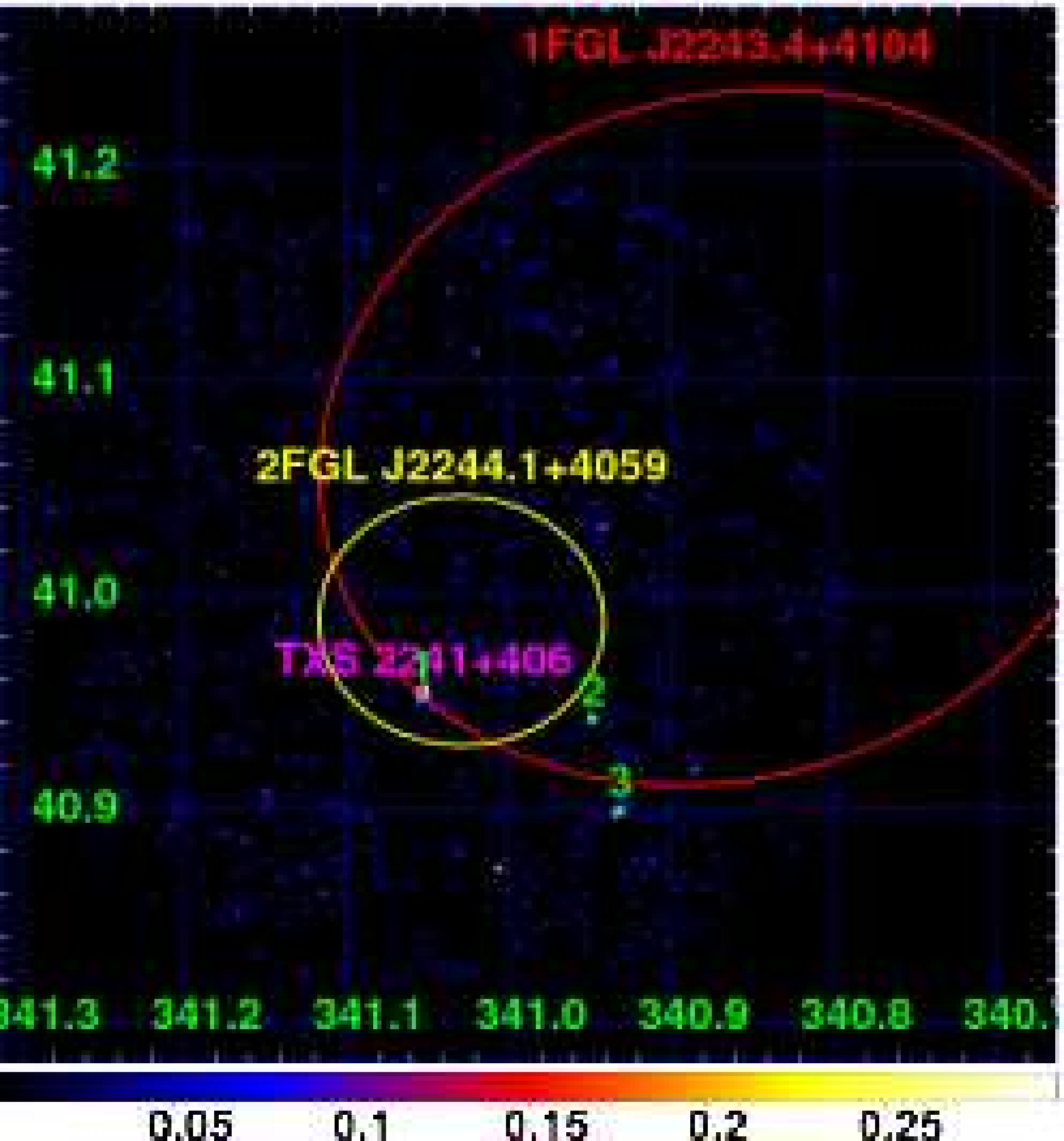}
    \end{center}
  \end{minipage}
  \begin{minipage}{0.32\hsize}
    \begin{center}
      {\small (124) 1FGL\,J2251.2--4928} \\
      \includegraphics[width=52mm]{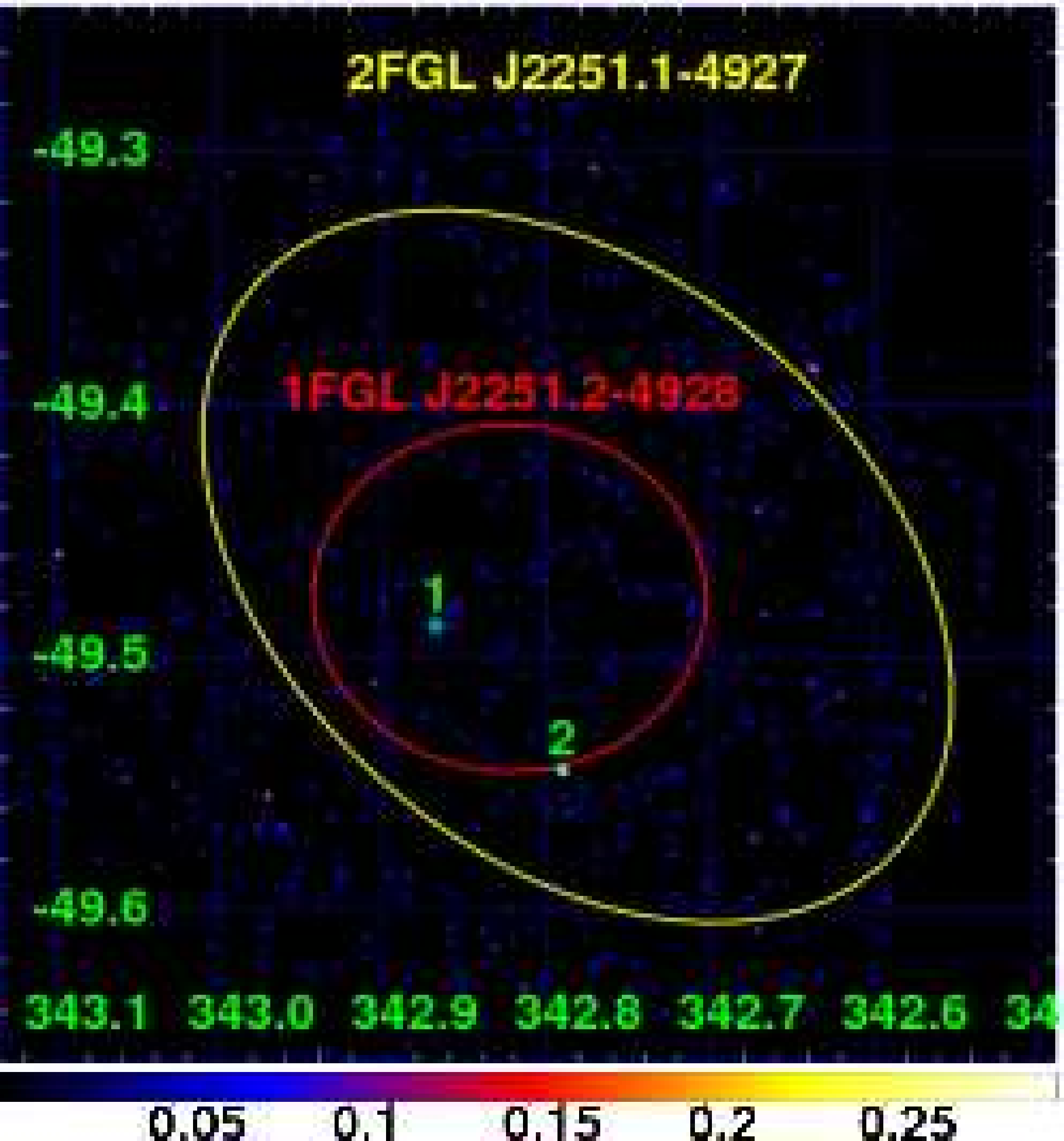}
    \end{center}
  \end{minipage}
  \begin{minipage}{0.32\hsize}
    \begin{center}
      {\small (125) 1FGL\,J2256.9--1024} \\
      \includegraphics[width=52mm]{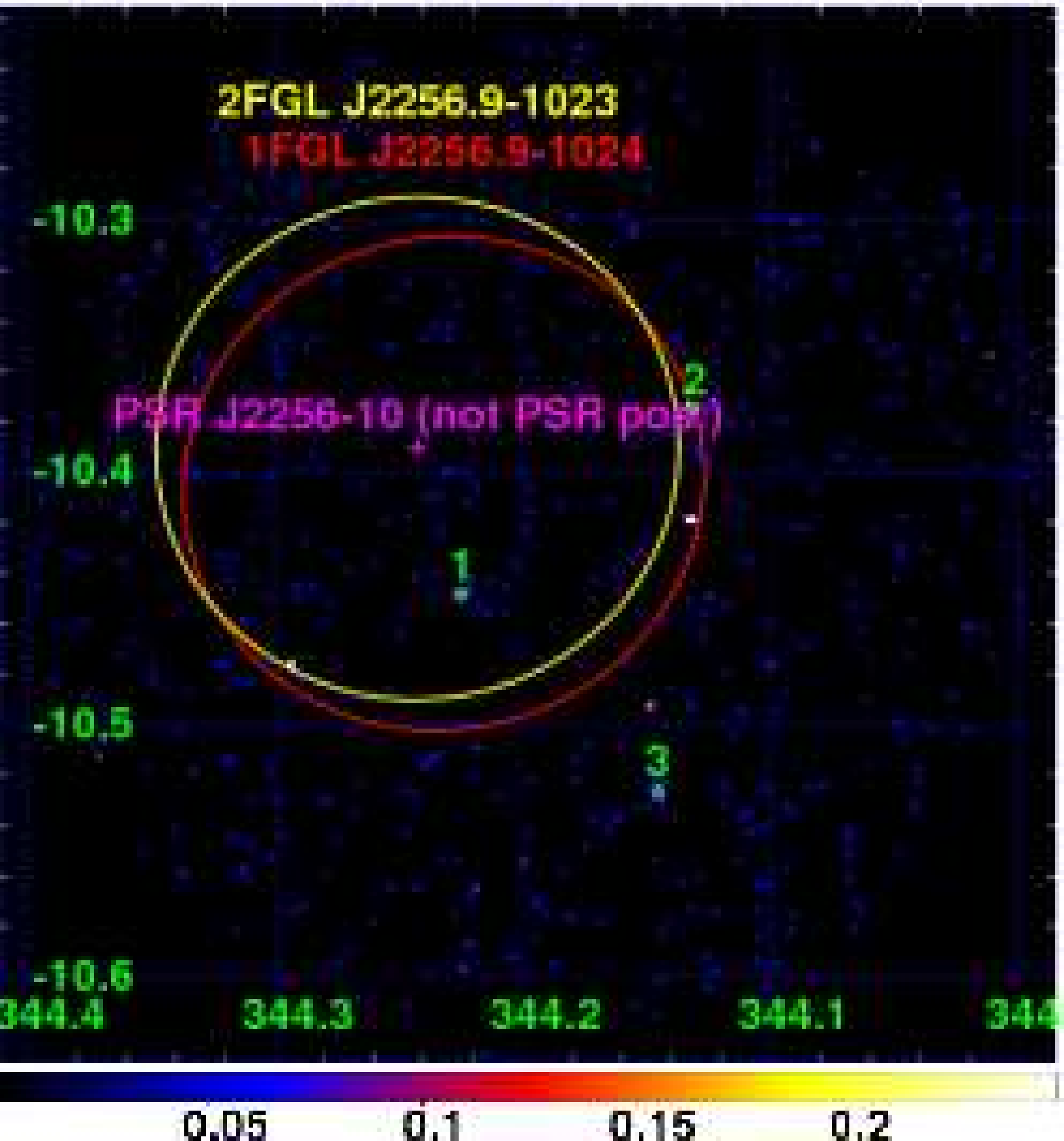}
    \end{center} 
  \end{minipage}
  \begin{minipage}{0.32\hsize}
    \begin{center}
      {\small (126) 1FGL\,J2257.9--3643} \\
      \includegraphics[width=52mm]{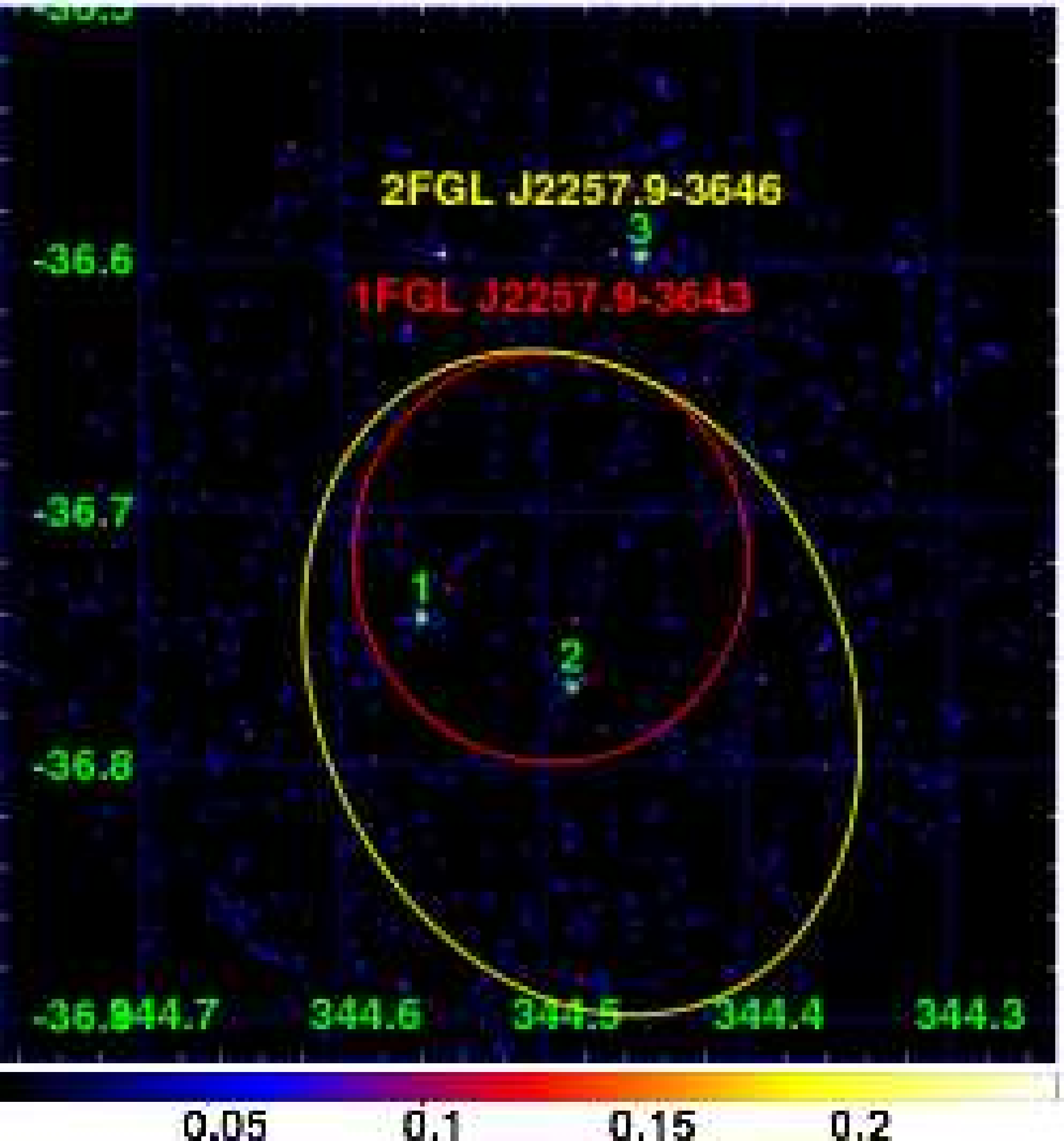}
    \end{center}
  \end{minipage}
  \begin{minipage}{0.32\hsize}
    \begin{center}
      {\small (127) 1FGL\,J2259.9--8255} \\
      \includegraphics[width=52mm]{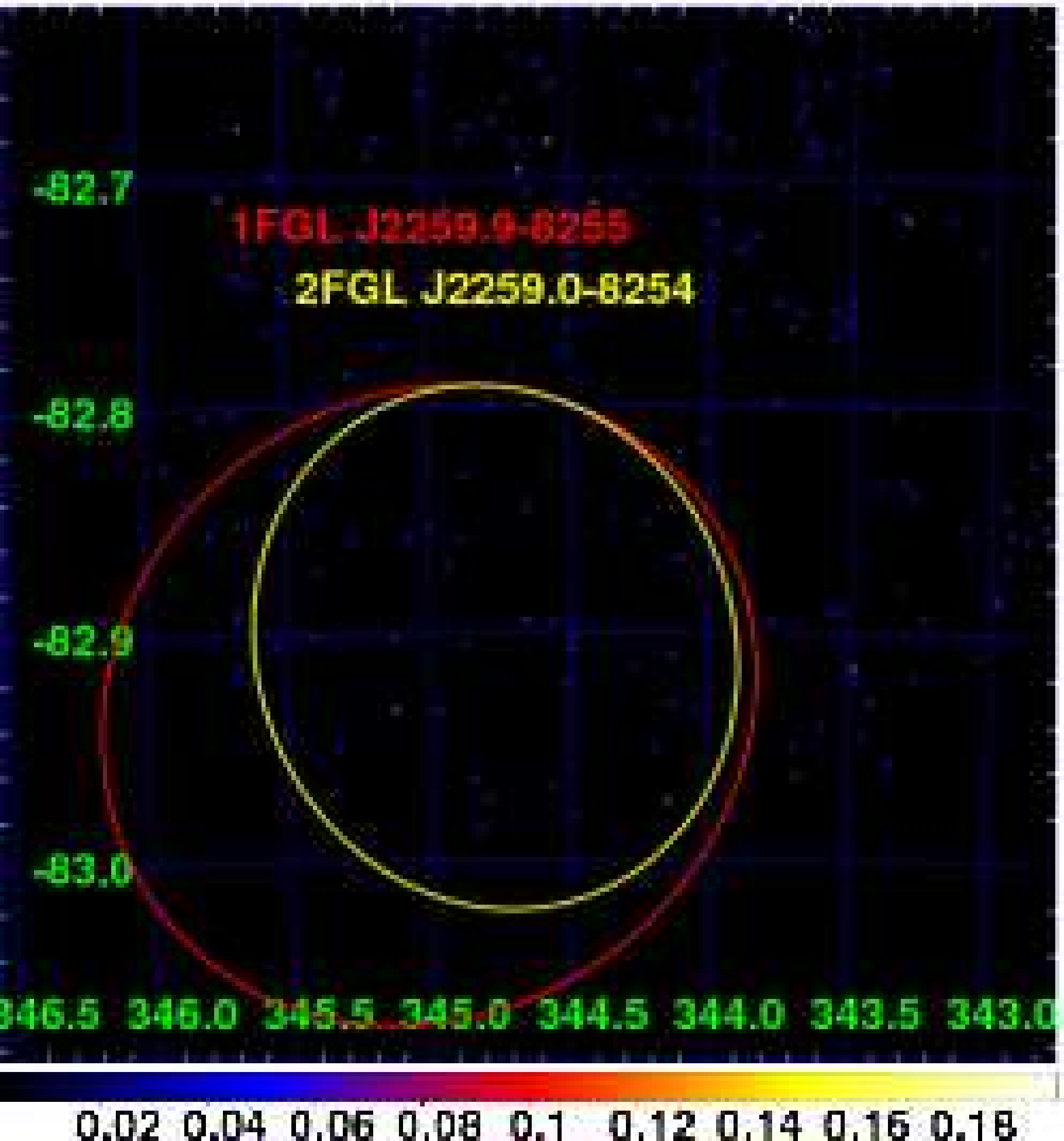}
    \end{center}
  \end{minipage}
  \begin{minipage}{0.32\hsize}
    \begin{center}
      {\small (128) 1FGL\,J2310.0--3627} \\
      \includegraphics[width=52mm]{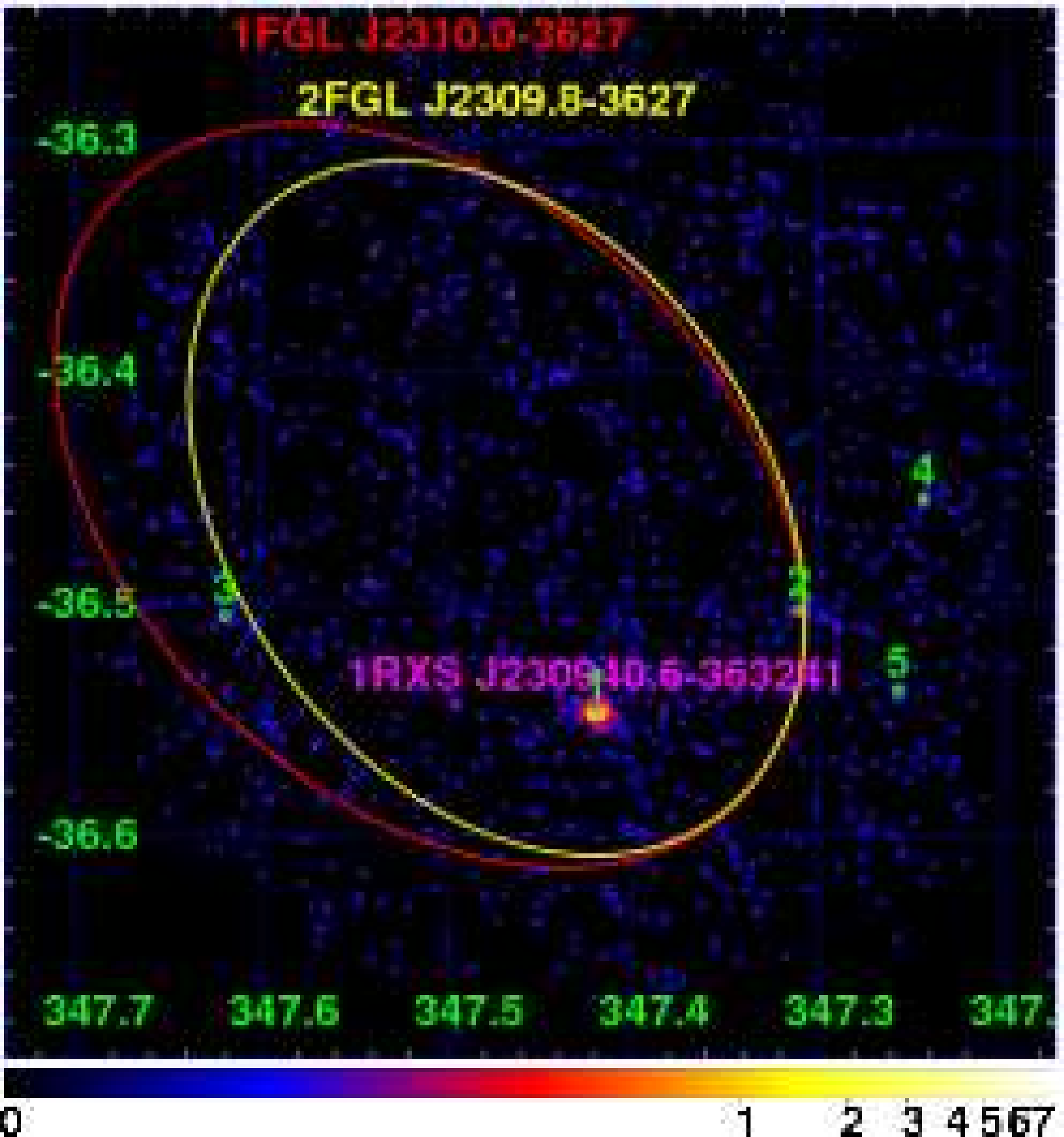}
    \end{center} 
  \end{minipage}
  \begin{minipage}{0.32\hsize}
    \begin{center}
      {\small (129) 1FGL\,J2323.0--4919} \\
      \includegraphics[width=52mm]{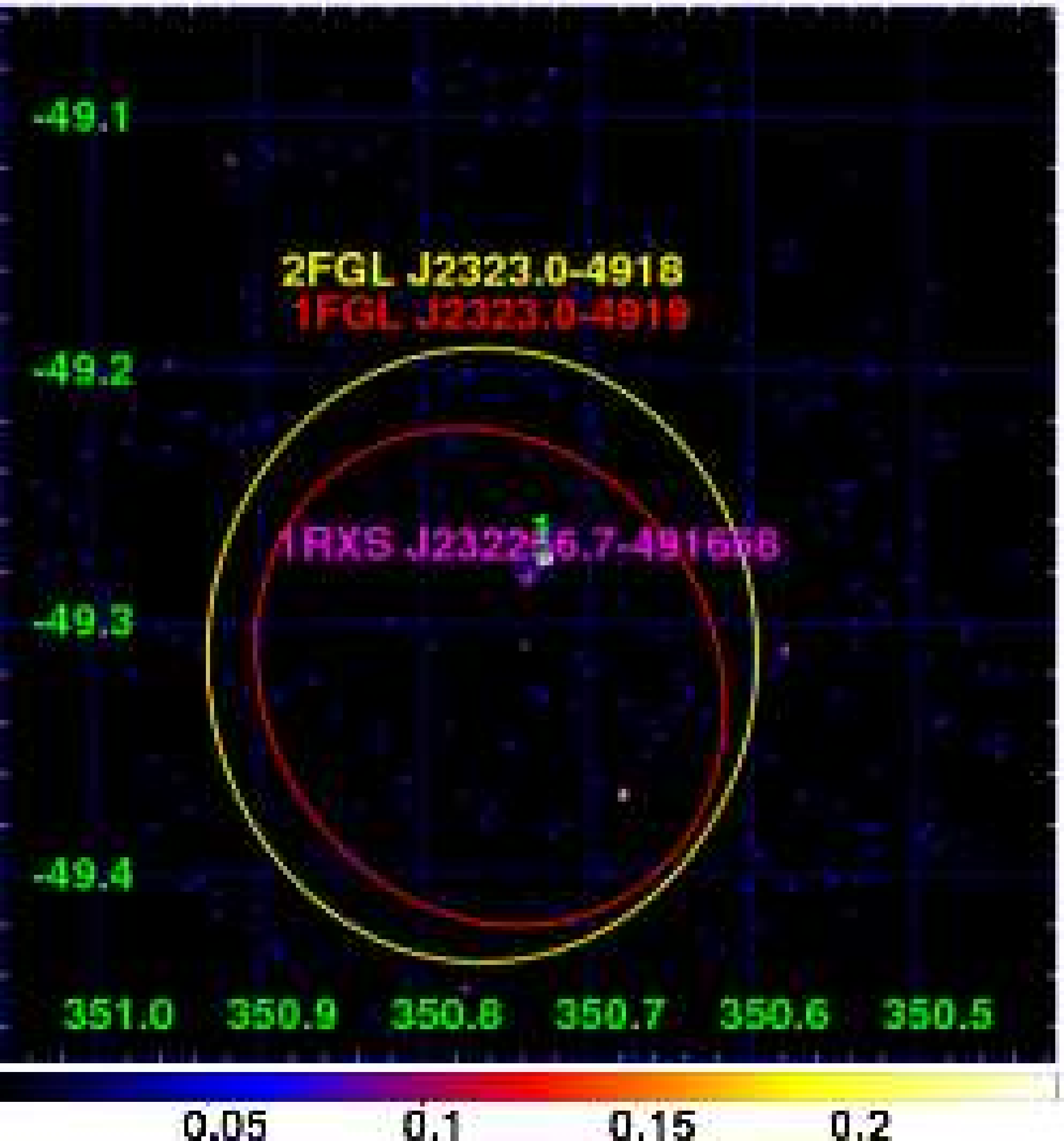}
    \end{center}
  \end{minipage}
  \begin{minipage}{0.32\hsize}
    \begin{center}
      {\small (130) 1FGL\,J2330.3--4745} \\
      \includegraphics[width=52mm]{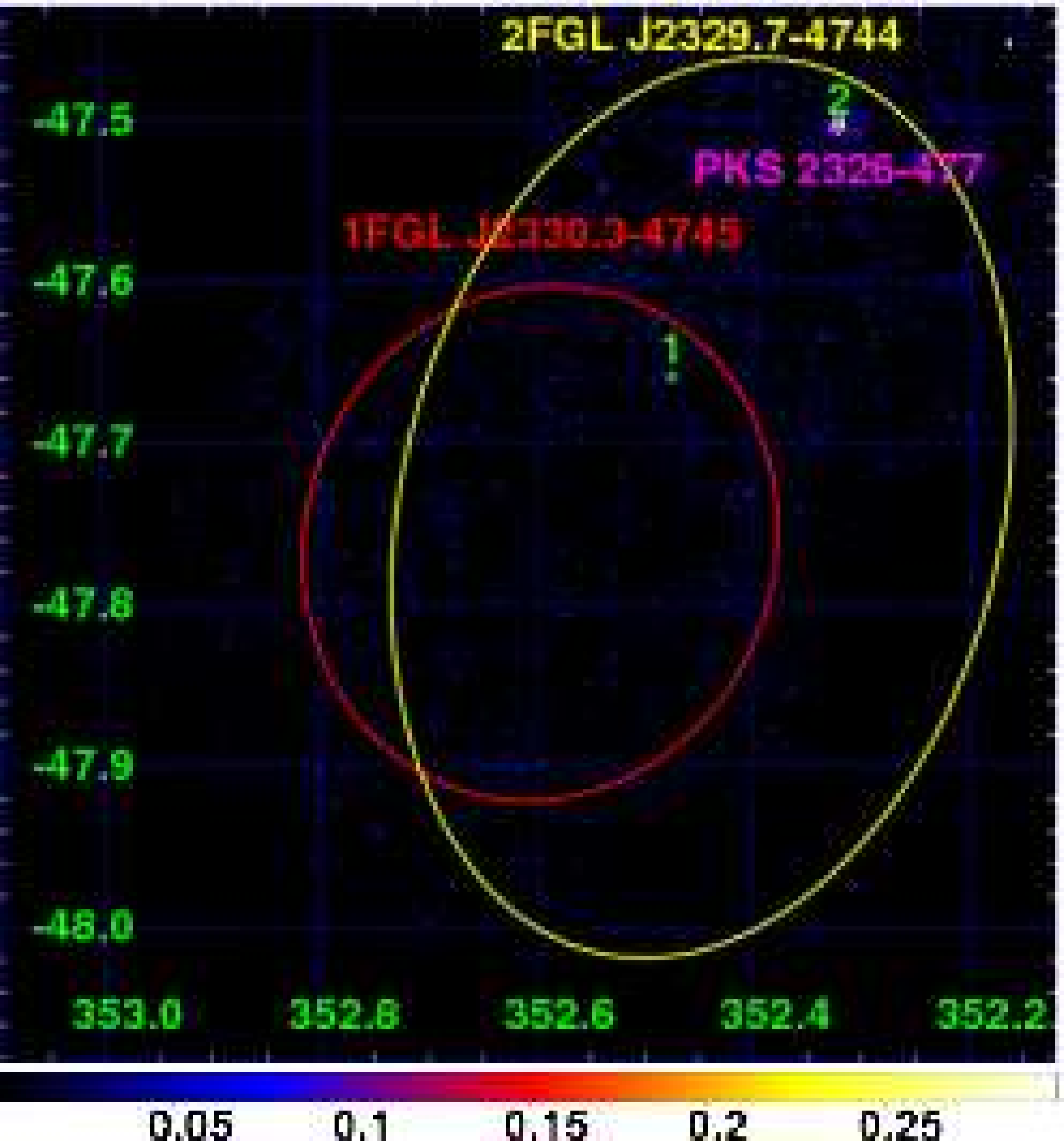}
    \end{center}
  \end{minipage}
  \begin{minipage}{0.32\hsize}
    \begin{center}
      {\small (131) 1FGL\,J2339.7--0531} \\
      \includegraphics[width=52mm]{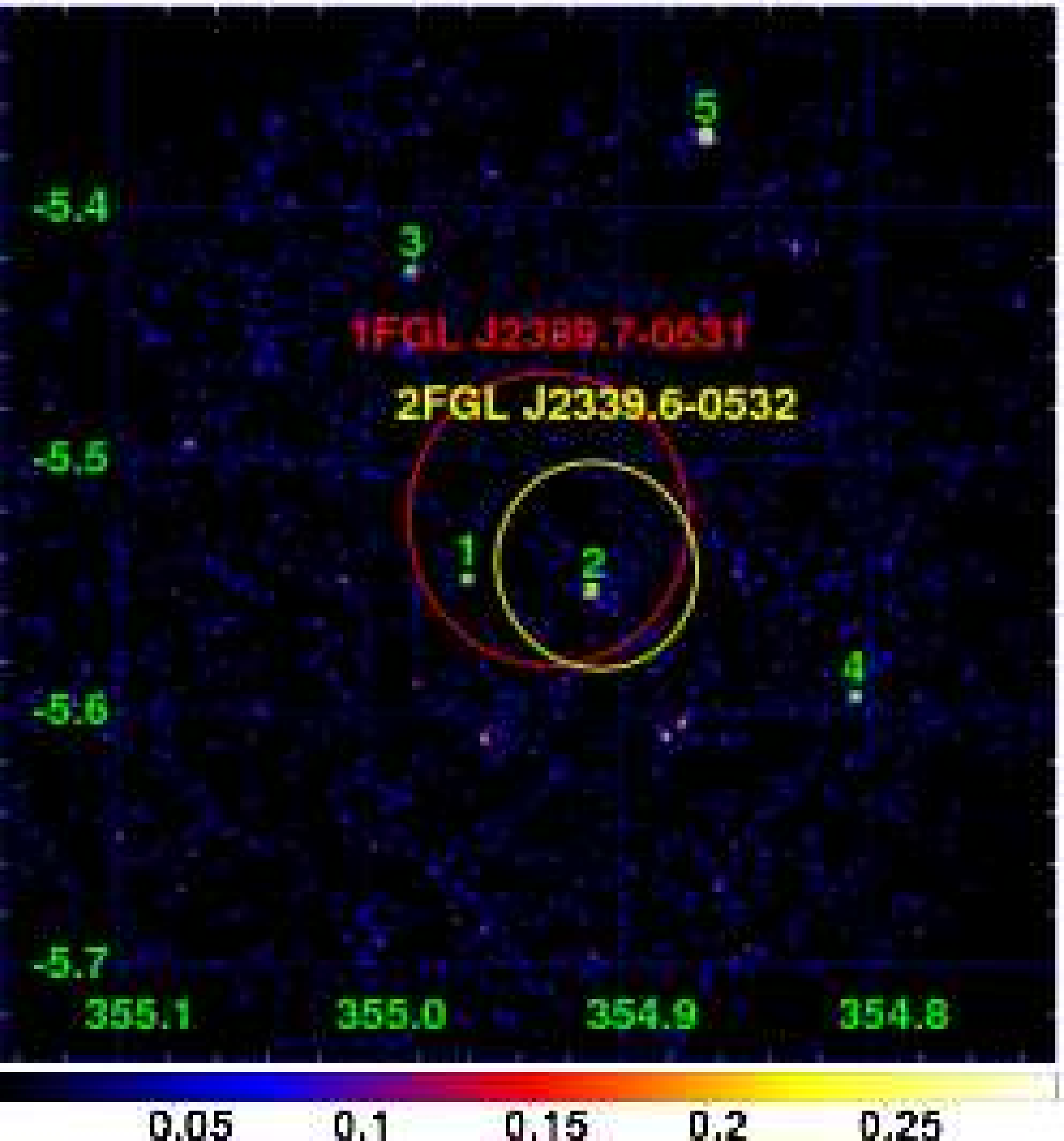}
    \end{center} 
  \end{minipage}
  \begin{minipage}{0.32\hsize}
    \begin{center}
      {\small (132) 1FGL\,J2347.3$+$0710} \\
      \includegraphics[width=52mm]{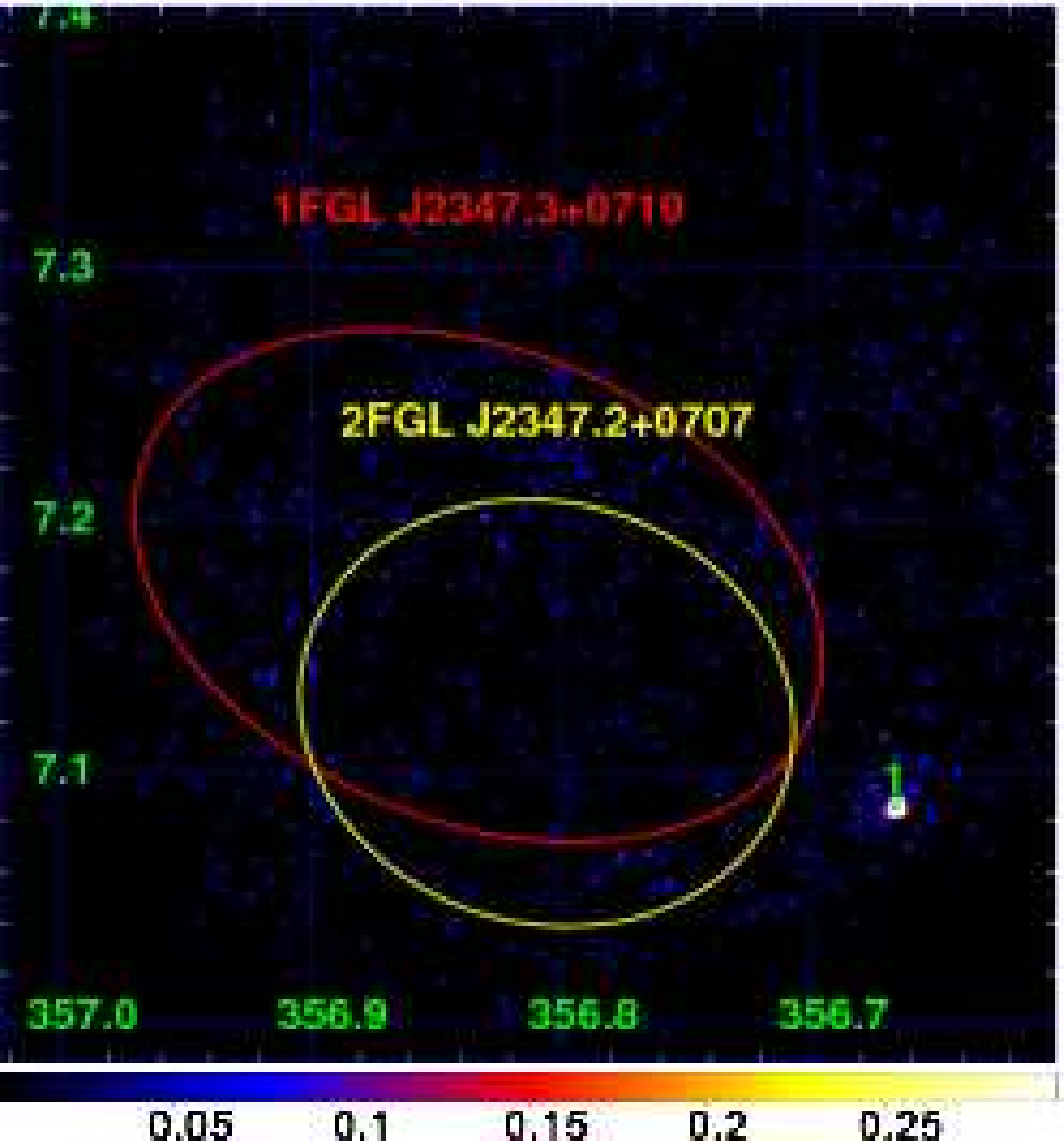}
    \end{center}
  \end{minipage}
 \end{center}
\end{figure}
\clearpage
\begin{figure}[m]
 \begin{center}
  \begin{minipage}{0.32\hsize}
    \begin{center}
      {\small (133) 1FGL\,J2350.1--3005} \\
      \includegraphics[width=52mm]{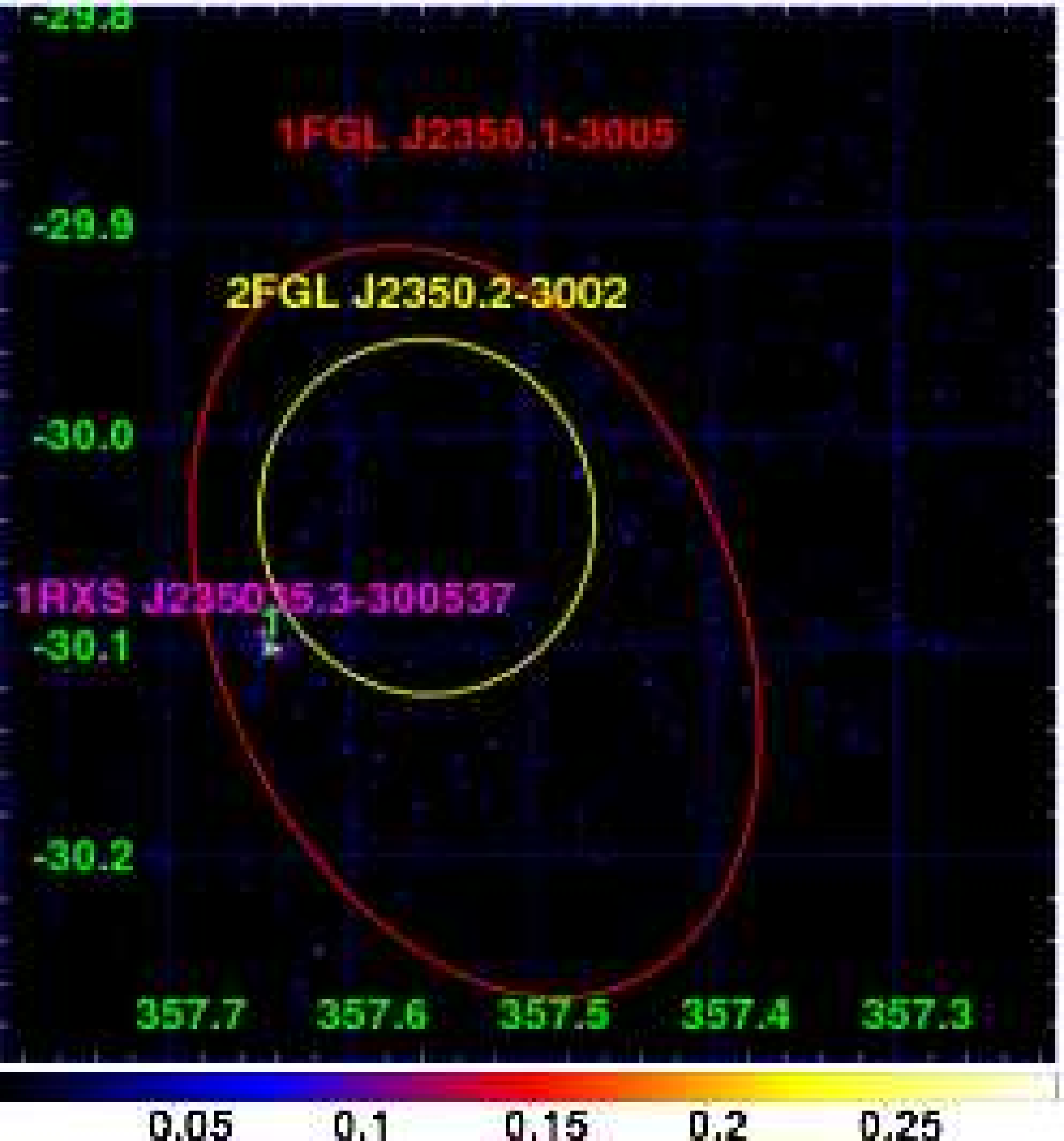}
    \end{center}
  \end{minipage}
  \begin{minipage}{0.32\hsize}
      {\small (134) 1FGL\,J2352.1$+$1752} \\
      \includegraphics[width=52mm]{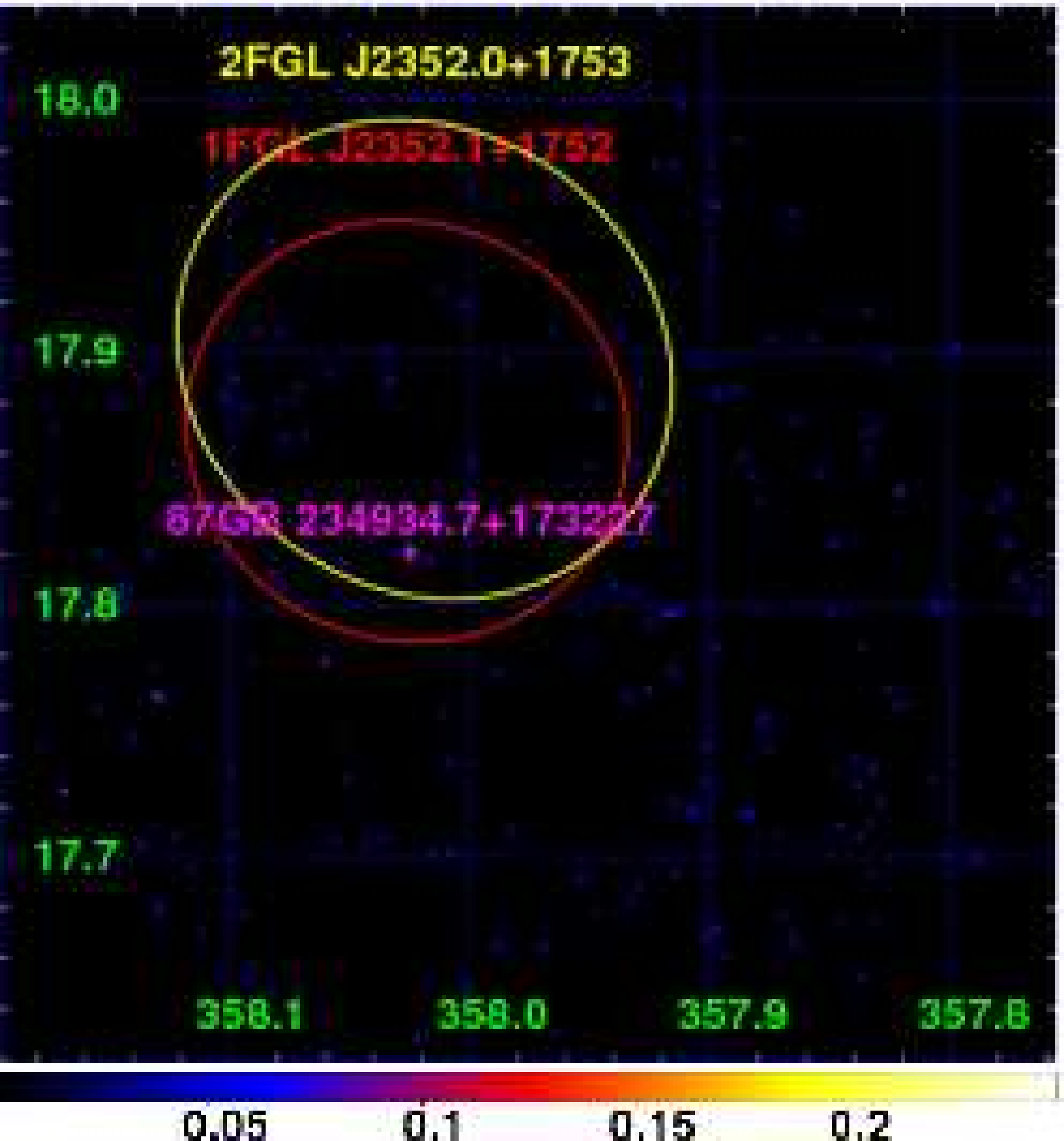}
  \end{minipage}
  \caption{
    {\it Swift}/XRT images of the 134 1FGL unID catolog sources
    of the $25'\times 25'$ FoV. 
    One or more sources are detected within the {\it Swift}/XRT FoV.
    Unfortunately, 2FGL error regions of some sources run off the edge of {\it Swift}/XRT FoV.
    The signal-to-noise acceptance threshold is set to $3 \sigma$.
    The yellow ellipses show $95\%$ error regions of 2FGL catalog
    gamma-ray sources, and red ellipses show that of 1FGL catalog
    sources. 
    If there are the radio and bright X-ray sources associated with gamma-ray source,
    we show those sources as magenta
    crosses.}
 \end{center}
\end{figure}
\clearpage
\begin{table}[m]
\small
\caption{{\it Swift}/XRT Results for 1FGL objects.
         This table only show the details of high signal-to-noise ($3 \sigma$) 
         sources detected within the {\it Swift}/XRT FoV. 
         The radius of 90\% congfidence error circles are described by $r_{90\%}$.
         Count rates are shown in the energy range of 0.3--10 keV.}
\scalebox{0.9}{
\begin{tabular}{ccccccc}
\hline \hline
1FGL Src & XRT Src\# & R.A.(J2000) [deg] & Dec.(J2000) [deg] & $r_{90\%}$ [$''$] & Count Rates [1E-03 cts/s] & S/N ratio \\
\hline
1FGL\,J0001.9--4158 & 1 & 0.3864 & -41.9227 & 3.73 & 6.95E+01+/-3.6 & 19.5 \\
 & 2 & 0.3671 & -42.0492 & 6.31 & 2.22+/-7.3E-01 & 3.0 \\
 & 3 & 0.2813 & -41.8854 & 5.70 & 3.36+/-9.2E-01 & 3.7 \\
\hline
1FGL\,J0009.1$+$5031 & 1 & 2.3453 & 50.5071 & 4.90 & 1.00E+01+/-1.6 & 6.2 \\
 & 2 & 2.3150 & 50.5960 & 5.64 & 3.05+/-9.3E-01 & 3.3 \\
 & 3 & 2.1369 & 50.5176 & 6.31 & 2.62+/-8.8E-01 & 3.0 \\
 & 4 & 2.4958 & 50.6476 & 6.41 & 3.16+/-1.0 & 3.0 \\
 & 5 & 2.0160 & 50.4936 & 6.12 & 3.39+/-1.1 & 3.1 \\
 & 6 & 1.9824 & 50.6048 & 7.09 & 4.18+/-1.3 & 3.2 \\
\hline
1FGL\,J0022.2--1850 & 1 & 5.5385 & -18.8928 & 5.53 & 3.18E+01+/-6.1 & 5.2 \\
\hline
1FGL\,J0023.5$+$0930 & 1 & 5.9788 & 9.6547 & 5.14 & 6.63+/-1.3 & 5.0 \\
 & 2 & 5.9083 & 9.3570 & 5.64 & 5.42+/-1.2 & 4.4 \\
 & 3 & 5.8325 & 9.3485 & 5.05 & 7.99+/-1.5 & 5.4 \\
\hline
1FGL\,J0030.7$+$0724 & 1 &  7.5928 &  7.4397 & 6.12 & 1.52+/-4.7E-01 & 3.2 \\
 & 2 &  7.7076 &  7.3877 & 6.12 & 1.37+/-4.6E-01 & 3.0 \\
 & 3 &  7.5743 &  7.3616 & 4.88 & 4.10+/-7.4E-01 & 5.6 \\
 & 4 &  7.6248 &  7.3369 & 5.11 & 3.86+/-7.1E-01 & 5.4 \\
 & 5 &  7.5010 &  7.3836 & 5.59 & 2.04+/-5.6E-01 & 3.7 \\
 & 6 &  7.8316 &  7.4155 & 4.64 & 7.20+/-1.1 & 6.7 \\
\hline
1FGL\,J0032.7--5519 & 1 & 8.3719 & -55.3270 & 6.41 & 4.49+/-1.3 & 3.5 \\
 & 2 & 7.9552 & -55.4314 & 7.09 & 3.61+/-1.2 & 3.0 \\
\hline
1FGL\,J0038.0$+$1236 & 1 & 9.4627 & 12.6391 & 4.55 & 1.46E+01+/-1.9 & 7.8 \\
 & 2 & 9.5753 & 12.4588 & 4.51 & 2.30E+01+/-2.9 & 7.9 \\
 & 3 & 9.4125 & 12.3414 & 6.12 & 7.87+/-2.2 & 3.5 \\
\hline
1FGL\,J0039.2$+$4331 & 1 & 9.7439 & 43.4974 & 5.14 & 7.46+/-1.6 & 4.7 \\
 & 2 & 9.5962 & 43.4438 & 5.35 & 3.93+/-1.2 & 3.2 \\
 & 3 & 9.9055 & 43.6382 & 5.89 & 5.12+/-1.4 & 3.5 \\
\hline
1FGL\,J0051.4--6242 & 1 & 12.8194 & -62.7007 & 3.60 & 1.53E-01+/-5.1 & 30.2 \\
 & 2 & 12.9464 & -62.5974 & 6.52 & 2.71+/-7.7E-01 & 3.5 \\
 & 3 & 13.0755 & -62.6899 & 6.65 & 2.25+/-7.3E-01 & 3.1 \\
 & 4 & 12.6349 & -62.7895 & 5.82 & 3.51+/-9.1E-01 & 3.8 \\
\hline
1FGL\,J0054.9--2455 & 1 & 13.6953 & -24.9250 & 4.22 & 1.04E-01+/-1.0E+01 & 10.4 \\
\hline
1FGL\,J0101.0--6423 & 1 & 15.1107 & -64.3823 & 7.27 & 3.66+/-1.2 & 3.0 \\
& 2 & 15.3836 & -64.5598 & 6.41 & 7.85+/-2.0 & 4.0 \\
\hline
1FGL\,J0102.3$+$0942 & 1 & 15.5715 & 9.7362 & 6.31 & 3.23+/-1.0 & 3.1 \\
\hline
1FGL\,J0103.1$+$4840 & 1 & 15.5295 & 48.6014 & 5.35 & 4.83+/-1.3 & 3.7 \\
 & 2 & 15.7110 & 48.5053 & 6.04 & 5.20+/-1.3 & 3.9 \\
 & 3 & 15.9266 & 48.5087 & 4.63 & 2.00E+01+/-2.7 & 7.5 \\
 & 4 & 15.7360 & 48.4724 & 4.81 & 4.93+/-1.4 & 3.6 \\
\hline
1FGL\,J0115.7$+$0357 & 1 & 18.9199 & 3.9463 & 6.65 & 1.88E+01+/-4.7 & 4.0 \\
\hline
1FGL\,J0143.9--5845 & 1 & 25.9472 & -58.7643 & 3.55 & 5.30E-01+/-1.1E+01 & 47.1 \\
 & 2 & 26.0424 & -58.6789 & 6.12 & 5.24+/-1.2 & 4.2 \\
 & 3 & 25.6207 & -58.7648 & 4.56 & 1.99E+01+/-2.7 & 7.4 \\
\hline
1FGL\,J0157.0--5259 & 1 & 29.2405 & -53.0334 & 3.59 & 5.11E-01+/-1.5E+01 & 33.2 \\
\hline
\end{tabular}
}
\end{table}
\clearpage
\begin{table}[m]
\small
\scalebox{0.9}{
\begin{tabular}{ccccccc}
\hline \hline
1FGL Src & XRT Src\# & R.A.(J2000) [deg] & Dec.(J2000) [deg] & $r_{90\%}$ [$''$] & Count Rates [1E-03 cts/s] & S/N ratio \\
\hline
1FGL\,J0217.9--6630 & 1 & 34.2113 & -66.6116 & 3.77 & 5.65E+01+/-3.2 & 17.5 \\
 & 2 & 34.2072 & -66.5820 & 5.08 & 5.89+/-1.1 & 5.2 \\
 & 3 & 34.3547 & -66.6033 & 4.95 & 6.40+/-1.2 & 5.3 \\
 & 4 & 34.3879 & -66.7521 & 5.59 & 5.72+/-1.2 & 4.7 \\
\hline
1FGL\,J0223.0--1118 & 1 & 35.8097 & -11.2939 & 3.98 & 4.36E+01+/-3.4 & 12.7 \\
\hline
1FGL\,J0239.5$+$1324 & 1 & 39.7734 & 13.4551 & 5.44 & 5.50+/-1.3 & 4.2 \\
 & 2 & 39.7275 & 13.2637 & 5.44 & 1.11E+01+/-2.2 & 5.1 \\
\hline
1FGL\,J0305.2--1601 & 1 & 46.3123 & -16.1387 & 3.95 & 5.76E+01+/-4.4 & 13.2 \\
 & 2 & 46.3527 & -16.1529 & 5.96 & 3.89+/-1.3 & 3.1 \\
 & 3 & 46.3041 & -16.1613 & 4.86 & 1.15E+01+/-2.1 & 5.5 \\
\hline
1FGL\,J0316.3--6438 & 1 & 49.0593 & -64.6252 & 3.70 & 1.14E-01+/-5.5 & 20.8 \\
 & 2 & 48.7945 & -64.4430 & 7.27 & 3.70+/-1.2 & 3.0 \\
\hline
1FGL\,J0335.5--4501 & 1 & 53.8329 & -45.0084 & 5.82 & 3.90+/-1.2 & 3.2 \\
 & 2 & 53.9449 & -44.9757 & 6.31 & 3.45+/-1.1 & 3.0 \\
 & 3 & 53.8073 & -44.9947 & 4.81 & 1.65E+01+/-2.5 & 6.7 \\
 & 4 & 53.7894 & -45.1661 & 6.65 & 5.41+/-1.5 & 3.6 \\
\hline
1FGL\,J0340.4$+$4130 & 1 & 55.0355 & 41.4795 & 7.09 & 1.92+/-6.1E-01 & 3.2 \\
 & 2 & 55.2222 & 41.6003 & 6.65 & 2.12+/-6.6E-01 & 3.2 \\
 & 3 & 55.2058 & 41.3476 & 6.65 & 2.44+/-7.1E-01 & 3.4 \\
 & 4 & 55.1090 & 41.3302 & 5.53 & 2.45+/-7.2E-01 & 3.4 \\
 & 5 & 54.8538 & 41.3795 & 5.64 & 5.60+/-1.2 & 4.6 \\
\hline
1FGL\,J0345.2--2355 & 1 & 56.3261 & -23.8726 & 5.76 & 5.09+/-1.3 & 4.1 \\
 & 2 & 56.1862 & -23.9965 & 5.82 & 6.88+/-1.6 & 4.4 \\
\hline
1FGL\,J0409.9--0357 & 1 & 62.4433 & -4.0006 & 5.11 & 7.89+/-1.5 & 5.3 \\
\hline
1FGL\,J0439.8--1857 & 1 & 69.9570 & -19.0170 & 6.52 & 4.75+/-1.3 & 3.8 \\
 & 2 & 70.1243 & -19.0763 & 7.09 & 4.09+/-1.3 & 3.2 \\
\hline
1FGL\,J0505.9$+$6121 & 1 & 76.4858 & 61.4165 & 6.41 & 1.91+/-6.3E-01 & 3.0 \\
 & 2 & 76.4943 & 61.2259 & 3.89 & 3.55E+01+/-2.5 & 14.3 \\
 & 3 & 76.2685 & 61.2885 & 3.69 & 8.08E+01+/-3.7 & 21.8 \\
 & 4 & 76.5497 & 61.5288 & 5.64 & 3.30+/-9.0E-01 & 3.6 \\
\hline
1FGL\,J0506.9--5435 & 1 & 76.7410 & -54.5859 & 3.95 & 1.84E-01+/-1.3E+01 & 13.9 \\
\hline
1FGL\,J0521.6$+$0103 & 1 & 80.4198 & 1.0495 & 6.12 & 4.03+/-1.2 & 3.4 \\
 & 2 & 80.5590 & 0.9602 & 5.11 & 1.01E+01+/-1.9 & 5.3 \\
 & 3 & 80.6260 & 1.0803 & 6.12 & 4.03+/-1.3 & 3.1 \\
\hline
1FGL\,J0523.5--2529 & 1 & 80.8216 & -25.4604 & 6.21 & 3.08+/-9.4E-01 & 3.3 \\
\hline
1FGL\,J0533.9$+$6758 & 1 & 83.2624 & 68.0123 & 5.40 & 8.03+/-1.5 & 5.4 \\
 & 2 & 83.3788 & 68.0480 & 6.41 & 3.49+/-1.0 & 3.5 \\
 & 3 & 83.4525 & 67.8272 & 6.04 & 4.49+/-1.2 & 3.8 \\
 & 4 & 83.3501 & 67.7908 & 5.44 & 4.90+/-1.3 & 3.8 \\
 & 5 & 83.4313 & 68.1494 & 6.31 & 5.06+/-1.3 & 3.8 \\
\hline
1FGL\,J0537.7--5717 & 1 & 84.4532 & -57.3083 & 4.55 & 3.13E+01+/-4.0 & 7.8 \\
 & 2 & 84.6011 & -57.3187 & 5.59 & 1.62E+01+/-3.2 & 5.1 \\
 & 3 & 84.4356 & -57.1650 & 4.17 & 6.13E+01+/-5.8 & 10.6 \\
\hline
1FGL\,J0538.4--3910 & 1 & 84.5432 & -39.1450 & 3.64 & 1.35E-01+/-5.3 & 25.4 \\
 & 2 & 84.3436 & -39.1091 & 6.65 & 3.40+/-1.0 & 3.3 \\
\hline
1FGL\,J0545.6$+$6022 & 1 & 86.4117 & 60.4666 & 6.78 & 3.57+/-1.2 & 3.0 \\
 & 2 & 85.9778 & 60.4020 & 5.82 & 6.72+/-1.8 & 3.8 \\
\hline
\end{tabular}
}
\end{table}
\clearpage
\begin{table}[m]
\small
\scalebox{0.9}{
\begin{tabular}{ccccccc}
\hline \hline
1FGL Src & XRT Src\# & R.A.(J2000) [deg] & Dec.(J2000) [deg] & $r_{90\%}$ [$''$] & Count Rates [1E-03 cts/s] & S/N ratio \\
\hline
1FGL\,J0600.5--2006 & 1 & 90.0660 & -20.1167 & 5.40 & 2.07E+01+/-5.4 & 3.9 \\
 & 2 & 90.0324 & -20.1368 & 5.02 & 3.36E+01+/-6.7 & 5.0 \\
\hline
1FGL\,J0603.0--4012 & 1 & 90.7134 & -40.3133 & 5.00 & 2.54E+01+/-4.0 & 6.4 \\
\hline
1FGL\,J0604.2--4817 & 1 & 91.0367 & -48.2904 & 3.59 & 1.91E-01+/-6.0 & 32.1 \\
 & 2 & 90.9685 & -48.2119 & 5.59 & 3.77+/-9.4E-01 & 4.0 \\
 & 3 & 90.9612 & -48.3352 & 6.41 & 2.25+/-7.5E-01 & 3.0 \\
 & 4 & 91.2562 & -48.1729 & 5.20 & 6.70+/-1.3 & 5.1 \\
 & 5 & 91.2618 & -48.3670 & 6.04 & 4.50+/-1.1 & 4.1 \\
 & 6 & 90.9115 & -48.4518 & 6.12 & 4.13+/-1.1 & 3.7 \\
 & 7 & 91.1940 & -48.4578 & 3.57 & 3.70E-01+/-9.8 & 37.6 \\
\hline
1FGL\,J0605.1$+$0005 & 1 & 91.1991 & 0.2213 & 4.92 & 4.97E+01+/-7.2 & 6.9 \\
\hline
1FGL\,J0609.3--0244 & 1 & 92.3127 & -2.7977 & 6.12 & 1.52E+01+/-3.6 & 4.2 \\
\hline
1FGL\,J0648.6--6052 & 1 & 101.9201 & -60.9673 & 4.88 & 5.22+/-9.4E-01 & 5.6 \\
 & 2 & 102.1083 & -60.9551 & 6.52 & 1.90+/-5.8E-01 & 3.3 \\
 & 3 & 101.8282 & -60.8258 & 5.89 & 2.73+/-6.9E-01 & 4.0 \\
 & 4 & 102.0740 & -61.0352 & 6.12 & 1.74+/-5.8E-01 & 3.0 \\
 & 5 & 101.6416 & -61.0087 & 6.31 & 2.56+/-6.9E-01 & 3.7 \\
 & 6 & 101.5771 & -60.9332 & 6.31 & 1.83+/-6.0E-01 & 3.0 \\
\hline
1FGL\,J0707.3$+$7742 & 1 & 106.7108 & 77.6937 & 5.59 & 7.04+/-1.5 & 4.6 \\
 & 2 & 107.3337 & 77.5734 & 6.12 & 3.91+/-1.2 & 3.2 \\
\hline
1FGL\,J0718.8--4958 & 1 & 109.8163 & -49.9788 & 7.27 & 2.89+/-9.5E-01 & 3.0 \\
\hline
1FGL\,J0803.1--0339 & 1 & 120.8008 & -3.6000 & 4.26 & 2.69E+01+/-2.8 & 9.6 \\
 & 2 & 120.7911 & -3.5159 & 6.65 & 3.72+/-1.2 & 3.2 \\
 & 3 & 120.7895 & -3.8270 & 5.53 & 1.10E+01+/-2.2 & 5.0 \\
\hline
1FGL\,J0814.5--1011 & 1 & 123.5479 & -10.2019 & 4.90 & 3.63E+01+/-5.3 & 6.8 \\
\hline
1FGL\,J0848.6$+$0504 & 1 & 132.1648 & 5.1057 & 3.90 & 5.29E+01+/-3.7 & 14.2 \\
 & 2 & 132.1449 & 5.2360 & 5.31 & 1.09E+01+/-2.0 & 5.5 \\
\hline
1FGL\,J0902.4$+$2050 & 1 & 135.6110 & 20.8456 & 5.14 & 6.25+/-1.2 & 5.4 \\
\hline
1FGL\,J0906.4--0903 & 1 & 136.5734 & -9.0954 & 7.09 & 1.17E+01+/-3.7 & 3.1 \\
\hline
1FGL\,J0908.7--2119 & 1 & 137.2438 & -21.3149 & 4.28 & 2.72E+01+/-2.8 & 9.6 \\
 & 2 & 137.3365 & -21.3181 & 6.52 & 4.02+/-1.2 & 3.4 \\
 & 3 & 137.0984 & -21.1945 & 5.14 & 1.02E+01+/-2.1 & 5.0 \\
\hline
1FGL\,J0922.0$+$2337 & 1 & 140.4383 & 23.5963 & 5.82 & 6.62+/-1.5 & 4.3 \\
 & 2 & 140.3968 & 23.5262 & 7.27 & 3.87+/-1.3 & 3.0 \\
 & 3 & 140.3262 & 23.5672 & 5.96 & 6.43+/-1.7 & 3.9 \\
\hline
1FGL\,J0955.2--3949 & 1 & 148.8656 & -39.7969 & 5.96 & 5.08+/-1.4 & 3.6 \\
 & 2 & 148.8290 & -39.6696 & 5.35 & 9.49+/-1.9 & 4.9 \\
\hline
1FGL\,J1040.5$+$0616 & 1 & 160.1313 & 6.2904 & 5.82 & 2.99+/-8.0E-01 & 3.7 \\
 & 2 & 160.1639 & 6.2566 & 5.20 & 5.75+/-1.1 & 5.2 \\
 & 3 & 160.2016 & 6.3061 & 5.76 & 3.76+/-8.9E-01 & 4.2 \\
 & 4 & 160.0322 & 6.4196 & 6.93 & 2.32+/-7.7E-01 & 3.0 \\
 & 5 & 160.3053 & 6.3220 & 6.65 & 2.61+/-8.2E-01 & 3.2 \\
\hline
1FGL\,J1119.9--2205 & 1 & 169.9925 & -22.0822 & 5.08 & 2.67+/-6.1E-01 & 4.4 \\
 & 2 & 170.0034 & -22.0245 & 5.82 & 1.80+/-5.1E-01 & 3.5 \\
 & 3 & 170.0897 & -22.1876 & 5.49 & 1.63+/-4.8E-01 & 3.4 \\
 & 4 & 170.1708 & -22.1321 & 3.97 & 1.89E+01+/-1.5 & 12.4 \\
 & 5 & 169.8795 & -21.9436 & 6.04 & 1.88+/-6.0E-01 & 3.1 \\
\hline 
\end{tabular}
}
\end{table}
\clearpage
\begin{table}[m]
\small
\scalebox{0.9}{
\begin{tabular}{ccccccc}
\hline \hline
1FGL Src & XRT Src\# & R.A.(J2000) [deg] & Dec.(J2000) [deg] & $r_{90\%}$ [$''$] & Count Rates [1E-03 cts/s] & S/N ratio \\
\hline
1FGL\,J1124.4--3654 & 1 & 171.0039 & -36.8891 & 6.21 & 1.71+/-5.5E-01 & 3.1 \\
 & 2 & 171.1585 & -36.9404 & 5.76 & 2.51+/-6.5E-01 & 3.8 \\
 & 3 & 171.1690 & -36.8199 & 5.40 & 2.23+/-6.3E-01 & 3.5 \\
\hline
1FGL\,J1129.3$+$3757 & 1 & 172.2650 & 37.9494 & 6.65 & 3.07+/-9.5E-01 & 3.2 \\
 & 2 & 172.3798 & 38.0324 & 6.52 & 2.74+/-9.1E-01 & 3.0 \\
\hline
1FGL\,J1141.8--1403 & 1 & 175.4239 & -14.1322 & 3.74 & 7.76E+01+/-4.1 & 19.0 \\
 & 2 & 175.3233 & -14.0799 & 6.78 & 2.88+/-9.1E-01 & 3.1 \\
\hline
1FGL\,J1218.4--0128 & 1 & 184.6469 & -1.3324 & 8.27 & 9.26+/-3.1 & 3.0 \\
\hline
1FGL\,J1221.4--0635 & 1 & 185.3635 & -6.4784 & 4.78 & 8.44+/-1.5 & 5.7 \\
 & 2 & 185.2748 & -6.6540 & 5.59 & 3.69+/-1.0 & 3.5 \\
 & 3 & 185.4747 & -6.5225 & 5.96 & 2.93+/-9.4E-01 & 3.1 \\
 & 4 & 185.3609 & -6.4182 & 5.14 & 4.90+/-1.2 & 4.1 \\
 & 5 & 185.4733 & -6.6844 & 4.97 & 9.29+/-1.7 & 5.5 \\
\hline
1FGL\,J1226.0$+$2954 & 1 & 186.5352 & 29.9957 & 5.00 & 9.19+/-1.7 & 5.4 \\
 & 2 & 186.4046 & 29.9057 & 5.59 & 5.17+/-1.4 & 3.8 \\
\hline
1FGL\,J1232.2--5118 & 1 & 187.9632 & -51.3279 & 6.21 & 4.83+/-1.3 & 3.6 \\
\hline
1FGL\,J1249.3--2812 & 1 & 192.3304 & -28.1422 & 3.95 & 4.76E+01+/-3.6 & 13.1 \\
 & 2 & 192.2482 & -28.2060 & 6.12 & 5.05+/-1.3 & 3.8 \\
 & 3 & 192.5229 & -28.2054 & 4.74 & 1.67E+01+/-2.4 & 7.1 \\
\hline
1FGL\,J1251.3$+$1044 & 1 & 192.8250 & 10.6513 & 4.09 & 5.94E+01+/-5.3 & 11.1 \\
\hline
1FGL\,J1254.4$+$2209 & 1 & 193.6389 & 22.1844 & 5.59 & 1.94E+01+/-3.7 & 5.2 \\
 & 2 & 193.6657 & 22.1065 & 4.92 & 3.42E+01+/-5.0 & 6.8 \\
\hline
1FGL\,J1256.9$+$3650 & 1 & 194.3924 & 36.7638 & 6.52 & 2.24E+01+/-5.4 & 4.1 \\
 & 2 & 194.3487 & 36.7376 & 6.52 & 2.02E+01+/-5.2 & 3.9 \\
\hline
1FGL\,J1301.8$+$0837 & 1 & 195.3958 & 8.4991 & 6.93 & 3.76+/-1.2 & 3.0 \\
\hline
1FGL\,J1302.3--3255 & 1 & 195.6096 & -33.0154 & 6.12 & 1.55+/-4.9E-01 & 3.2 \\
 & 2 & 195.5295 & -32.7874 & 5.59 & 1.85+/-5.4E-01 & 3.5 \\
 & 3 & 195.7604 & -32.8232 & 5.17 & 2.32+/-6.0E-01 & 3.8 \\
\hline
1FGL\,J1304.3--4352 & 1 & 196.0869 & -43.8864 & 3.86 & 1.22E-01+/-8.0 & 15.3 \\
 & 2 & 196.1849 & -43.8558 & 7.27 & 5.79+/-1.9 & 3.0 \\
\hline
1FGL\,J1307.0--4030 & 1 & 196.6086 & -40.4145 & 3.64 & 1.23E-01+/-4.9 & 25.2 \\
 & 2 & 196.8008 & -40.4080 & 3.61 & 1.74E-01+/-6.1 & 28.6 \\
 & 3 & 196.7351 & -40.5895 & 5.31 & 3.65+/-1.1 & 3.4 \\
\hline
1FGL\,J1307.6--4259 & 1 &  196.9083 &  -42.9943 & 3.65 & 5.46E-01+/-2.1E+01 & 25.5 \\
\hline
1FGL\,J1311.7--3429 & 1 & 197.8769 & -34.3037 & 5.08 & 8.00+/-1.8 & 4.3 \\
 & 2 & 197.9380 & -34.5076 & 5.96 & 4.90+/-1.6 & 3.1 \\
\hline
1FGL\,J1312.7$+$0051 & 1 & 198.2429 & 0.8348 & 6.78 & 3.72+/-1.2 & 3.0 \\
\hline
1FGL\,J1315.6--0729 & 1 &  198.9712 &  -7.5508 & 3.69 & 8.16E+01+/-3.9 & 21.2 \\
 & 2 &  198.9320 &  -7.6155 & 4.71 & 8.51+/-1.4 & 6.2 \\
\hline
1FGL\,J1328.2--4729 & 1 &  202.1693 &  -47.4641 & 5.49 & 5.77+/-1.1 & 5.1 \\
 & 2 &  202.3257 &  -47.3795 & 5.82 & 4.94+/-1.1 & 4.6 \\
 & 3 &  202.0373 &  -47.5586 & 6.21 & 3.76+/-9.8E-01 & 3.8 \\
 & 4 &  201.7203 &  -47.4907 & 5.53 & 1.02E+01+/-2.7 & 3.8 \\
\hline
1FGL\,J1340.5--0413 & 1 & 205.1756 & -4.1685 & 3.97 & 2.48E+01+/-2.0 & 12.4 \\
 & 2 & 205.1773 & -4.2541 & 5.59 & 2.79+/-7.6E-01 & 3.7 \\
 & 3 & 205.0287 & -4.1842 & 5.49 & 3.75+/-8.7E-01 & 4.3 \\
 & 4 & 205.0739 & -4.2852 & 4.41 & 6.19+/-1.1 & 5.6 \\
\hline
1FGL\,J1406.2--2510 & 1 & 211.5423 & -25.1342 & 5.96 & 2.49E+01+/-5.4 & 4.6 \\
\hline
1FGL\,J1419.7$+$7731 & 1 & 214.7543 & 77.5418 & 3.77 & 7.17E+01+/-4.1 & 17.7 \\
\hline
\end{tabular}
}
\end{table}
\clearpage
\begin{table}[m]
\small
\scalebox{0.9}{
\begin{tabular}{ccccccc}
\hline \hline
1FGL Src & XRT Src\# & R.A.(J2000) [deg] & Dec.(J2000) [deg] & $r_{90\%}$ [$''$] & Count Rates [1E-03 cts/s] & S/N ratio \\
\hline
1FGL\,J1511.8--0513 & 1 & 227.9530 & -5.2297 & 3.55 & 4.50E-01+/-1.1E+01 & 42.5 \\
 & 2 & 228.0008 & -5.2577 & 5.02 & 9.17+/-1.7 & 5.5 \\
\hline
1FGL\,J1521.0--0350 & 1 & 230.2047 & -3.8140 & 4.01 & 1.94E+01+/-1.7 & 11.6 \\
 & 2 & 230.2114 & -3.6863 & 4.92 & 2.31+/-6.6E-01 & 3.5 \\
\hline
1FGL\,J1544.5--1127 & 1 & 236.1636 & -11.4676 & 3.70 & 7.55E+01+/-3.6 & 20.9 \\
 & 2 & 236.2932 & -11.6010 & 5.96 & 3.57+/-9.2E-01 & 3.9 \\
\hline
1FGL\,J1549.7--0659 & 1 & 237.4678 & -6.9857 & 3.70 & 1.17E-01+/-5.6 & 20.8 \\
\hline
1FGL\,J1627.6$+$3218 & 1 & 246.9290 & 32.3502 & 5.49 & 1.00E+01+/-1.9 & 5.3 \\
 & 2 & 247.0013 & 32.4037 & 5.96 & 7.85+/-1.8 & 4.4 \\
\hline
1FGL\,J1653.6--0158 & 1 & 253.4077 & -1.9764 & 6.04 & 2.84+/-9.5E-01 & 3.0 \\
 & 2 & 253.3149 & -1.9732 & 4.11 & 2.93E+01+/-2.7 & 10.8 \\
 & 3 & 253.4946 & -2.0543 & 5.02 & 9.17+/-1.7 & 5.4 \\
 & 4 & 253.4024 & -1.8221 & 5.49 & 6.99+/-1.6 & 4.4 \\
\hline
1FGL\,J1738.9$+$8716 & 1 & 265.4297 & 87.2447 & 6.93 & 3.23+/-1.0 & 3.1 \\
 & 2 & 265.4339 & 87.4113 & 5.76 & 4.98+/-1.4 & 3.7 \\
\hline
1FGL\,J1744.1$+$7620 & 1 & 266.2855 & -76.1771 & 6.12 & 3.73+/-9.2E-01 & 4.0 \\
\hline
1FGL\,J1745.5$+$1018 & 1 & 266.4193 & 10.2526 & 6.65 & 4.03+/-1.2 & 3.2 \\
 & 2 & 266.3164 & 10.3072 & 5.64 & 7.42+/-1.7 & 4.4 \\
\hline
1FGL\,J1754.0--5002 & 1 & 268.5392 & -50.1507 & 5.49 & 6.18+/-1.5 & 4.0 \\
 & 2 & 268.3887 & -50.1235 & 5.70 & 5.30+/-1.4 & 3.7 \\
 & 3 & 268.6812 & -50.1930 & 7.27 & 4.14+/-1.4 & 3.0 \\
\hline
1FGL\,J1806.2$+$0609 & 1 & 271.7333 & 6.1765 & 5.96 & 7.81+/-2.1 & 3.8 \\
\hline
1FGL\,J1810.3$+$1741 & 1 & 272.6161 & 17.8358 & 5.44 & 8.85+/-1.8 & 4.8 \\
\hline
1FGL\,J1816.7$+$4509 & 1 & 274.2201 & 45.1247 & 6.04 & 5.57+/-1.3 & 4.4 \\
\hline
1FGL\,J1841.9$+$3220 & 1 & 280.4803 & 32.3805 & 5.82 & 4.01+/-9.7E-01 & 4.2 \\
 & 2 & 280.4459 & 32.3103 & 3.75 & 6.11E+01+/-3.4 & 18.2 \\
 & 3 & 280.5522 & 32.3770 & 5.82 & 2.66+/-8.0E-01 & 3.3 \\
\hline
1FGL\,J1842.3--5845 & 1 & 280.6243 & -58.7002 & 3.58 & 4.37E-01+/-1.3E+01 & 34.1 \\
\hline
1FGL\,J1916.9--3028 & 1 & 289.2676 & -30.3227 & 6.31 & 9.91+/-2.8 & 3.5 \\
 & 2 & 289.0549 & -30.4573 & 5.76 & 1.78E+01+/-4.0 & 4.4 \\
\hline
1FGL\,J1926.8$+$6153 & 1 & 291.7104 & 61.9111 & 4.18 & 1.27E-01+/-1.2E+01 & 10.6 \\
\hline
1FGL\,J1956.2--0238 & 1 & 299.0734 & -2.7197 & 6.21 & 4.68+/-1.3 & 3.7 \\
\hline
1FGL\,J1959.7--4730 & 1 & 299.9399 & -47.4216 & 3.94 & 5.51E+01+/-4.1 & 13.5 \\
 & 2 & 299.8476 & -47.4352 & 6.52 & 3.32+/-1.1 & 3.0 \\
\hline
1FGL\,J2004.8$+$7004 & 1 & 301.2739 & 70.0766 & 3.69 & 1.09E-01+/-5.0 & 22.0 \\
 & 2 & 301.1617 & 70.1349 & 5.11 & 7.68+/-1.4 & 5.4 \\
 & 3 & 300.7818 & 70.0746 & 7.27 & 3.32+/-1.0 & 3.2 \\
 & 4 & 300.8928 & 70.2259 & 3.73 & 1.15E-01+/-5.9 & 19.4 \\
\hline
1FGL\,J2014.4$+$0647 & 1 & 303.5461 & 6.7328 & 5.76 & 4.85+/-1.1 & 4.4 \\
 & 2 & 303.4967 & 6.7707 & 6.41 & 2.44+/-8.1E-01 & 3.0 \\
 & 3 & 303.5626 & 6.5772 & 4.39 & 1.79E+01+/-2.1 & 8.3 \\
 & 4 & 303.6298 & 6.8144 & 3.78 & 7.48E+01+/-4.3 & 17.4 \\
\hline
1FGL\,J2034.6--4202 & 1 & 308.7116 & -42.0101 & 3.93 & 4.18E+01+/-3.2 & 13.0 \\
 & 2 & 308.5874 & -41.9693 & 5.89 & 4.20+/-1.1 & 3.7 \\
\hline
1FGL\,J2039.4--5621 & 1 & 310.0106 & -56.4619 & 4.97 & 1.12E+01+/-1.8 & 6.2 \\
\hline
1FGL\,J2129.8--0427 & 1 & 322.4393 & -4.4849 & 4.90 & 4.25+/-7.6E-01 & 5.6 \\
 & 2 & 322.5741 & -4.3742 & 5.76 & 1.60+/-4.9E-01 & 3.3 \\
 & 3 & 322.5339 & -4.5812 & 4.38 & 1.05E+01+/-1.3 & 8.3 \\
\hline
1FGL\,J2134.5--2130 & 1 & 323.6404 & -21.5176 & 5.64 & 3.01+/-8.2E-01 & 3.7 \\
\hline
\end{tabular}
}
\end{table}
\clearpage
\begin{table}[m]
\small
\scalebox{0.9}{
\begin{tabular}{ccccccc}
\hline \hline
1FGL Src & XRT Src\# & R.A.(J2000) [deg] & Dec.(J2000) [deg] & $r_{90\%}$ [$''$] & Count Rates [1E-03 cts/s] & S/N ratio \\
\hline
1FGL\,J2223.3$+$0103 & 1 & 335.8724 & 1.0403 & 5.82 & 3.93+/-1.1 & 3.6 \\
 & 2 & 335.6256 & 1.0416 & 5.31 & 9.29+/-1.8 & 5.2 \\
\hline
1FGL\,J2228.5--1633 & 1 & 337.0186 & -16.6058 & 5.59 & 4.19+/-1.2 & 3.6 \\
 & 2 & 336.9564 & -16.4692 & 6.21 & 5.33+/-1.3 & 4.0 \\
 & 3 & 337.0684 & -16.3948 & 3.73 & 1.07E-01+/-5.6 & 19.3 \\
\hline
1FGL\,J2243.4$+$4104 & 1 & 341.0528 & 40.9538 & 4.57 & 1.52E+01+/-2.0 & 7.5 \\
 & 2 & 340.9480 & 40.9421 & 6.04 & 3.60+/-1.0 & 3.5 \\
 & 3 & 340.9314 & 40.8992 & 5.96 & 5.16+/-1.2 & 4.2 \\
\hline
1FGL\,J2251.2--4928 & 1 & 342.8677 & -49.4876 & 6.93 & 3.65+/-1.1 & 3.2 \\
 & 2 & 342.7909 & -49.5447 & 5.96 & 4.44+/-1.3 & 3.5 \\
\hline
1FGL\,J2256.9--1024 & 1 & 344.2248 & -10.4482 & 6.65 & 3.80+/-1.2 & 3.2 \\
 & 2 & 344.1310 & -10.3750 & 5.44 & 9.14+/-1.8 & 5.1 \\
 & 3 & 344.1457 & -10.5265 & 6.65 & 3.89+/-1.2 & 3.2 \\
\hline
1FGL\,J2257.9--3643 & 1 & 344.5606 & -36.7415 & 6.04 & 5.85+/-1.5 & 3.9 \\
 & 2 & 344.4865 & -36.7684 & 6.52 & 5.28+/-1.4 & 3.7 \\
 & 3 & 344.4524 & -36.5985 & 5.17 & 1.11E+01+/-2.1 & 5.3 \\
\hline
1FGL\,J2310.0--3627 & 1 & 347.4192 & -36.5466 & 3.73 & 7.12E+01+/-3.7 & 19.4 \\
 & 2 & 347.3100 & -36.5028 & 5.31 & 5.10+/-1.1 & 4.7 \\
 & 3 & 347.6192 & -36.5051 & 6.12 & 3.31+/-9.1E-01 & 3.6 \\
 & 4 & 347.2436 & -36.4546 & 6.12 & 3.20+/-9.4E-01 & 3.4 \\
 & 5 & 347.2568 & -36.5376 & 6.21 & 2.56+/-8.4E-01 & 3.0 \\
\hline
1FGL\,J2323.0--4919 & 1 & 350.7275 & -49.2745 & 4.92 & 4.03E+01+/-6.0 & 6.8 \\
\hline
1FGL\,J2330.3--4745 & 1 & 352.4761 & -47.6597 & 6.21 & 1.63+/-5.5E-01 & 3.0 \\
 & 2 & 352.3234 & -47.5036 & 3.94 & 7.65E+01+/-5.8 & 13.1 \\
\hline
1FGL\,J2339.7--0531 & 1 & 354.9608 & -5.5473 & 5.31 & 4.19+/-7.9E-01 & 5.3 \\
 & 2 & 354.9106 & -5.5524 & 4.90 & 4.58+/-8.1E-01 & 5.6 \\
 & 3 & 354.9828 & -5.4250 & 5.76 & 2.48+/-6.5E-01 & 3.8 \\
 & 4 & 354.8064 & -5.5937 & 5.82 & 2.17+/-6.3E-01 & 3.5 \\
 & 5 & 354.8657 & -5.3717 & 4.81 & 8.36+/-1.3 & 6.5 \\
\hline
1FGL\,J2347.3$+$0710 & 1 & 356.6665 & 7.0863 & 4.13 & 3.03E+01+/-2.9 & 10.3 \\
\hline
1FGL\,J2350.1--3005 & 1 & 357.6430 & -30.1017 & 4.60 & 2.39E+01+/-3.1 & 7.7 \\
\hline
\end{tabular}
}
\end{table}
\clearpage
\begin{figure}[m]
 \begin{center}
  \begin{minipage}{0.32\hsize}
    \begin{center}
      \includegraphics[width=55mm]{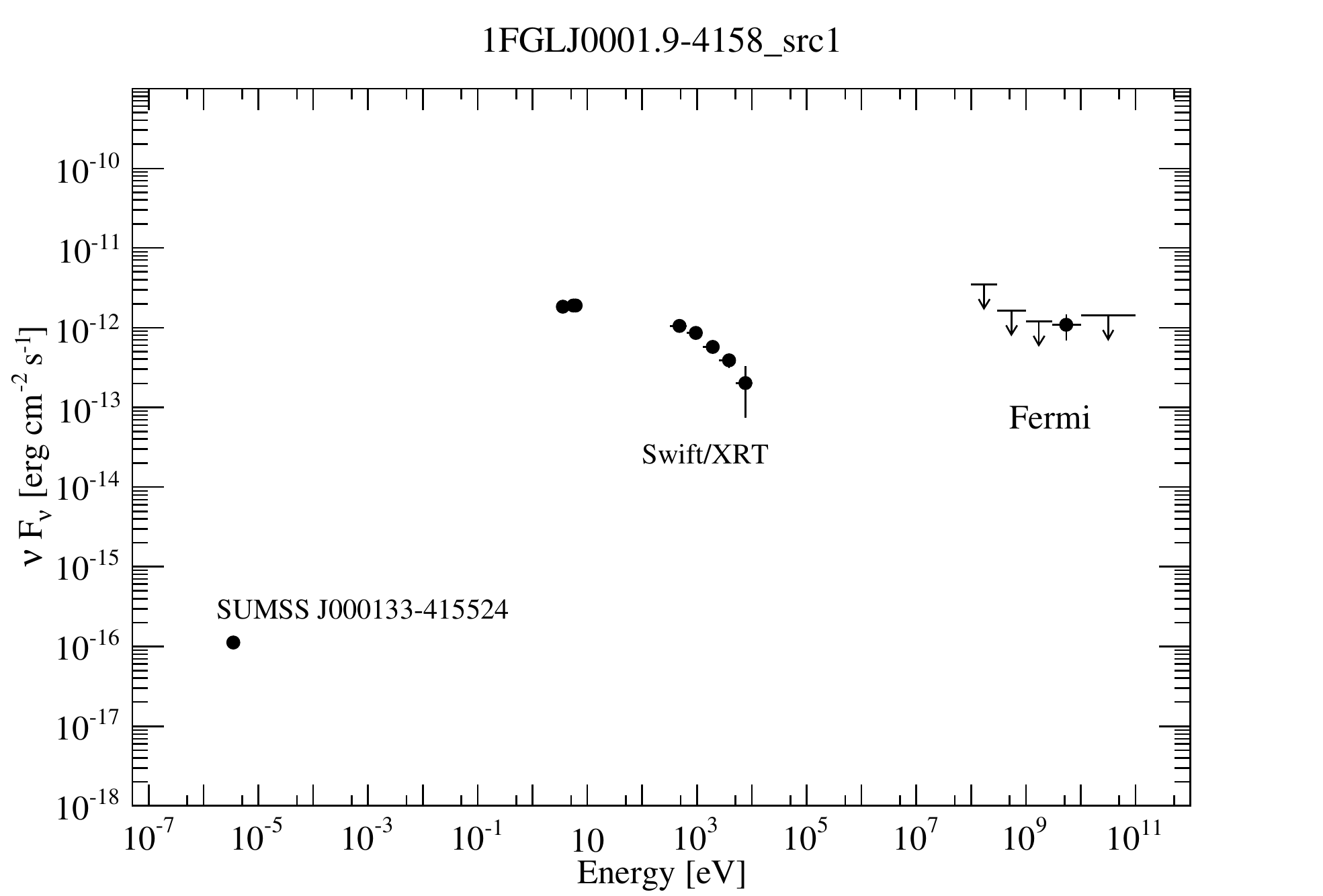}
    \end{center}
  \end{minipage}
  \begin{minipage}{0.32\hsize}
    \begin{center}
      \includegraphics[width=55mm]{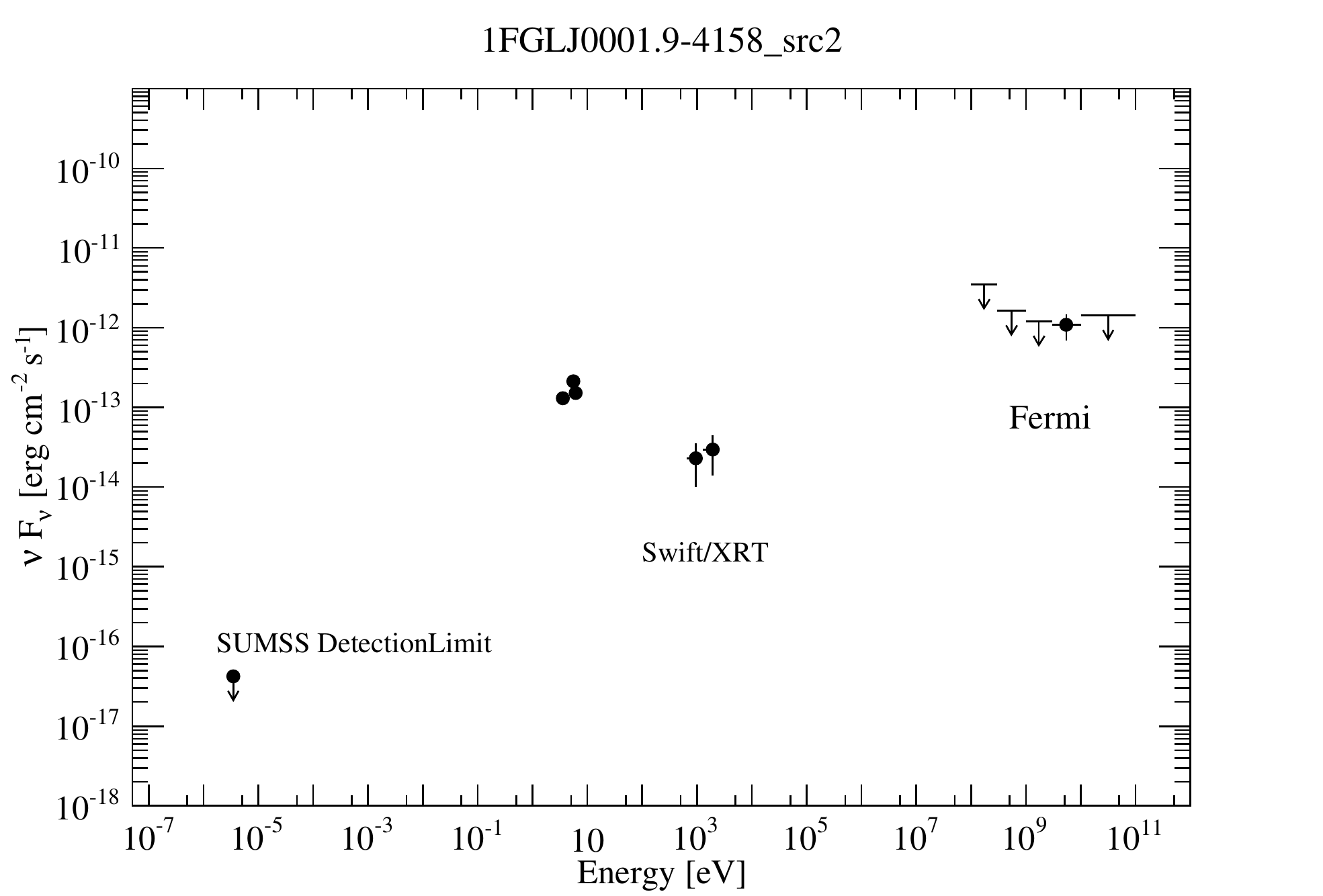}
    \end{center}
  \end{minipage}
  \begin{minipage}{0.32\hsize}
    \begin{center}
      \includegraphics[width=55mm]{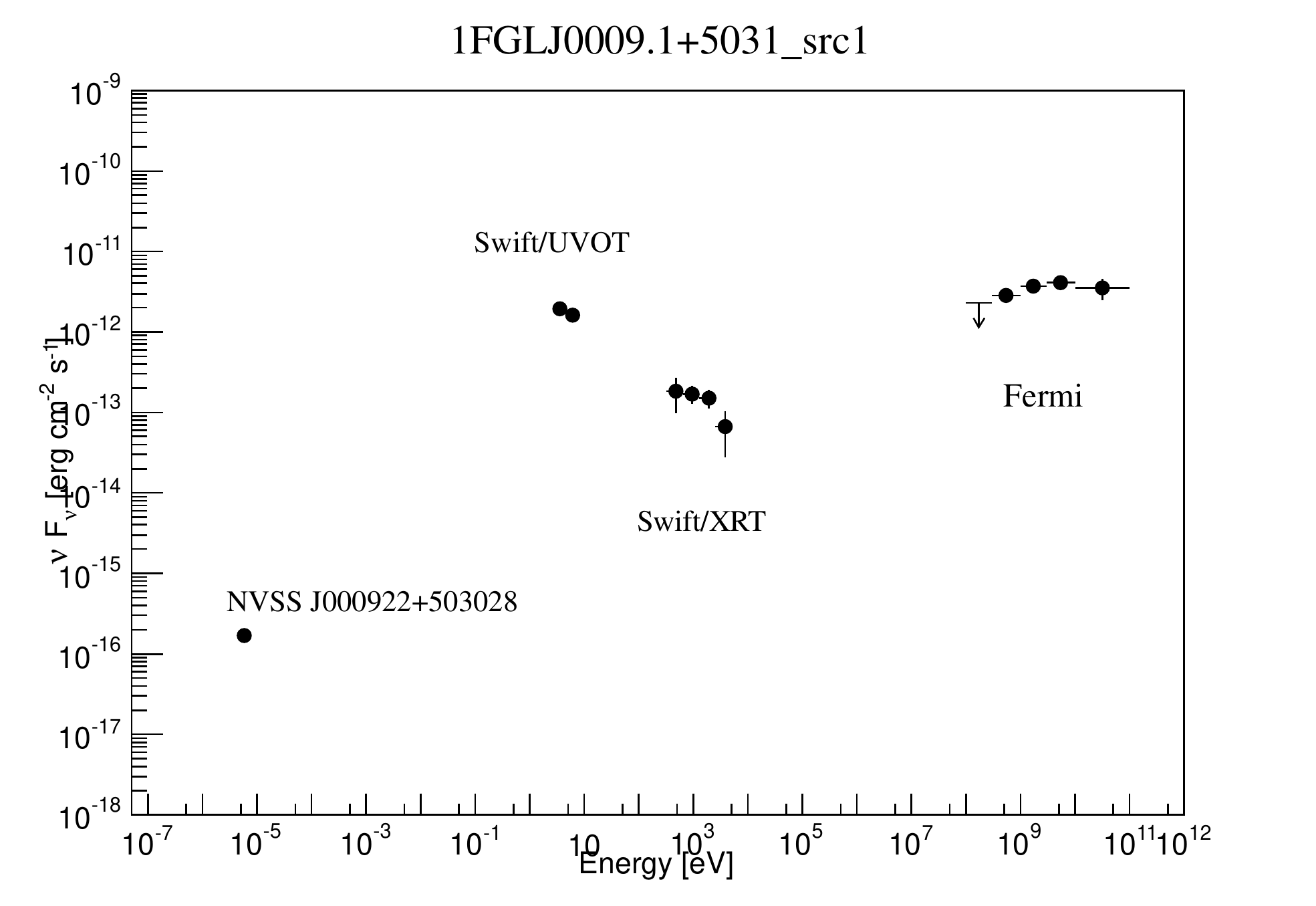}
    \end{center}
  \end{minipage}
  \begin{minipage}{0.32\hsize}
    \begin{center}
      \includegraphics[width=55mm]{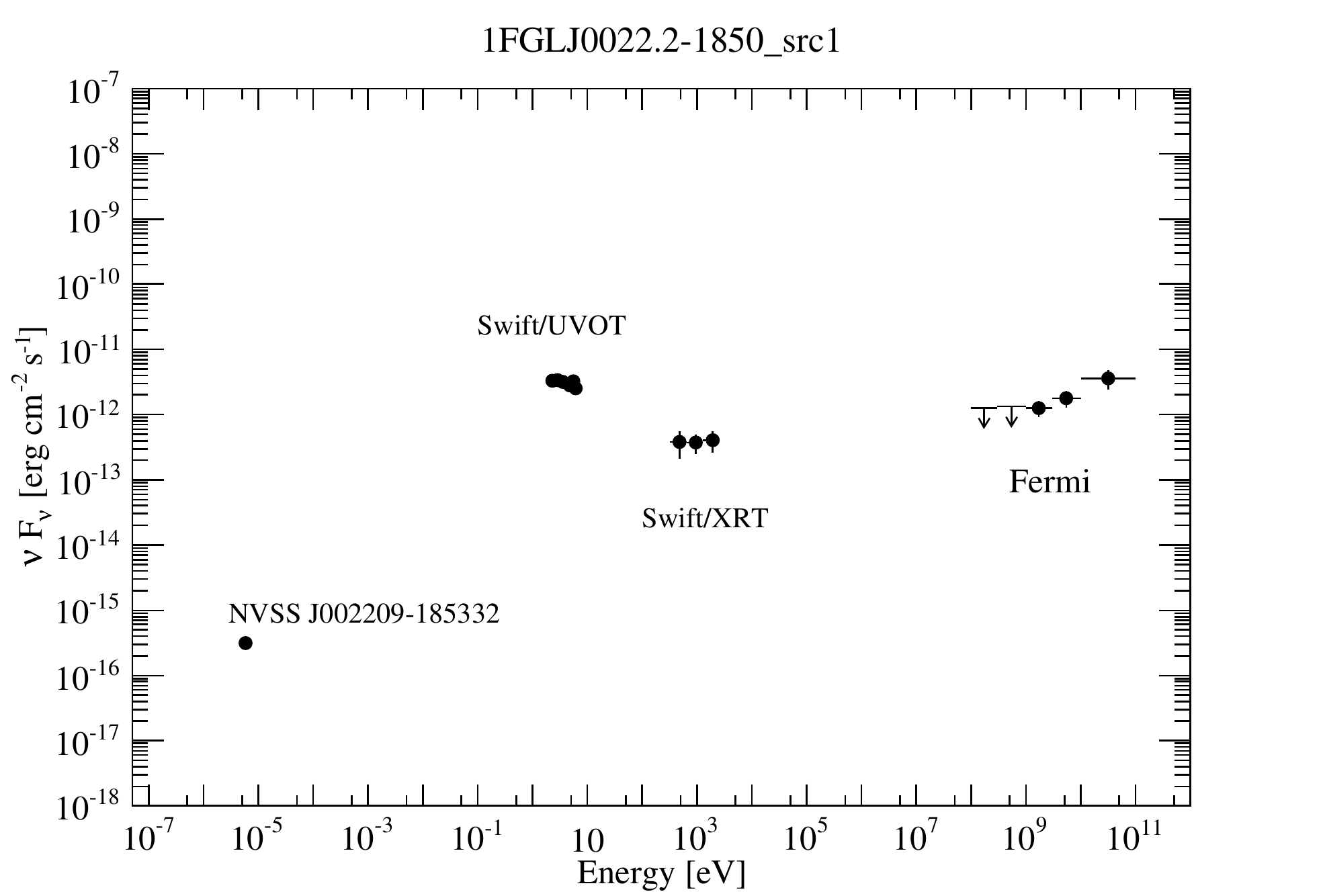}
    \end{center}
  \end{minipage}
  \begin{minipage}{0.32\hsize}
    \begin{center}
      \includegraphics[width=55mm]{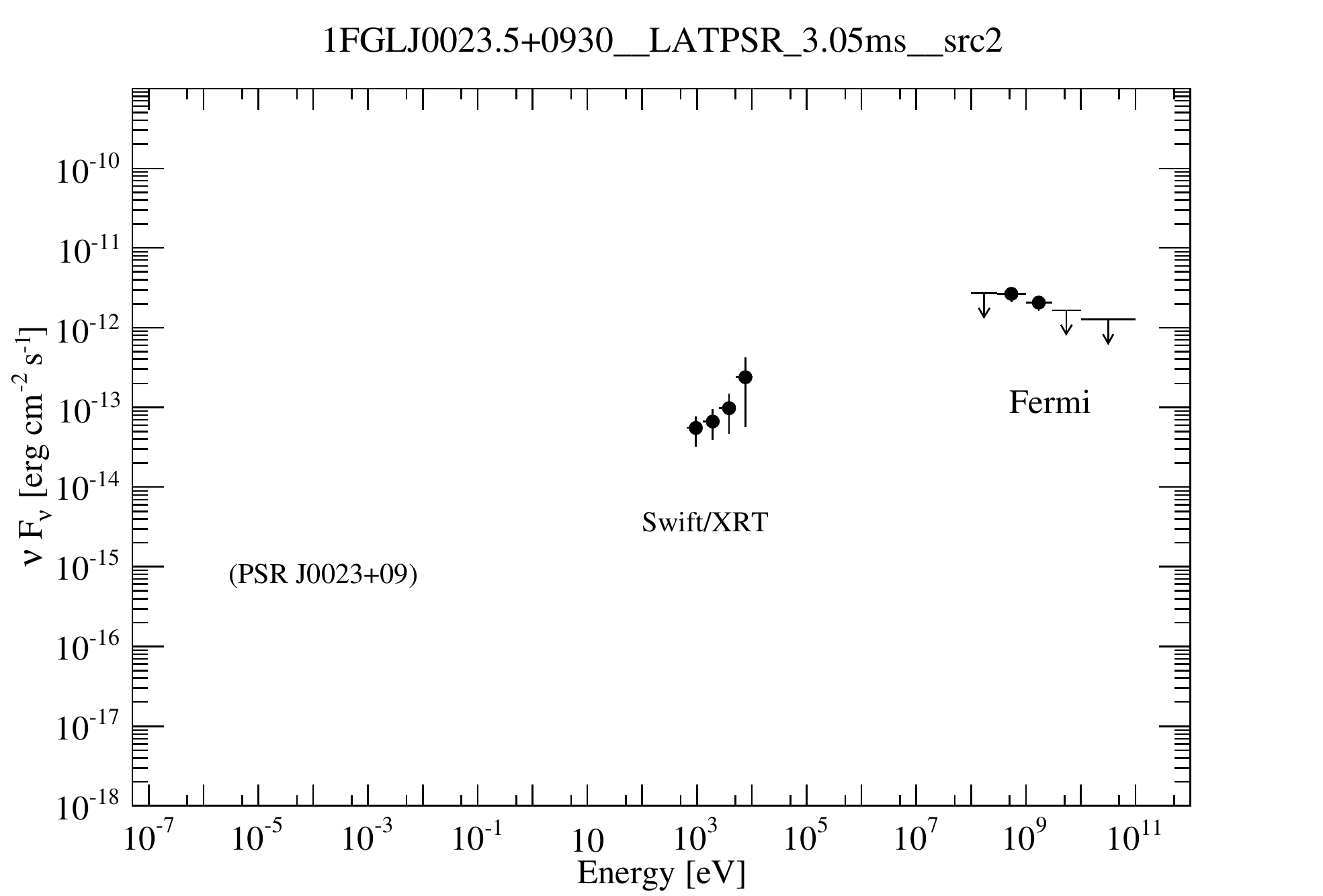}
    \end{center}
  \end{minipage}
  \begin{minipage}{0.32\hsize}
    \begin{center}
      \includegraphics[width=55mm]{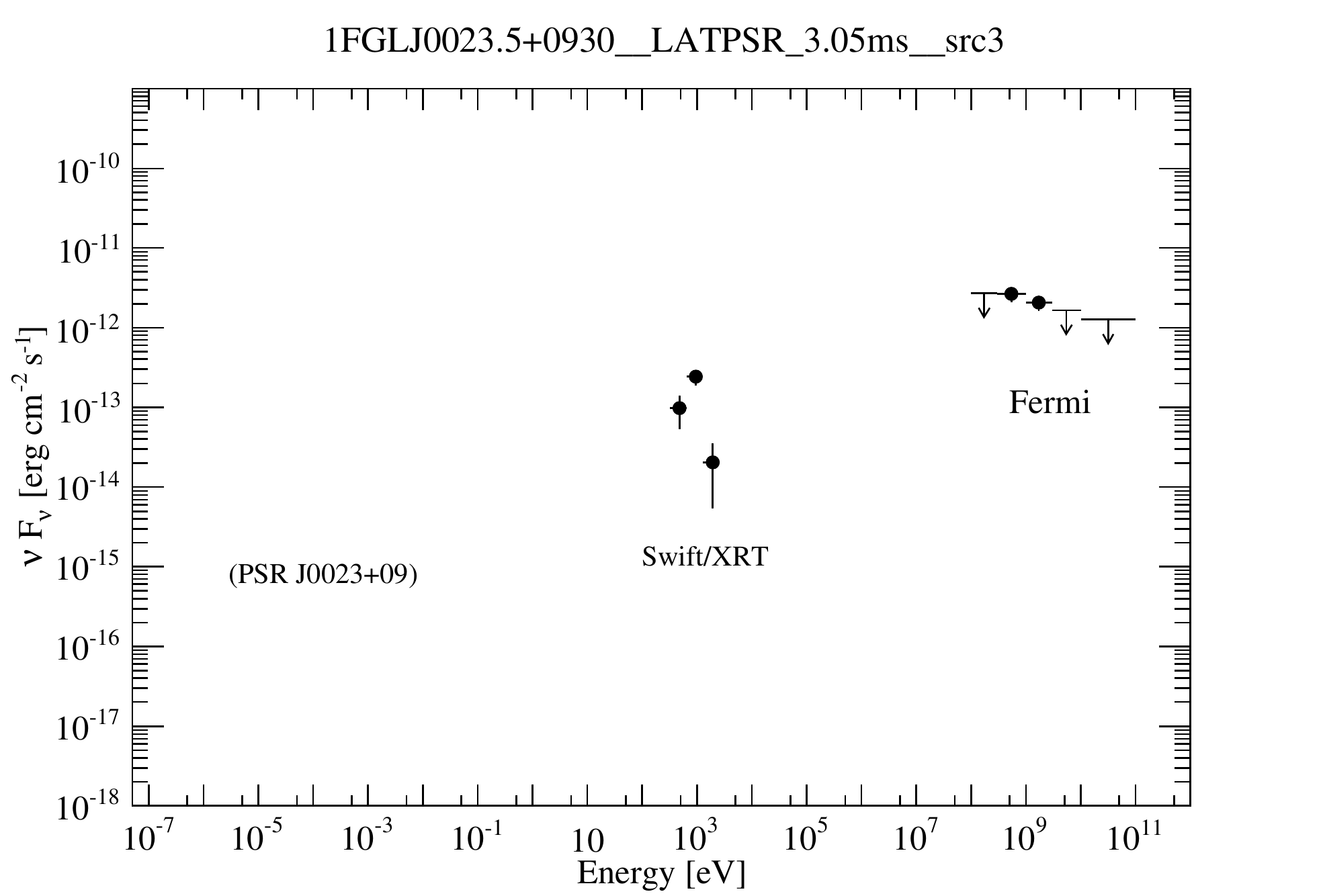}
    \end{center}
  \end{minipage}
  \begin{minipage}{0.32\hsize}
    \begin{center}
      \includegraphics[width=55mm]{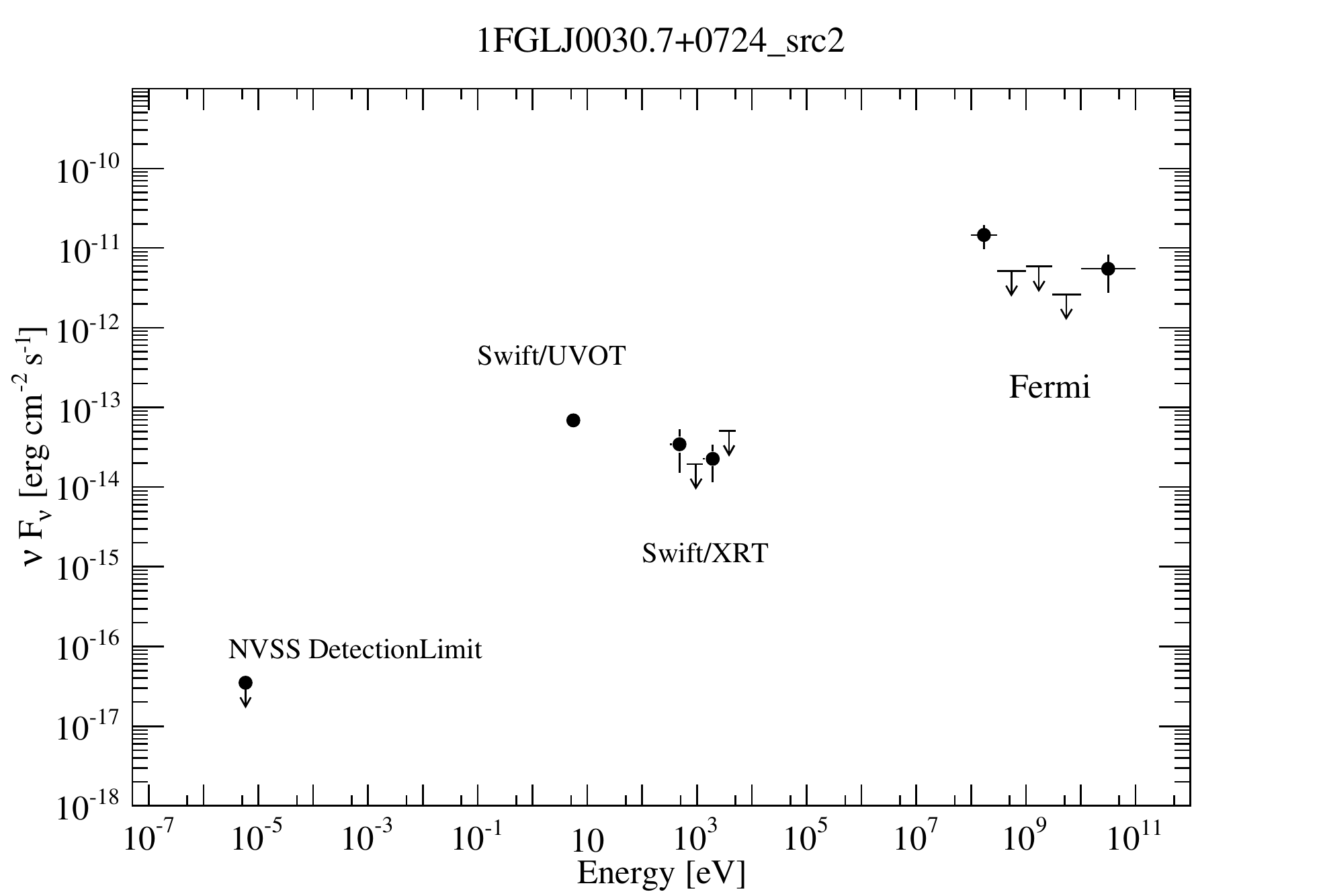}
    \end{center}
  \end{minipage}
  \begin{minipage}{0.32\hsize}
    \begin{center}
      \includegraphics[width=55mm]{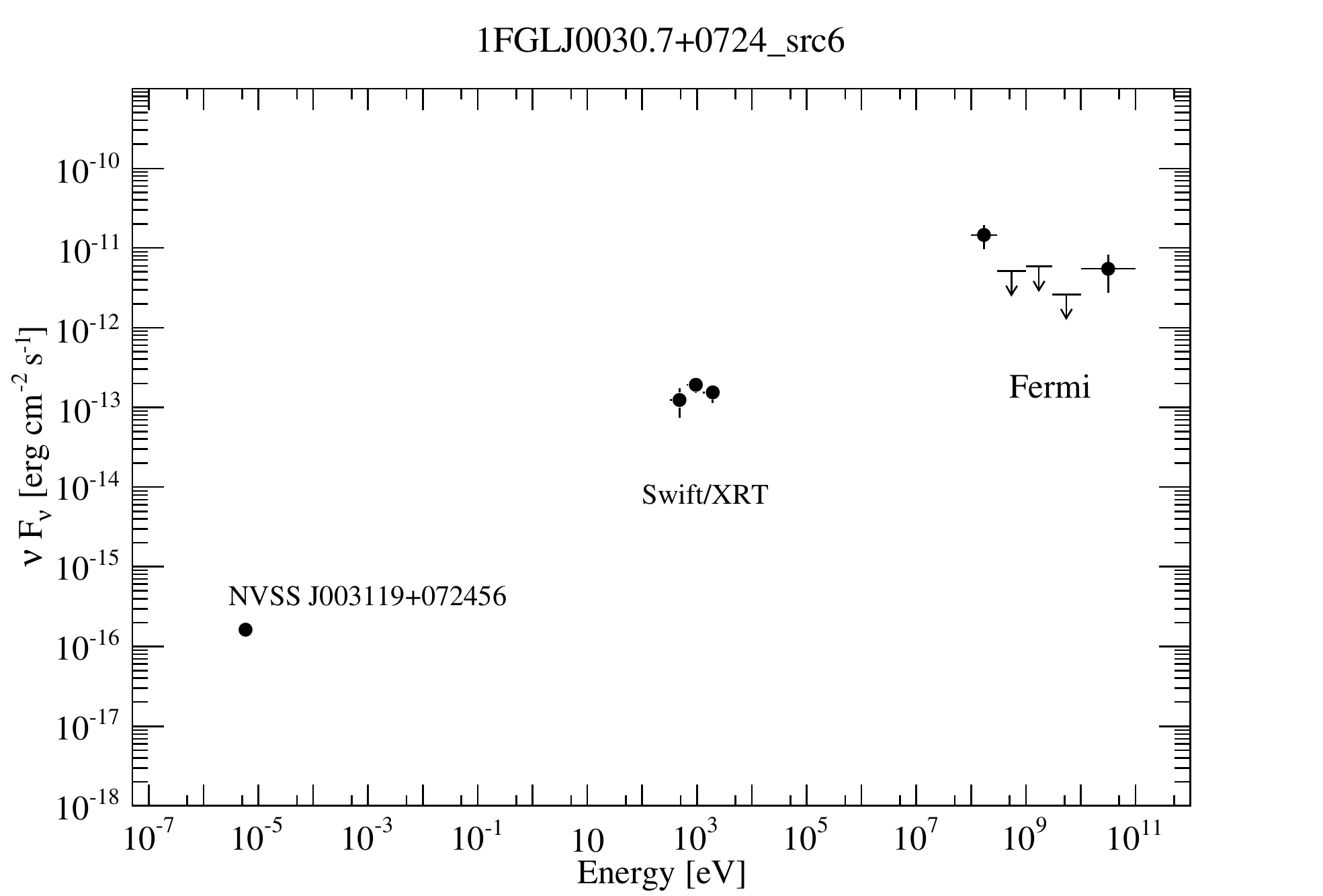}
    \end{center}
  \end{minipage}
  \begin{minipage}{0.32\hsize}
    \begin{center}
      \includegraphics[width=55mm]{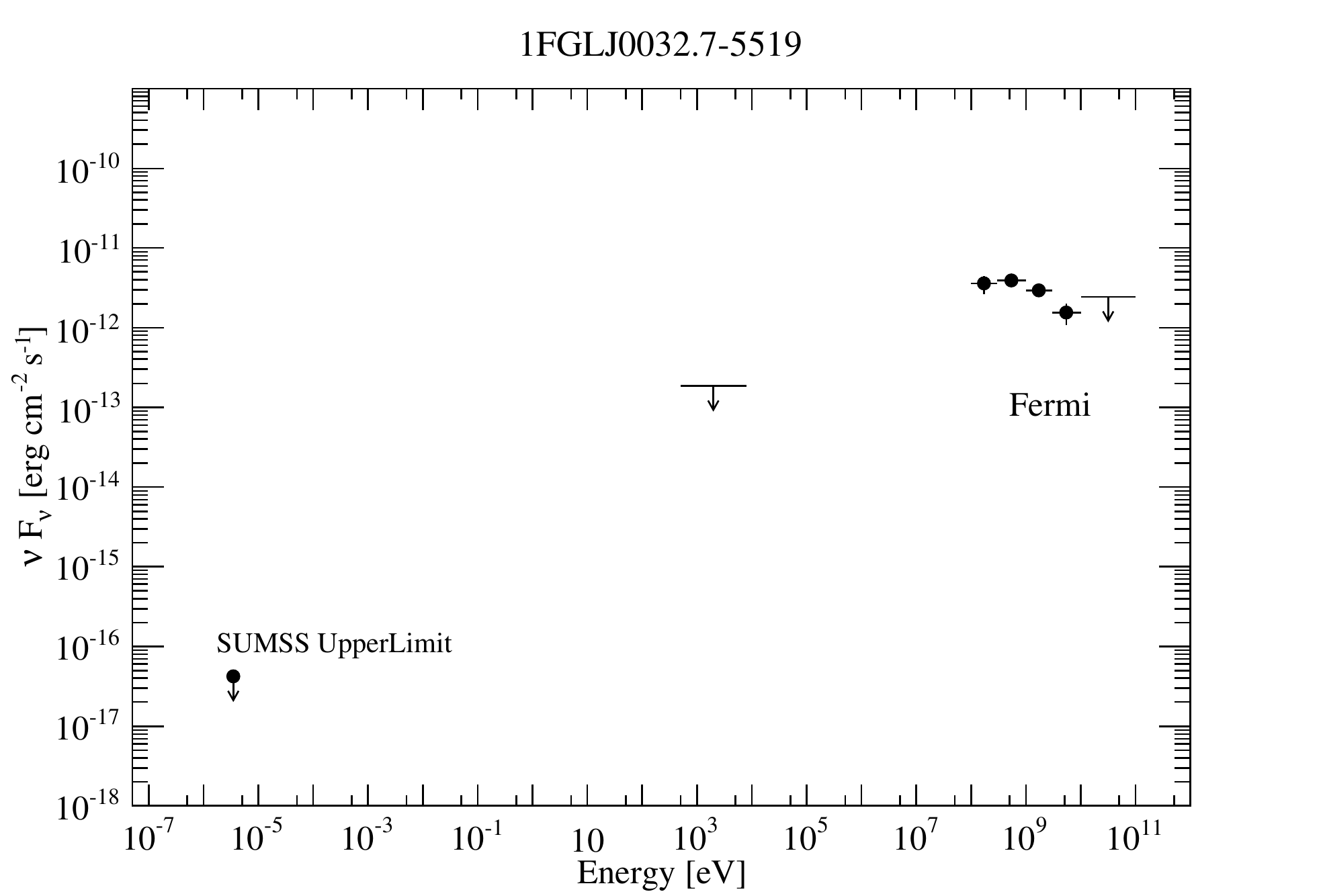}
    \end{center}
  \end{minipage}
  \begin{minipage}{0.32\hsize}
    \begin{center}
      \includegraphics[width=55mm]{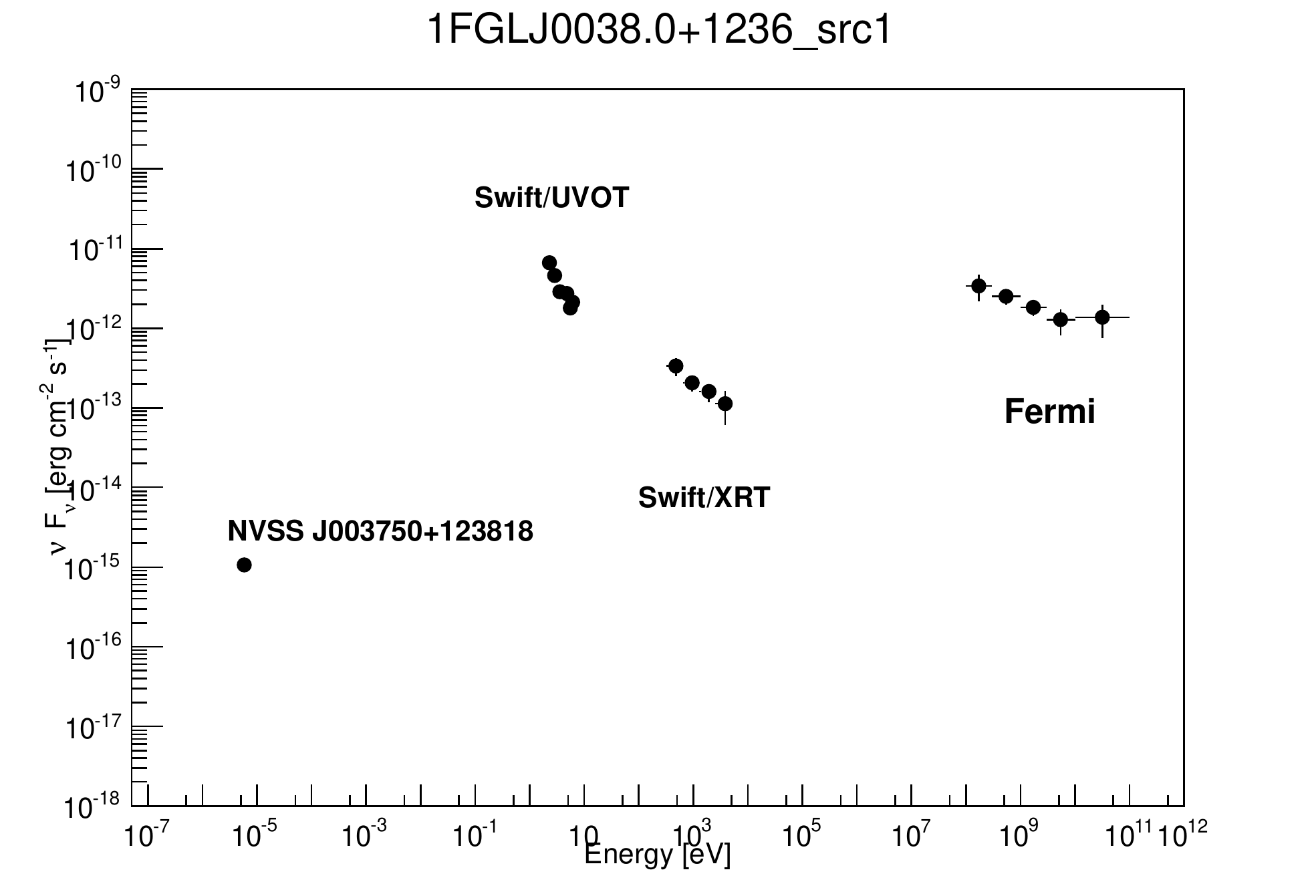}
    \end{center}
  \end{minipage}
  \begin{minipage}{0.32\hsize}
    \begin{center}
      \includegraphics[width=55mm]{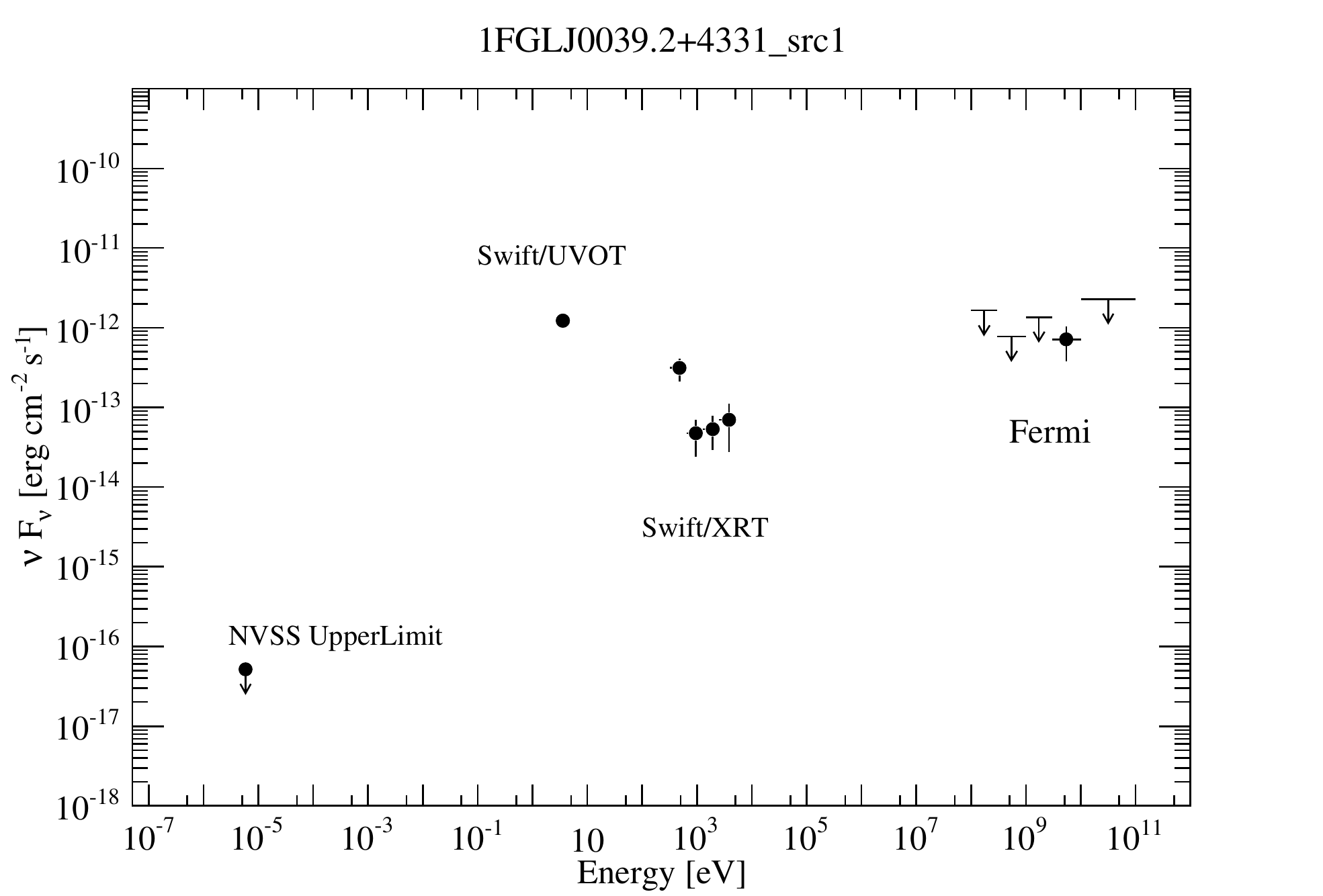}
    \end{center}
  \end{minipage}
  \begin{minipage}{0.32\hsize}
    \begin{center}
      \includegraphics[width=55mm]{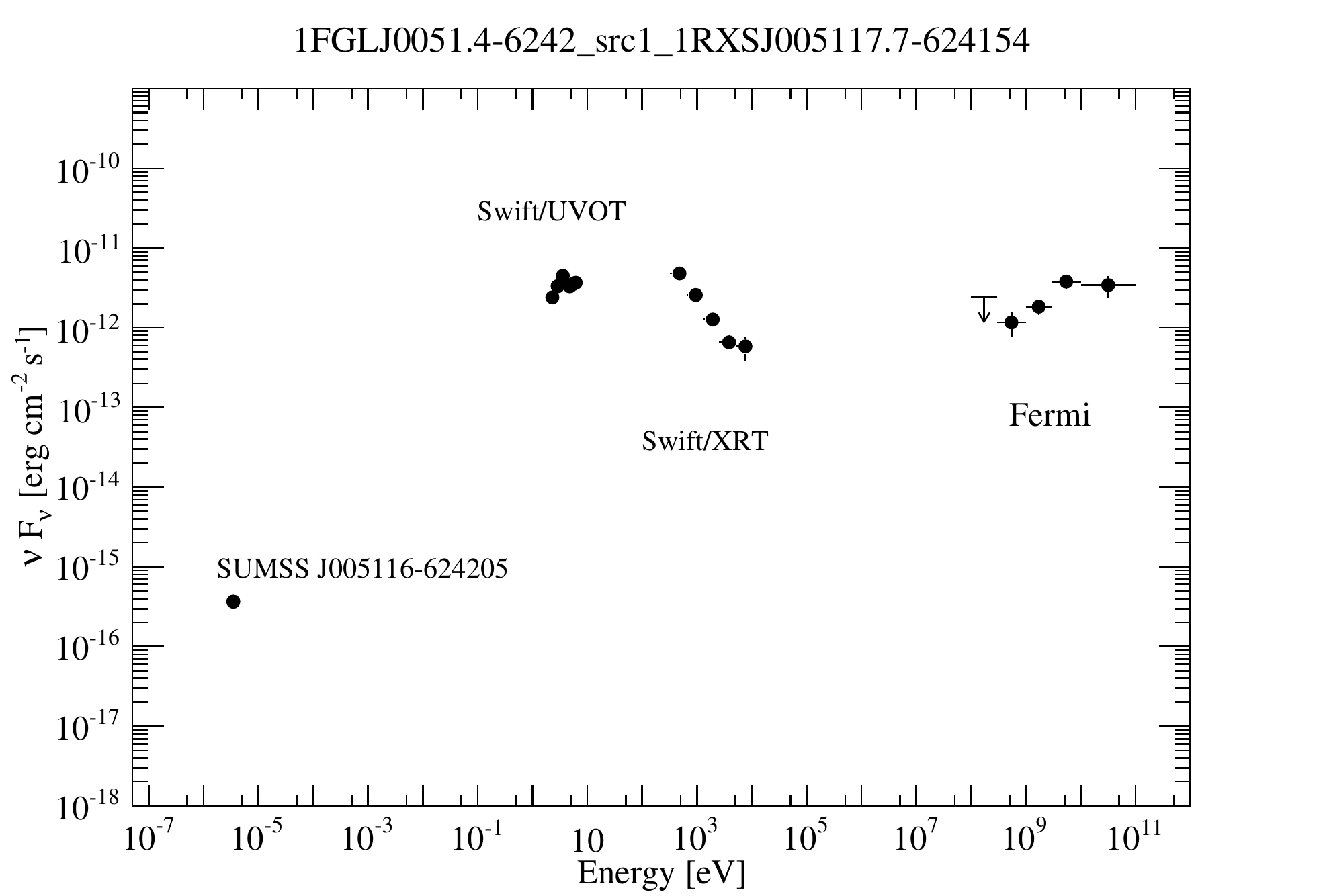}
    \end{center}
  \end{minipage}
  \begin{minipage}{0.32\hsize}
    \begin{center}
      \includegraphics[width=55mm]{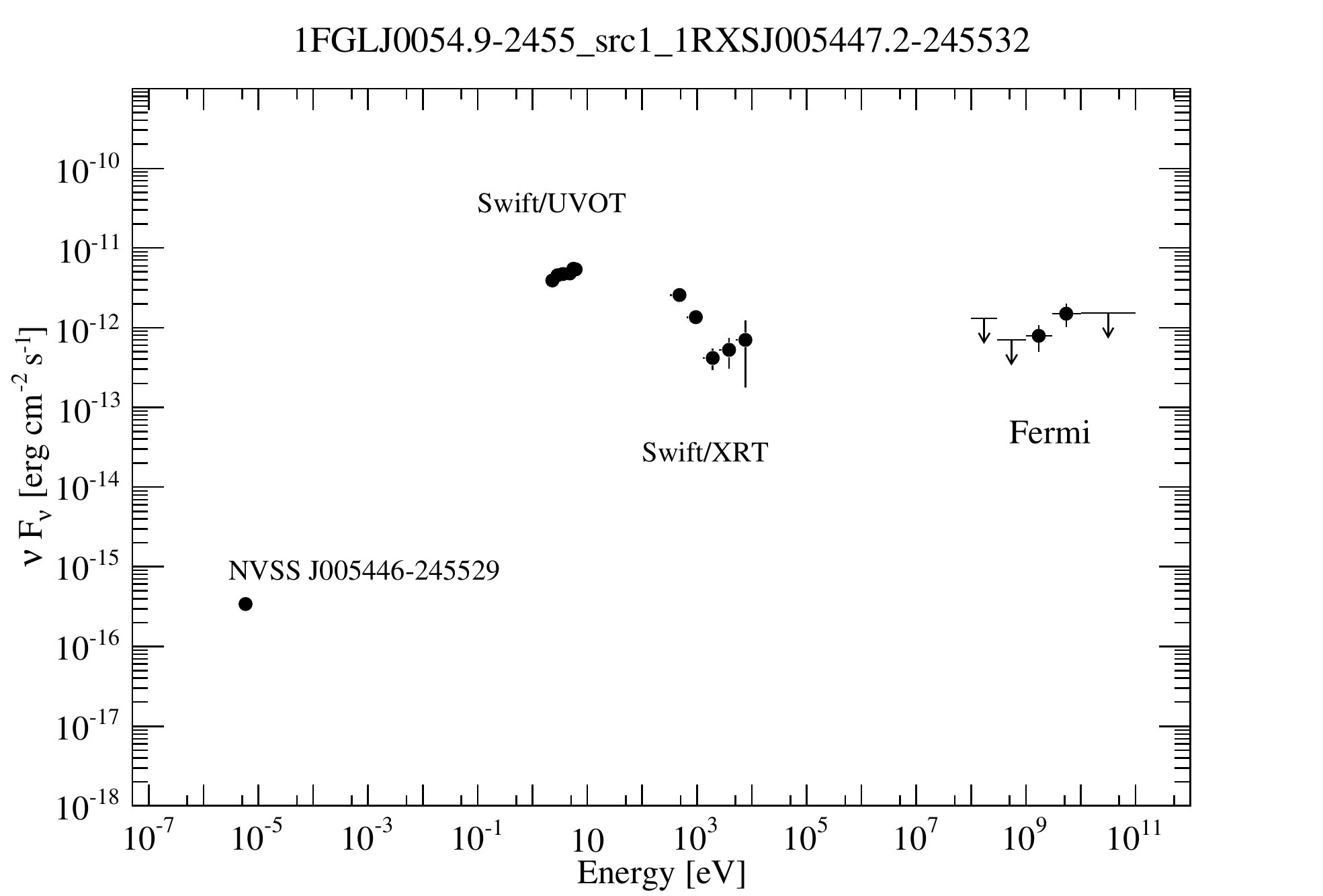}
    \end{center}
  \end{minipage}
  \begin{minipage}{0.32\hsize}
    \begin{center}
      \includegraphics[width=55mm]{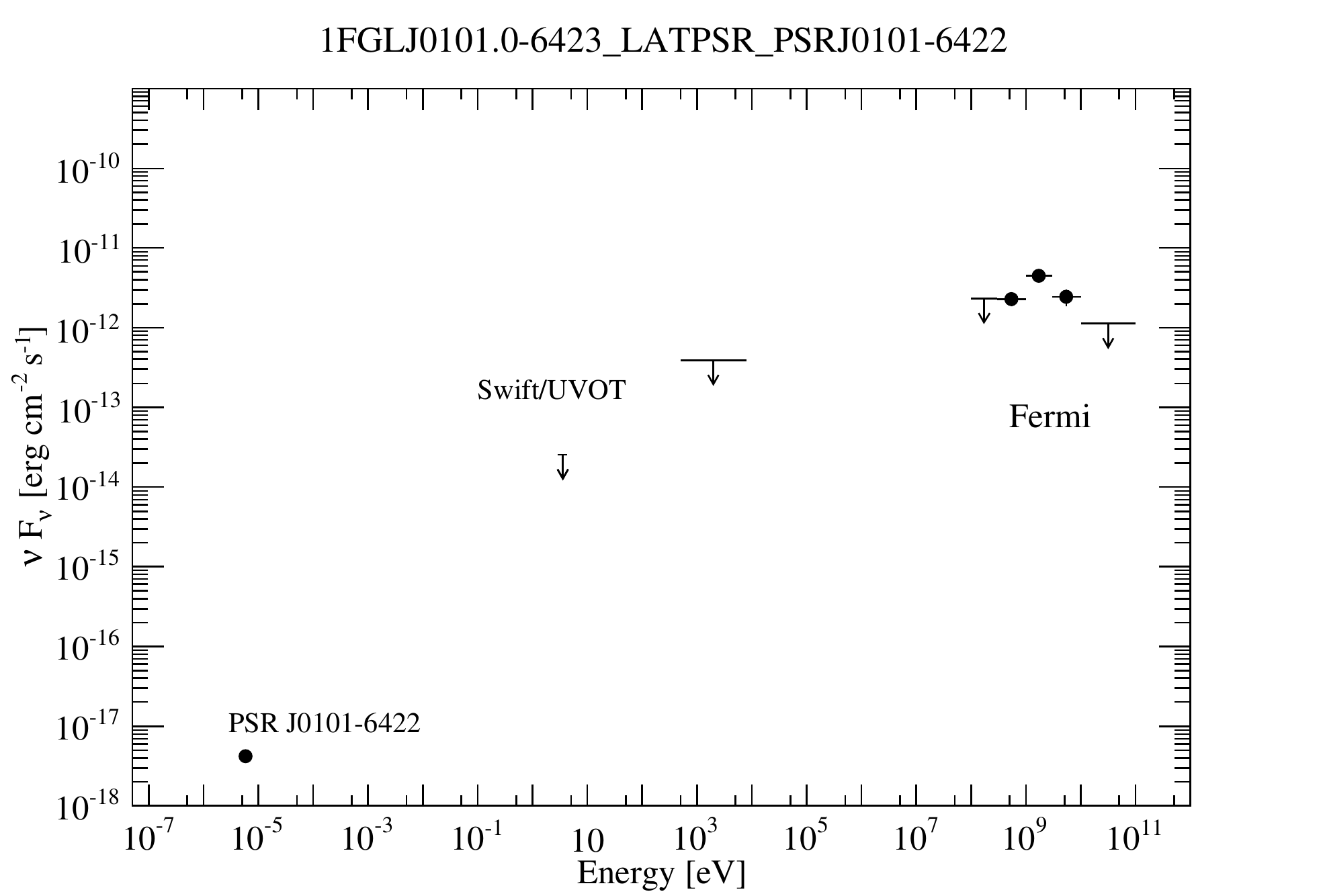}
    \end{center}
  \end{minipage}
  \begin{minipage}{0.32\hsize}
    \begin{center}
      \includegraphics[width=55mm]{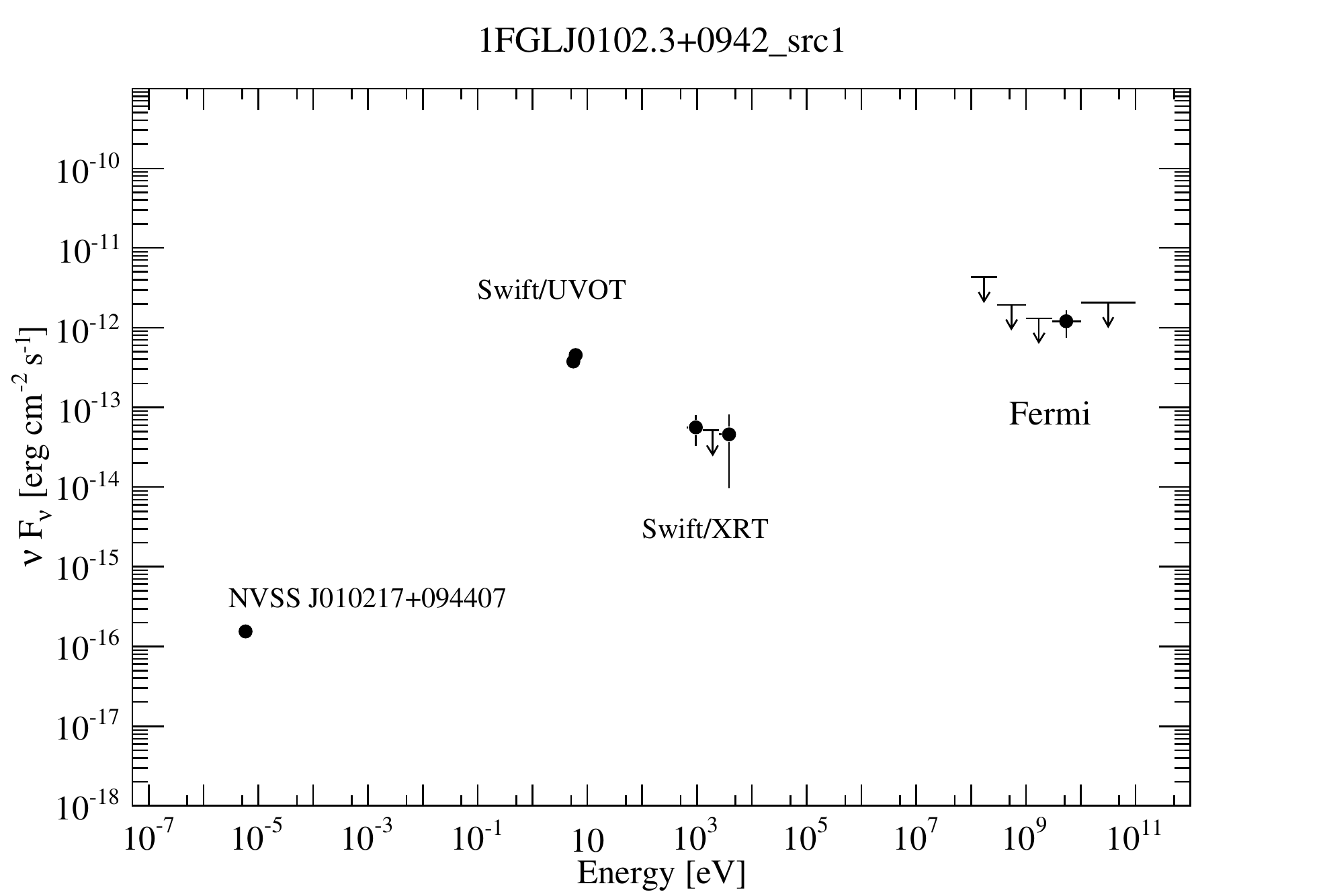}
    \end{center}
  \end{minipage}
  \begin{minipage}{0.32\hsize}
    \begin{center}
      \includegraphics[width=55mm]{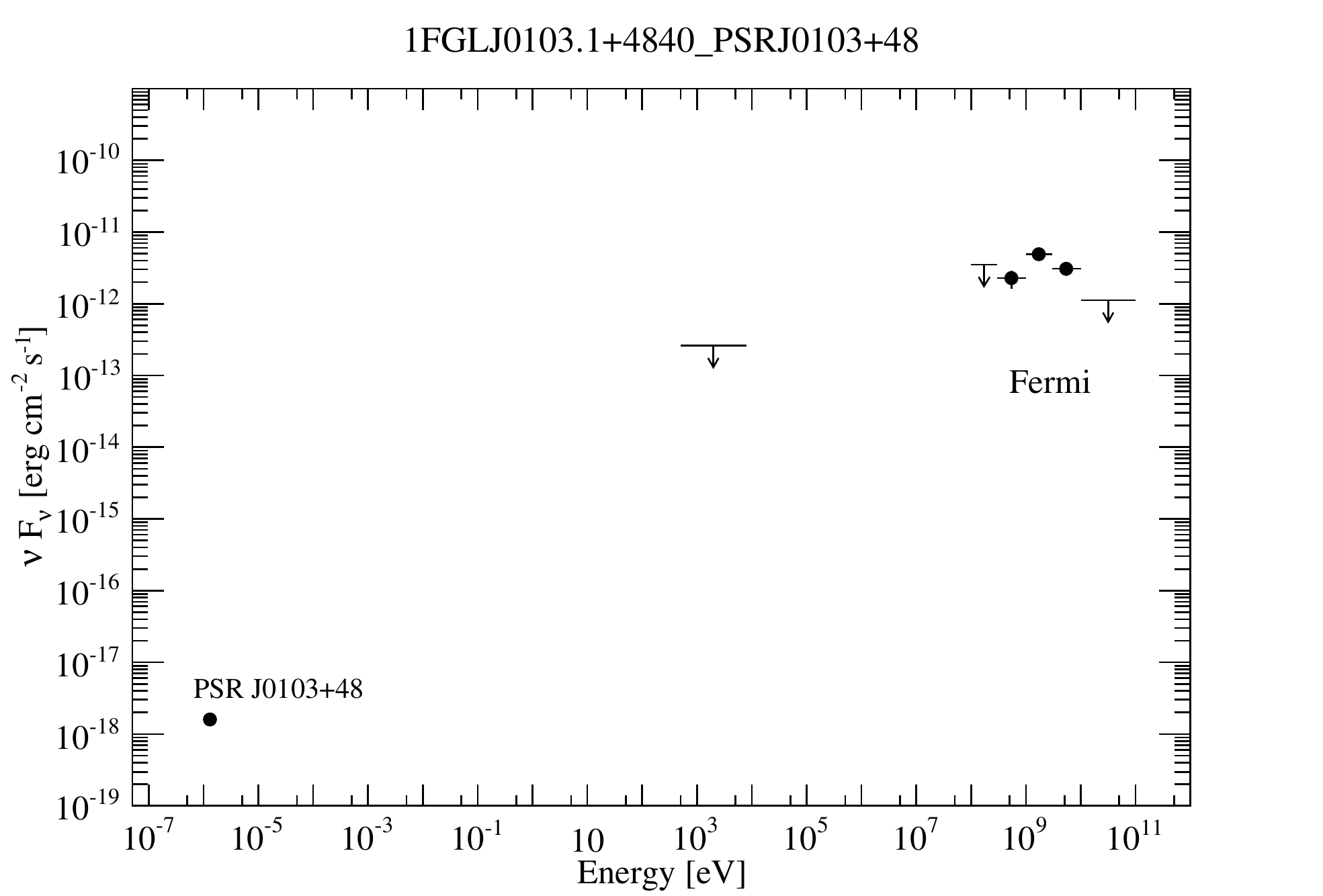}
    \end{center}
  \end{minipage}
  \begin{minipage}{0.32\hsize}
    \begin{center}
      \includegraphics[width=55mm]{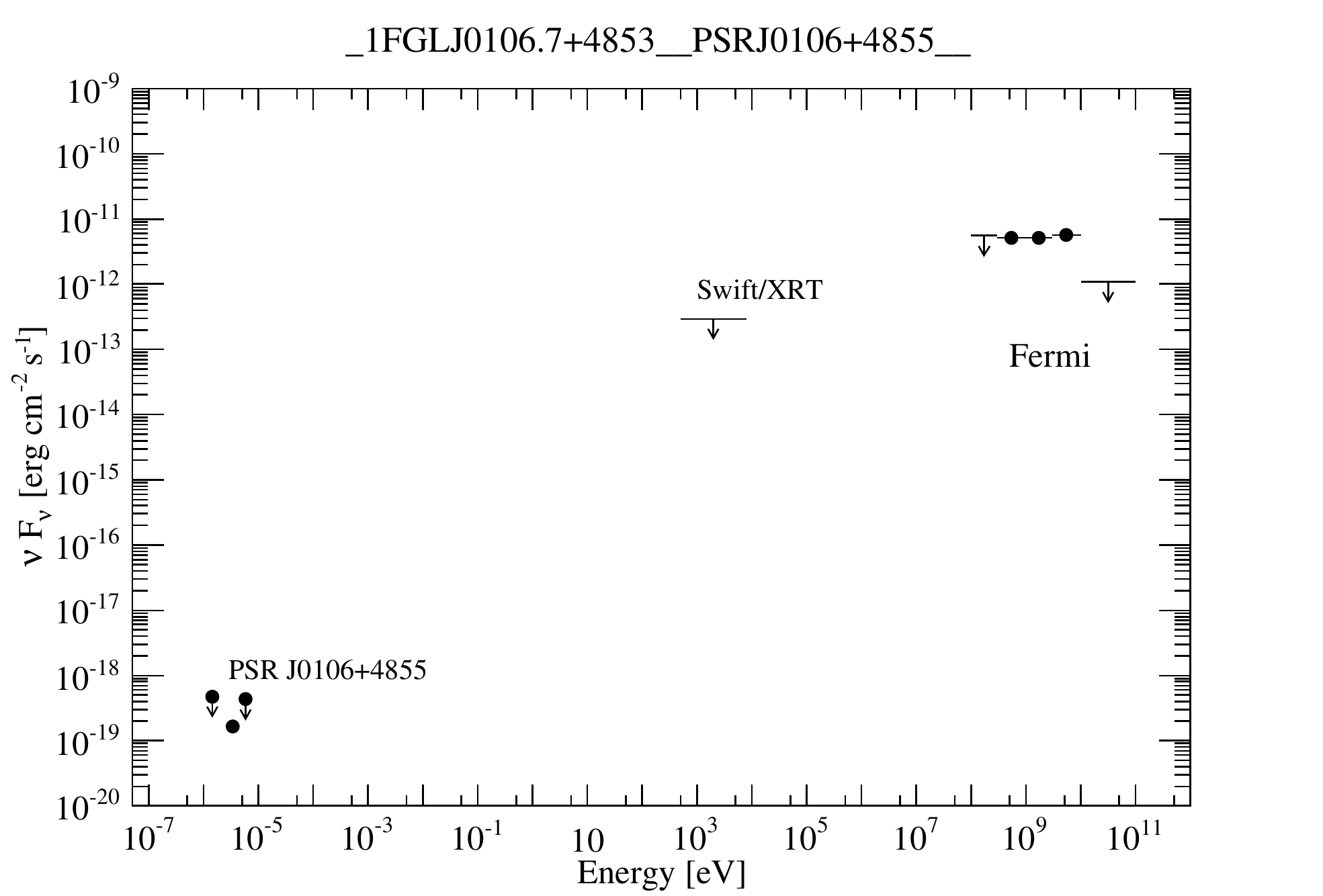}
    \end{center}
  \end{minipage} 
  \begin{minipage}{0.32\hsize}
    \begin{center}
      \includegraphics[width=55mm]{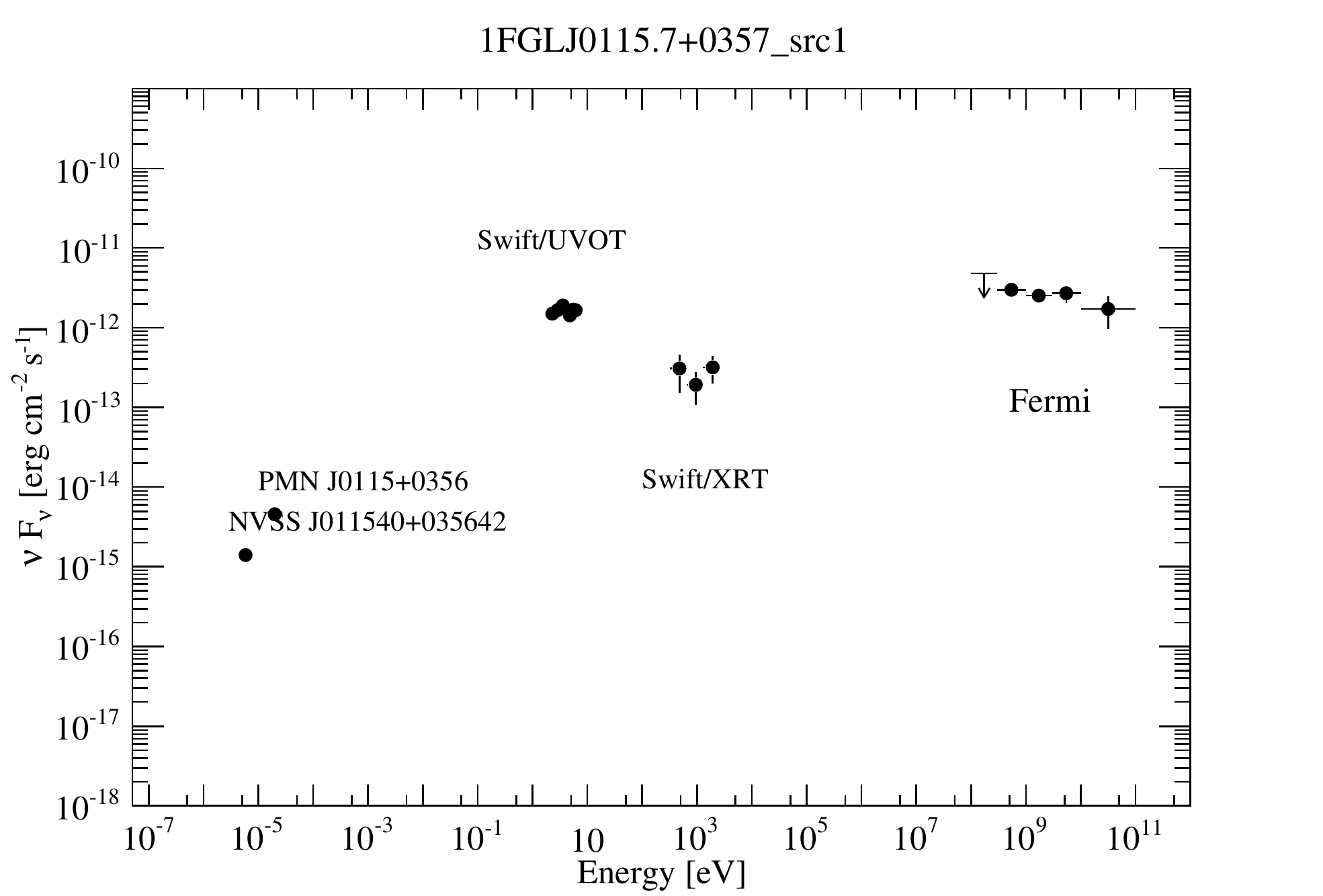}
    \end{center}
  \end{minipage}
 \end{center}
\end{figure}
\clearpage
\begin{figure}[m]
 \begin{center}
  \begin{minipage}{0.32\hsize}
    \begin{center}
      \includegraphics[width=55mm]{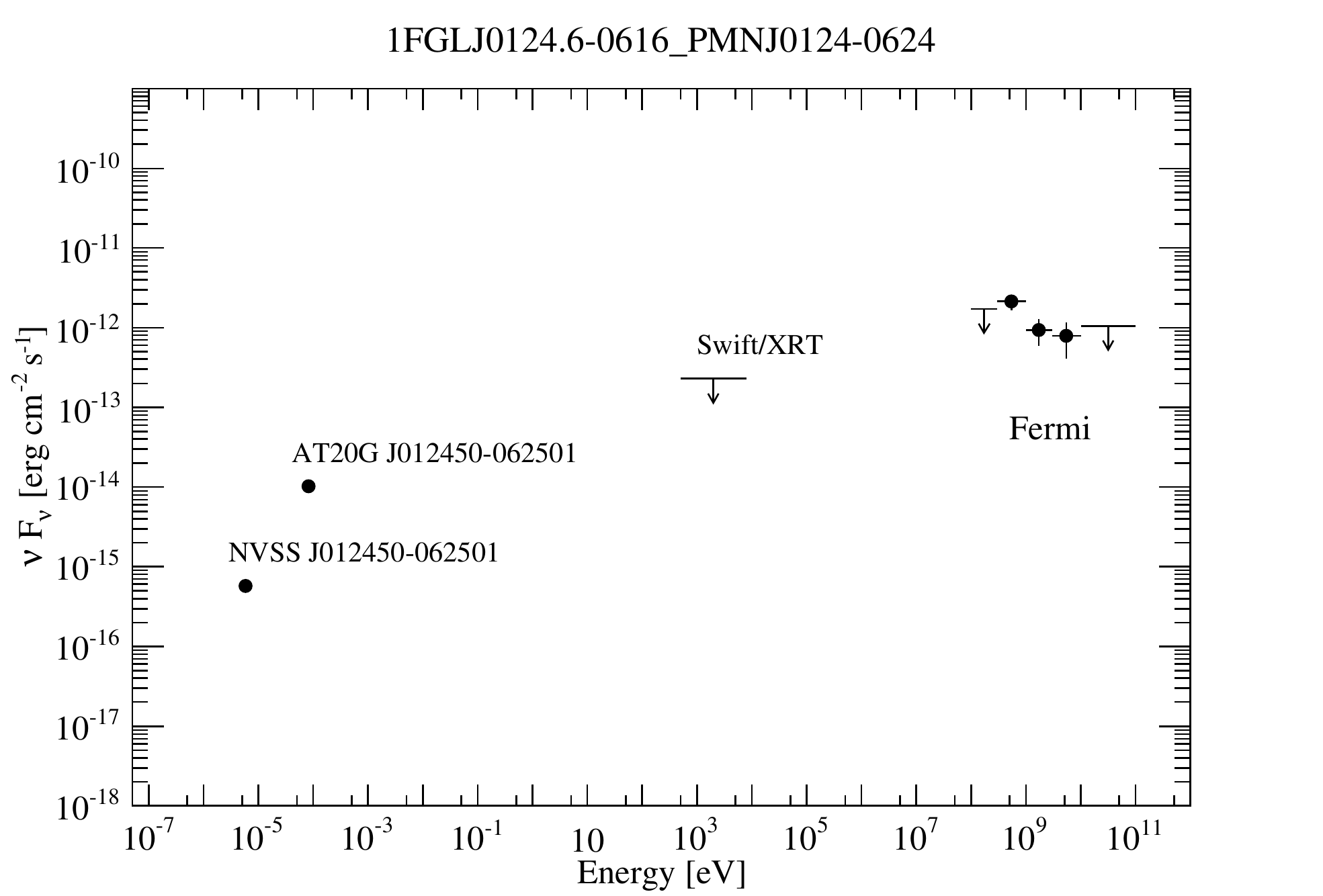}
    \end{center}
  \end{minipage}
  \begin{minipage}{0.32\hsize}
    \begin{center}
      \includegraphics[width=55mm]{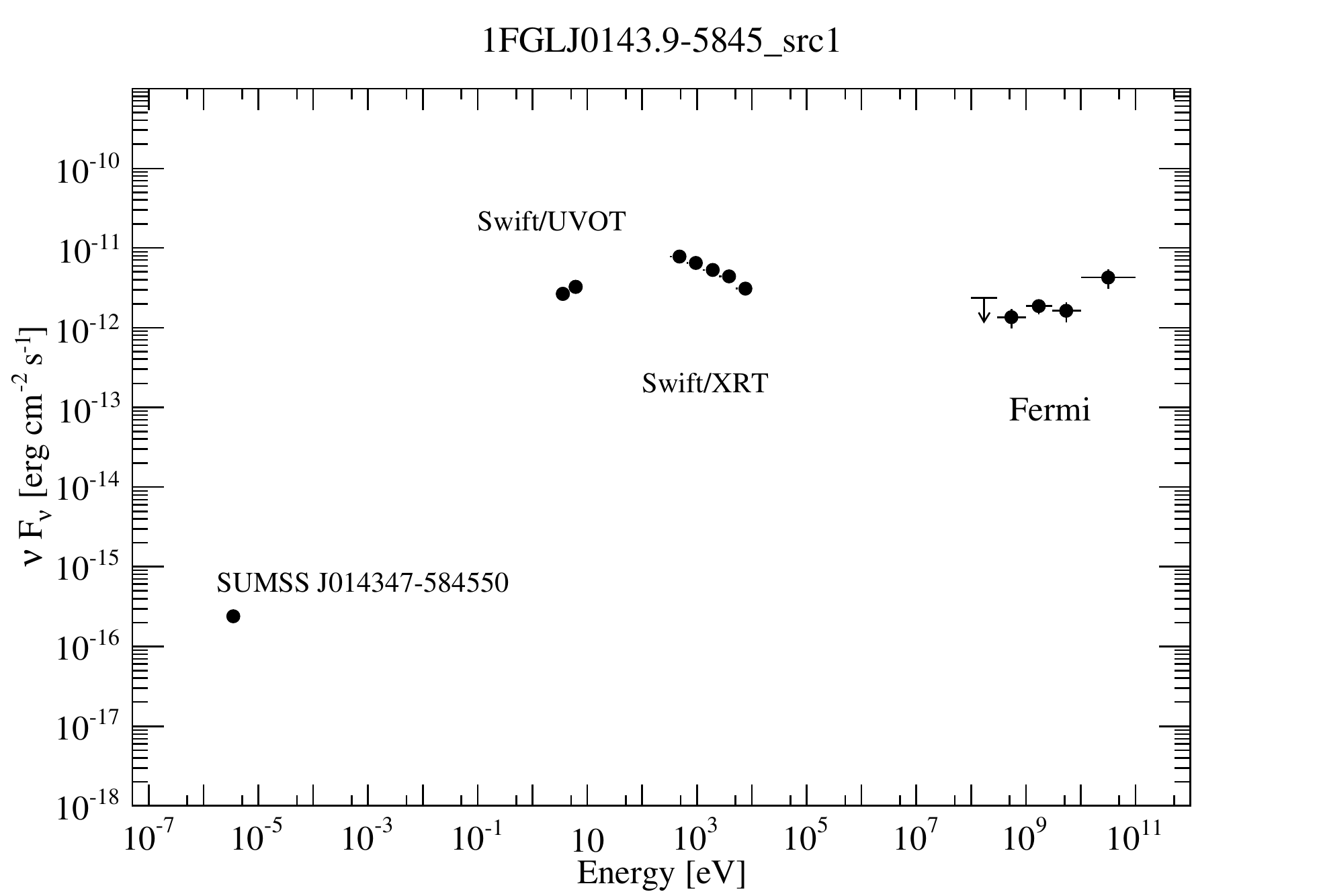}
    \end{center}
  \end{minipage}
  \begin{minipage}{0.32\hsize}
    \begin{center}
      \includegraphics[width=55mm]{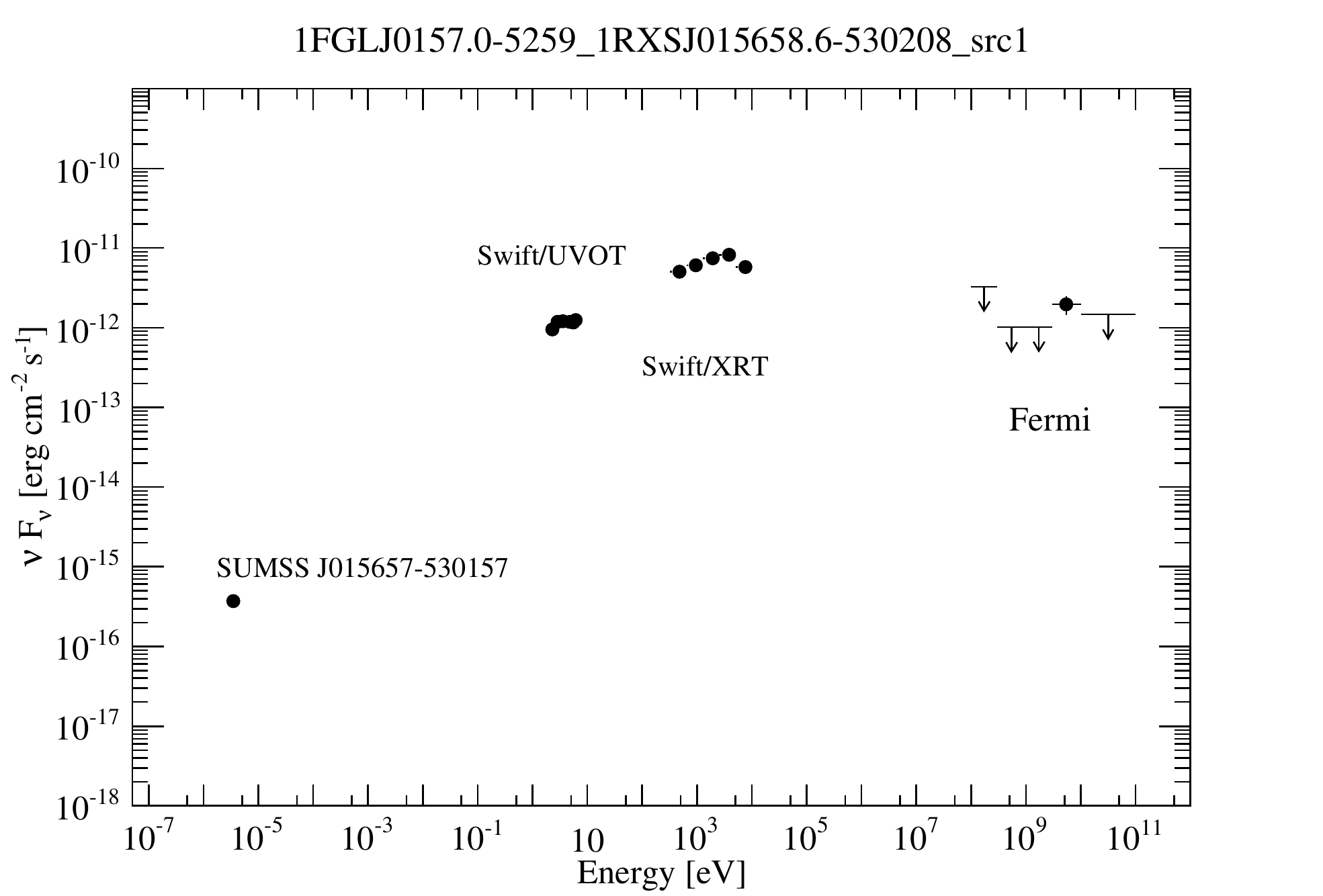}
    \end{center}
  \end{minipage}
  \begin{minipage}{0.32\hsize}
    \begin{center}
      \includegraphics[width=55mm]{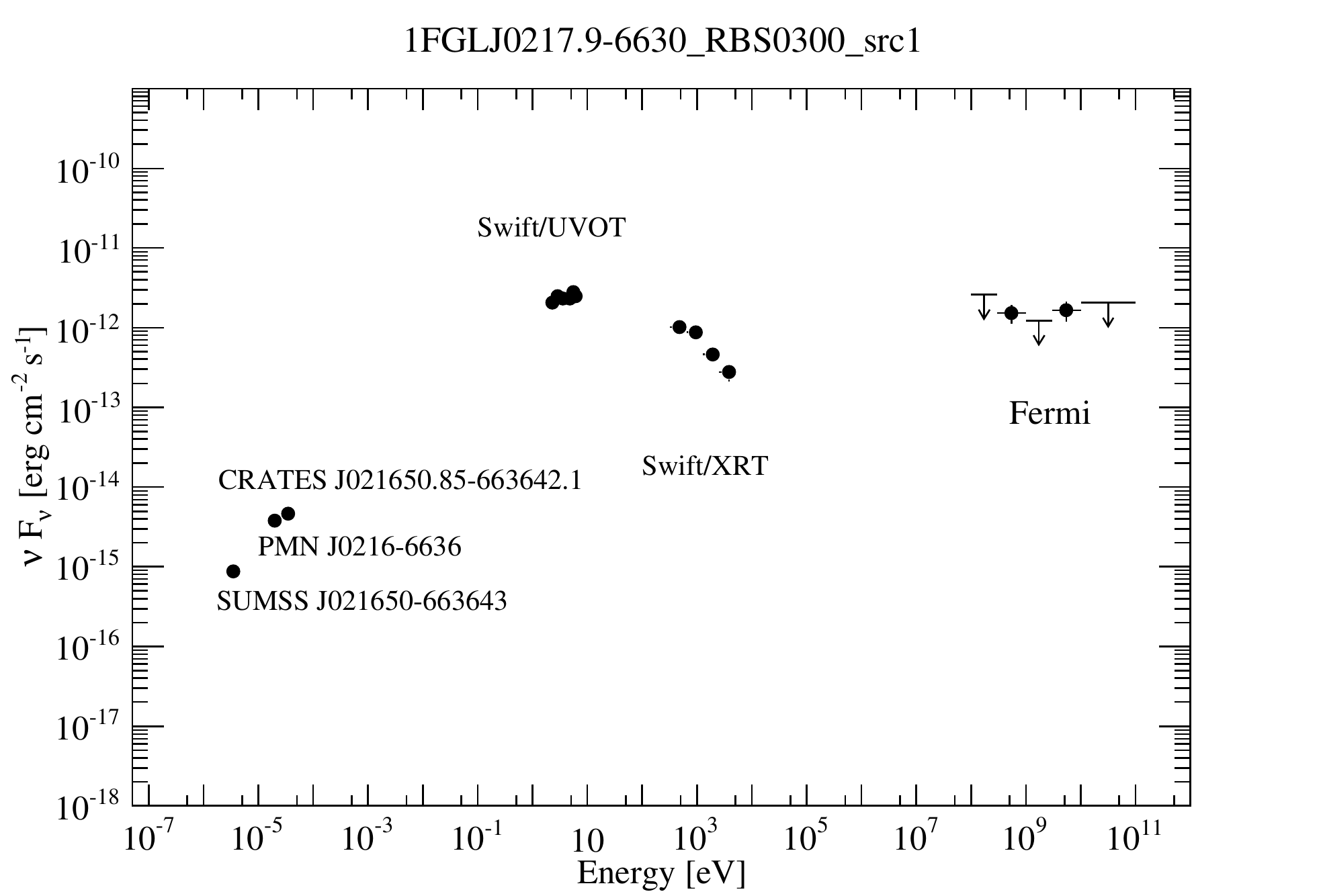}
    \end{center}
  \end{minipage}
  \begin{minipage}{0.32\hsize}
    \begin{center}
      \includegraphics[width=55mm]{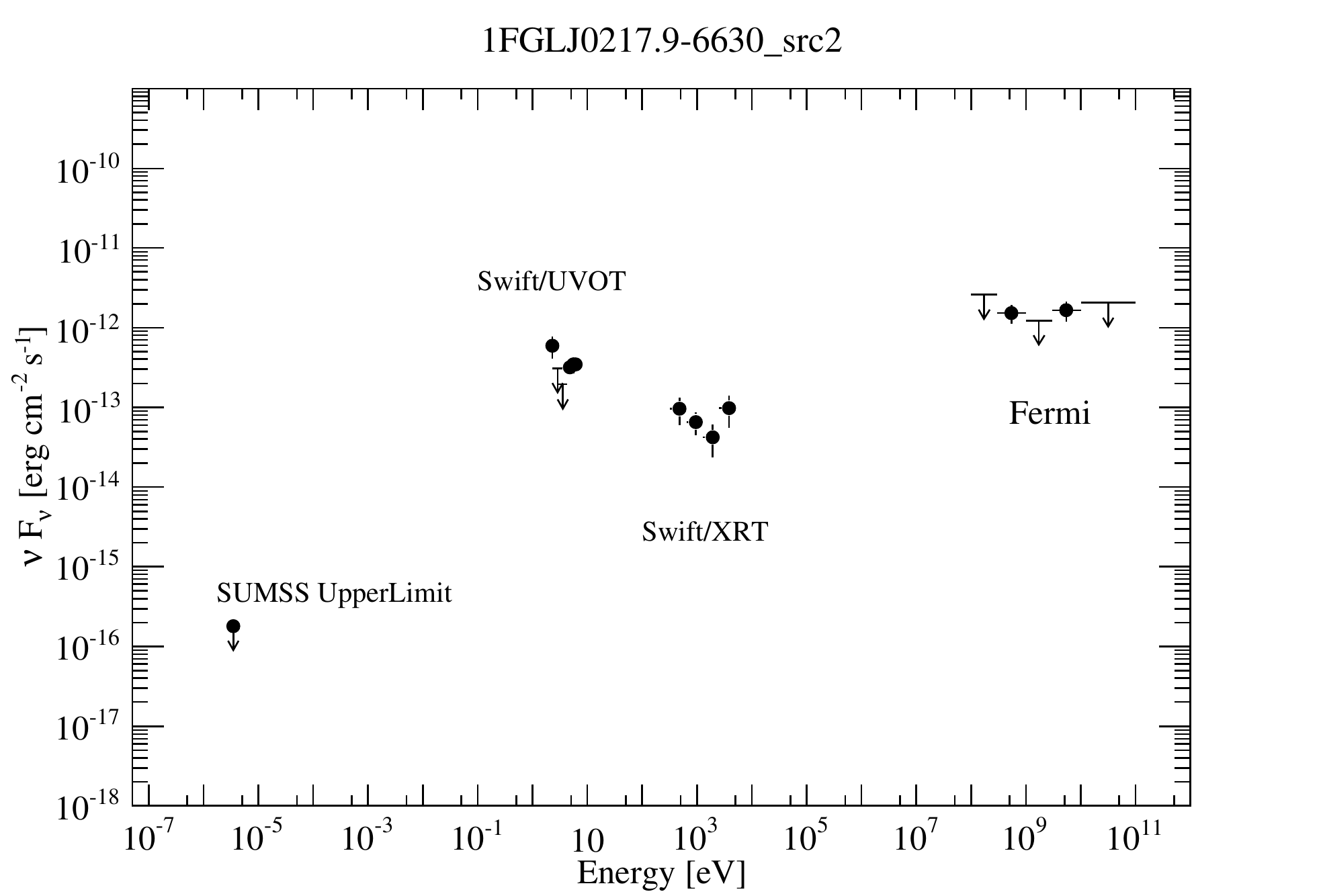}
    \end{center}
  \end{minipage}
  \begin{minipage}{0.32\hsize}
    \begin{center}
      \includegraphics[width=55mm]{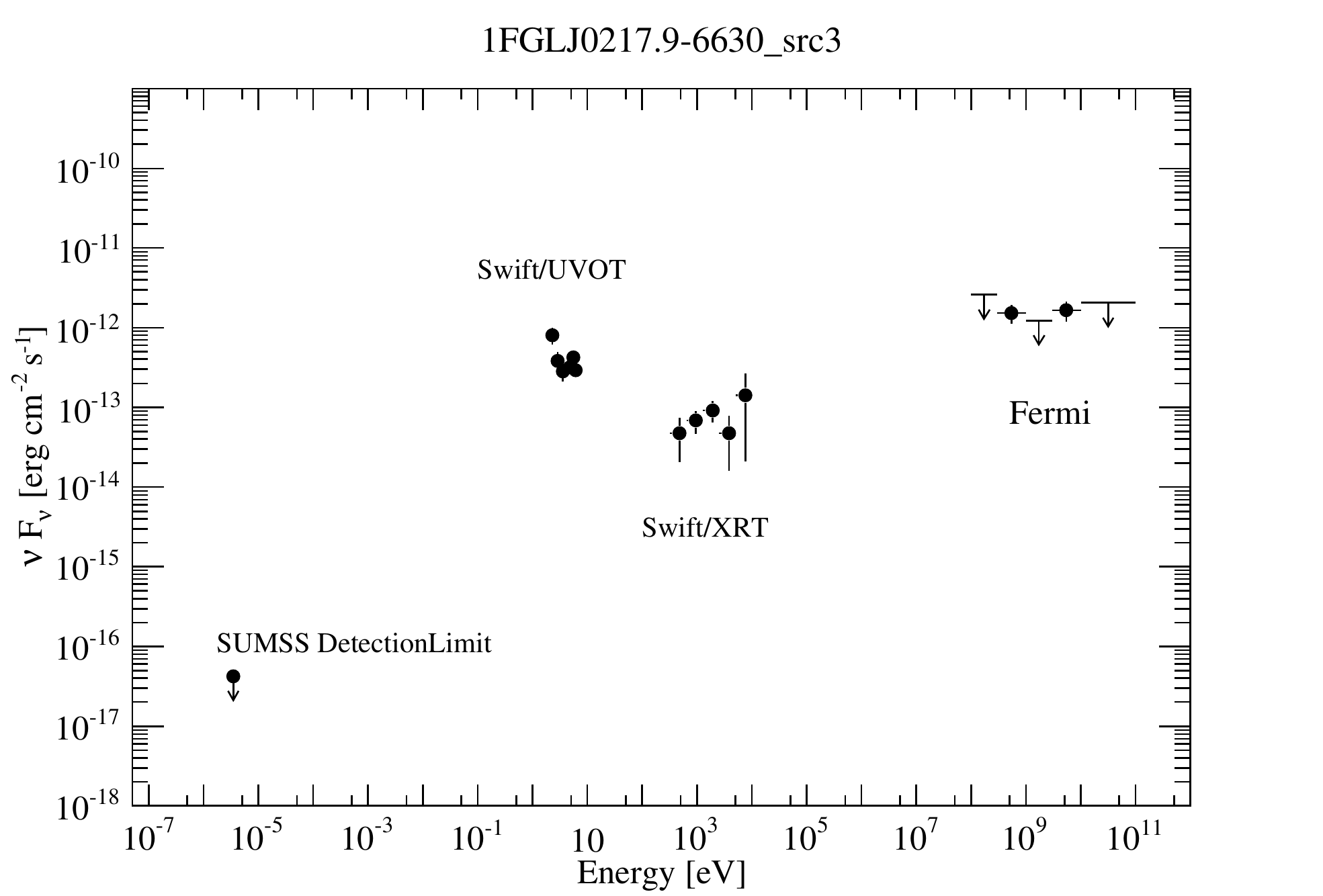}
    \end{center}
  \end{minipage}
  \begin{minipage}{0.32\hsize}
    \begin{center}
      \includegraphics[width=55mm]{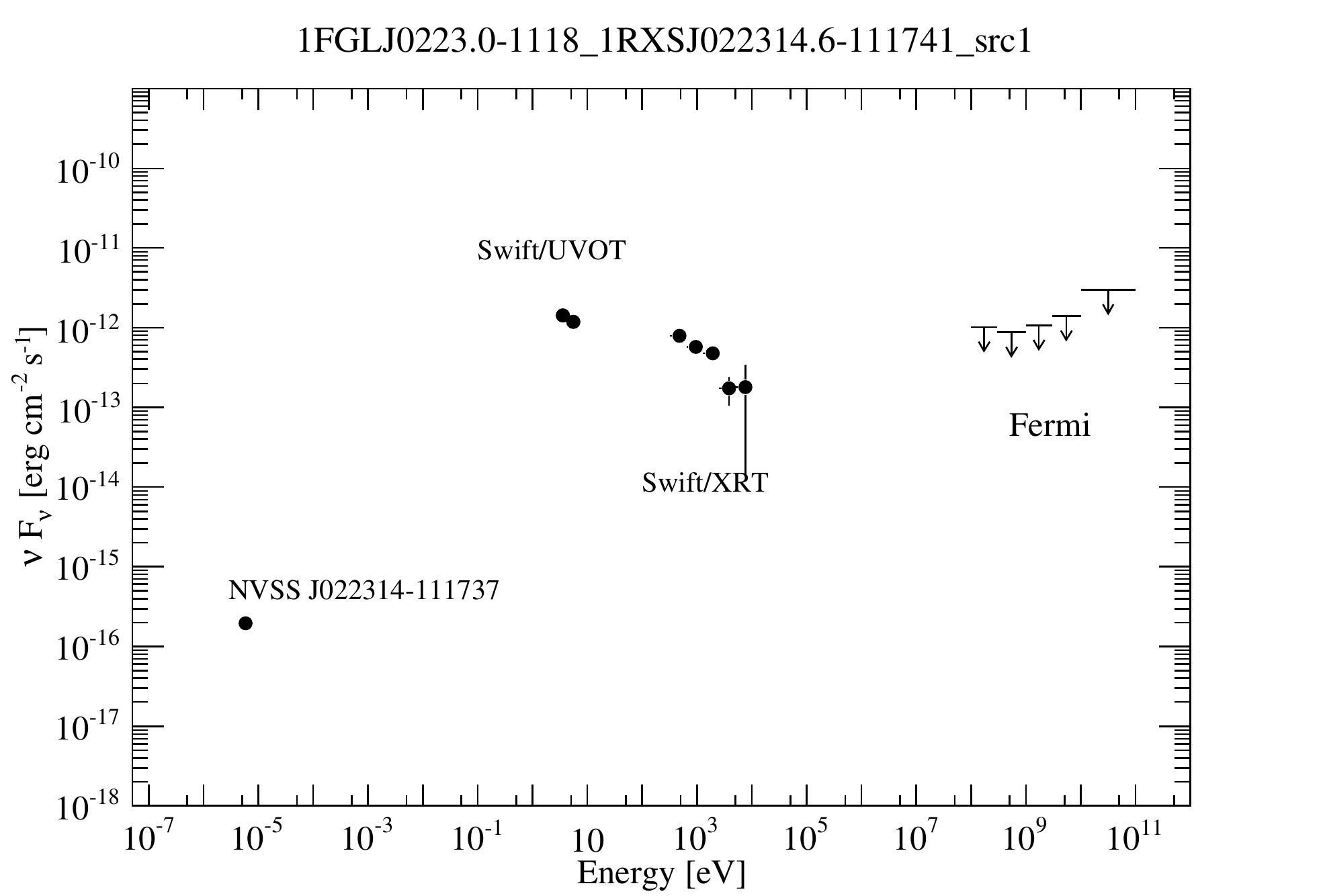}
    \end{center}
  \end{minipage}
  \begin{minipage}{0.32\hsize}
    \begin{center}
      \includegraphics[width=55mm]{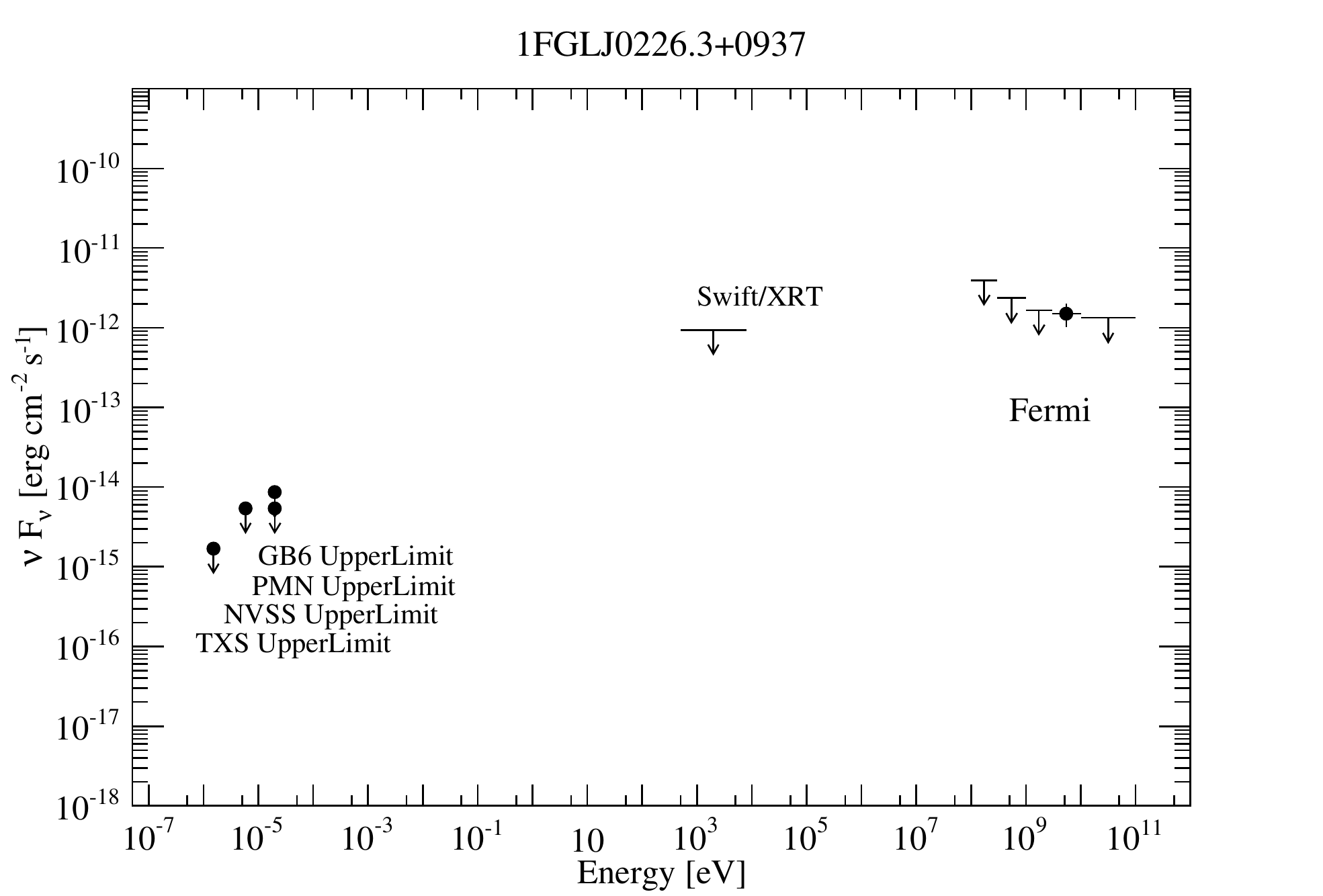}
    \end{center}
  \end{minipage}
  \begin{minipage}{0.32\hsize}
    \begin{center}
      \includegraphics[width=55mm]{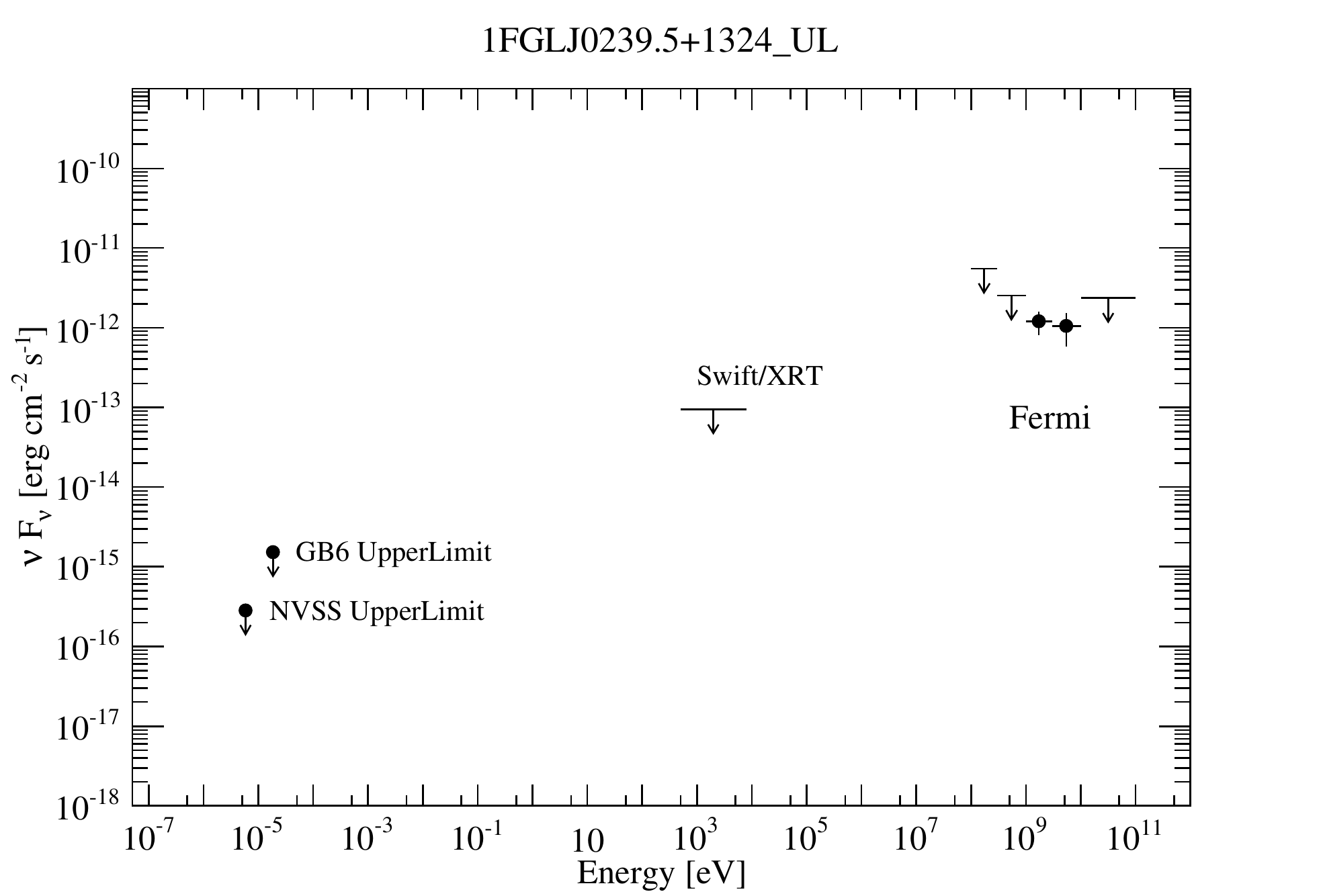}
    \end{center}
  \end{minipage}
  \begin{minipage}{0.32\hsize}
    \begin{center}
      \includegraphics[width=55mm]{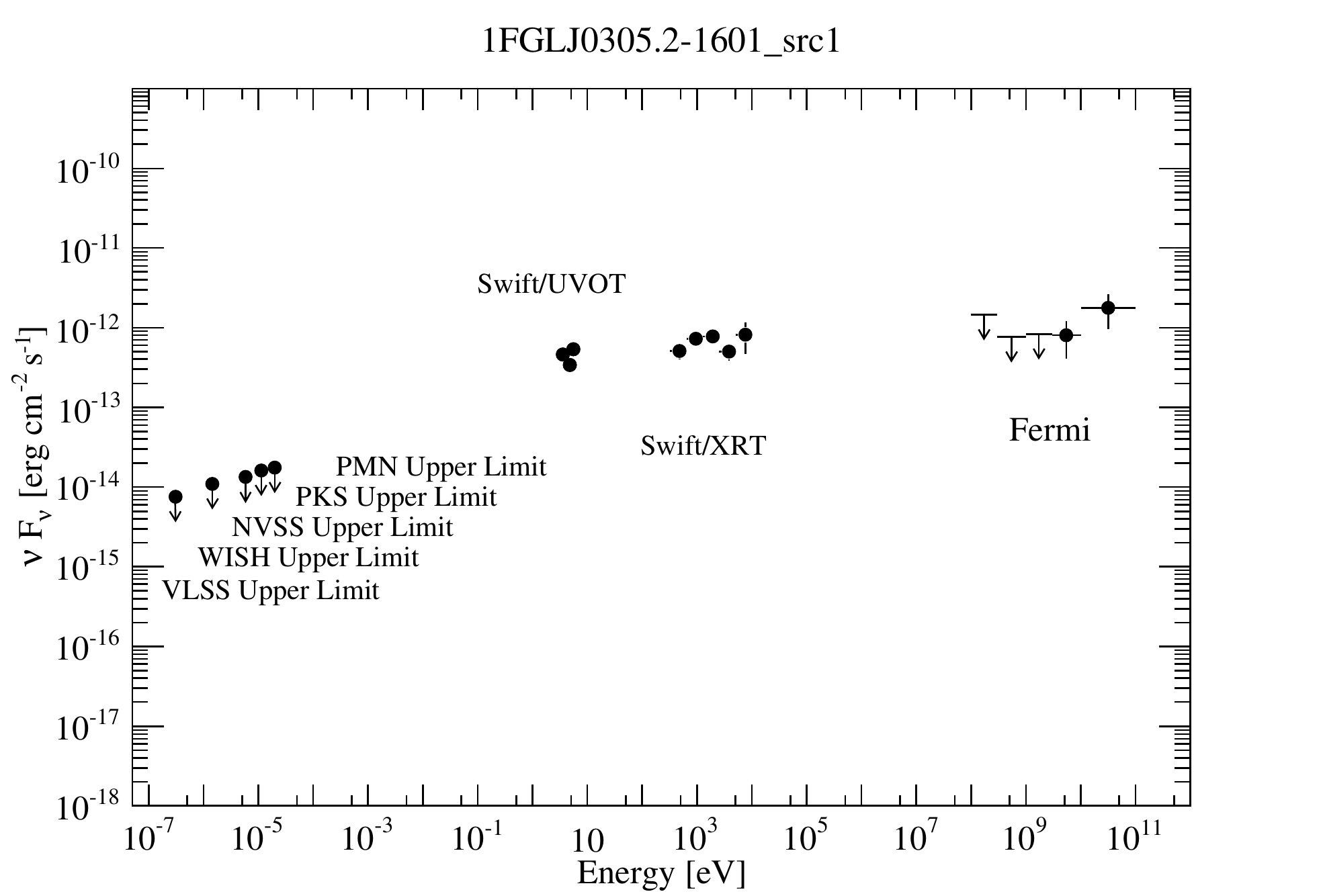}
    \end{center}
  \end{minipage}
  \begin{minipage}{0.32\hsize}
    \begin{center}
      \includegraphics[width=55mm]{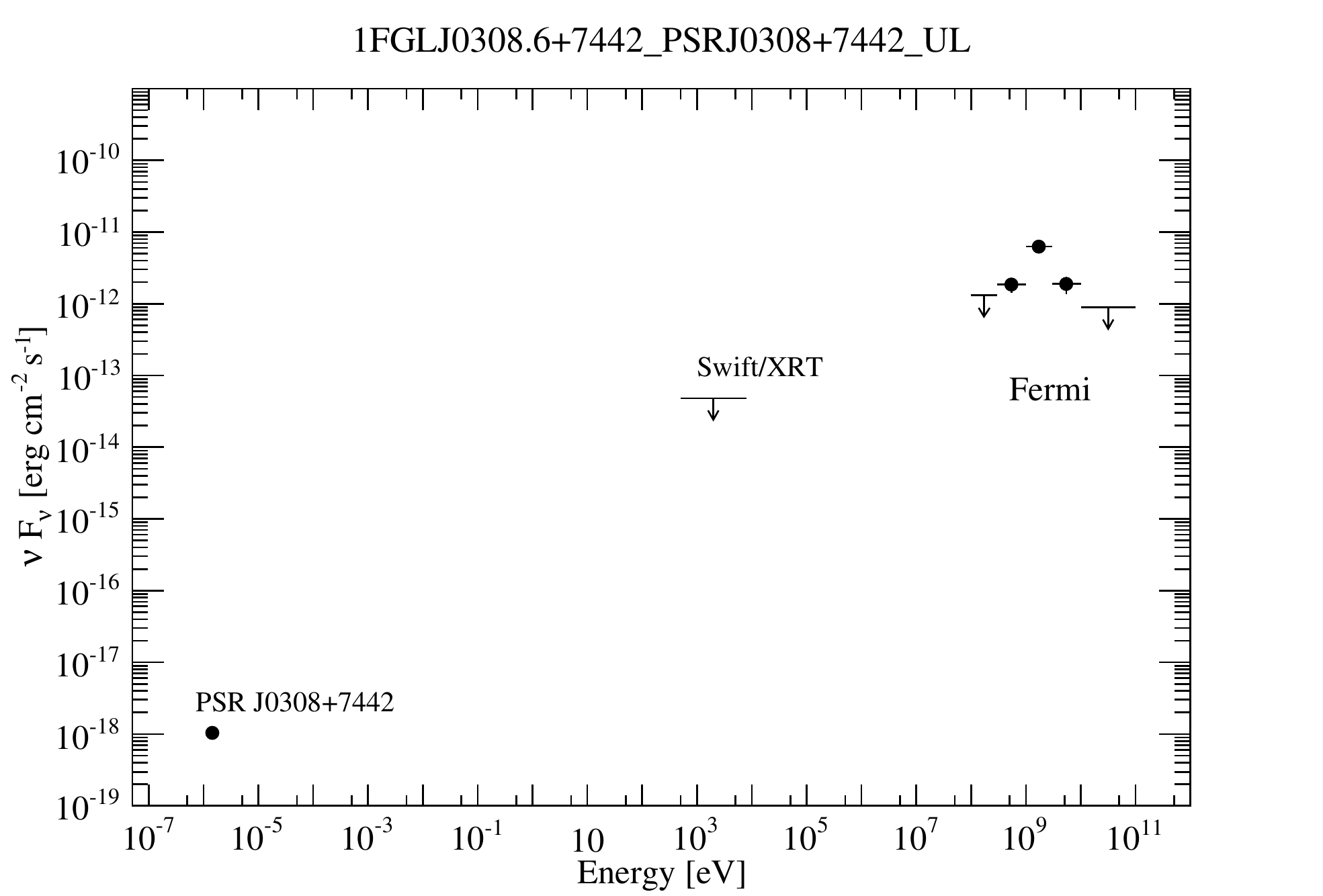}
    \end{center}
  \end{minipage}
  \begin{minipage}{0.32\hsize}
    \begin{center}
      \includegraphics[width=55mm]{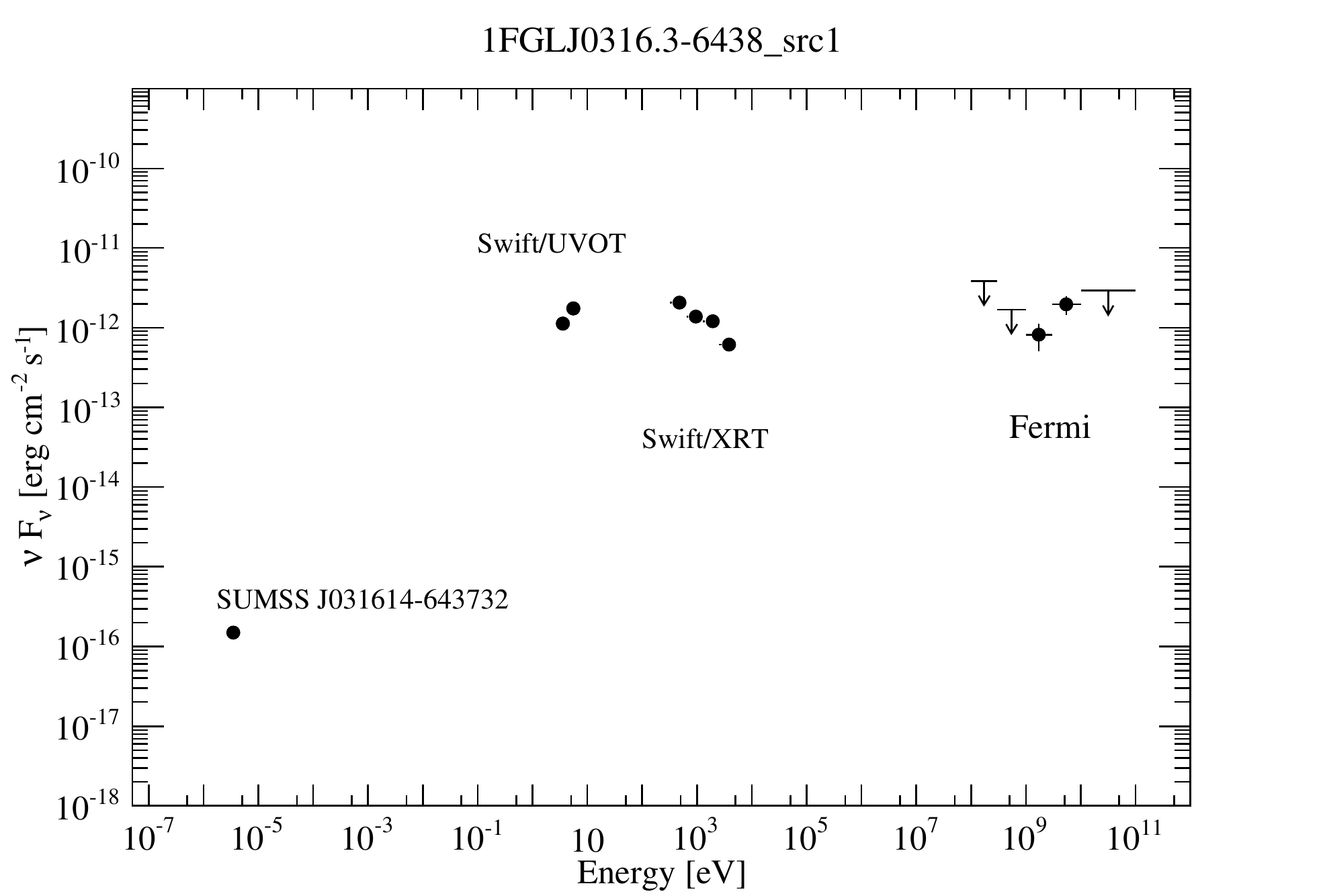}
    \end{center}
  \end{minipage}
  \begin{minipage}{0.32\hsize}
    \begin{center}
      \includegraphics[width=55mm]{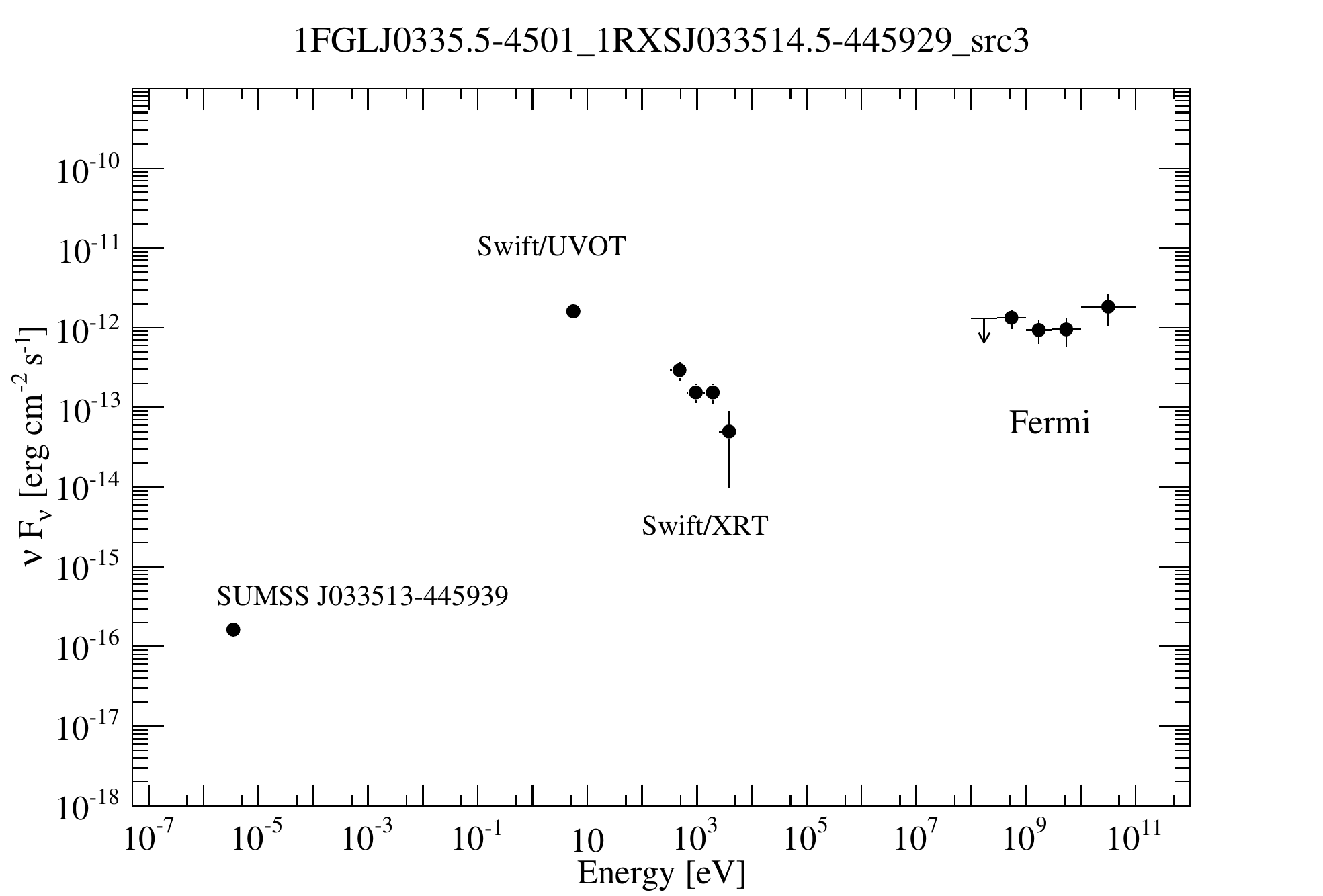}
    \end{center}
  \end{minipage}
  \begin{minipage}{0.32\hsize}
    \begin{center}
      \includegraphics[width=55mm]{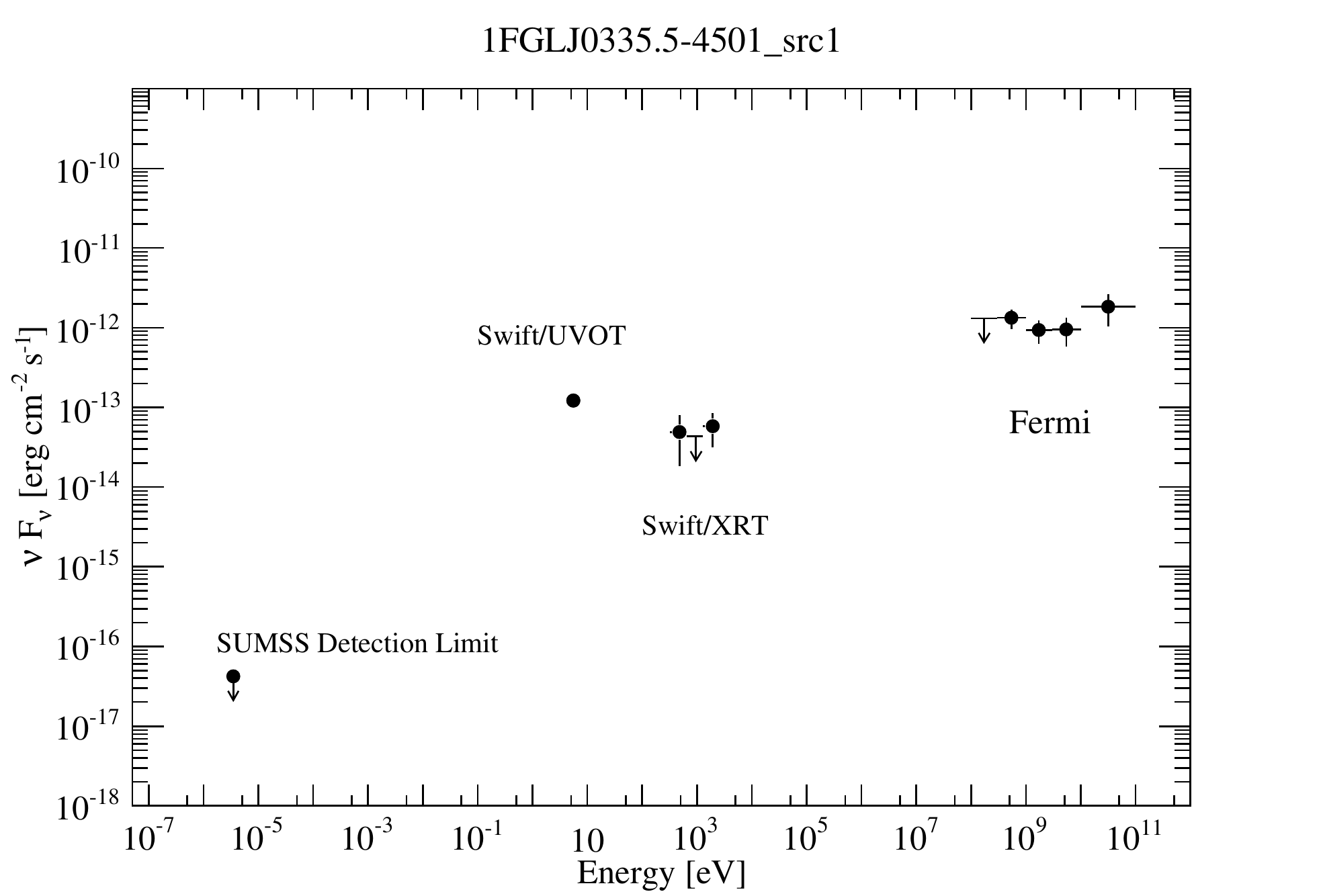}
    \end{center}
  \end{minipage}
  \begin{minipage}{0.32\hsize}
    \begin{center}
      \includegraphics[width=55mm]{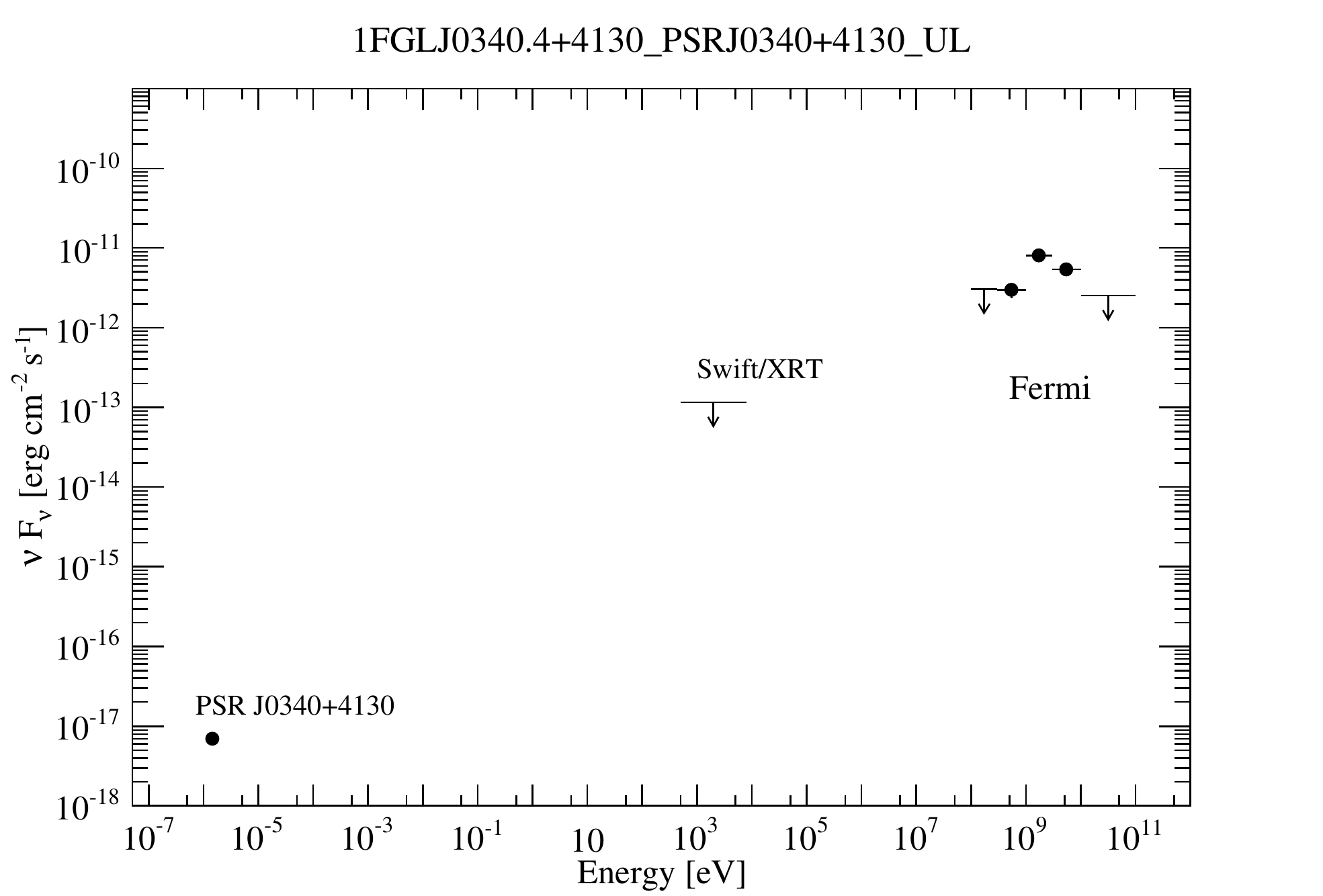}
    \end{center}
  \end{minipage}
  \begin{minipage}{0.32\hsize}
    \begin{center}
      \includegraphics[width=55mm]{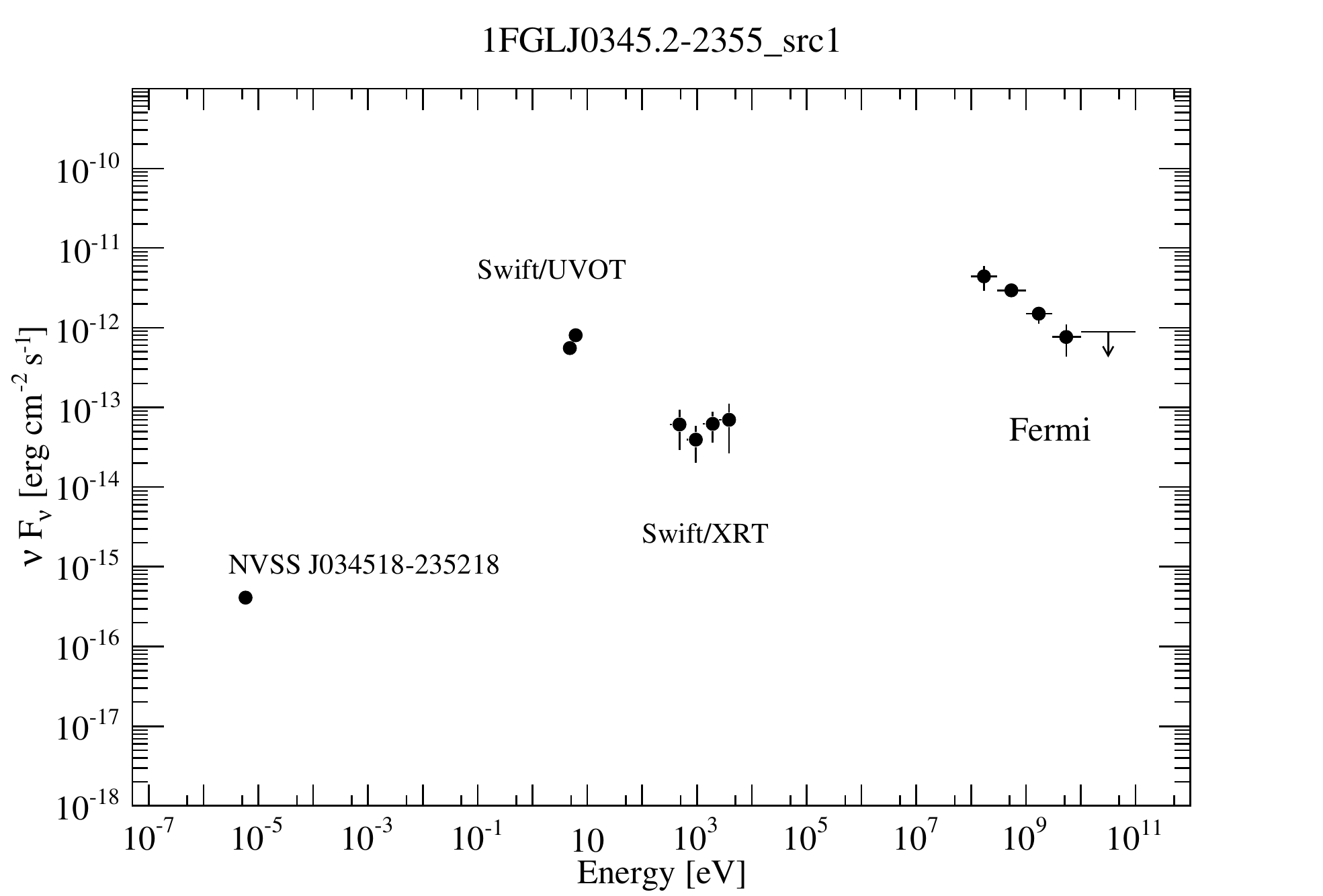}
    \end{center}
  \end{minipage}
  \begin{minipage}{0.32\hsize}
    \begin{center}
      \includegraphics[width=55mm]{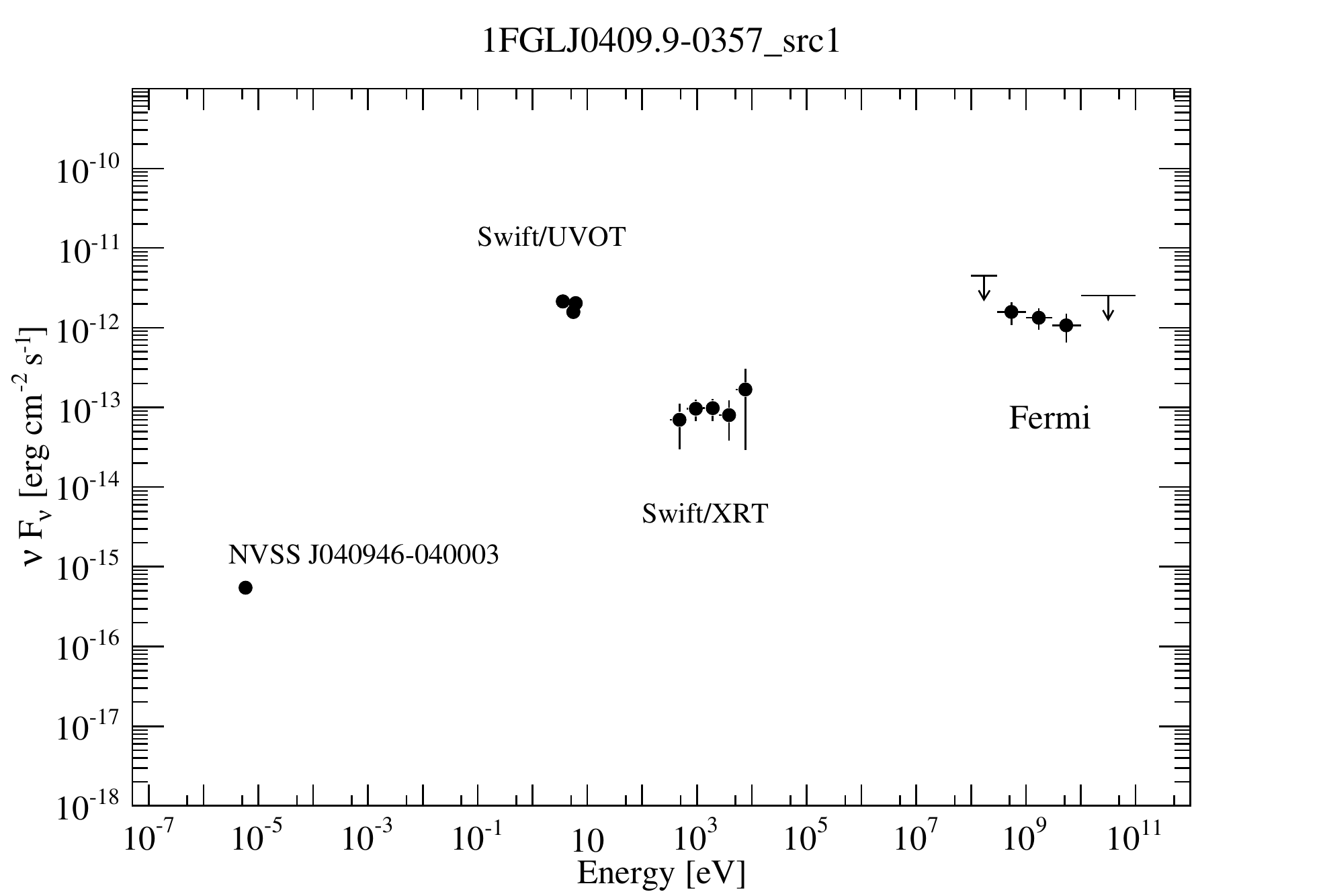}
    \end{center}
  \end{minipage}
  \begin{minipage}{0.32\hsize}
    \begin{center}
      \includegraphics[width=55mm]{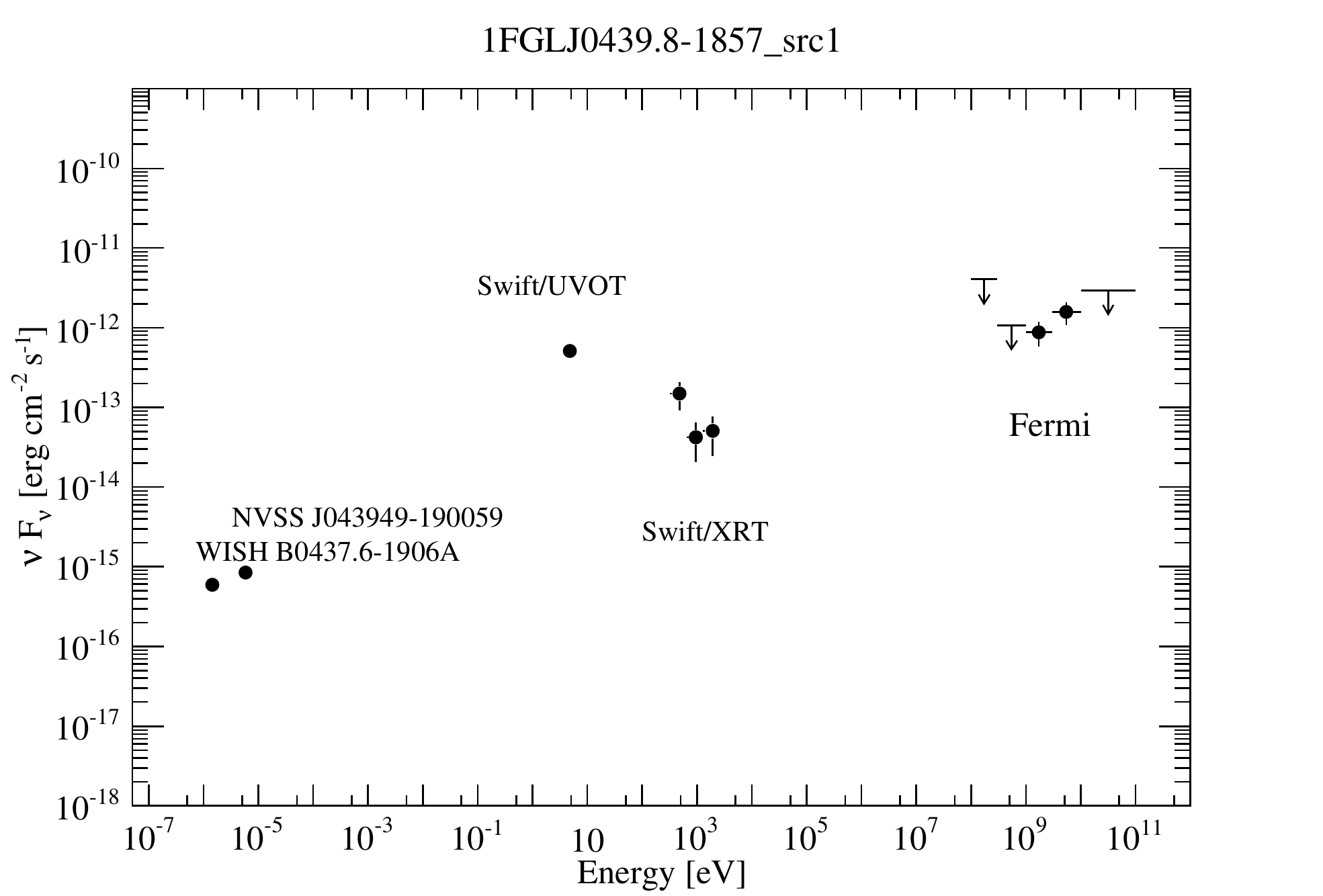}
    \end{center}
  \end{minipage}
 \end{center}
\end{figure}
\clearpage
\begin{figure}[m]
 \begin{center}
  \begin{minipage}{0.32\hsize}
    \begin{center}
      \includegraphics[width=55mm]{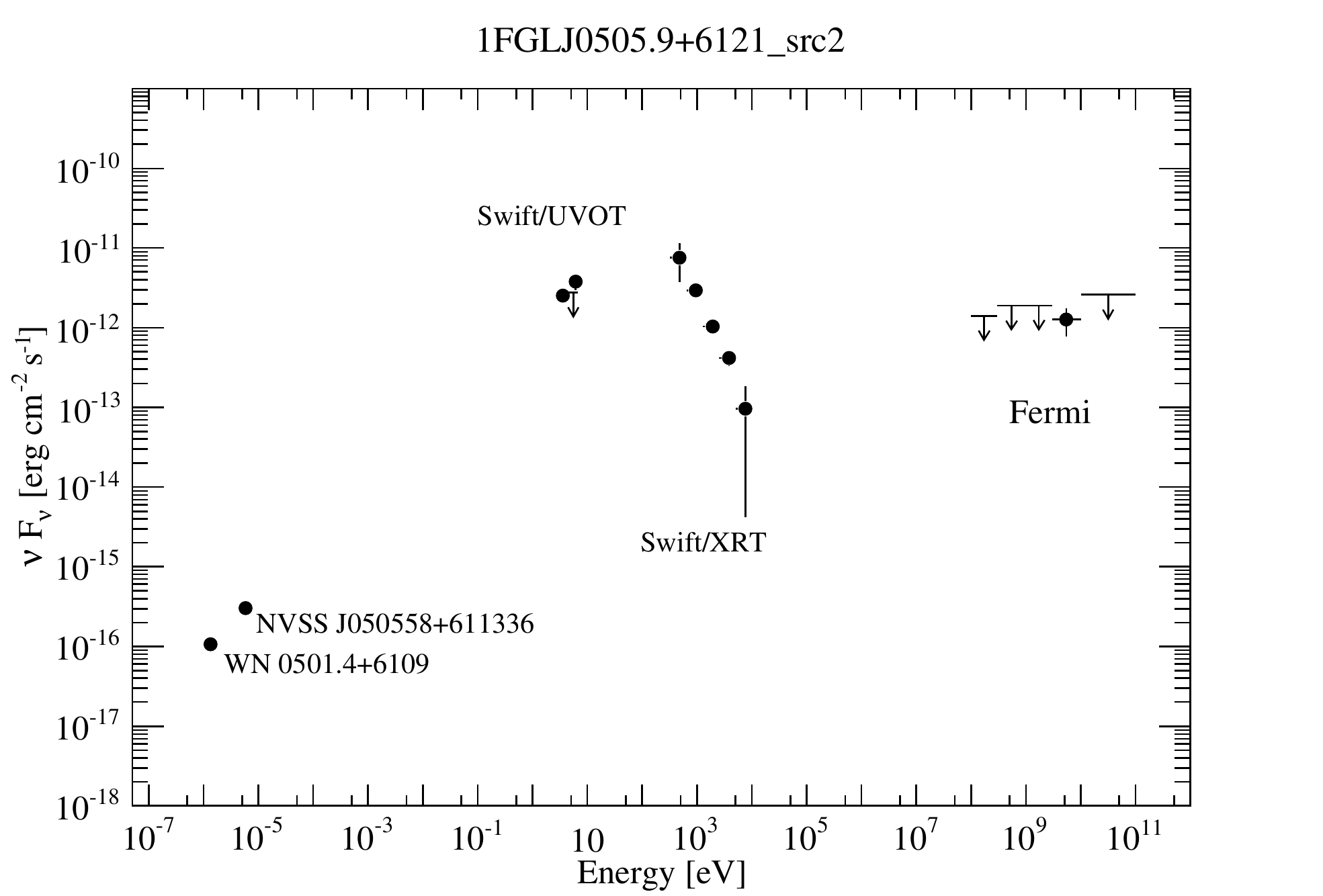}
    \end{center}
  \end{minipage}
  \begin{minipage}{0.32\hsize}
    \begin{center}
      \includegraphics[width=55mm]{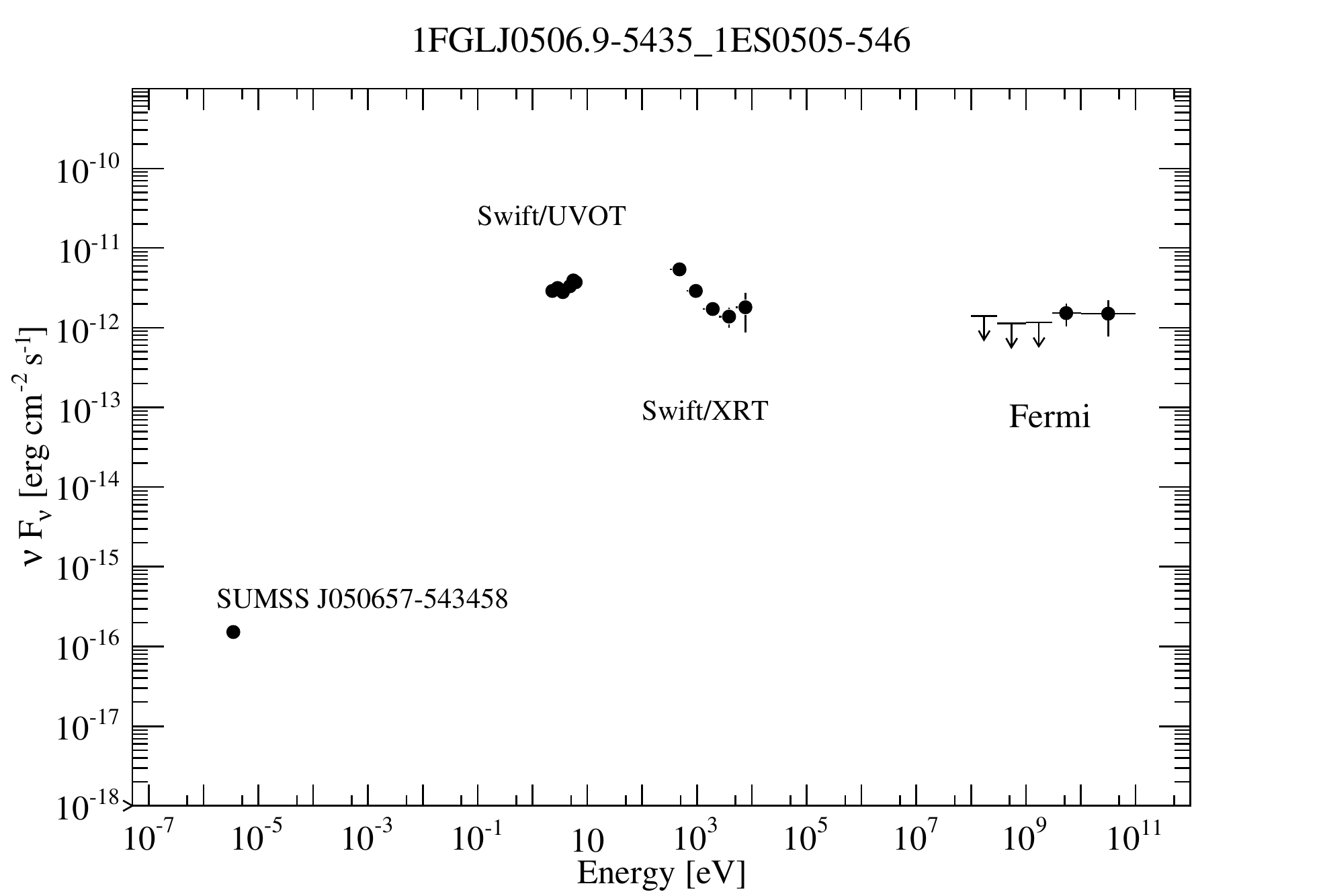}
    \end{center}
  \end{minipage}
  \begin{minipage}{0.32\hsize}
    \begin{center}
      \includegraphics[width=55mm]{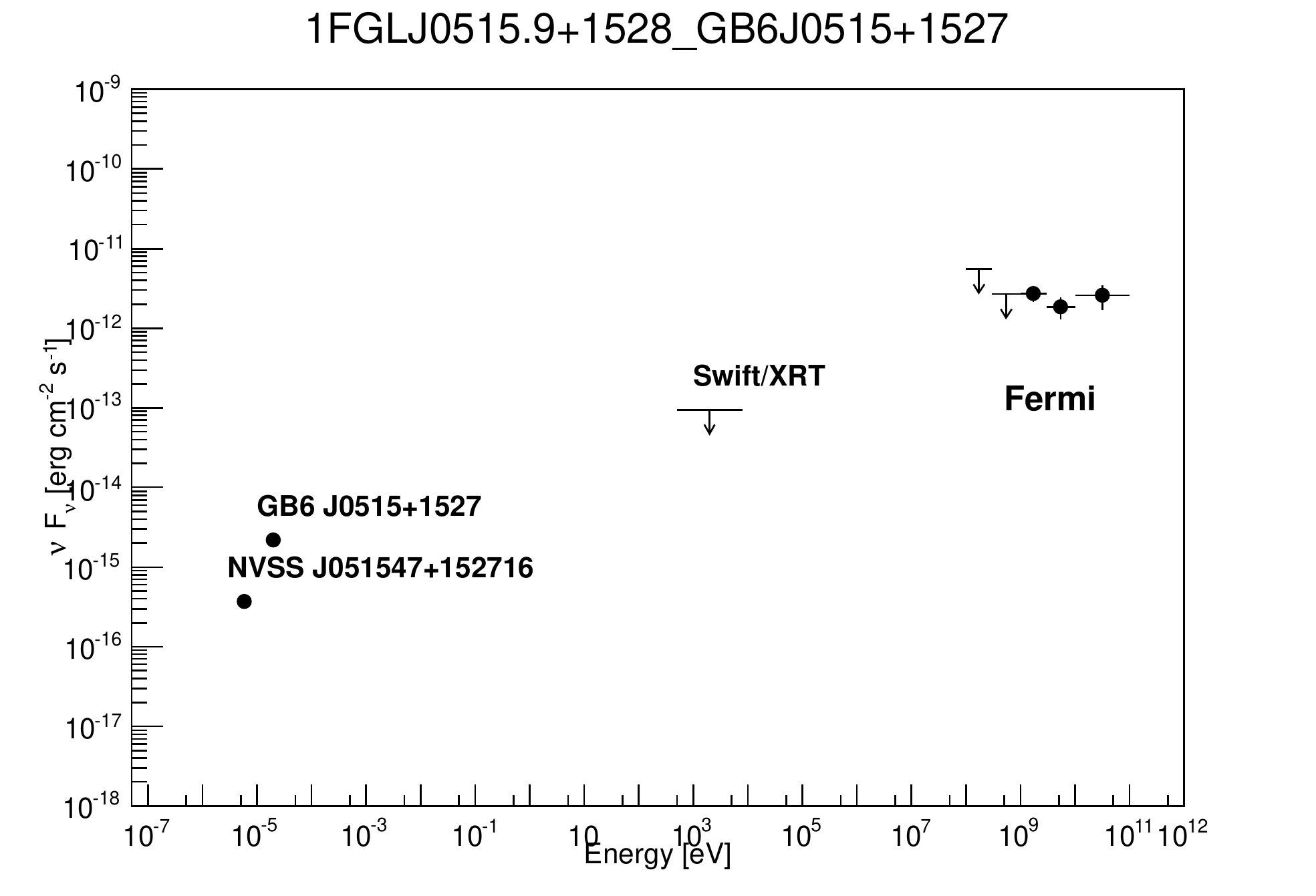}
    \end{center}
  \end{minipage}
  \begin{minipage}{0.32\hsize}
    \begin{center}
      \includegraphics[width=55mm]{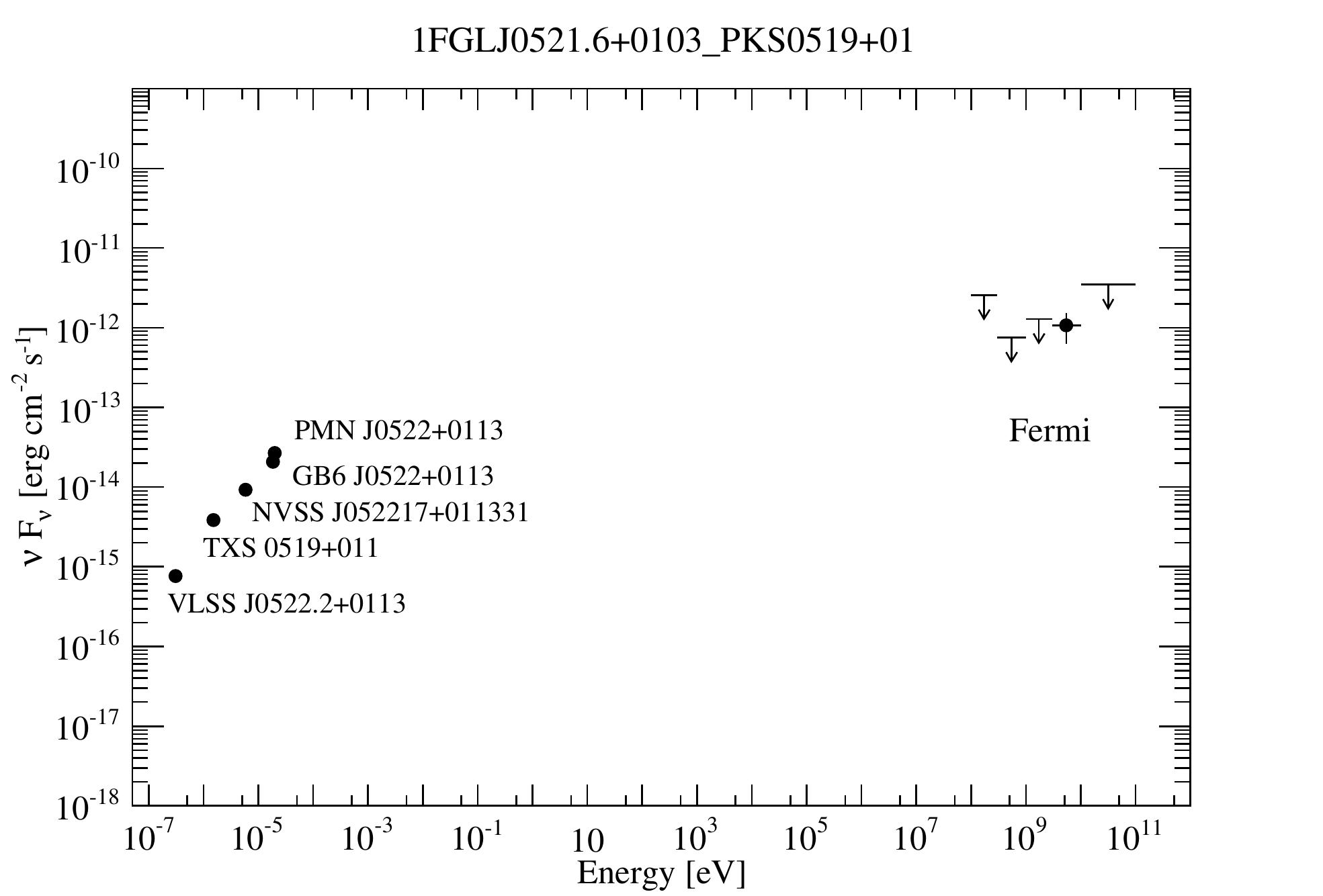}
    \end{center}
  \end{minipage}
  \begin{minipage}{0.32\hsize}
    \begin{center}
      \includegraphics[width=55mm]{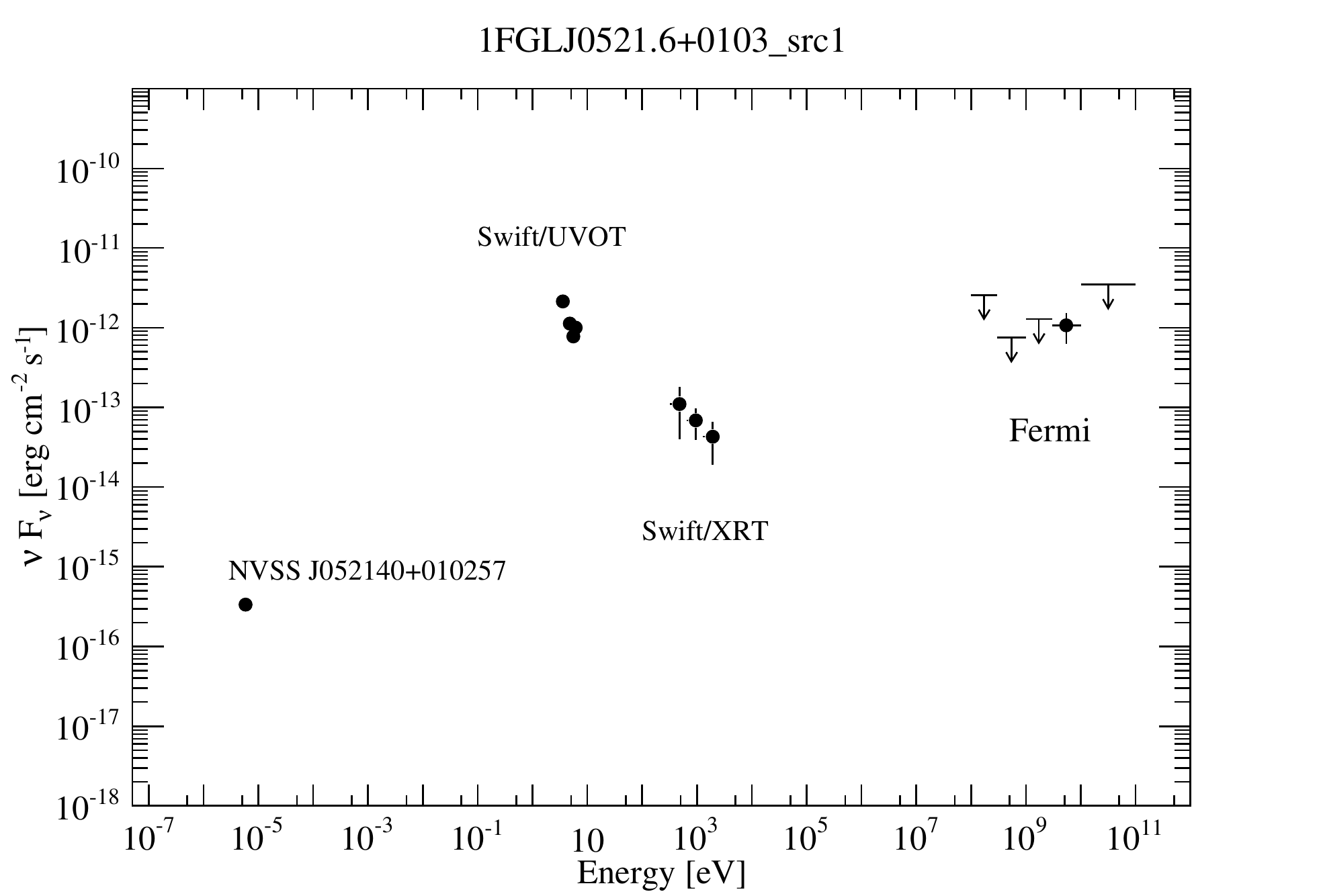}
    \end{center}
  \end{minipage}
  \begin{minipage}{0.32\hsize}
    \begin{center}
      \includegraphics[width=55mm]{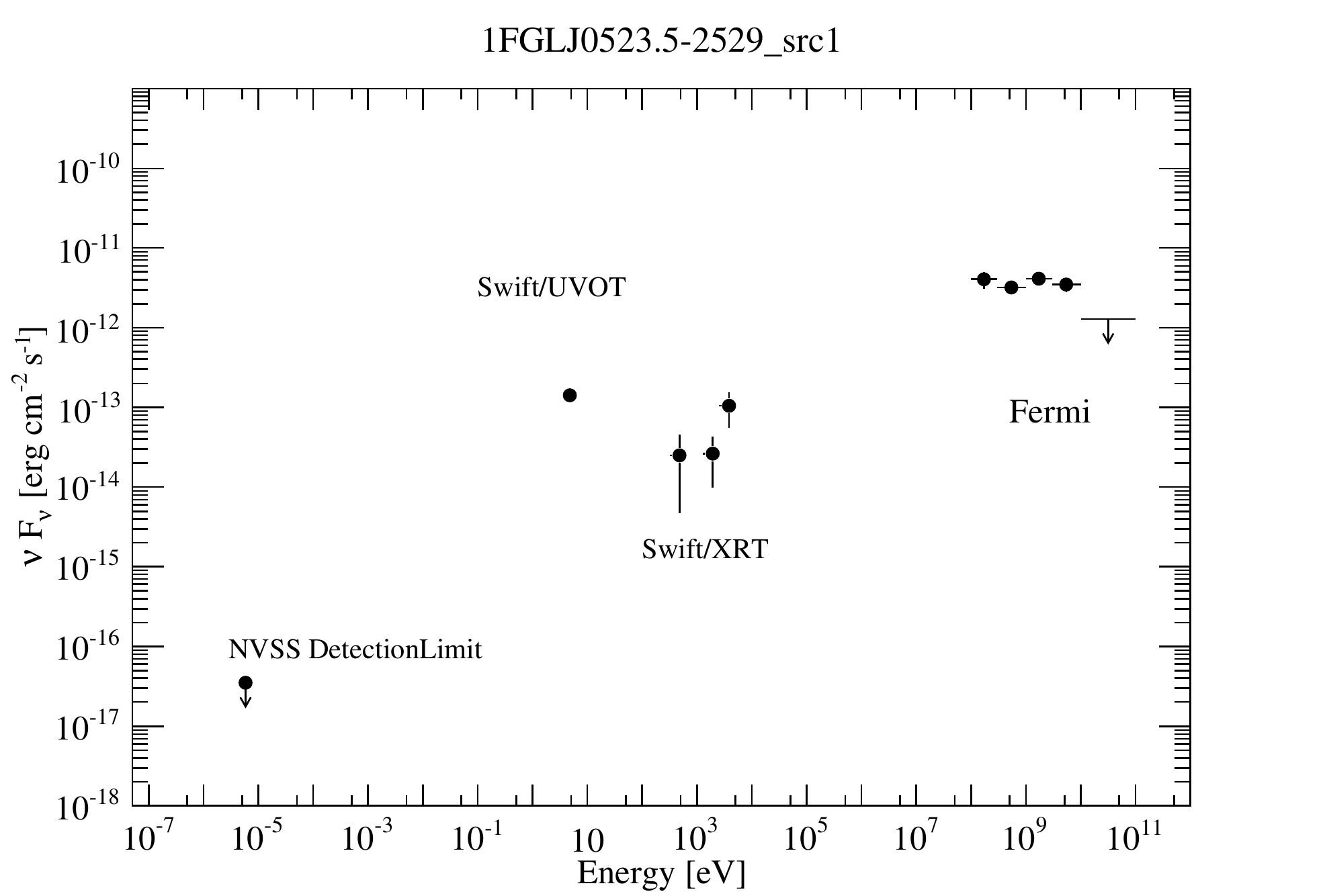}
    \end{center}
  \end{minipage}
  \begin{minipage}{0.32\hsize}
    \begin{center}
      \includegraphics[width=55mm]{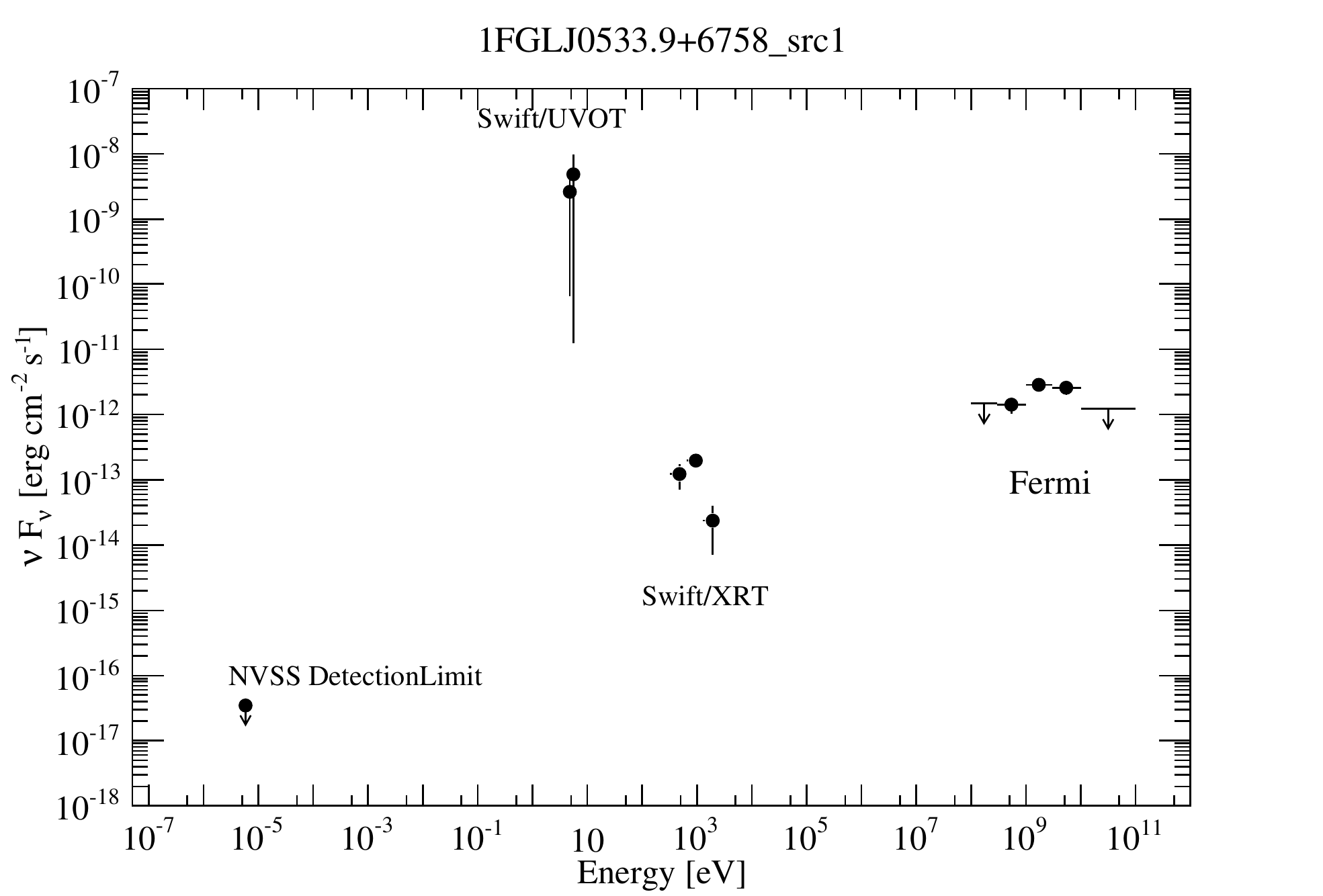}
    \end{center}
  \end{minipage}
  \begin{minipage}{0.32\hsize}
    \begin{center}
      \includegraphics[width=55mm]{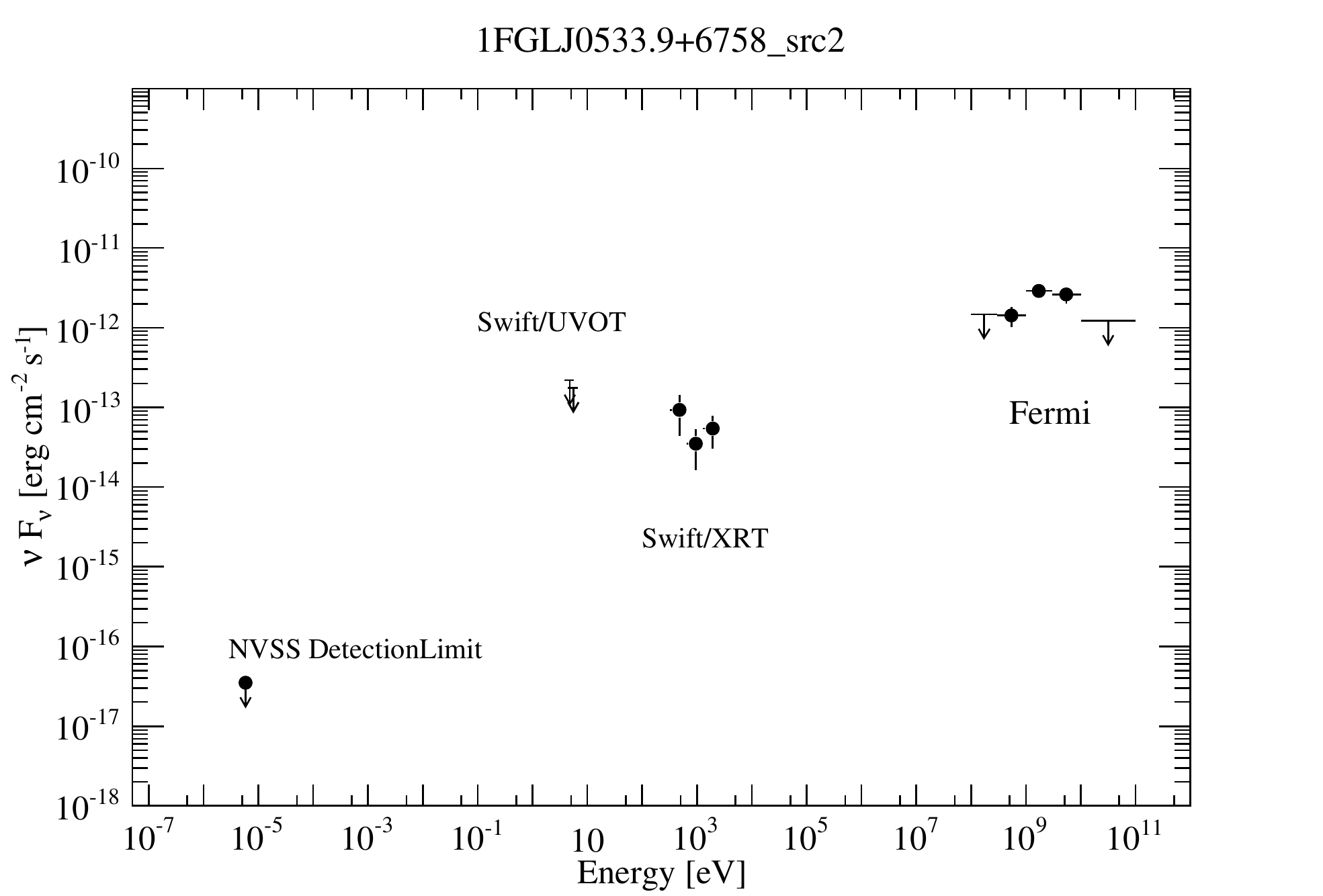}
    \end{center}
  \end{minipage}
  \begin{minipage}{0.32\hsize}
    \begin{center}
      \includegraphics[width=55mm]{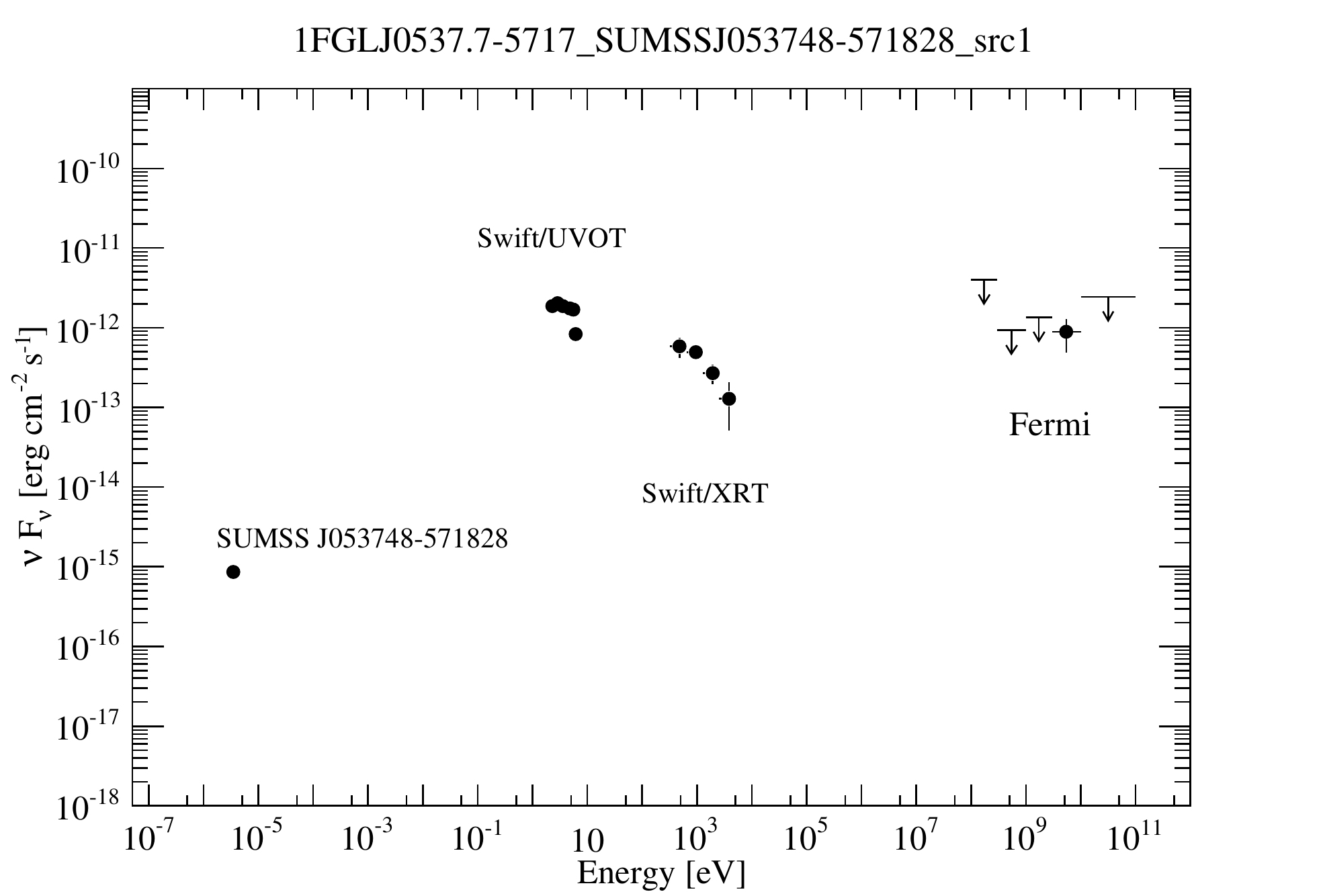}
    \end{center}
  \end{minipage}
  \begin{minipage}{0.32\hsize}
    \begin{center}
      \includegraphics[width=55mm]{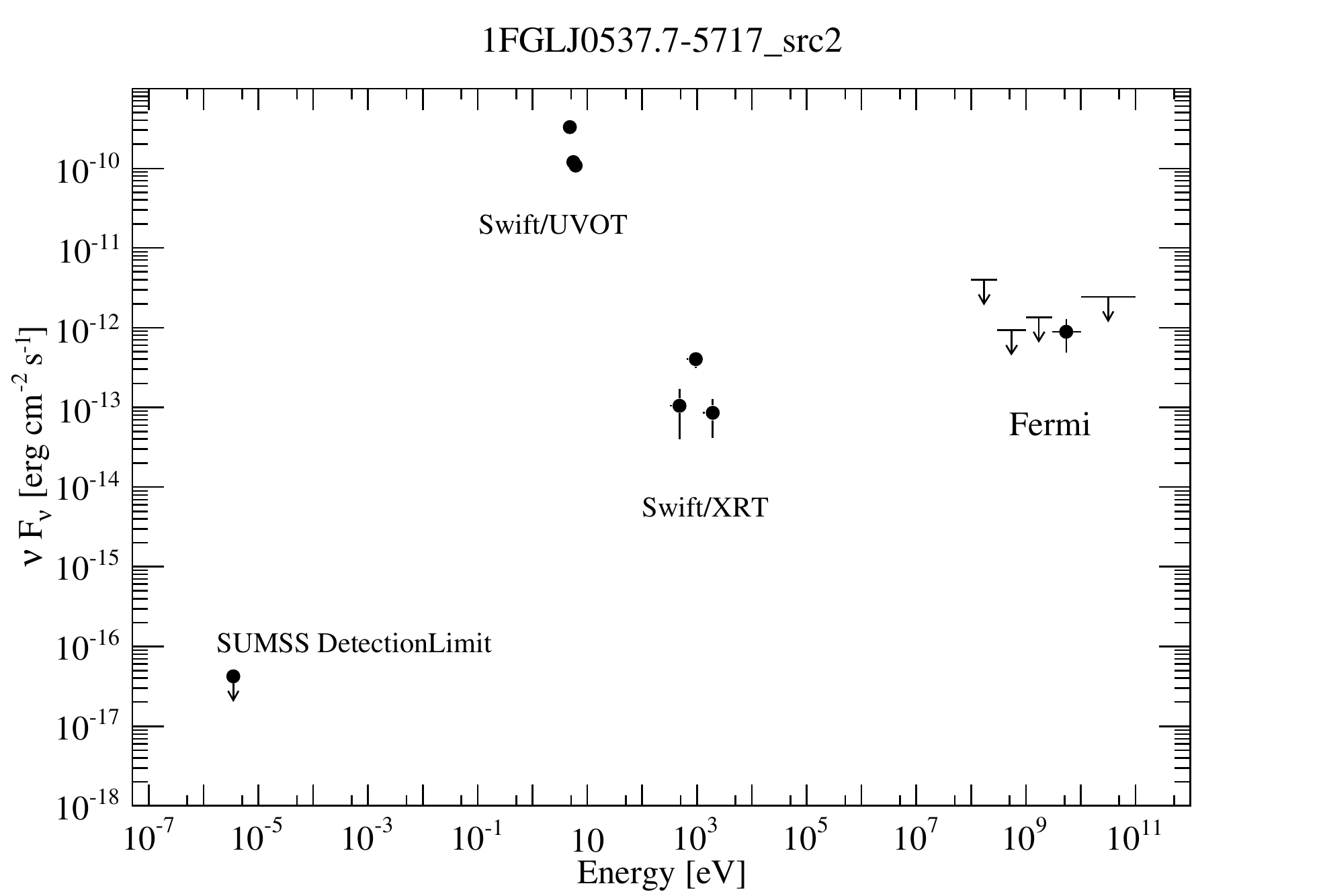}
    \end{center}
  \end{minipage}
  \begin{minipage}{0.32\hsize}
    \begin{center}
      \includegraphics[width=55mm]{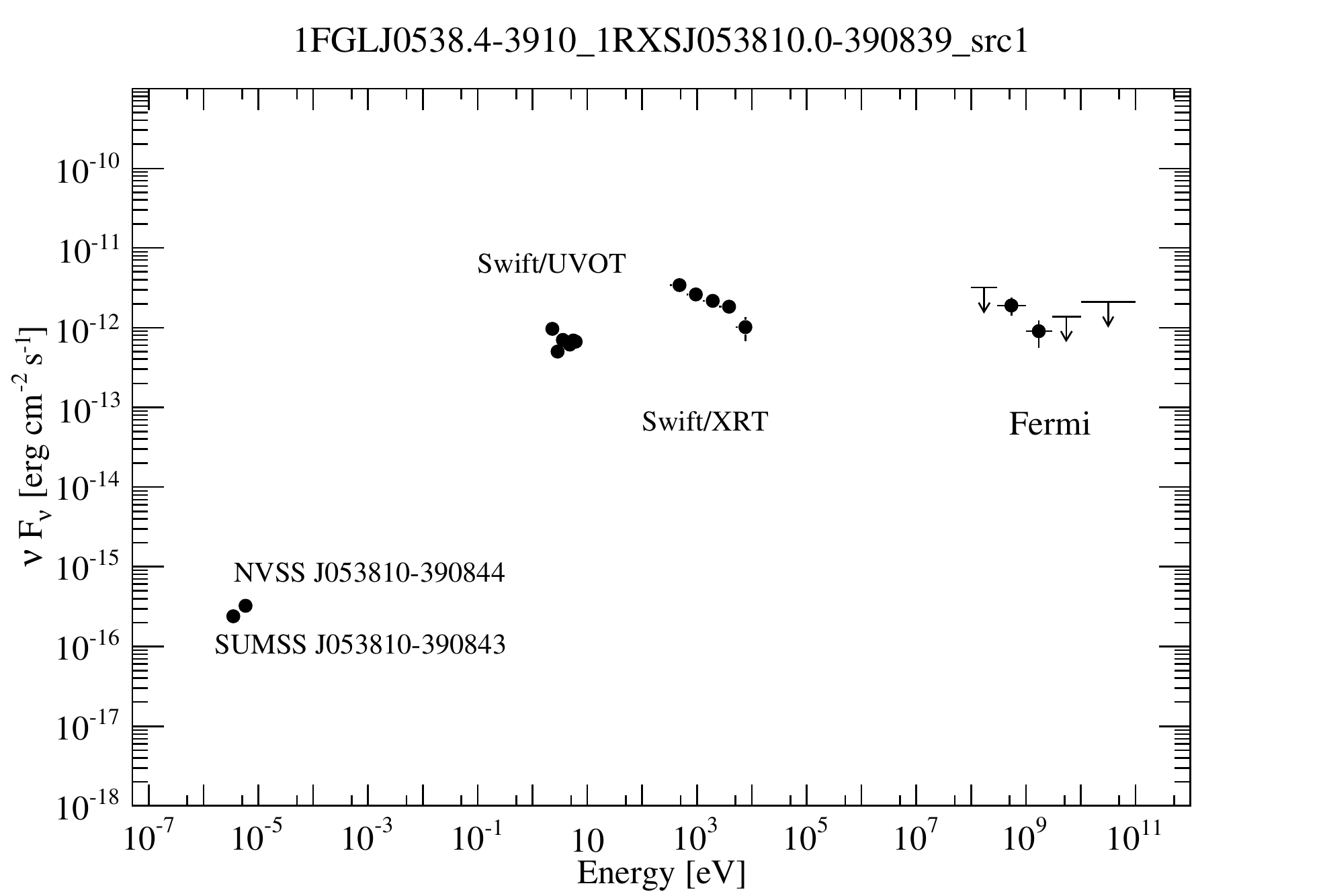}
    \end{center}
  \end{minipage}
  \begin{minipage}{0.32\hsize}
    \begin{center}
      \includegraphics[width=55mm]{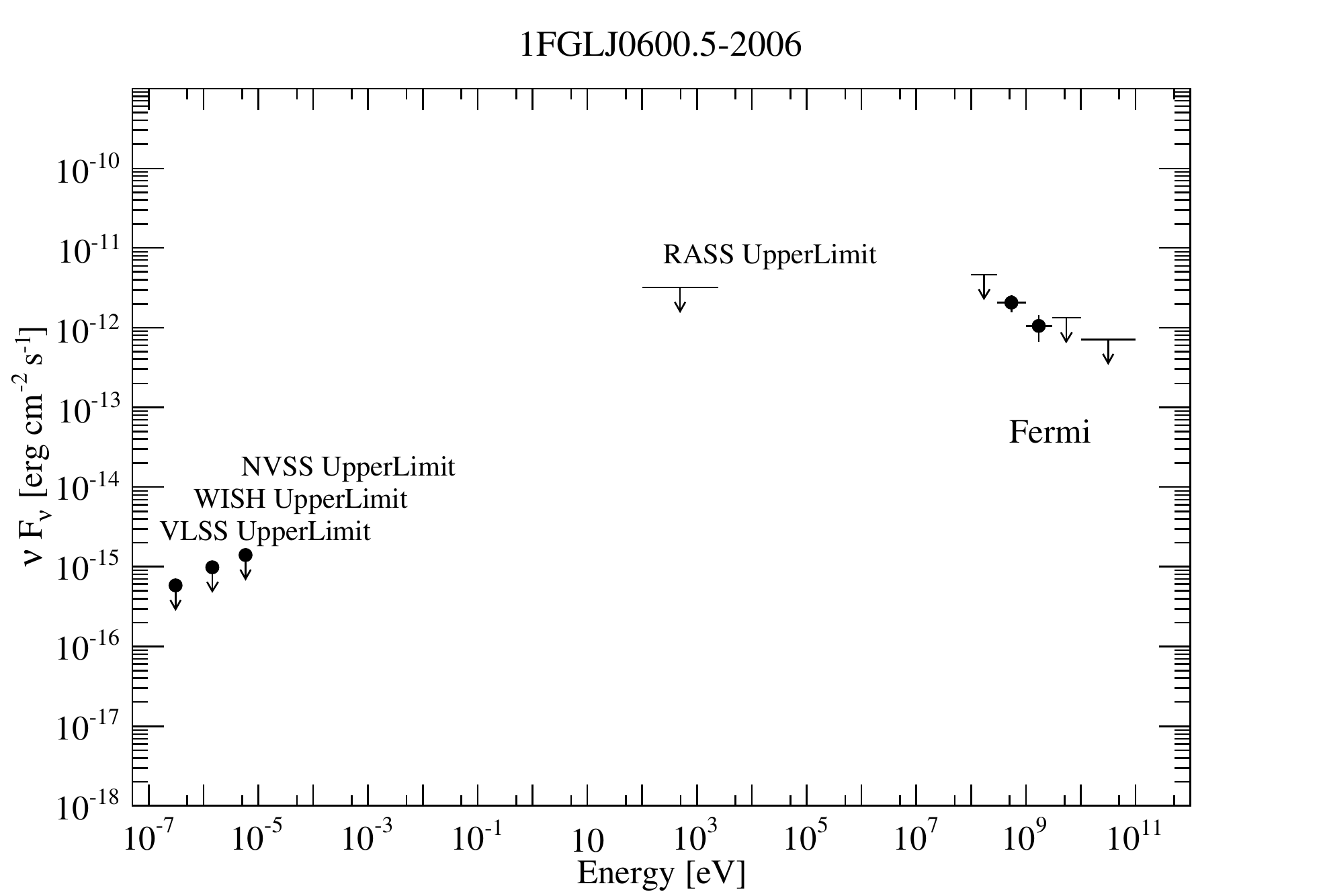}
    \end{center}
  \end{minipage}
  \begin{minipage}{0.32\hsize}
    \begin{center}
      \includegraphics[width=55mm]{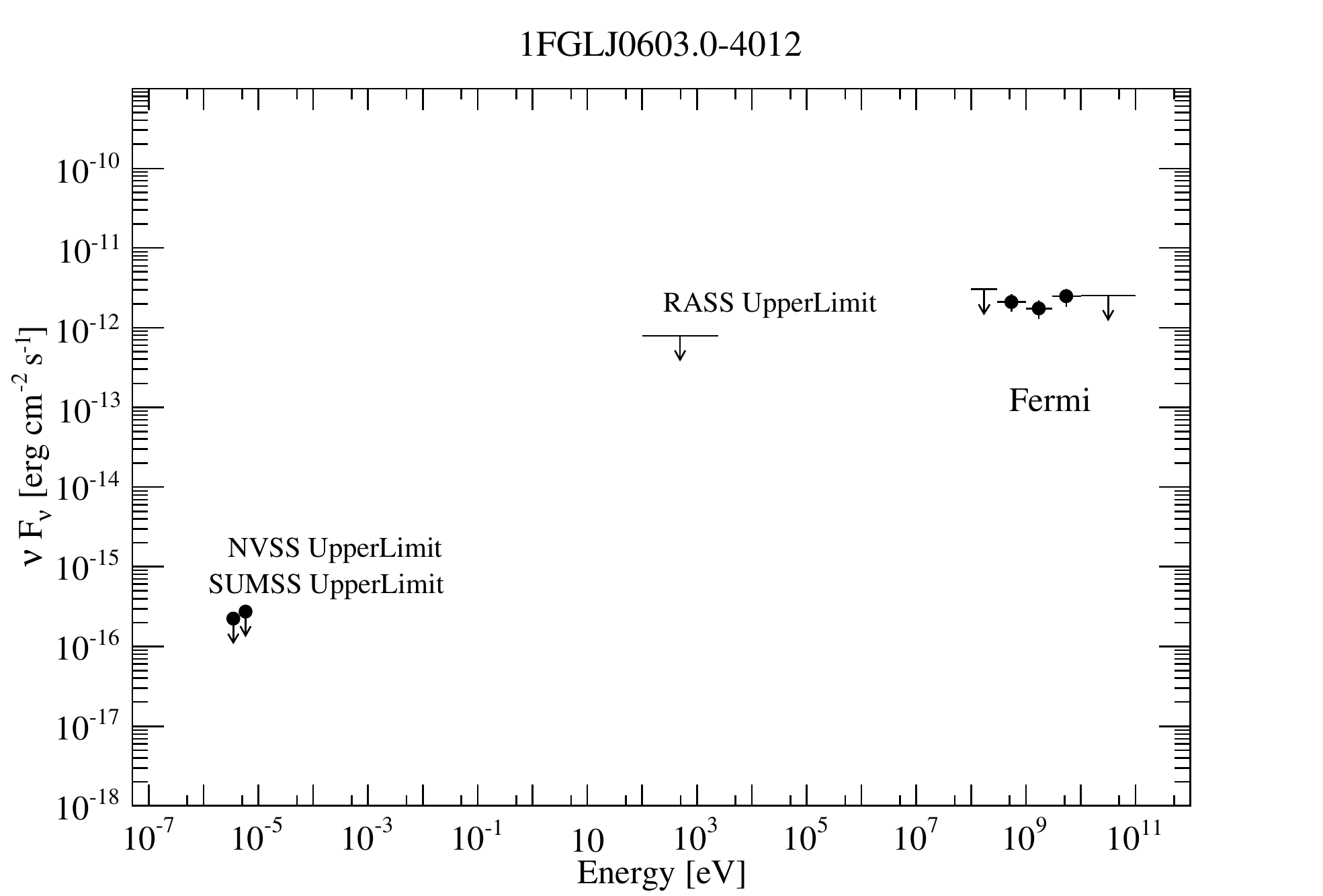}
    \end{center}
  \end{minipage}
  \begin{minipage}{0.32\hsize}
    \begin{center}
      \includegraphics[width=55mm]{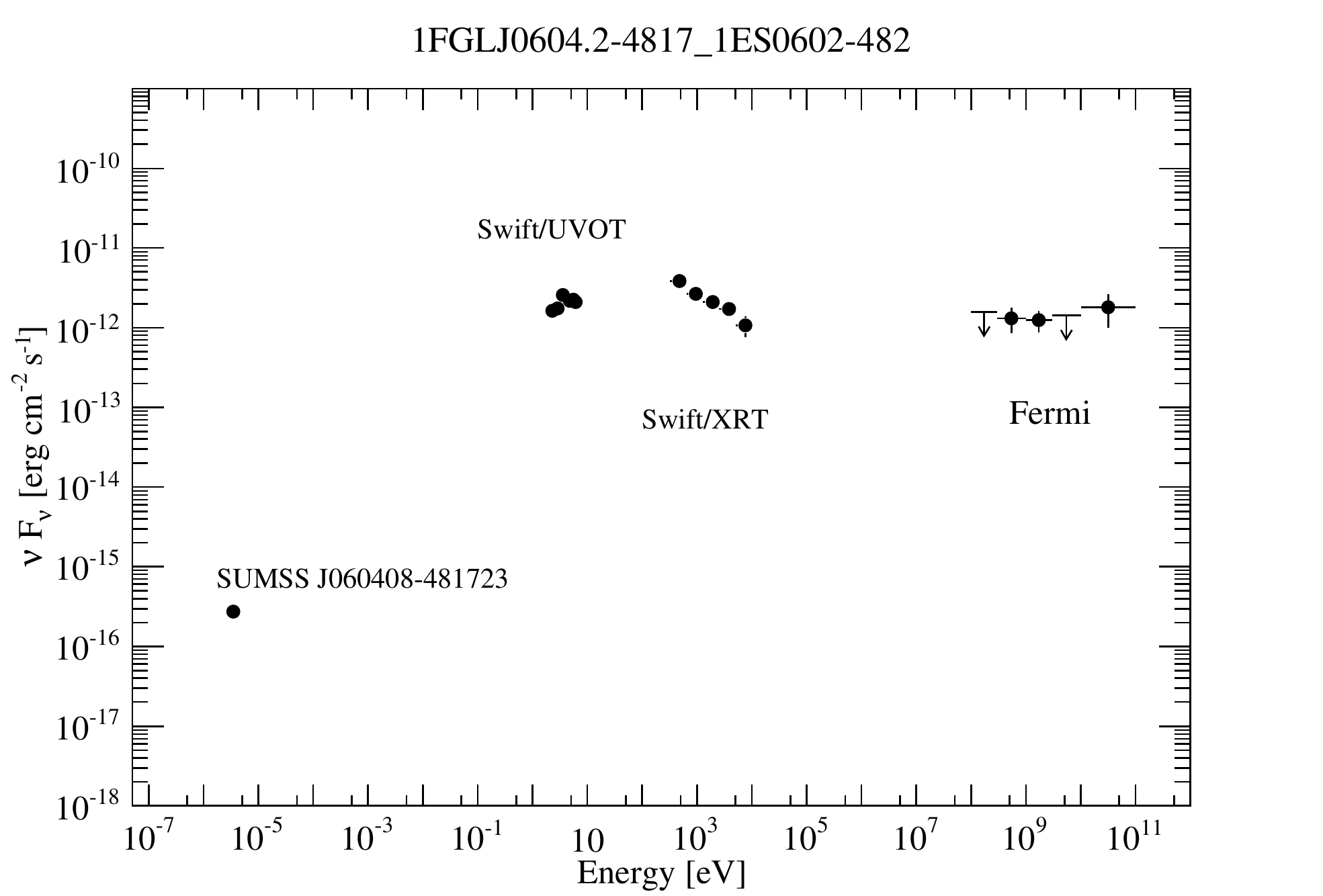}
    \end{center}
  \end{minipage}
  \begin{minipage}{0.32\hsize}
    \begin{center}
      \includegraphics[width=55mm]{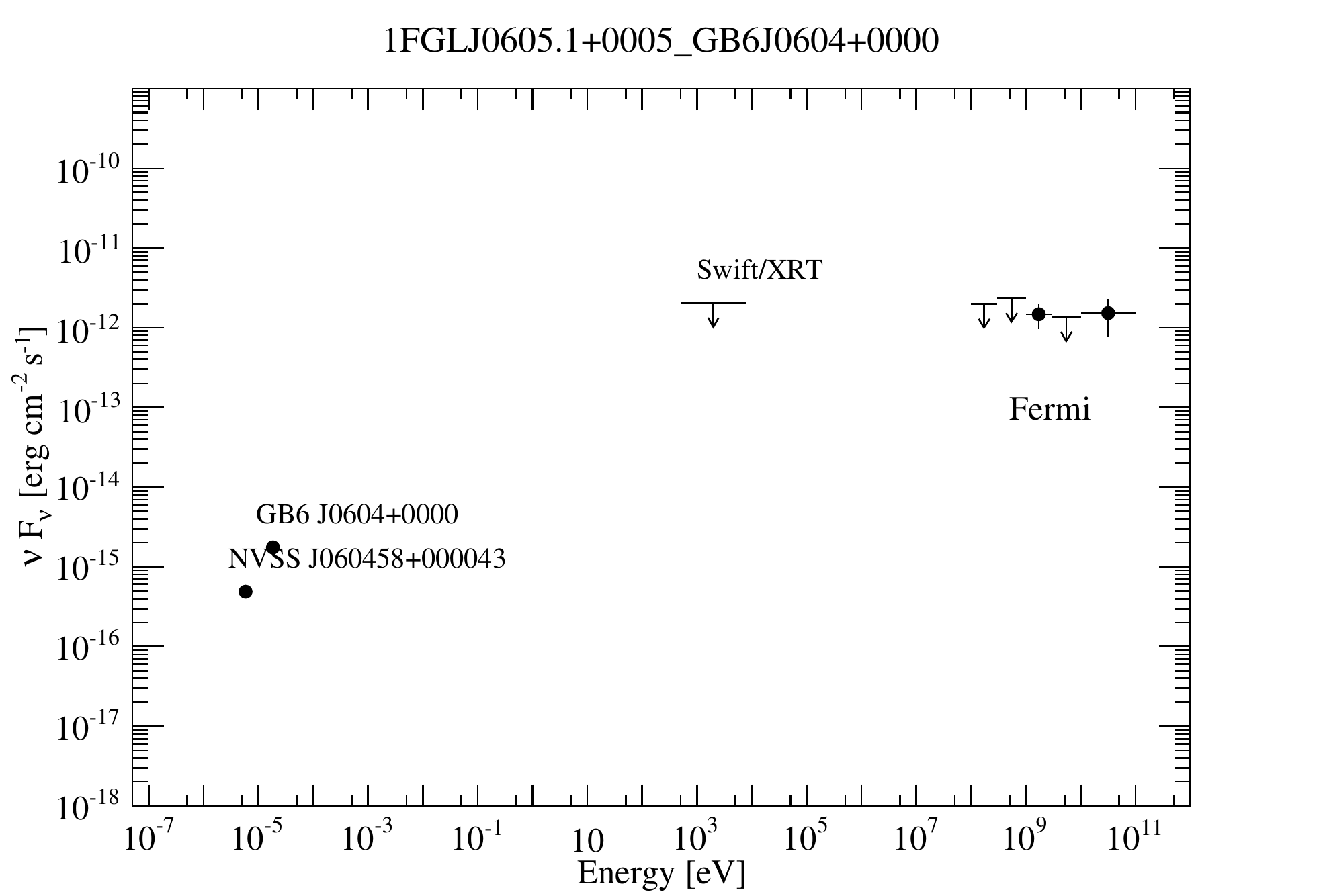}
    \end{center}
  \end{minipage}
  \begin{minipage}{0.32\hsize}
    \begin{center}
      \includegraphics[width=55mm]{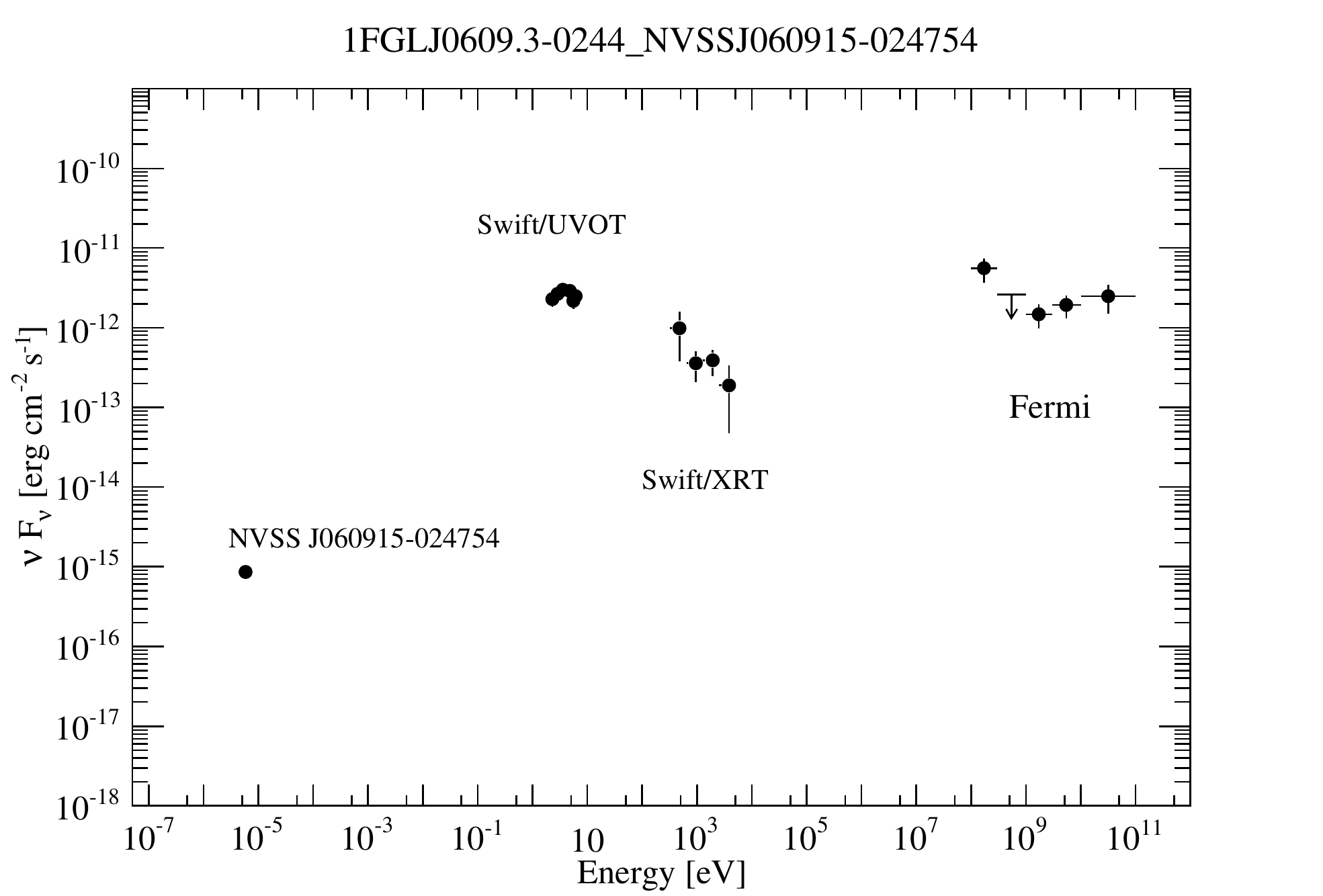}
    \end{center}
  \end{minipage}
  \begin{minipage}{0.32\hsize}
    \begin{center}
      \includegraphics[width=55mm]{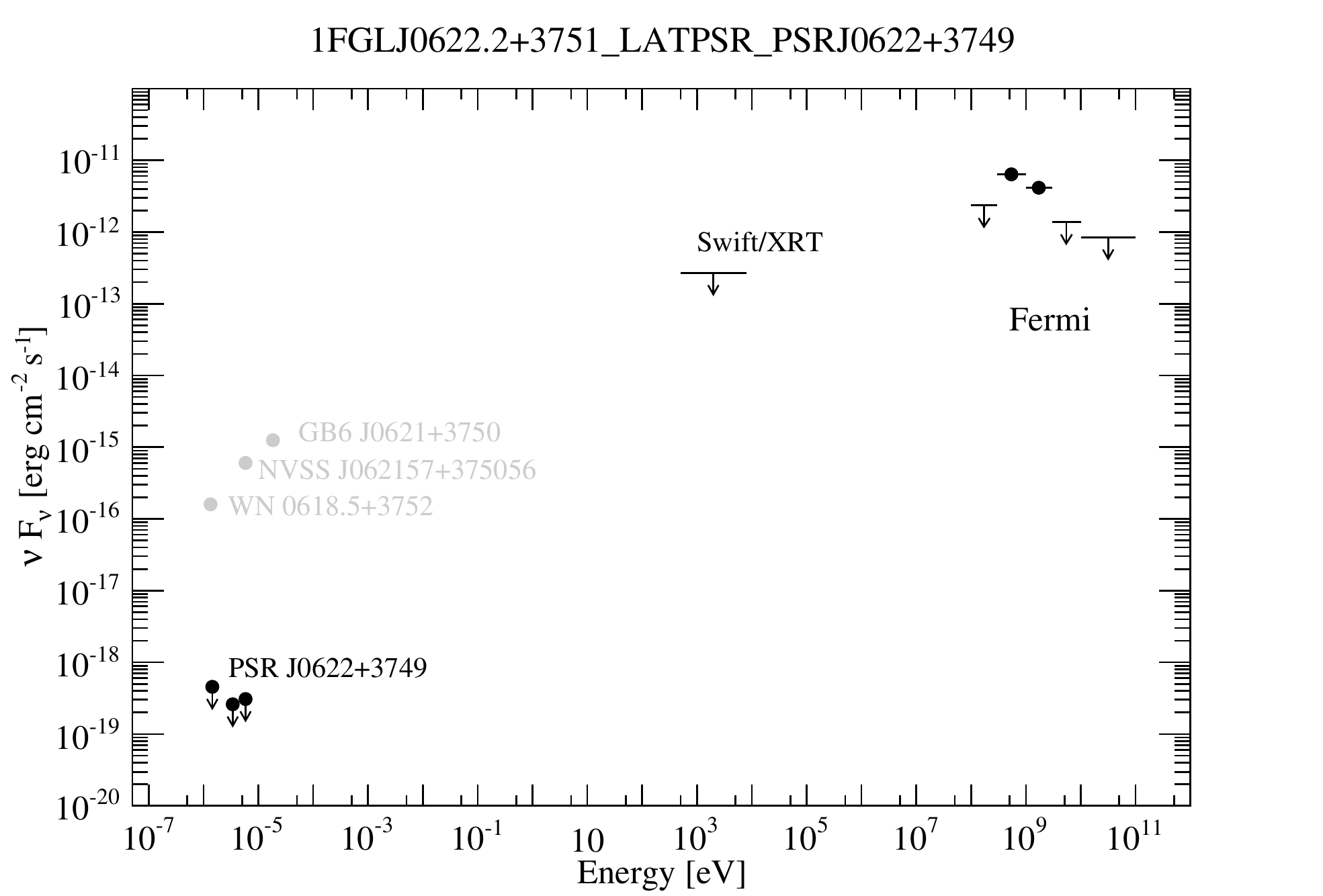}
    \end{center}
  \end{minipage}
  \begin{minipage}{0.32\hsize}
    \begin{center}
      \includegraphics[width=55mm]{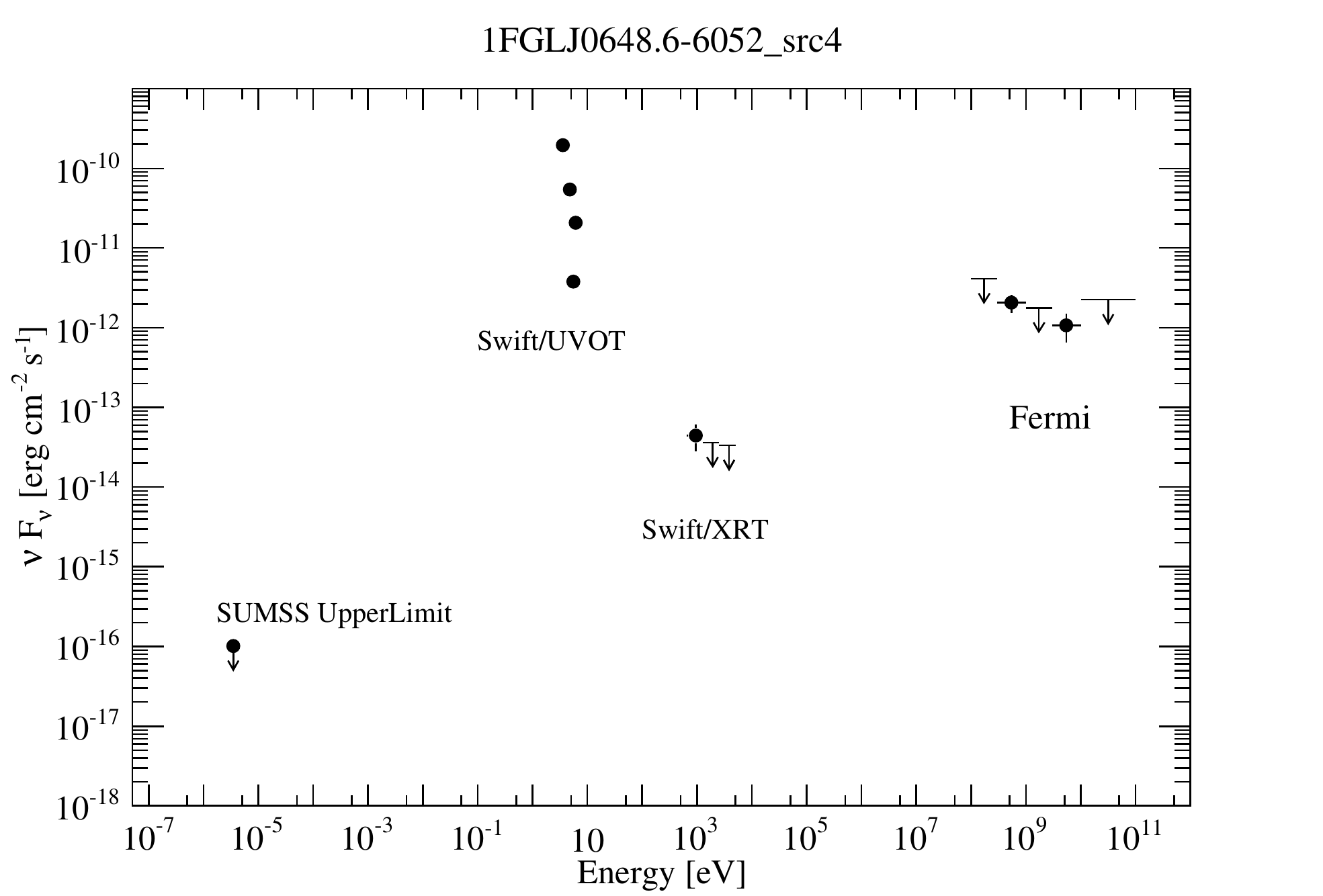}
    \end{center}
  \end{minipage}
 \end{center}
\end{figure}
\clearpage
\begin{figure}[m]
 \begin{center}
  \begin{minipage}{0.32\hsize}
    \begin{center}
      \includegraphics[width=55mm]{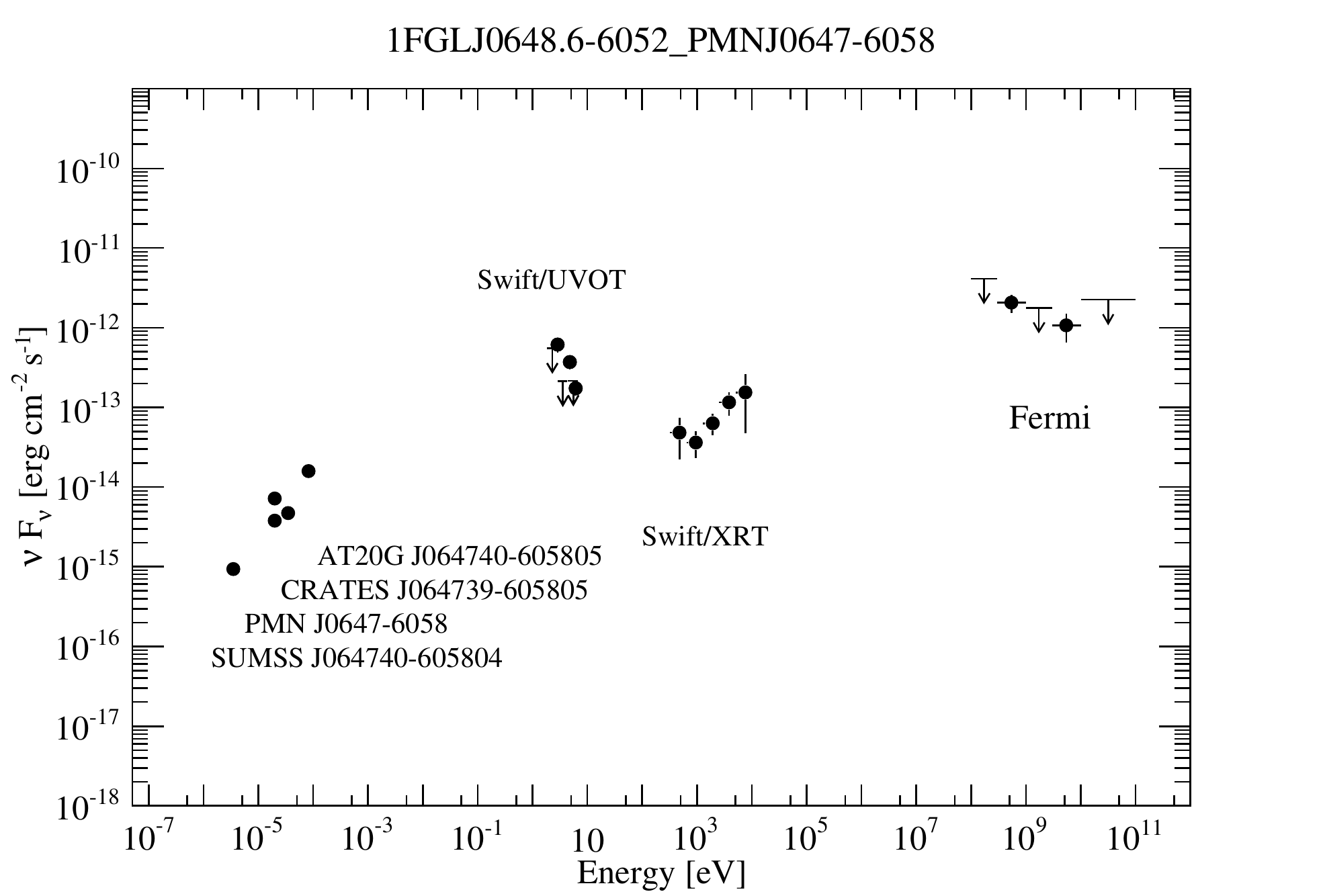}
    \end{center}
  \end{minipage}
  \begin{minipage}{0.32\hsize}
    \begin{center}
      \includegraphics[width=55mm]{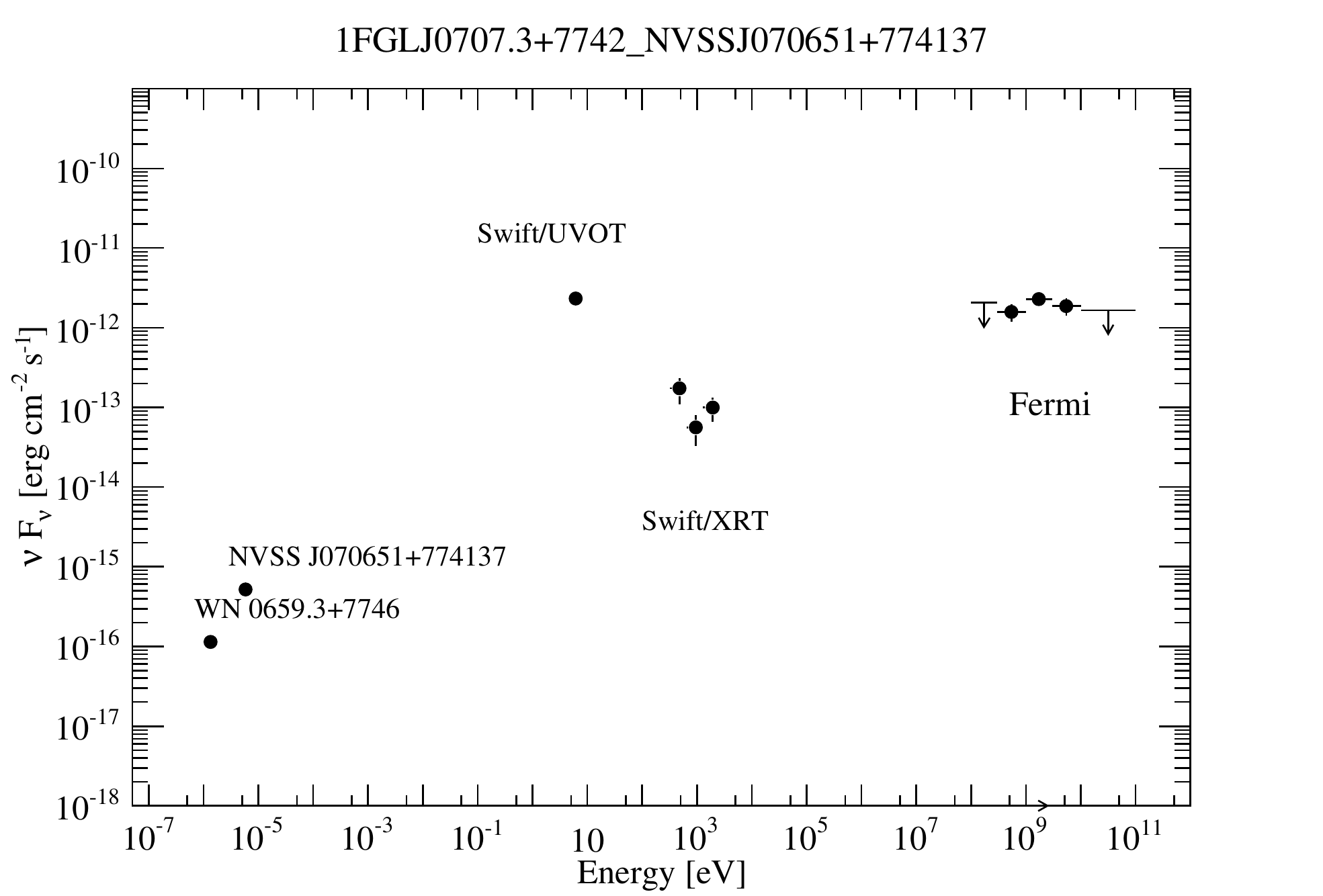}
    \end{center}
  \end{minipage}
  \begin{minipage}{0.32\hsize}
    \begin{center}
      \includegraphics[width=55mm]{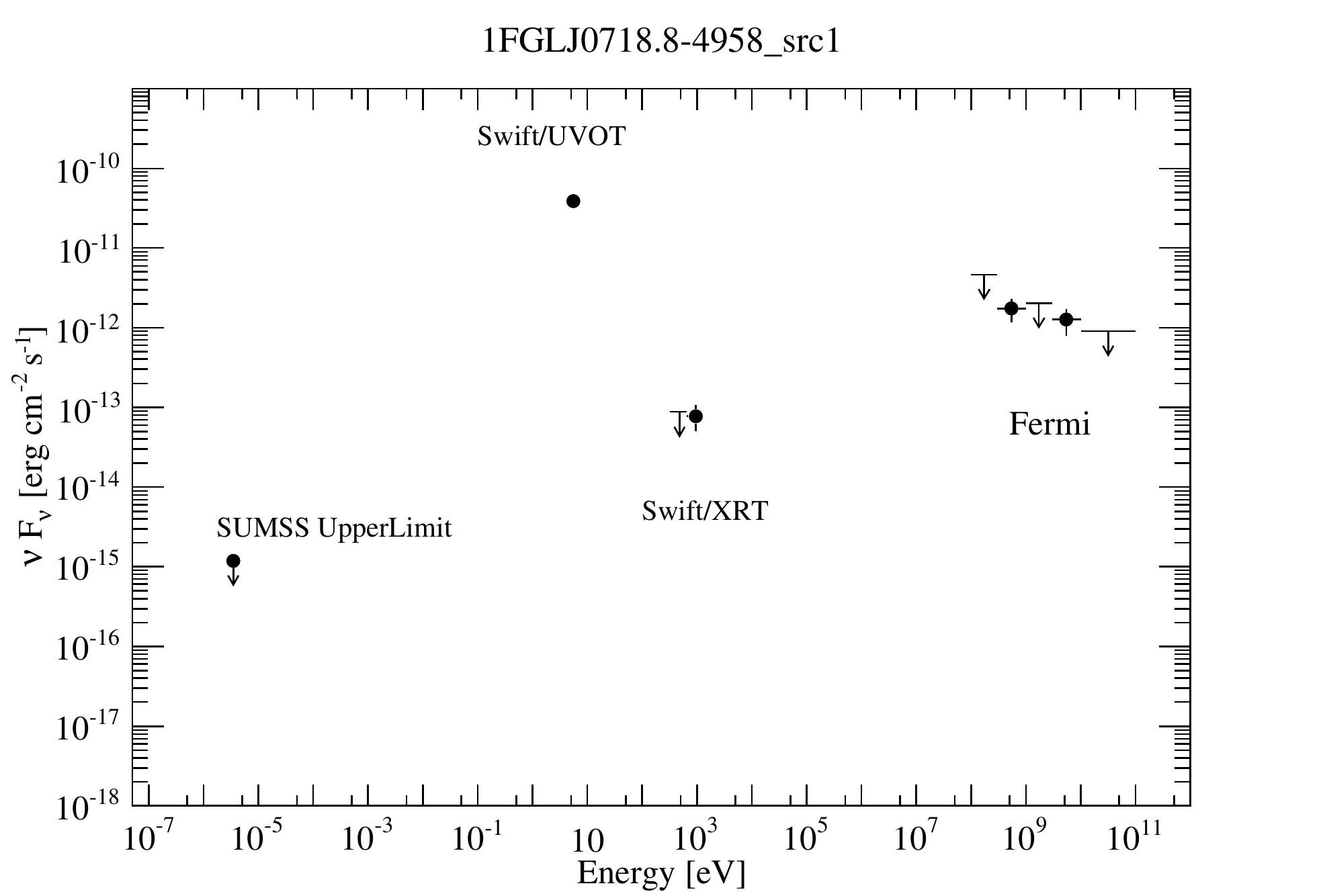}
    \end{center}
  \end{minipage}
  \begin{minipage}{0.32\hsize}
    \begin{center}
      \includegraphics[width=55mm]{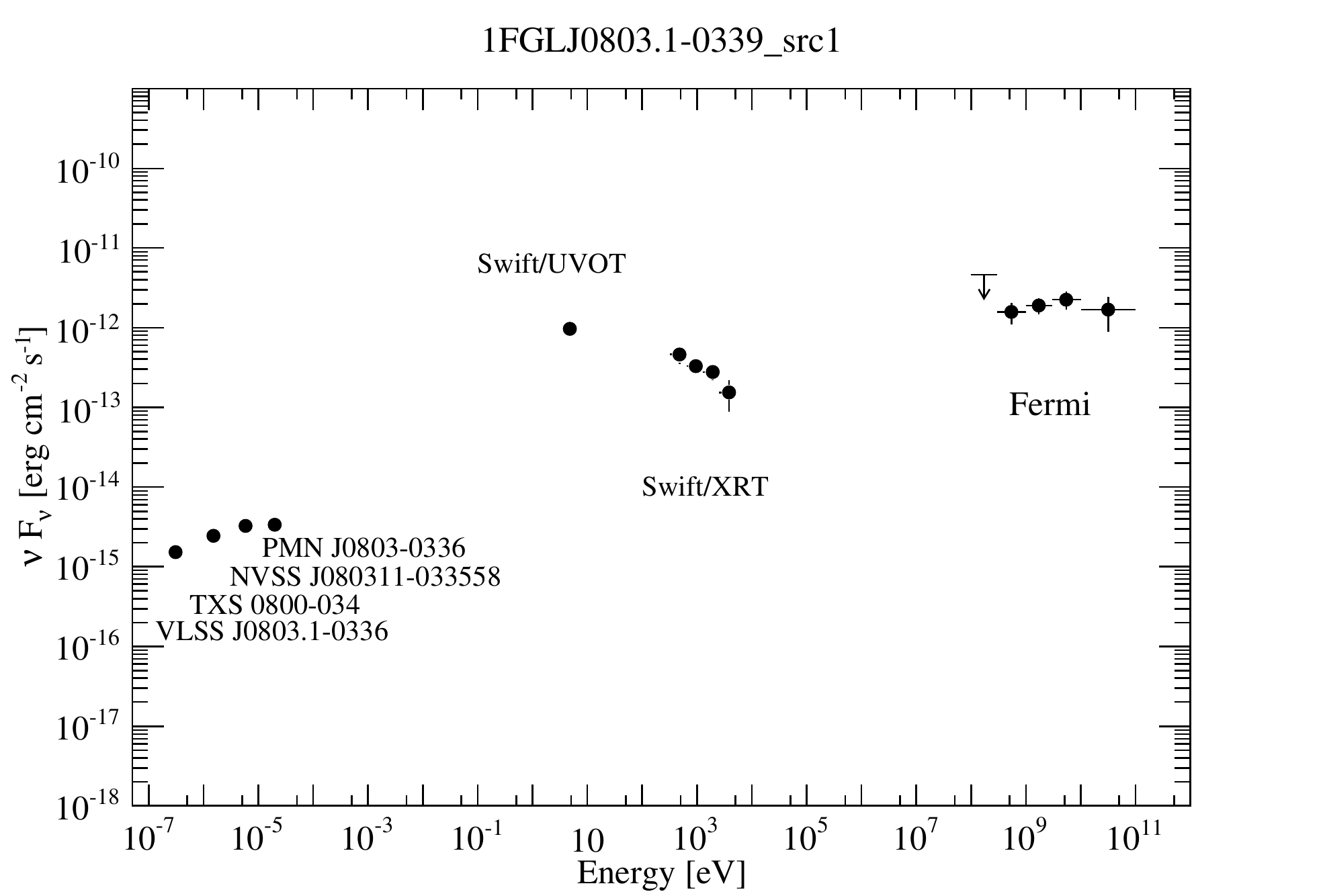}
    \end{center}
  \end{minipage}
  \begin{minipage}{0.32\hsize}
    \begin{center}
      \includegraphics[width=55mm]{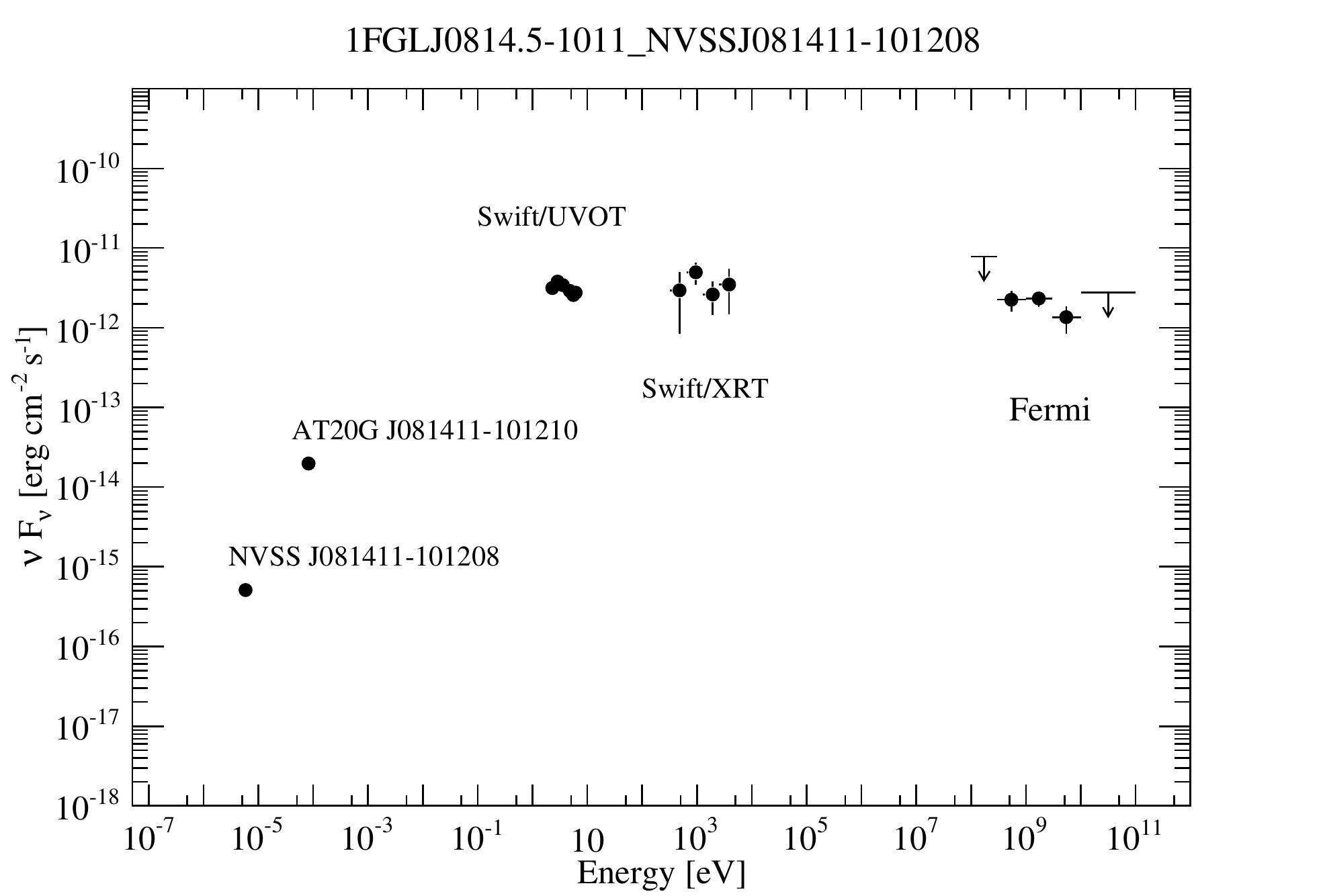}
    \end{center}
  \end{minipage}
  \begin{minipage}{0.32\hsize}
    \begin{center}
      \includegraphics[width=55mm]{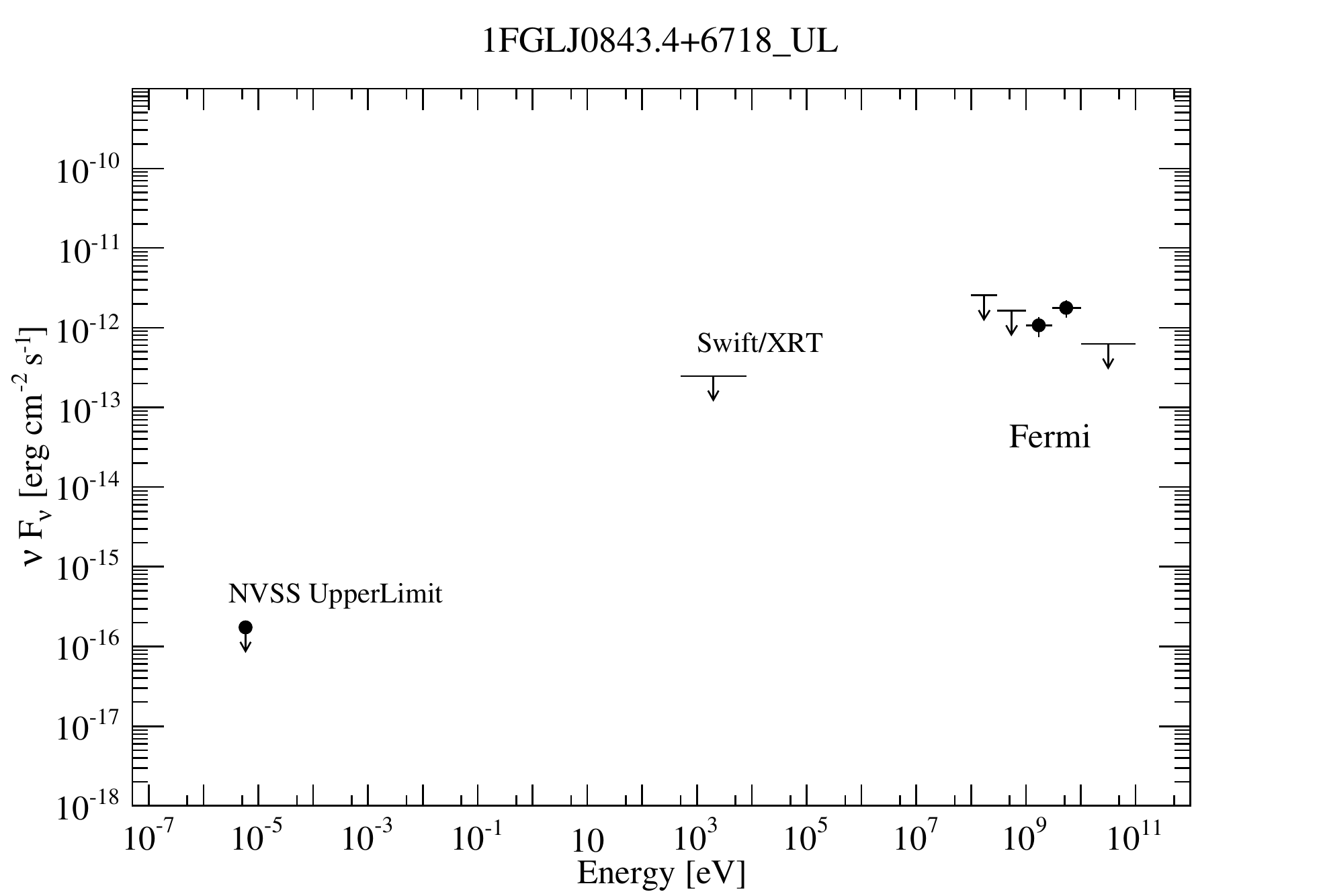}
    \end{center}
  \end{minipage}
  \begin{minipage}{0.32\hsize}
    \begin{center}
      \includegraphics[width=55mm]{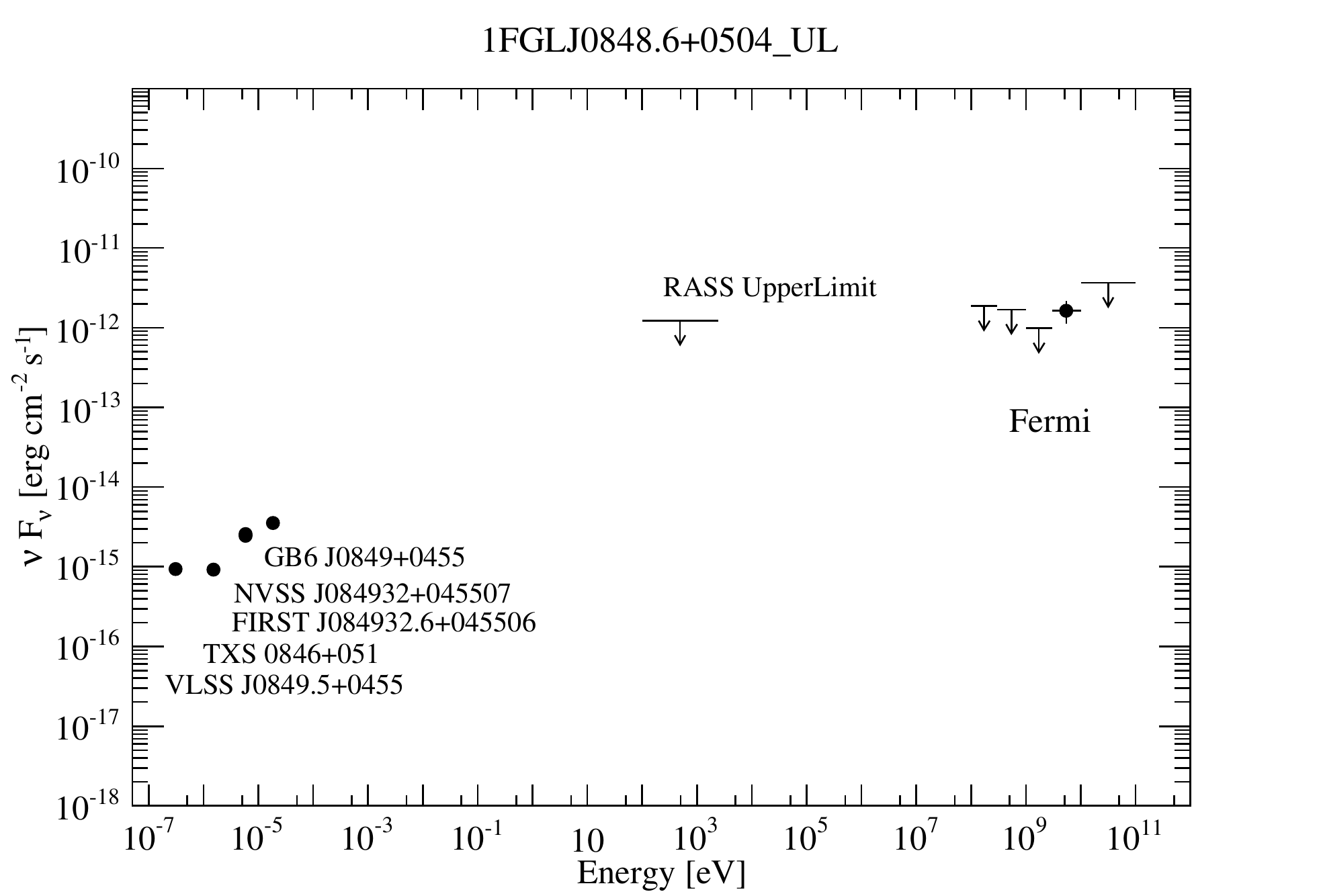}
    \end{center}
  \end{minipage}
  \begin{minipage}{0.32\hsize}
    \begin{center}
      \includegraphics[width=55mm]{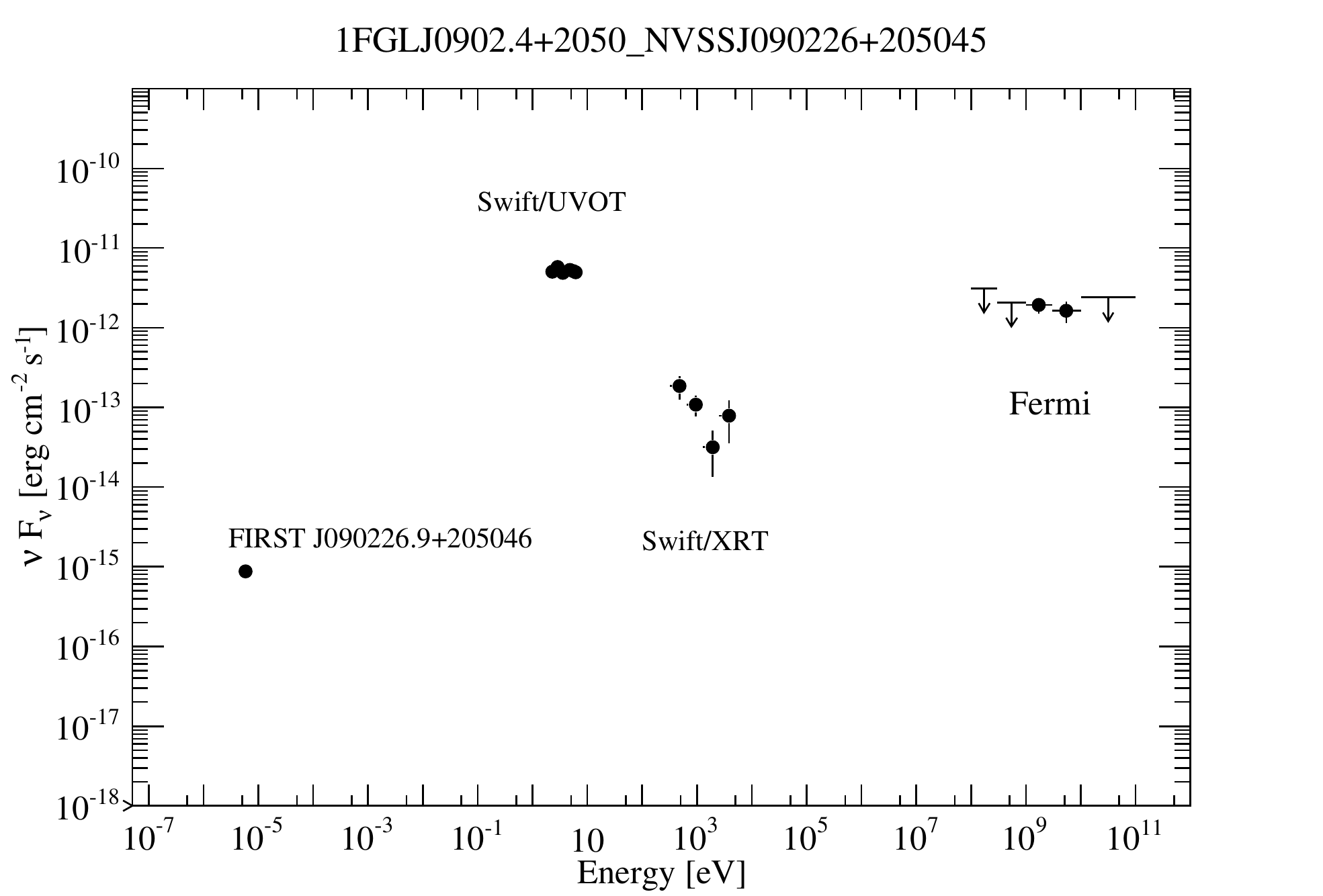}
    \end{center}
  \end{minipage}
  \begin{minipage}{0.32\hsize}
    \begin{center}
      \includegraphics[width=55mm]{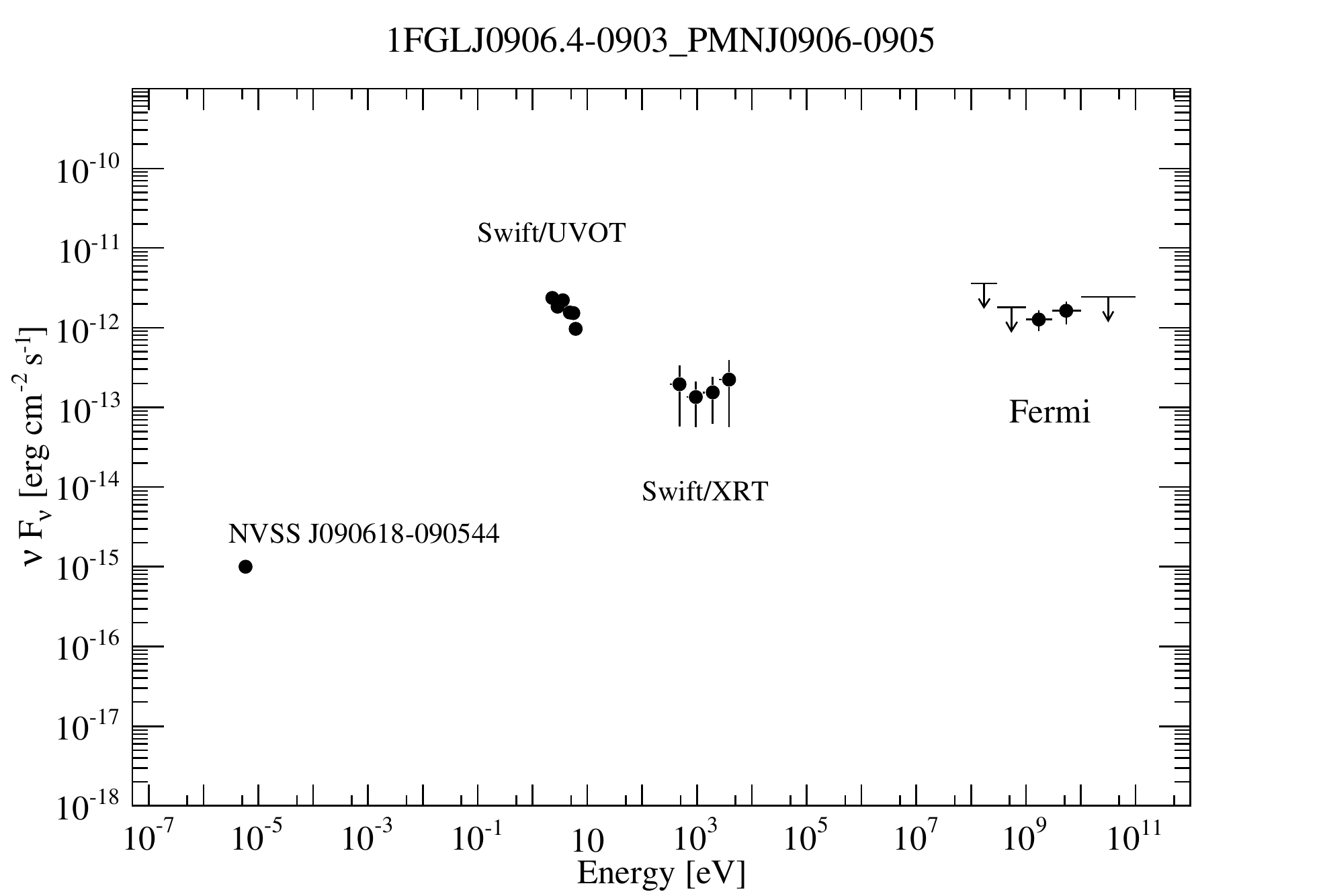}
    \end{center}
  \end{minipage}
  \begin{minipage}{0.32\hsize}
    \begin{center}
      \includegraphics[width=55mm]{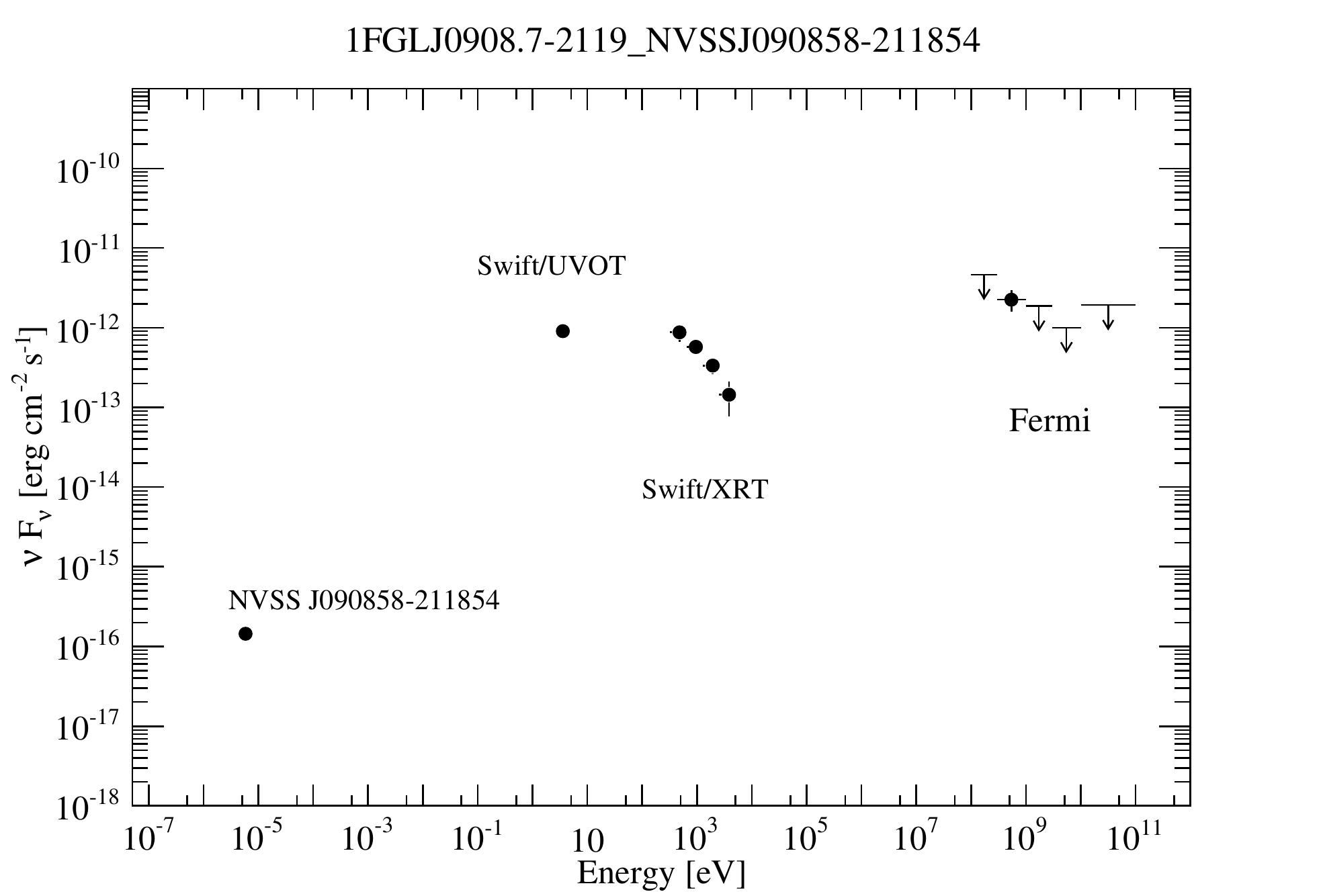}
    \end{center}
  \end{minipage}
  \begin{minipage}{0.32\hsize}
    \begin{center}
      \includegraphics[width=55mm]{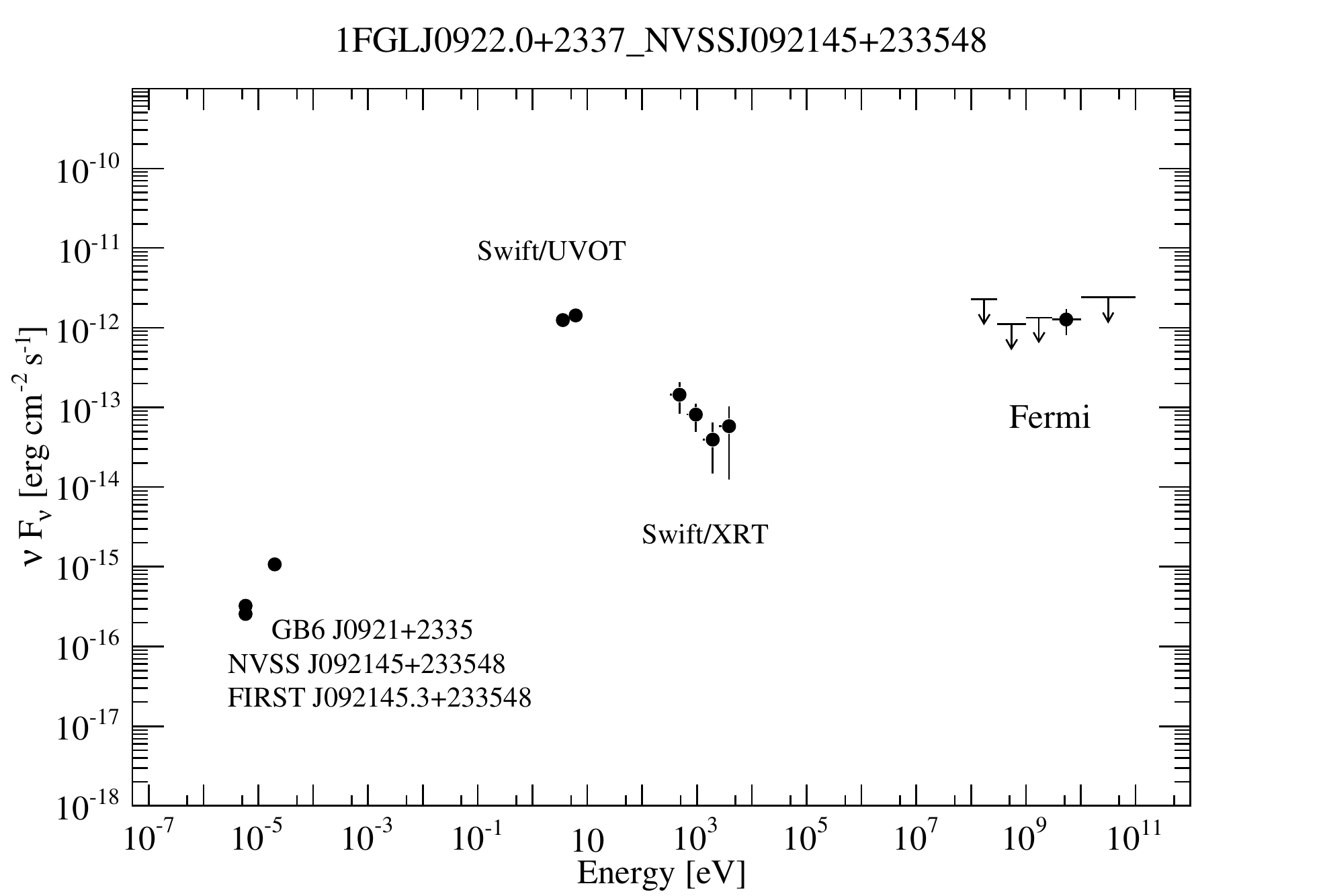}
    \end{center}
  \end{minipage}
  \begin{minipage}{0.32\hsize}
    \begin{center}
      \includegraphics[width=55mm]{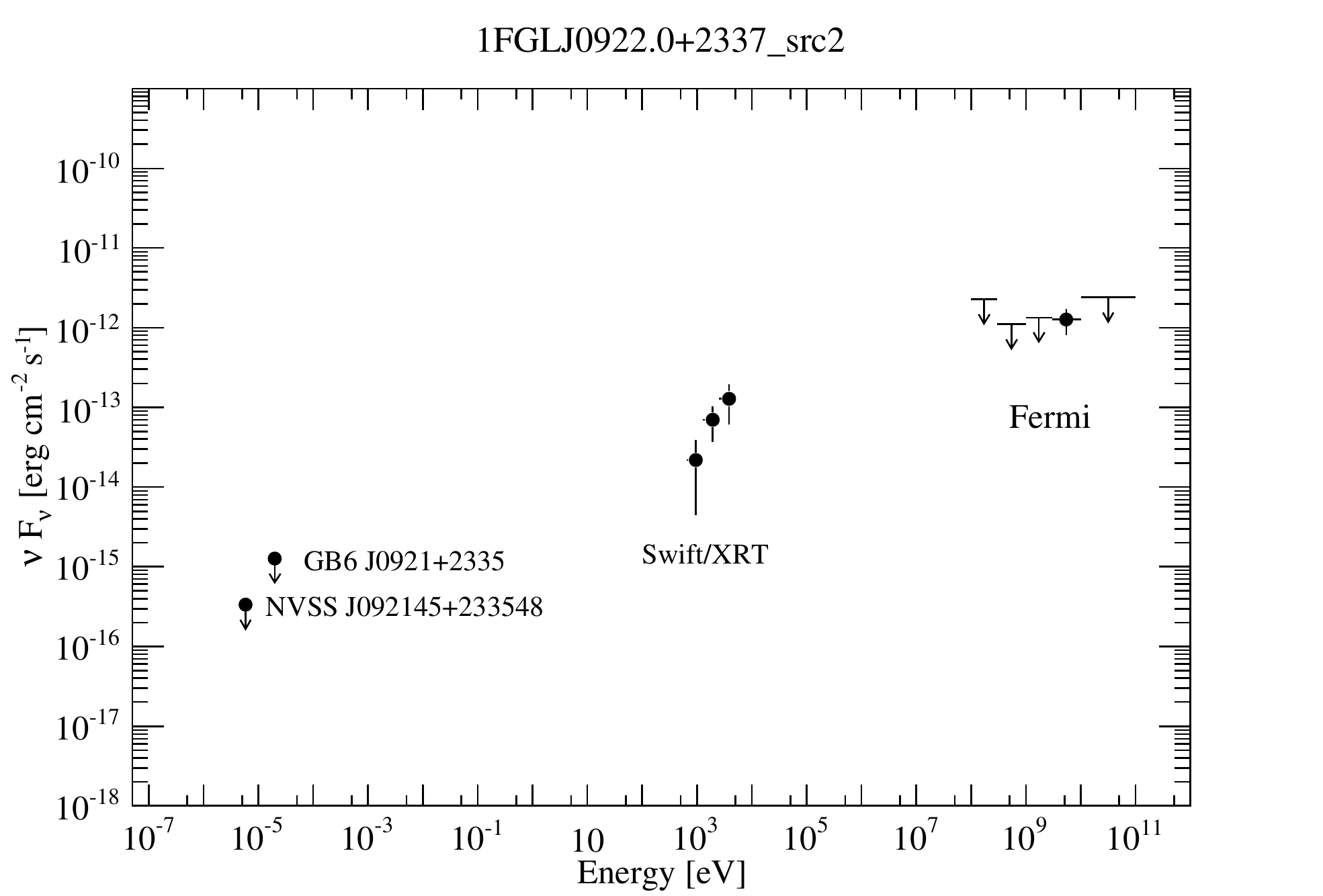}
    \end{center}
  \end{minipage}
  \begin{minipage}{0.32\hsize}
    \begin{center}
      \includegraphics[width=55mm]{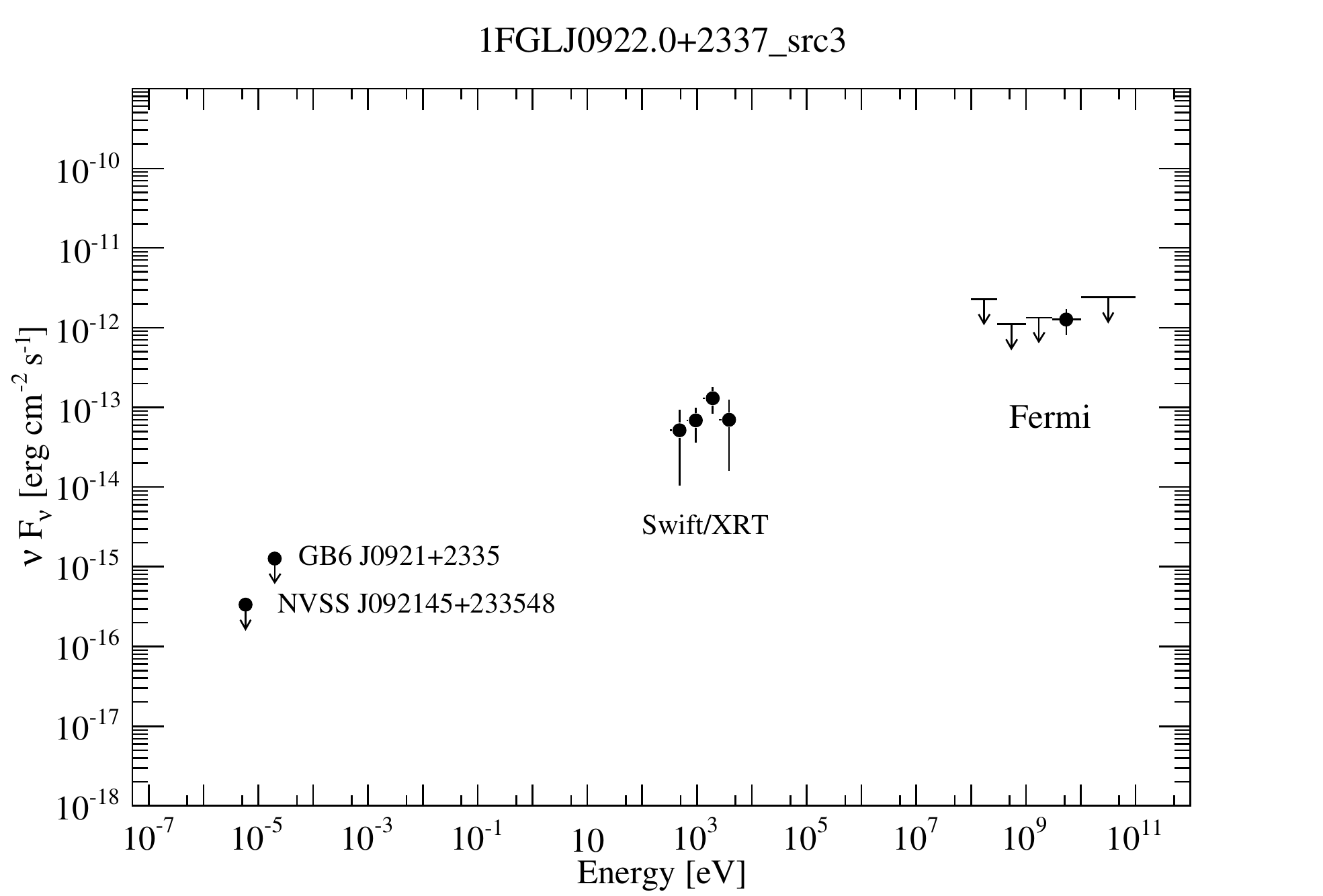}
    \end{center}
  \end{minipage}
  \begin{minipage}{0.32\hsize}
    \begin{center}
      \includegraphics[width=55mm]{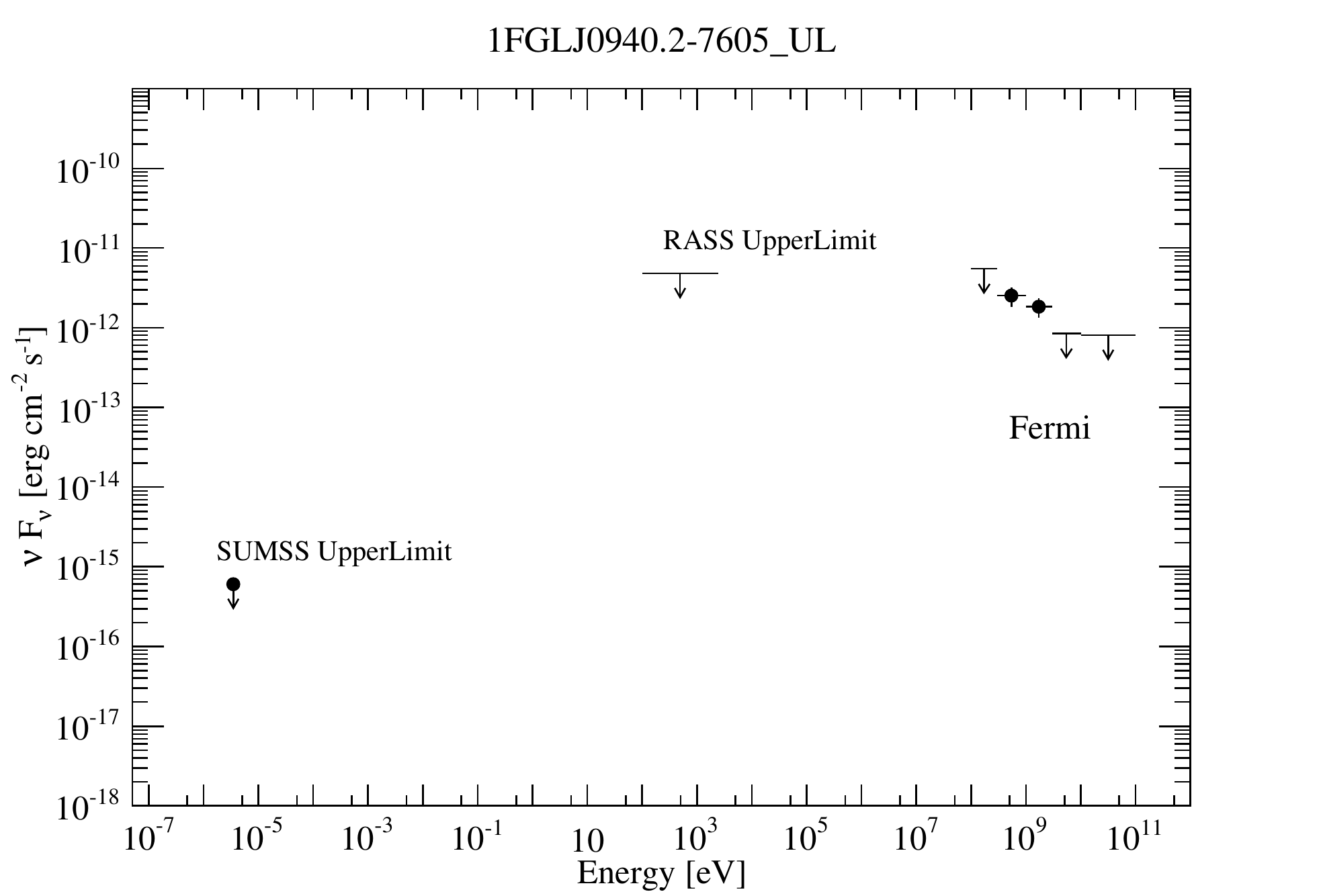}
    \end{center}
  \end{minipage}
  \begin{minipage}{0.32\hsize}
    \begin{center}
      \includegraphics[width=55mm]{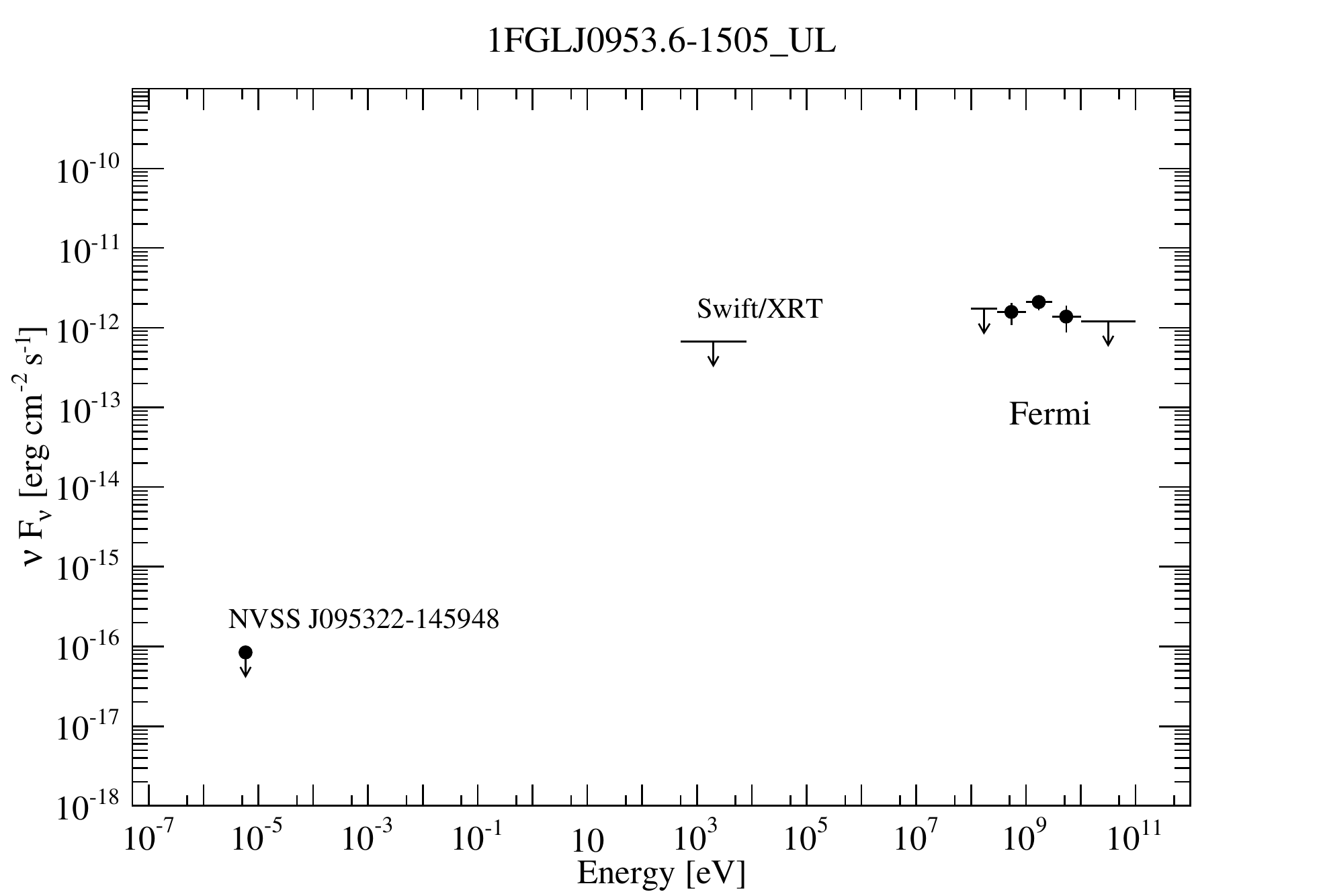}
    \end{center}
  \end{minipage}
  \begin{minipage}{0.32\hsize}
    \begin{center}
      \includegraphics[width=55mm]{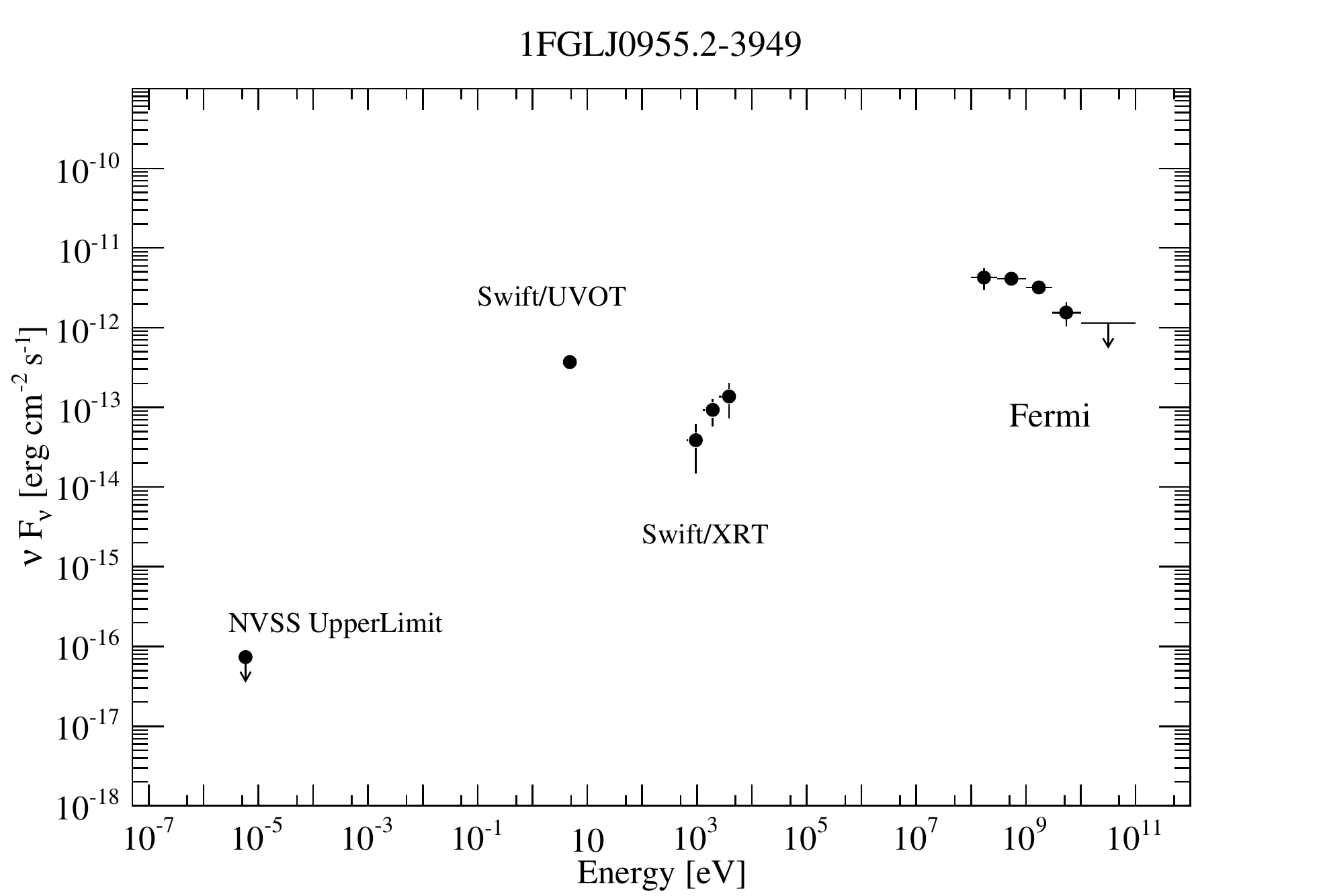}
    \end{center}
  \end{minipage}
  \begin{minipage}{0.32\hsize}
    \begin{center}
      \includegraphics[width=55mm]{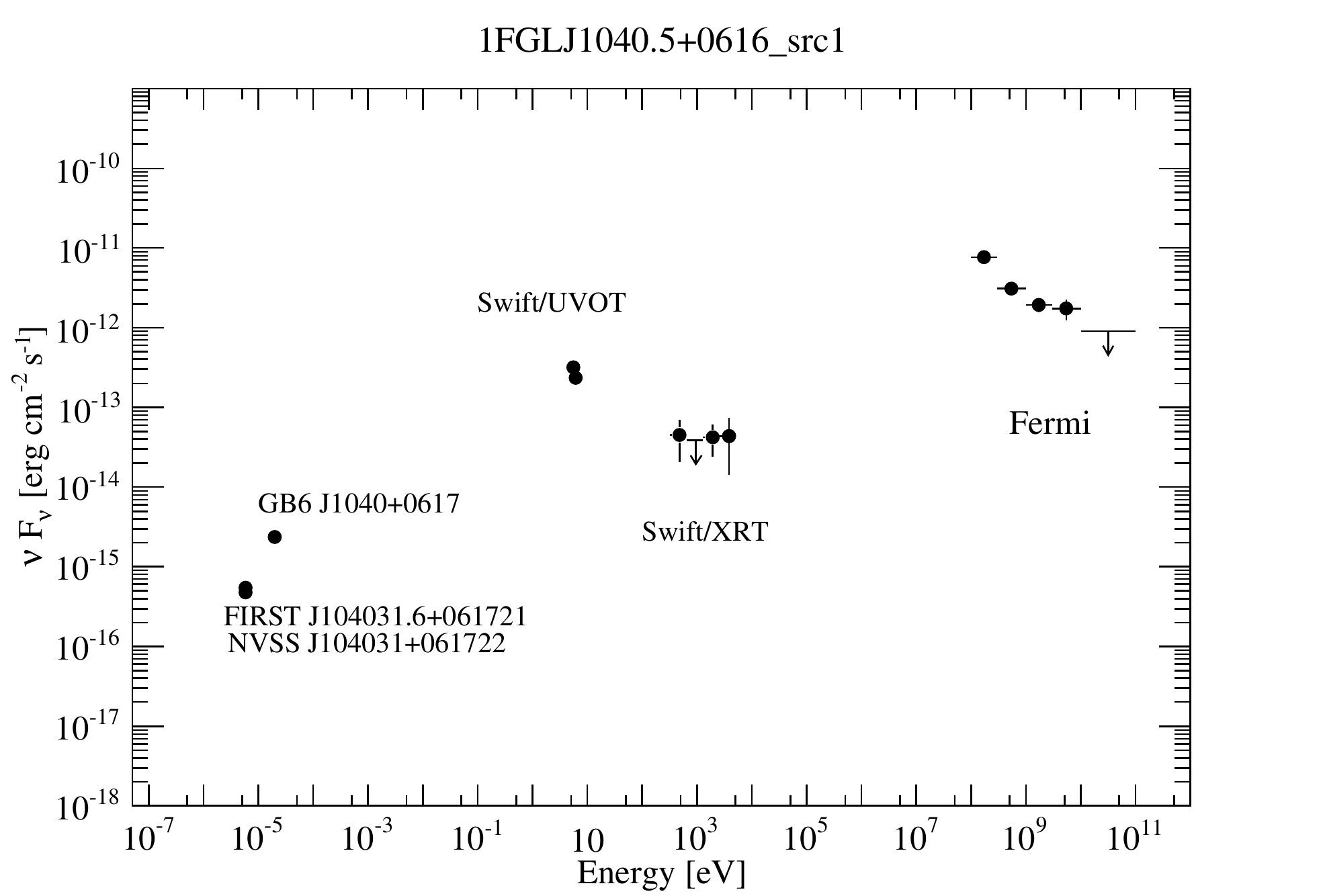}
    \end{center}
  \end{minipage}
  \begin{minipage}{0.32\hsize}
    \begin{center}
      \includegraphics[width=55mm]{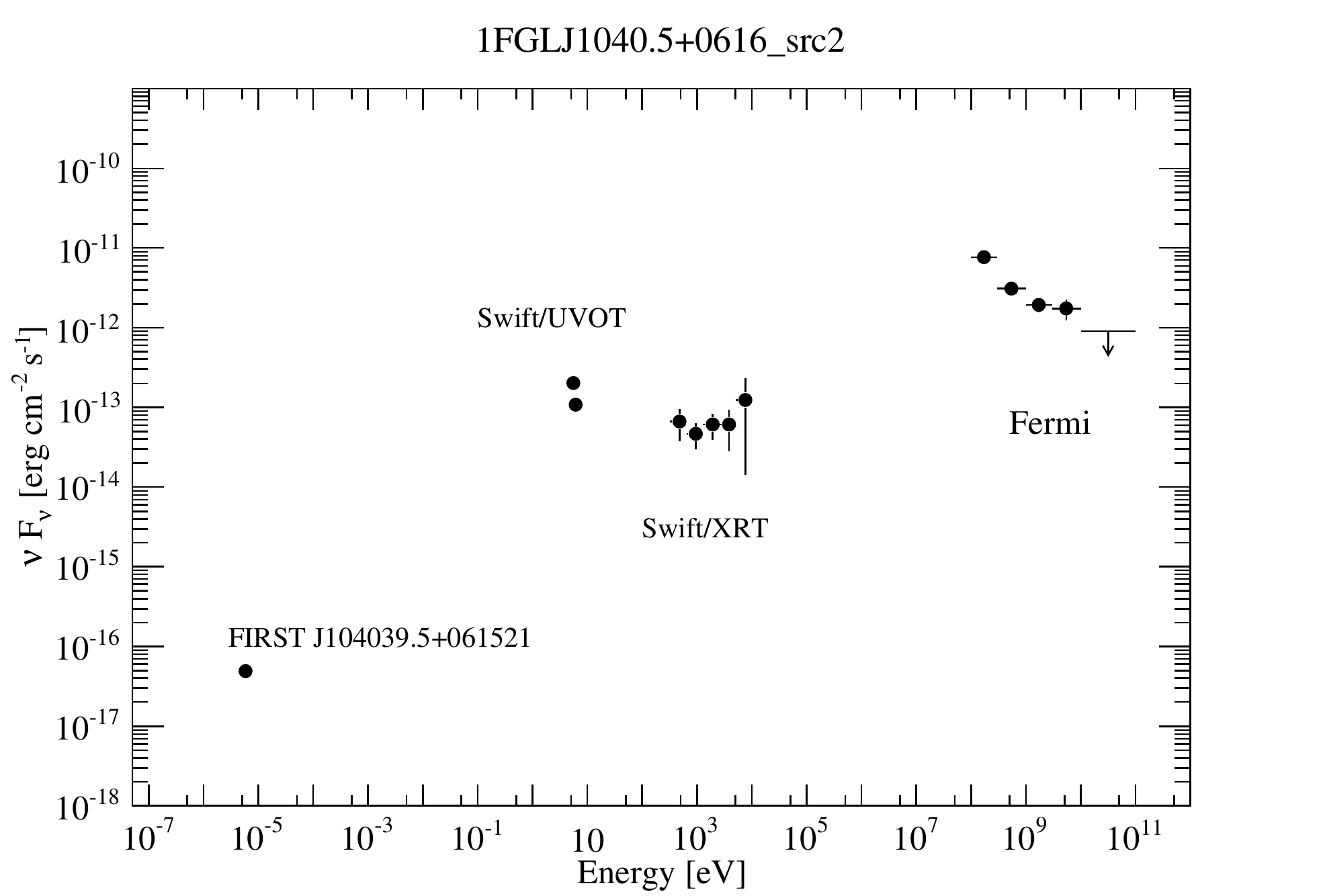}
    \end{center}
  \end{minipage}
 \end{center}
\end{figure}
\clearpage
\begin{figure}[m]
 \begin{center}
  \begin{minipage}{0.32\hsize}
    \begin{center}
      \includegraphics[width=55mm]{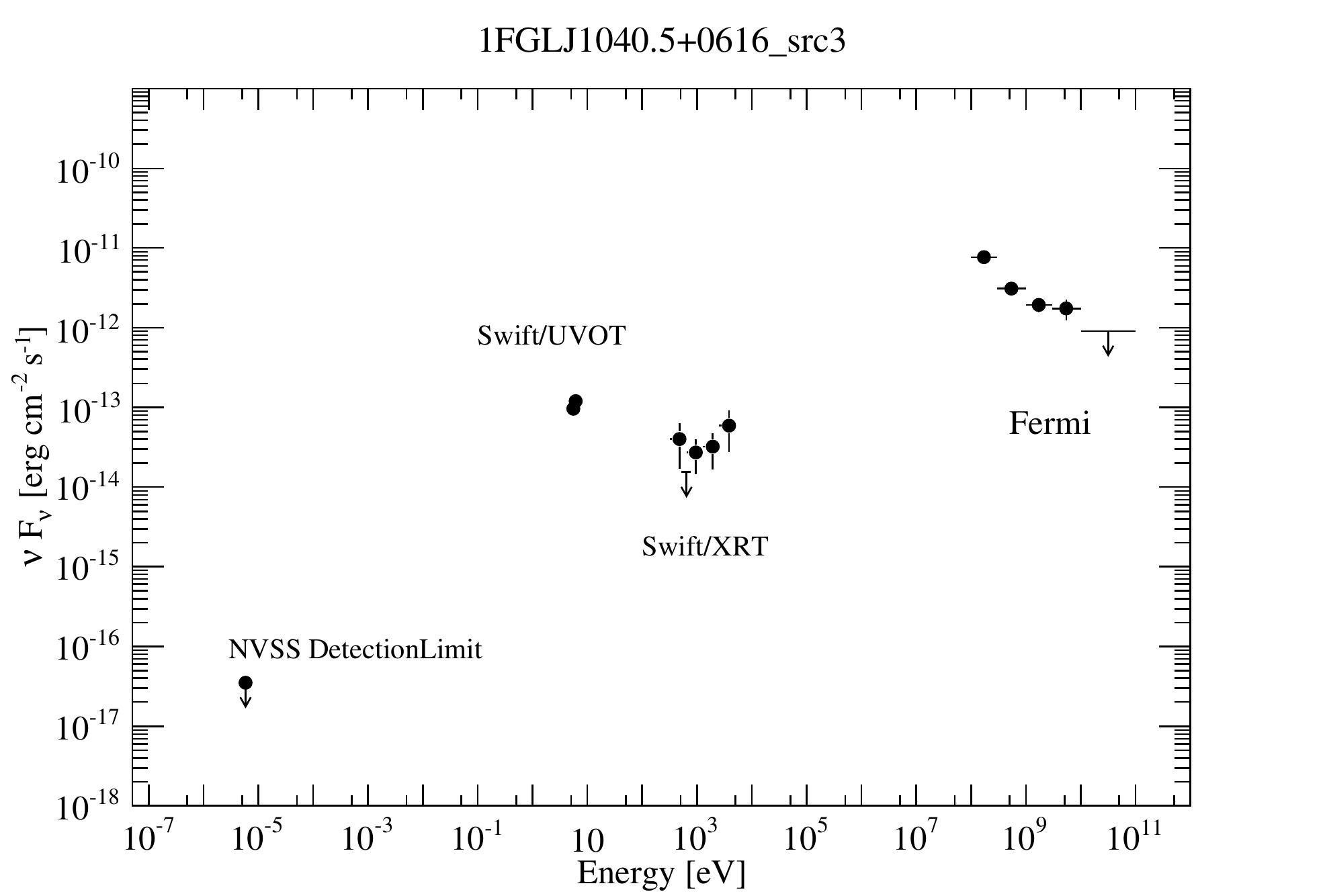}
    \end{center}
  \end{minipage}
  \begin{minipage}{0.32\hsize}
    \begin{center}
      \includegraphics[width=55mm]{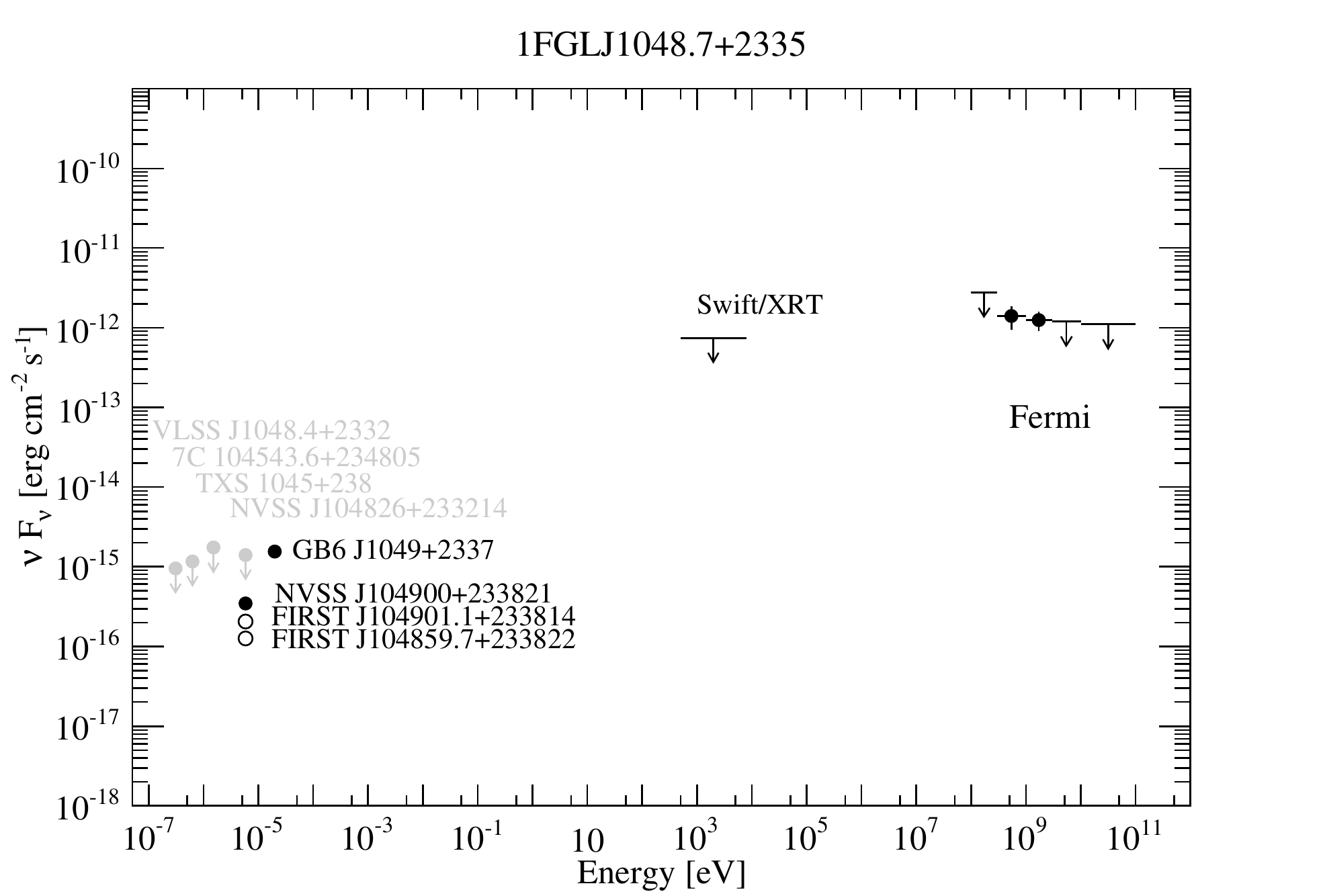}
    \end{center}
  \end{minipage}
  \begin{minipage}{0.32\hsize}
    \begin{center}
      \includegraphics[width=55mm]{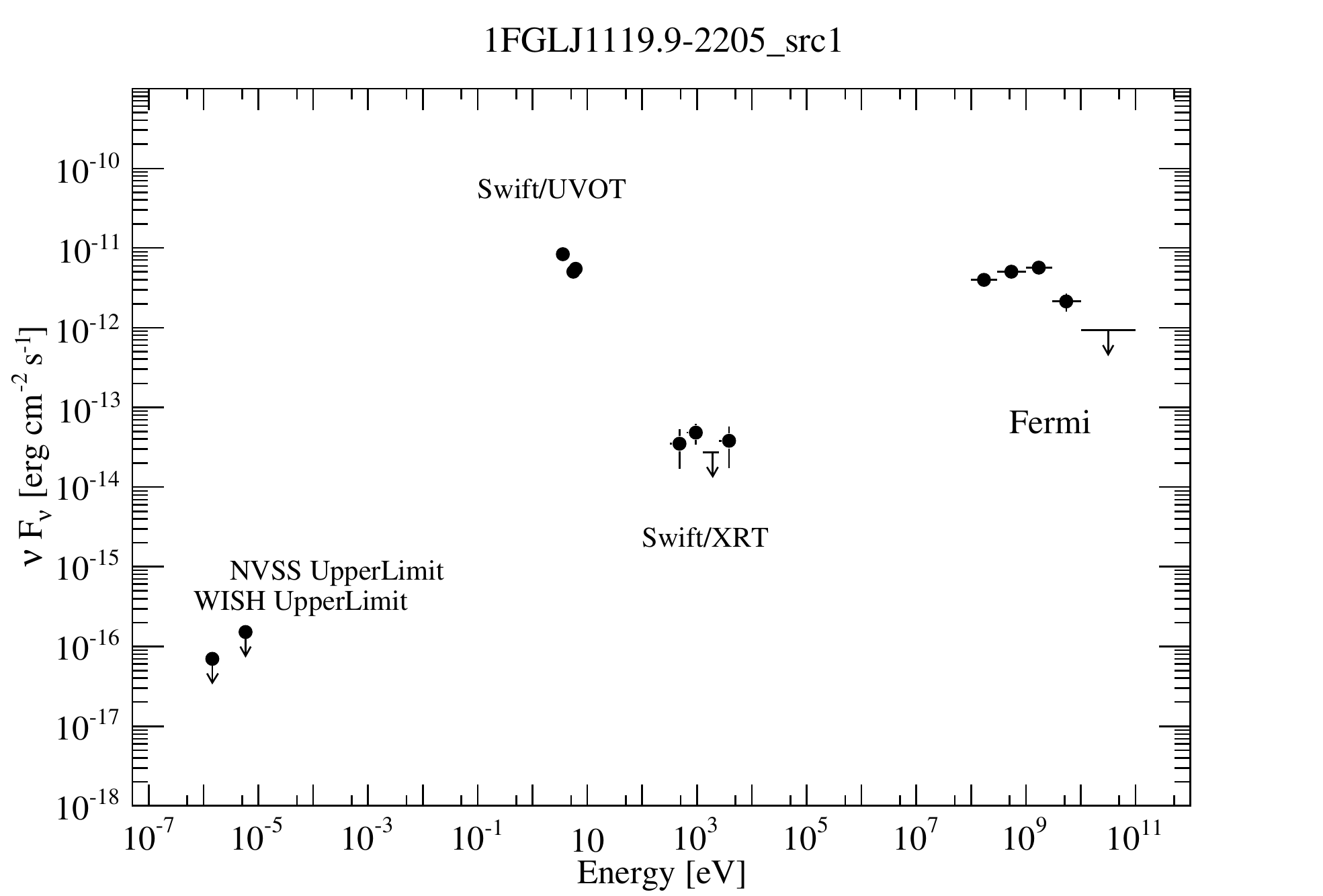}
    \end{center}
  \end{minipage}
  \begin{minipage}{0.32\hsize}
    \begin{center}
      \includegraphics[width=55mm]{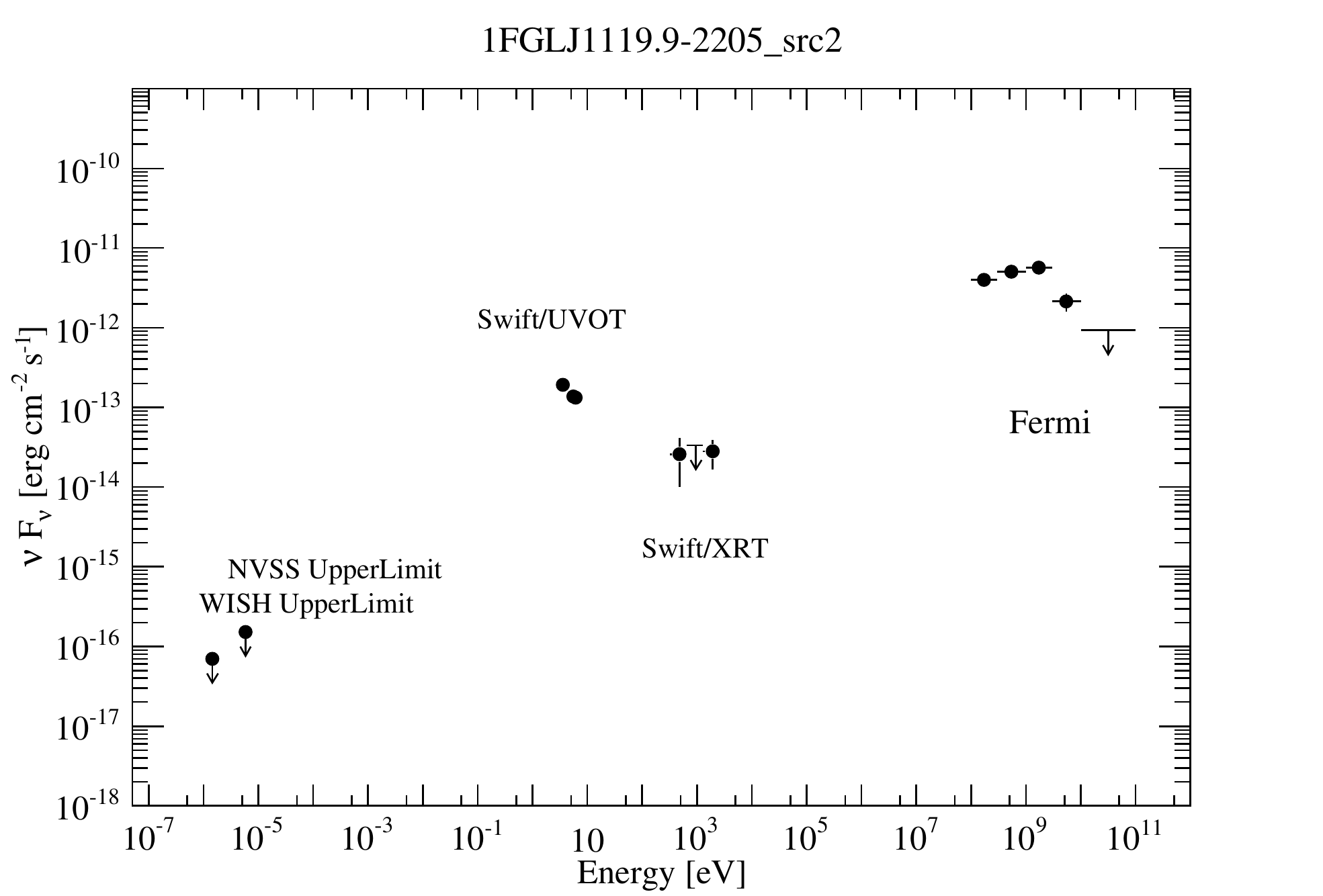}
    \end{center}
  \end{minipage}
  \begin{minipage}{0.32\hsize}
    \begin{center}
      \includegraphics[width=55mm]{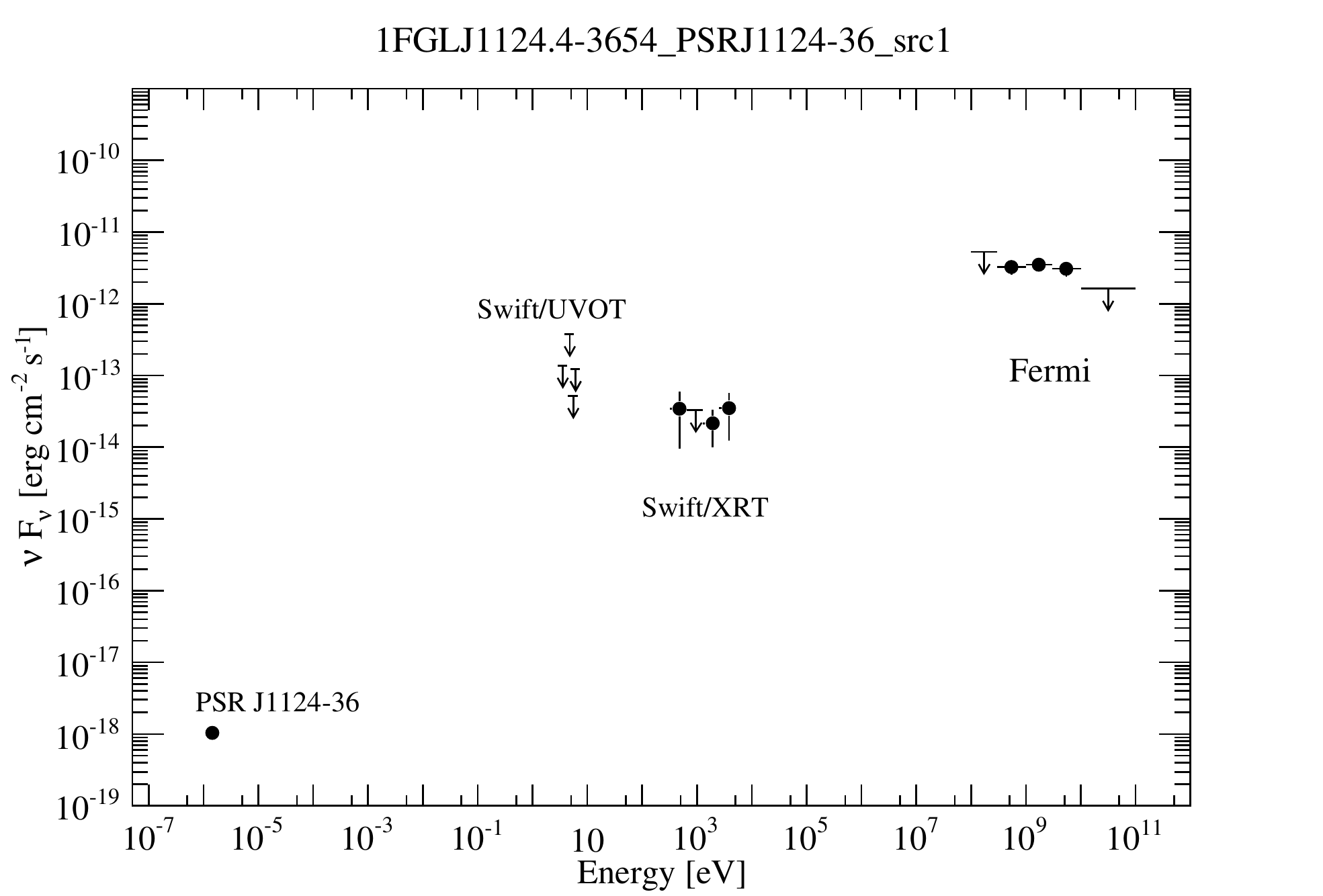}
    \end{center}
  \end{minipage}
  \begin{minipage}{0.32\hsize}
    \begin{center}
      \includegraphics[width=55mm]{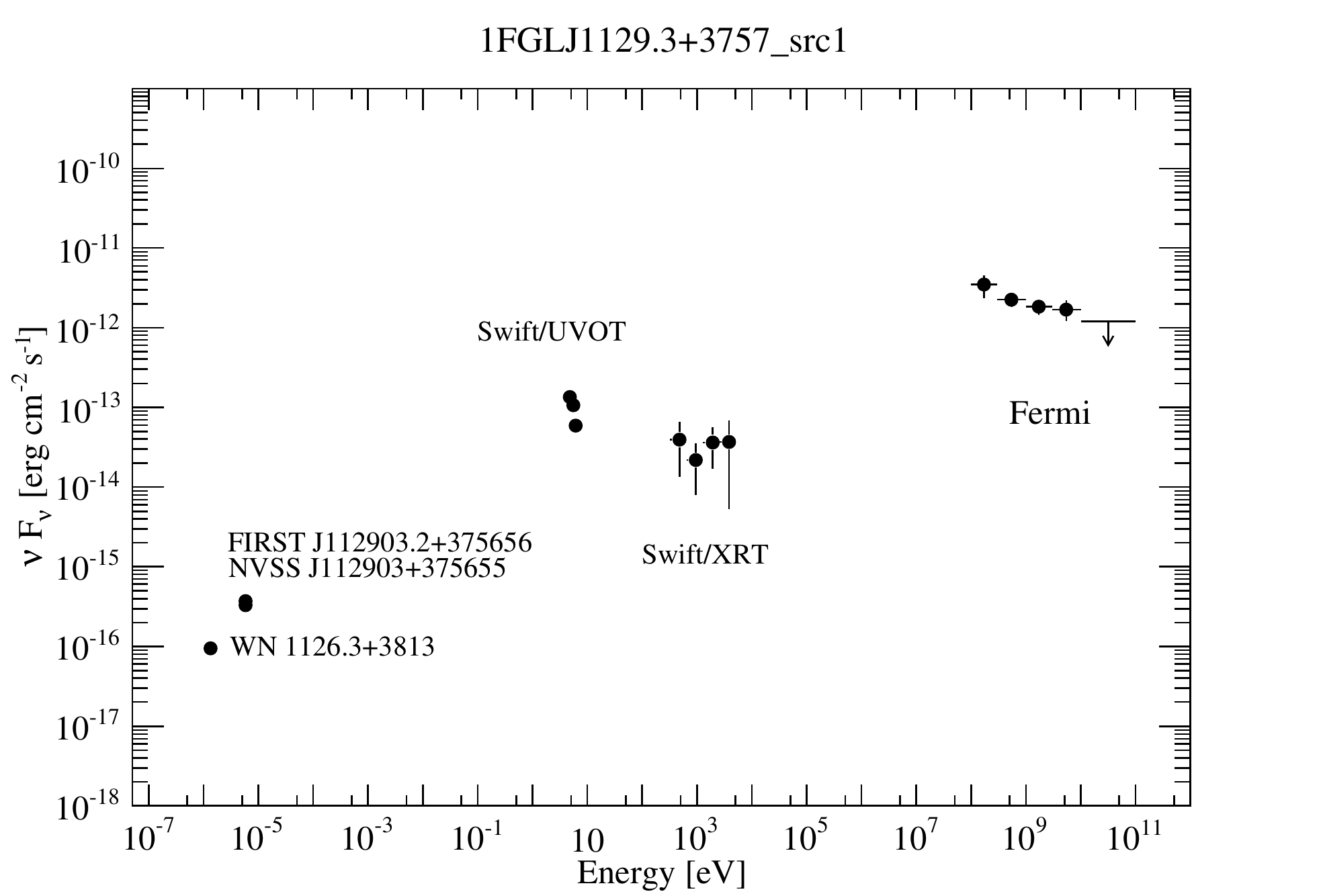}
    \end{center}
  \end{minipage}
  \begin{minipage}{0.32\hsize}
    \begin{center}
      \includegraphics[width=55mm]{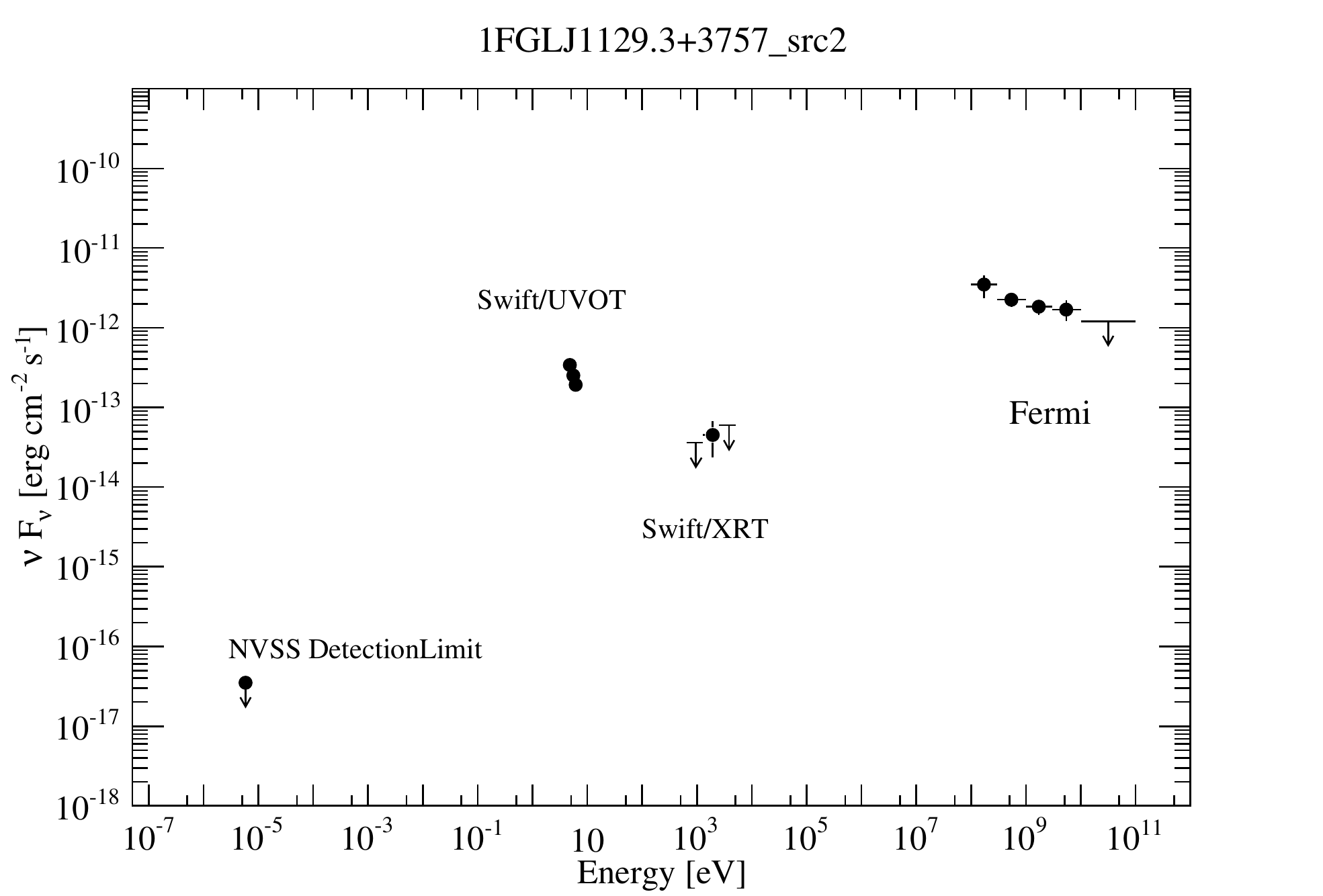}
    \end{center}
  \end{minipage}
  \begin{minipage}{0.32\hsize}
    \begin{center}
      \includegraphics[width=55mm]{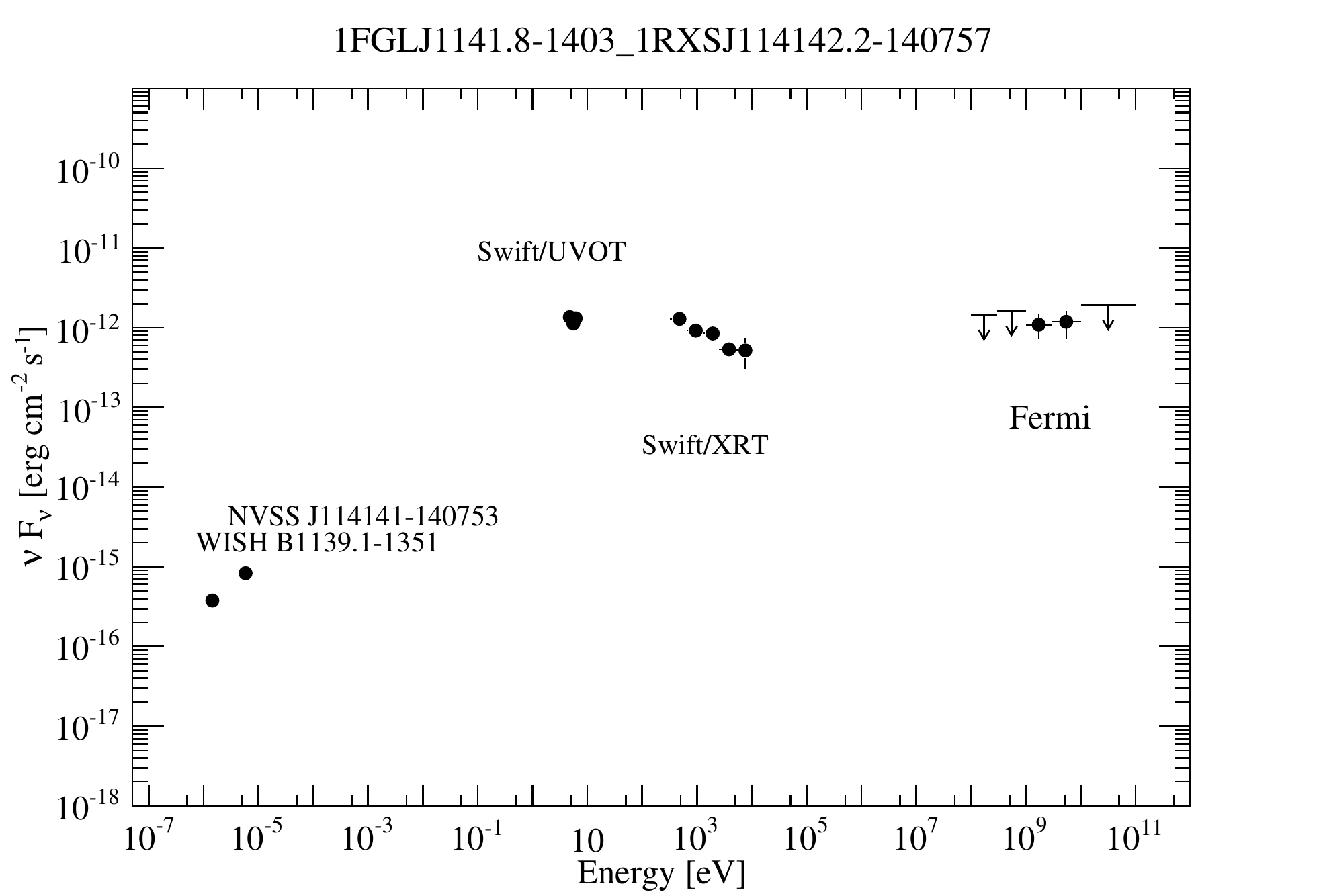}
    \end{center}
  \end{minipage}
  \begin{minipage}{0.32\hsize}
    \begin{center}
      \includegraphics[width=55mm]{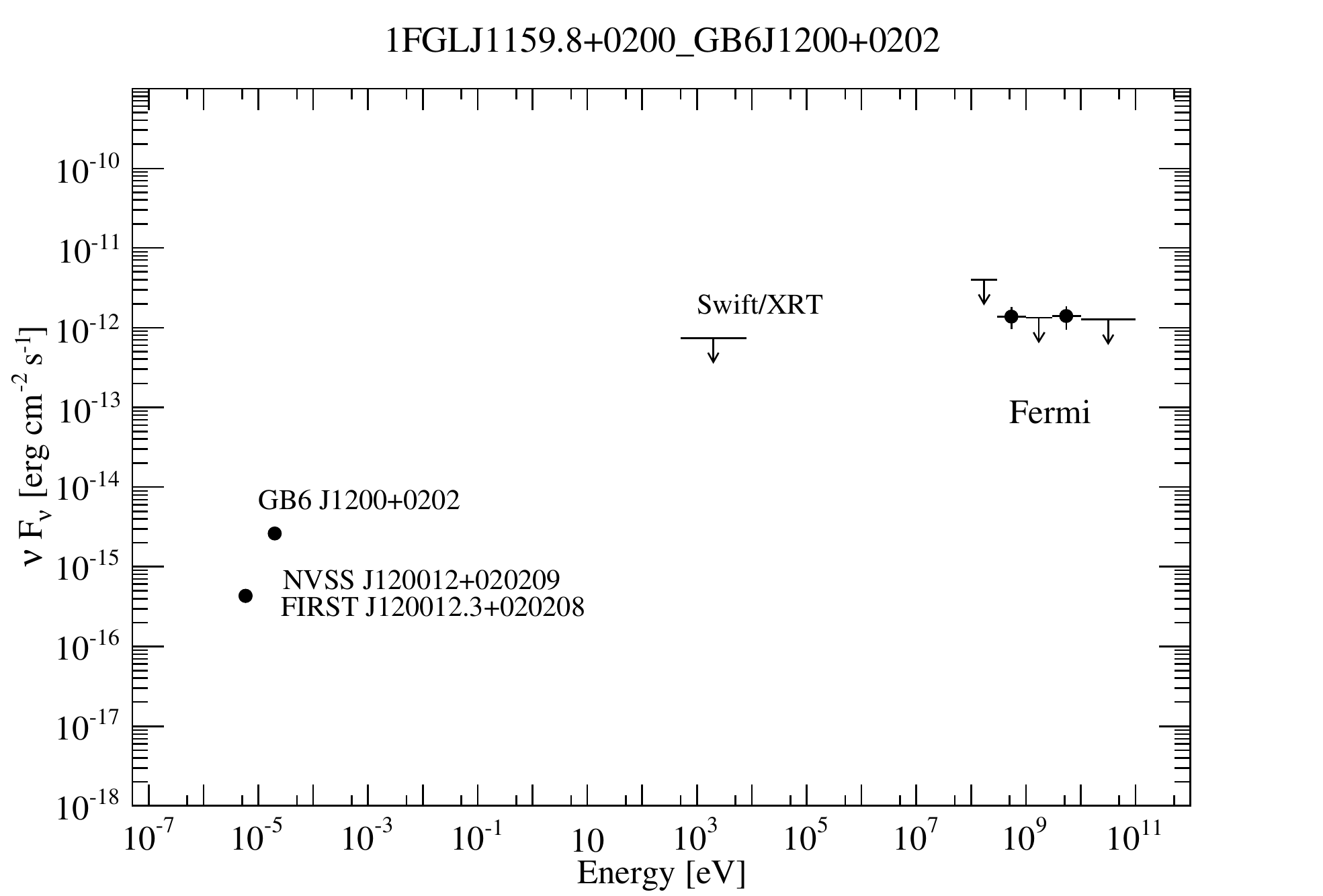}
    \end{center}
  \end{minipage}
  \begin{minipage}{0.32\hsize}
    \begin{center}
      \includegraphics[width=55mm]{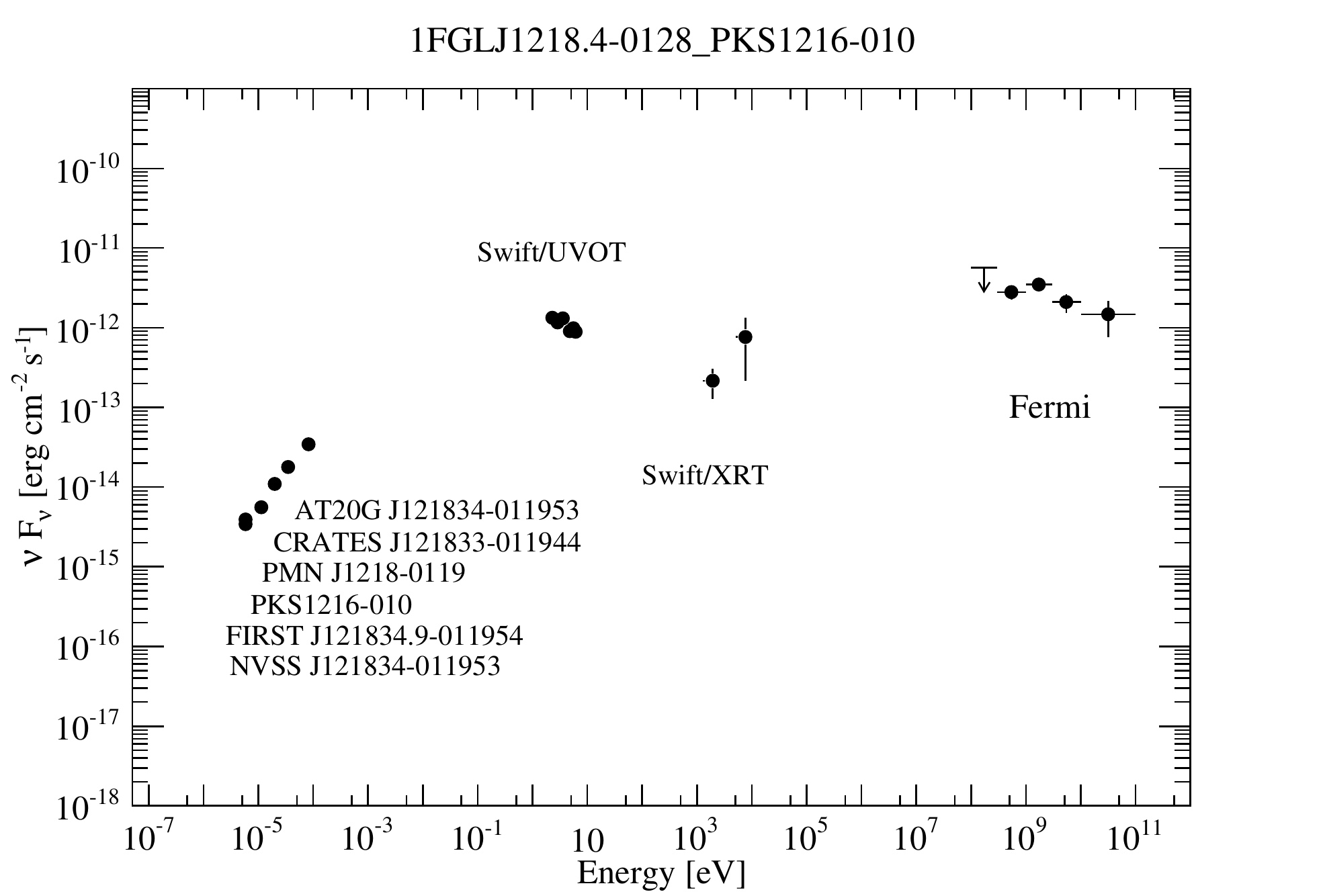}
    \end{center}
  \end{minipage}
  \begin{minipage}{0.32\hsize}
    \begin{center}
      \includegraphics[width=55mm]{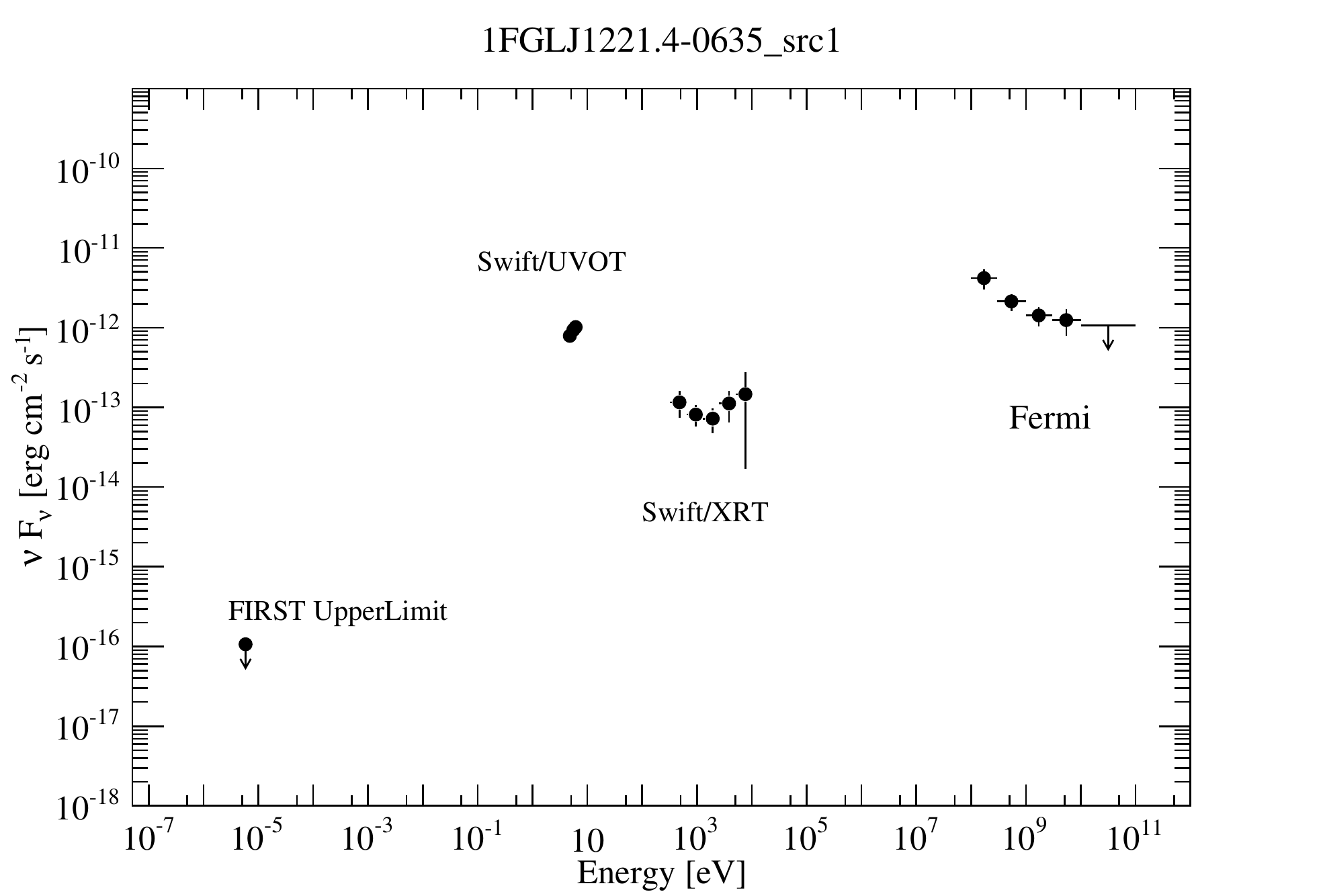}
    \end{center}
  \end{minipage}
  \begin{minipage}{0.32\hsize}
    \begin{center}
      \includegraphics[width=55mm]{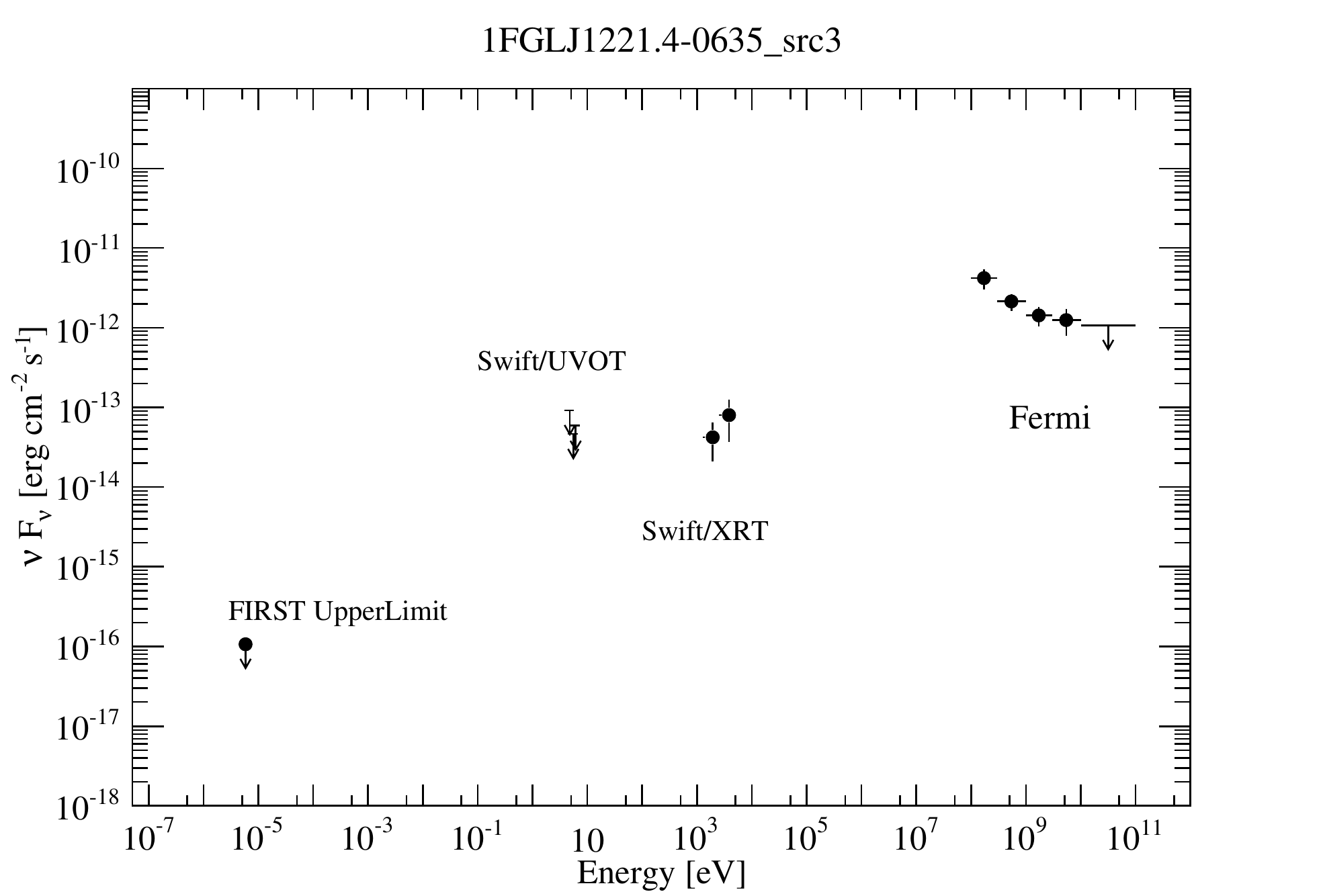}
    \end{center}
  \end{minipage}
  \begin{minipage}{0.32\hsize}
    \begin{center}
      \includegraphics[width=55mm]{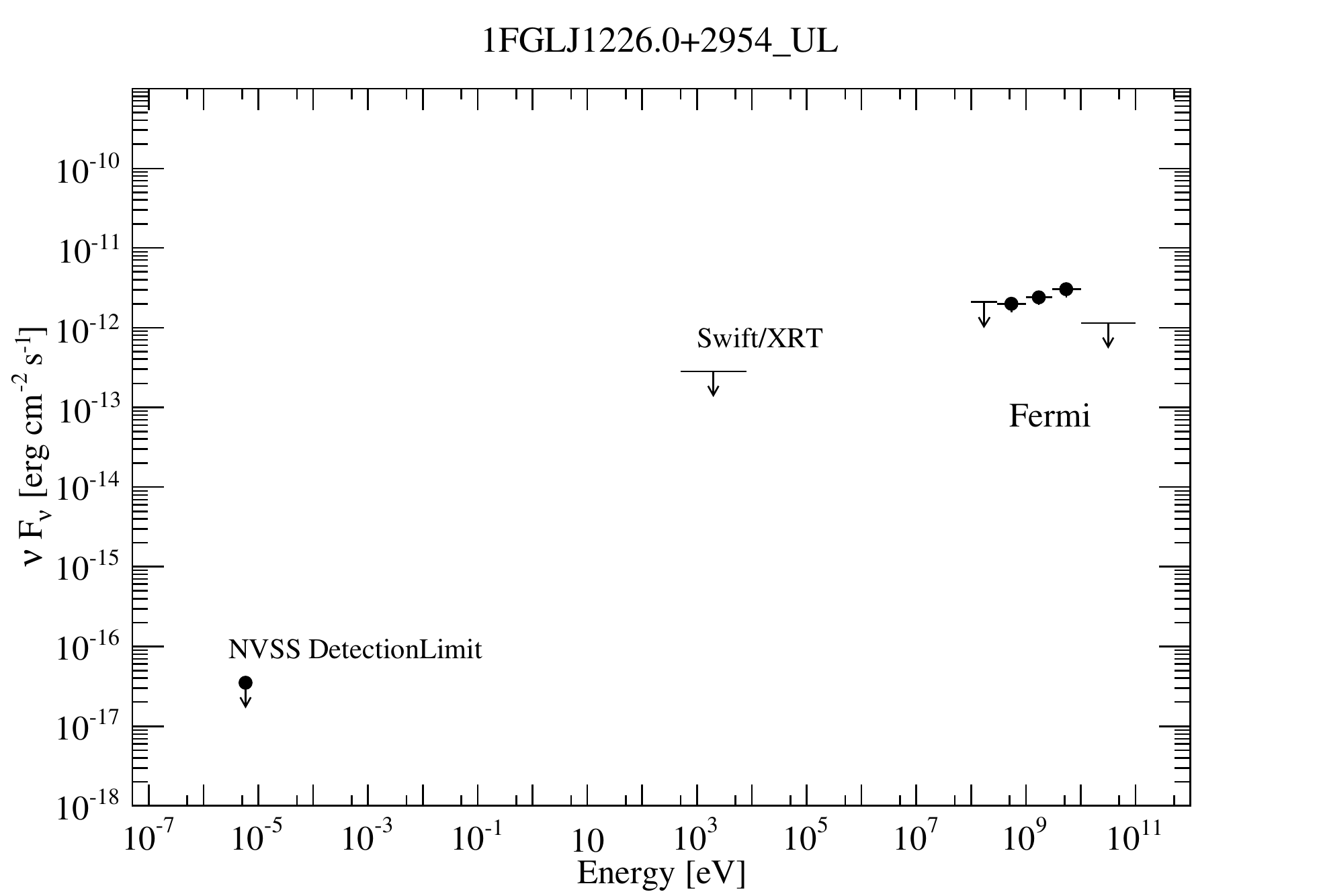}
    \end{center}
  \end{minipage}
  \begin{minipage}{0.32\hsize}
    \begin{center}
      \includegraphics[width=55mm]{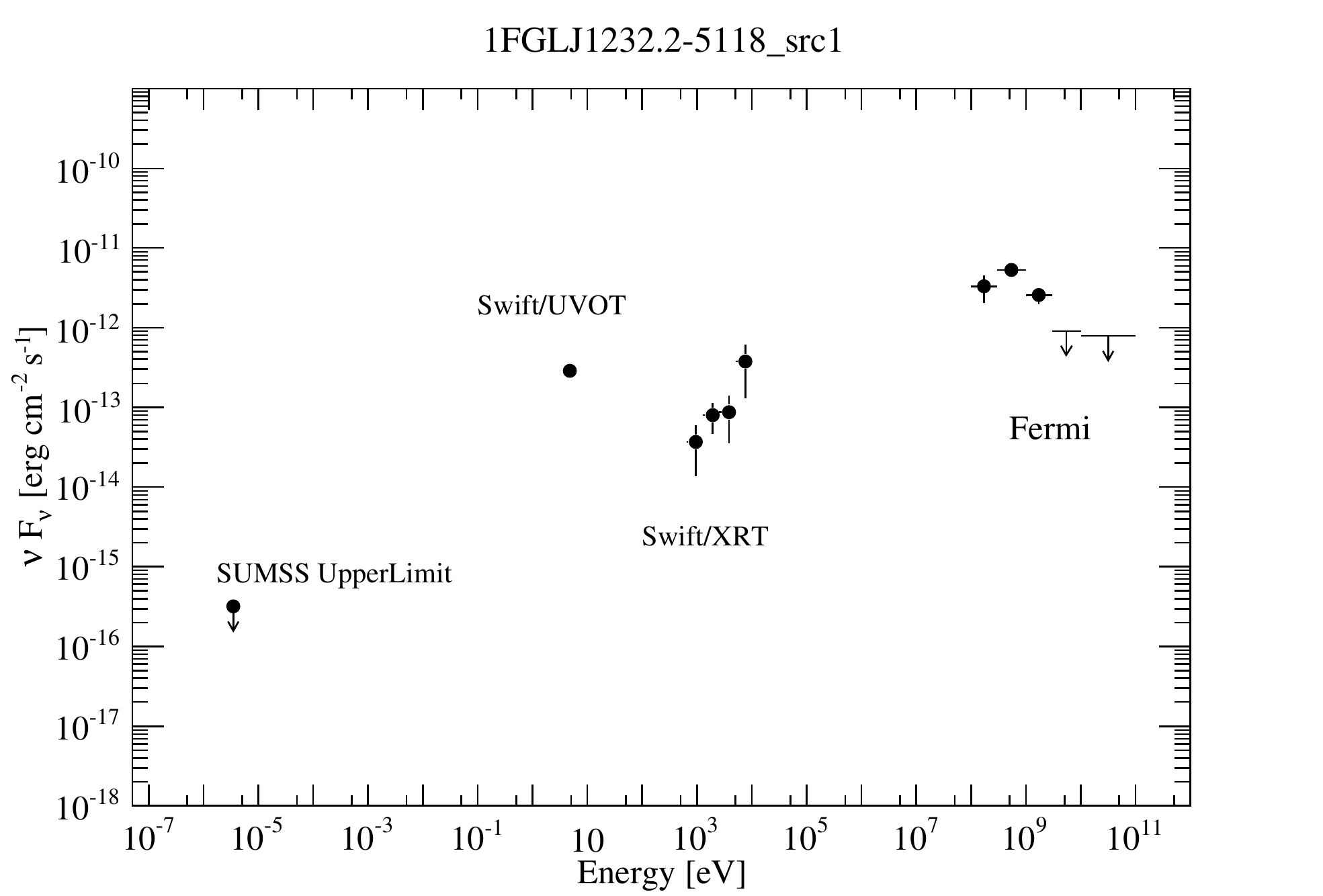}
    \end{center}
  \end{minipage}
  \begin{minipage}{0.32\hsize}
    \begin{center}
      \includegraphics[width=55mm]{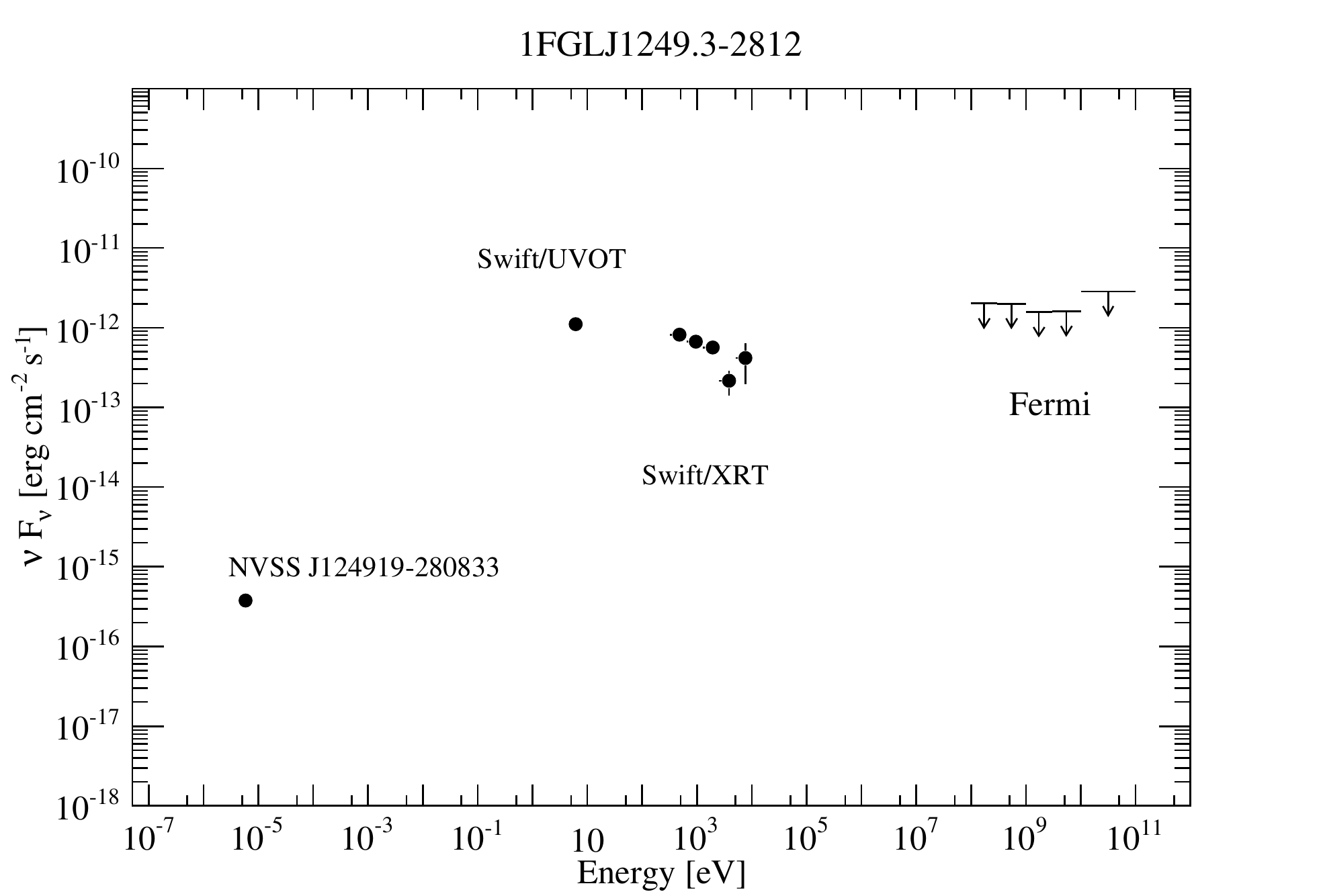}
    \end{center}
  \end{minipage}
  \begin{minipage}{0.32\hsize}
    \begin{center}
      \includegraphics[width=55mm]{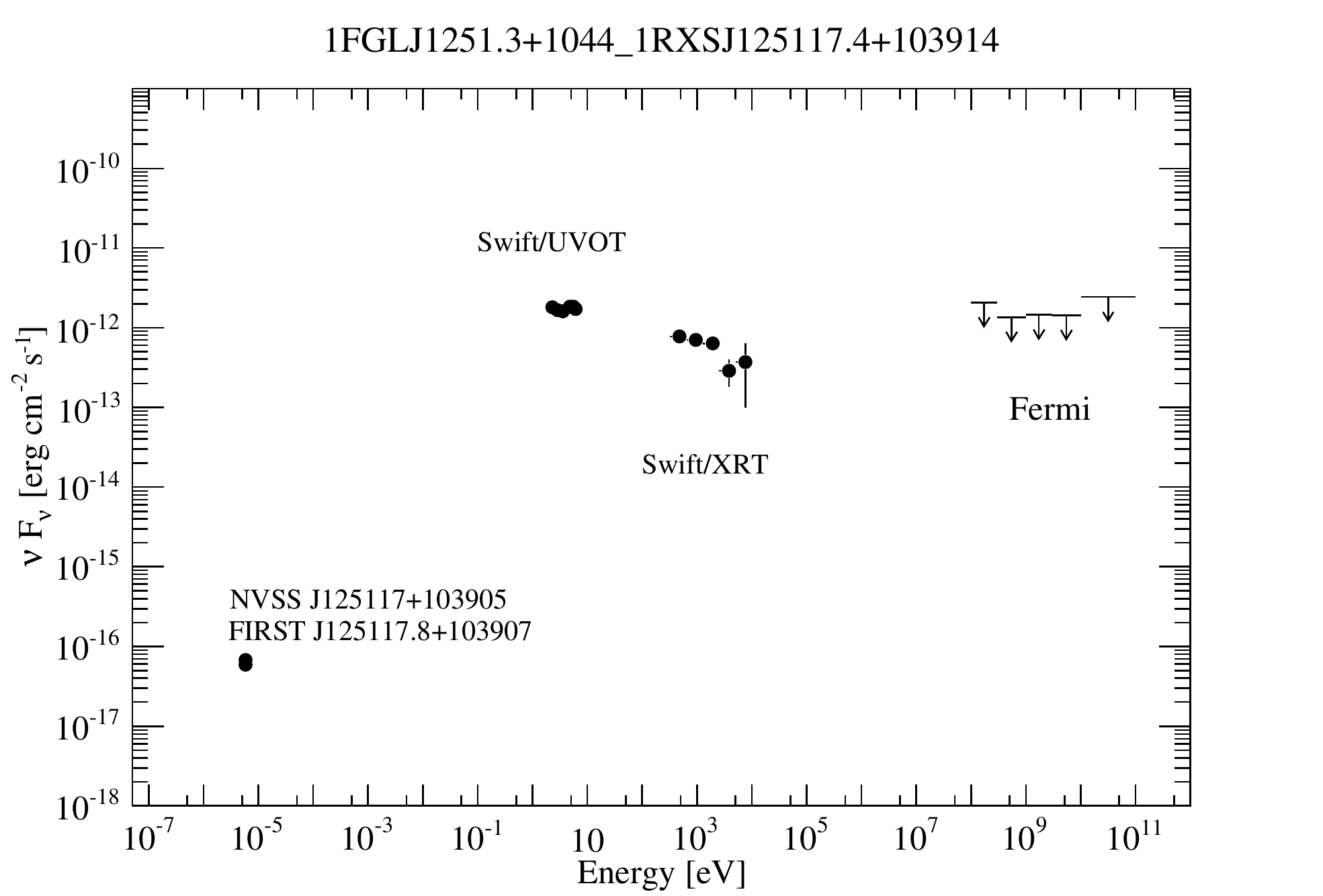}
    \end{center}
  \end{minipage}
  \begin{minipage}{0.32\hsize}
    \begin{center}
      \includegraphics[width=55mm]{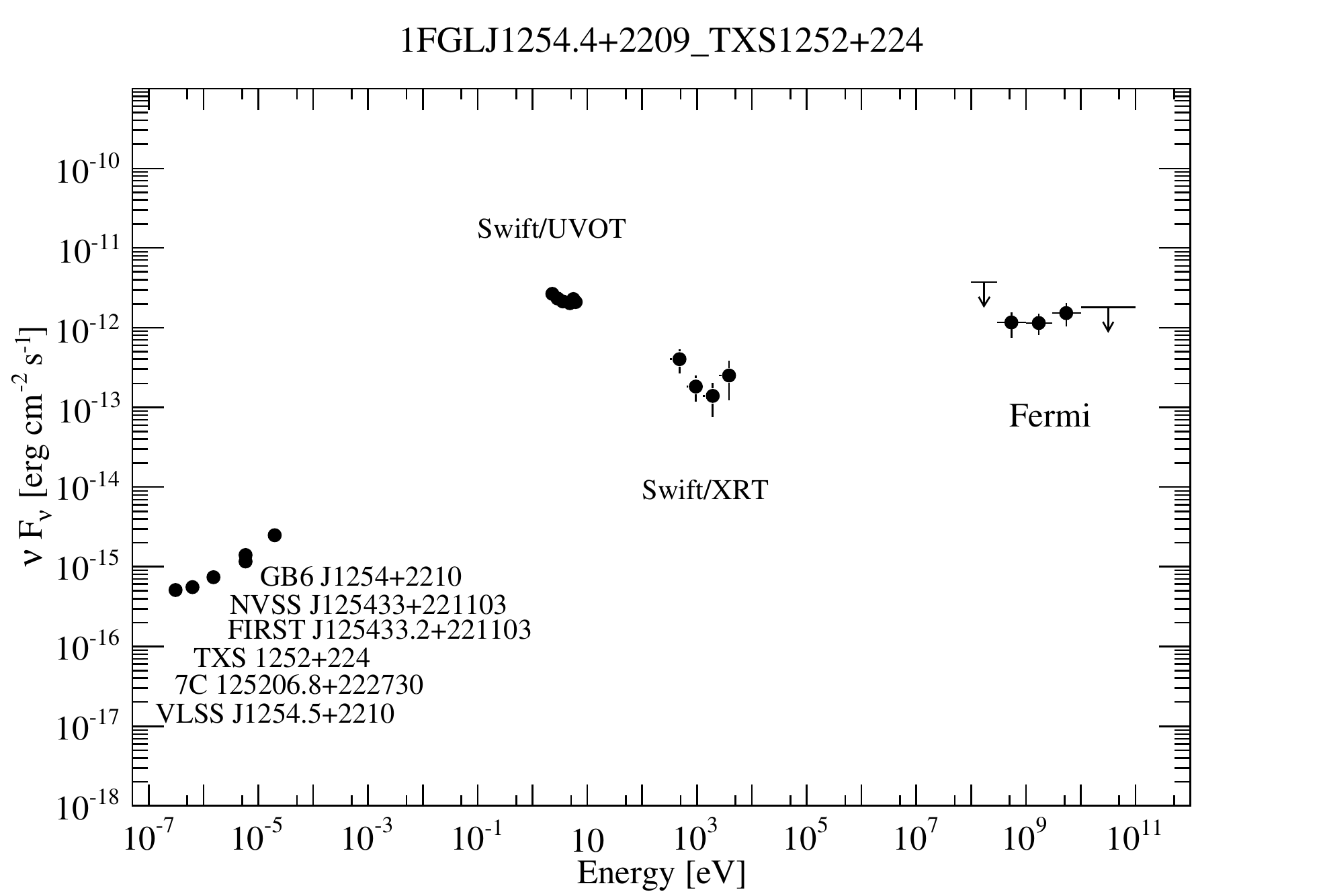}
    \end{center}
  \end{minipage}
  \begin{minipage}{0.32\hsize}
    \begin{center}
      \includegraphics[width=55mm]{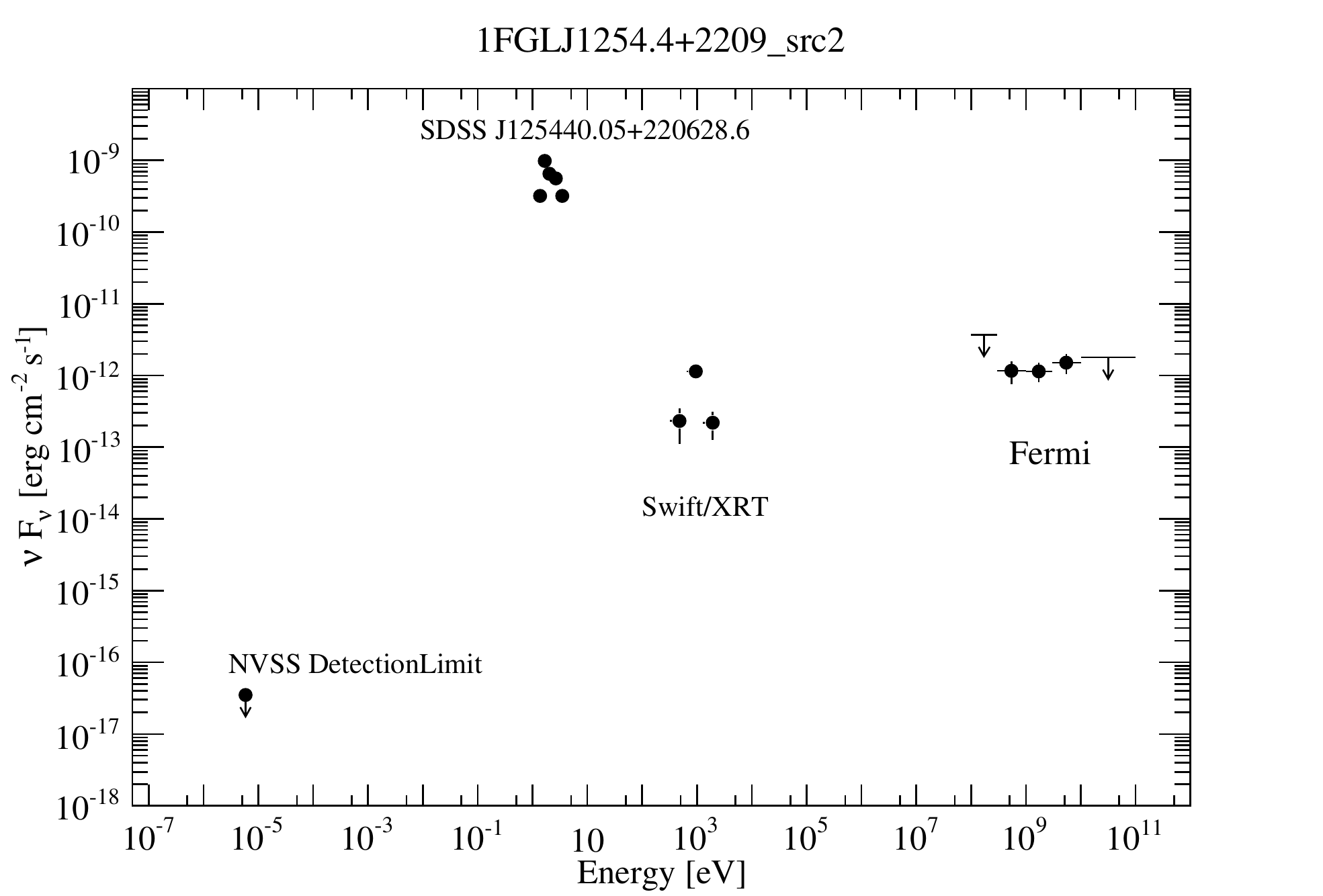}
    \end{center}
  \end{minipage}
 \end{center}
\end{figure}
\clearpage
\begin{figure}[m]
 \begin{center}
  \begin{minipage}{0.32\hsize}
    \begin{center}
      \includegraphics[width=55mm]{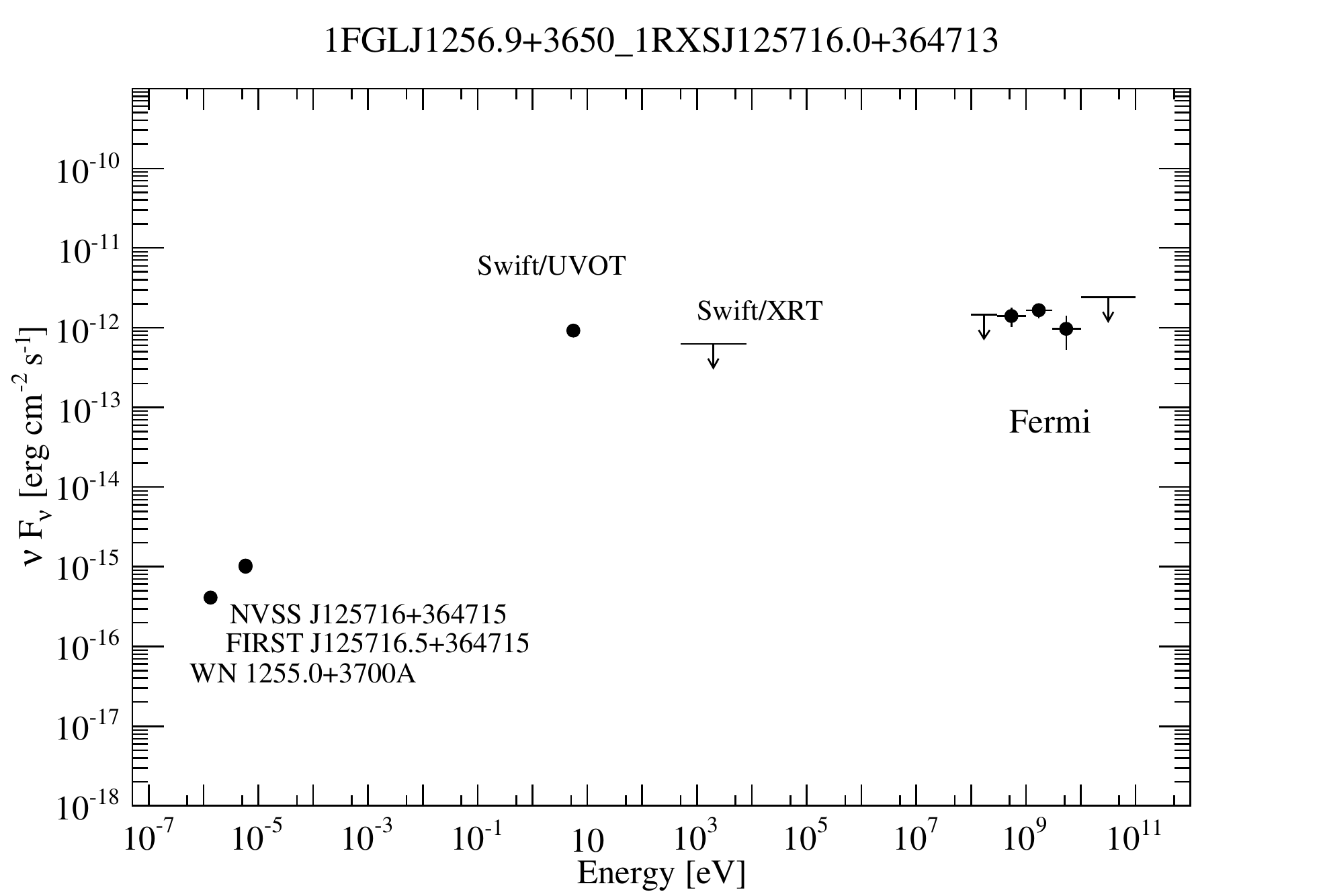}
    \end{center}
  \end{minipage}
  \begin{minipage}{0.32\hsize}
    \begin{center}
      \includegraphics[width=55mm]{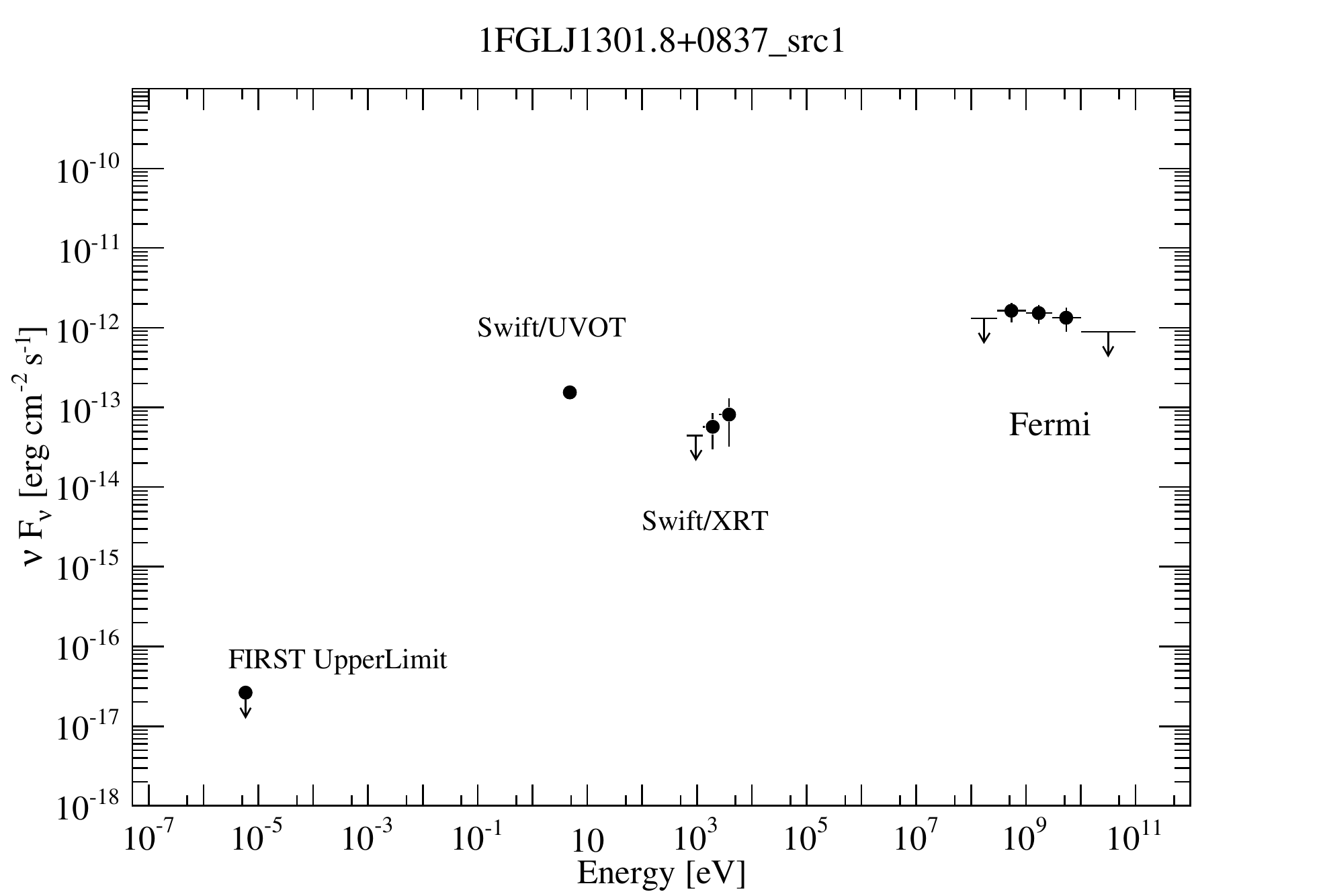}
    \end{center}
  \end{minipage}
  \begin{minipage}{0.32\hsize}
    \begin{center}
      \includegraphics[width=55mm]{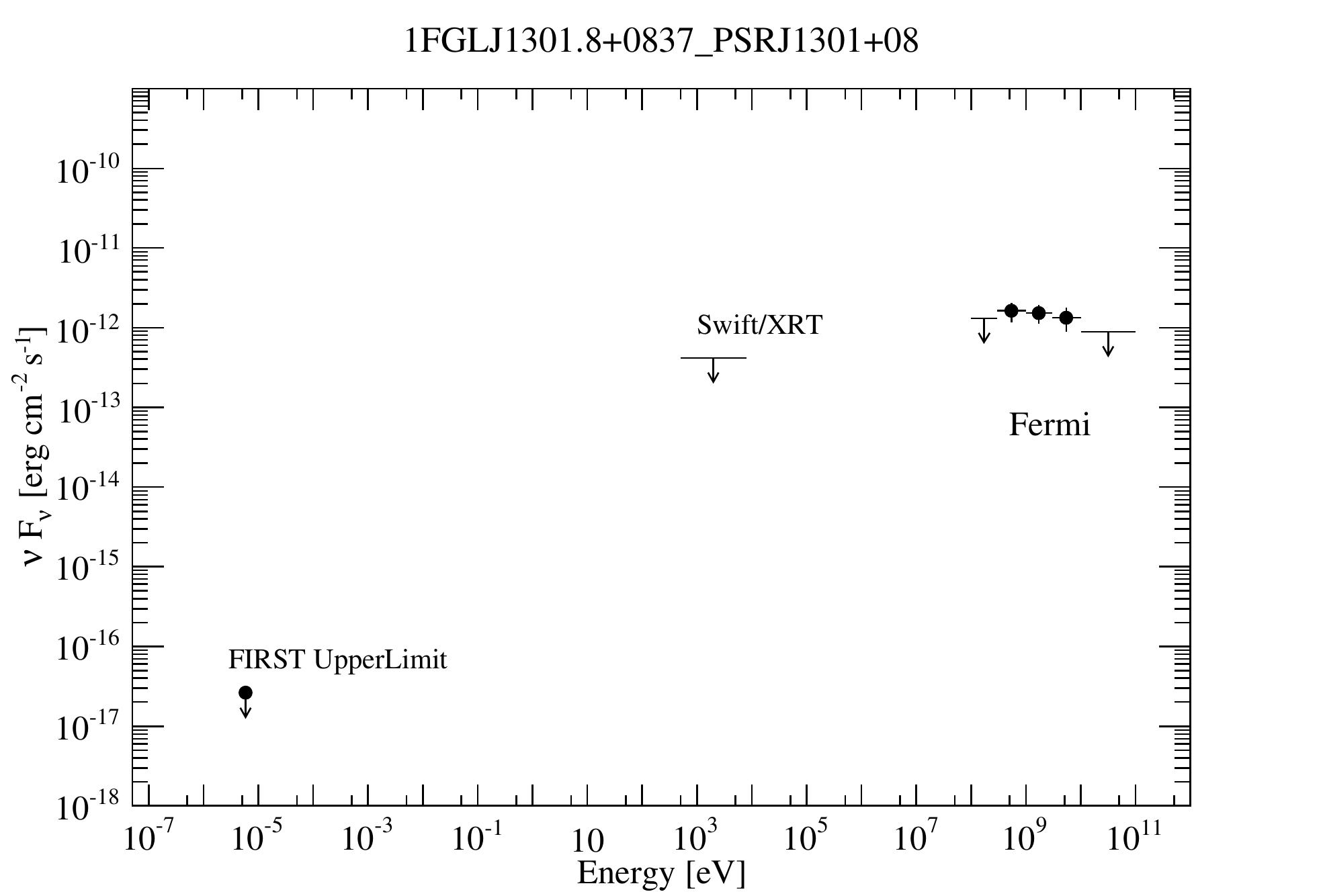}
    \end{center}
  \end{minipage}
  \begin{minipage}{0.32\hsize}
    \begin{center}
      \includegraphics[width=55mm]{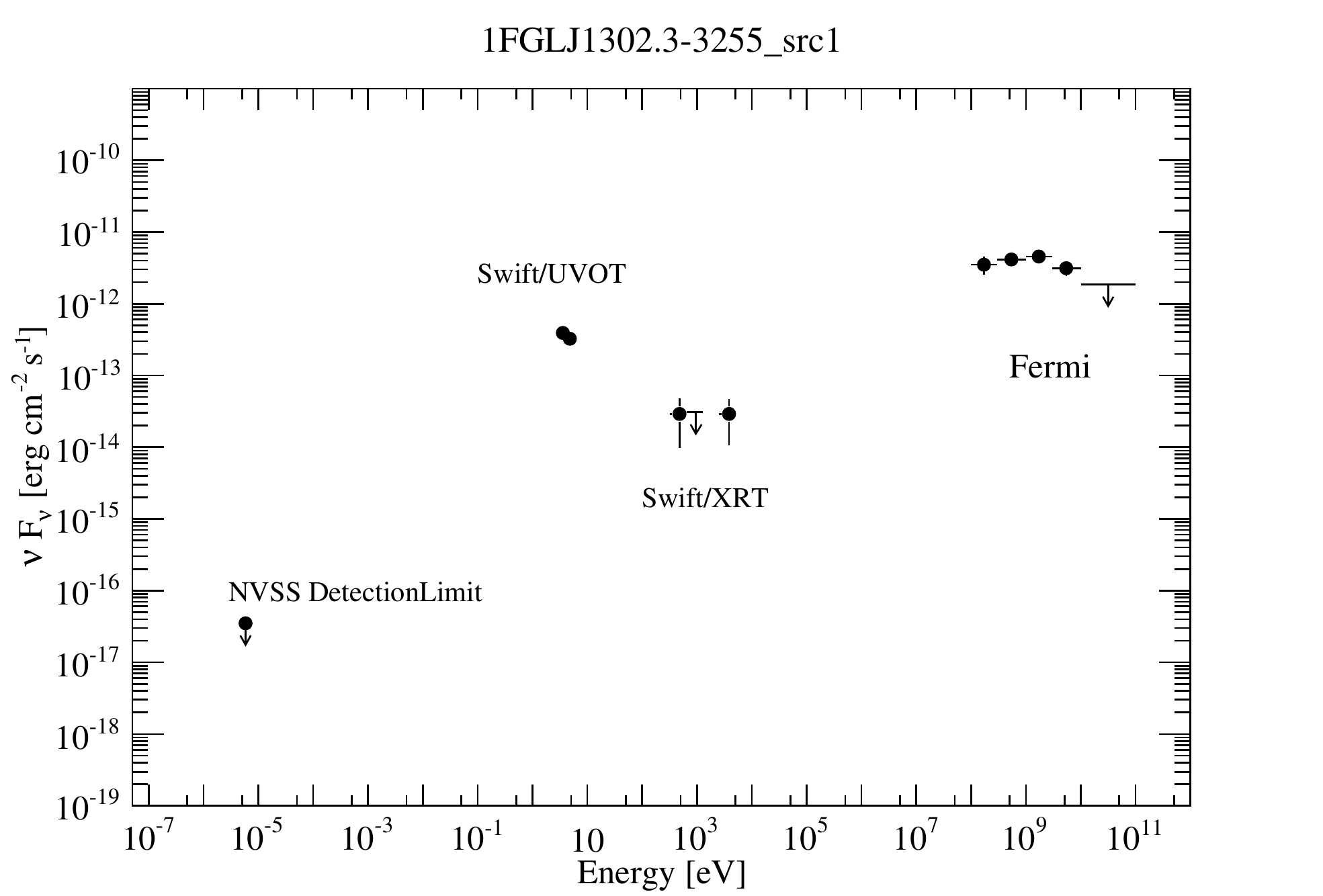}
    \end{center}
  \end{minipage}
  \begin{minipage}{0.32\hsize}
    \begin{center}
      \includegraphics[width=55mm]{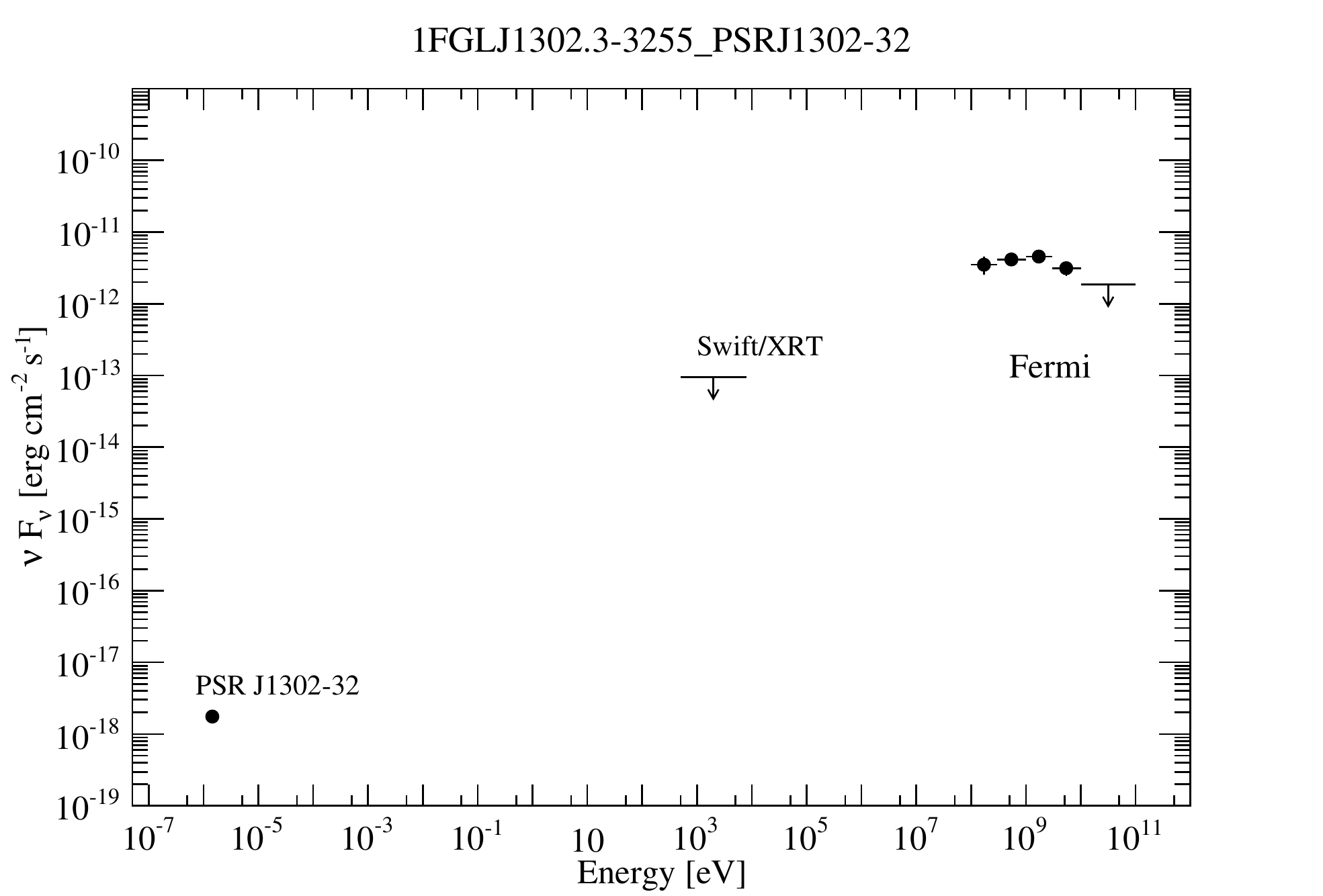}
    \end{center}
  \end{minipage}
  \begin{minipage}{0.32\hsize}
    \begin{center}
      \includegraphics[width=55mm]{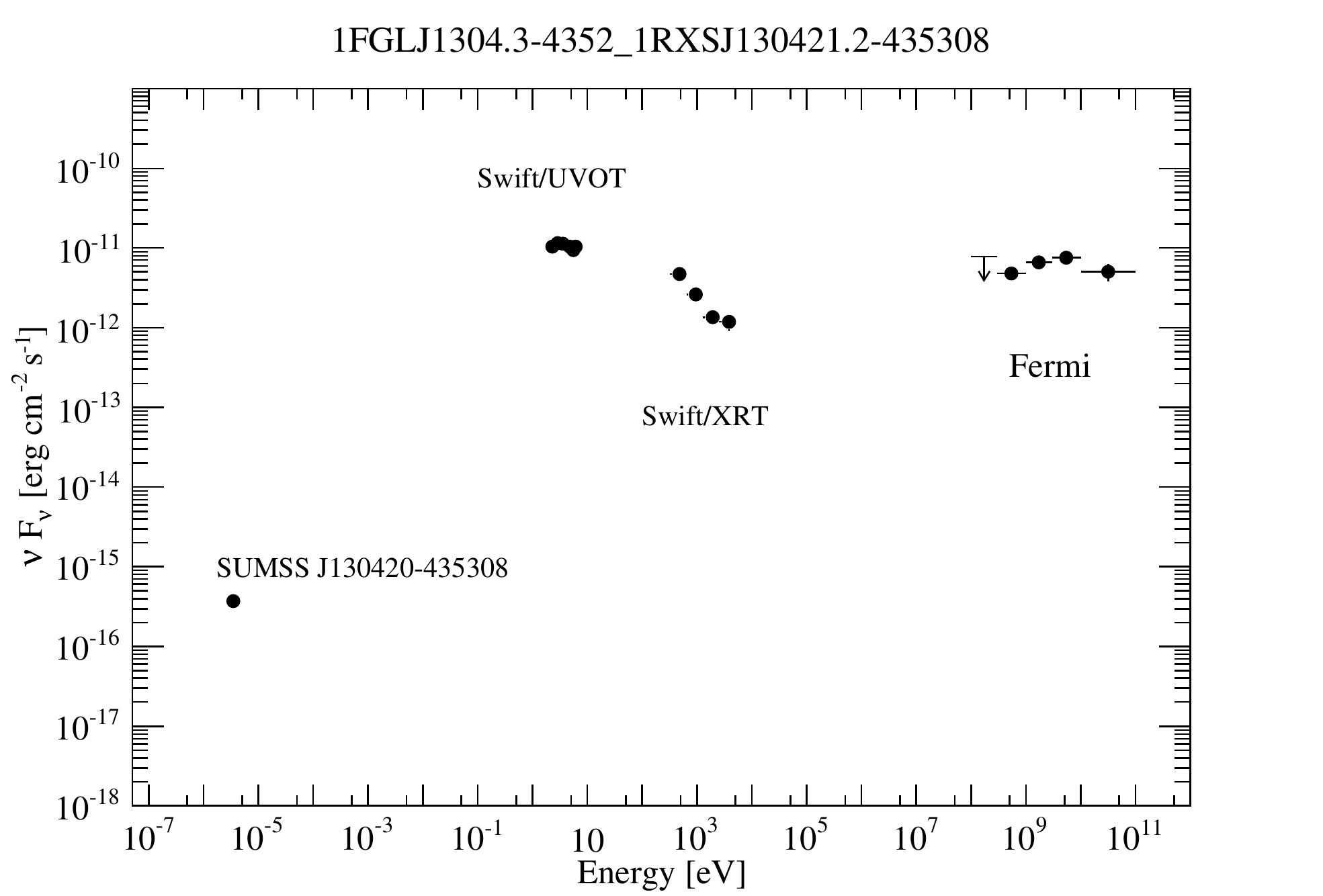}
    \end{center}
  \end{minipage}
  \begin{minipage}{0.32\hsize}
    \begin{center}
      \includegraphics[width=55mm]{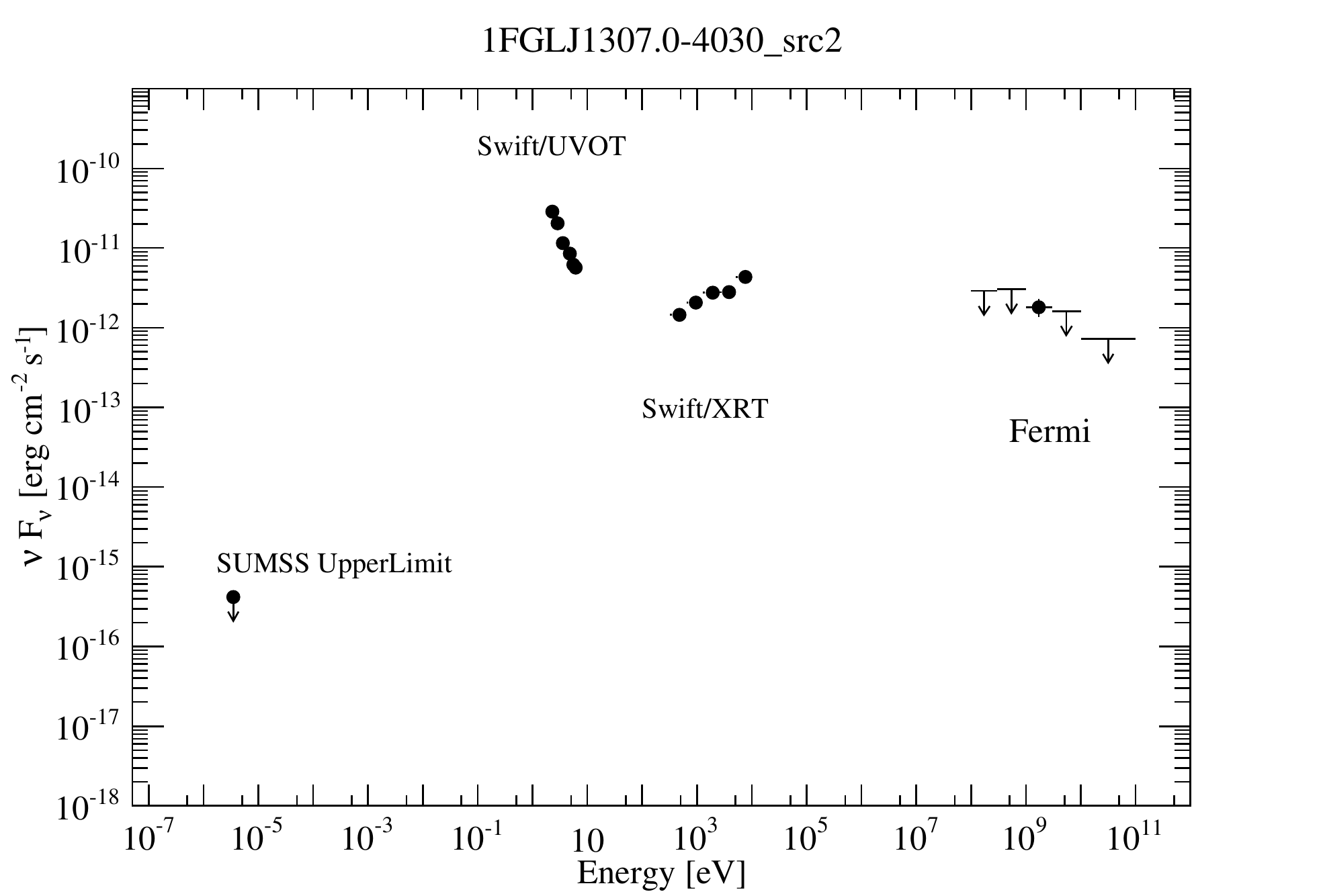}
    \end{center}
  \end{minipage}
  \begin{minipage}{0.32\hsize}
    \begin{center}
      \includegraphics[width=55mm]{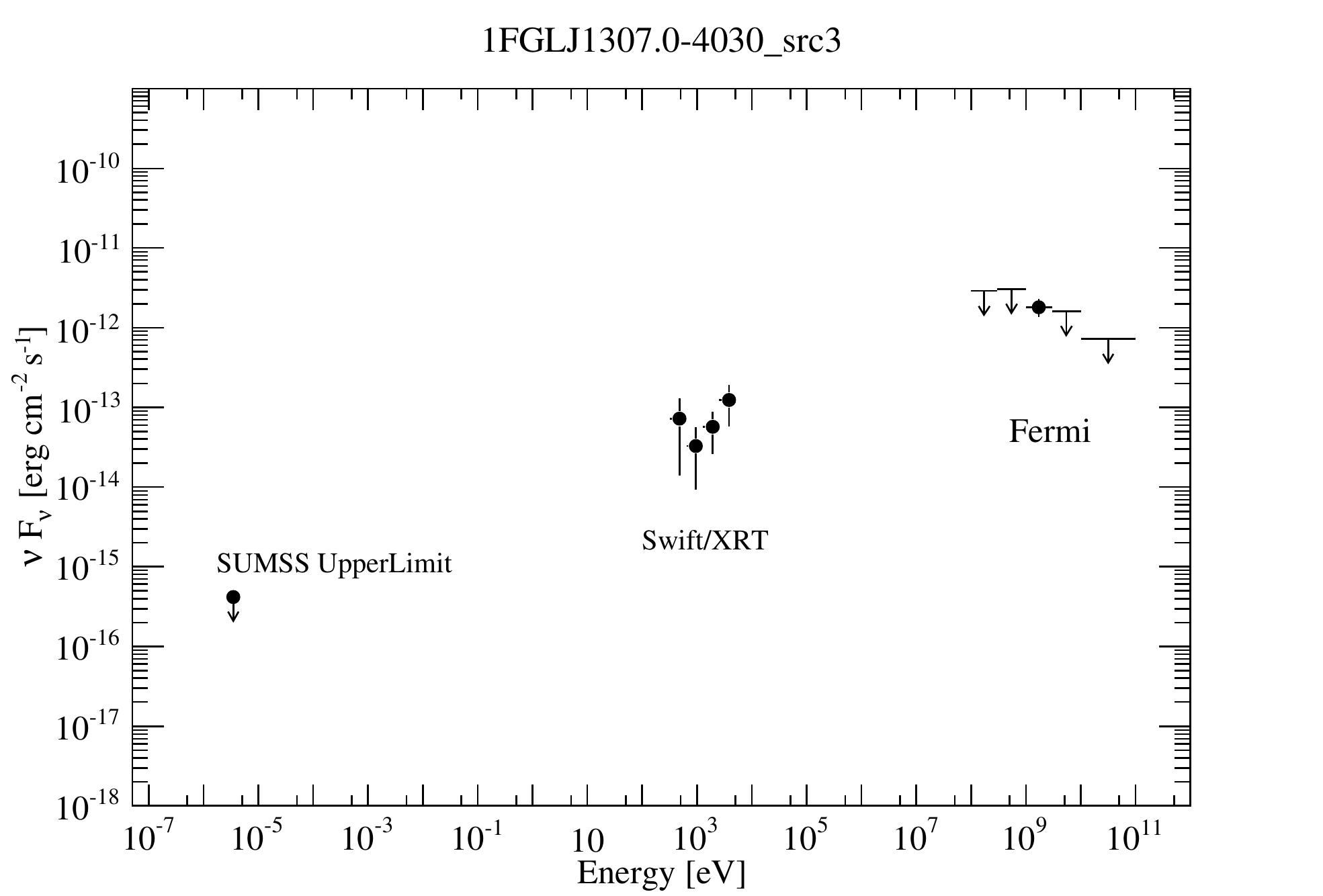}
    \end{center}
  \end{minipage}
  \begin{minipage}{0.32\hsize}
    \begin{center}
      \includegraphics[width=55mm]{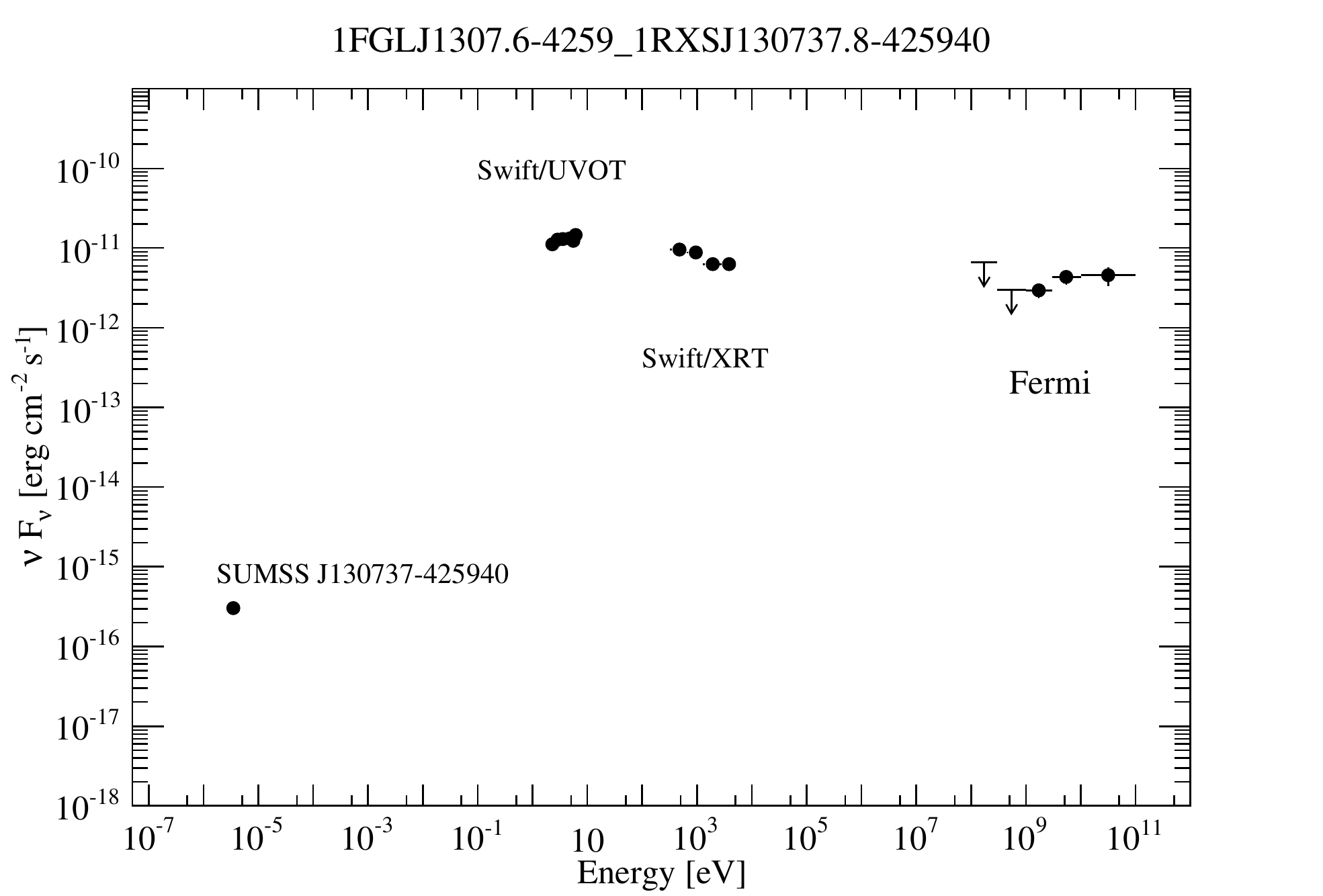}
    \end{center}
  \end{minipage}
  \begin{minipage}{0.32\hsize}
    \begin{center}
      \includegraphics[width=55mm]{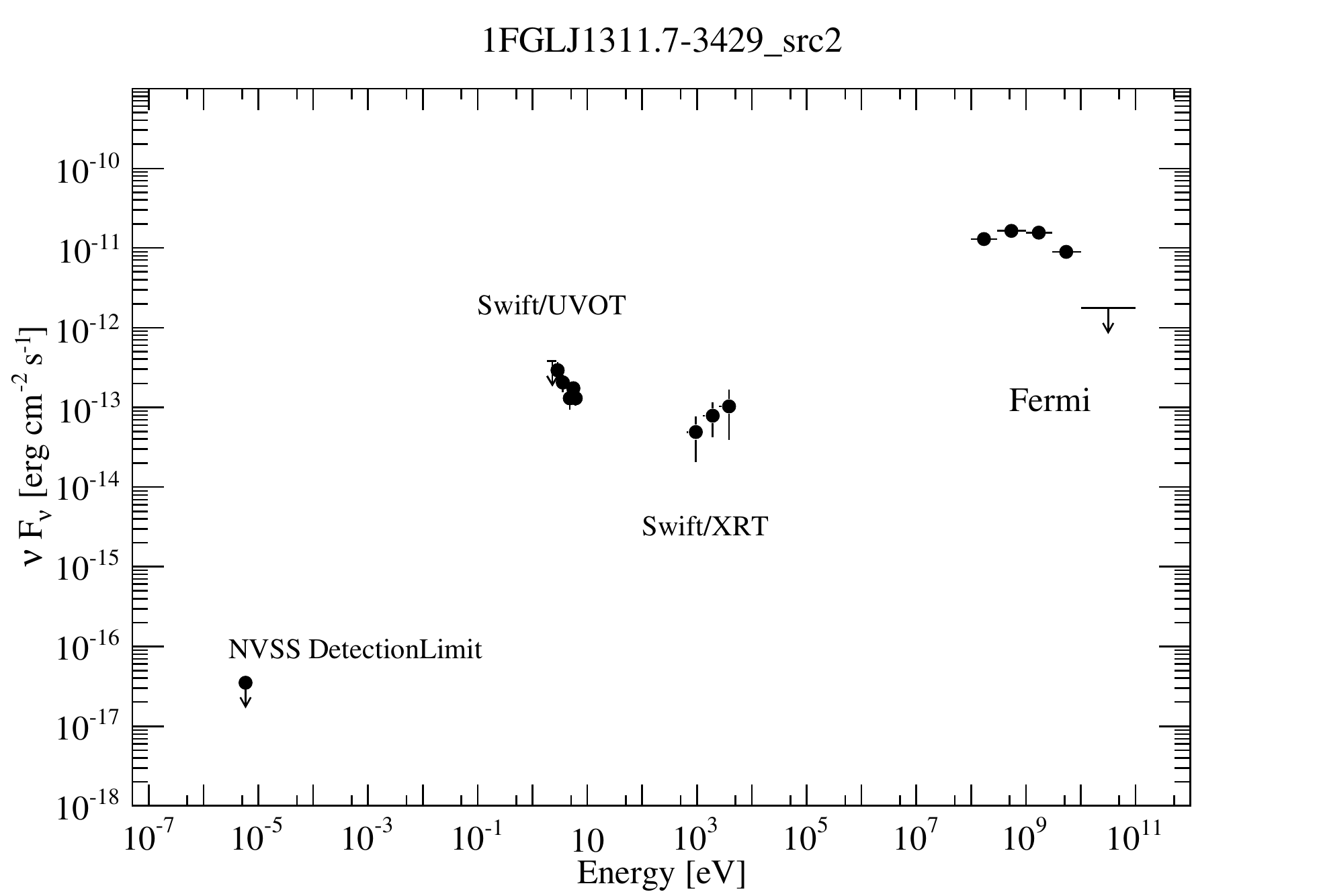}
    \end{center}
  \end{minipage}
  \begin{minipage}{0.32\hsize}
    \begin{center}
      \includegraphics[width=55mm]{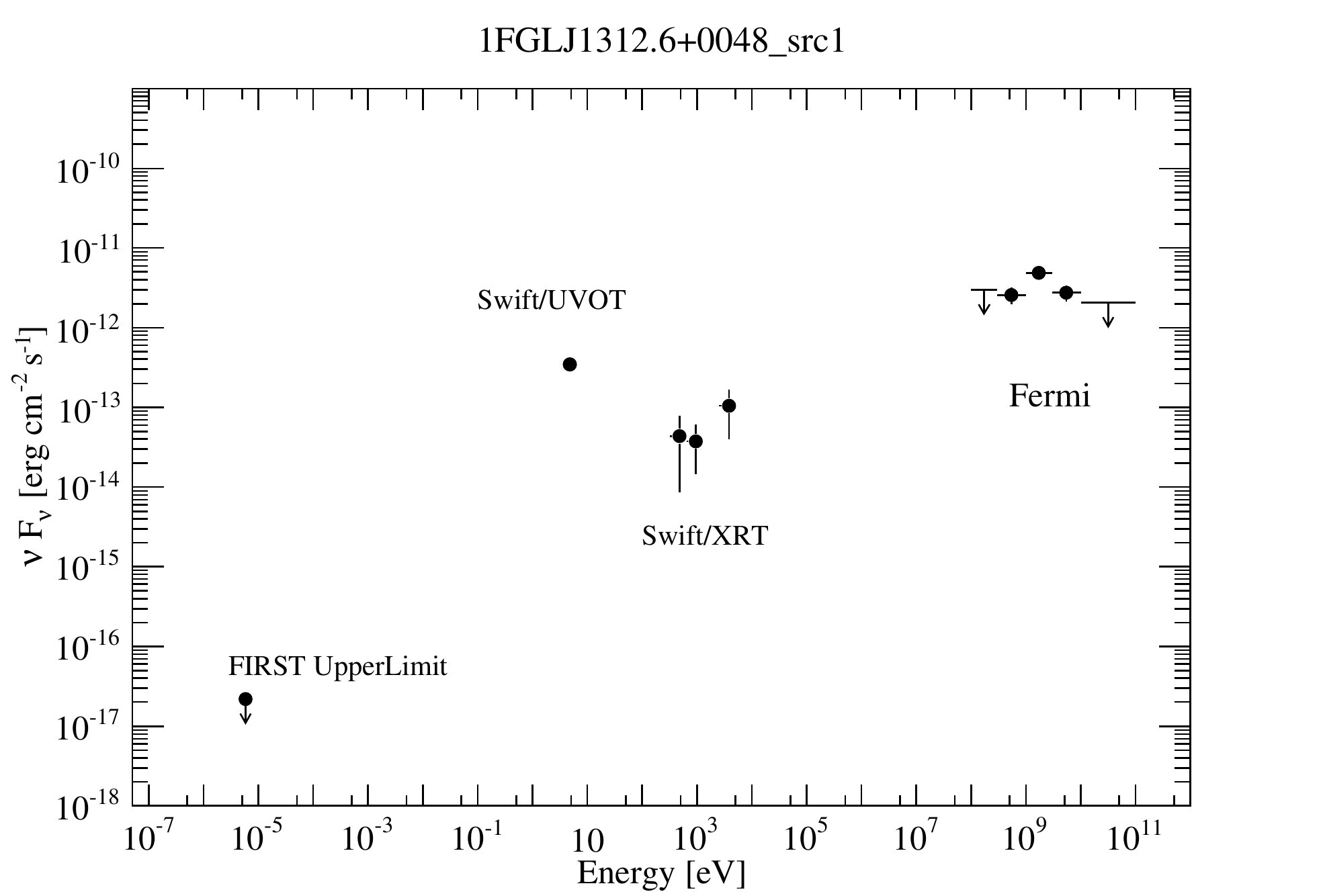}
    \end{center}
  \end{minipage}
  \begin{minipage}{0.32\hsize}
    \begin{center}
      \includegraphics[width=55mm]{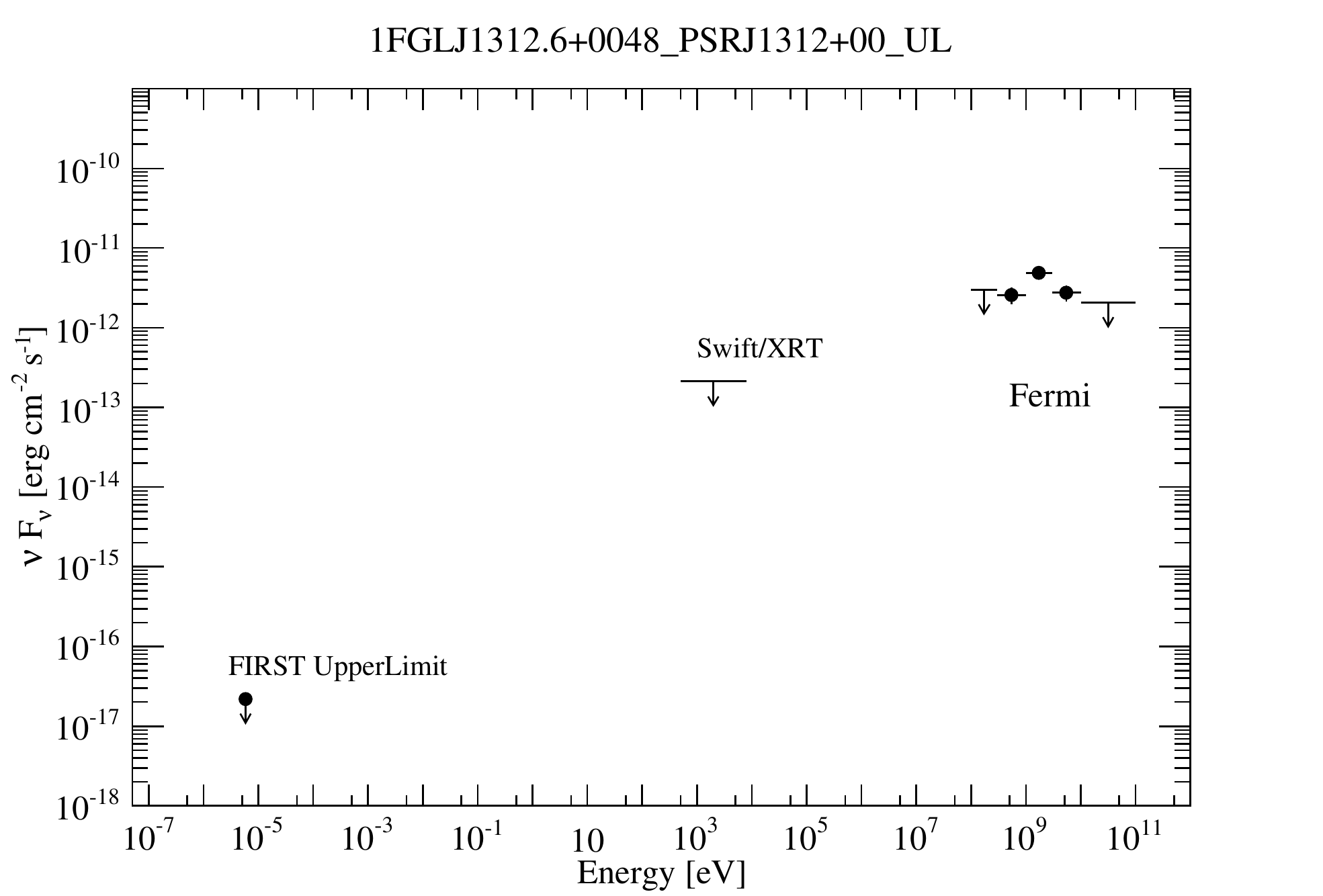}
    \end{center}
  \end{minipage}
  \begin{minipage}{0.32\hsize}
    \begin{center}
      \includegraphics[width=55mm]{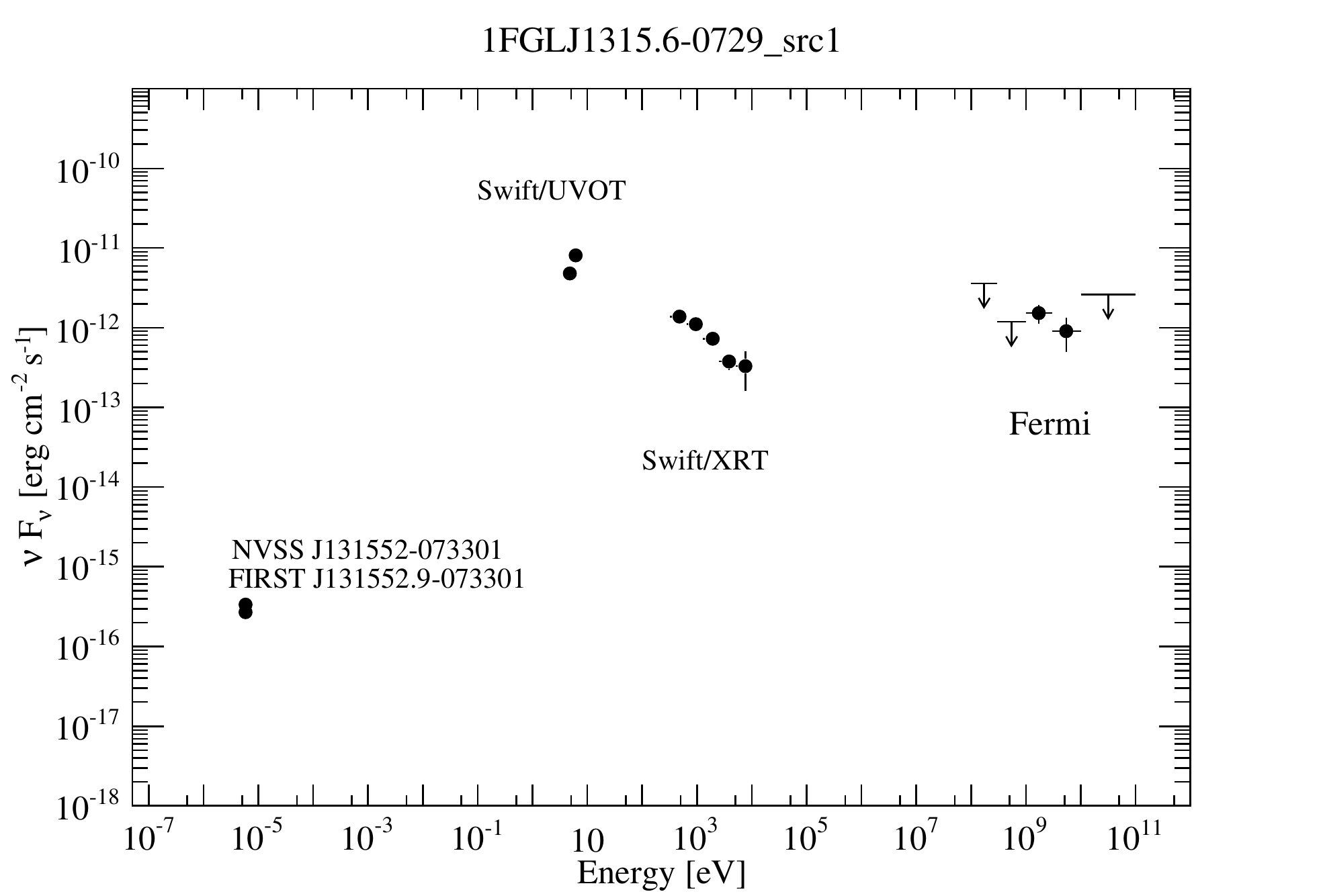}
    \end{center}
  \end{minipage}
  \begin{minipage}{0.32\hsize}
    \begin{center}
      \includegraphics[width=55mm]{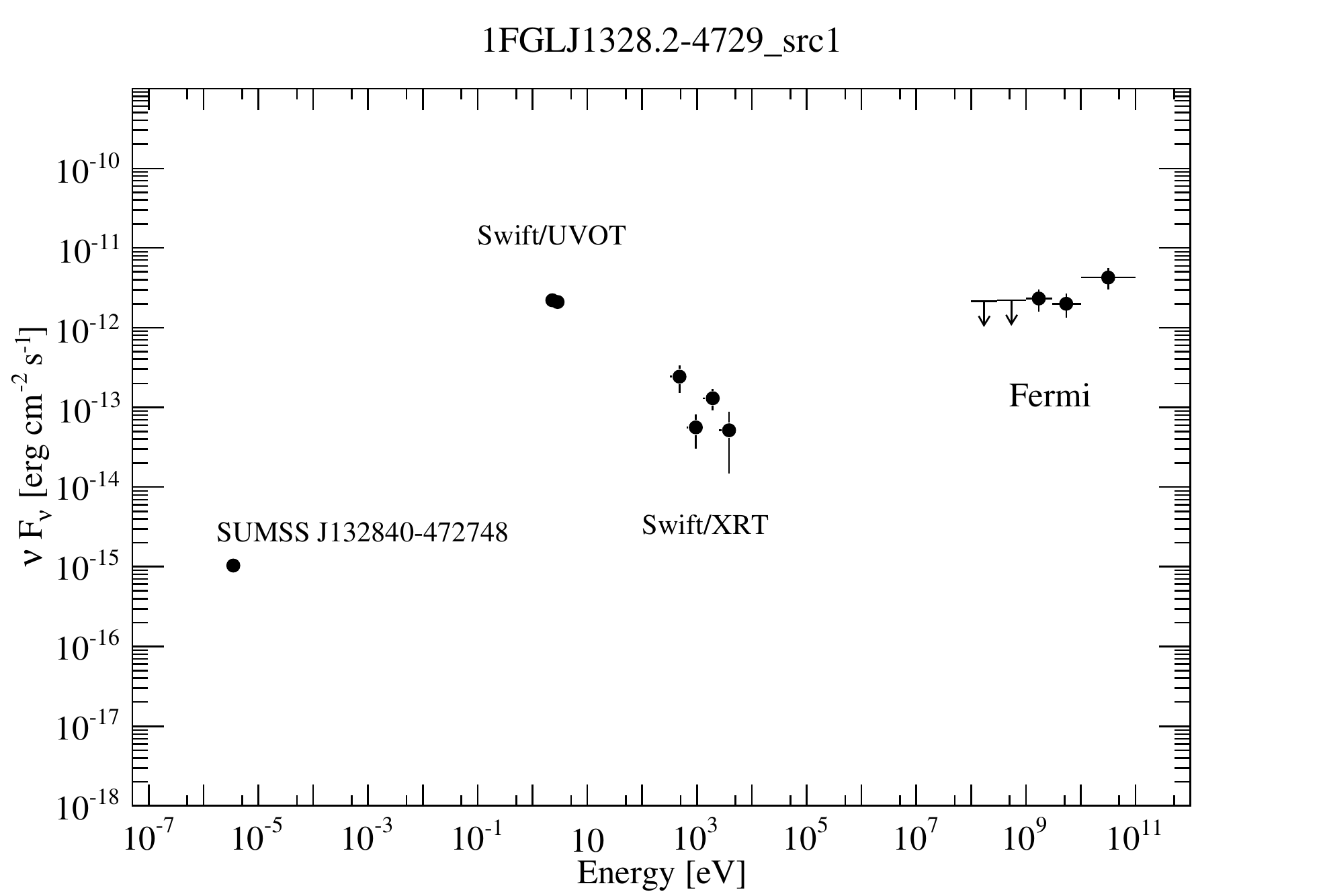}
    \end{center}
  \end{minipage}
  \begin{minipage}{0.32\hsize}
    \begin{center}
      \includegraphics[width=55mm]{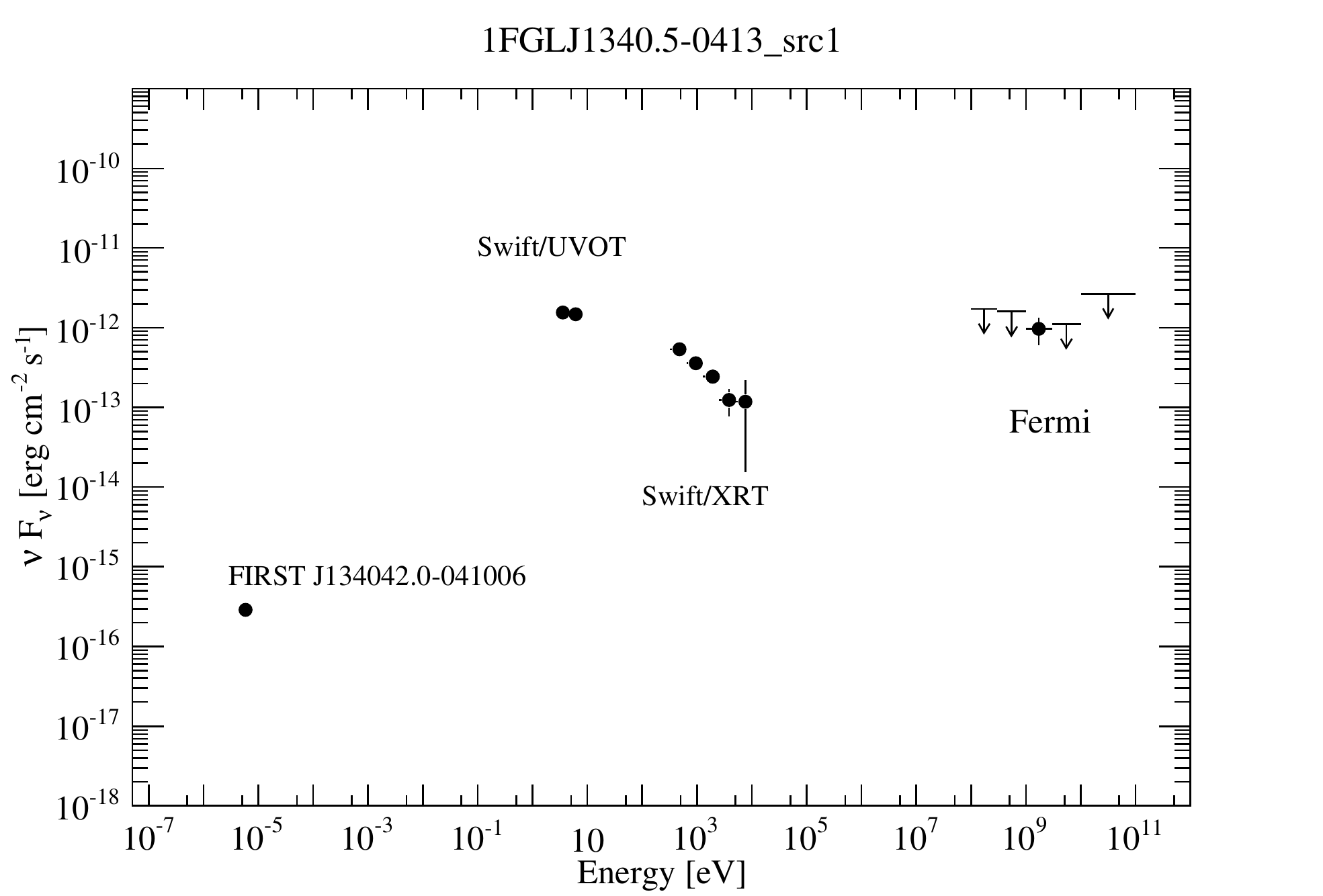}
    \end{center}
  \end{minipage}
  \begin{minipage}{0.32\hsize}
    \begin{center}
      \includegraphics[width=55mm]{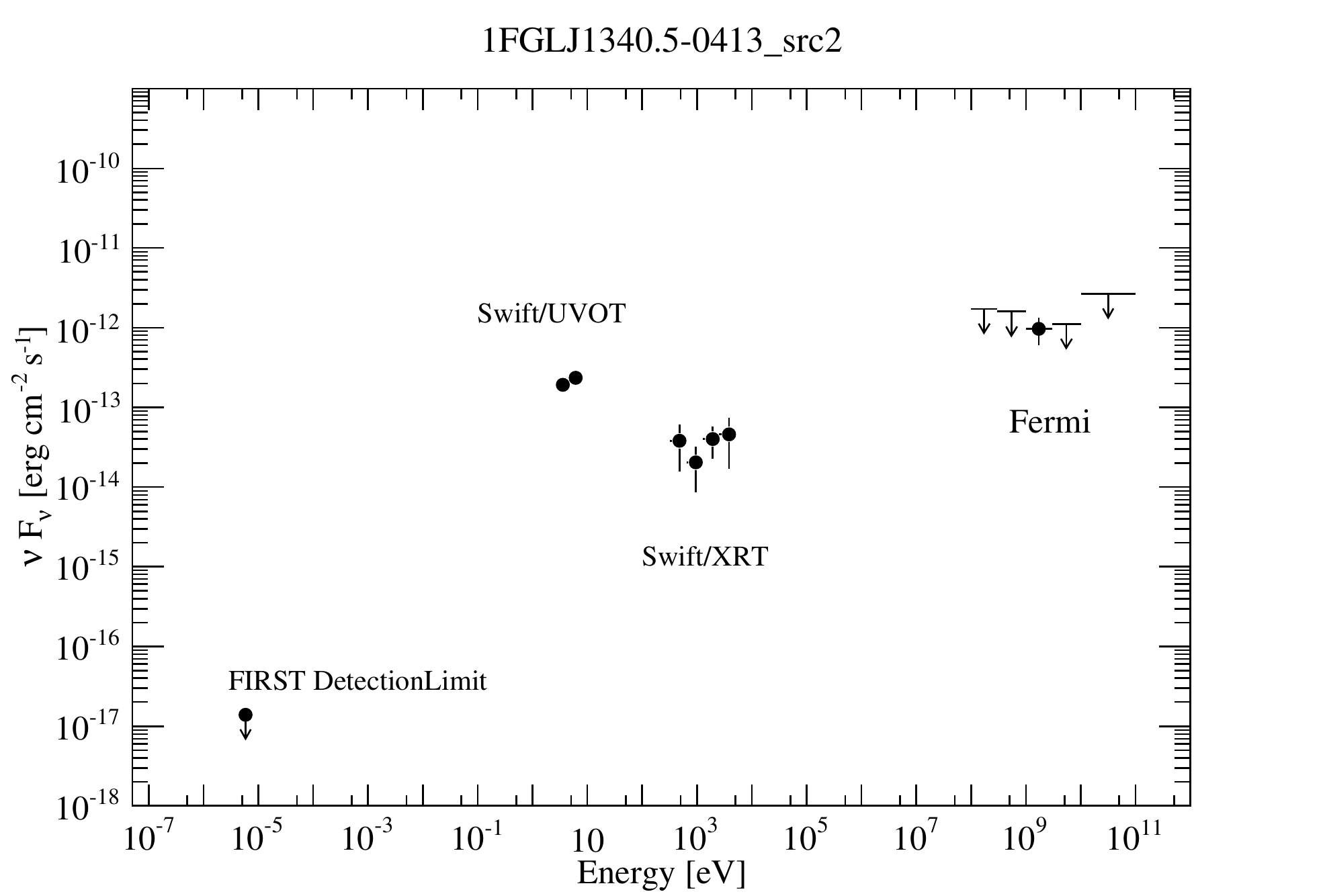}
    \end{center}
  \end{minipage}
  \begin{minipage}{0.32\hsize}
    \begin{center}
      \includegraphics[width=55mm]{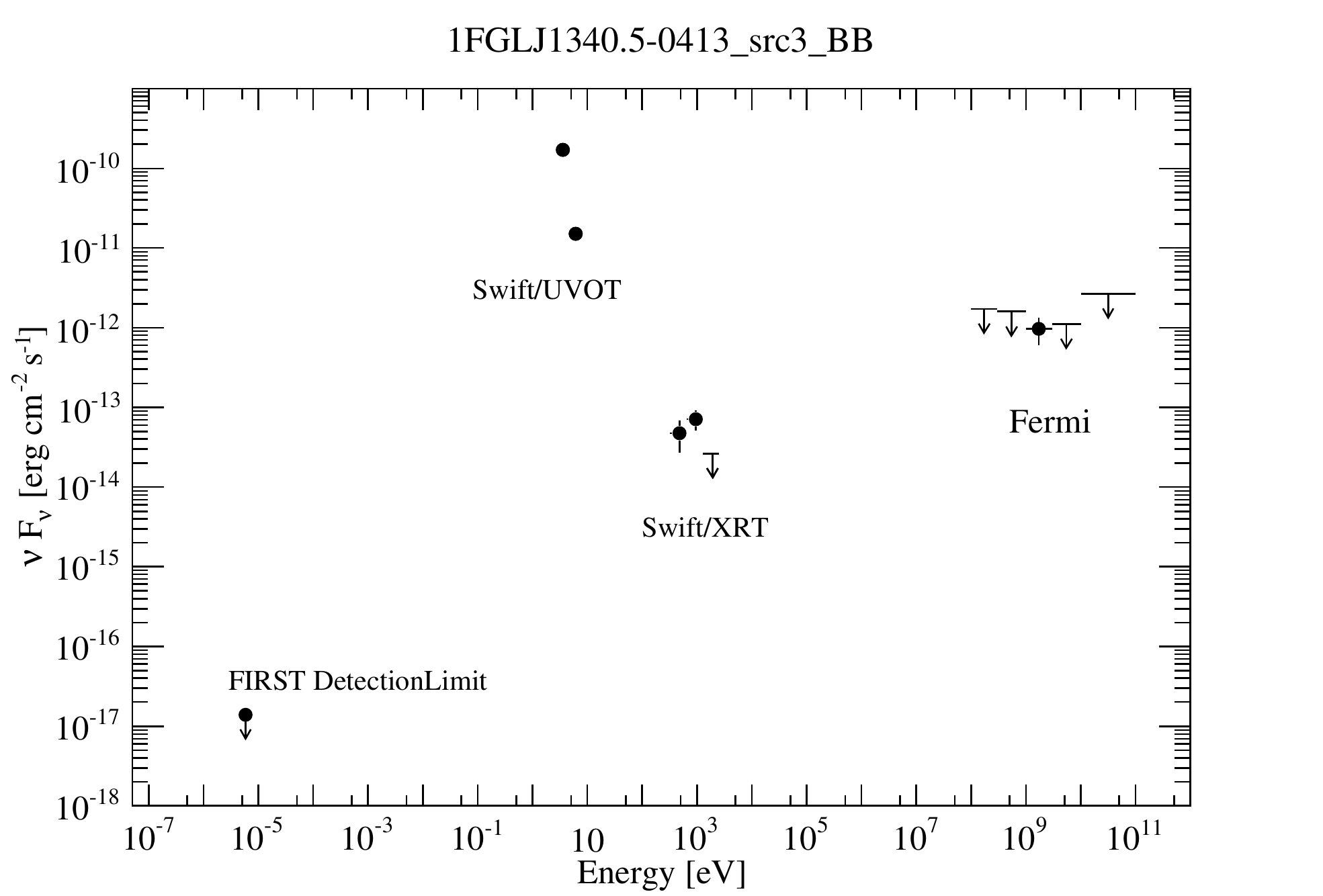}
    \end{center}
  \end{minipage}
  \begin{minipage}{0.32\hsize}
    \begin{center}
      \includegraphics[width=55mm]{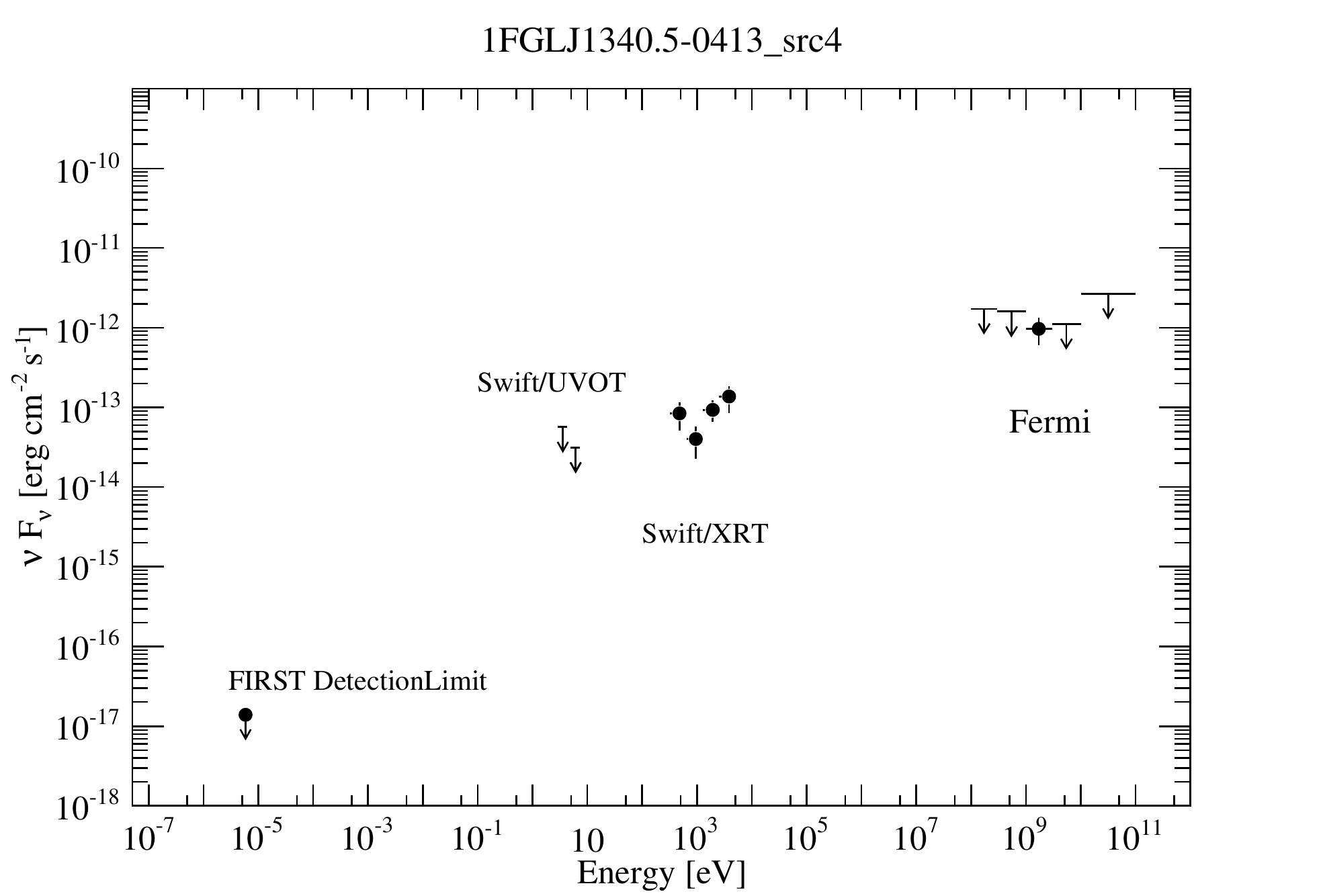}
    \end{center}
  \end{minipage}
 \end{center}
\end{figure}
\clearpage
\begin{figure}[m]
 \begin{center}
  \begin{minipage}{0.32\hsize}
    \begin{center}
      \includegraphics[width=55mm]{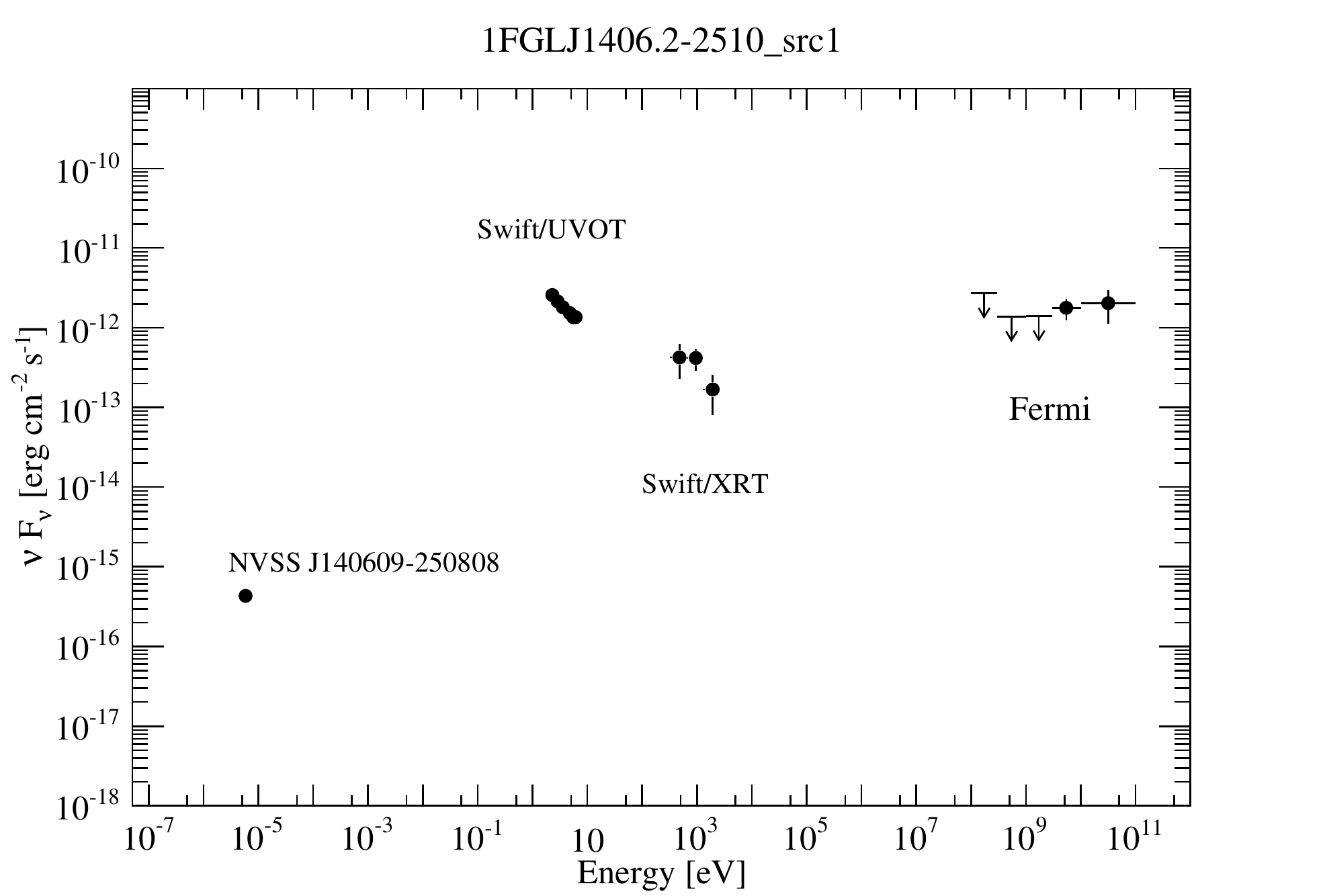}
    \end{center}
  \end{minipage}
  \begin{minipage}{0.32\hsize}
    \begin{center}
      \includegraphics[width=55mm]{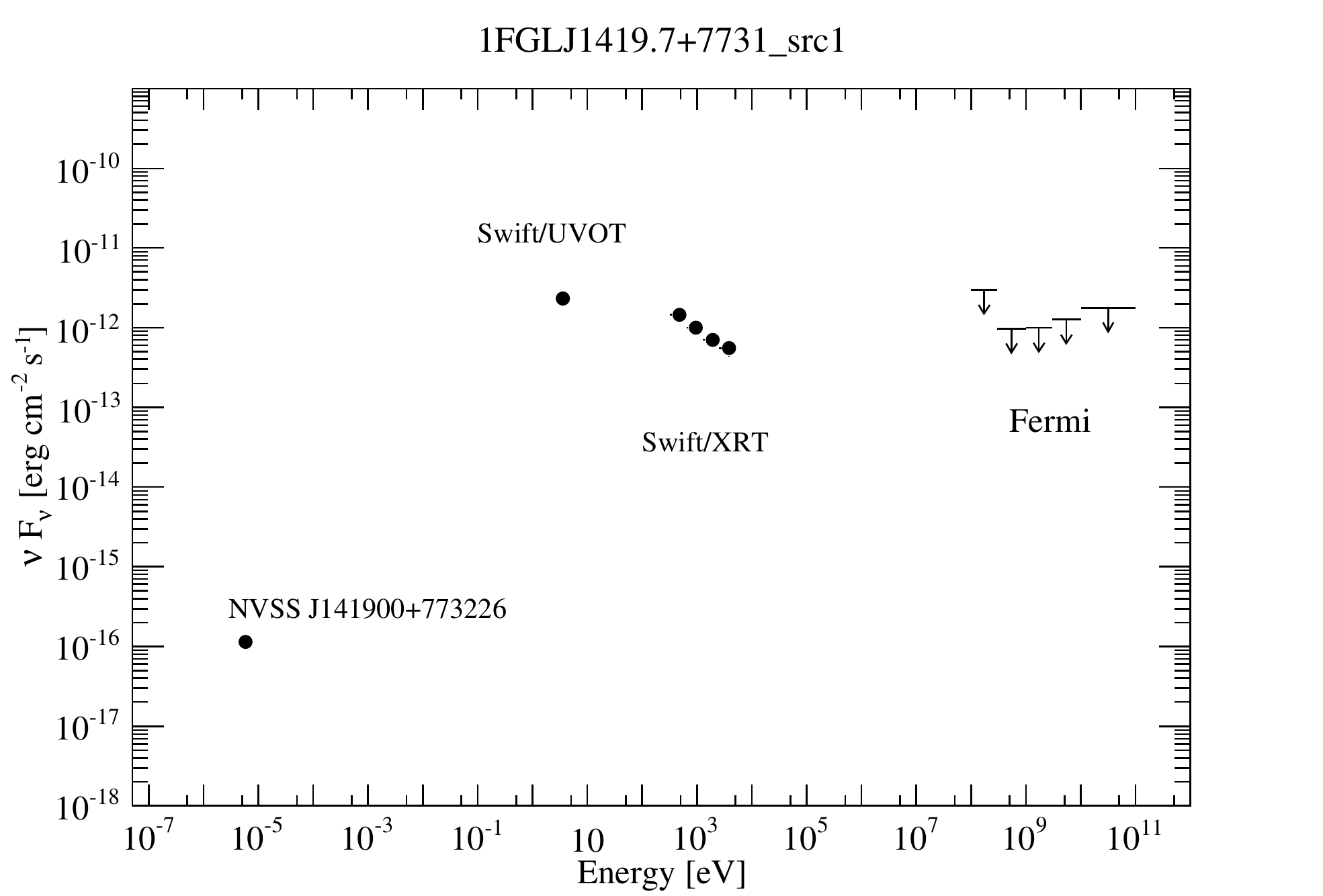}
    \end{center}
  \end{minipage}
  \begin{minipage}{0.32\hsize}
    \begin{center}
      \includegraphics[width=55mm]{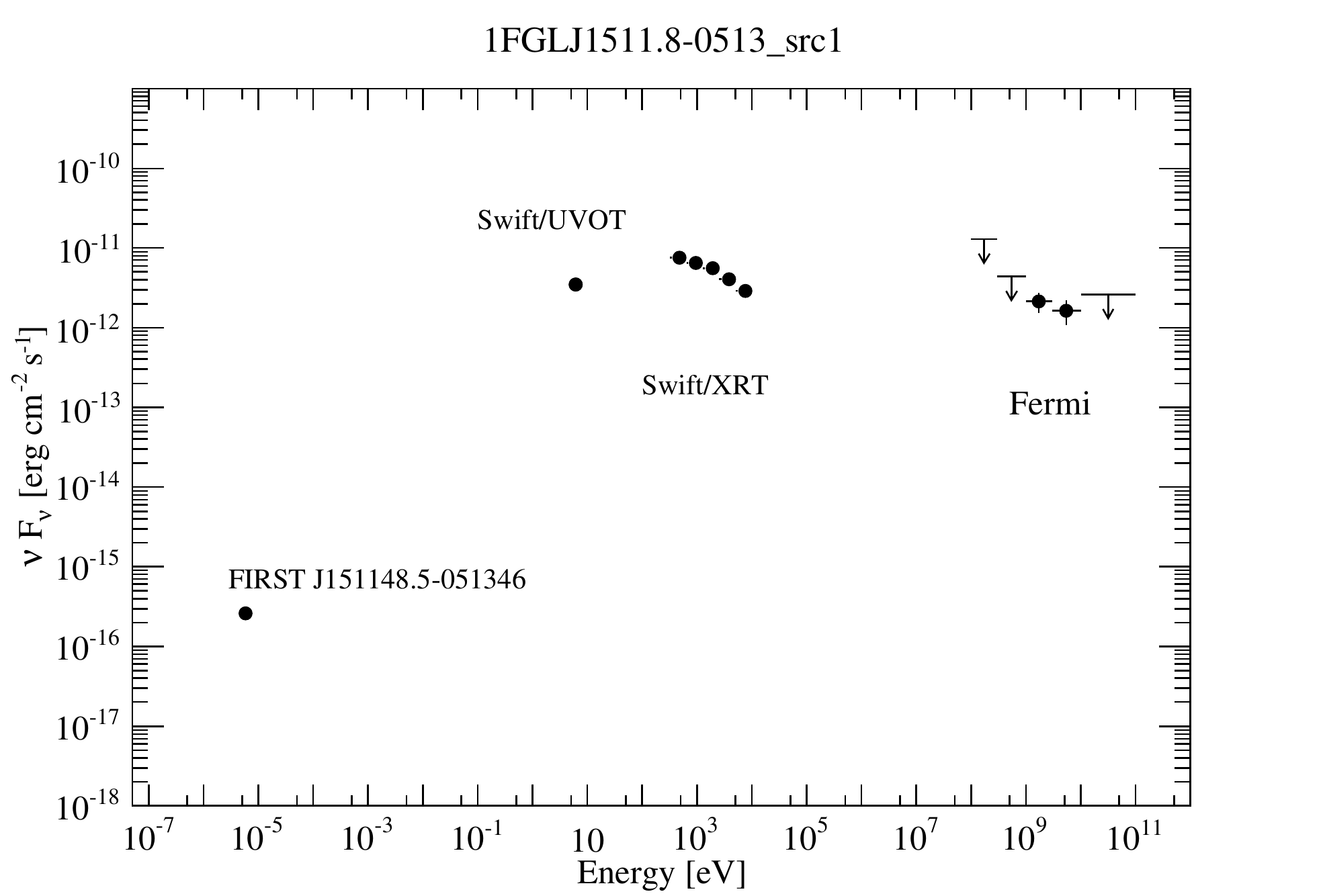}
    \end{center}
  \end{minipage}
  \begin{minipage}{0.32\hsize}
    \begin{center}
      \includegraphics[width=55mm]{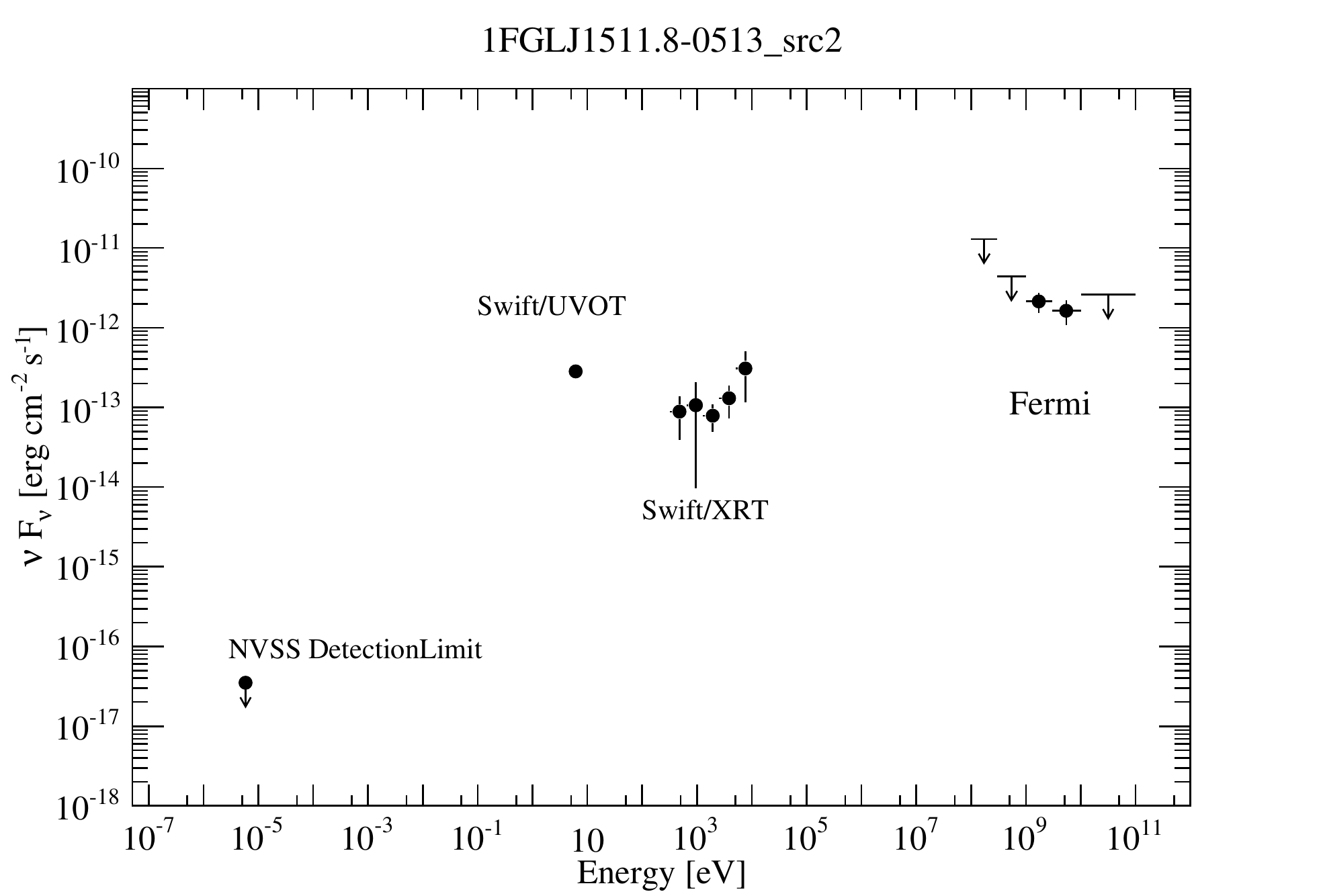}
    \end{center}
  \end{minipage}
  \begin{minipage}{0.32\hsize}
    \begin{center}
      \includegraphics[width=55mm]{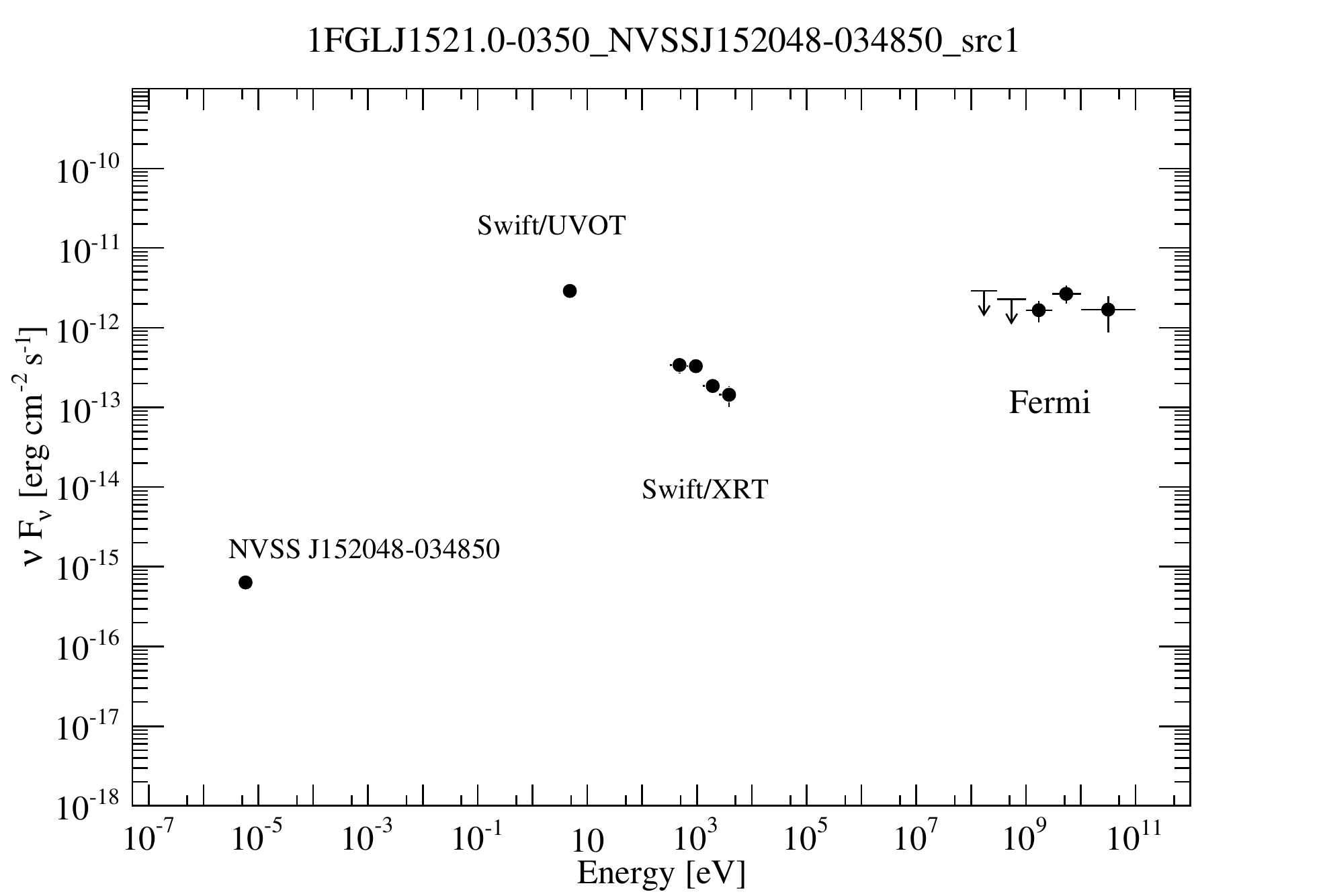}
    \end{center}
  \end{minipage}
  \begin{minipage}{0.32\hsize}
    \begin{center}
      \includegraphics[width=55mm]{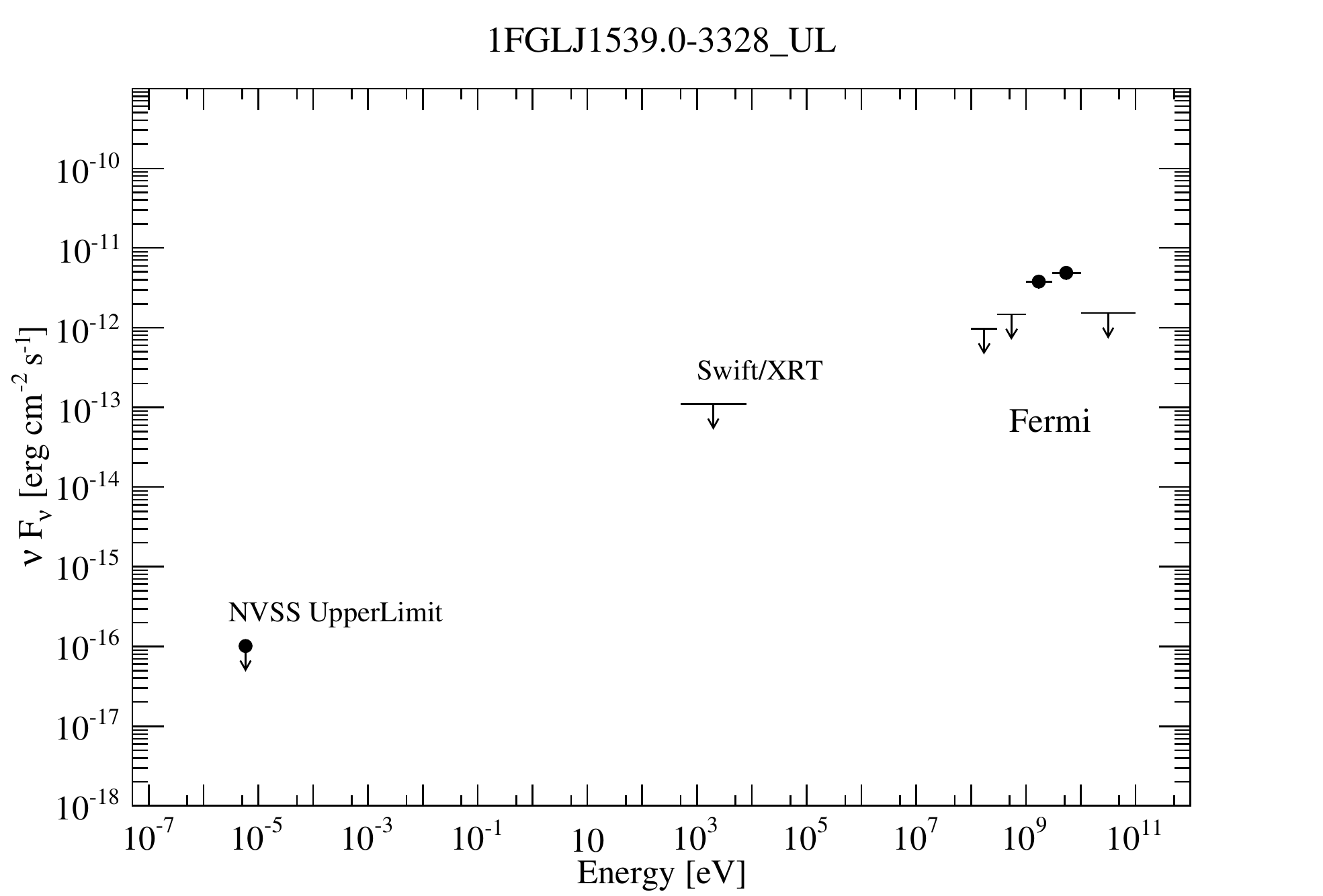}
    \end{center}
  \end{minipage}
  \begin{minipage}{0.32\hsize}
    \begin{center}
      \includegraphics[width=55mm]{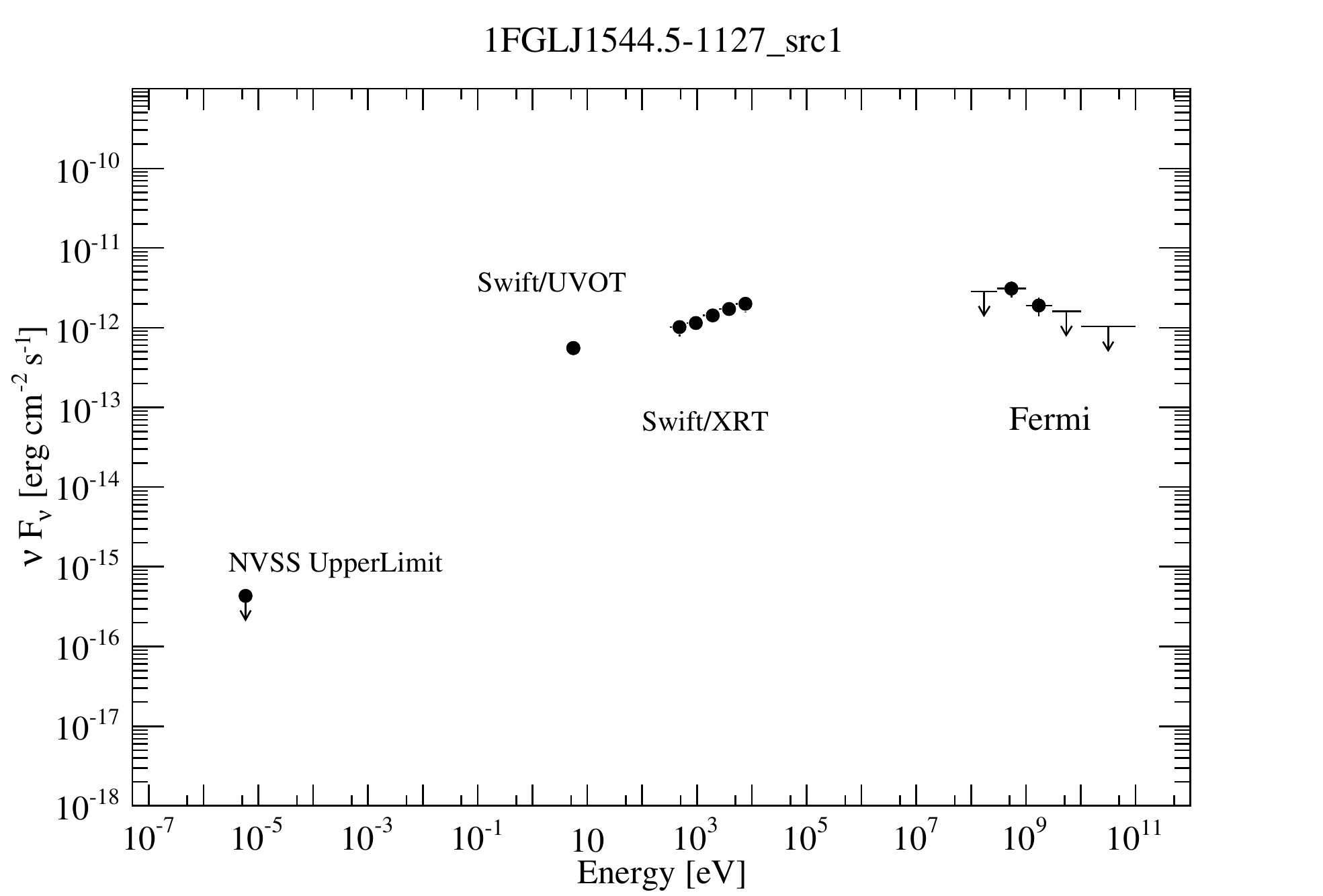}
    \end{center}
  \end{minipage}
  \begin{minipage}{0.32\hsize}
    \begin{center}
      \includegraphics[width=55mm]{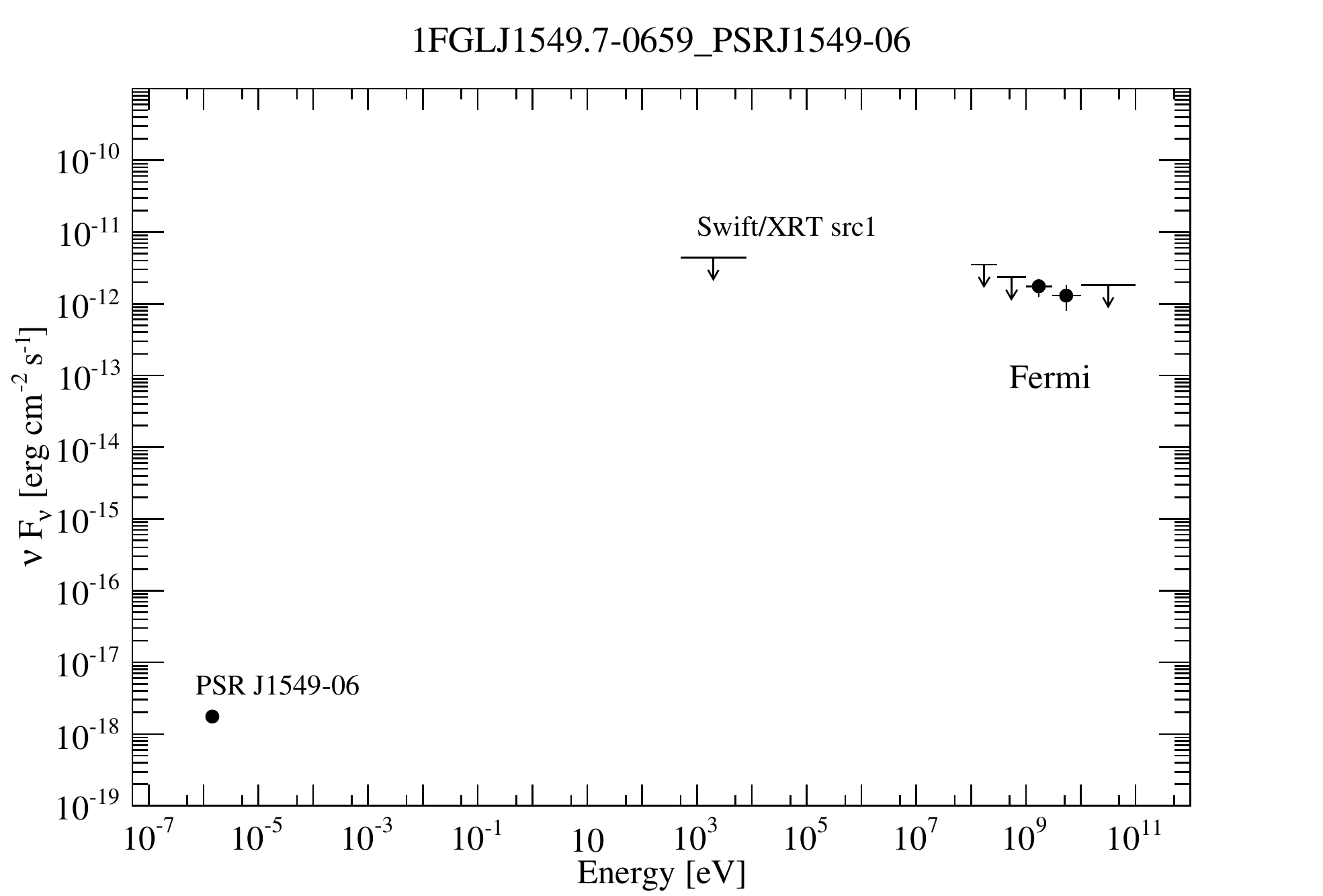}
    \end{center}
  \end{minipage}
  \begin{minipage}{0.32\hsize}
    \begin{center}
      \includegraphics[width=55mm]{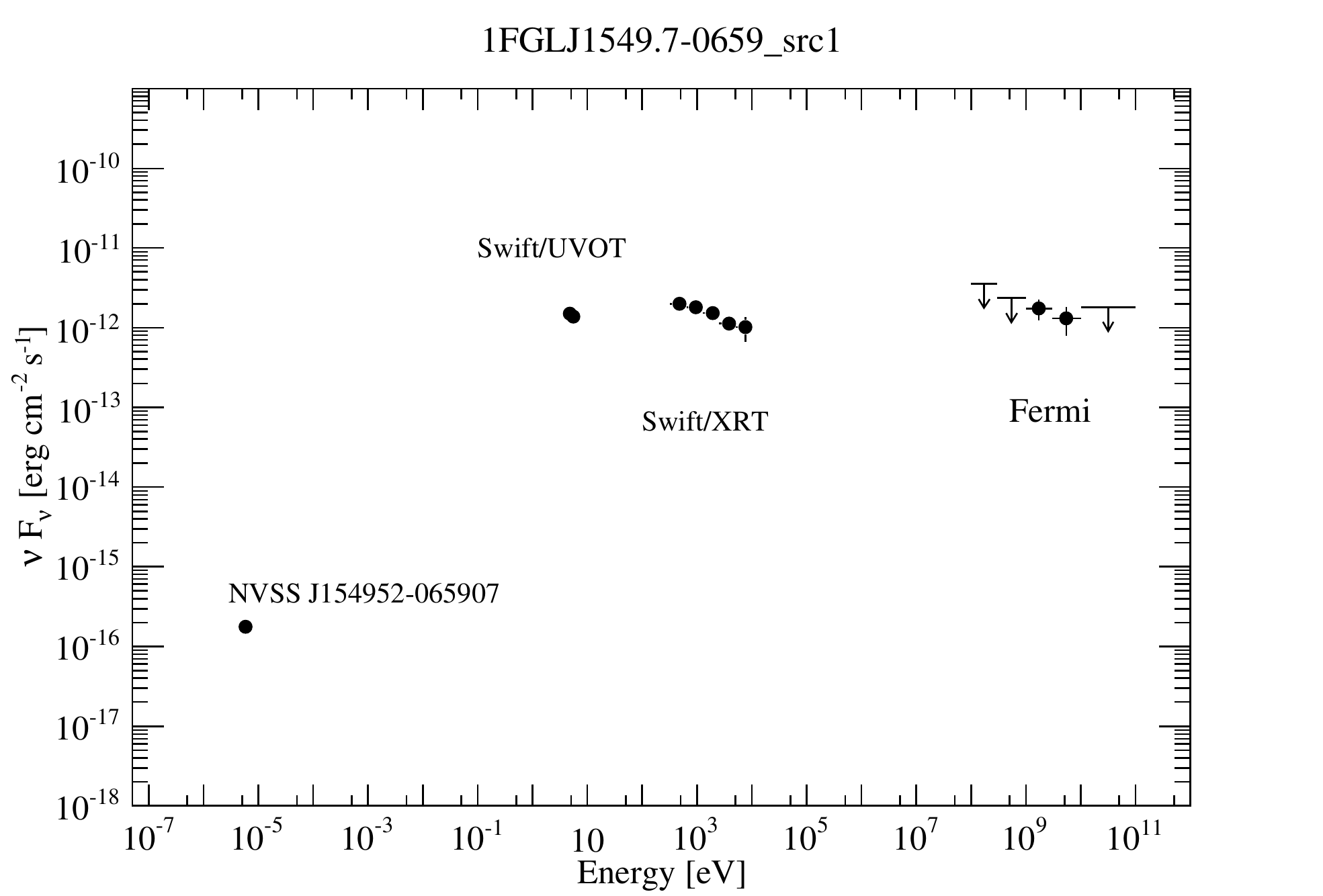}
    \end{center}
  \end{minipage}
  \begin{minipage}{0.32\hsize}
    \begin{center}
      \includegraphics[width=55mm]{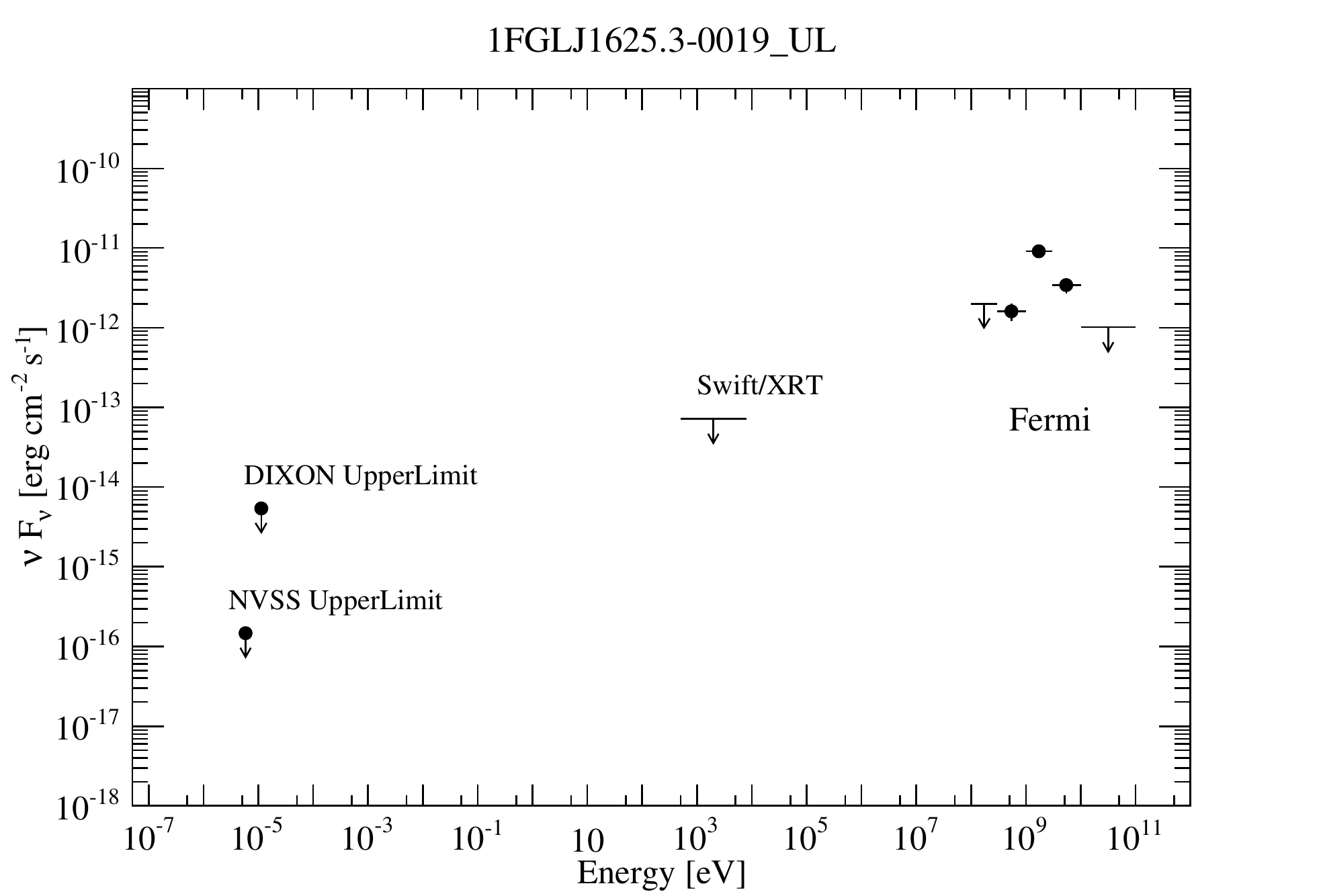}
    \end{center}
  \end{minipage}
  \begin{minipage}{0.32\hsize}
    \begin{center}
      \includegraphics[width=55mm]{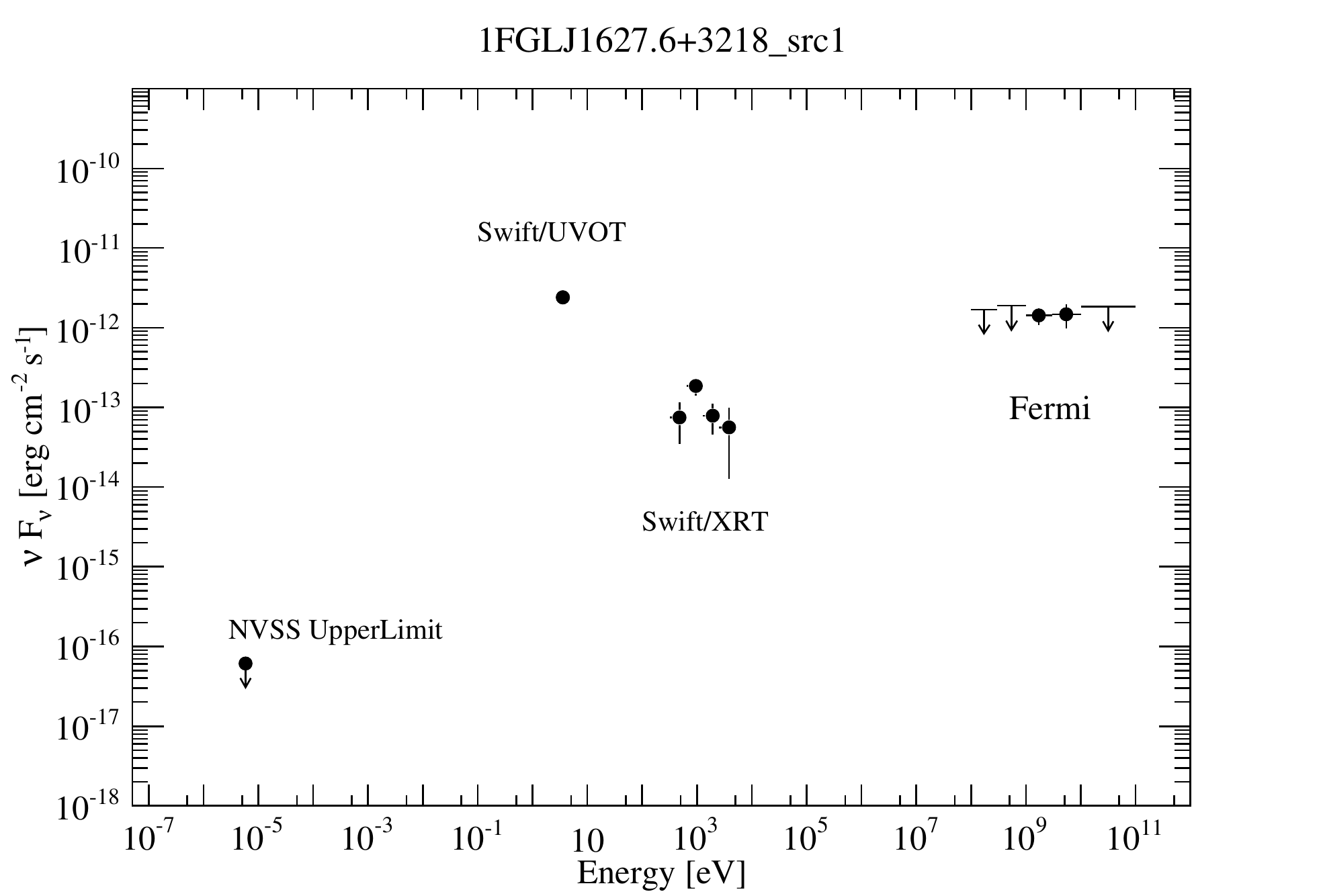}
    \end{center}
  \end{minipage}
  \begin{minipage}{0.32\hsize}
    \begin{center}
      \includegraphics[width=55mm]{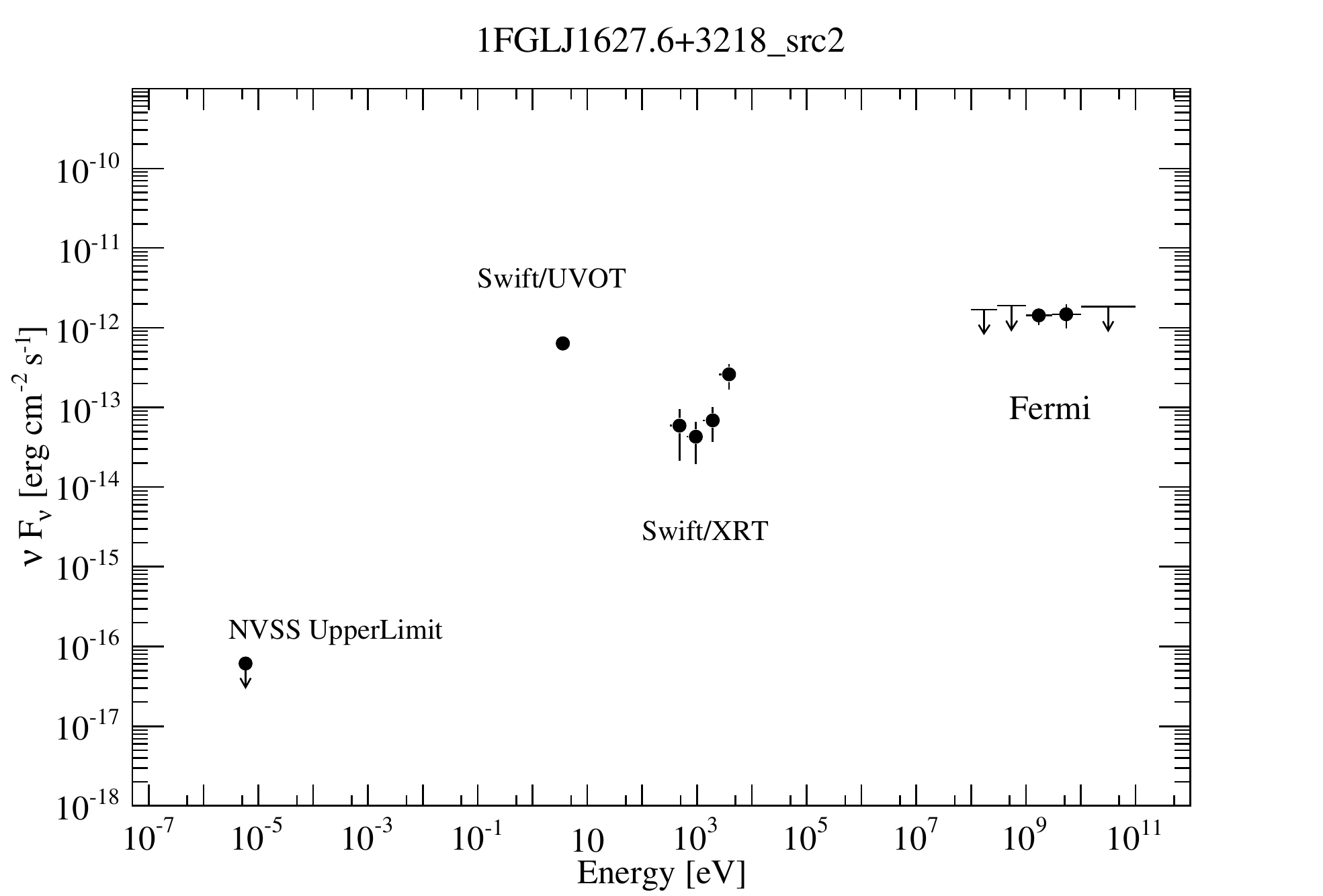}
    \end{center}
  \end{minipage}
  \begin{minipage}{0.32\hsize}
    \begin{center}
      \includegraphics[width=55mm]{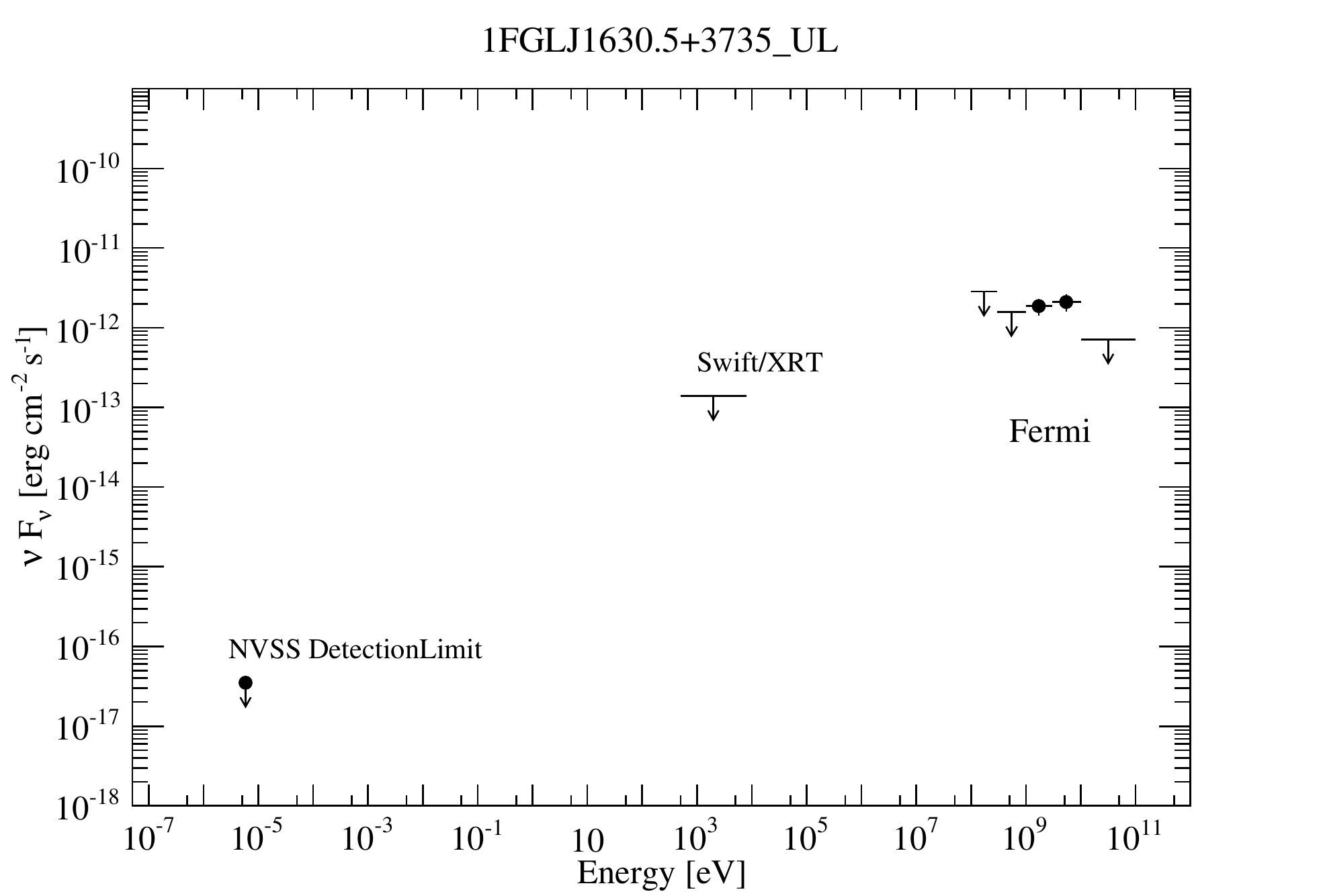}
    \end{center}
  \end{minipage}
  \begin{minipage}{0.32\hsize}
    \begin{center}
      \includegraphics[width=55mm]{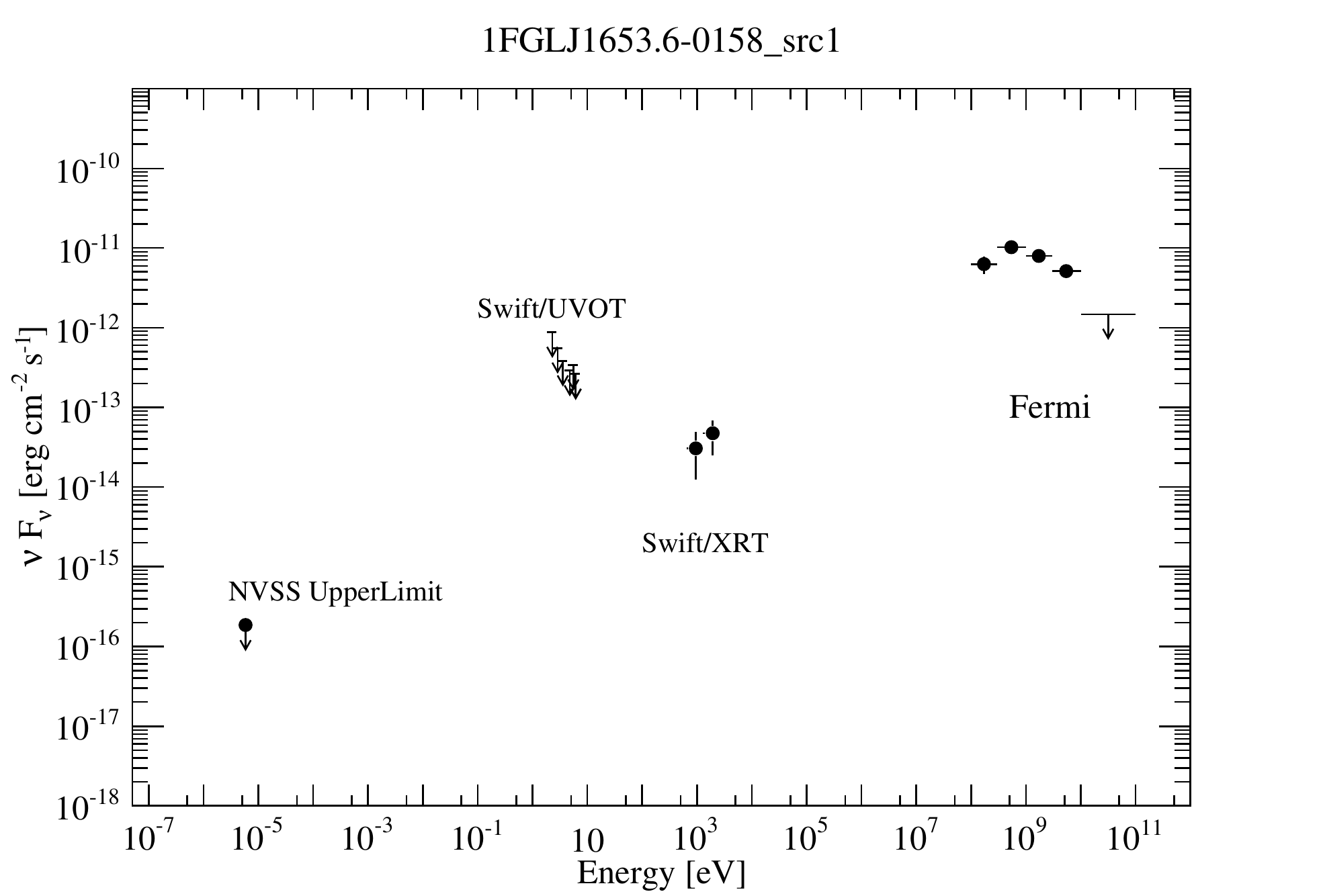}
    \end{center}
  \end{minipage}
  \begin{minipage}{0.32\hsize}
    \begin{center}
      \includegraphics[width=55mm]{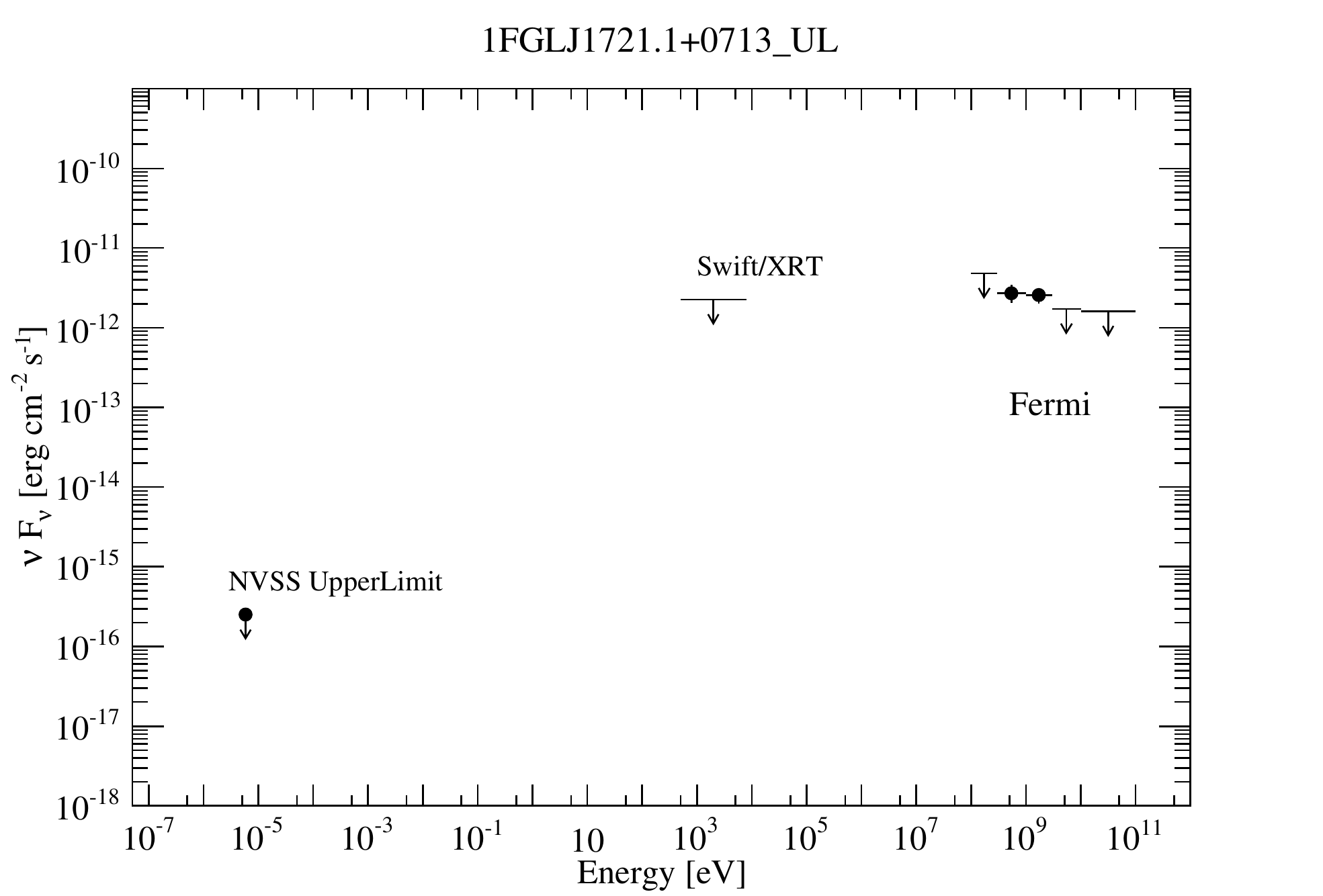}
    \end{center}
  \end{minipage}
  \begin{minipage}{0.32\hsize}
    \begin{center}
      \includegraphics[width=55mm]{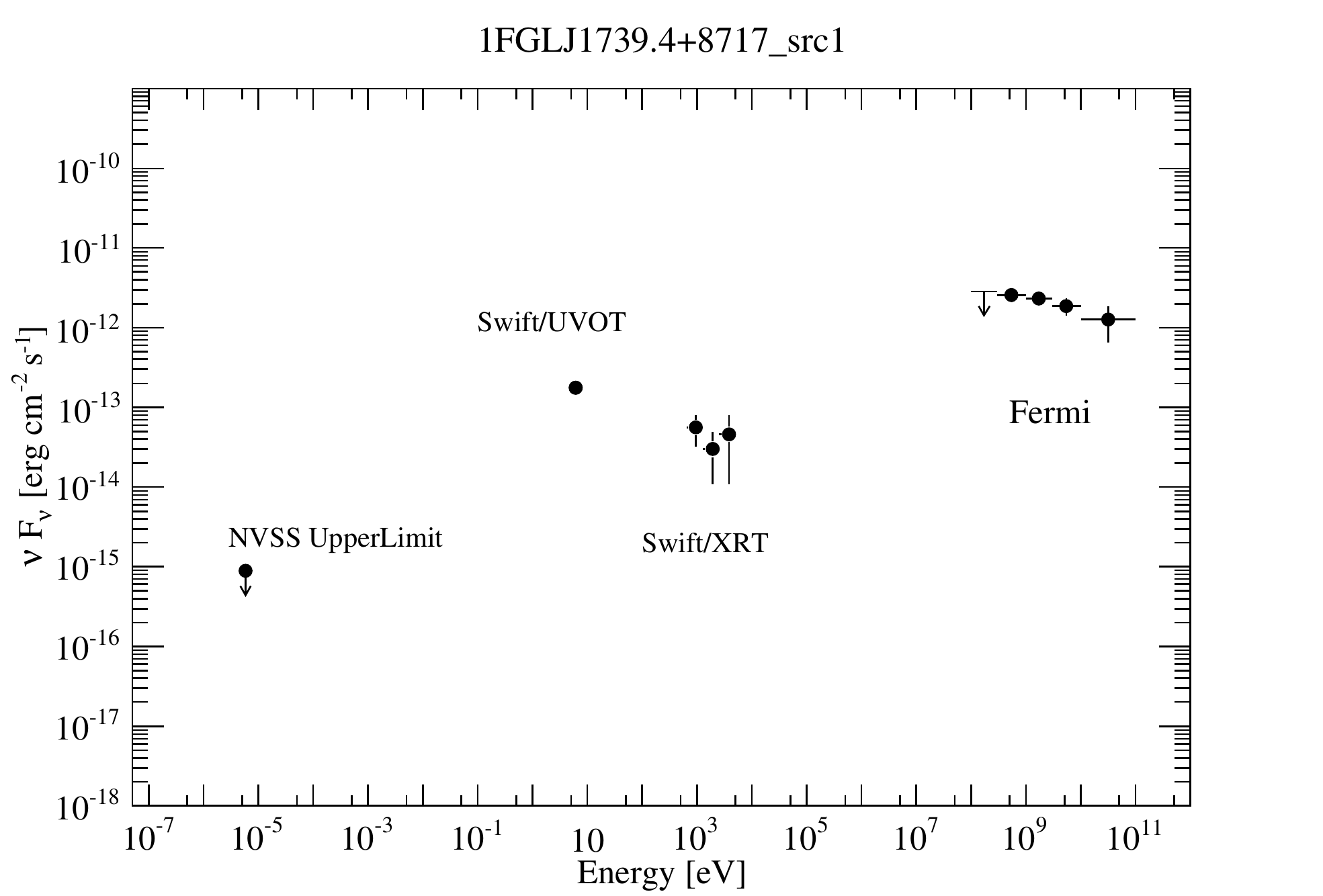}
    \end{center}
  \end{minipage}
  \begin{minipage}{0.32\hsize}
    \begin{center}
      \includegraphics[width=55mm]{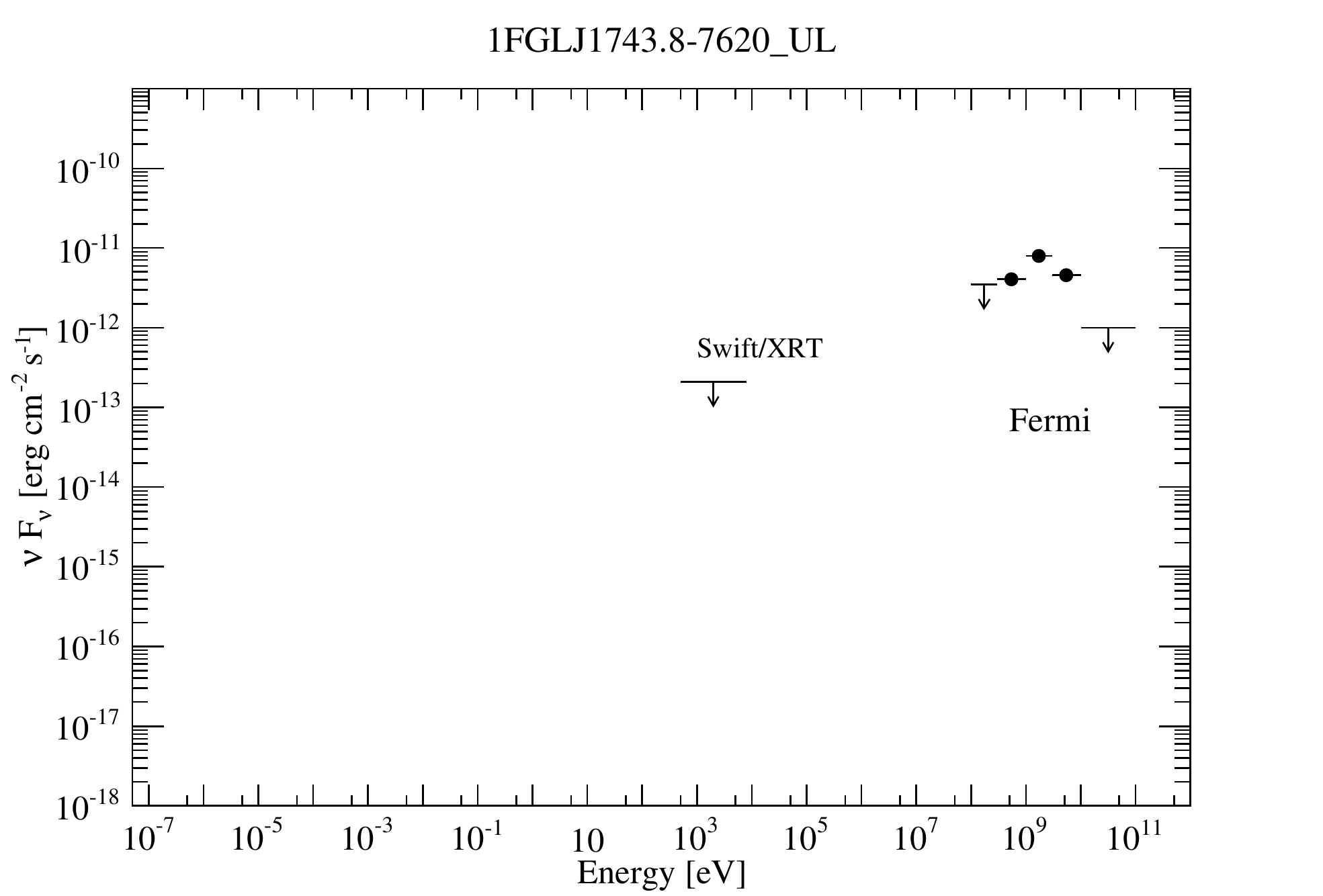}
    \end{center}
  \end{minipage}
  \begin{minipage}{0.32\hsize}
    \begin{center}
      \includegraphics[width=55mm]{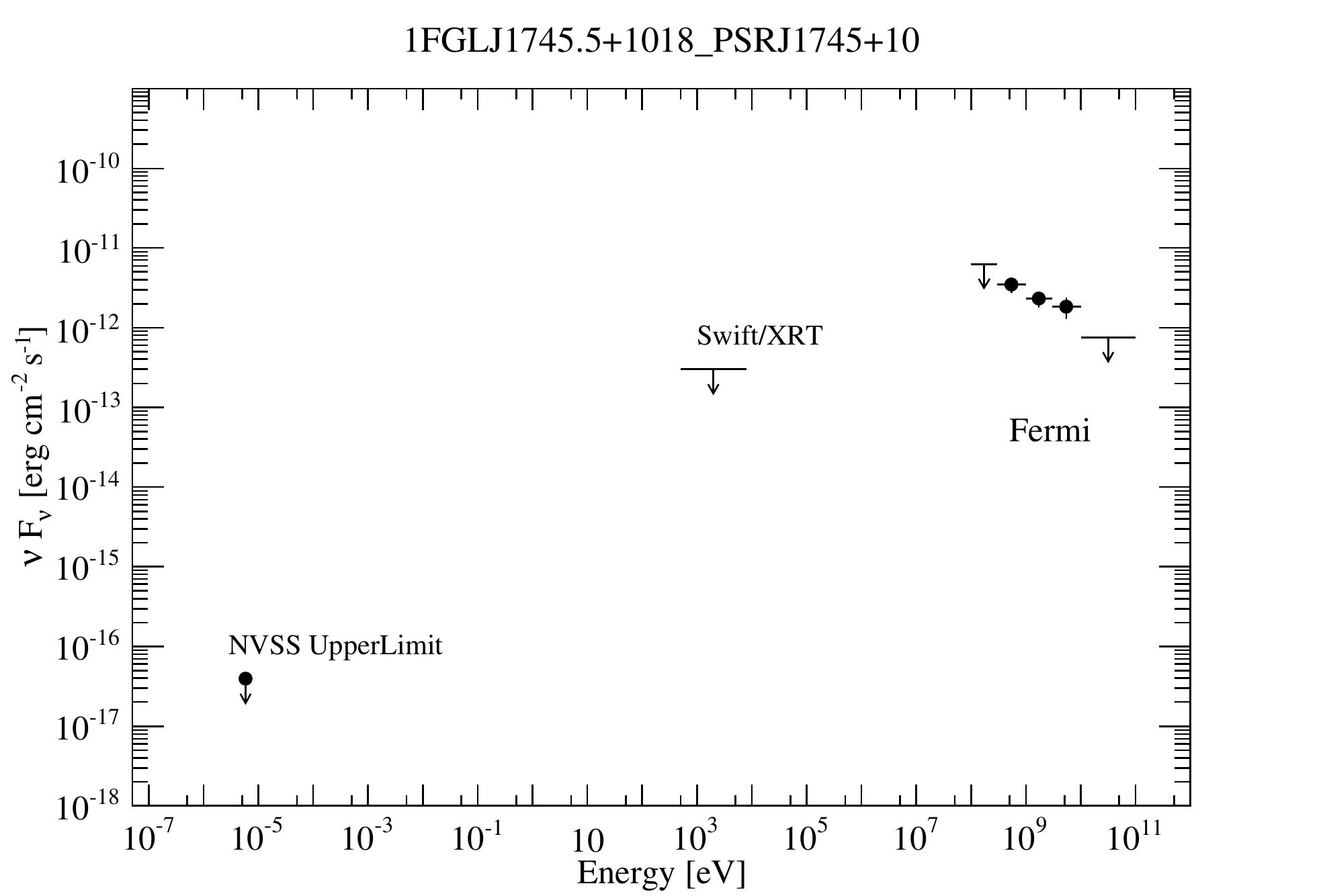}
    \end{center}
  \end{minipage}
 \end{center}
\end{figure}
\clearpage
\begin{figure}[m]
 \begin{center}
  \begin{minipage}{0.32\hsize}
    \begin{center}
      \includegraphics[width=55mm]{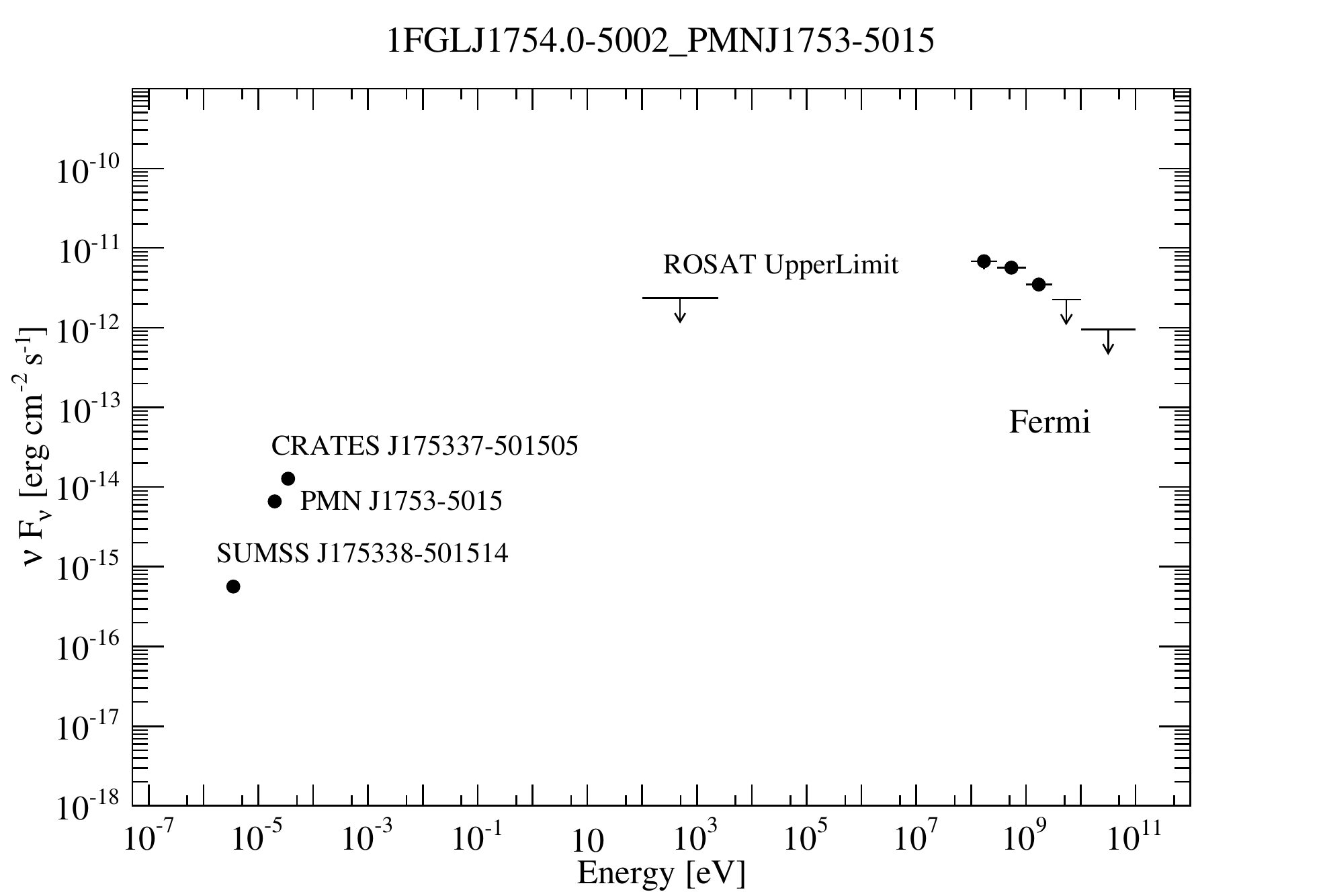}
    \end{center}
  \end{minipage}
  \begin{minipage}{0.32\hsize}
    \begin{center}
      \includegraphics[width=55mm]{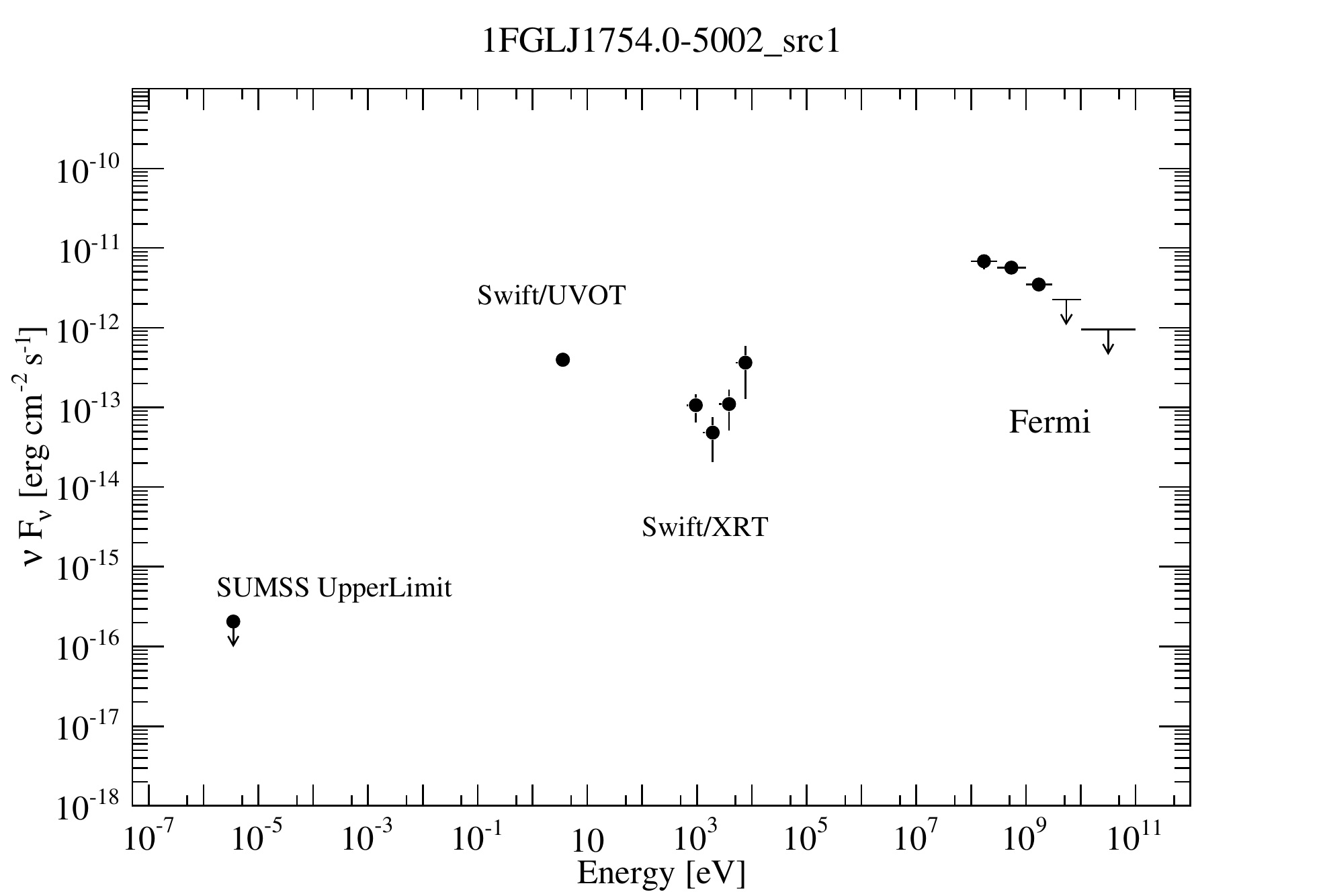}
    \end{center}
  \end{minipage}
  \begin{minipage}{0.32\hsize}
    \begin{center}
      \includegraphics[width=55mm]{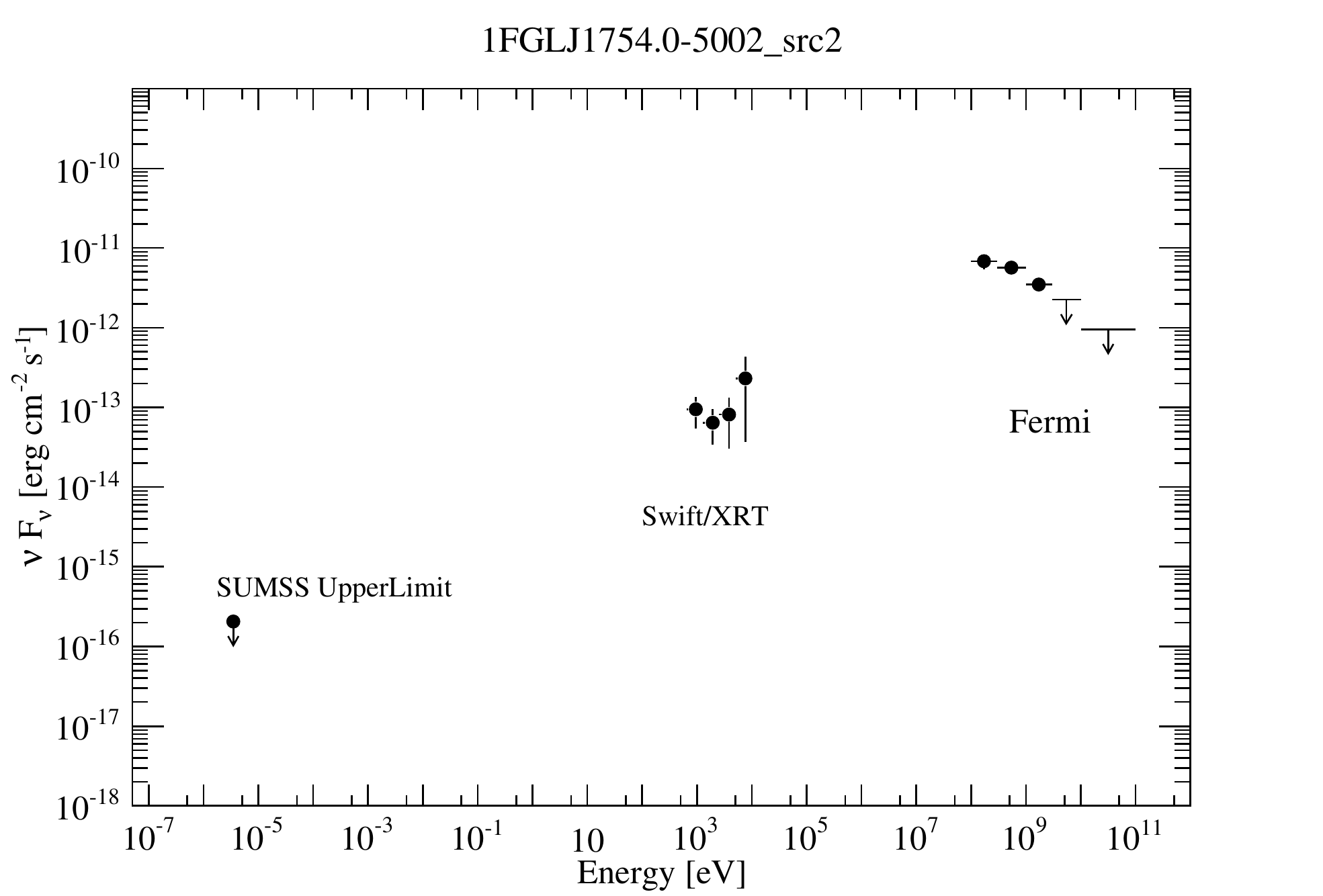}
    \end{center}
  \end{minipage}
  \begin{minipage}{0.32\hsize}
    \begin{center}
      \includegraphics[width=55mm]{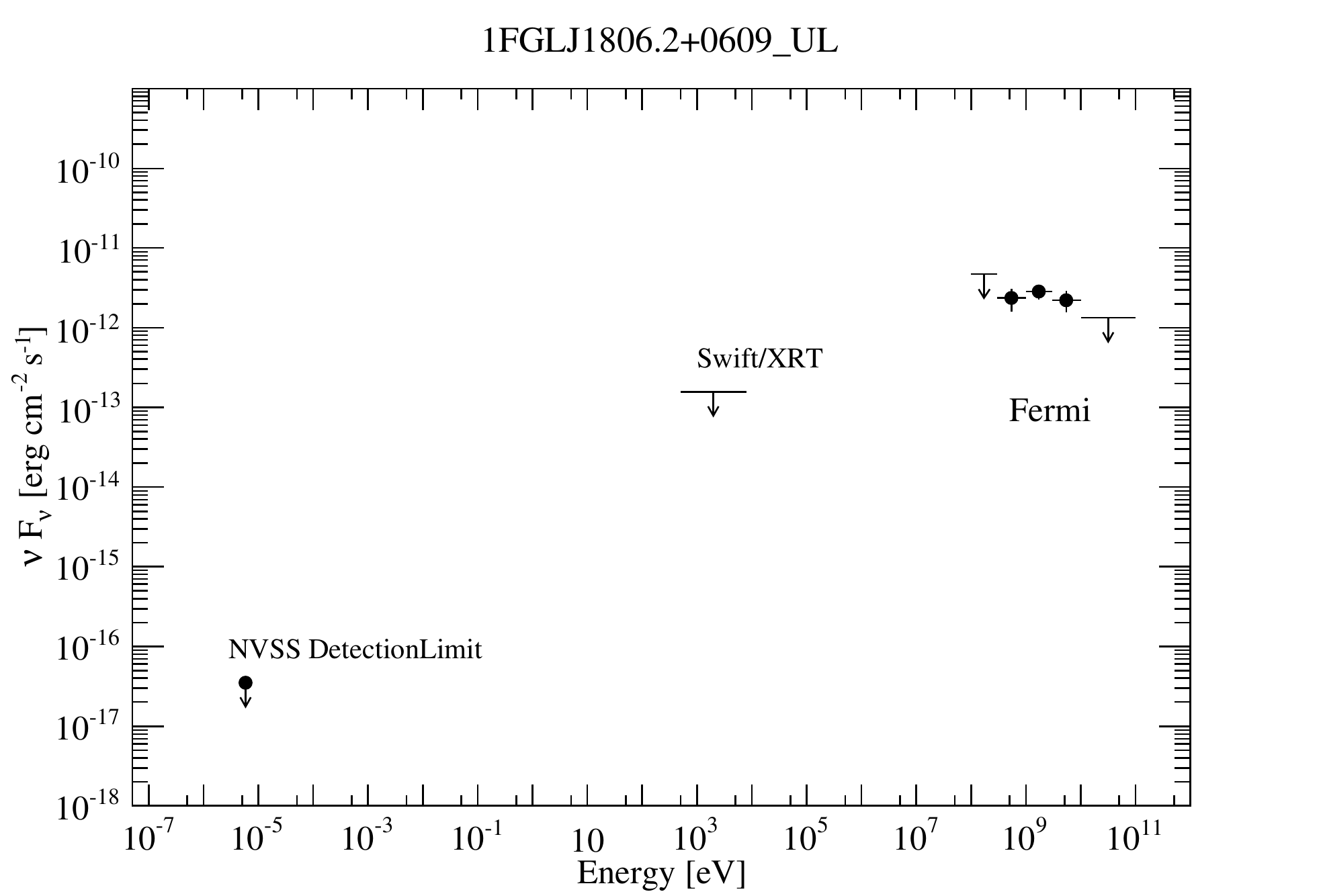}
    \end{center}
  \end{minipage}
  \begin{minipage}{0.32\hsize}
    \begin{center}
      \includegraphics[width=55mm]{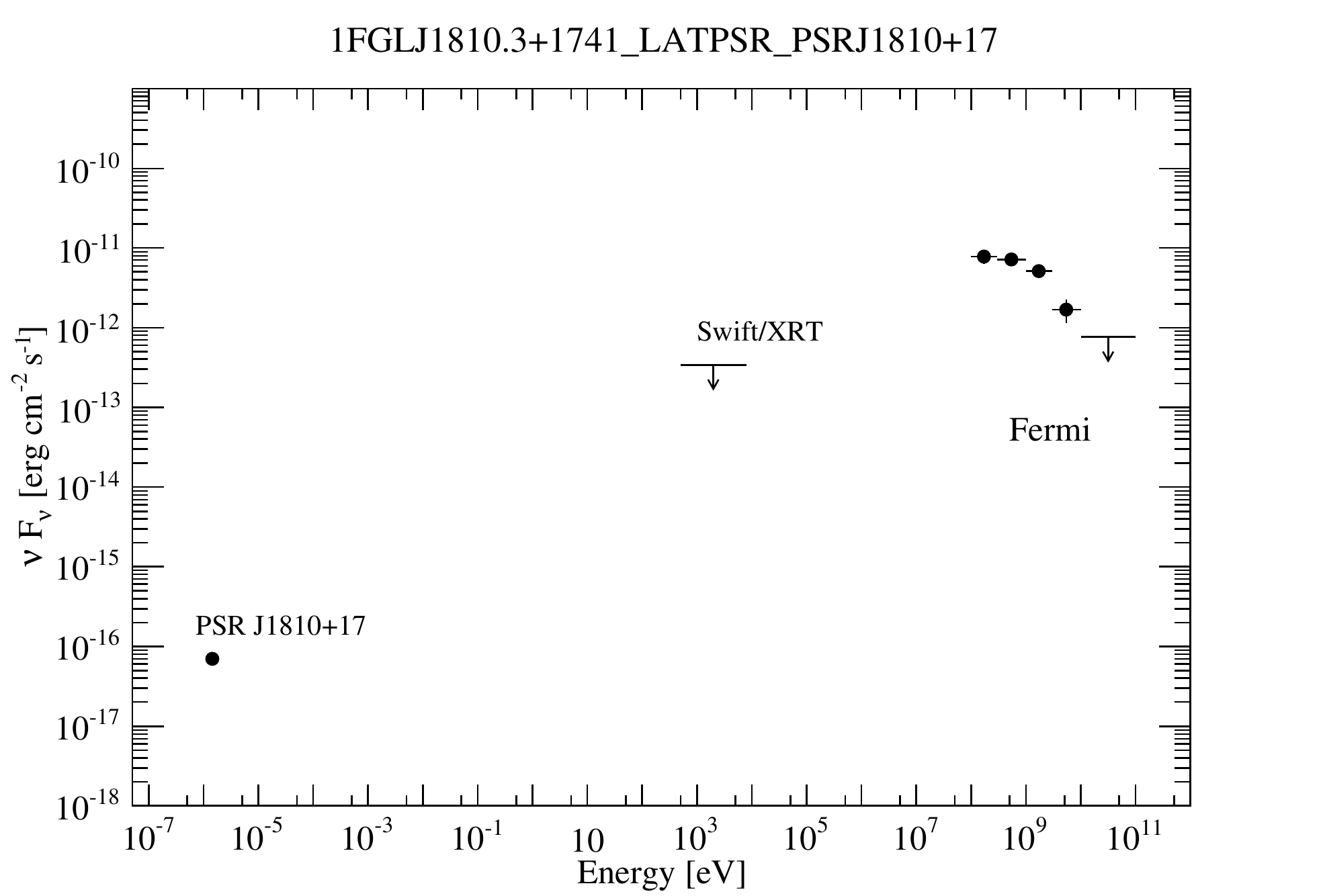}
    \end{center}
  \end{minipage}
  \begin{minipage}{0.32\hsize}
    \begin{center}
      \includegraphics[width=55mm]{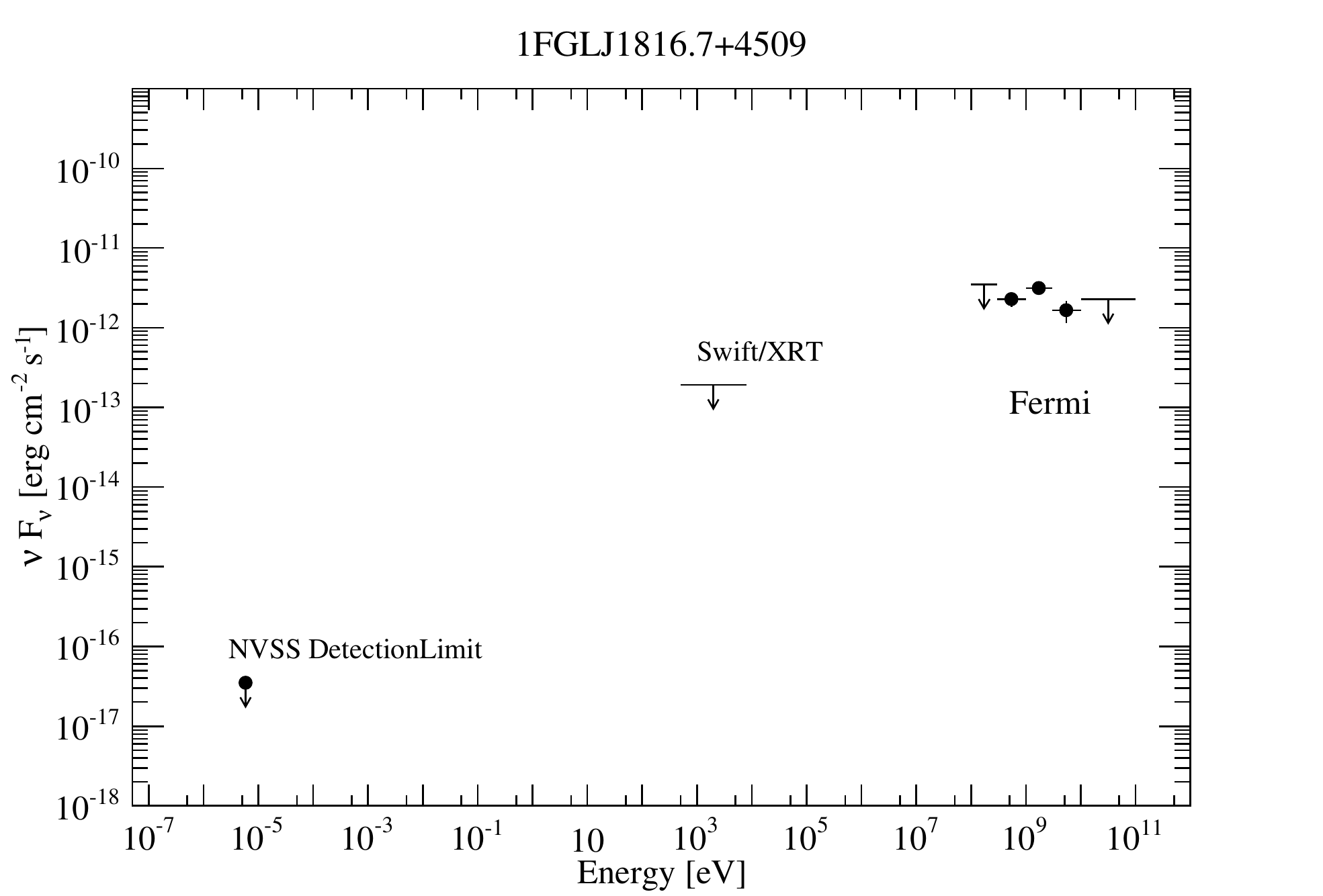}
    \end{center}
  \end{minipage}
  \begin{minipage}{0.32\hsize}
    \begin{center}
      \includegraphics[width=55mm]{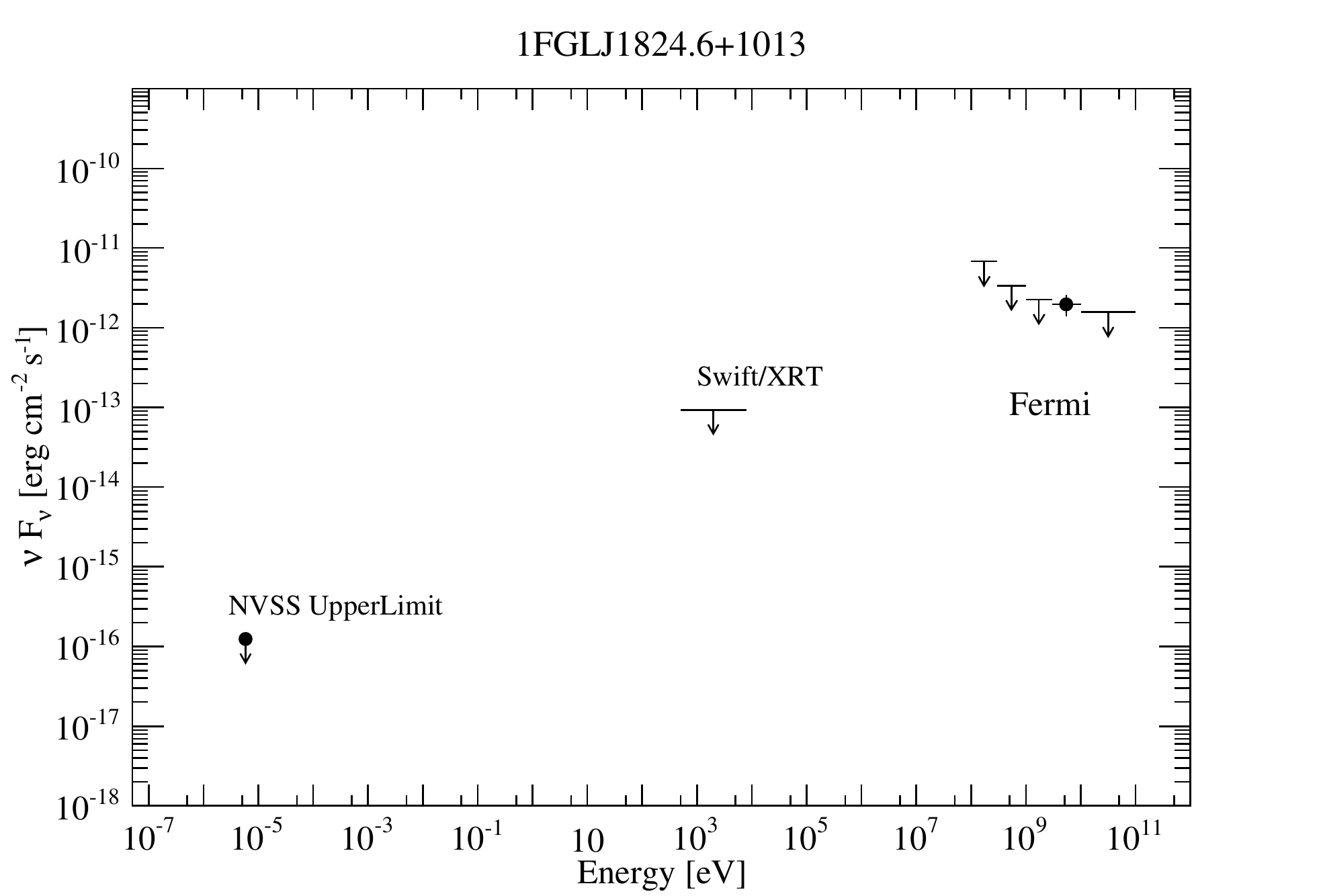}
    \end{center}
  \end{minipage}
  \begin{minipage}{0.32\hsize}
    \begin{center}
      \includegraphics[width=55mm]{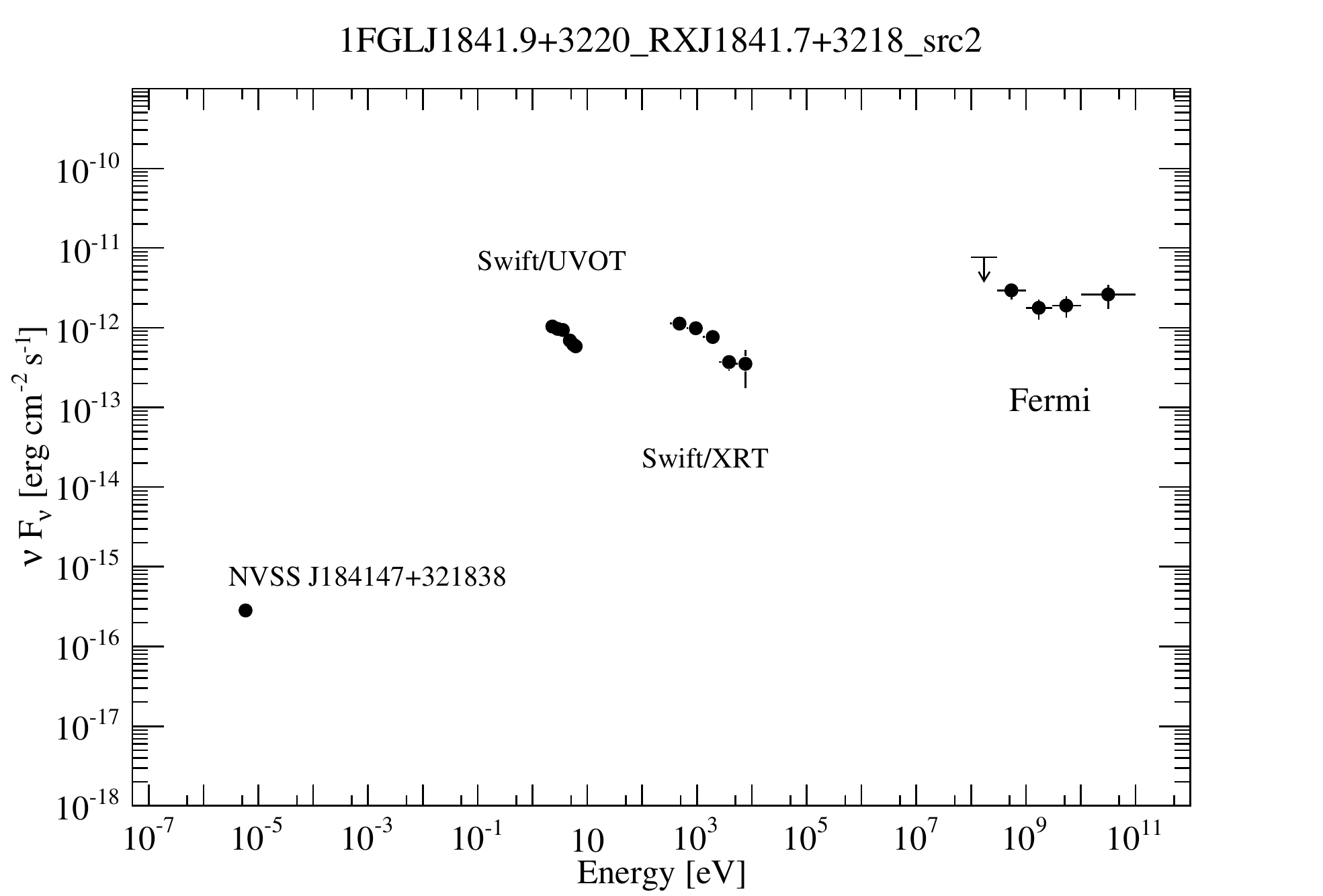}
    \end{center}
  \end{minipage}
  \begin{minipage}{0.32\hsize}
    \begin{center}
      \includegraphics[width=55mm]{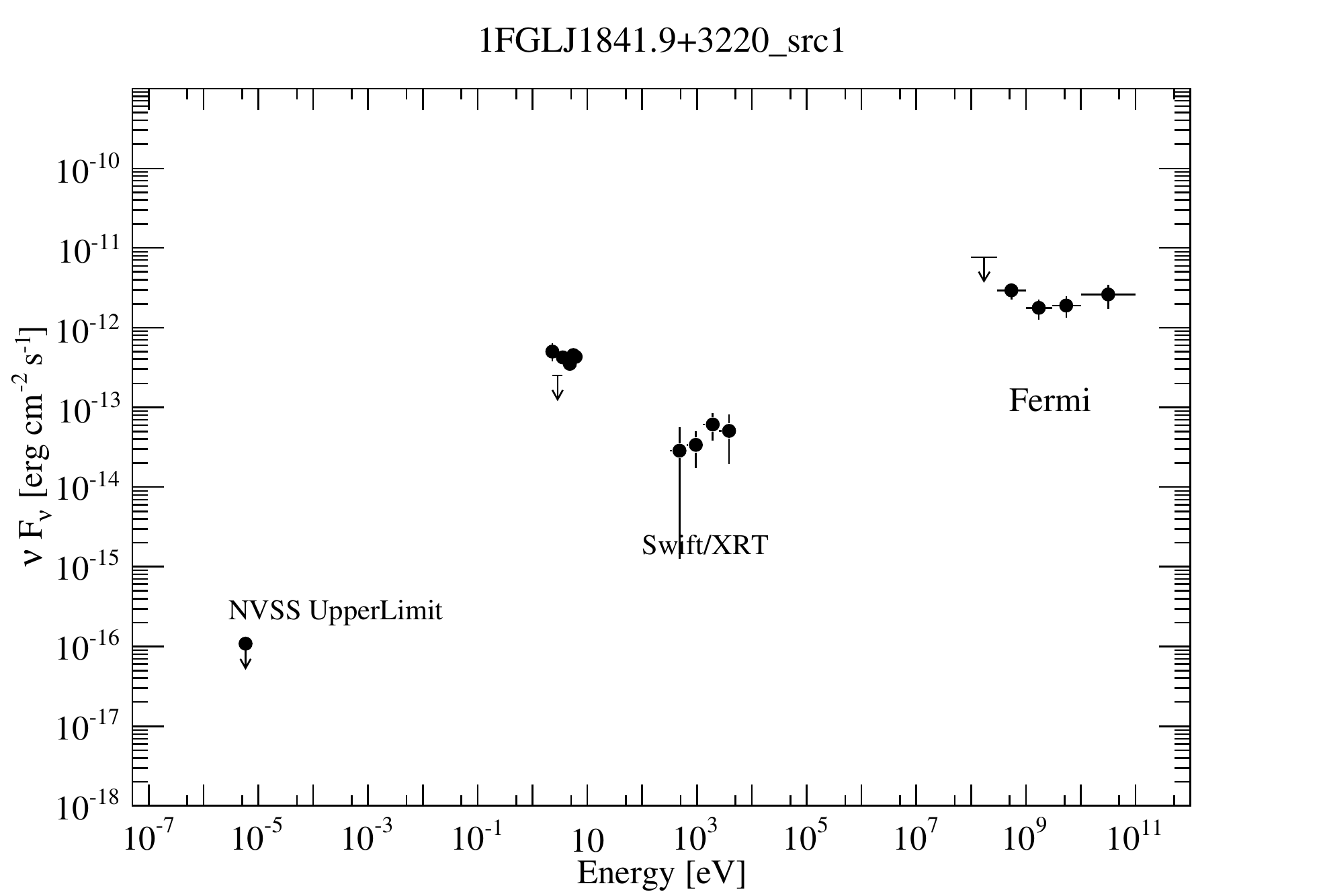}
    \end{center}
  \end{minipage}
  \begin{minipage}{0.32\hsize}
    \begin{center}
      \includegraphics[width=55mm]{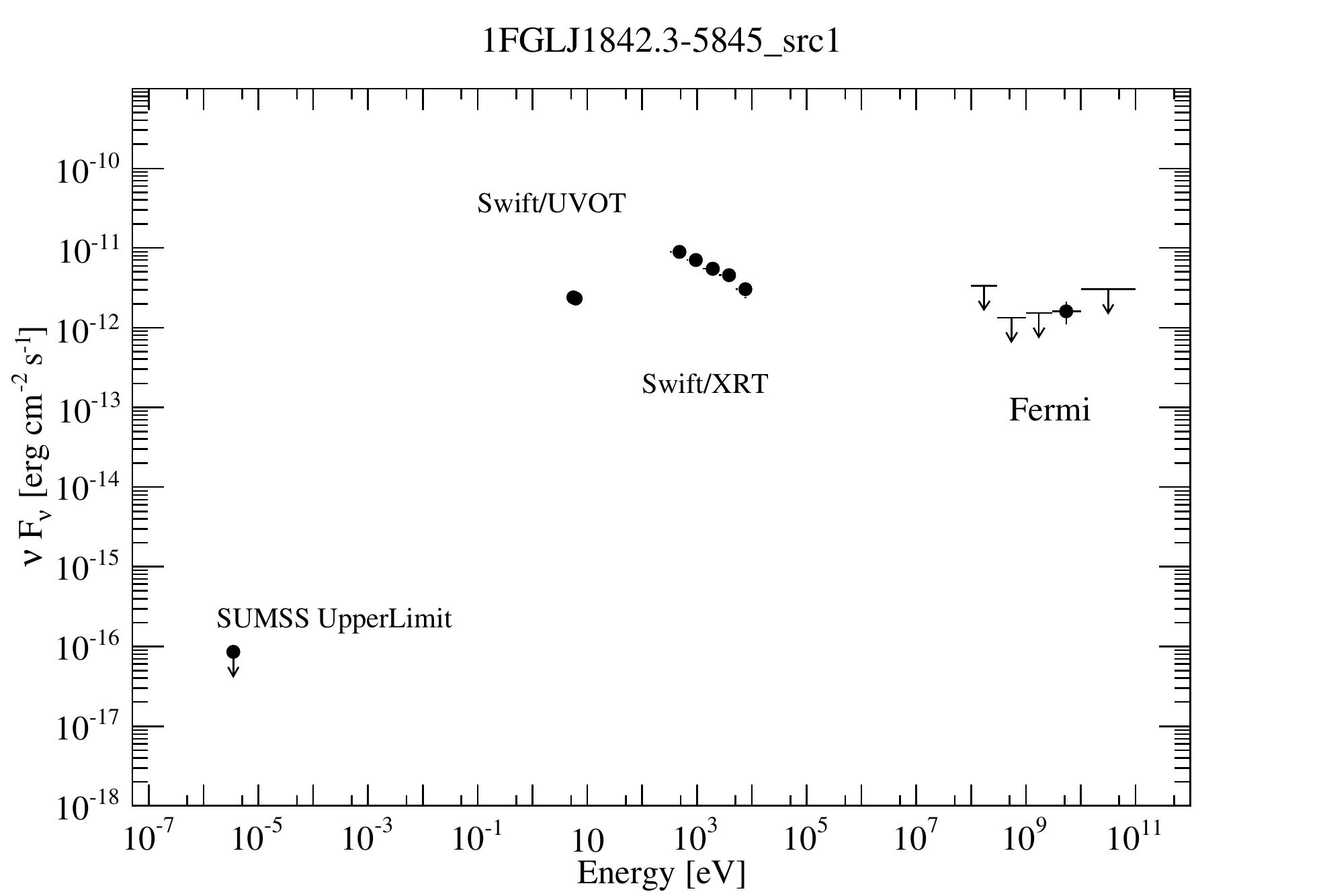}
    \end{center}
  \end{minipage}
  \begin{minipage}{0.32\hsize}
    \begin{center}
      \includegraphics[width=55mm]{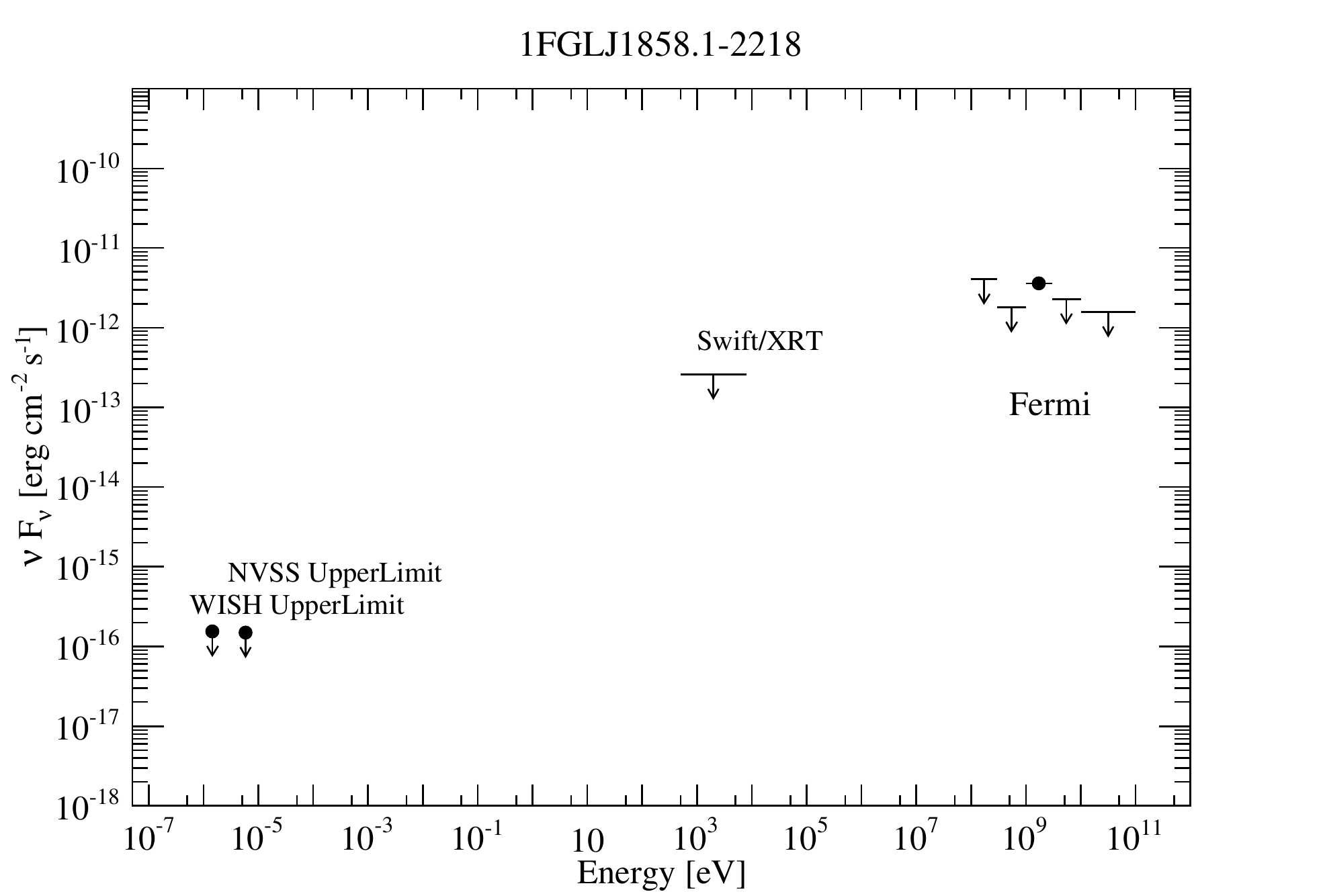}
    \end{center}
  \end{minipage}
  \begin{minipage}{0.32\hsize}
    \begin{center}
      \includegraphics[width=55mm]{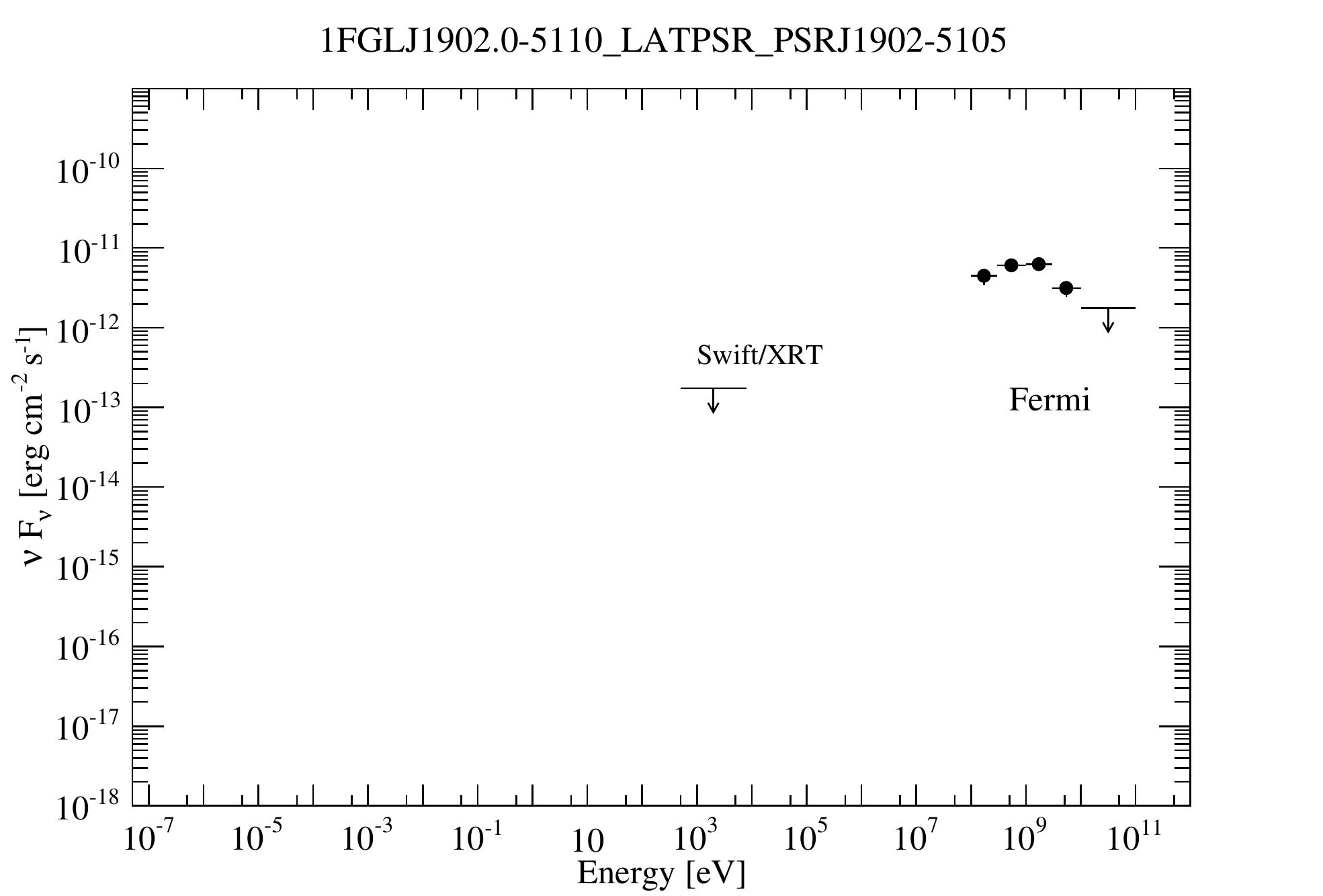}
    \end{center}
  \end{minipage}
  \begin{minipage}{0.32\hsize}
    \begin{center}
      \includegraphics[width=55mm]{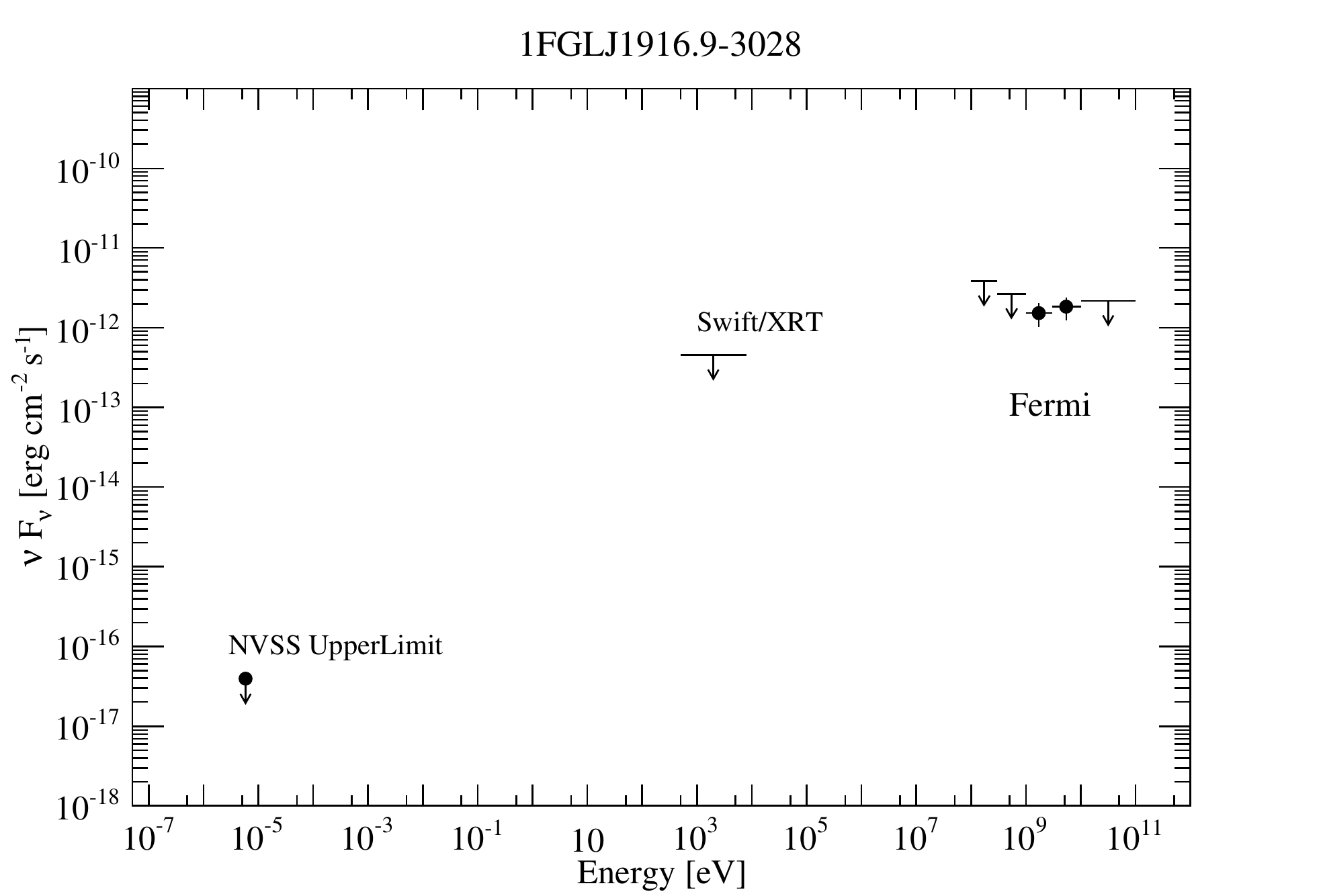}
    \end{center}
  \end{minipage}
  \begin{minipage}{0.32\hsize}
    \begin{center}
      \includegraphics[width=55mm]{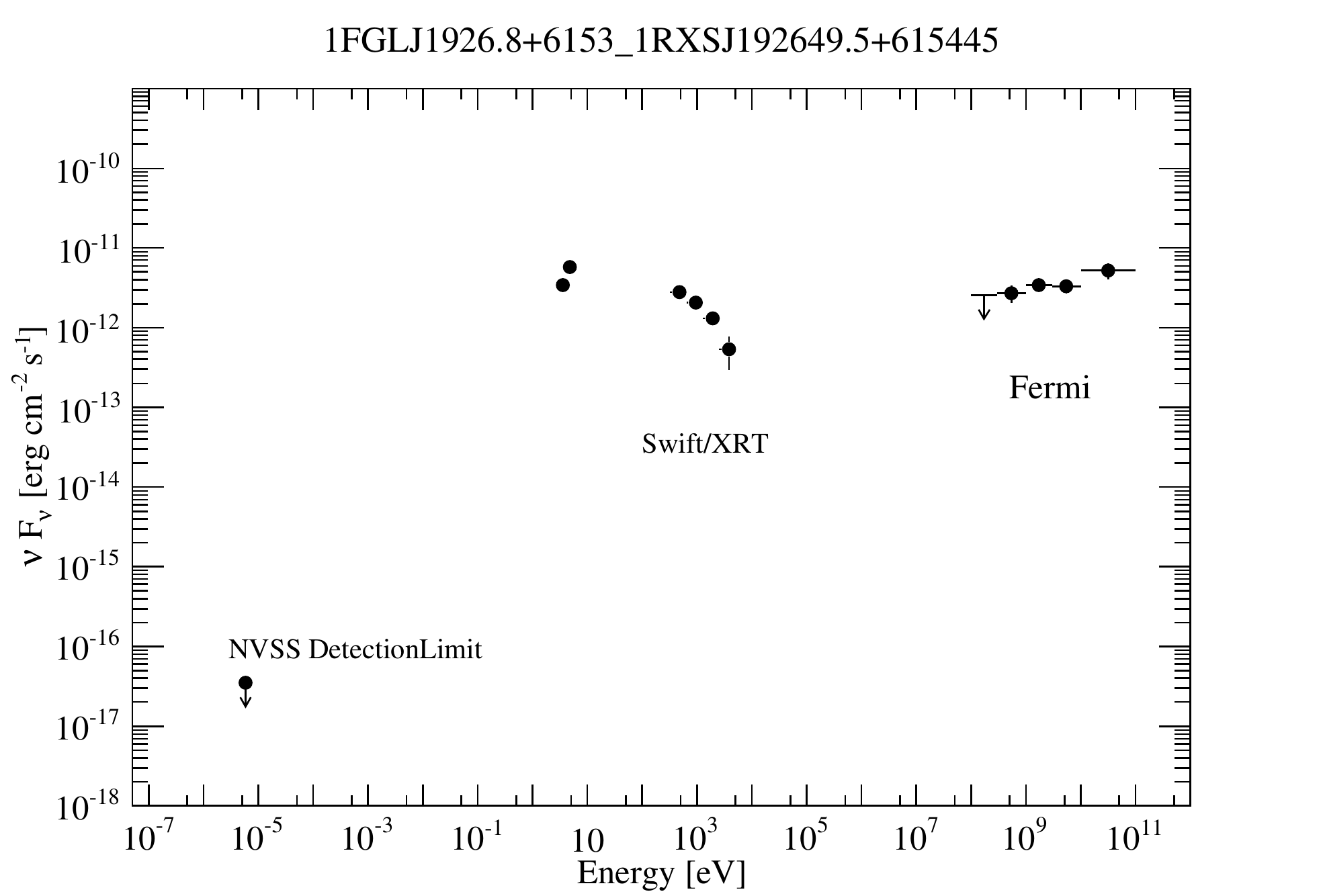}
    \end{center}
  \end{minipage}
  \begin{minipage}{0.32\hsize}
    \begin{center}
      \includegraphics[width=55mm]{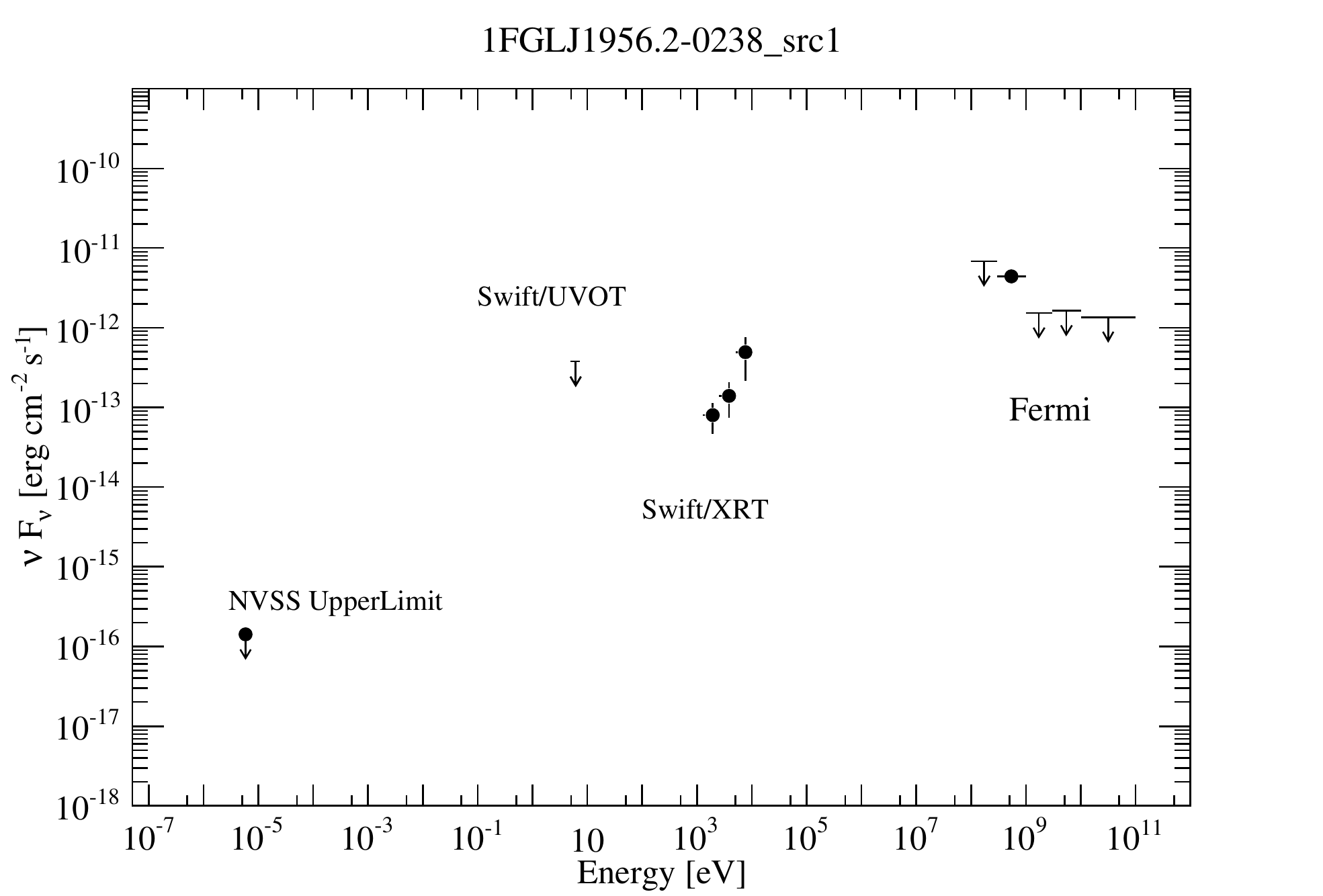}
    \end{center}
  \end{minipage}
  \begin{minipage}{0.32\hsize}
    \begin{center}
      \includegraphics[width=55mm]{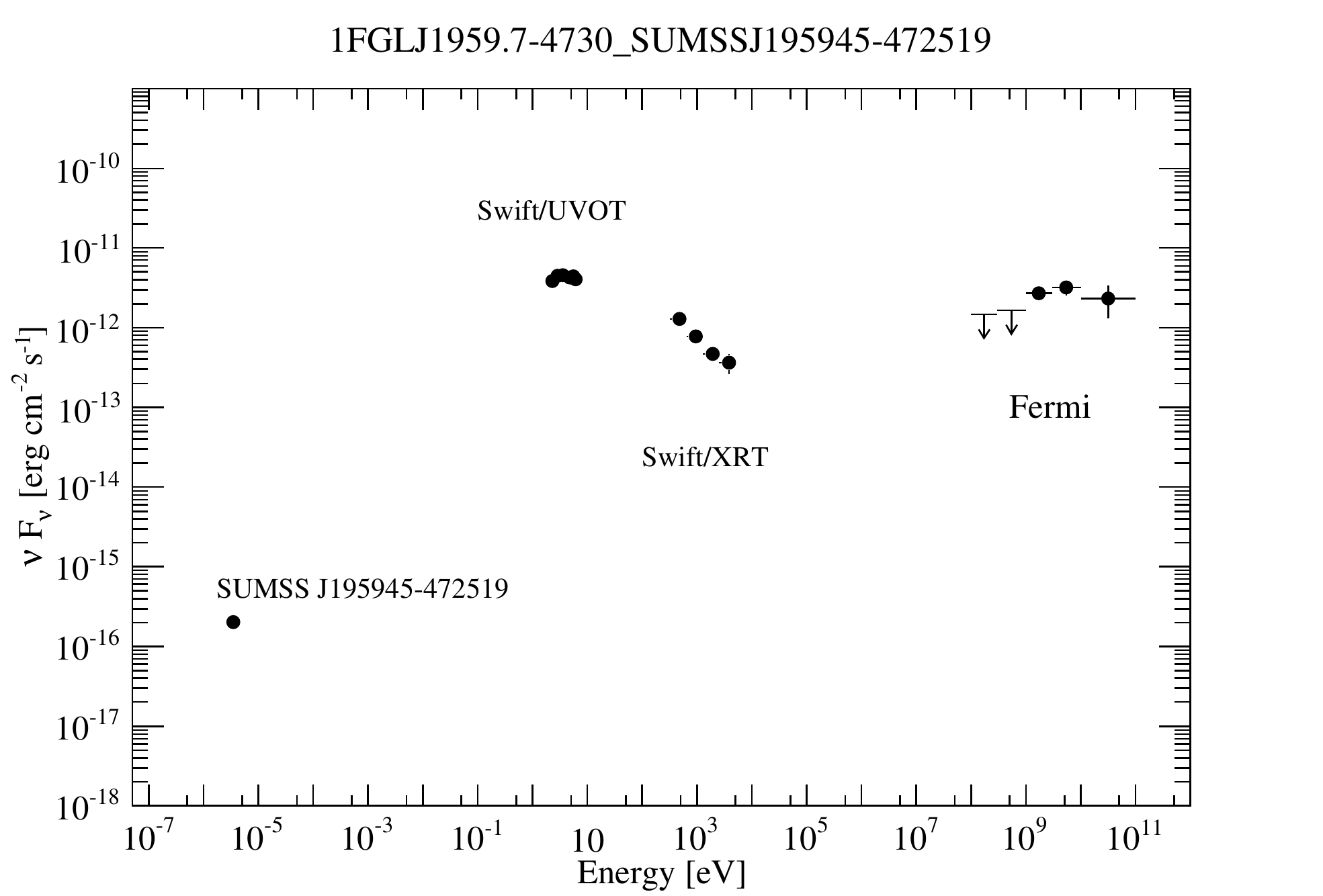}
    \end{center}
  \end{minipage}
  \begin{minipage}{0.32\hsize}
    \begin{center}
      \includegraphics[width=55mm]{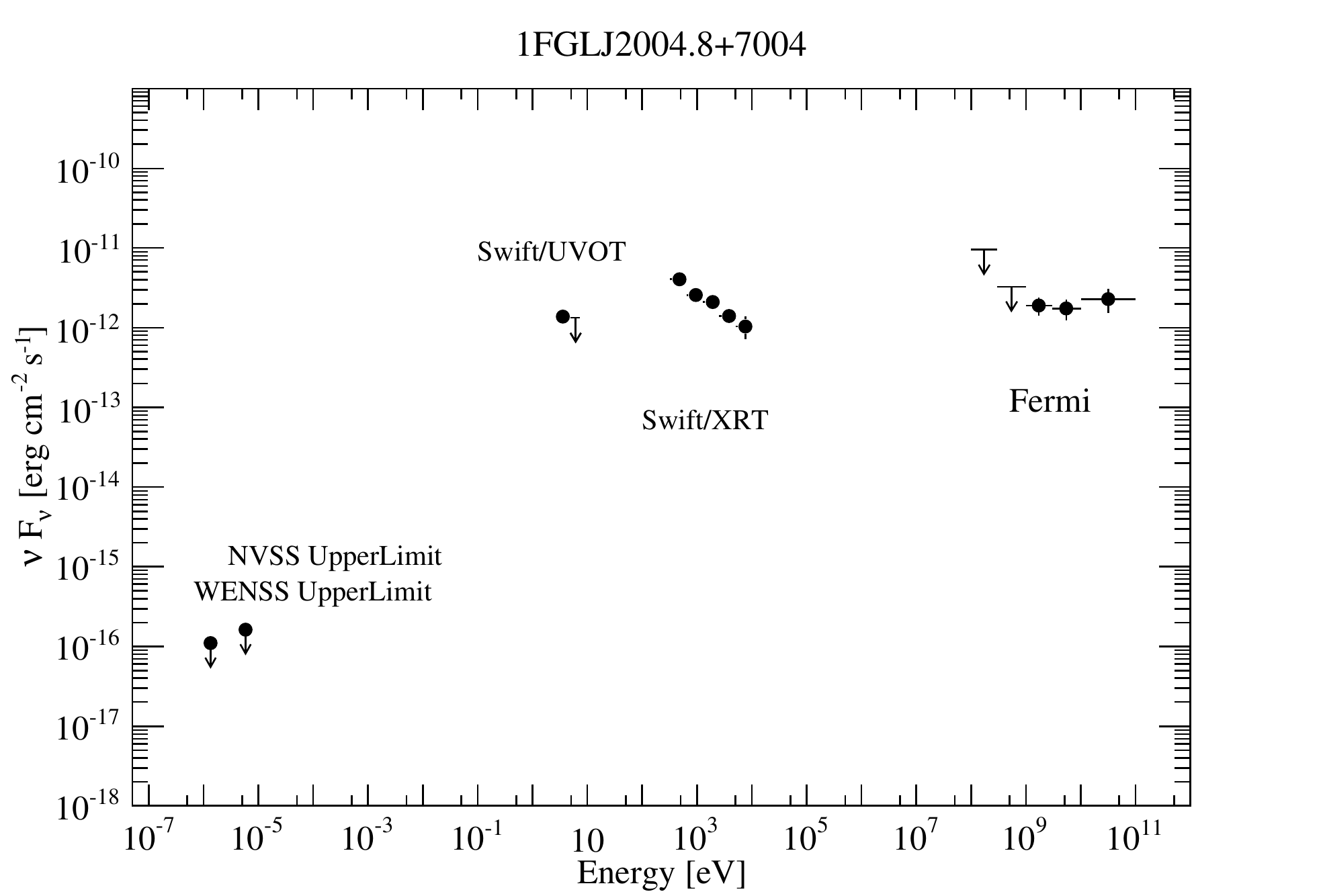}
    \end{center}
  \end{minipage}
  \begin{minipage}{0.32\hsize}
    \begin{center}
      \includegraphics[width=55mm]{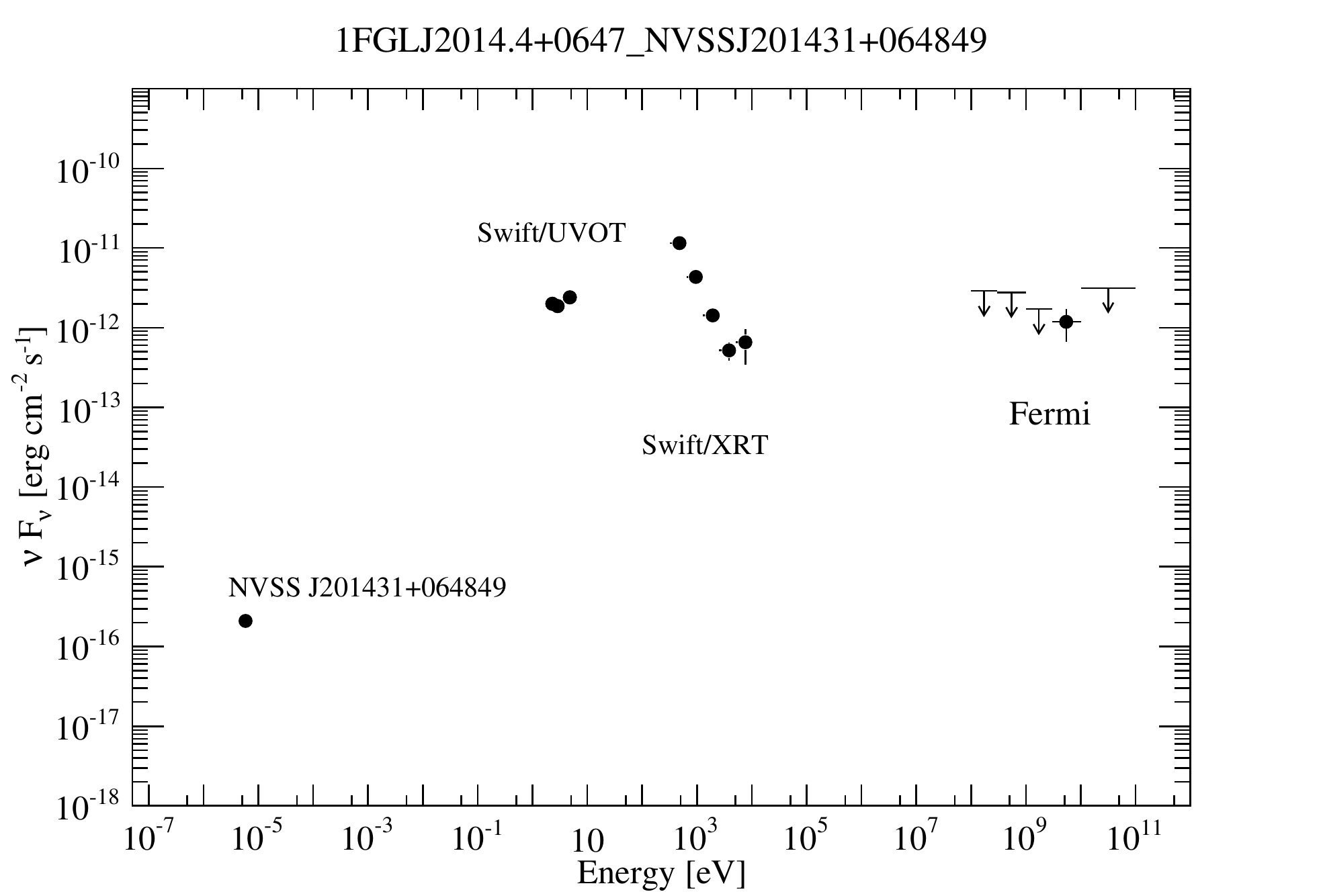}
    \end{center}
  \end{minipage}
 \end{center}
\end{figure}
\clearpage
\begin{figure}[m]
 \begin{center}
  \begin{minipage}{0.32\hsize}
    \begin{center}
      \includegraphics[width=55mm]{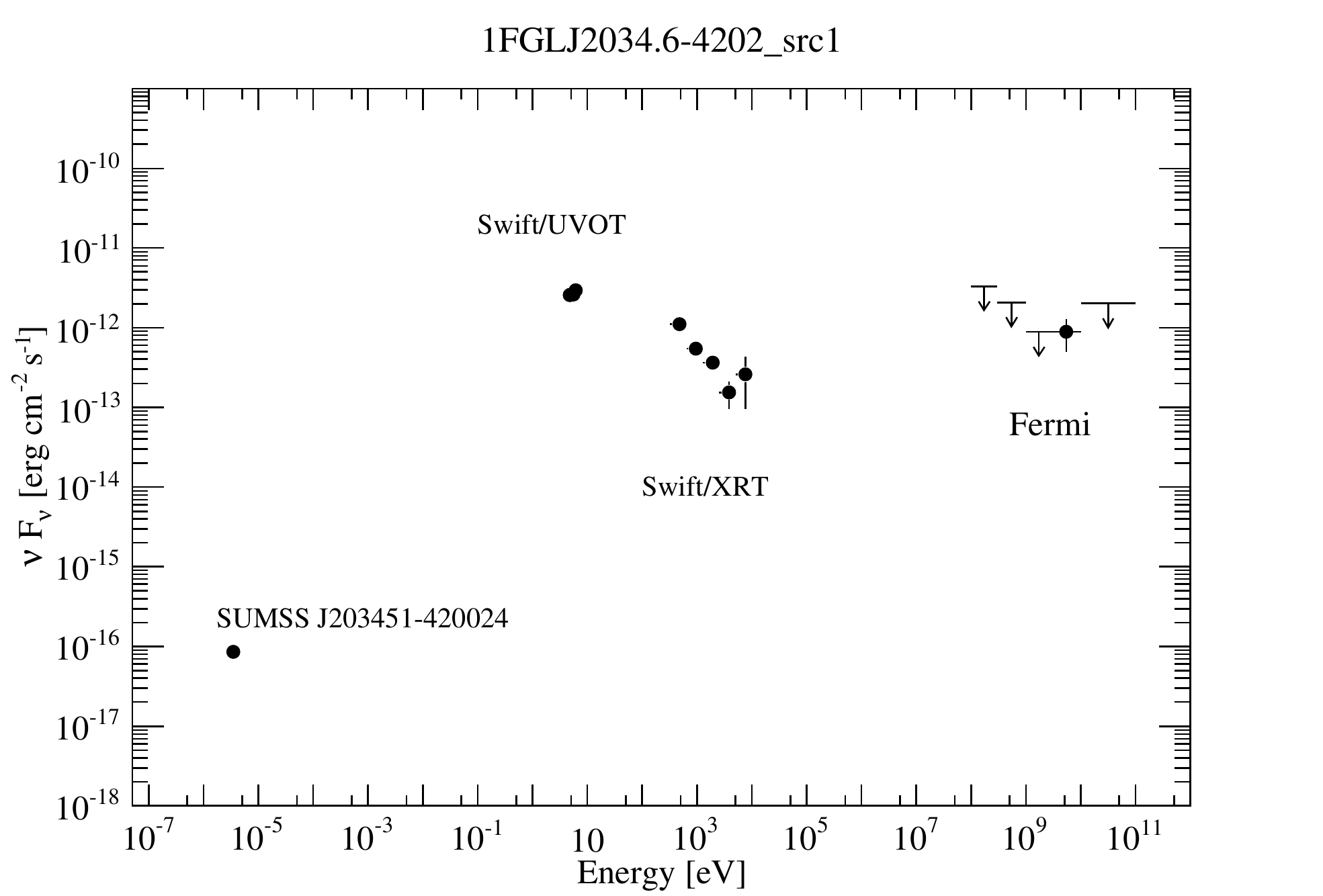}
    \end{center}
  \end{minipage}
  \begin{minipage}{0.32\hsize}
    \begin{center}
      \includegraphics[width=55mm]{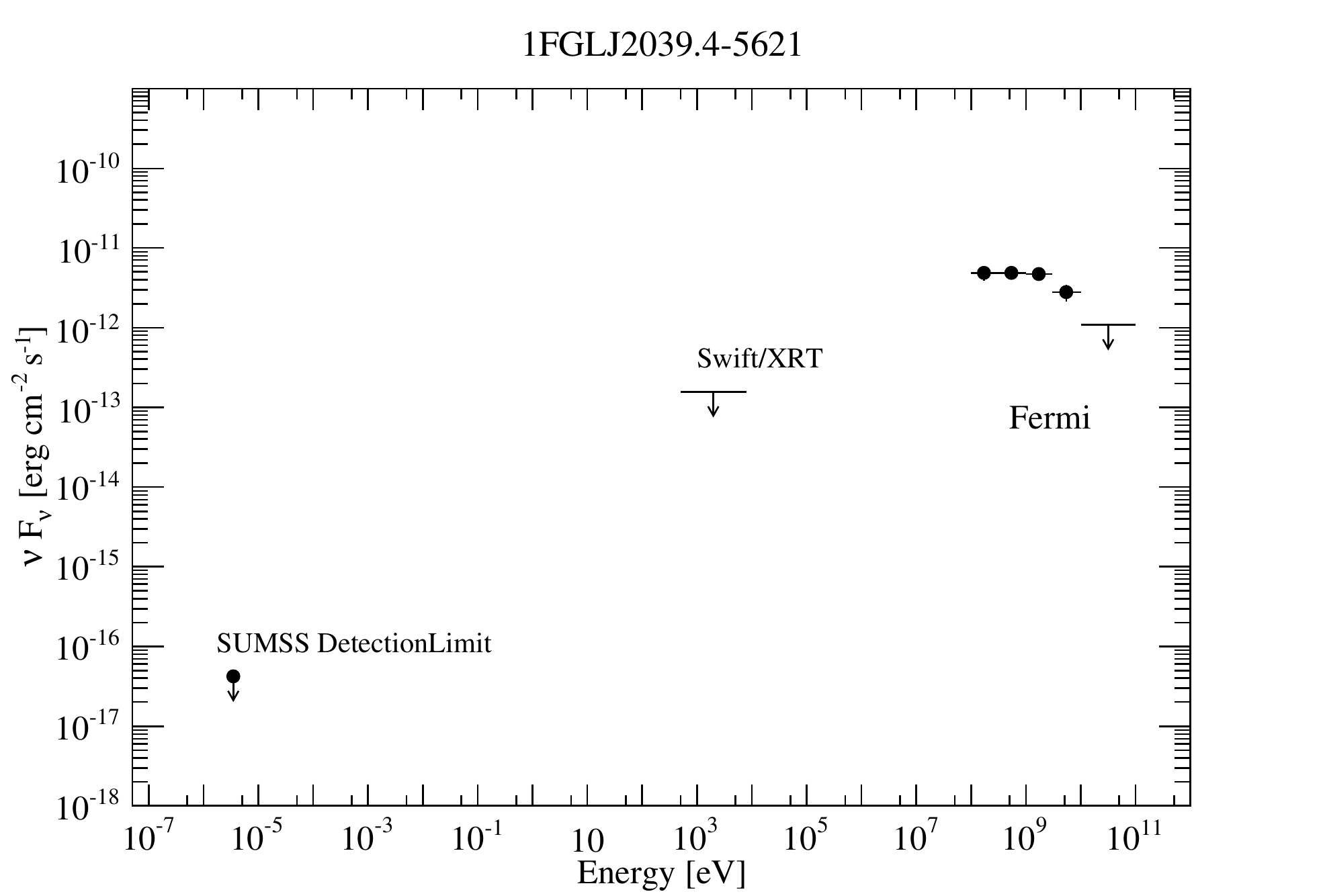}
    \end{center}
  \end{minipage}
  \begin{minipage}{0.32\hsize}
    \begin{center}
      \includegraphics[width=55mm]{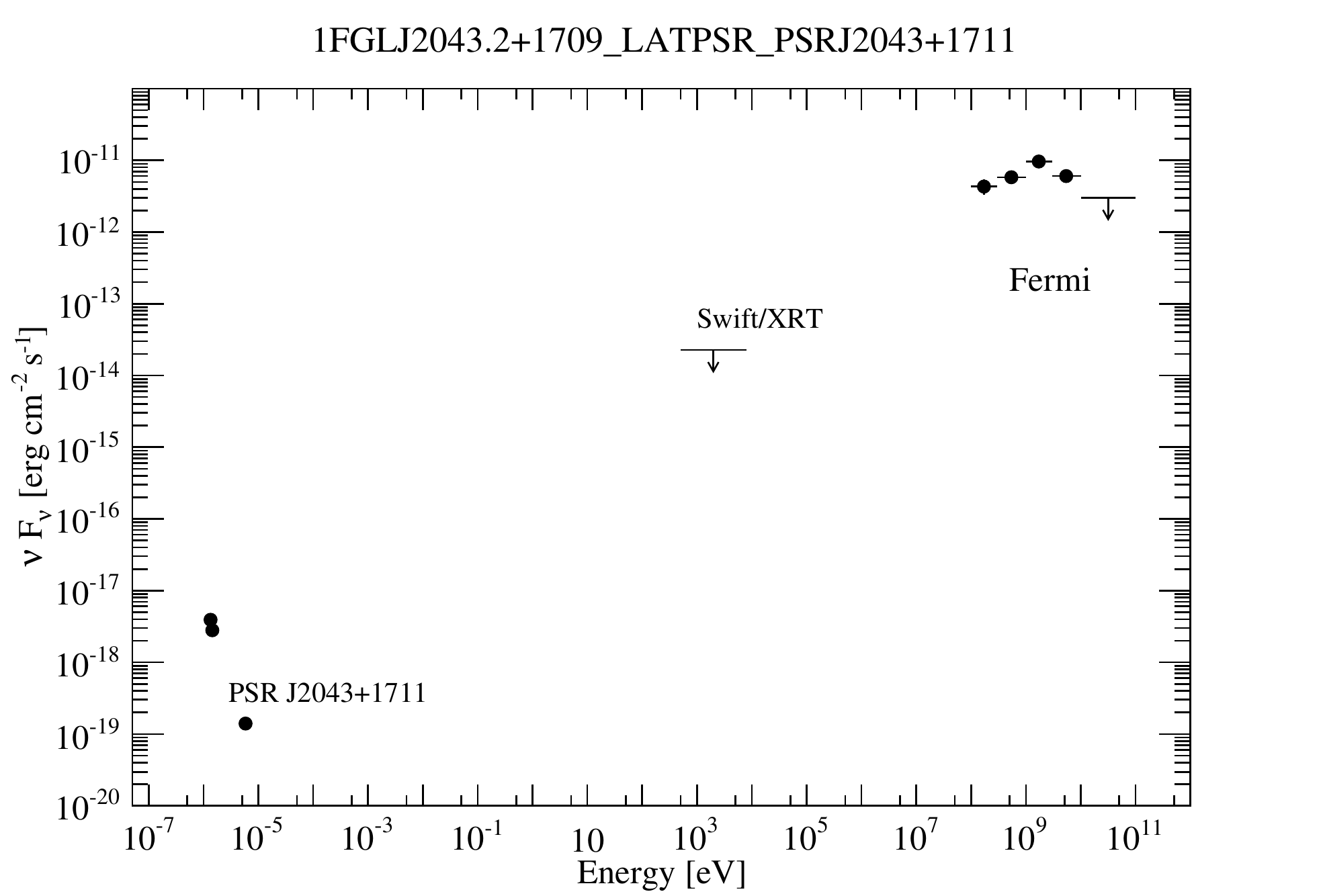}
    \end{center}
  \end{minipage}
  \begin{minipage}{0.32\hsize}
    \begin{center}
      \includegraphics[width=55mm]{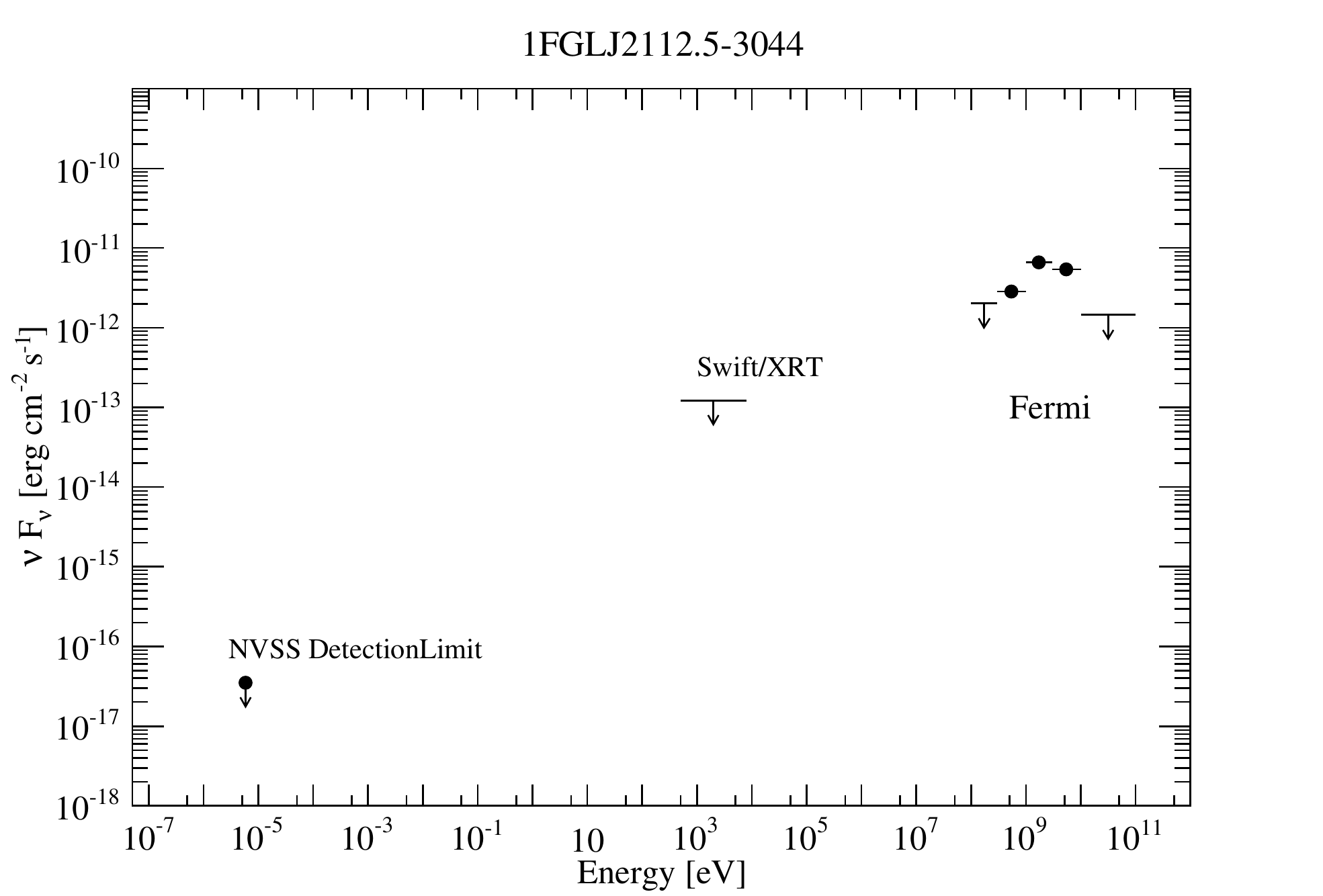}
    \end{center}
  \end{minipage}
  \begin{minipage}{0.32\hsize}
    \begin{center}
      \includegraphics[width=55mm]{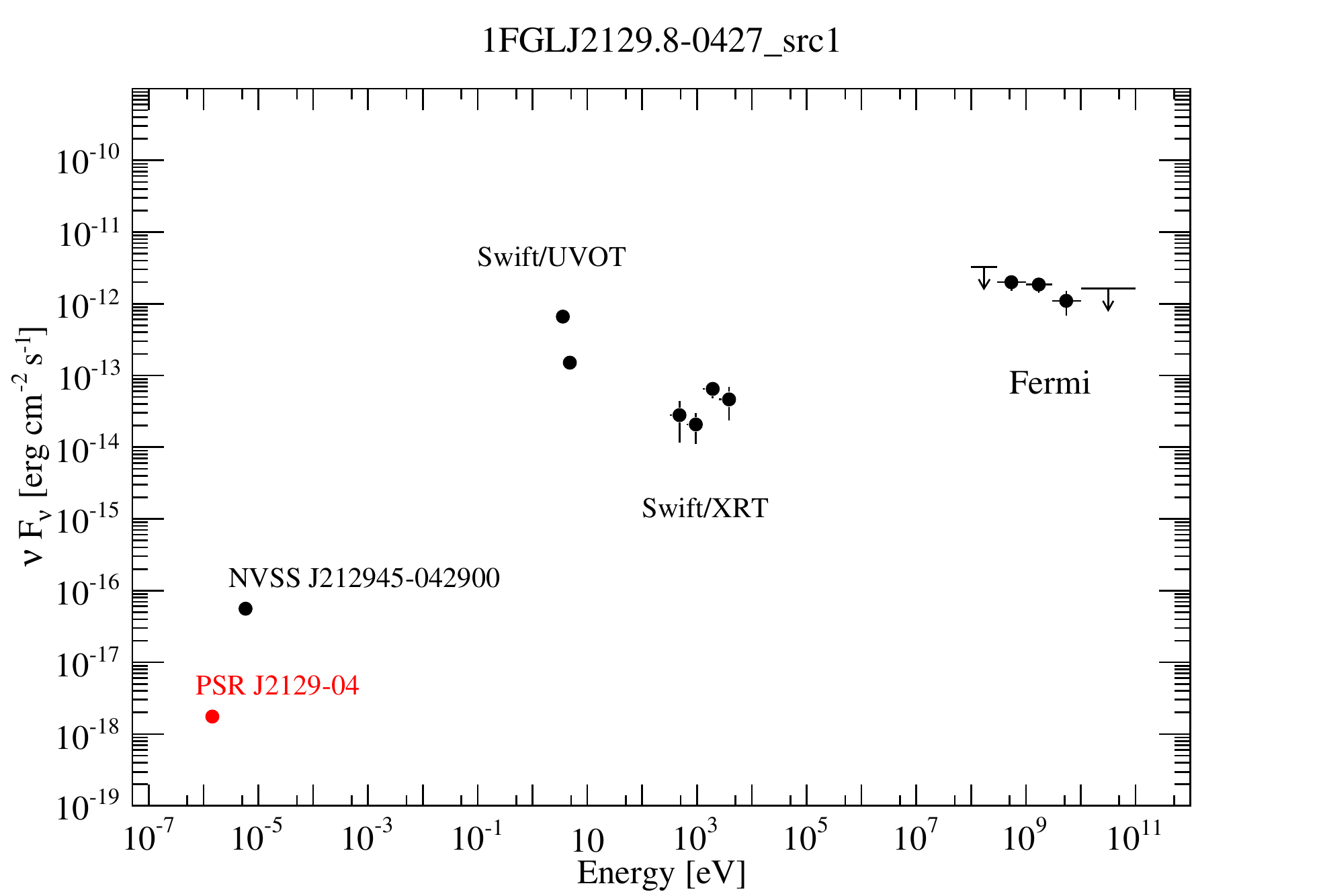}
    \end{center}
  \end{minipage}
  \begin{minipage}{0.32\hsize}
    \begin{center}
      \includegraphics[width=55mm]{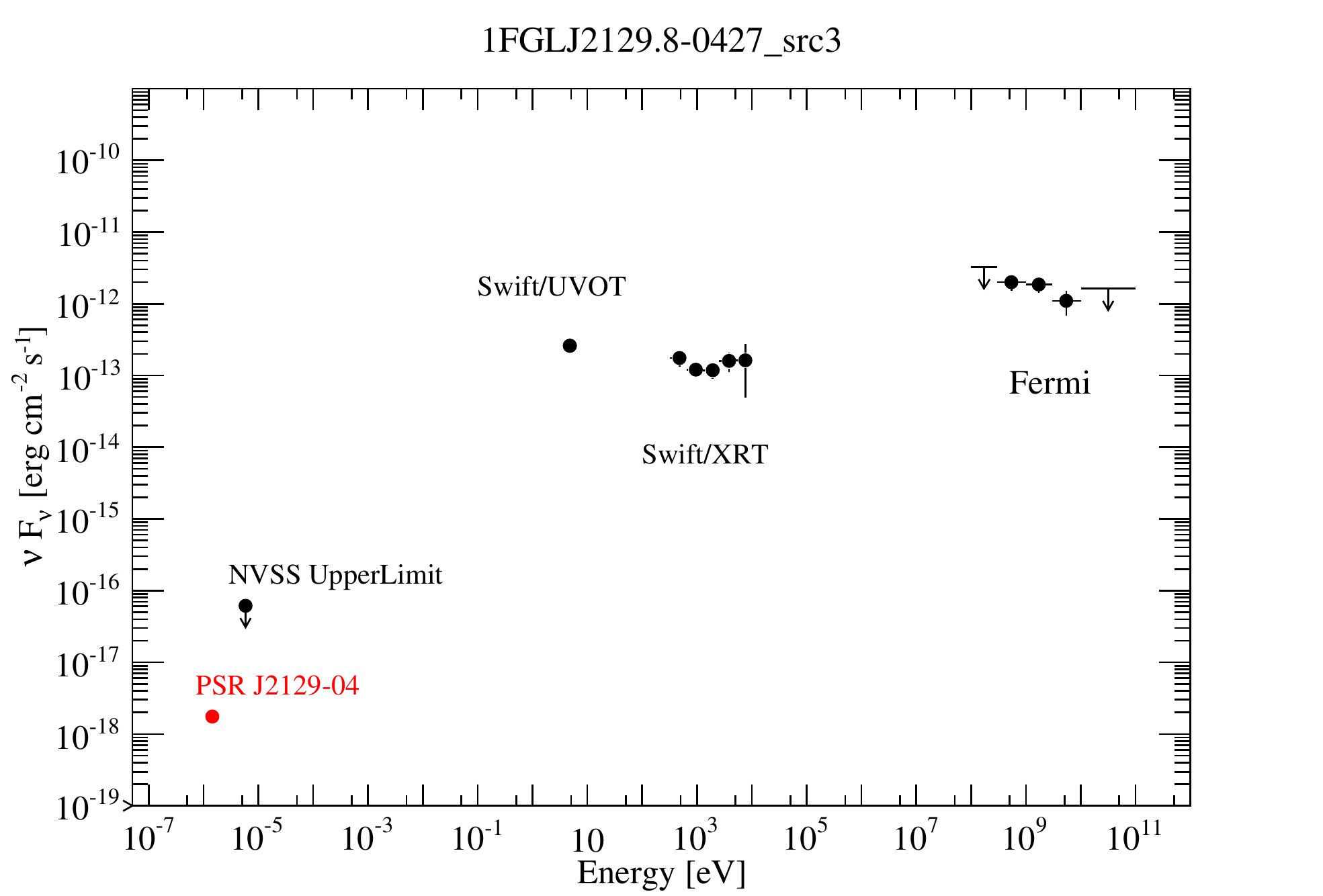}
    \end{center}
  \end{minipage}
  \begin{minipage}{0.32\hsize}
    \begin{center}
      \includegraphics[width=55mm]{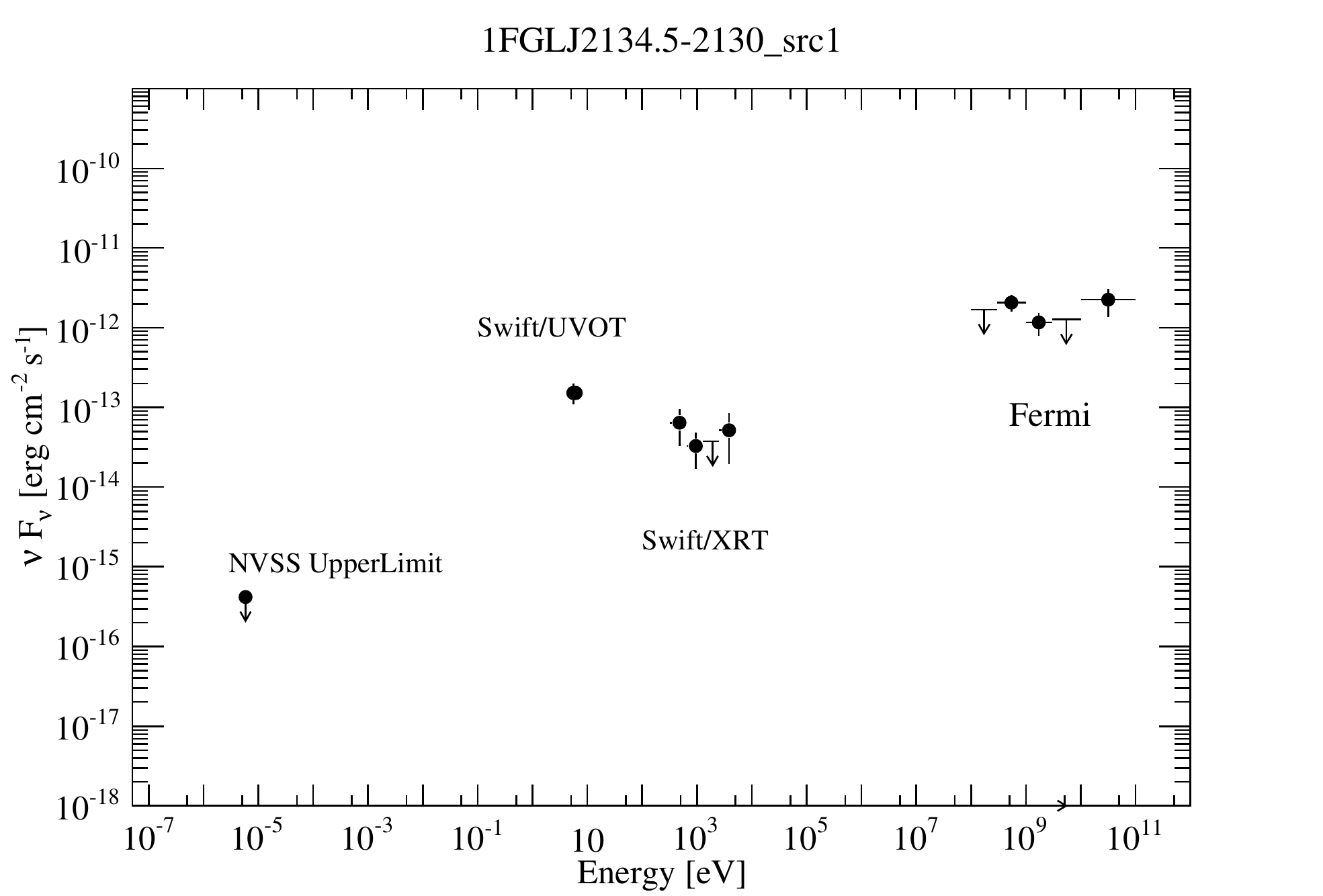}
    \end{center}
  \end{minipage}
  \begin{minipage}{0.32\hsize}
    \begin{center}
      \includegraphics[width=55mm]{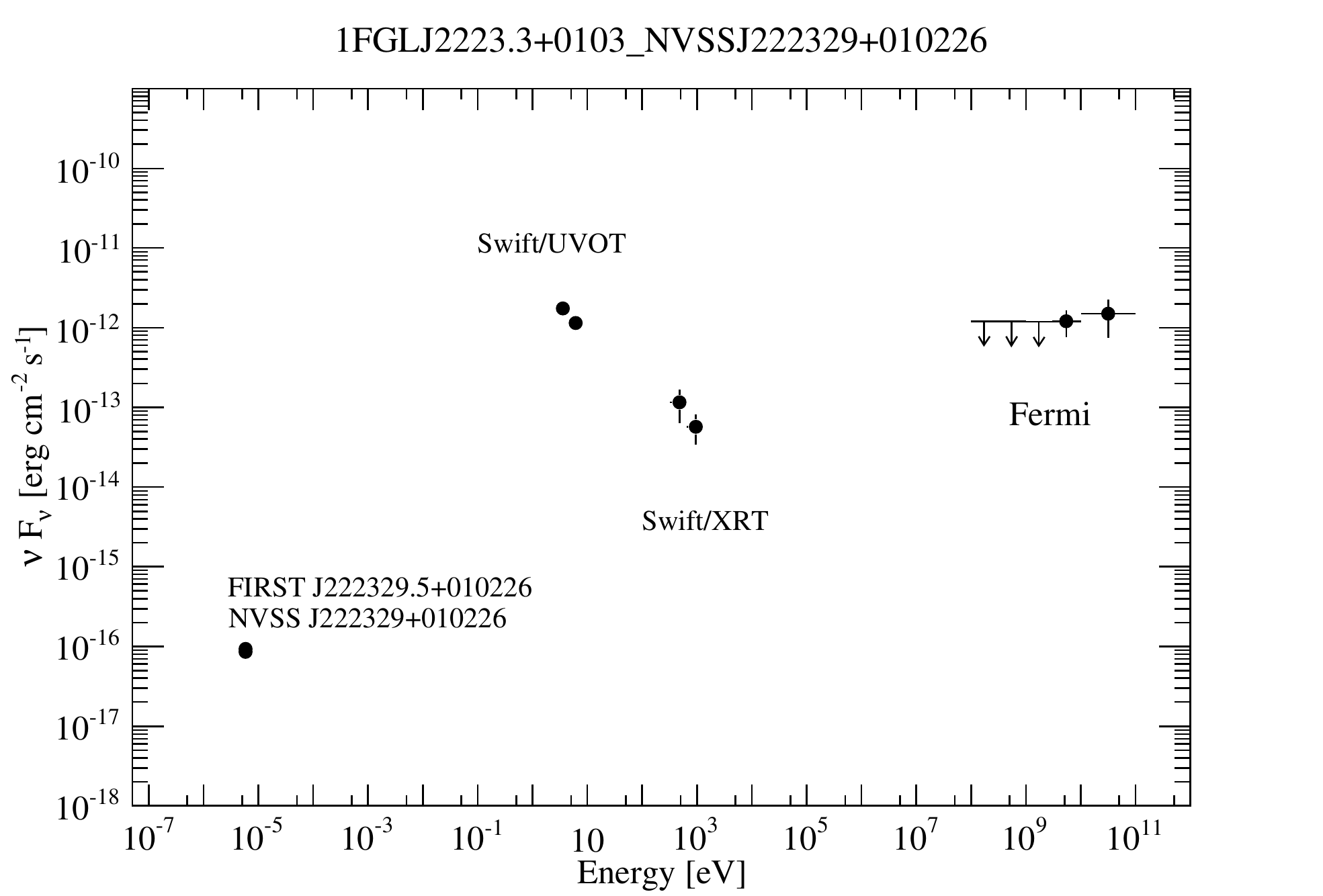}
    \end{center}
  \end{minipage}
  \begin{minipage}{0.32\hsize}
    \begin{center}
      \includegraphics[width=55mm]{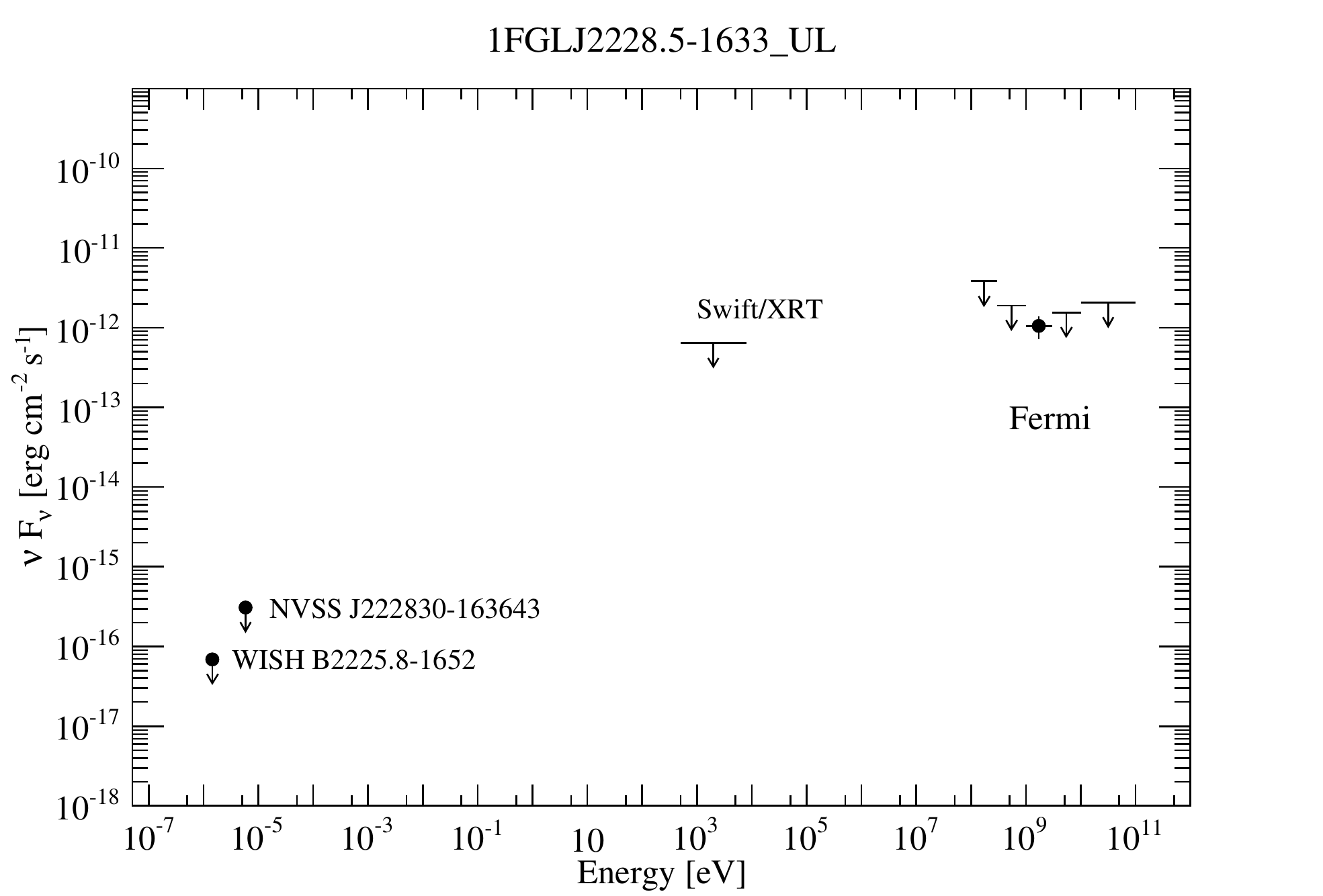}
    \end{center}
  \end{minipage}
  \begin{minipage}{0.32\hsize}
    \begin{center}
      \includegraphics[width=55mm]{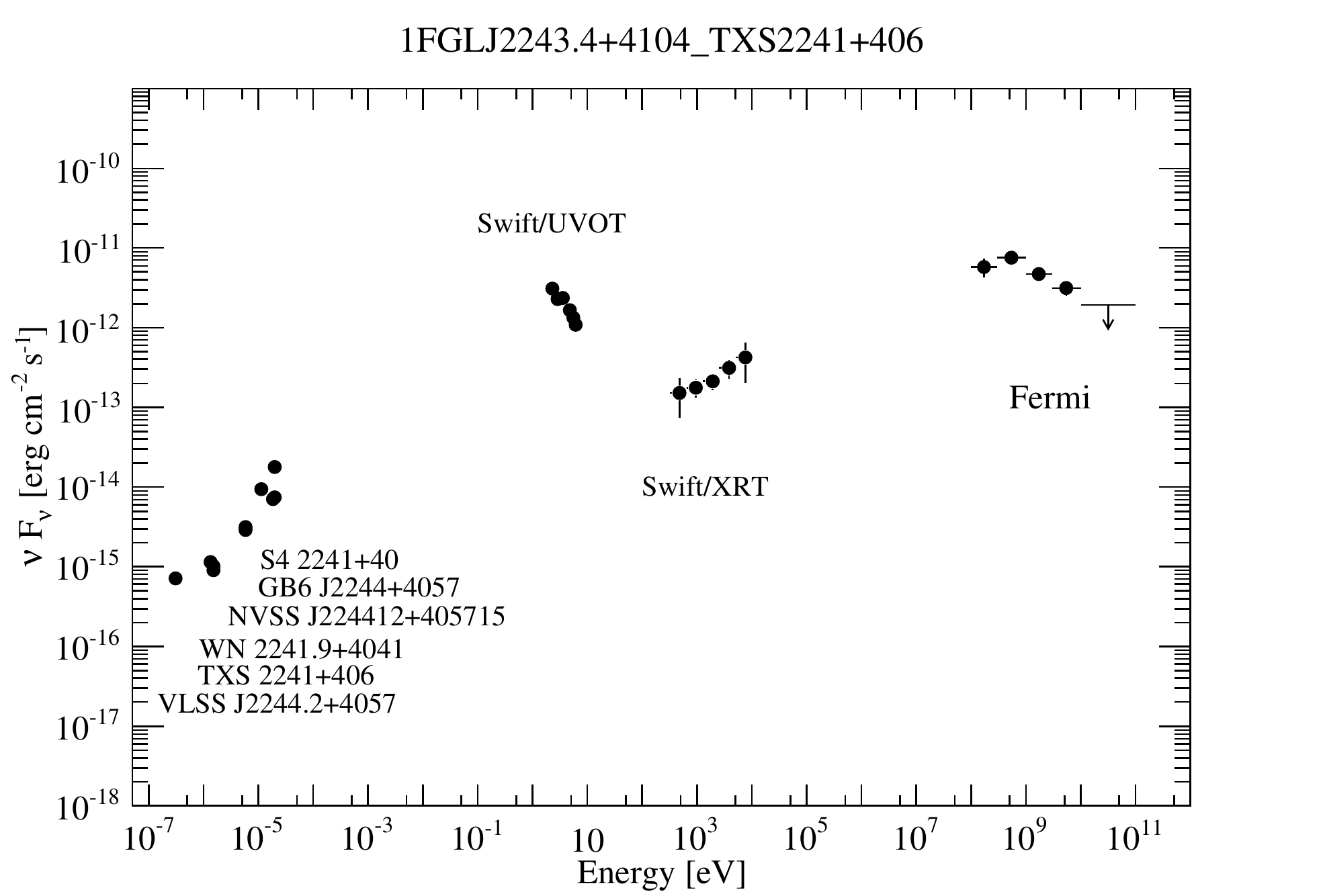}
    \end{center}
  \end{minipage}
  \begin{minipage}{0.32\hsize}
    \begin{center}
      \includegraphics[width=55mm]{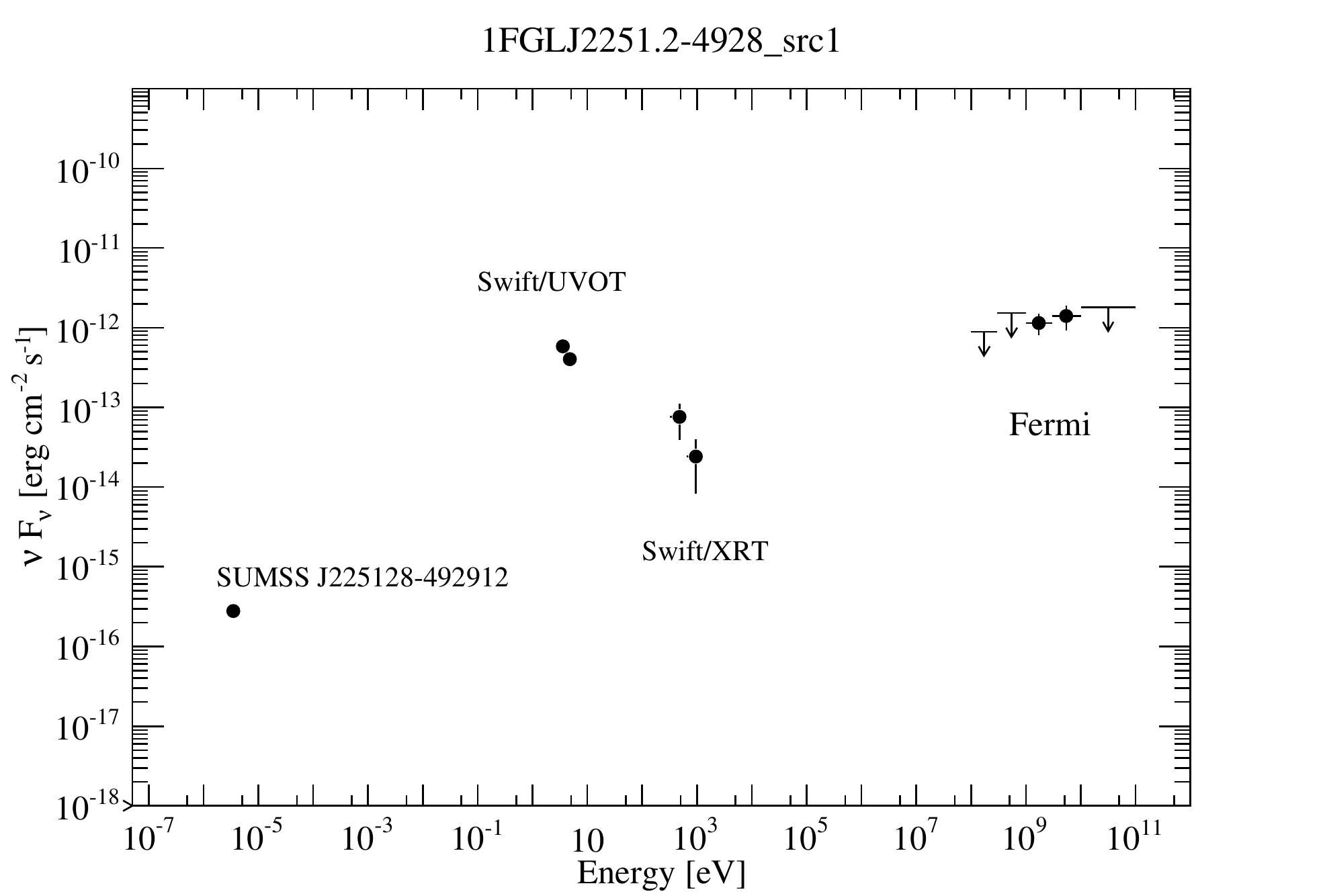}
    \end{center}
  \end{minipage}
  \begin{minipage}{0.32\hsize}
    \begin{center}
      \includegraphics[width=55mm]{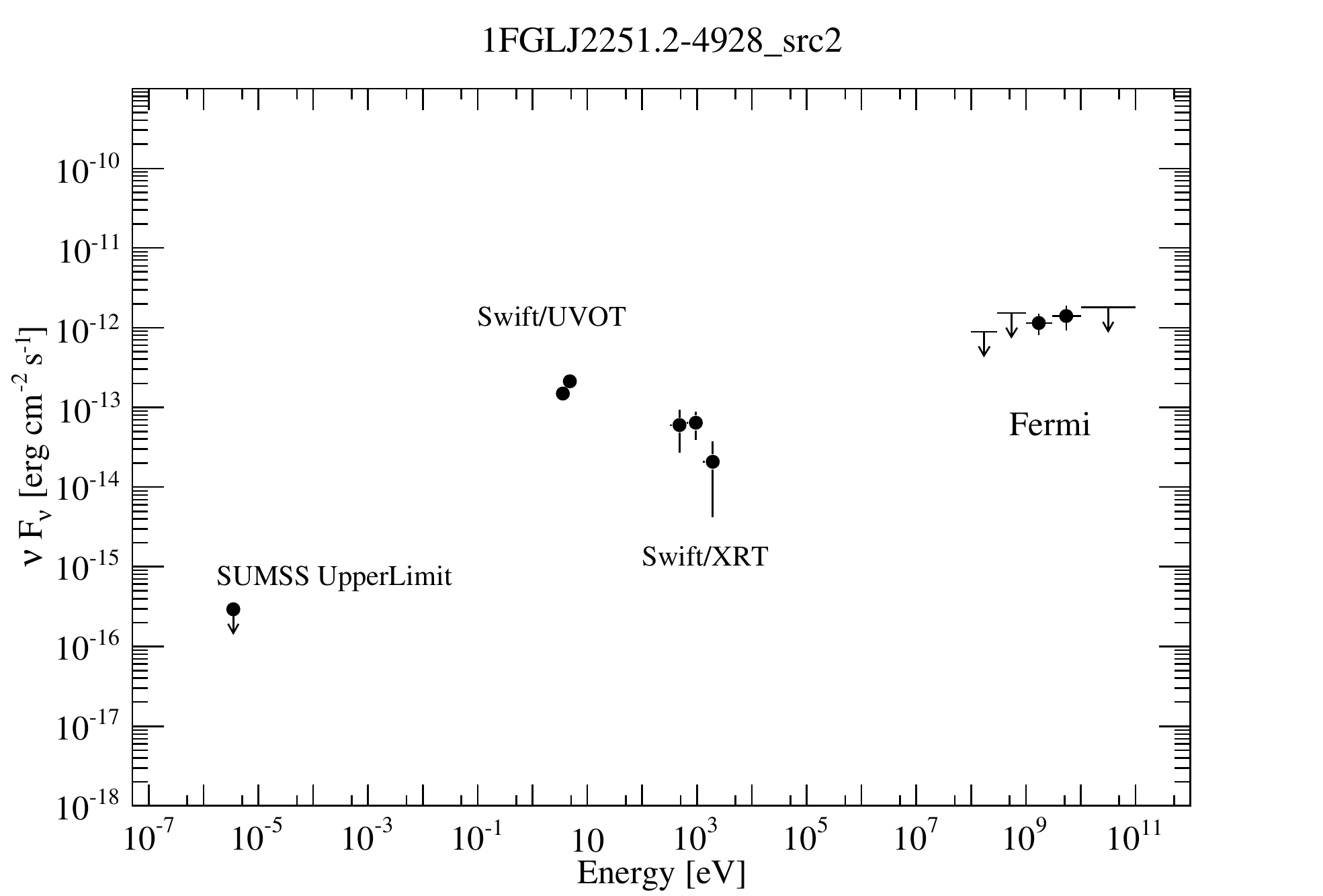}
    \end{center}
  \end{minipage}
  \begin{minipage}{0.32\hsize}
    \begin{center}
      \includegraphics[width=55mm]{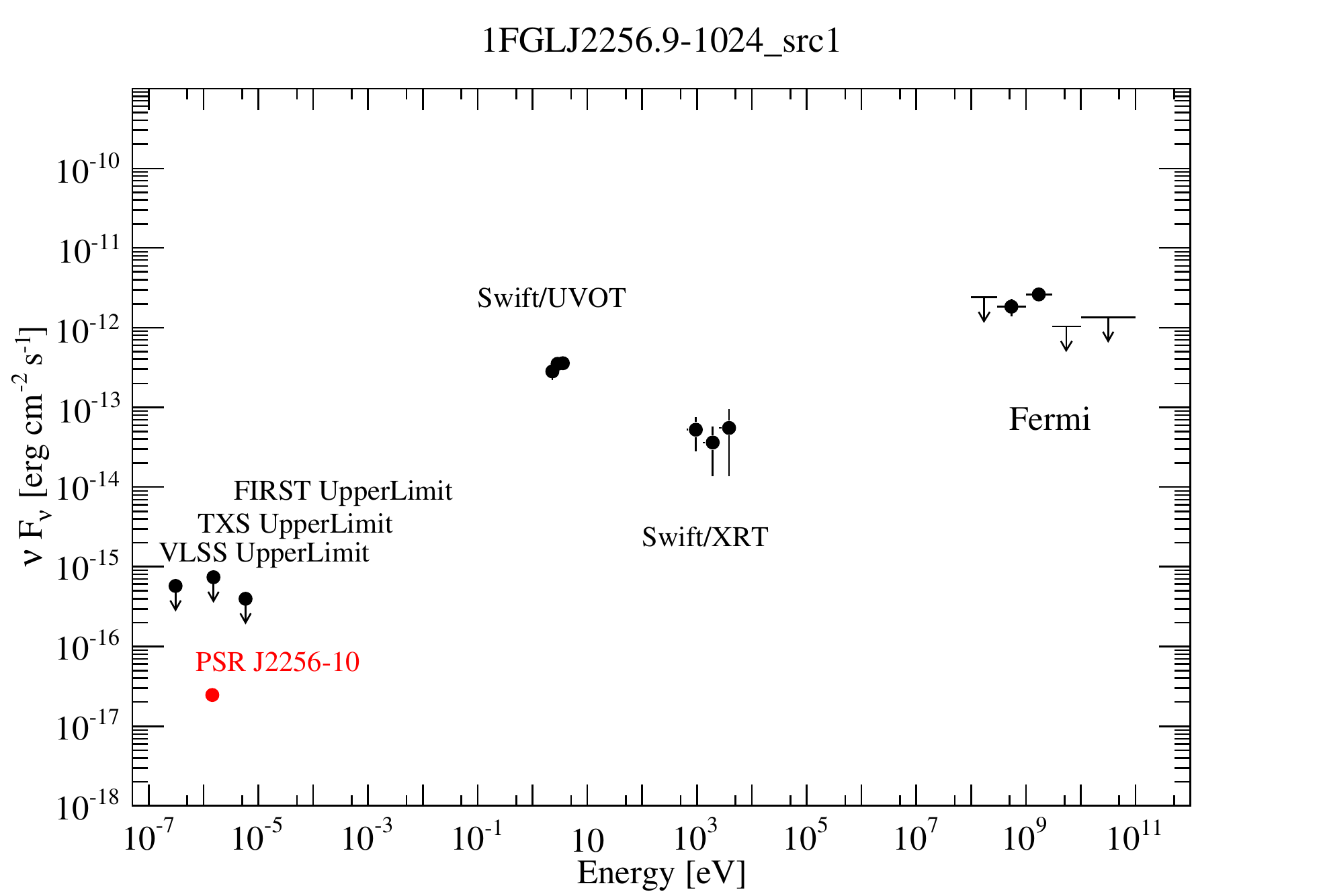}
    \end{center}
  \end{minipage}
  \begin{minipage}{0.32\hsize}
    \begin{center}
      \includegraphics[width=55mm]{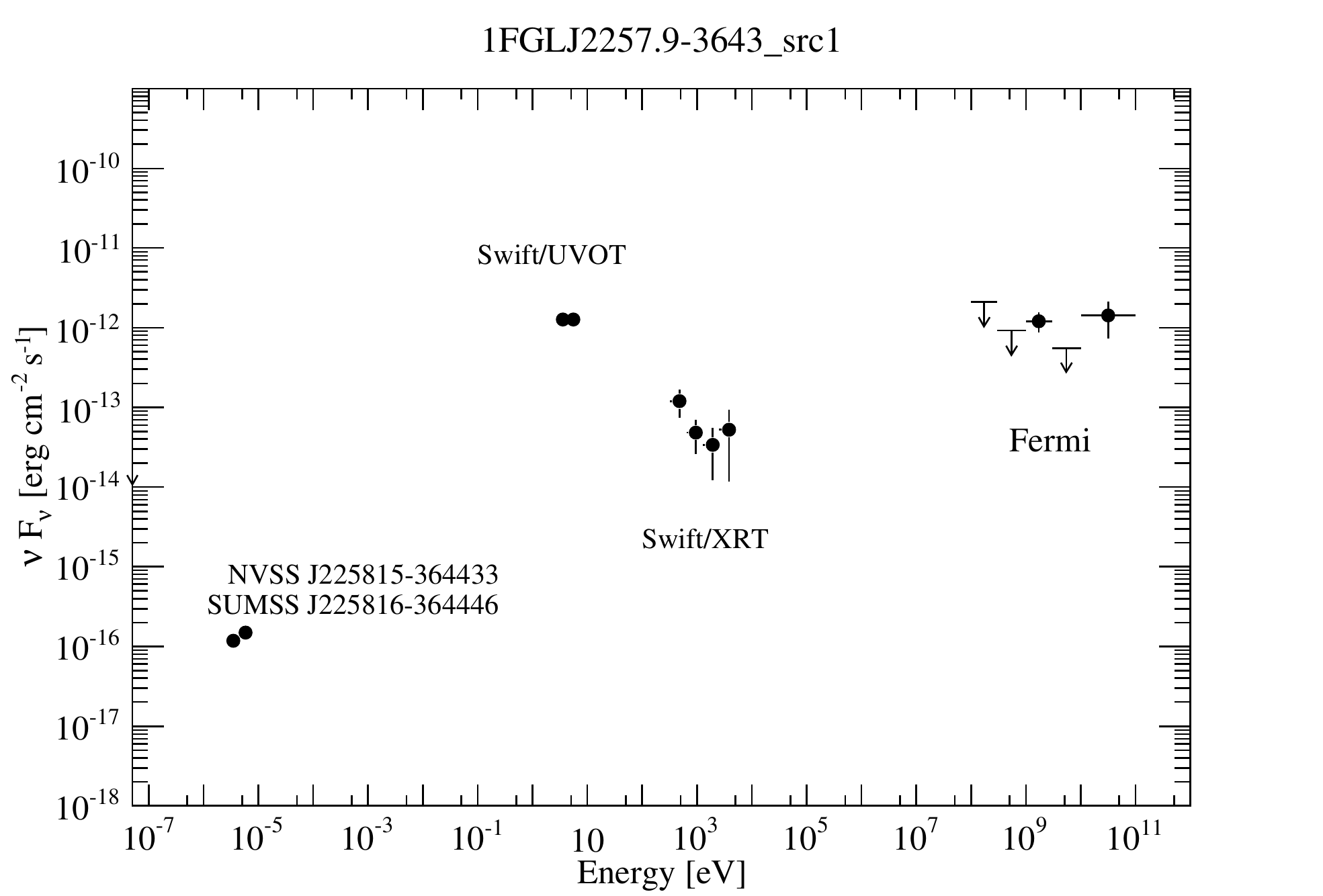}
    \end{center}
  \end{minipage}
  \begin{minipage}{0.32\hsize}
    \begin{center}
      \includegraphics[width=55mm]{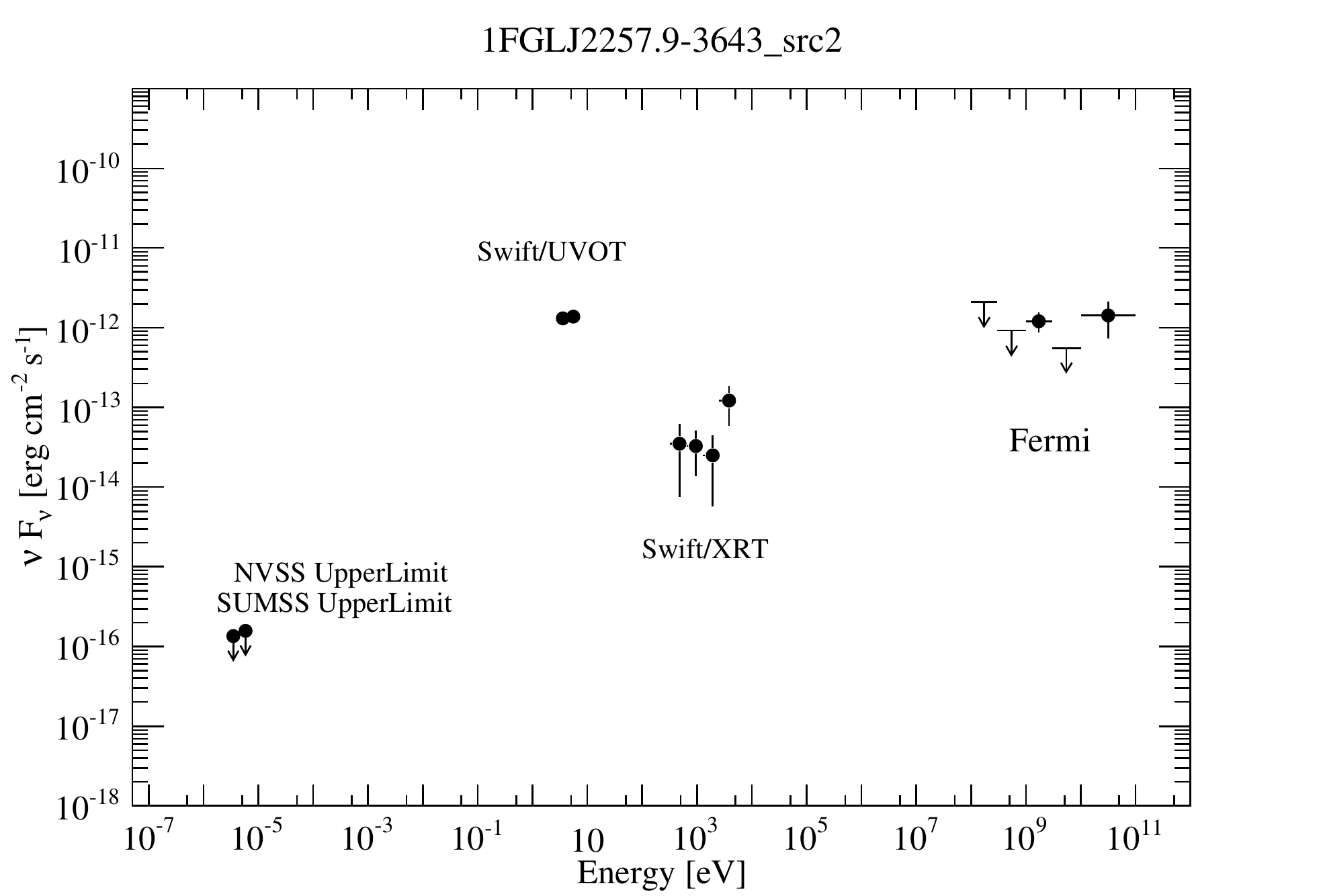}
    \end{center}
  \end{minipage}
  \begin{minipage}{0.32\hsize}
    \begin{center}
      \includegraphics[width=55mm]{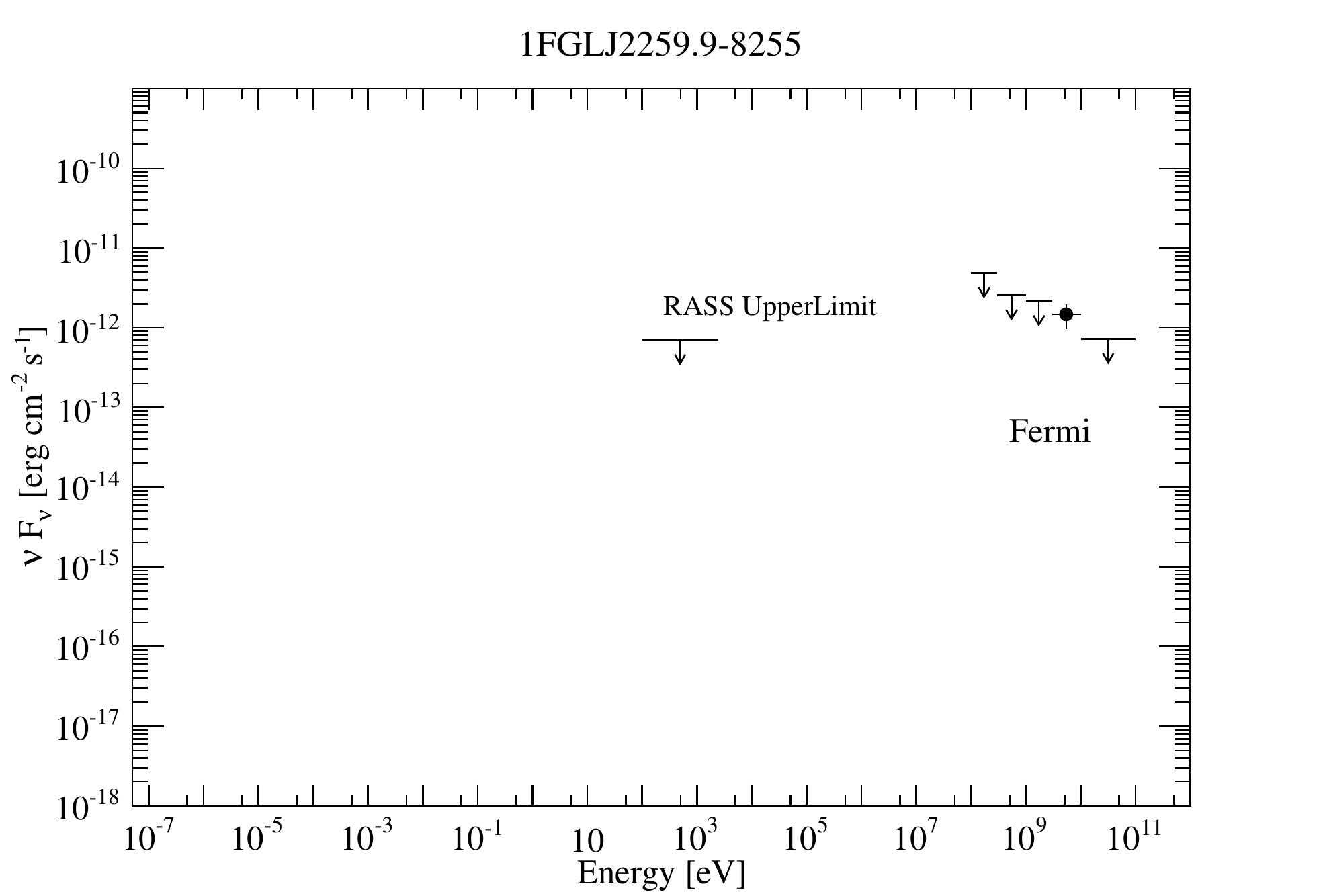}
    \end{center}
  \end{minipage}
  \begin{minipage}{0.32\hsize}
    \begin{center}
      \includegraphics[width=55mm]{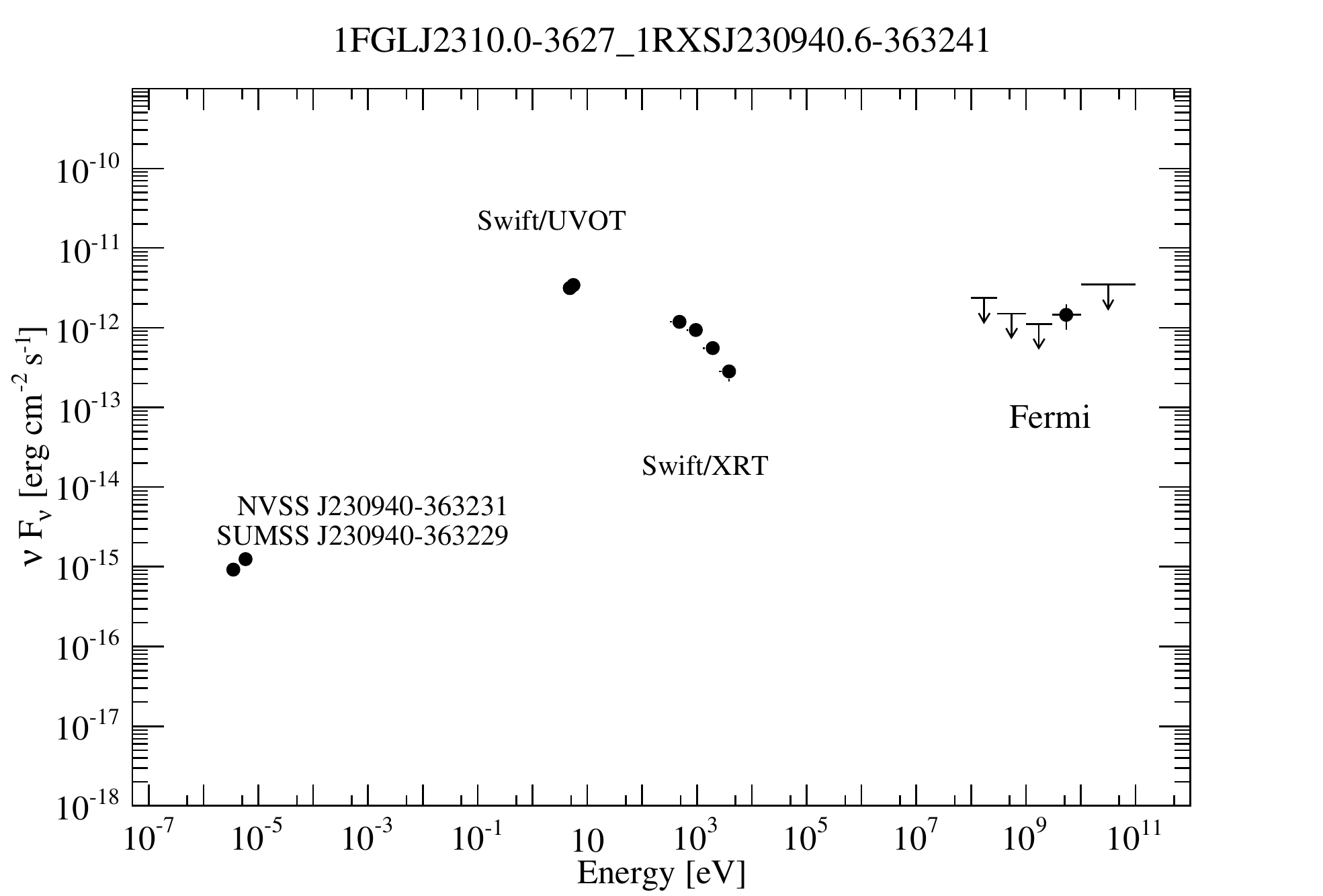}
    \end{center}
  \end{minipage}
  \begin{minipage}{0.32\hsize}
    \begin{center}
      \includegraphics[width=55mm]{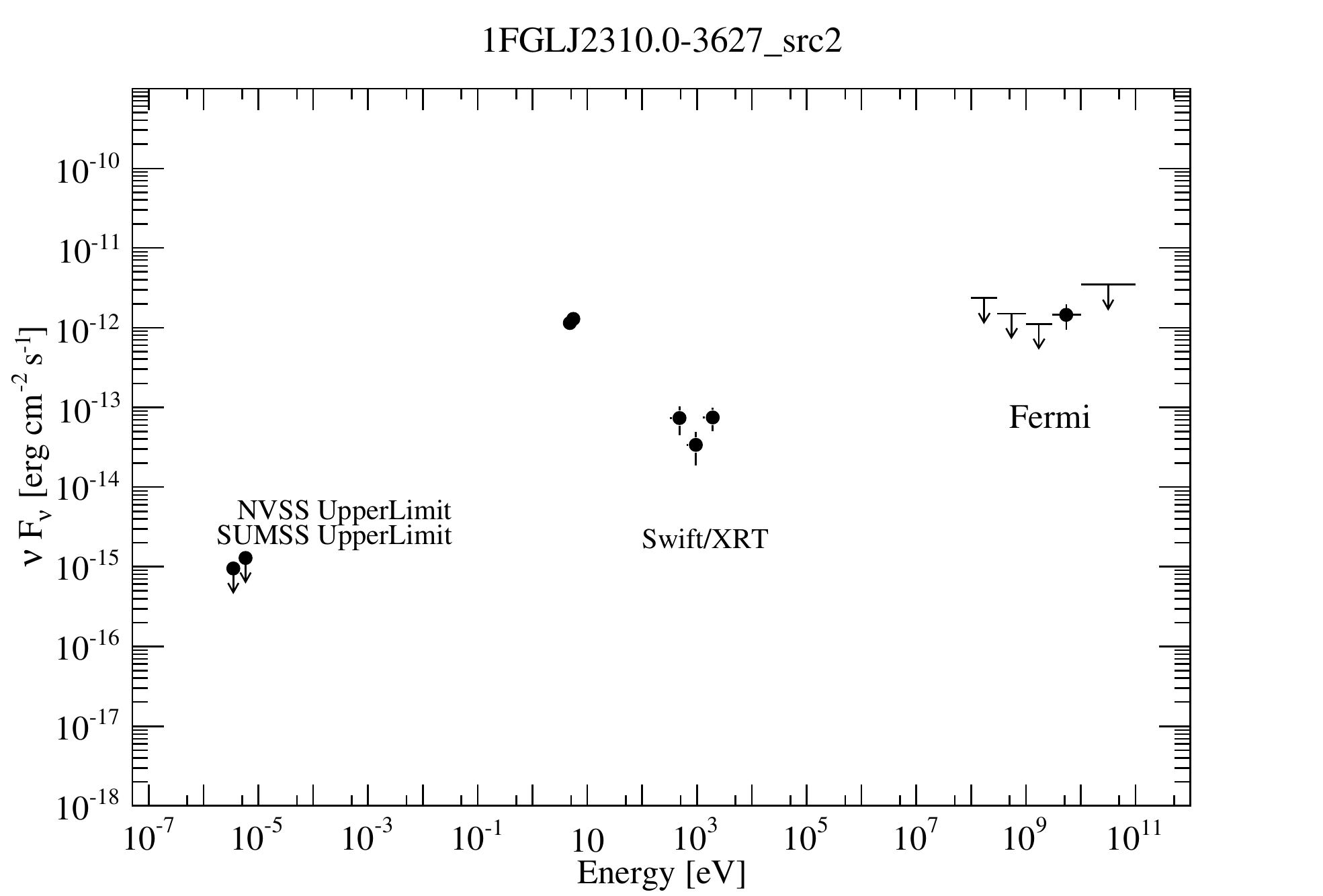}
    \end{center}
  \end{minipage}
 \end{center}
\end{figure}
\clearpage
\begin{figure}[m]
 \begin{center}
  \begin{minipage}{0.32\hsize}
    \begin{center}
      \includegraphics[width=55mm]{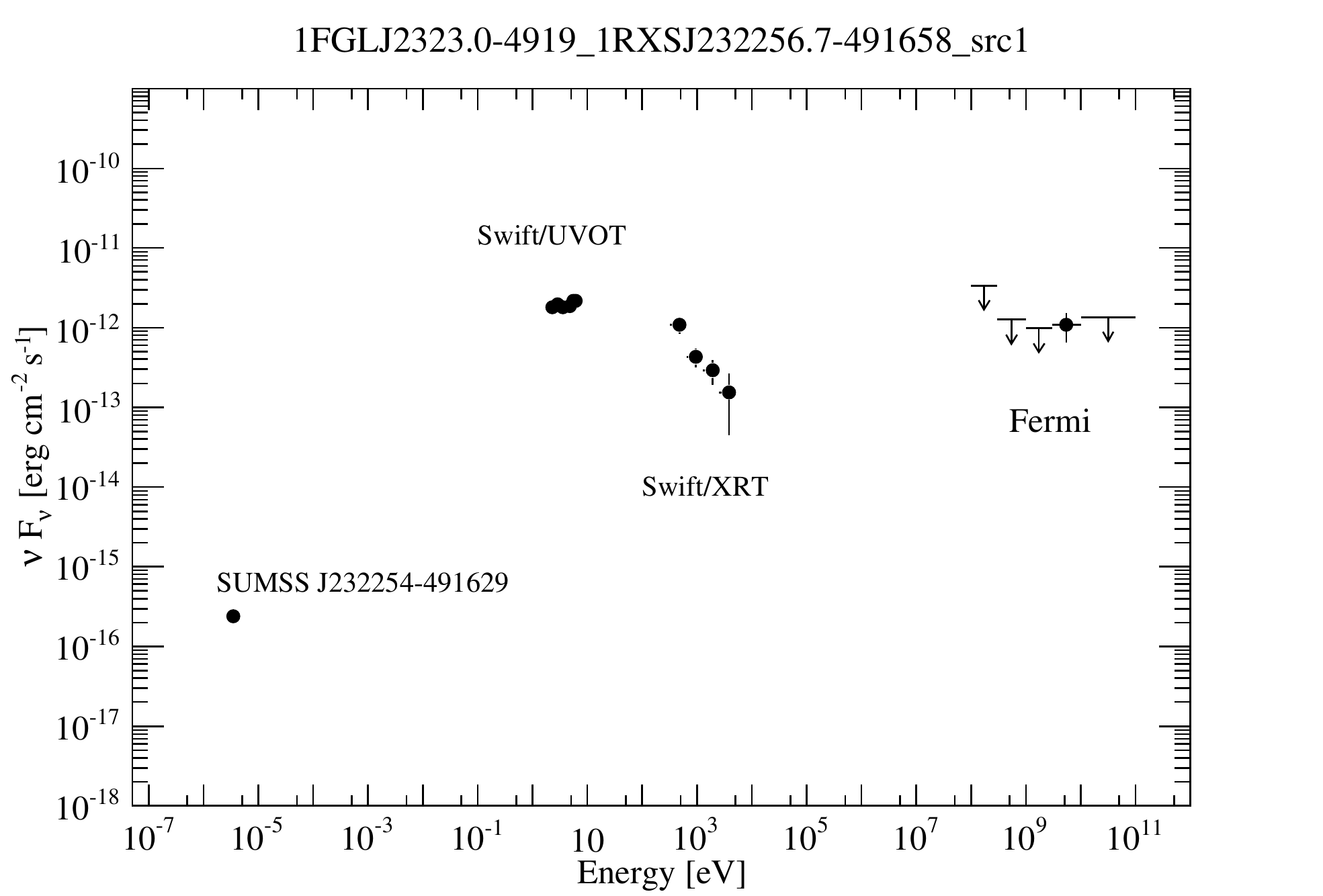}
    \end{center}
  \end{minipage}
  \begin{minipage}{0.32\hsize}
    \begin{center}
      \includegraphics[width=55mm]{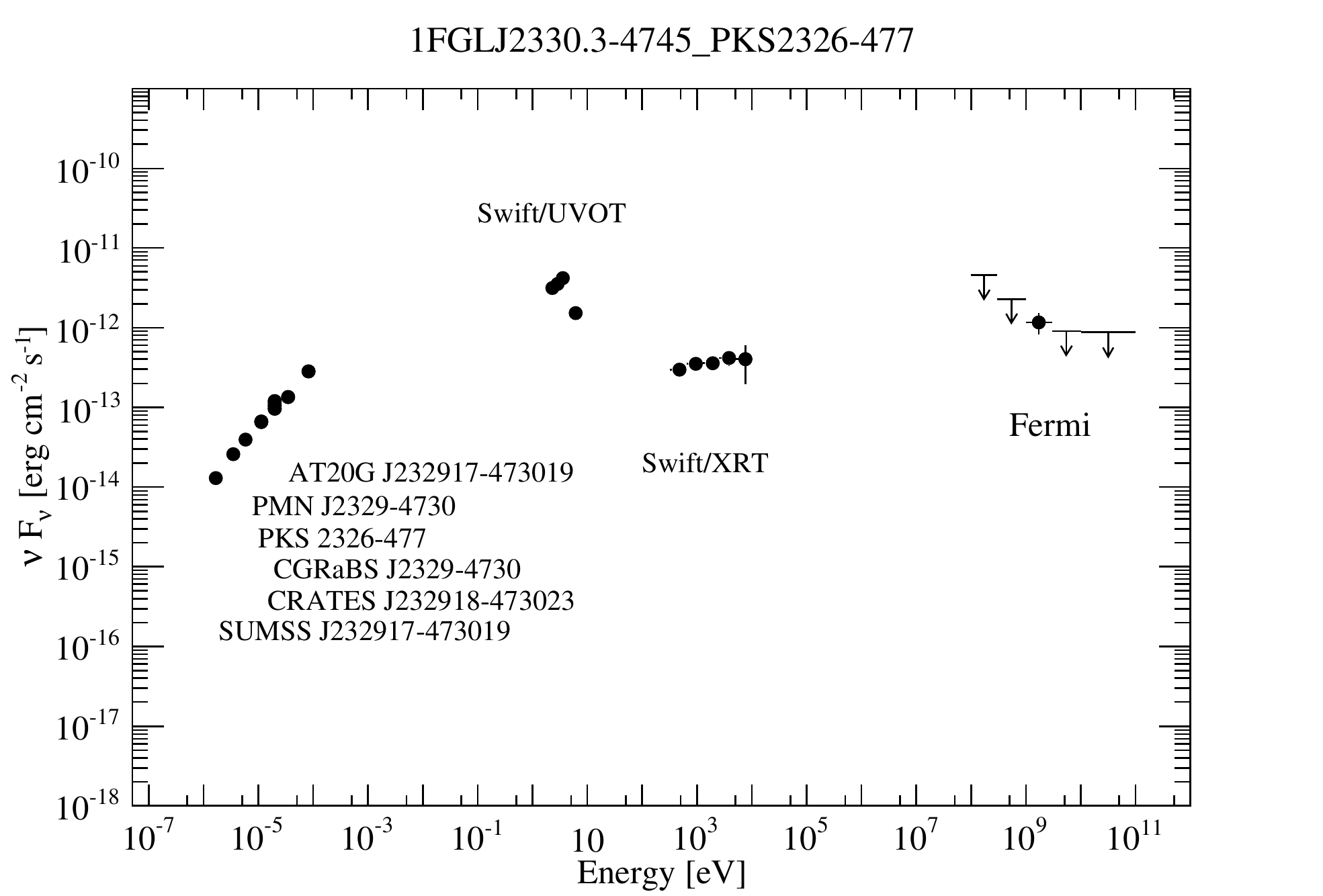}
    \end{center}
  \end{minipage}
  \begin{minipage}{0.32\hsize}
    \begin{center}
      \includegraphics[width=55mm]{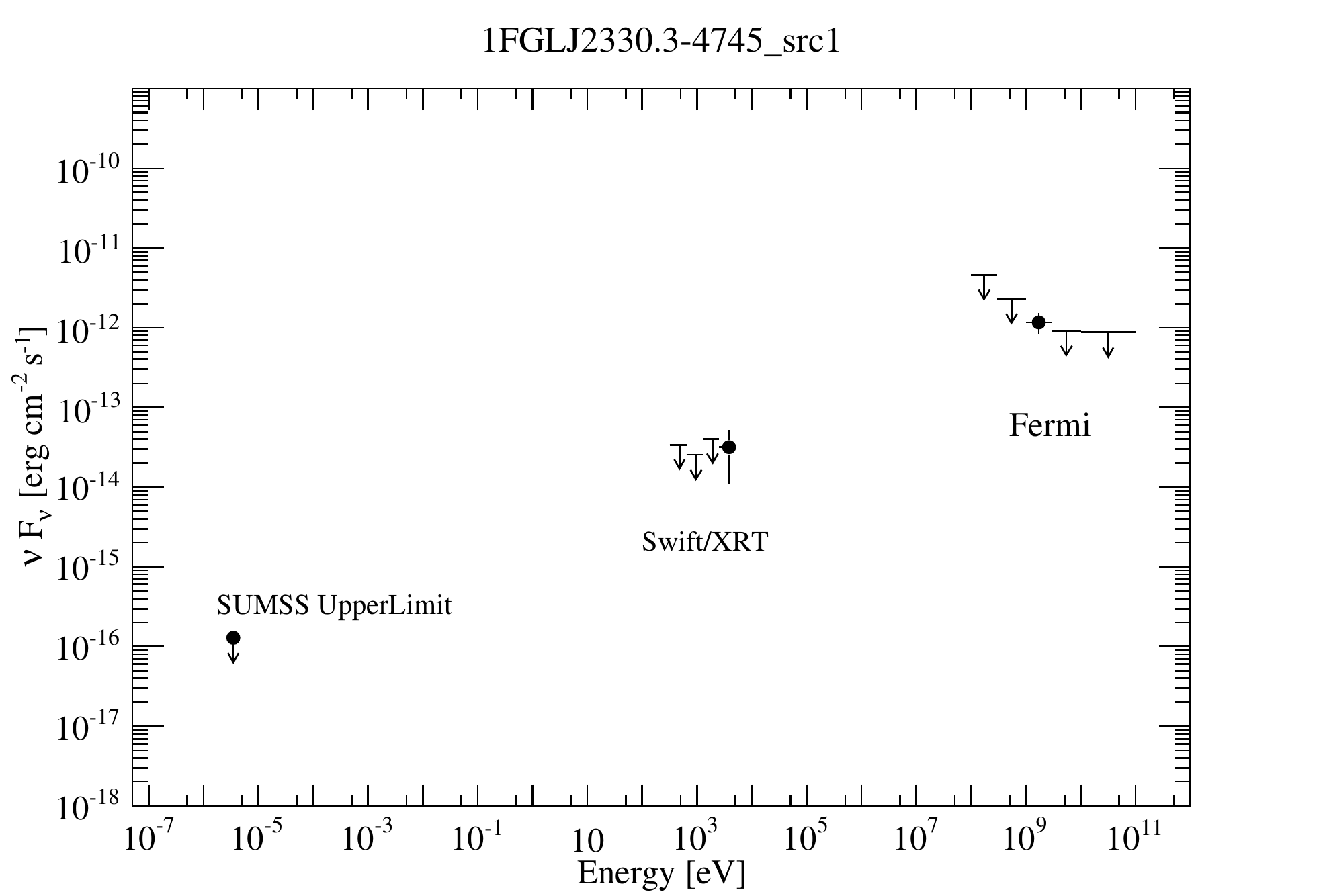}
    \end{center}
  \end{minipage}
  \begin{minipage}{0.32\hsize}
    \begin{center}
      \includegraphics[width=55mm]{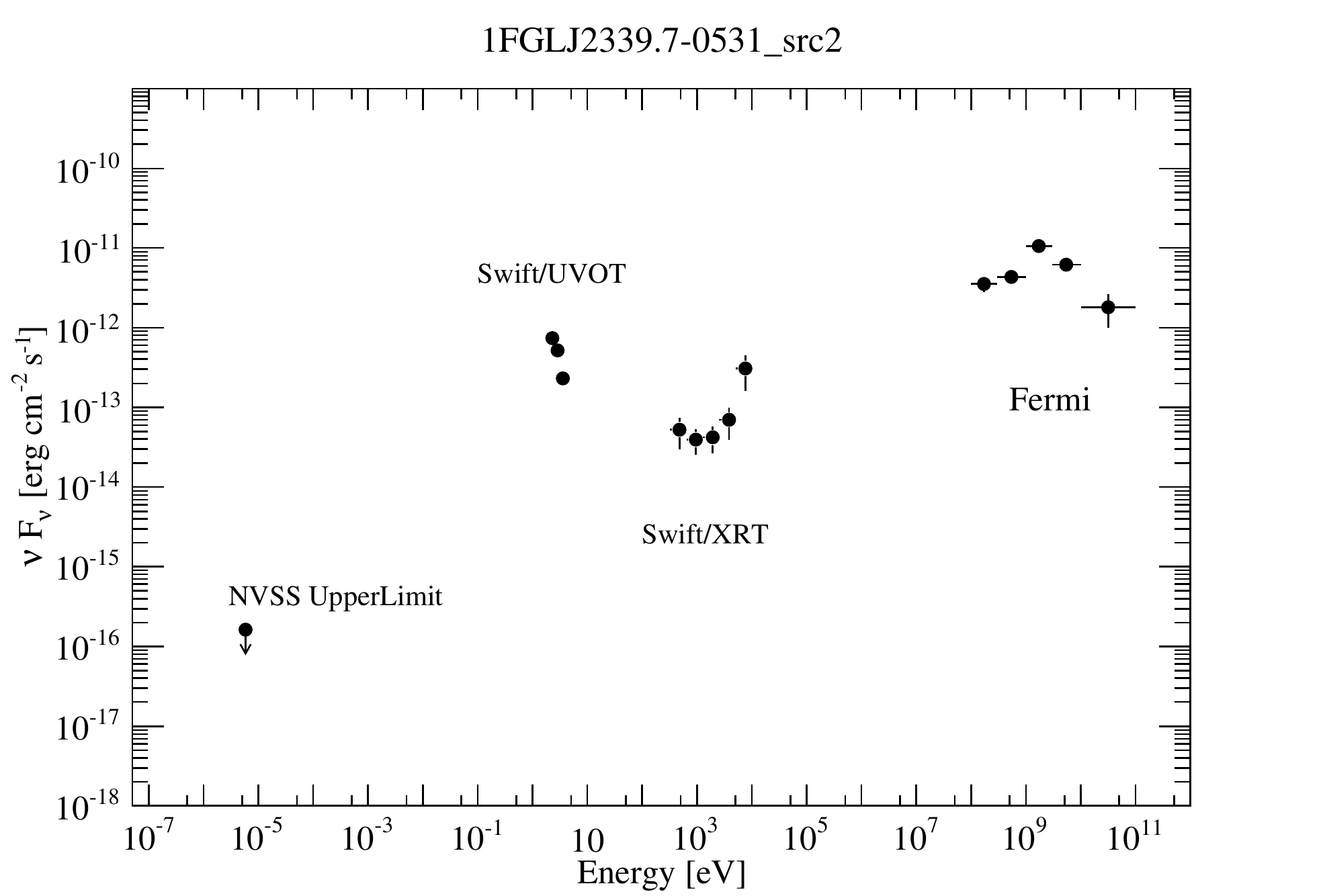}
    \end{center}
  \end{minipage}
  \begin{minipage}{0.32\hsize}
    \begin{center}
      \includegraphics[width=55mm]{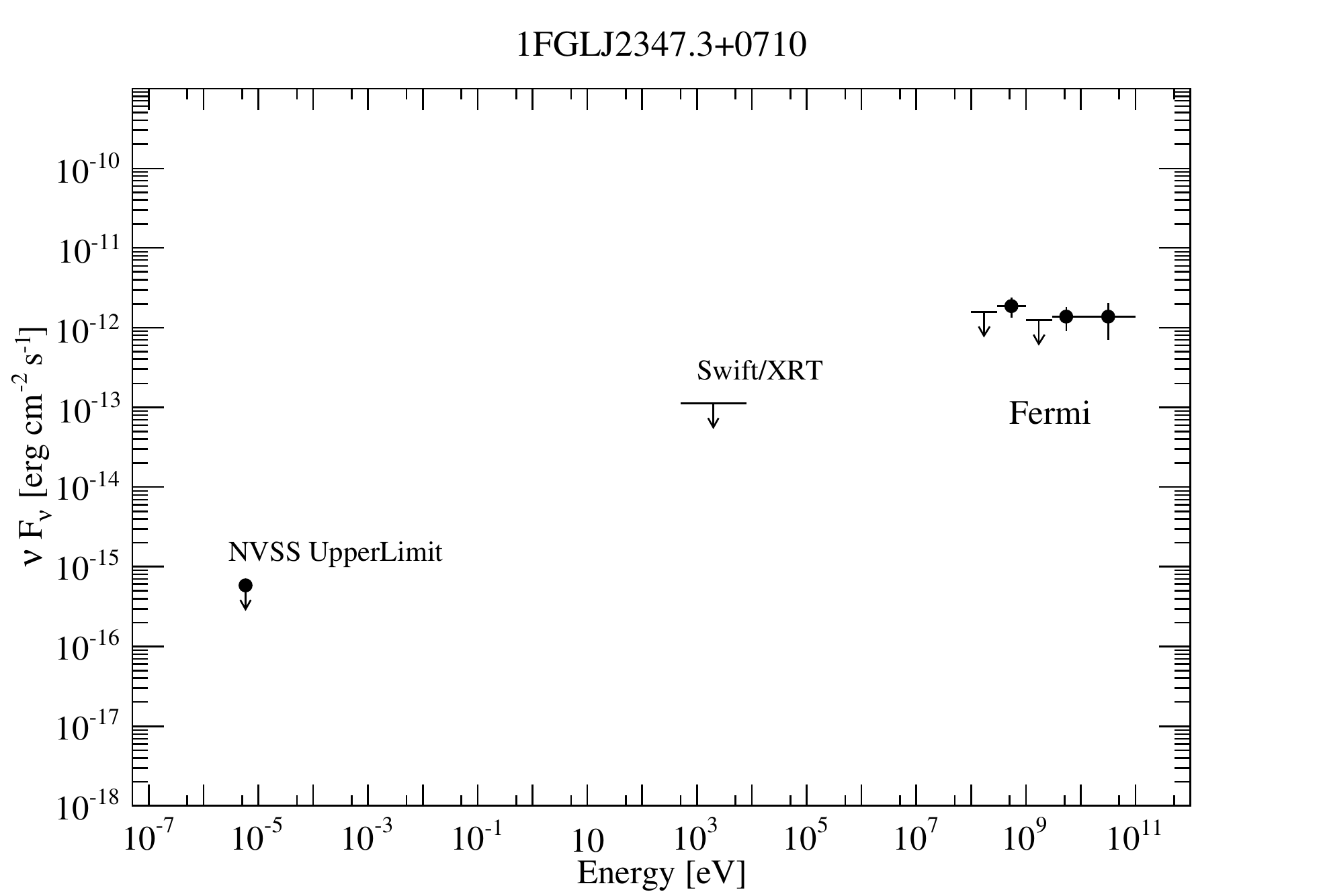}
    \end{center}
  \end{minipage}
  \begin{minipage}{0.32\hsize}
    \begin{center}
      \includegraphics[width=55mm]{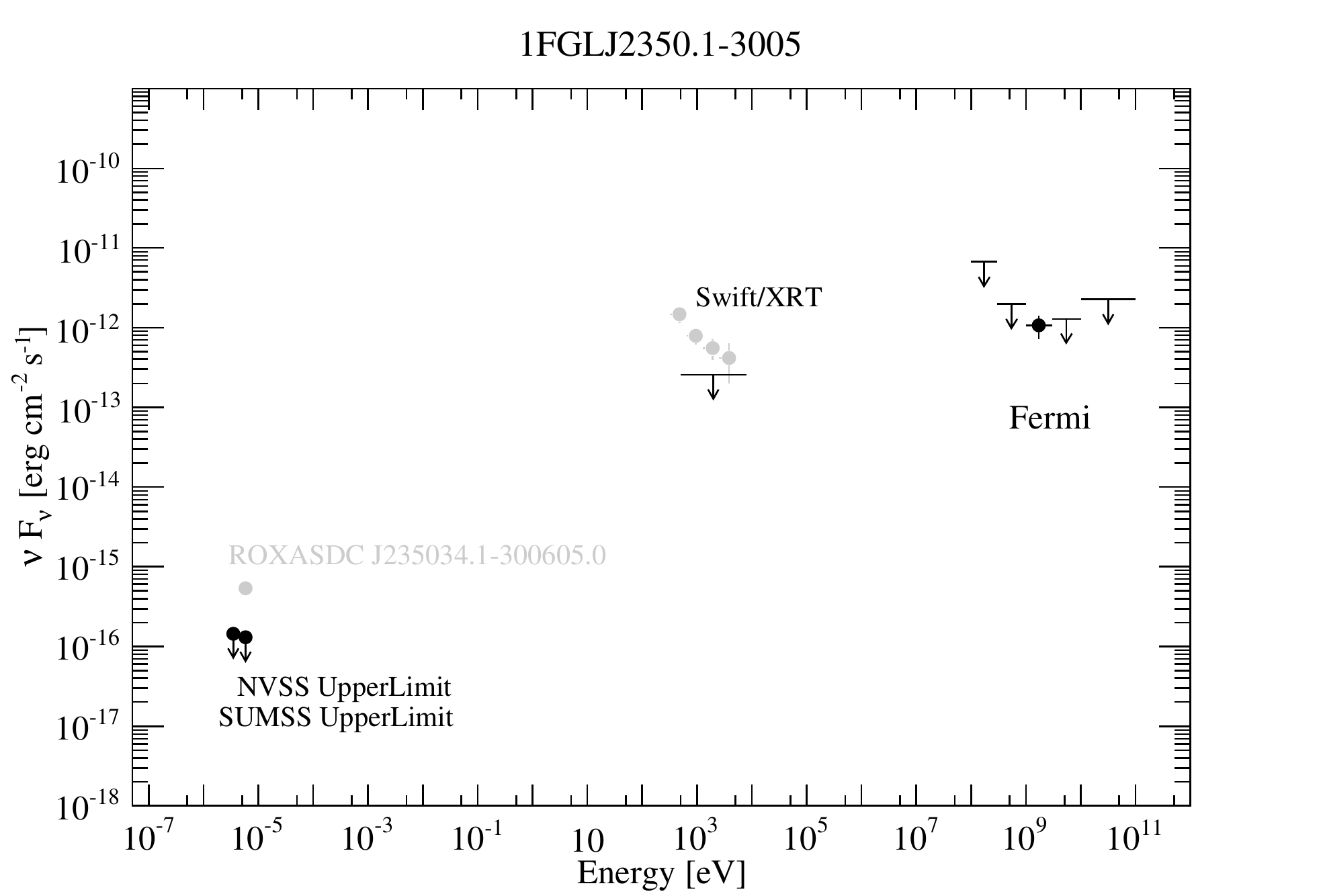}
    \end{center}
  \end{minipage}
  \begin{minipage}{0.32\hsize}
    \begin{center}
      \includegraphics[width=55mm]{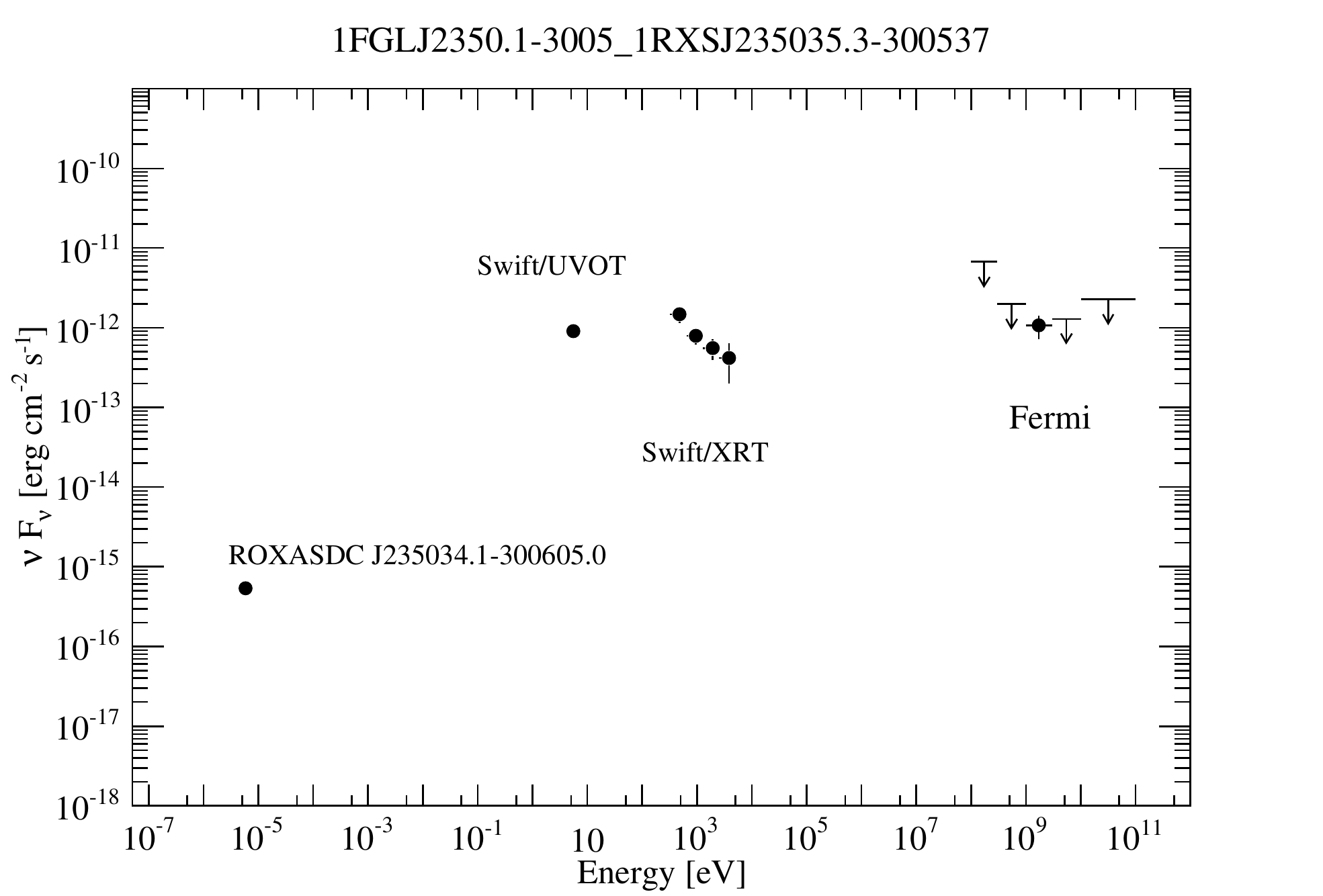}
    \end{center}
  \end{minipage}
  \begin{minipage}{0.32\hsize}
    \begin{center}
      \includegraphics[width=55mm]{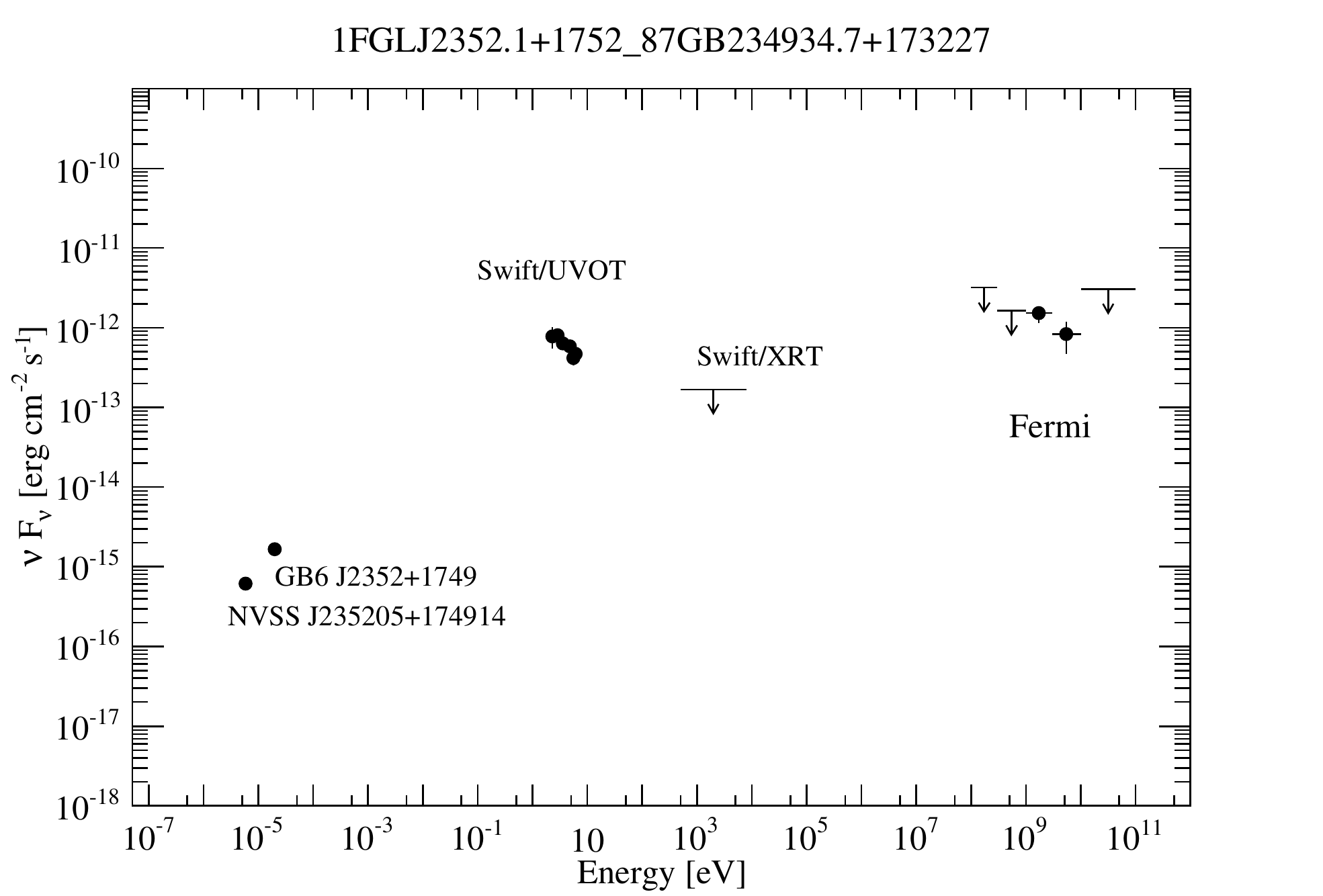}
    \end{center}
  \end{minipage}
 \caption{Broadband SEDs of the 134 1FGL unidentified sources we selected in this paper. The radio fluxes of these sources are described in Section\,2.3. The X-ray fluxes for these sources are given from {\it Swift}/XRT analysis (see Section\,2.1). Finally, the LAT fluxes of these sources are taken from the 2FGL catalog \citep{2FGL}.}
 \end{center}
\end{figure}
\clearpage

{
\setlength{\textheight}{40\baselineskip}
\begin{landscape}
\begin{table}
\label{UV&Radio}
\caption{All Radio fluxes and {\it Swift}/UVOT fluxes plotted in SEDs(Figure 12) are listed here.}
\begin{center}
\scalebox{0.7}{
\begin{tabular}{lccccccccccc}
\hline \hline
 &  \multicolumn{4}{c}{Radio [mJy]} & \multicolumn{6}{c}{UV (Swift/UVOT) [E-12 erg/cm$^2$/s]}\\ \cline{2-5} \cline{7-12}
\multicolumn{1}{c}{Name} & $F_{350MHz}$ & $F_{843MHz}$ & $F_{1.4GHz}$ & $F_{4.85GHz}$  && V-band & B-band & U-band & UVW1-band & UVM2-band & UVW2-band\\ \hline
J0001.9-4158\_src1	&-	&$13.2\pm1.2$	&-	&-	&	&-	&-	&-	&-	&-	&-\\
J0001.9-4158\_src2	&-	&$<5$	&-	&-	&	&-	&-	&$0.131\pm0.017$	&-	&$0.213\pm0.030$	&$0.152\pm0.010$\\
J0009.1+5031	&-	&-	&$12.1\pm0.6 $	&-	&	&-	&-	&$1.94\pm0.06$	&-	&-	&$1.61\pm0.063$\\
J0022.2-1850	&-	&-	&$22.4\pm0.8 $	&-	&	&$3.31\pm0.33$	&$3.39\pm0.23$	&$3.19\pm0.19$	&$2.83\pm0.17$	&$3.26\pm0.21$	&$2.51\pm0.12$\\
J0023.5+0930\_PSRJ0023+48\_src2	&$2.00\pm0.01$	&-	&-	&-	&	&-	&-	&-	&-	&-	&-\\
J0023.5+0930\_PSRJ0023+48\_src3	&$2.00\pm0.01$	&-	&-	&-	&	&-	&-	&-	&-	&-	&-\\
J0030.7+0724\_src2	&-	&-	&$<2.5$	&-	&	&-	&-	&-	&-	&$0.0685\pm0.0067$	&-\\
J0030.7+0724\_src6	&-	&-	&$12.2$	&-	&	&-	&-	&-	&-	&-	&-\\
J0032.7-5519	&-	&$<5$	&-	&-	&	&-	&-	&-	&-	&-	&-\\
J0038.0+1236	&-	&-	&$75.6\pm2.3$	&-	&	&$6.63\pm0.51$	&$4.59\pm0.31$	&$2.87\pm0.22$	&$2.71\pm0.21$	&$1.79\pm0.18$	&$2.14\pm0.15$\\
J0039.2+4331	&-	&-	&$<3.7$	&-	&	&-	&-	&$1.21\pm0.04$	&-	&-	&-\\
J0051.4-6242	&-	&$43.2\pm1.5$	&-	&-	&	&$2.39\pm0.19$	&$3.31\pm0.16$	&$4.45\pm0.10$	&$3.29\pm0.13$	&$3.53\pm0.13$	&$3.67\pm0.12222$\\
J0054.9-2455	&-	&-	&$24.3\pm0.9$	&-	&	&$3.92\pm0.32$	&$4.55\pm0.23$	&$4.70\pm0.21$	&$4.75\pm0.20$	&$5.51\pm0.26$	&$5.43\pm0.18$\\
J0101.0-6423\_PSRJ0101-6422	&-	&-	&$0.20\pm0.01$	&-	&	&-	&-	&$<0.0253$	&-	&-	&-\\
J0102.3+0942	&-	&-	&$11.0\pm0.5$	&-	&	&-	&-	&-	&-	&$0.375\pm0.028$	&$0.454\pm0.026$\\
J0103.1+4840	&$0.50\pm0.01$	&-	&-	&-	&	&-	&-	&-	&-	&-	&-\\
J0106.7+4853	&-	&-	&-	&-	&	&-	&-	&-	&-	&-	&-\\
J0115.7+0357	&-	&-	&$99.4\pm3.0$	&-	&	&$1.51\pm0.27$	&$1.65\pm0.18$	&$1.90\pm0.15$	&$1.42\pm0.12$	&$1.68\pm0.14$	&$1.65\pm0.10$\\
J0124.6-0616\_PMNJ0124-0624	&-	&-	&$40.9\pm1.3$	&-	&	&-	&-	&-	&-	&-	&-\\
J0143.9-5845	&-	&$28.3\pm1.2$	&-	&-	&	&-	&-	&$2.66\pm0.07$	&-	&-	&$3.27\pm0.08$\\
J0157.0-5259	&-	&$44.2\pm1.6$	&-	&-	&	&$0.946\pm0.139$	&$1.17\pm0.10$	&$1.21\pm0.08$	&$1.19\pm0.07$	&$1.17\pm0.08$	&$1.24\pm0.06$\\
J0217.9-6630\_RBS0300\_src1	&-	&-	&-	&$78.0\pm8.0$	&	&$2.05\pm0.25$	&$2.48\pm0.18$	&$2.32\pm0.15$	&$2.31\pm0.10$	&$2.79\pm0.20$	&$2.50\pm0.05$\\
J0217.9-6630\_src2	&-	&$<21.4$	&-	&-	&	&$0.592\pm0.185$	&$<0.307$	&$<0.194$	&$0.319\pm0.047$	&$0.347\pm0.079$	&$0.348\pm0.015$\\
J0217.9-6630\_src3	&-	&$<5$	&-	&-	&	&$0.807\pm0.192$	&$0.383\pm0.106$	&$0.282\pm0.069$	&$0.318\pm0.045$	&$0.424\pm0.085$	&$0.292\pm0.014$\\
J0223.0-1118\_1RXSJ022314.6-111741	&-	&-	&$14.0\pm0.6$	&-	&	&-	&-	&$1.43\pm0.04$	&-	&$1.18\pm0.06$	&-\\
J0226.3+0937	&-	&-	&$<385.8$	&-	&	&-	&-	&-	&-	&-	&-\\
J0239.5+1324	&-	&-	&$<20.4$	&-	&	&-	&-	&-	&-	&-	&-\\
J0305.2-1601	&-	&-	&$<951.6$	&-	&	&-	&-	&$0.464\pm0.046$	&$0.342\pm0.051$	&$0.538\pm0.046$	&-\\
J0308.6+7442	&$0.30\pm0.01$	&-	&-	&-	&	&-	&-	&-	&-	&-	&-\\
J0316.3-6438	&-	&$17.8\pm1.3$	&-	&-	&	&-	&-	&$1.13\pm0.04$	&-	&$1.74\pm0.06$	&-\\
J0335.5-4501\_src1	&-	&$<20.5$	&-	&-	&	&-	&-	&-	&-	&$0.122\pm0.012$	&-\\
J0335.5-4501\_src3	&-	&$19.3\pm1.2$	&-	&-	&	&-	&-	&-	&-	&$1.61\pm0.05$	&-\\
J0340.4+4130\_PSRJ0340+4130	&$2.0\pm0.1$	&-	&-	&-	&	&-	&-	&-	&-	&-	&-\\
J0345.2-2355	&-	&-	&$29.2\pm1.0$	&-	&	&-	&-	&-	&$0.552\pm0.023$	&-	&$0.805\pm0.031$\\
J0409.9-0357	&-	&-	&$39.1\pm1.2$	&-	&	&-	&-	&$2.12\pm0.08$	&-	&$1.57\pm0.07$	&$2.02\pm0.09$\\
J0439.8-1857	&-	&-	&$60\pm2.3$	&-	&	&-	&-	&-	&$0.513\pm0.020$	&-	&-\\
J0505.9+6121\_NVSSJ050558+611336	&-	&-	&$21.8\pm0.8$	&-	&	&-	&-	&$2.53\pm0.36$	&-	&$<2.74$	&$3.79\pm0.87$\\
J0506.9-5435\_1ES0505-546	&-	&$18.0\pm1.1$	&-	&-	&	&$2.91\pm0.28$	&$3.16\pm0.20$	&$2.80\pm0.17$	&$3.29\pm0.16$	&$3.89\pm0.20$	&$3.72\pm0.15$\\
J0515.9+1528\_GB6J0515+1527	&-	&-	&$26.5\pm0.9$	&-	&	&$5.91\pm0.95$	&$6.47\pm0.95$	&$6.41\pm0.93$	&$4.78\pm1.36$	&$<4.53$	&$<6.36$\\
J0521.6+0103\_src1	&-	&-	&$24.1\pm0.8$	&-	&	&-	&-	&$2.12\pm0.17$	&$1.13\pm0.064$	&$0.778\pm0.082$	&$1.01\pm0.15$\\ \hline

\end{tabular}
}
\end{center}
\end{table}
\end{landscape}
}

{
\setlength{\textheight}{40\baselineskip}
\begin{landscape}
\begin{table}
\label{UV&Radio}
\begin{center}
\scalebox{0.7}{
\begin{tabular}{lccccccccccc}
\hline \hline
 &  \multicolumn{4}{c}{Radio [mJy]} & \multicolumn{6}{c}{UV (Swift/UVOT) [E-12 erg/cm$^2$/s]}\\ \cline{2-5} \cline{7-12}
\multicolumn{1}{c}{Name} & $F_{350MHz}$ & $F_{843MHz}$ & $F_{1.4GHz}$ & $F_{4.85GHz}$  && V-band & B-band & U-band & UVW1-band & UVM2-band & UVW2-band\\ \hline
J0523.5-2529	&-	&-	&$<2.5$	&-	&	&-	&-	&-	&$0.142\pm0.010$	&-	&-\\
J0533.9+6758\_src1	&-	&-	&$<2.5$	&-	&	&-	&-	&-	&$2615\pm2550$	&$4886\pm4874$	&-\\
J0533.9+6758\_src2	&-	&-	&$<2.5$	&-	&	&-	&-	&-	&$<0.219$	&$<0.176$	&-\\
J0537.7-5717\_src2	&-	&$<5$	&-	&-	&	&$1.87\pm0.18$	&$2.05\pm0.13$	&$1.88\pm0.11$	&$1.75\pm0.10$	&$1.70\pm0.11$	&$0.829\pm0.053$\\
J0537.7-5717\_SUMSSJ053748-571828	&-	&$102.6\pm3.3$	&-	&-	&	&-	&-	&-	&$327\pm7$	&$120\pm3$	&$108\pm2$\\
J0538.4-3910\_1RXSJ053810.0-390839	&-	&-	&$23\pm0.8$	&-	&	&$0.970\pm0.157$	&$0.503\pm0.087$	&$0.699\pm0.023$	&$0.610\pm0.060$	&$0.691\pm0.065$	&$0.669\pm0.073$\\
J0600.5-2006	&-	&-	&$<100$	&-	&	&-	&-	&-	&-	&-	&-\\
J0603.0-4012	&-	&-	&$<19.6$	&-	&	&-	&-	&-	&-	&-	&-\\
J0604.2-4817\_1ES0602-482	&-	&$32.3\pm1.3$	&-	&-	&	&$1.64\pm0.17$	&$1.75\pm0.12$	&$2.57\pm0.06$	&$2.16\pm0.11$	&$2.23\pm0.13$	&$2.12\pm0.07$\\
J0605.1+0005\_GB6J0604+0000	&-	&-	&$34.4\pm1.1$	&-	&	&-	&-	&-	&-	&-	&-\\
J0609.3-0244\_NVSSJ060915-024754	&-	&-	&$61.3\pm1.9$	&-	&	&$2.30\pm0.49$	&$2.67\pm0.41$	&$3.00\pm0.28$	&$2.87\pm0.31$	&$2.17\pm0.46$	&$2.46\pm0.38$\\
J0622.2+3751\_LATPSR\_PSRJ0622+3749	&-	&-	&$<0.022$	&-	&	&-	&-	&-	&-	&-	&-\\
J0648.6-6052\_PMNJ0647-6058	&-	&-	&-	&$78.0\pm4.0$	&	&$<0.552$	&$0.610\pm0.122$	&$<0.213$	&$0.372\pm0.073$	&$<0.216$	&$0.173\pm0.030$\\
J0648.6-6052\_src4	&-	&$<12$	&-	&-	&	&-	&-	&$193\pm4$	&$54.6\pm1.2$	&$3.75\pm0.21$	&$20.9\pm0.4$\\
J0707.3+7742\_NVSSJ070651+774137	&-	&-	&$37.0\pm1.2$	&-	&	&-	&-	&-	&-	&-	&$2.32\pm0.05$\\
J0718.8-4958\_src1	&-	&$<141.4$	&-	&-	&	&-	&-	&-	&-	&$38.9\pm0.88$	&-\\
J0803.1-0339\_src1	&-	&-	&$231.1\pm8.2$	&-	&	&-	&-	&-	&$0.971\pm0.031$	&-	&-\\
J0814.5-1011\_NVSSJ081411-101208	&-	&-	&$36.4\pm1.2$	&-	&	&$3.13\pm0.36$	&$3.75\pm0.28$	&$3.42\pm0.23$	&$2.90\pm0.23$	&$2.56\pm0.22$	&$2.77\pm0.18$\\
J0843.4+6718	&-	&-	&$<12.5$	&-	&	&-	&-	&-	&-	&-	&-\\
J0848.6+0504\_TXS0846+051	&-	&-	&$174.09\pm0.14$	&-	&	&-	&-	&-	&-	&-	&-\\
J0902.4+2050\_NVSSJ090226+205045	&-	&-	&$62.85\pm0.15$	&-	&	&$5.03\pm0.34$	&$5.79\pm0.25$	&$4.86\pm0.11$	&$5.29\pm0.21$	&$5.17\pm0.23$	&$4.94\pm0.16$\\
J0906.4-0903\_PMNJ0906-0905	&-	&-	&$71.3\pm2.2$	&-	&	&$2.38\pm0.30$	&$1.83\pm0.19$	&$2.20\pm0.16$	&$1.56\pm0.13$	&$1.51\pm0.15$	&$0.960\pm0.079$\\
J0908.7-2119\_NVSSJ090858-211854	&-	&-	&$10.4\pm0.6$	&-	&	&-	&-	&$0.909\pm0.040$	&-	&-	&-\\
J0922.0+2337\_NVSSJ092145+233548	&-	&-	&$18.4\pm0.14$	&-	&	&-	&-	&$1.25\pm0.05$	&-	&-	&$1.42\pm0.05$\\
J0922.0+2337\_src2	&-	&-	&$<23.8$	&-	&	&-	&-	&-	&-	&-	&-\\
J0922.0+2337\_src3	&-	&-	&$<23.8$	&-	&	&-	&-	&-	&-	&-	&-\\
J0940.2-7605\_UL	&-	&$<71.7$	&-	&-	&	&-	&-	&-	&-	&-	&-\\
J0953.6-1505	&-	&-	&$<6$	&-	&	&-	&-	&-	&-	&-	&-\\
J0955.2-3949\_src1	&-	&-	&$<5.3$	&-	&	&-	&-	&-	&$0.373\pm0.045$	&-	&-\\
J1040.5+0616\_src1	&-	&-	&$39.02\pm0.17$	&-	&	&-	&-	&-	&-	&$0.316\pm0.019$	&$0.234\pm0.024$\\
J1040.5+0616\_src2	&-	&-	&$3.53\pm0.19$	&-	&	&-	&-	&-	&-	&$0.201\pm0.016$	&$0.109\pm0.020$\\
J1040.5+0616\_src3	&-	&-	&$<2.5$	&-	&	&-	&-	&-	&-	&$0.0960\pm0.0134$	&$0.120\pm0.020$\\
J1048.7+2335	&-	&-	&-	&-	&	&-	&-	&-	&-	&-	&-\\
J1119.9-2205\_src1	&-	&-	&$<10.9$	&-	&	&-	&-	&$8.42\pm0.19$	&-	&$5.04\pm0.12$	&$5.52\pm0.13$\\
J1119.9-2205\_src2	&-	&-	&$<10.9$	&-	&	&-	&-	&$0.193\pm0.024$	&-	&$0.136\pm0.013$	&$0.132\pm0.019$\\
J1124.4-3654\_PSRJ1124-36	&$<0.3$	&-	&-	&-	&	&-	&-	&$<0.137$	&$<0.377$	&$<0.0523$	&$<0.124$\\
J1129.3+3757\_src1	&-	&-	&$26.5\pm1.2$	&-	&	&-	&-	&-	&$0.136\pm0.017$	&$0.107\pm0.018$	&$0.0595\pm0.0109$\\
J1129.3+3757\_src2	&-	&-	&$<2.5$	&-	&	&-	&-	&-	&$0.343\pm0.023$	&$0.249\pm0.023$	&$0.192\pm0.016$\\
J1141.8-1403\_1RXSJ114142.2-140757	&-	&-	&$59.2\pm2.2$	&-	&	&-	&-	&-	&$1.36\pm0.11$	&$1.13\pm0.04$	&$1.30\pm0.04$\\
J1159.8+0200\_GB6J1200+0202	&-	&-	&$30.82\pm0.14$	&-	&	&-	&-	&-	&-	&-	&-\\
J1218.4-0128\_PKS1216-010	&-	&-	&$280.11\pm0.15$	&-	&	&$1.34\pm0.25$	&$1.16\pm0.15$	&$1.30\pm0.12$	&$0.897\pm0.090$	&$0.982\pm0.098$	&$0.896\pm0.067$\\
J1221.4-0635\_src1	&-	&-	&$<7.59$	&-	&	&-	&-	&-	&$0.788\pm0.042$	&$0.936\pm0.038$	&$1.02\pm0.04$\\
J1221.4-0635\_src3	&-	&-	&$<7.59$	&-	&	&-	&-	&-	&$<0.0913$	&$<0.0464$	&$<0.0588$\\ \hline
\end{tabular}
}
\end{center}
\end{table}
\end{landscape}
}

{
\setlength{\textheight}{40\baselineskip}
\begin{landscape}
\begin{table}
\label{UV&Radio}
\begin{center}
\scalebox{0.7}{
\begin{tabular}{lccccccccccc}
\hline \hline
 &  \multicolumn{4}{c}{Radio [mJy]} & \multicolumn{6}{c}{UV (Swift/UVOT) [E-12 erg/cm$^2$/s]}\\ \cline{2-5} \cline{7-12}
\multicolumn{1}{c}{Name} & $F_{350MHz}$ & $F_{843MHz}$ & $F_{1.4GHz}$ & $F_{4.85GHz}$  && V-band & B-band & U-band & UVW1-band & UVM2-band & UVW2-band\\ \hline
J1226.0+2954\_UL	&-	&-	&$<2.5$	&-	&	&-	&-	&-	&-	&-	&-\\
J1232.2-5118\_src1	&-	&$<37.5$	&-	&-	&	&-	&-	&-	&$0.288\pm0.040$	&-	&-\\
J1249.3-2812	&-	&-	&$26.8\pm0.9$	&-	&	&-	&-	&-	&-	&-	&$1.11\pm0.03$\\
J1251.3+1044\_1RXSJ125117.4+103914	&-	&-	&$4.21\pm0.15$	&-	&	&$1.79\pm0.18$	&$1.66\pm0.11$	&$1.60\pm0.10$	&$1.83\pm0.09$	&$1.82\pm0.10$	&$1.71\pm0.07$\\
J1254.4+2209\_src2	&-	&-	&$<2.5$	&-	&	&-	&-	&-	&-	&-	&-\\
J1254.4+2209\_TXS1252+224	&-	&-	&$83.43\pm0.15$	&-	&	&$2.66\pm0.32$	&$2.30\pm0.14$	&$2.13\pm0.12$	&$2.04\pm0.11$	&$2.30\pm0.18$	&$2.11\pm0.09$\\
J1256.9+3650\_1RXSJ125716.0+364713	&-	&-	&$73.86\pm0.17$	&-	&	&-	&-	&-	&-	&$0.914\pm0.052$	&-\\
J1301.8+0837\_PSRJ1301+08	&-	&-	&$<1.87$	&-	&	&-	&-	&-	&-	&-	&-\\
J1301.8+0837\_src1	&-	&-	&$<1.87$	&-	&	&-	&-	&-	&$0.153\pm0.018$	&-	&-\\
J1302.3-3255\_PSRJ1302-32	&$0.500\pm0.001$	&-	&-	&-	&	&-	&-	&-	&-	&-	&-\\
J1302.3-3255\_src1	&-	&-	&$<2.5$	&-	&	&-	&-	&$0.389\pm0.016$	&$0.325\pm0.024$	&-	&-\\
J1304.3-4352\_1RXSJ130421.2-435308	&-	&$43.8\pm1.7$	&-	&-	&	&$10.5\pm0.5$	&$11.5\pm0.4$	&$11.2\pm0.4$	&$10.5\pm0.3$	&$9.39\pm0.38$	&$10.4\pm0.3$\\
J1307.0-4030\_ESO323-77	&-	&$48.0\pm1.9$	&-	&-	&	&$83.2\pm1.9$	&$57.9\pm1.1$	&$32.7\pm0.7$	&$15.6\pm0.3$	&$7.20\pm0.22$	&$7.40\pm0.23$\\
J1307.0-4030\_src2	&-	&$<49.9$	&-	&-	&	&$28.6\pm0.7$	&$20.3\pm0.5$	&$11.4\pm0.3$	&$8.46\pm0.22$	&$6.16\pm0.21$	&$5.71\pm0.21$\\
J1307.0-4030\_src3	&-	&$<49.9$	&-	&-	&	&-	&-	&-	&-	&-	&-\\
J1307.6-4259\_1RXSJ130737.8-425940	&-	&$36.2\pm1.6$	&-	&-	&	&$11.1\pm0.56$	&$12.7\pm0.4$	&$12.9\pm0.5$	&$13.1\pm0.5$	&$12.3\pm0.5$	&$14.5\pm0.4$\\
J1311.7-3429	&-	&-	&-	&-	&	&-	&-	&-	&-	&-	&-\\
J1312.6+0048	&-	&-	&-	&-	&	&-	&-	&-	&-	&-	&-\\
J1315.6-0729\_src1	&-	&-	&$19.21\pm0.15$	&-	&	&-	&-	&-	&$4.76\pm0.10$	&-	&$8.03\pm0.17$\\
J1328.2-4729\_src1	&-	&$123.6\pm3.8$	&-	&-	&	&$2.20\pm0.33$	&$2.09\pm0.23$	&-	&-	&-	&-\\
J1340.5-0413\_src1	&-	&-	&$20.74\pm0.14$	&-	&	&-	&-	&$1.56\pm0.04$	&-	&-	&$1.48\pm0.04$\\
J1340.5-0413\_src2	&-	&-	&$<1$	&-	&	&-	&-	&$0.190\pm0.022$	&-	&-	&$0.233\pm0.016$\\
J1340.5-0413\_src3	&-	&-	&$<1$	&-	&	&-	&-	&$170\pm4$	&-	&-	&$15.1\pm0.3$\\
J1340.5-0413\_src4	&-	&-	&$<1$	&-	&	&-	&-	&$<0.0567$	&-	&-	&$<0.0313$\\
J1406.2-2510\_NVSSJ140609-250808	&-	&-	&$30.7\pm1.0$	&-	&	&$2.57\pm0.34$	&$2.13\pm0.21$	&$1.81\pm0.16$	&$1.53\pm0.14$	&$1.36\pm0.15$	&$1.36\pm0.10$\\
J1419.7+7731\_1RXSJ141901.8+773229	&-	&-	&$8.1\pm0.5$	&-	&	&-	&-	&$2.32\pm0.05$	&-	&-	&-\\
J1511.8-0513\_src1	&-	&-	&$18.72\pm0.14$	&-	&	&-	&-	&-	&-	&-	&$3.45\pm0.08$\\
J1511.8-0513\_src2	&-	&-	&$<2.5$	&-	&	&-	&-	&-	&-	&-	&$0.284\pm0.021$\\
J1521.0-0350\_NVSSJ152048-034850	&-	&-	&$45.4\pm1.4$	&-	&	&-	&-	&-	&$2.91\pm0.06$	&-	&-\\
J1539.0-3328\_UL	&-	&-	&$<7.3$	&-	&	&-	&-	&-	&-	&-	&-\\
J1544.5-1127\_src1	&-	&-	&$<31$	&-	&	&-	&-	&-	&-	&$0.556\pm0.090$	&-\\
J1549.7-0659\_PSRJ1549-06	&$<0.5$	&-	&-	&-	&	&-	&-	&-	&-	&-	&-\\
J1549.7-0659\_src1	&-	&-	&-	&-	&	&-	&-	&-	&$1.50\pm0.08$	&$1.39\pm0.07$	&-\\
J1625.3-0019\_UL	&-	&-	&$<10.5$	&-	&	&-	&-	&-	&-	&-	&-\\
J1627.6+3218\_src1	&-	&-	&$<4.4$	&-	&	&-	&-	&$2.41\pm0.06$	&-	&-	&-\\
J1627.6+3218\_src2	&-	&-	&$<4.4$	&-	&	&-	&-	&$0.629\pm0.020$	&-	&-	&-\\
J1630.5+3735\_UL	&-	&-	&$<2.5$	&-	&	&-	&-	&-	&-	&-	&-\\
J1653.6-0158\_src1	&-	&-	&$<13.2$	&-	&	&$<0.875$	&$<0.554$	&$<0.382$	&$<0.290$	&$<0.342$	&$<0.262$\\
J1721.1+0713\_UL	&-	&-	&$<17.9$	&-	&	&-	&-	&-	&-	&-	&-\\
J1739.4+8717	&-	&-	&-	&-	&	&-	&-	&-	&-	&-	&-\\
J1221.4-0635\_src1	&-	&-	&$<7.59$	&-	&	&-	&-	&-	&$0.788\pm0.042$	&$0.936\pm0.038$	&$1.02\pm0.04$\\
J1743.8-7620	&-	&-	&-	&-	&	&-	&-	&-	&-	&-	&-\\
J1745.5+1018\_PSRJ1745+10	&-	&-	&$<2.8$	&-	&	&-	&-	&-	&-	&-	&-\\ \hline
\end{tabular}
}
\end{center}
\end{table}
\end{landscape}
}

{
\setlength{\textheight}{40\baselineskip}
\begin{landscape}
\begin{table}
\label{UV&Radio}
\begin{center}
\scalebox{0.7}{
\begin{tabular}{lccccccccccc}
\hline \hline
 &  \multicolumn{4}{c}{Radio [mJy]} & \multicolumn{6}{c}{UV (Swift/UVOT) [E-12 erg/cm$^2$/s]}\\ \cline{2-5} \cline{7-12}
\multicolumn{1}{c}{Name} & $F_{350MHz}$ & $F_{843MHz}$ & $F_{1.4GHz}$ & $F_{4.85GHz}$  && V-band & B-band & U-band & UVW1-band & UVM2-band & UVW2-band\\ \hline

J1754.0-5002\_PMNJ1753-5015	&-	&-	&$137.0\pm11.0$	&-	&	&-	&-	&-	&-	&-	&-\\
J1754.0-5002\_src1	&-	&$<14.8$	&-	&-	&	&-	&-	&$0.394\pm0.038$	&-	&-	&-\\
J1754.0-5002\_src2	&-	&$<14.8$	&-	&-	&	&-	&-	&-	&-	&-	&-\\
J1806.2+0609\_UL	&-	&-	&$<2.5$	&-	&	&-	&-	&-	&-	&-	&-\\
J1810.3+1741\_LATPSR\_PSRJ1810+17	&$20.0\pm0.1$	&-	&-	&-	&	&-	&-	&-	&-	&-	&-\\
J1816.7+4509	&-	&-	&$<2.5$	&-	&	&-	&-	&-	&-	&-	&-\\
J1824.6+1013	&-	&-	&$<8.9$	&-	&	&-	&-	&-	&-	&-	&-\\
J1841.9+3220\_RXJ1841.7+3218\_src2	&-	&-	&$20.4\pm0.7$	&-	&	&$1.04\pm0.13$	&$0.965\pm0.095$	&$0.931\pm0.073$	&$0.688\pm0.060$	&$0.611\pm0.067$	&$0.588\pm0.042$\\
J1841.9+3220\_src1	&-	&-	&$<7.7$	&-	&	&$0.503\pm0.127$	&$<0.252$	&$0.422\pm0.062$	&$0.351\pm0.051$	&$0.457\pm0.061$	&$0.430\pm0.037$\\
J1842.3-5845\_src1	&-	&$<10.1$	&-	&-	&	&-	&-	&-	&-	&$2.38\pm0.10$	&$2.32\pm0.09$\\
J1858.1-2218	&-	&-	&$<10.7$	&-	&	&-	&-	&-	&-	&-	&-\\
J1902.0-5110	&-	&-	&-	&-	&	&-	&-	&-	&-	&-	&-\\
J1916.9-3028	&-	&-	&$<2.8$	&-	&	&-	&-	&-	&-	&-	&-\\
J1926.8+6153\_1RXSJ192649.5+615445	&-	&-	&$<2.5$	&-	&	&-	&-	&$3.43\pm0.11$	&$5.73\pm0.17$	&-	&-\\
J1956.2-0238\_src1	&-	&-	&$<10.1$	&-	&	&-	&-	&-	&-	&-	&$0.37458$\\
J1959.7-4730\_SUMSSJ195945-472519	&-	&$23.8\pm1.1$	&-	&-	&	&$3.82\pm0.32$	&$4.44\pm0.23$	&$4.55\pm0.21$	&$4.28\pm0.20$	&$4.40\pm0.22$	&$4.02\pm0.10$\\
J2004.8+7004	&-	&-	&$<11.6$	&-	&	&-	&-	&$1.37\pm0.13$	&-	&-	&$<1.33$\\
J2014.4+0647\_NVSSJ201431+064849	&-	&-	&$15.0\pm0.9$	&-	&	&$2.01\pm0.11$	&$1.87\pm0.09$	&-	&$2.40\pm0.29$	&-	&-\\
J2034.6-4202	&-	&$10.2\pm1.6$	&-	&-	&	&-	&-	&-	&$2.59\pm0.08$	&$2.61\pm0.11$	&$2.96\pm0.08$\\
J2039.4-5621	&-	&$<5$	&-	&-	&	&-	&-	&-	&-	&-	&-\\
J2043.2+1709	&-	&-	&-	&-	&	&-	&-	&-	&-	&-	&-\\
J2112.5-3044	&-	&-	&$<2.5$	&-	&	&-	&-	&-	&-	&-	&-\\
J2129.8-0427\_PSRJ2129-04\_src1	&-	&-	&$<4.4$	&-	&	&-	&-	&$0.657\pm0.027$	&$0.151\pm0.012$	&-	&-\\
J2129.8-0427\_PSRJ2129-04\_src3	&$<0.5$	&-	&-	&-	&	&-	&-	&-	&$0.261\pm0.010$	&-	&-\\
J2134.5-2130\_src1	&-	&-	&$<30$	&-	&	&-	&-	&-	&-	&$0.153\pm0.045$	&$0.152\pm0.022$\\
J2223.3+0103\_NVSSJ222329+010226	&-	&-	&$6.1\pm0.5$	&-	&	&-	&-	&$1.73\pm0.04$	&-	&-	&$1.14\pm0.07$\\
J2228.5-1633\_UL	&-	&-	&$<22$	&-	&	&-	&-	&-	&-	&-	&-\\
J2243.4+4104\_TXS2241+406	&-	&-	&$226.2\pm6.8$	&-	&	&$3.10\pm0.42$	&$2.27\pm0.28$	&$2.37\pm0.23$	&$1.65\pm0.20$	&$1.34\pm0.19$	&$1.08\pm0.10$\\
J2251.2-4928\_src1	&-	&$33.3\pm1.4$	&-	&-	&	&-	&-	&$0.582\pm0.020$	&$0.403\pm0.032$	&-	&-\\
J2251.2-4928\_src2	&-	&$<34.7$	&-	&-	&	&-	&-	&$0.150\pm0.012$	&$0.211\pm0.026$	&-	&-\\
J2256.9-1024	&-	&-	&-	&-	&	&$0.284\pm0.064$	&$0.351\pm0.043$	&$0.359\pm0.027$	&-	&-	&-\\
J2257.9-3643\_src1	&-	&-	&$10.6\pm0.6$	&-	&	&-	&-	&$1.26\pm0.04$	&-	&$1.26\pm0.05$	&-\\
J2257.9-3643\_src2	&-	&-	&$<11.2$	&-	&	&-	&-	&$1.30\pm0.04$	&-	&$1.37\pm0.05$	&-\\
J2259.9-8255	&-	&-	&-	&-	&	&-	&-	&-	&-	&-	&-\\
J2310.0-3627\_1RXSJ230940.6-363241	&-	&-	&$89.8\pm3.1$	&-	&	&-	&-	&-	&$3.14\pm0.07$	&$3.41\pm0.09$	&-\\
J2310.0-3627\_src2	&-	&-	&$<92.9$	&-	&	&-	&-	&-	&$1.15\pm0.03$	&$1.29\pm0.05$	&-\\
J2323.0-4919\_1RXSJ232256.7-491658	&-	&$28.3\pm1.3$	&-	&-	&	&$1.79\pm0.21$	&$1.97\pm0.14$	&$1.81\pm0.12$	&$1.87\pm0.11$	&$2.18\pm0.14$	&$2.17\pm0.09$\\
J2330.3-4745\_PKS2326-477	&-	&$2820\pm1$	&-	&-	&	&-	&-	&-	&-	&-	&-\\
J2330.3-4745\_src1	&-	&$<15.1$	&-	&-	&	&$3.13\pm0.14$	&$3.53\pm0.11$	&$4.17\pm0.12$	&-	&-	&$1.52\pm0.04$\\
J2339.7-0531\_src2	&-	&-	&$<11.6$	&-	&	&$0.73\pm0.07$	&$0.52\pm0.05$	&$0.23\pm0.02$	&-	&-	&-\\
J2347.3+0710\_UL	&-	&-	&$<41.6$	&-	&	&-	&-	&-	&-	&-	&-\\
J2350.1-3005	&-	&-	&$<9.3$	&-	&	&-	&-	&-	&-	&-	&-\\
J2352.1+1752\_87GB234934.7+173227	&-	&-	&$44.2\pm1.4$	&-	&	&$0.779\pm0.234$	&$0.801\pm0.152$	&$0.631\pm0.101$	&$0.581\pm0.083$	&$0.416\pm0.080$	&$0.469\pm0.055$\\ \hline
\end{tabular}
}
\end{center}
\end{table}
\end{landscape}
}

{
\setlength{\textheight}{40\baselineskip}
\begin{landscape}
\begin{table}
\label{X-ray}
\caption{{\it Swift}/XRT X-ray fluxes and 2FGL Gamma-ray fluxes plotted in SEDs(Figure 12) are shown here. $U.L._{0.5-8keV}$ shows the upper-limit of X-ray fluxes. Note that all X-ray fluxes are in [E-14 erg/cm$^2$/s], while Gamma-ray fluxes in [E-12 erg/cm$^2$/s].}
\begin{center}
\scalebox{0.65}{
\begin{tabular}{lcccccccccccc}
\hline \hline 
 &  \multicolumn{6}{c}{X-ray (Swift/XRT) [E-14 erg/cm$^2$/s]}  &&  \multicolumn{5}{c}{Gamma-ray (2FGL Catalog) [E-12 erg/cm$^2$/s]}\\ \cline{2-7} \cline{9-13}
\multicolumn{1}{c}{Name} & $F_{0.32-.64 keV}$ & $F_{0.64-1.28 keV}$ & $F_{1.28-2.56 keV}$ & $F_{2.56-5.12 keV}$ & $F_{5.12-10.24 keV}$ & $U.L._{0.5-8keV}$ && $F_{100-300MeV}$ & $F_{0.3-1GeV}$ & $F_{1-3GeV}$ & $F_{3-10GeV}$ & $F_{10-100GeV}$\\ \hline
J0001.9-4158\_src1	&$104\pm11$	&$86.2\pm7.3$	&$57.5\pm6.4$	&$39.2\pm8.13$	&$20.2\pm12.9$	&-	&	&$<3.47$	&$<1.63$	&$<1.21$	&$1.09\pm0.39$	&$<1.42$\\
J0001.9-4158\_src2	&-	&$2.29\pm1.29$	&$2.96\pm1.58$	&-	&-	&-	&	&$<3.47$	&$<1.63$	&$<1.21$	&$1.09\pm0.39$	&$<1.42$\\
J0009.1+5031	&$18.3\pm8.45$	&$17.0\pm4.3$	&$15.1\pm3.9$	&$6.62\pm3.86$	&-	&-	&	&$<2.29$	&$2.85\pm0.53$	&$3.70\pm0.55$	&$4.10\pm0.73$	&$3.53\pm1.04$\\
J0022.2-1850	&$38.2\pm17.4$	&$37.2\pm12.5$	&$40.7\pm14.5$	&-	&-	&-	&	&$<1.27$	&$<1.34$	&$1.27\pm0.35$	&$1.80\pm0.52$	&$3.61\pm1.16$\\
J0023.5+0930\_PSRJ0023+48\_src2	&-	&$5.49\pm2.27$	&$6.68\pm2.77$	&$9.78\pm5.08$	&$24.1\pm18.5$	&-	&	&$<2.70$	&$2.64\pm0.53$	&$2.08\pm0.44$	&$<1.67$	&$<1.27$\\
J0023.5+0930\_PSRJ0023+48\_src3	&$9.82\pm4.45$	&$24.1\pm5.2$	&$2.06\pm1.52$	&-	&-	&-	&	&$<2.70$	&$2.64\pm0.53$	&$2.08\pm0.44$	&$<1.67$	&$<1.27$\\
J0030.7+0724\_src2	&$3.42\pm1.91$	&$0.745\pm0.606$	&$2.27\pm1.11$	&$1.96\pm1.57$	&-	&-	&	&$3.41\pm1.13$	&$<1.20$	&$<1.38$	&$<0.609$	&$1.28\pm0.64$\\
J0030.7+0724\_src6	&$12.3\pm5.0$	&$19.2\pm4.1$	&$15.4\pm3.9$	&-	&-	&-	&	&$3.41\pm1.13$	&$<1.20$	&$<1.38$	&$<0.609$	&$1.28\pm0.64$\\
J0032.7-5519	&-	&-	&-	&-	&-	&$<18.7$	&	&$3.58\pm0.91$	&$3.94\pm0.46$	&$2.92\pm0.44$	&$1.54\pm0.45$	&$<2.46$\\
J0038.0+1236	&$33.4\pm8.7$	&$20.5\pm4.4$	&$16.0\pm4.0$	&$11.2\pm5.14$	&-	&-	&	&$3.42\pm1.24$	&$2.51\pm0.52$	&$1.84\pm0.42$	&$1.28\pm0.46$	&$1.38\pm0.62$\\
J0039.2+4331	&$31.3\pm9.9$	&$4.74\pm2.30$	&$5.38\pm2.47$	&$6.98\pm4.23$	&-	&-	&	&$<1.64$	&$<0.776$	&$<1.36$	&$0.709\pm0.334$	&$<2.28$\\
J0051.4-6242	&$481\pm32$	&$255\pm13$	&$127\pm9$	&$65.6\pm10.2$	&$57.9\pm20.0$	&-	&	&$<2.38$	&$1.17\pm0.39$	&$1.82\pm0.38$	&$3.79\pm0.68$	&$3.43\pm1.05$\\
J0054.9-2455	&$254\pm41$	&$136\pm21$	&$41.7\pm12.6$	&$52.9\pm21.9$	&$70.7\pm53.1$	&-	&	&$<1.31$	&$<0.701$	&$0.790\pm0.299$	&$1.51\pm0.50$	&$<1.54$\\
J0101.0-6423\_PSRJ0101-6422	&-	&-	&-	&-	&-	&$<39.1$	&	&$<2.23$	&$2.15\pm0.36$	&$4.28\pm0.51$	&$2.30\pm0.55$	&$<0.925$\\
J0102.3+0942	&-	&$5.62\pm2.36$	&$1.93\pm1.59$	&$4.58\pm3.60$	&-	&-	&	&$<4.30$	&$<1.94$	&$<1.31$	&$1.21\pm0.46$	&$<2.08$\\
J0103.1+4840	&-	&-	&-	&-	&-	&$<26.1$	&	&$<3.44$	&$2.21\pm0.63$	&$4.82\pm0.67$	&$3.00\pm0.65$	&$<1.01$\\
J0106.7+4853	&-	&-	&-	&-	&-	&-	&	&$<5.47$	&$5.03\pm0.77$	&$5.11\pm0.69$	&$5.60\pm0.84$	&$<1.05$\\
J0115.7+0357	&$30.7\pm15.4$	&$19.3\pm8.6$	&$32.0\pm12.3$	&-	&-	&-	&	&$<4.80$	&$2.98\pm0.50$	&$2.54\pm0.45$	&$2.69\pm0.62$	&$1.73\pm0.76$\\
J0124.6-0616\_PMNJ0124-0624	&-	&-	&-	&-	&-	&$<22.9$	&	&$<1.73$	&$2.12\pm0.46$	&$0.943\pm0.353$	&$0.786\pm0.377$	&$<1.06$\\
J0143.9-5845	&$780\pm37$	&$650\pm22$	&$533\pm22$	&$441\pm30$	&$311\pm56$	&-	&	&$<2.38$	&$1.36\pm0.37$	&$1.86\pm0.37$	&$1.62\pm0.46$	&$4.25\pm1.15$\\
J0157.0-5259	&$500\pm40$	&$607\pm32$	&$737\pm39$	&$825\pm63$	&$579\pm118$	&-	&	&$<3.26$	&$<1.01$	&$<1.02$	&$1.98\pm0.52$	&$<1.46$\\
J0217.9-6630\_RBS0300\_src1	&$101\pm11$	&$86.8\pm7.5$	&$45.7\pm5.8$	&$28.0\pm7.0$	&-	&-	&	&$<2.60$	&$1.52\pm0.40$	&$<1.23$	&$1.67\pm0.48$	&$<2.05$\\
J0217.9-6630\_src2	&$9.70\pm3.65$	&$6.58\pm2.12$	&$4.20\pm1.88$	&$9.79\pm4.28$	&-	&-	&	&$<2.60$	&$1.52\pm0.40$	&$<1.23$	&$1.67\pm0.48$	&$<2.05$\\
J0217.9-6630\_src3	&$4.72\pm2.68$	&$6.83\pm2.20$	&$9.13\pm2.76$	&$4.72\pm3.13$	&$14.2\pm12.1$	&-	&	&$<2.60$	&$1.52\pm0.40$	&$<1.23$	&$1.67\pm0.48$	&$<2.05$\\
J0223.0-1118\_1RXSJ022314.6-111741	&$78.6\pm11.9$	&$57.6\pm7.5$	&$47.5\pm7.4$	&$17.4\pm7.0$	&$18.0\pm16.6$	&-	&	&$<1.02$	&$<0.873$	&$<1.08$	&$<1.40$	&$<3.01$\\
J0226.3+0937	&-	&-	&-	&-	&-	&$<93.9$	&	&$<3.92$	&$<2.35$	&$<1.66$	&$1.50\pm0.48$	&$<1.32$\\
J0239.5+1324	&-	&-	&-	&-	&-	&$<9.42$	&	&$<5.46$	&$<2.35$	&$1.20\pm0.40$	&$1.06\pm0.47$	&$<2.35$\\
J0305.2-1601	&$51.3\pm11.2$	&$72.4\pm9.0$	&$77.3\pm10.2$	&$50.4\pm12.3$	&$81.7\pm34.7$	&-	&	&$<1.45$	&$<0.765$	&$<0.838$	&$0.808\pm0.395$	&$1.78\pm0.81$\\
J0308.6+7442	&-	&-	&-	&-	&-	&$<4.86$	&	&$<1.31$	&$1.84\pm0.40$	&$6.19\pm0.65$	&$1.87\pm0.50$	&$<0.876$\\
J0316.3-6438	&$206\pm20$	&$138\pm11$	&$120\pm11$	&$61.3\pm12.4$	&-	&-	&	&$<3.88$	&$<1.69$	&$0.813\pm0.315$	&$1.96\pm0.52$	&$<2.92$\\
J0335.5-4501\_src1	&$4.94\pm3.10$	&$1.57\pm1.38$	&$5.84\pm2.69$	&-	&-	&-	&	&$<1.31$	&$1.33\pm0.37$	&$0.930\pm0.302$	&$0.953\pm0.378$	&$1.83\pm0.79$\\
J0335.5-4501\_src3	&$29.4\pm7.7$	&$15.4\pm4.1$	&$15.5\pm4.4$	&$4.98\pm3.99$	&-	&-	&	&$<1.31$	&$1.33\pm0.37$	&$0.930\pm0.302$	&$0.953\pm0.378$	&$1.83\pm0.79$\\
J0340.4+4130\_PSRJ0340+4130	&-	&-	&-	&-	&-	&$<11.6$	&	&$<2.89$	&$2.81\pm0.57$	&$7.74\pm0.75$	&$5.13\pm0.83$	&$<2.07$\\
J0345.2-2355	&$6.12\pm3.20$	&$3.97\pm1.96$	&$6.25\pm2.65$	&$6.94\pm4.30$	&-	&-	&	&$4.44\pm1.53$	&$2.96\pm0.52$	&$1.49\pm0.36$	&$0.768\pm0.337$	&$<0.886$\\
J0409.9-0357	&$7.04\pm4.08$	&$9.58\pm2.86$	&$9.78\pm3.07$	&$7.97\pm4.13$	&$16.7\pm13.8$	&-	&	&$<4.47$	&$1.58\pm0.51$	&$1.34\pm0.40$	&$1.08\pm0.43$	&$<2.53$\\
J0439.8-1857	&$14.9\pm5.8$	&$4.22\pm2.17$	&$5.06\pm2.61$	&-	&-	&-	&	&$<4.02$	&$<1.07$	&$0.879\pm0.306$	&$1.58\pm0.50$	&$<2.93$\\
J0505.9+6121\_NVSSJ050558+611336	&$755\pm386$	&$295\pm34$	&$103\pm10$	&$41.7\pm8.6$	&$9.56\pm9.14$	&-	&	&$<1.41$	&$<1.89$	&$<1.90$	&$1.27\pm0.49$	&$<2.60$\\
J0506.9-5435\_1ES0505-546	&$539\pm78$	&$290\pm36$	&$172\pm28$	&$138\pm38$	&$181\pm94$	&-	&	&$<1.41$	&$<1.12$	&$<1.16$	&$1.51\pm0.48$	&$1.49\pm0.71$\\
J0515.9+1528\_GB6J0515+1527	&-	&$12.1\pm6.2$	&$4.81\pm3.45$	&$9.15\pm6.97$	&-	&-	&	&$<5.51$	&$<2.68$	&$2.74\pm0.61$	&$1.86\pm0.58$	&$2.58\pm0.89$\\
J0521.6+0103\_src1	&$11.0\pm7.0$	&$6.87\pm2.93$	&$4.30\pm2.39$	&-	&-	&-	&	&$<2.58$	&$<0.754$	&$<1.28$	&$1.07\pm0.44$	&$<3.49$\\
J0523.5-2529	&$2.52\pm2.05$	&-	&$2.64\pm1.66$	&$10.4\pm4.9$	&-	&-	&	&$4.03\pm0.93$	&$3.18\pm0.49$	&$4.14\pm0.53$	&$3.49\pm0.69$	&$<1.29$\\ \hline
\end{tabular}
}
\end{center}
\end{table}
\end{landscape}
}

{
\setlength{\textheight}{40\baselineskip}
\begin{landscape}
\begin{table}
\label{X-ray}
\begin{center}
\scalebox{0.65}{
\begin{tabular}{lcccccccccccc}
\hline \hline 
 &  \multicolumn{6}{c}{X-ray (Swift/XRT) [E-14 erg/cm$^2$/s]}  &&  \multicolumn{5}{c}{Gamma-ray (2FGL Catalog) [E-12 erg/cm$^2$/s]}\\ \cline{2-7} \cline{9-13}
\multicolumn{1}{c}{Name} & $F_{0.32-.64 keV}$ & $F_{0.64-1.28 keV}$ & $F_{1.28-2.56 keV}$ & $F_{2.56-5.12 keV}$ & $F_{5.12-10.24 keV}$ & $U.L._{0.5-8keV}$ && $F_{100-300MeV}$ & $F_{0.3-1GeV}$ & $F_{1-3GeV}$ & $F_{3-10GeV}$ & $F_{10-100GeV}$\\ \hline
J0533.9+6758\_src1	&$12.3\pm5.3$	&$19.6\pm4.6$	&$2.36\pm1.66$	&-	&-	&-	&	&$<1.44$	&$1.38\pm0.38$	&$2.81\pm0.46$	&$2.51\pm0.56$	&$<1.09$\\
J0533.9+6758\_src2	&$9.35\pm4.97$	&$3.48\pm1.83$	&$5.41\pm2.43$	&-	&-	&-	&	&$<1.44$	&$1.38\pm0.38$	&$2.81\pm0.46$	&$2.51\pm0.56$	&$<1.09$\\
J0537.7-5717\_src2	&$58.3\pm16.5$	&$49.4\pm9.6$	&$27.0\pm7.3$	&$12.9\pm7.7$	&-	&-	&	&$<4.00$	&$<0.931$	&$<1.36$	&$0.888\pm0.403$	&$<2.46$\\
J0537.7-5717\_SUMSSJ053748-571828	&$10.5\pm6.5$	&$40.4\pm9.2$	&$8.53\pm4.36$	&-	&-	&-	&	&$<4.00$	&$<0.931$	&$<1.36$	&$0.888\pm0.403$	&$<2.46$\\
J0538.4-3910\_1RXSJ053810.0-390839	&$344\pm39$	&$261\pm17$	&$216\pm15$	&$183\pm20$	&$102\pm34$	&-	&	&$<3.18$	&$1.90\pm0.51$	&$0.901\pm0.348$	&$<1.37$	&$<2.11$\\
J0600.5-2006	&-	&-	&-	&-	&-	&$<321$	&	&$<4.67$	&$2.07\pm0.52$	&$1.06\pm0.39$	&$<1.33$	&$<0.718$\\
J0603.0-4012	&-	&-	&-	&-	&-	&$<79.4$	&	&$<3.02$	&$2.12\pm0.52$	&$1.76\pm0.46$	&$2.46\pm0.61$	&$<2.53$\\
J0604.2-4817\_1ES0602-482	&$384\pm26$	&$263\pm14$	&$208\pm13$	&$170\pm17$	&$107\pm31$	&-	&	&$<1.57$	&$1.32\pm0.45$	&$1.24\pm0.37$	&$<1.44$	&$1.80\pm0.79$\\
J0605.1+0005\_GB6J0604+0000	&-	&-	&-	&-	&-	&$<203$	&	&$<1.99$	&$<2.35$	&$1.48\pm0.53$	&$<1.37$	&$1.53\pm0.76$\\
J0609.3-0244\_NVSSJ060915-024754	&$98.0\pm60.0$	&$35.8\pm14.9$	&$38.9\pm14.0$	&$19.0\pm14.2$	&-	&-	&	&$5.58\pm1.89$	&$<2.62$	&$1.47\pm0.49$	&$1.93\pm0.61$	&$2.48\pm0.97$\\
J0622.2+3751\_LATPSR\_PSRJ0622+3749	&-	&-	&-	&-	&-	&$<27.1$	&	&$<2.25$	&$6.09\pm0.64$	&$3.95\pm0.60$	&$<1.32$	&$<0.712$\\
J0648.6-6052\_PMNJ0647-6058	&$4.80\pm2.59$	&$3.64\pm1.36$	&$6.32\pm1.92$	&$11.6\pm3.8$	&$15.5\pm10.7$	&-	&	&$<4.13$	&$2.05\pm0.52$	&$<1.78$	&$1.07\pm0.43$	&$<2.26$\\
J0648.6-6052\_src4	&-	&$4.45\pm1.63$	&$1.52\pm1.06$	&$0.663\pm1.343$	&-	&-	&	&$<4.13$	&$2.05\pm0.52$	&$<1.78$	&$1.07\pm0.43$	&$<2.26$\\
J0707.3+7742\_NVSSJ070651+774137	&$17.2\pm6.2$	&$5.65\pm2.39$	&$9.89\pm3.39$	&-	&-	&-	&	&$<2.05$	&$1.57\pm0.38$	&$2.30\pm0.38$	&$1.85\pm0.45$	&$<1.67$\\
J0718.8-4958\_src1	&$1.22\pm3.78$	&$7.76\pm2.80$	&-	&-	&-	&-	&	&$<4.64$	&$1.73\pm0.57$	&$<2.04$	&$1.26\pm0.47$	&$<0.902$\\
J0803.1-0339\_src1	&$45.7\pm10.3$	&$33.1\pm5.9$	&$27.8\pm5.8$	&$15.5\pm6.6$	&-	&-	&	&$<4.67$	&$1.58\pm0.48$	&$1.89\pm0.42$	&$2.25\pm0.58$	&$1.68\pm0.78$\\
J0814.5-1011\_NVSSJ081411-101208	&$292\pm206$	&$496\pm157$	&$261\pm117$	&$349\pm201$	&-	&-	&	&$<7.85$	&$2.24\pm0.67$	&$2.33\pm0.53$	&$1.36\pm0.52$	&$<2.75$\\
J0843.4+6718	&-	&-	&-	&-	&-	&$<24.7$	&	&$<2.55$	&$<1.64$	&$1.06\pm0.30$	&$1.78\pm0.45$	&$<0.623$\\
J0848.6+0504\_TXS0846+051	&-	&-	&-	&-	&-	&$<123$	&	&$<1.86$	&$<1.69$	&$<0.986$	&$1.64\pm0.52$	&$<3.69$\\
J0902.4+2050\_NVSSJ090226+205045	&$18.7\pm6.0$	&$10.9\pm3.1$	&$3.19\pm1.85$	&$7.81\pm4.27$	&-	&-	&	&$<3.11$	&$<2.05$	&$1.92\pm0.41$	&$1.63\pm0.49$	&$<2.39$\\
J0906.4-0903\_PMNJ0906-0905	&$19.6\pm13.9$	&$13.4\pm7.8$	&$15.3\pm9.1$	&$22.4\pm16.8$	&-	&-	&	&$<3.58$	&$<1.82$	&$1.28\pm0.38$	&$1.62\pm0.51$	&$<2.44$\\
J0908.7-2119\_NVSSJ090858-211854	&-	&$2.67\pm1.97$	&-	&$11.9\pm6.2$	&-	&-	&	&$<4.65$	&$2.25\pm0.68$	&$<1.87$	&$<1.01$	&$<1.92$\\
J0922.0+2337\_NVSSJ092145+233548	&$14.4\pm6.0$	&$8.10\pm3.18$	&$3.96\pm2.48$	&$5.81\pm4.57$	&-	&-	&	&$<2.27$	&$<1.10$	&$<1.33$	&$1.26\pm0.45$	&$<2.38$\\
J0922.0+2337\_src2	&-	&$2.19\pm1.75$	&$6.97\pm3.31$	&$12.9\pm6.7$	&-	&-	&	&$<2.27$	&$<1.10$	&$<1.33$	&$1.26\pm0.45$	&$<2.38$\\
J0922.0+2337\_src3	&$5.18\pm4.14$	&$6.82\pm3.18$	&$13.1\pm4.8$	&$7.04\pm5.45$	&-	&-	&	&$<2.27$	&$<1.10$	&$<1.33$	&$1.26\pm0.45$	&$<2.38$\\
J0940.2-7605\_UL	&-	&-	&-	&-	&-	&$<480$	&	&$<5.45$	&$2.53\pm0.69$	&$1.84\pm0.50$	&$<0.841$	&$<0.805$\\
J0953.6-1505	&-	&-	&-	&-	&-	&$<66.8$	&	&$<1.73$	&$1.57\pm0.48$	&$2.08\pm0.44$	&$1.38\pm0.51$	&$<1.21$\\
J0955.2-3949\_src1	&-	&$3.85\pm2.36$	&$9.25\pm3.56$	&$13.7\pm6.5$	&-	&-	&	&$4.28\pm1.33$	&$4.14\pm0.67$	&$3.20\pm0.58$	&$1.56\pm0.52$	&$<1.14$\\
J1040.5+0616\_src1	&$4.51\pm2.46$	&$1.71\pm1.07$	&$4.23\pm1.83$	&$4.39\pm2.95$	&-	&-	&	&$7.67\pm1.05$	&$3.10\pm0.51$	&$1.95\pm0.40$	&$1.75\pm0.49$	&$<0.912$\\
J1040.5+0616\_src2	&$6.64\pm2.91$	&$4.64\pm1.69$	&$6.13\pm2.15$	&$6.07\pm3.26$	&$12.3\pm10.9$	&-	&	&$7.67\pm1.05$	&$3.10\pm0.51$	&$1.95\pm0.40$	&$1.75\pm0.49$	&$<0.912$\\
J1040.5+0616\_src3	&$4.00\pm2.28$	&$2.74\pm1.28$	&$3.22\pm1.55$	&$5.94\pm3.16$	&-	&-	&	&$7.67\pm1.05$	&$3.10\pm0.51$	&$1.95\pm0.40$	&$1.75\pm0.49$	&$<0.912$\\
J1048.7+2335	&-	&-	&-	&-	&-	&$<73.7$	&	&$<2.73$	&$1.41\pm0.47$	&$1.25\pm0.35$	&$<1.20$	&$<1.10$\\
J1119.9-2205\_src1	&$3.52\pm1.84$	&$4.80\pm1.39$	&$1.14\pm0.79$	&$3.78\pm2.05$	&-	&-	&	&$3.90\pm0.74$	&$4.88\pm0.53$	&$5.55\pm0.61$	&$2.06\pm0.55$	&$<0.828$\\
J1119.9-2205\_src2	&$2.59\pm1.58$	&$1.70\pm0.83$	&$2.81\pm1.15$	&-	&-	&-	&	&$3.90\pm0.74$	&$4.88\pm0.53$	&$5.55\pm0.61$	&$2.06\pm0.55$	&$<0.828$\\
J1124.4-3654\_PSRJ1124-36	&$3.45\pm2.50$	&$1.43\pm0.93$	&$2.17\pm1.17$	&$3.48\pm2.25$	&-	&-	&	&$<5.27$	&$3.24\pm0.68$	&$3.53\pm0.57$	&$3.08\pm0.70$	&$<1.63$\\
J1129.3+3757\_src1	&$3.93\pm2.59$	&$2.18\pm1.38$	&$3.65\pm1.95$	&$3.69\pm3.16$	&-	&-	&	&$3.45\pm1.14$	&$2.26\pm0.45$	&$1.83\pm0.37$	&$1.69\pm0.50$	&$<1.21$\\
J1129.3+3757\_src2	&-	&$1.35\pm1.12$	&$4.54\pm2.17$	&$1.51\pm2.22$	&-	&-	&	&$3.45\pm1.14$	&$2.26\pm0.45$	&$1.83\pm0.37$	&$1.69\pm0.50$	&$<1.21$\\
J1141.8-1403\_1RXSJ114142.2-140757	&$128\pm14$	&$91.5\pm8.4$	&$84.5\pm8.7$	&$53.6\pm10.4$	&$52.1\pm22.4$	&-	&	&$<1.42$	&$<1.60$	&$1.09\pm0.38$	&$1.18\pm0.45$	&$<1.91$\\
J1159.8+0200\_GB6J1200+0202	&-	&-	&-	&-	&-	&$<73.8$	&	&$<3.99$	&$1.39\pm0.44$	&$<1.33$	&$1.41\pm0.47$	&$<1.26$\\
J1218.4-0128\_PKS1216-010	&-	&-	&$21.6\pm8.8$	&-	&$77.0\pm55.3$	&-	&	&$<5.70$	&$2.79\pm0.54$	&$3.47\pm0.54$	&$2.09\pm0.56$	&$1.46\pm0.70$\\
J1221.4-0635\_src1	&$11.6\pm4.2$	&$8.17\pm2.46$	&$7.29\pm2.57$	&$11.2\pm4.8$	&$14.8\pm13.1$	&-	&	&$4.21\pm1.15$	&$2.14\pm0.51$	&$1.43\pm0.39$	&$1.25\pm0.46$	&$<1.08$\\
J1221.4-0635\_src3	&-	&-	&$4.24\pm2.13$	&$7.99\pm4.32$	&-	&-	&	&$4.21\pm1.15$	&$2.14\pm0.51$	&$1.43\pm0.39$	&$1.25\pm0.46$	&$<1.07$\\
J1226.0+2954\_UL	&-	&-	&-	&-	&-	&$<28.4$	&	&$<2.0738$	&$1.98\pm0.44$	&$2.38\pm0.46$	&$3.02\pm0.63$	&$<1.13$\\
J1232.2-5118\_src1	&-	&$3.68\pm2.33$	&$8.05\pm3.37$	&$8.72\pm5.18$	&$37.4\pm24.4$	&-	&	&$3.19\pm1.20$	&$5.11\pm0.72$	&$2.45\pm0.56$	&$<0.868$	&$<0.68$\\
J1249.3-2812	&$81.2\pm14.6$	&$67.0\pm8.3$	&$56.4\pm7.9$	&$21.5\pm7.4$	&$41.5\pm22.0$	&-	&	&$<2.01$	&$<1.99$	&$<1.57$	&$<1.60$	&$<2.86$\\
J1251.3+1044\_1RXSJ125117.4+103914	&$77.2\pm15.1$	&$70.6\pm10.4$	&$63.2\pm10.7$	&$29.0\pm11.0$	&$37.0\pm27.0$	&-	&	&$<2.08$	&$<1.37$	&$<1.45$	&$<1.43$	&$<2.43$\\ \hline
\end{tabular}
}
\end{center}
\end{table}
\end{landscape}
}

{
\setlength{\textheight}{40\baselineskip}
\begin{landscape}
\begin{table}
\label{X-ray}
\begin{center}
\scalebox{0.65}{
\begin{tabular}{lcccccccccccc}
\hline \hline 
 &  \multicolumn{6}{c}{X-ray (Swift/XRT) [E-14 erg/cm$^2$/s]}  &&  \multicolumn{5}{c}{Gamma-ray (2FGL Catalog) [E-12 erg/cm$^2$/s]}\\ \cline{2-7} \cline{9-13}
\multicolumn{1}{c}{Name} & $F_{0.32-.64 keV}$ & $F_{0.64-1.28 keV}$ & $F_{1.28-2.56 keV}$ & $F_{2.56-5.12 keV}$ & $F_{5.12-10.24 keV}$ & $U.L._{0.5-8keV}$ && $F_{100-300MeV}$ & $F_{0.3-1GeV}$ & $F_{1-3GeV}$ & $F_{3-10GeV}$ & $F_{10-100GeV}$\\ \hline
J1254.4+2209\_src2	&$40.3\pm13.7$	&$18.4\pm6.6$	&$14.0\pm6.4$	&$25.2\pm13.0$	&-	&-	&	&$<3.74$	&$1.15\pm0.40$	&$1.15\pm0.35$	&$1.52\pm0.49$	&$<1.80$\\
J1254.4+2209\_TXS1252+224	&$23.1\pm12.9$	&$114\pm18$	&$22.0\pm9.2$	&-	&-	&-	&	&$<3.74$	&$1.15\pm0.40$	&$1.15\pm0.35$	&$1.52\pm0.49$	&$<1.80$\\
J1256.9+3650\_1RXSJ125716.0+364713	&-	&-	&-	&-	&-	&$<62.4$	&	&$<1.44$	&$1.40\pm0.39$	&$1.65\pm0.35$	&$0.967\pm0.437$	&$<2.41$\\
J1301.8+0837\_PSRJ1301+08	&-	&$1.65\pm1.41$	&$5.69\pm2.74$	&$8.09\pm4.86$	&-	&$<41.7$	&	&$<1.31$	&$1.62\pm0.45$	&$1.52\pm0.39$	&$1.33\pm0.44$	&$<0.89$\\
J1301.8+0837\_src1	&$2.91\pm1.93$	&$1.40\pm0.83$	&-	&$2.89\pm1.83$	&-	&$<41.7$	&	&$<1.31$	&$1.62\pm0.45$	&$1.52\pm0.39$	&$1.33\pm0.44$	&$<0.89$\\
J1302.3-3255\_PSRJ1302-32	&$2.91\pm1.93$	&$1.40\pm0.83$	&-	&$2.89\pm1.83$	&-	&$<9.42$	&	&$3.48\pm0.98$	&$4.04\pm0.60$	&$4.45\pm0.62$	&$3.06\pm0.69$	&$<1.73$\\
J1302.3-3255\_src1	&$473\pm64$	&$262\pm28$	&$134\pm20$	&$118\pm28$	&-	&$<9.42$	&	&$3.48\pm0.98$	&$4.04\pm0.60$	&$4.45\pm0.62$	&$3.06\pm0.69$	&$<1.73$\\
J1304.3-4352\_1RXSJ130421.2-435308	&$473\pm64$	&$262\pm28$	&$134\pm20$	&$118\pm28$	&-	&-	&	&$<7.81$	&$4.80\pm0.73$	&$6.54\pm0.71$	&$7.55\pm1.03$	&$5.02\pm1.26$\\
J1307.0-4030\_ESO323-77	&$0.756\pm0.267$	&$6.65\pm1.64$	&$69.1\pm7.9$	&$653\pm35$	&$1510\pm134$	&-	&	&$<2.88$	&$<3.06$	&$1.81\pm0.47$	&$<1.61$	&$<0.722$\\
J1307.0-4030\_src2	&$146\pm16$	&$206\pm13$	&$274\pm17$	&$281\pm25$	&$435\pm73$	&-	&	&$<2.878$	&$<3.06$	&$1.81\pm0.47$	&$<1.61$	&$<0.722$\\
J1307.0-4030\_src3	&$7.21\pm5.82$	&$3.26\pm2.33$	&$5.75\pm3.14$	&$12.5\pm6.7$	&-	&-	&	&$<2.878$	&$<3.06$	&$1.81\pm0.47$	&$<1.61$	&$<0.722$\\
J1307.6-4259\_1RXSJ130737.8-425940	&$952\pm99$	&$872\pm58$	&$628\pm51$	&$624\pm76$	&-	&-	&	&$<6.61$	&$<3.01$	&$2.95\pm0.57$	&$4.32\pm0.82$	&$4.54\pm1.23$\\
J1311.7-3429	&-	&-	&-	&-	&-	&-	&	&$12.8\pm1.6$	&$15.9\pm0.9$	&$15.2\pm1.0$	&$8.75\pm1.05$	&$<1.63$\\
J1312.6+0048	&-	&-	&-	&-	&-	&-	&	&$<2.94$	&$2.53\pm0.61$	&$4.79\pm0.65$	&$2.69\pm0.62$	&$<1.92$\\
J1315.6-0729\_src1	&$138\pm13$	&$110\pm8$	&$72.4\pm7.4$	&$37.4\pm8.1$	&$33.1\pm17.2$	&-	&	&$<3.62$	&$<1.18$	&$1.53\pm0.41$	&$0.908\pm0.417$	&$<2.63$\\
J1328.2-4729\_src1	&$24.3\pm9.3$	&$5.60\pm2.56$	&$13.0\pm3.9$	&$5.16\pm3.67$	&-	&-	&	&$<2.13$	&$<2.21$	&$2.32\pm0.73$	&$2.01\pm0.68$	&$4.27\pm1.25$\\
J1340.5-0413\_src1	&$54.0\pm8.2$	&$36.0\pm4.8$	&$24.2\pm4.2$	&$12.3\pm4.6$	&$11.7\pm10.2$	&-	&	&$<1.72$	&$<1.59$	&$0.965\pm0.365$	&$<1.10$	&$<2.67$\\
J1340.5-0413\_src2	&$3.82\pm2.27$	&$2.05\pm1.19$	&$4.04\pm1.78$	&$4.57\pm2.86$	&-	&-	&	&$<1.72$	&$<1.59$	&$0.965\pm0.365$	&$<1.10$	&$<2.67$\\
J1340.5-0413\_src3	&$8.36\pm3.26$	&$3.98\pm1.71$	&$9.35\pm2.76$	&$13.6\pm5.1$	&-	&-	&	&$<1.72$	&$<1.59$	&$0.965\pm0.365$	&$<1.10$	&$<2.67$\\
J1340.5-0413\_src4	&$42.5\pm19.7$	&$41.4\pm12.6$	&$16.7\pm8.7$	&-	&-	&-	&	&$<1.72$	&$<1.59$	&$0.965\pm0.365$	&$<1.10$	&$<2.67$\\
J1406.2-2510\_NVSSJ140609-250808	&$42.5\pm19.7$	&$41.4\pm12.6$	&$16.7\pm8.7$	&-	&-	&-	&	&$<2.72$	&$<1.38$	&$<1.40$	&$1.76\pm0.54$	&$2.02\pm0.89$\\
J1419.7+7731\_1RXSJ141901.8+773229	&$145\pm16$	&$99.6\pm9.7$	&$70.3\pm8.7$	&$55.2\pm11.9$	&-	&-	&	&$<2.98$	&$<0.975$	&$<0.994$	&$<1.26$	&$<1.78$\\
J1511.8-0513\_src1	&$753\pm43.6$	&$649\pm25$	&$559\pm24$	&$401\pm30$	&$289\pm56$	&-	&	&$<13.0$	&$<4.42$	&$2.14\pm0.62$	&$1.63\pm0.57$	&$<2.61$\\
J1511.8-0513\_src2	&$8.83\pm4.89$	&$10.7\pm3.3$	&$7.89\pm3.00$	&$13.0\pm5.8$	&$30.9\pm19.3$	&-	&	&$<13.0$	&$<4.42$	&$2.14\pm0.62$	&$1.63\pm0.57$	&$<2.61$\\
J1521.0-0350\_NVSSJ152048-034850	&$33.8\pm7.2$	&$33.1\pm4.2$	&$18.5\pm3.2$	&$14.4\pm4.3$	&-	&-	&	&$<2.89$	&$<2.28$	&$1.67\pm0.50$	&$2.68\pm0.67$	&$1.67\pm0.80$\\
J1539.0-3328\_UL	&-	&-	&-	&-	&-	&$<11.0$	&	&$<0.967$	&$<1.46$	&$3.75\pm0.67$	&$4.87\pm0.92$	&$<1.48$\\
J1544.5-1127\_src1	&$101\pm24$	&$114\pm11$	&$143\pm11$	&$171\pm17$	&$199\pm44$	&-	&	&$<2.86$	&$3.08\pm0.69$	&$1.89\pm0.51$	&$<1.61$	&$<1.03$\\
J1549.7-0659\_PSRJ1549-06	&-	&-	&-	&-	&-	&$<436$	&	&$<3.52$	&$<2.34$	&$1.75\pm0.50$	&$1.31\pm0.52$	&$<1.81$\\
J1549.7-0659\_src1	&$199\pm26$	&$179\pm14$	&$152\pm13$	&$112\pm16$	&$102\pm35$	&$<436$	&	&$<3.52$	&$<2.34$	&$1.75\pm0.50$	&$1.31\pm0.52$	&$<1.81$\\
J1625.3-0019\_UL	&-	&-	&-	&-	&-	&$<7.17$	&	&$<1.99$	&$1.57\pm0.39$	&$8.95\pm0.79$	&$3.40\pm0.71$	&$<0.970$\\
J1627.6+3218\_src1	&$7.47\pm4.03$	&$18.4\pm4.5$	&$7.91\pm3.29$	&$5.63\pm4.36$	&-	&-	&	&$<1.68$	&$<1.88$	&$1.42\pm0.34$	&$1.47\pm0.48$	&$<1.83$\\
J1627.6+3218\_src2	&$5.86\pm3.71$	&$4.26\pm2.32$	&$6.85\pm3.20$	&$25.9\pm9.3$	&-	&-	&	&$<1.68$	&$<1.88$	&$1.42\pm0.34$	&$1.47\pm0.48$	&$<1.83$\\
J1630.5+3735\_UL	&-	&-	&-	&-	&-	&$<13.9$	&	&$<2.80$	&$<1.54$	&$1.81\pm0.41$	&$2.05\pm0.50$	&$<0.657$\\
J1653.6-0158\_src1	&-	&$3.06\pm1.84$	&$4.70\pm2.19$	&-	&-	&-	&	&$6.16\pm1.54$	&$9.96\pm0.88$	&$7.79\pm0.79$	&$4.99\pm0.89$	&$<1.37$\\
J1721.1+0713\_UL	&-	&-	&-	&-	&-	&$<225$	&	&$<4.79$	&$2.72\pm0.70$	&$2.57\pm0.56$	&$<1.72$	&$<1.60$\\
J1739.4+8717	&-	&-	&-	&-	&-	&-	&	&$<2.85$	&$2.57\pm0.47$	&$2.32\pm0.41$	&$1.88\pm0.45$	&$1.26\pm0.60$\\
J1743.8-7620	&-	&-	&-	&-	&-	&-	&	&$<3.38$	&$3.85\pm0.55$	&$7.63\pm0.75$	&$4.40\pm0.79$	&$<0.862$\\
J1745.5+1018\_PSRJ1745+10	&-	&-	&-	&-	&-	&$<30.1$	&	&$<6.32$	&$3.49\pm0.76$	&$2.32\pm0.54$	&$1.84\pm0.56$	&$<0.749$\\
J1754.0-5002\_PMNJ1753-5015	&-	&-	&-	&-	&-	&$<235$	&	&$6.69\pm1.43$	&$5.57\pm0.78$	&$3.42\pm0.62$	&$<2.21$	&$<0.890$\\
J1754.0-5002\_src1	&-	&$10.6\pm4.2$	&$4.82\pm2.72$	&$10.9\pm5.8$	&$36.2\pm23.4$	&-	&	&$6.69\pm1.43$	&$5.57\pm0.78$	&$3.42\pm0.62$	&$<2.21$	&$<0.890$\\
J1754.0-5002\_src2	&-	&$9.44\pm4.00$	&$6.47\pm3.11$	&$8.15\pm5.10$	&$23.2\pm19.6$	&-	&	&$6.69\pm1.43$	&$5.57\pm0.78$	&$3.42\pm0.62$	&$<2.21$	&$<0.890$\\
J1806.2+0609\_UL	&-	&-	&-	&-	&-	&$<15.7$	&	&$<4.69$	&$2.36\pm0.74$	&$2.85\pm0.60$	&$2.21\pm0.66$	&$<1.33$\\
J1810.3+1741\_LATPSR\_PSRJ1810+17	&-	&$4.48\pm2.42$	&$2.15\pm1.78$	&-	&$35.4\pm22.4$	&-	&	&$7.55\pm1.43$	&$6.97\pm0.73$	&$5.00\pm0.65$	&$1.63\pm0.52$	&$<0.675$\\
J1816.7+4509	&-	&-	&-	&-	&-	&$<19.3$	&	&$<3.48$	&$2.28\pm0.46$	&$3.14\pm0.48$	&$1.66\pm0.52$	&$<2.29$\\
J1824.6+1013	&-	&-	&-	&-	&-	&$<9.31$	&	&$<6.78$	&$<3.36$	&$<2.24$	&$1.97\pm0.60$	&$<1.56$\\
J1841.9+3220\_RXJ1841.7+3218\_src2	&$2.85\pm2.73$	&$3.37\pm1.64$	&$6.11\pm2.27$	&$5.04\pm3.09$	&-	&-	&	&$<7.68$	&$2.92\pm0.67$	&$1.76\pm0.48$	&$1.90\pm0.56$	&$2.60\pm0.89$\\ \hline
\end{tabular}
}
\end{center}
\end{table}
\end{landscape}
}

{
\setlength{\textheight}{40\baselineskip}
\begin{landscape}
\begin{table}
\label{X-ray}
\begin{center}
\scalebox{0.65}{
\begin{tabular}{lcccccccccccc}
\hline \hline 
 &  \multicolumn{6}{c}{X-ray (Swift/XRT) [E-14 erg/cm$^2$/s]}  &&  \multicolumn{5}{c}{Gamma-ray (2FGL Catalog) [E-12 erg/cm$^2$/s]}\\ \cline{2-7} \cline{9-13}
\multicolumn{1}{c}{Name} & $F_{0.32-.64 keV}$ & $F_{0.64-1.28 keV}$ & $F_{1.28-2.56 keV}$ & $F_{2.56-5.12 keV}$ & $F_{5.12-10.24 keV}$ & $U.L._{0.5-8keV}$ && $F_{100-300MeV}$ & $F_{0.3-1GeV}$ & $F_{1-3GeV}$ & $F_{3-10GeV}$ & $F_{10-100GeV}$\\ \hline
J1841.9+3220\_src1	&$113\pm15$	&$98.2\pm8.7$	&$76.1\pm7.8$	&$36.9\pm8.2$	&$35.1\pm17.9$	&-	&	&$<7.68$	&$2.92\pm0.67$	&$1.76\pm0.48$	&$1.90\pm0.56$	&$2.60\pm0.89$\\ \hline
J1842.3-5845\_src1	&$887\pm75$	&$701\pm35$	&$548\pm30$	&$459\pm40$	&$303\pm71$	&-	&	&$<3.35$	&$<1.34$	&$<1.53$	&$1.61\pm0.51$	&$<3.02$\\
J1858.1-2218	&-	&-	&-	&-	&-	&$<26.1$	&	&$<4.08$	&$<1.81$	&$3.61\pm0.66$	&$<2.29$	&$<1.58$\\
J1902.0-5110	&-	&-	&-	&-	&-	&-	&	&$4.25\pm0.98$	&$5.74\pm0.62$	&$5.93\pm0.67$	&$2.95\pm0.64$	&$<1.43$\\
J1916.9-3028	&-	&-	&-	&-	&-	&$<45.4$	&	&$<3.86$	&$<2.66$	&$1.54\pm0.52$	&$1.83\pm0.58$	&$<2.18$\\
J1926.8+6153\_1RXSJ192649.5+615445	&$278\pm55$	&$206\pm30$	&$130\pm24$	&$53.3\pm23.8$	&-	&-	&	&$<2.58$	&$2.70\pm0.66$	&$3.41\pm0.50$	&$3.32\pm0.62$	&$5.20\pm1.16$\\
J1956.2-0238\_src1	&$401\pm78$	&$255\pm21$	&$208\pm15$	&$140\pm17$	&$104\pm32$	&$<9.56$	&	&$<6.80$	&$4.38\pm0.79$	&$<1.52$	&$<1.64$	&$<1.35$\\
J1959.7-4730\_SUMSSJ195945-472519	&$128\pm18$	&$77.9\pm9.6$	&$47.3\pm7.8$	&$36.4\pm10.4$	&-	&-	&	&$<1.47$	&$<1.66$	&$2.68\pm0.49$	&$3.21\pm0.68$	&$2.34\pm1.04$\\
J2004.8+7004	&$401\pm78$	&$255\pm21$	&$208\pm15$	&$140\pm17$	&$104\pm32$	&-	&	&$<9.48$	&$<3.26$	&$1.90\pm0.51$	&$1.74\pm0.50$	&$2.30\pm0.77$\\
J2014.4+0647\_NVSSJ201431+064849	&$1151\pm182$	&$434\pm36$	&$143\pm16$	&$52.1\pm13.7$	&$65.9\pm31.3$	&-	&	&$<2.87$	&$<2.76$	&$<1.70$	&$1.19\pm0.53$	&$<3.14$\\
J2034.6-4202	&$111\pm14$	&$54.7\pm7.0$	&$36.4\pm6.0$	&$15.4\pm5.9$	&$26.1\pm16.7$	&-	&	&$<3.29$	&$<2.06$	&$<0.885$	&$0.890\pm0.397$	&$<2.01$\\
J2039.4-5621	&-	&-	&-	&-	&-	&$<15.6$	&	&$4.86\pm1.02$	&$4.83\pm0.58$	&$4.68\pm0.59$	&$2.79\pm0.64$	&$<1.09$\\
J2043.2+1709	&-	&-	&-	&-	&-	&-	&	&$4.14\pm0.99$	&$5.52\pm0.62$	&$9.14\pm0.76$	&$5.70\pm0.87$	&$<2.49$\\
J2112.5-3044	&-	&-	&-	&-	&-	&$<12.1$	&	&$<1.95$	&$2.72\pm0.49$	&$6.36\pm0.69$	&$5.21\pm0.86$	&$<1.27$\\
J2129.8-0427\_PSRJ2129-04\_src1	&$2.78\pm1.62$	&$2.06\pm0.94$	&$6.56\pm1.76$	&$4.68\pm2.30$	&-	&-	&	&$<3.25$	&$2.01\pm0.49$	&$1.85\pm0.42$	&$1.10\pm0.42$	&$<1.63$\\
J2129.8-0427\_PSRJ2129-04\_src3	&$17.7\pm4.46$	&$12.0\pm2.5$	&$11.8\pm2.7$	&$16.0\pm4.8$	&$16.3\pm11.4$	&-	&	&$<3.25$	&$2.01\pm0.49$	&$1.85\pm0.42$	&$1.10\pm0.42$	&$<1.63$\\
J2134.5-2130\_src1	&$6.47\pm3.21$	&$3.25\pm1.55$	&$1.44\pm1.17$	&$5.18\pm3.25$	&-	&-	&	&$<1.70$	&$2.07\pm0.49$	&$1.15\pm0.36$	&$<1.26$	&$2.23\pm0.87$\\
J2223.3+0103\_NVSSJ222329+010226	&$11.6\pm5.3$	&$5.73\pm2.34$	&-	&-	&-	&-	&	&$<1.20$	&$<1.20$	&$<1.18$	&$1.21\pm0.44$	&$1.49\pm0.73$\\
J2228.5-1633\_UL	&-	&-	&-	&-	&-	&$<64.3$	&	&$<3.82$	&$<1.90$	&$1.05\pm0.33$	&$<1.55$	&$<2.07$\\
J2243.4+4104\_TXS2241+406	&$15.3\pm7.9$	&$17.8\pm4.4$	&$21.3\pm4.8$	&$31.2\pm8.5$	&$42.4\pm22.3$	&-	&	&$5.79\pm1.52$	&$7.51\pm0.72$	&$4.70\pm0.60$	&$3.14\pm0.66$	&$<1.94$\\
J2251.2-4928\_src1	&$7.56\pm3.61$	&$2.40\pm1.59$	&-	&-	&-	&-	&	&$<0.894$	&$<1.52$	&$1.15\pm0.35$	&$1.41\pm0.48$	&$<1.80$\\
J2251.2-4928\_src2	&$6.02\pm3.29$	&$6.42\pm2.47$	&$2.08\pm1.66$	&-	&-	&-	&	&$<0.894$	&$<1.52$	&$1.15\pm0.35$	&$1.41\pm0.48$	&$<1.80$\\
J2256.9-1024	&$6.39\pm3.80$	&$4.01\pm2.11$	&-	&-	&-	&-	&	&$<2.42$	&$1.84\pm0.46$	&$2.61\pm0.44$	&$<1.03$	&$<1.36$\\
J2257.9-3643\_src1	&$12.0\pm4.6$	&$4.81\pm2.22$	&$3.38\pm2.16$	&$5.23\pm4.05$	&-	&-	&	&$<2.09$	&$<0.914$	&$1.21\pm0.34$	&$<0.56$	&$1.43\pm0.70$\\
J2257.9-3643\_src2	&$3.50\pm2.75$	&$3.27\pm1.90$	&$2.52\pm1.95$	&$12.2\pm6.3$	&-	&-	&	&$<2.09$	&$<0.914$	&$1.21\pm0.34$	&$<0.56$	&$1.43\pm0.70$\\
J2259.9-8255	&-	&-	&-	&-	&-	&$<70.8$	&	&$<4.89$	&$<2.58$	&$<2.18$	&$1.46\pm0.51$	&$<0.725$\\
J2310.0-3627\_1RXSJ230940.6-363241	&$118\pm11$	&$93.3\pm7.7$	&$55.2\pm6.4$	&$28.4\pm7.0$	&-	&-	&	&$<2.34$	&$<1.50$	&$<1.11$	&$1.44\pm0.50$	&$<3.46$\\
J2310.0-3627\_src2	&$7.39\pm2.95$	&$3.40\pm1.52$	&$7.46\pm2.45$	&-	&-	&-	&	&$<2.34$	&$<1.50$	&$<1.11$	&$1.44\pm0.50$	&$<3.46$\\
J2323.0-4919\_1RXSJ232256.7-491658	&$109\pm24$	&$43.3\pm11.3$	&$29.1\pm9.8$	&$15.4\pm10.9$	&-	&-	&	&$<3.35$	&$<1.26$	&$<0.979$	&$1.08\pm0.43$	&$<1.35$\\
J2330.3-4745\_PKS2326-477	&$1.16\pm1.11$	&$1.04\pm0.74$	&$1.90\pm1.06$	&$3.17\pm2.10$	&-	&-	&	&$<4.58$	&$<2.28$	&$1.17\pm0.35$	&$<0.910$	&$<0.87$\\
J2330.3-4745\_src1	&$30.0\pm5.90$	&$35.3\pm4.7$	&$35.8\pm5.3$	&$41.9\pm8.9$	&$40.1\pm20.4$	&-	&	&$<4.58$	&$<2.28$	&$1.17\pm0.35$	&$<0.910$	&$<0.87$\\
J2339.7-0531\_src2	&$5.22\pm2.27$	&$3.92\pm1.38$	&$4.18\pm1.55$	&$6.96\pm3.05$	&$30.6\pm14.4$	&-	&	&$3.42\pm0.69$	&$4.21\pm0.48$	&$10.3\pm0.8$	&$6.02\pm0.89$	&$1.63\pm0.73$\\
J2347.3+0710\_UL	&-	&-	&-	&-	&-	&$<11.2$	&	&$<1.58$	&$1.87\pm0.53$	&$<1.25$	&$1.37\pm0.47$	&$1.38\pm0.67$\\
J2350.1-3005	&$147\pm33$	&$78.9\pm18.0$	&$55.0\pm16.1$	&$41.4\pm21.5$	&-	&-	&	&$<6.67$	&$<1.98$	&$1.06\pm0.34$	&$<1.29$	&$<2.27$\\
J2352.1+1752\_87GB234934.7+173227	&-	&-	&-	&-	&-	&$<16.8$	&	&$<3.19$	&$<1.64$	&$1.52\pm0.36$	&$0.830\pm0.363$	&$<3.06$\\ \hline
\end{tabular}
}
\end{center}
\end{table}
\end{landscape}
}

\end{appendix}

\end{document}